\documentclass[11pt]{ucthesis}
\usepackage{graphicx}
\usepackage{dcolumn}
\usepackage{bm, amsmath, amssymb}
\usepackage{subfigure}
\usepackage{caption}	
\usepackage[displaymath]{lineno}

\usepackage[final,		
	colorlinks,			
	linktocpage,		
	linkcolor=blue,		
	citecolor=blue,		
	urlcolor=blue,		
	breaklinks=true,	
	]{hyperref}
\usepackage[all]{hypcap} 

\newcommand{\comment}[1]{}

\begin{document}


\title{Acoustic detection of astrophysical neutrinos in South Pole ice}
\author{Justin Arthur Vandenbroucke}
\degreeyear{2009}
\degreesemester{Fall}
\degree{Doctor of Philosophy}
\chair{Professor P. Buford Price}
\othermembers{Professor Don Backer\\
Professor Bernard Sadoulet}
\numberofmembers{3}
\prevdegrees{B. S. (Stanford University) 2002\\
M. A. (University of California, Berkeley) 2006}
\field{Physics}
\campus{Berkeley}

\maketitle
\approvalpage
\copyrightpage

\begin{abstract}


When high-energy particles interact in dense media to produce a particle shower, most of the shower energy is deposited in the medium as heat.  This causes the medium to expand locally and emit a shock wave with a medium-dependent peak frequency on the order of 10~kHz.  In South Pole ice in particular, the elastic properties of the medium have been theorized to provide good coupling of particle energy to acoustic energy.  The acoustic attenuation length has been theorized to be several km, which could enable a sparsely instrumented large-volume detector to search for rare signals from high-energy astrophysical neutrinos.  We simulated a hybrid optical/radio/acoustic extension to the IceCube array, specifically intended to detect cosmogenic (GZK) neutrinos with multiple methods simultaneously in order to achieve high confidence in a discovered signal and to measure angular, temporal, and spectral distributions of GZK neutrinos.  Detecting $\sim$100 GZK events could help resolve the question of ultra-high cosmic ray acceleration and allow us to measure the total neutrino-nucleon cross section at $\sim$100 TeV center-of-mass energy.  Our simulation showed that such a hybrid array could detect 10-20 GZK neutrinos per year, half of which would be detected by both the radio and acoustic methods.

This work motivated the design, deployment, and operation of the South Pole Acoustic Test Setup (SPATS).  The main purpose of SPATS is to measure the acoustic attenuation length, sound speed profile, noise floor, and transient noise sources \emph{in situ} at the South Pole.  We describe the design, performance, and results from SPATS.  We measured the sound speed in the fully dense ice between 200~m and 500~m depth to be 3878~$\pm$~12~m/s for pressure waves and 1975.8~$\pm$~8.0~m/s for shear waves.  We measured the acoustic amplitude attenuation length to be 316~$\pm$~105~m.  We measured the background noise floor to be Gaussian and very stable on all time scales from one second to two years.  Finally, we have detected an interesting set of well-reconstructed transient events in over one year of high quality transient data acquisition.  We conclude with a discussion of what is next for SPATS and of the prospects for acoustic neutrino detection in ice.

\abstractsignature

\end{abstract}

\begin{frontmatter}

\begin{dedication}
\null\vfil
{\large
\begin{center}
To Ardelle, Arthur, George, and Mary Louise.
\end{center}}
\vfil\null
\end{dedication}

\tableofcontents
\listoffigures
\listoftables

\end{frontmatter}

\begin{acknowledgements}

Working with the IceCube Collaboration has given me some of the best experiences of my life.  You've taught me how to be a physicist, and you've given me great friendships all around the world.  I'm especially grateful to the other grad students I worked with closely on SPATS: Sebastian B\"{oser}, Freija Descamps, and Delia Tosi.  SPATS has been a successful experiment, and more importantly a fun one, because of you.  I'll always remember the intense periods we spent together (on four different continents!), designing, building, testing, deploying, commissioning, and analyzing.  Thanks for the bottomless coffees and long nights in the ${\AA}$ngstr\"om and DESY labs; the shipwreck rescue in Uppsala; cooking in the BVG lab; climbing, fondue, and dolls in Berlin; the intercontinental DAQ development sessions with everything compiling in time for Abisko; the long arguments over clock lines and PCB design; convincing me of the wonders of CVS; teaching me how to be a hardware guy; swimming in the Zeuthen see; the shirts and ties in Lake Geneva; the canoeing in Utrecht; the woks and Qwaks in Ghent;  the blues club in Chicago; the melon in Madison; my first ski hut trip, in -60~$^{\circ}$C wind chill; leaving a dinner party on multiple occasions to log on to SPATS during a satellite pass; the Turkish restaurant in San Francisco in the rain; the frantic intercontinental debugging sessions when something was wrong with our baby; the South Pole tunnels and hero photos on the white sand ``beach''; and memories at the SPATS bar and especially at Jupiter.

DESY Zeuthen has been a second home throughout my time in graduate school, starting with a fun and productive visit in June 2004.  Thank you to Rolf Nahnhauer, Sebastian B\"{o}ser, Delia Tosi, Kalle Sulanke, Christian Spiering, Bernhard Voigt, Elisa Bernardini, Oxana Tarasova, and the whole IceCube group at DESY for being gracious hosts on several very productive visits.  Thanks also to the three restaurants in Zeuthen, where we spent many hours first thinking about neutrino detector simulations and doing calculations on napkins, then struggling with embedded Linux driver hacking for the SPATS DAQ, and finally discussing results from SPATS (after successful installation in the ice!) over German beers, Italian digestives, and Greek ouzos.

Thanks to Allan Hallgren, who hosted Sebastian and me for the first ``SPATS working weeks'' to integrate the whole system in the  ${\AA}$ngstr\"om Laboratory in Uppsala, and who worked harder and more joyfully than anyone else I've ever seen, during some tough and tiring seasons at the Pole.  We could not have built SPATS without Spider Man the sailor.

Thanks to the Berkeley guys: Andres Morey, David Hardtke, Ignacio Taboada, Kurt Woschnagg, Kirill Filomonov, Ryan Bay, Bobby Rohde, and Michelangelo D'Agostino.  I never thought a physics group could be so much fun.  Thanks for New Orleans, for nights at the kiddie table, for tie fridays and tie dinners, for coffee and lunch and coffee, for exploding stuff at the South Pole ten years ago, for the author list script, for the hooding and the champagne cooler, and for the wall of love.  As Kirill said, ``you can never have too much love.''  Michelangelo, if you ever need someone to hold your hand on a flight to France, I'm there.  Go easy on the tires when you're parking.

Thanks to my two physics ``fathers,'' Giorgio Gratta and Buford Price, for supporting me while trusting me and giving me space to be independent and take on large responsibilities, to make my own choices, and to grow into a mature scientist.

Thanks to my readers, Buford Price, Don Backer, and Bernard Sadoulet, and to my collaborators, Rolf Nahnhauer, Delia Tosi, Naoko Kurahashi, and Christopher Wiebusch for reading this dissertation closely and giving many valuable comments.

Finally, thank you to my family: Aynsley, Mary, Matt, and Russell; and my friends: Alex S., Ann D., Becky B., Brenna H., Daisy P. L., Dan K., Dave L., EJ B., Jeff M., Jen A., Jude S., Kate L., Kater M., Kimberly H., Lauren Q., Leah B., Mandeep G., Mary D., Maryam K., Mike R., Mike S., Nathan M., Pepe P., Phil B., Rahel W., Ray M., Ryan G. R., Sarah R., Scott M., Sumi N., Todd R., Val Z., Veronica Y., and Yossi F.  I've learned more from you than a PhD could ever give me.

\end{acknowledgements}
\section*{Foreword}

Like many projects in particle physics and, increasingly, in astronomy, the work I report here is the result of fruitful collaboration, both within the UC Berkeley Price group and with my collaborators in IceCube and in particular in SPATS.  For the benefit of my thesis committee I summarize my own particular role in the work, as well as other projects I worked on in graduate school.  I also give a brief account of the development of the SPATS project from my perspective, focusing especially on the early development of SPATS.

In the first year of graduate school, I finished my work on the Study of Acoustic Ultra-high energy Neutrino Detection (SAUND) project.  I collaborated on this work with Giorgio Gratta and Nikolai Lehtinen.  We developed a data acquisition system and installed it to read out seven underwater hydrophones at a naval array in the Bahamas.  The system was optimized to search for acoustic signals from ultra-high-energy neutrinos interacting in the water.  My personal responsibility in the project was to implement and install the data acquisition system, then to operate the experiment and analyze the data.  This analysis was published in the Astrophysical Journal~\cite{Vandenbroucke05}.  Although our neutrino flux limit was not competitive with existing radio limits, it was the first search for astrophysical particles with the acoustic method.  It has been the only such limit for several years, although the ACORNE group~\cite{Bevan07,Bevan09limit} has now also determined a similar limit.  In the Fall of 2003, Giorgio and I hosted the first acoustic neutrino detection workshop, which together with the RADHEP workshop led to a series of successful ARENA workshops held so far in 2005, 2006, and 2008, focusing on radio and acoustic detection of high energy particles.

I've continued contributing to the SAUND project as a side project in graduate school.  Naoko Kurahashi and Giorgio Gratta have been doing most of the work, which constitutes the SAUND-II project and is an order of magnitude more sensitive than SAUND-I.  In this thesis I will not describe the SAUND work but instead refer you to~\cite{Vandenbroucke05} and~\cite{Kurahashi07}.

From water I turned to acoustic neutrino detection in ice and salt.  This was based on the theoretical work Buford Price had done to estimate how well acoustic waves could propagate through both ice and salt.  Starting with a productive visit to DESY Zeuthen hosted by Rolf Nahnhauer in June 2004, I expanded the work I had done for SAUND to develop a software package for simulating acoustic neutrino detector arrays in water, ice, and salt.  The salt work was spurred by a SalSA workshop held at SLAC in February 2005.  In the spring of 2005, Buford, Sebastian B\"{o}ser, Rolf, Dave Besson and I completed the simulation work that would motivate SPATS and the idea of a hybrid radio/acoustic/neutrino extension to IceCube~\cite{Besson05}.  For the simulation project, I generated a common neutrino event set that we then fed to the three detector sub-arrays.  Sebastian ran the optical detector response simulation, Dave ran the radio simulation, and I ran the acoustic simulation.  I then combined the output of the three sub-detector simulations to determine the neutrino sensitivity (effective volume vs. energy) and expected GZK event rate.  These simulations indicated that we could detect more than 10 GZK events per year, with about half of them detected by both the radio and the acoustic method, assuming the theoretical acoustic attenuation model presented in~\cite{Price96},~\cite{Price93}, and~\cite{Price06}.

At the same time, we started designing an experimental setup to be deployed in IceCube holes at the South Pole, in order to measure the acoustic properties of the ice \emph{in situ} and determine the feasibility of acoustic neutrino detection there.  This next step was necessary to test the parameters we were assuming in our neutrino sensitivity simulations, because many of them were theoretical estimates without experimental ground truth.  At the spring IceCube collaboration meeting in Berkeley in 2005, we (Buford Price and myself from Berkeley, Rolf Nahnhauer and Sebastian B\"{o}ser from DESY, and Allan Hallgren from Uppsala) started sketching the design of SPATS, including the number of strings, what instrumentation should be on each string, and how we would control it.  Soon afterward Stephan Hundertmark, Per Olof Hulth, and Christian Bohm from Stockholm University joined.  The Berkeley group took responsibility for the data acquisition system including both the hardware and software.  The DESY group was responsible for the in-ice instrumentation, Uppsala for in-ice cables, and Stockholm for other surface equipment.

That summer (2005) we had regular phone calls to design what would be SPATS.  I was responsible for designing the DAQ system, which meant that I spent much of the summer researching various options for rugged computing to interface with the instrumentation in the ice, then ordering and testing the hardware.  The first String PC arrived at the end of Summer 2005, plunging me into months of hacking Linux embedded computing.  We could not get the solid state device that came with the systems to be readable, and furthermore we had problems getting a stable operating system to run at all.  Sebastian and I spent two caffeine-fueled months with Allan Hallgren in Uppsala University in the Fall of 2005, integrating the different SPATS components we had built at separate institutions, designing and building the Acoustic Junction Boxes, and testing everything.

In the winter of 2005-2006, Freija Descamps from Ghent University joined SPATS and took on a large responsibility in testing and debugging the hardware of the first three strings, for installation in the 2006-2007 season.  Delia Tosi from DESY Zeuthen joined in the summer of 2006 and also performed essential work on the final construction and testing of the first three strings.  In the spring of 2006, Delia and Freija led the development of the retrievable pinger and String D hardware design and construction projects.

Yasser Abdou from Ghent joined soon after Freija and has focused on simulation studies.  The University of Wuppertal (Klaus Helbing, Timo Karg, and Benjamin Semburg) and RWTH Aachen groups also joined in 2006 and contributed toward String D development.  The Wuppertal group has been responsible in particular for the HADES sensors.  The Aachen group (Christopher Wiebusch, Christian Vogt, Karim Laihem, Matthias Schunck, and Martin Bissok) developed the Aachen Acoustic Laboratory, featuring an IceTop tank in a freezer room, used to make large bubble-free ice blocks for sensor calibration and quantitative studies of the thermoacoustic effect in ice.

For the design, deployment, and commissioning of SPATS, my main responsibility was the data acquisition system.  In addition to selecting and designing the String PC systems, I wrote most of the data acquisition software for SPATS.  I started with a primitive set of software in the Fall of 2005, mostly featuring a single program that could read out a single sensor channel with the option of simultaneously pulsing a transmitter either once or repeatedly.  I continued expanding and improving the DAQ software during and after deployment and commissioning.  The software progressed from primitive to fully featured during the first year of operation of SPATS (2007), with several milestones including GPS time stamping, HV read-back readout, and long-duration threshold triggered acquisition to complement the forced-mode readout supported from the beginning.  

In addition to the data acquisition software, I wrote and maintained an offline analysis framework that several of us have used for SPATS analysis.

I worked at the South Pole during three consecutive austral summer seasons (2005-2006, 2006-2007, and 2007-2008), for several purposes: IceCube string installation, IceCube Standard Candle installation, IceTop installation, SPATS string installation, and SPATS pinger operation.  Together with IceCube collaborators, I installed all four SPATS strings and participated in six of the ten retrievable pinger deployments.  Those of us who had worked long hard hours in basement labs, slaving over tedious but interesting challenges to make SPATS a reality, were beaming during the successful deployment of each string and pinger.  One friend who saw a photo of us after the first SPATS string deployment said I looked like a new father, and indeed we were all happy fathers and mothers.  This excitement continued as we commissioned the array at the Pole and began the first data taking campaigns, as well as long after we left our new children behind in the ice and got to know them from the North.

I've contributed several pieces of analysis both at the science-results level and at the lower level of technical understanding of our detector.  Especially in the beginning, SPATS has benefitted from being a small enough project that analysis has been a collaborative process, often involving breakthroughs arising from discussions among multiple people.  Much of my own analysis effort was focused on using inter-string data to measure the attenuation length.  This proceeded through a series of challenges, many of which were overcome through improvements in data taking and analysis.  One important discovery in the process of this analysis was the presence of ADC clock drift in waveform averaging, which is now recognized to be an important feature in most of our analyses.  I subsequently developed several algorithms for correcting this clock drift, which have been used for many of our analyses.  Another piece of low-level analysis was my work to explain that the strange spikes in our otherwise Gaussian noise amplitude histograms are due to binning effects in our ADC's.

In analyzing 2007-2008 pinger data, I realized that the mysterious ``after-pulses'' we originally thought were reflections were actually shear waves.  This was the first realization that we detect shear waves (in addition to pressure waves) with SPATS, and we have subsequently identified them from nearly all of our detected sources: in-ice transmitters, the retrievable pinger operated in water, and ambient transients.  The discovery of shear wave signals led me to focus on measuring both the pressure and shear wave speed vs. depth, now our first SPATS measurement submitted for publication in a peer-reviewed journal.  Finally, in the last year several of us have been analyzing data from the transients data stream, our newest mode of SPATS operation and a rich data set.  I completed the first analysis of this data with automated event reconstruction, discovering that we saw a hot spot in our data which we subsequently identified as an IceCube ``Rodriguez'' (Rod) well.  These are deep wells used as water reservoirs for the IceCube hot water drill system.  Since then we have identified transient signals from most of the Rod wells drilled over the years for IceCube construction, as well as from IceCube holes themselves.  We have even detected acoustic emission from an AMANDA Rod well last heated nine years ago.

In addition to my focus on acoustic research and development for IceCube, I helped build a laser calibration device for IceCube (the ``Standard Candle'').  At Berkeley we built two of these devices, each of which features a pulsed Nitrogen laser pointed at the apex of a reflecting cone to generate a simulated Cherenkov cone in the ice.  The two devices were installed deep in the ice on two IceCube strings and have been used to validate and calibrate the response of IceCube, both for position reconstruction and (especially) for the absolute intensity calibration of IceCube, necessary to determine the energy of neutrino-induced cascades.  I was responsible for the optics in each of the two Standard Candles, which included mirrors, filters, fibers, and lenses to deliver the beam from the laser to the cone.  This included two weeks of long nights at the South Pole in the 2007-2008 season to solve alignment problems that arose during shipping, capped by deploying the second device in the middle of the night and running it a few days later to see that it had survived deployment to hundreds of atmospheres of pressure and was working well.
\chapter{Introduction}

\label{introductionChapter}

\noindent\emph{In this chapter we introduce the physics of extremely high energy neutrinos.  We then explain the mechanism of acoustic signal production by high energy particle showers and describe how well it is currently understood in theory, in simulation, and in laboratory studies.  We conclude with an overview of existing field studies exploring the use of acoustic techniques for high energy neutrino detection.}

\section{Astrophysical neutrinos at extremely high energy (EHE)}


The field of neutrino astronomy is progressing quickly.  While no astrophysical sources of neutrinos have yet been detected other than the sun and Supernova 1987a, the Baikal, AMANDA, ANTARES, and IceCube experiments have set increasingly stringent limits on neutrino fluxes over several orders of magnitude in energy, ruling out multiple theoretical models in the process.  Each of these experiments uses the optical Cherenkov technique to detect neutrinos interacting in a transparent medium (water or ice) to produce charged particles.  IceCube, now three-quarters constructed, is already by far the most sensitive neutrino ``telescope.''  It has a $\sim$1~km$^3$ instrumented volume and will operate for a decade after construction is complete in early 2011.

While these general-purpose neutrino telescopes have sensitivity between 10$^{11}$ and 10$^{18}$~eV and are particularly optimized for discovery in the TeV (10$^{12}$~eV) to PeV (10$^{15}$~eV) region, a second type of astrophysical neutrino detectors has also proliferated in the past decade, optimized for EeV (10$^{18}$~eV) energies.  This energy range is considered ``extremely high energy'' (EHE).  While the neutrino-nucleon interaction rate is higher in this energy range, the expected flux is lower, so very large instrumented volumes are necessary.  Experiments have therefore depended on creative instrumentation techniques to monitor large naturally occurring volumes of matter for neutrino interactions.

Experiments in the EeV range have relied mostly on the radio Cherenkov techniques, with additional contributions from the optical Cherenkov technique and the extensive air shower technique.  Although it is not yet competitive with these techniques, the acoustic technique has also been used to set neutrino flux limits, and has has been developed steadily over the past few years.

\subsection{Theories of EHE neutrino production}

For the past decade there have ben two outstanding questions concerning ultra-high-energy cosmic rays (UHECR): (1) Is there a GZK cutoff?  (2) What is the source of the cosmic rays?  The first question has now been resolved by the HiRes and Auger experiments.  The second is still one of the most important questions in astro-particle physics.  Both questions are intimately connected to extremely high energy (EHE) neutrinos.

The most important theoretical source of astrophysical neutrinos expected in the EeV (EHE) range is the ``cosmogenic'' or Greisen-Zatsepin-Kuzmin (GZK) neutrinos.  These neutrinos are produced when ultra-high-energy cosmic rays (UHECR, with energies in excess of 10$^{19}$~eV) interact with the cosmic microwave background at the delta resonance, producing pions and nucleons which decay to neutrinos.  The interaction length for this process is $\sim$50~Mpc at 10$^{19}$~eV.

In addition to producing neutrinos, this interaction is predicted to suppress the ultra-high-energy cosmic ray flux above $\sim$10$^{19.5}$~eV, an effect known as the GZK cutoff.  The UHECR spectrum is otherwise described well by a broken power law over many orders of magnitude.  The GZK mechanism is a solid prediction of basic particle physics and astrophysics and has therefore been an important focus of astro-particle physics.  Absence of a cutoff would indicate something new either in the physics or in the astrophysics of the process, and would yield insight into the source of these ultra-high-energy cosmic rays.  For several years there were indications from the AGASA experiment that the cutoff did not exist~\cite{AGASA98}, which sparked intense activity in the field and a proliferation of exotic models.

Theories of ultra-high energy neutrinos and charged cosmic rays fall for the most part into two categories.  ``Bottom-up'' models invoke acceleration of hadrons by astrophysical engines such as gamma-ray bursts (GRB's) and active galactic nuclei (AGN).  The accelerated hadrons provide detectable cosmic ray, neutrino, and photon fluxes.  ``Top-down'' models instead invoke the decay of heavy exotic particles, again producing detectable charged cosmic rays, neutrinos, and photons.  The energy in these models comes from the rest mass of the exotic particles rather than from astrophysical accelerators.

Improvements both in understanding detector systematics and in building larger, redundant detectors have improved our understanding of ultra-high energy cosmic rays in the past several years.  Both the HiRes experiment~\cite{HiRes08} and, with larger sensitivity, the Auger experiment~\cite{Auger08} have refuted the AGASA result and measured with high confidence a steepening of the UHECR spectrum around 4 x 10$^{19}$~eV.  The simplest explanation of this steepening is the GZK mechanism.

The confirmation of the GZK cutoff increases our confidence that the GZK mechanism is a ``guaranteed'' source of EHE neutrinos.  Other exotic theories, currently disfavored both by the confirmation of the GZK cutoff and by increasingly stringent experimental constraints on the neutrino flux, include topological defect models~\cite{Sigl98} and Z-burst models~\cite{Weiler82}.

Several groups have calculated the predicted GZK neutrino spectrum.  A baseline model, used to estimate the expected neutrino rate of many EHE neutrino experiments, was calculated by Engel, Seckel, and Stanev (ESS) in 2001~\cite{Engel01}.  The composition of UHECR is difficult to measure and is poorly known.  Most GZK neutrino calculations have assumed the composition is predominantly protons.  If the composition is heavy (iron-like) or mixed, the GZK neutrino flux is expected to be lower than that in the pure proton case, for a fixed UHECR spectrum~\cite{Hooper05}.  This is because if the energy of a primary UHECR is shared by multiple nucleons, the nuclei are first photodisintegrated by background photons.  The resulting individual nuclei can then undergo the GZK process, but they have reduced energy relative to the original nucleus.  The number of protons available at the delta resonance is lower than in the case of pure proton primaries.

\subsection{Experimental EHE neutrino results to date}

Three dense media have been instrumented (or considered for future instrumentation) for astrophysical neutrino searches: water, ice, and salt.  These particular media have been considered because they occur naturally in very large volumes, as is necessary to achieve the instrumented volumes on the order of 1-100~km$^3$ necessary for extremely high energy neutrino detection.

The Baikal, ANTARES, NESTOR, and NEMO projects have used lake and sea water, searching for optical Cherenkov signals.  The SAUND experiment has searched for acoustic signals in sea water, as have the ACORNE and AMADEUS projects.

The AMANDA and IceCube projects have searched for optical Cherenkov signals with instrumented ice.  The RICE, FORTE, and ANITA projects have searched for radio Cherenkov signals originating in the polar ice caps (Greenland and Antarctica).  SPATS is now searching for acoustic signals in South Pole ice.

Finally, salt occurs naturally in underground domes (\emph{diapirs}, or geological intrusions) that can be several km in width and height.  They have been considered for radio Cherenkov detection of neutrino interactions by the SalSA project, and measurements of site properties (particularly attenuation length) have been made at several domes.  These salt domes could also be used for acoustic detection of neutrinos, although they typically have layered heterogeneities that could significantly scatter the acoustic signals.

Most of these projects instrument a large volume with sparse array and then search for signals ``contained'' inside the array.  Exceptions to this include ANITA, which uses a high-altitude balloon to search for radio signals from the Antarctic ice sheet, and FORTE, which used a radio satellite to search for signals from the Greenland ice sheet.  Continuing to even further remote sensing of neutrinos, the GLUE project searched for radio Cherenkov signals produced by neutrinos interacting in the moon.

In addition to water, ice, salt, and moon rock, the atmosphere can also be used as a target medium for neutrino detection.  Neutrino-induced showers are distinguished from hadron (cosmic ray) showers by searching for highly inclined, slightly downgoing (nearly horizontal) air showers.  These showers must have traversed a large column of density of air before interacting and therefore must be weakly interacting (neutrinos or something exotic).  In addition to slightly down-going events, slightly-up going neutrino events can also be detected with air shower experiments.  This type of search requires that a tau neutrino skims the Earth (nearly horizontally) and interacts inside it via the charged-current interaction to produce a tau lepton, which then escapes the Earth and then decays to produce an air shower.  In addition to slightly up-going Earth-skimming searches, searches can be performed for the same phenomenon occurring via conversion inside a nearby mountain rather than the Earth.

The Pierre Auger Observatory has performed both down-going and up-going searches.~\cite{Auger09tau}.  While the down-going search is sensitive to both charged-current and neutral-current interactions and to all neutrino flavors, the up-going search is sensitive only to tau neutrino charged-current (CC) interactions.  This is because electron neutrino CC interactions produce an electron that showers before escaping the Earth, and muon neutrino CC interactions produce a single muon that can escape the Earth but then traverses the atmosphere without producing a significant signal detectable by an air shower array.  Despite being sensitive to only a single flavor and a single interaction type, the up-going search is more sensitive than the down-going search due to the large density of the Earth compared to the atmosphere.

\begin{figure}[tbp]
\begin{center}
\includegraphics[width = 0.5\textwidth]{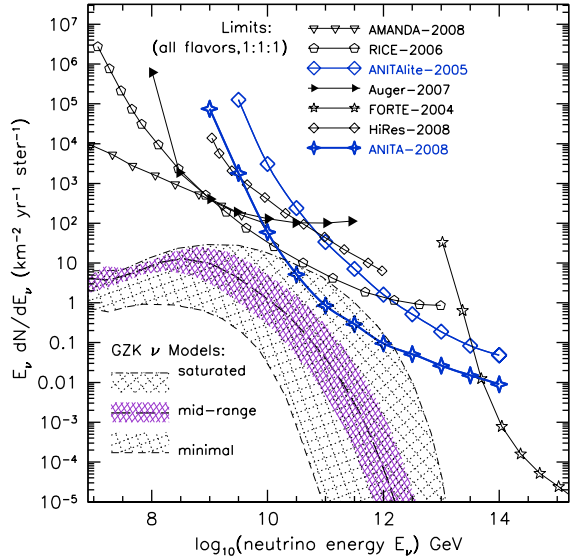}
\end{center}
\caption[Neutrino flux models and experimental limits]{Current flux limits from several experimental EHE neutrino searches: AMANDA, RICE, ANITA, Auger, FORTE, and HiRes.  A range of theoretical models for GZK neutrino fluxes is also shown.  Experiments have now excluded the more optimistic theoretical models.  In the next few years ANITA, Auger, and IceCube will either exclude the moderate models or detect GZK neutrinos if they are correct.  Figure taken from~\cite{ANITA09}.}
\label{fluxLimits}
\end{figure}

Current neutrino flux limits in the 10$^{18}$~eV energy range are summarized in Figure~\ref{fluxLimits}.

\section{Acoustic detection of particle showers}

\begin{figure}[tbp]
\begin{center}
\includegraphics[width = 0.5\textwidth]{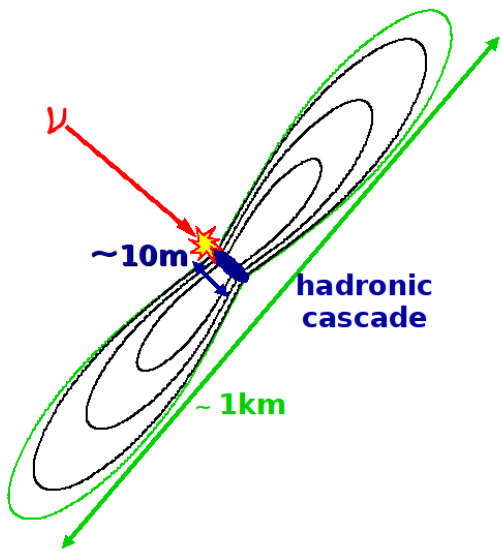}
\end{center}
\caption[Mechanism of acoustic signal production by neutrinos]{Acoustic signal production by a neutrino-induced particle shower.  A high-energy neutrino interacts to produce a hadronic (or electromagnetic) cascade, which heats a long, thin volume of the medium.  This filament is on the order of 10~cm diameter and 10~m length.  The thermal energy is deposited rapidly, faster than both the thermal and acoustic time scales.  This rapid heating causes rapid expansion, resulting in an outgoing shock wave.  Because the filament is long and thin, the radiation pattern is a wide, flat disk perpendicular to the filament.  The acoustic pulse expands outward from the filament in a ring shape, illuminating a disk of the medium perpendicular to the cascade.  Figure taken from~\cite{Richardt09}.}
\label{pancakeCartoon}
\end{figure}

When high-energy particles interact in dense (solid or liquid) media to produce secondary particles that eventually deposit most of the original particle energy as heat, the heat expands the medium locally causing a shock wave to be emitted (Figure~\ref{pancakeCartoon}).  This ``thermoacoustic'' effect, first proposed in the 1950's, can be used to detect very high energy astrophysical particles.  It is particularly well suited for neutrinos, which pass for the most part through the atmosphere and interact only when traversing a denser medium.

\subsection{Theory and simulation}

The theory of the thermoacoustic effect was first established in the 1950's by Gurgen Askarian~\cite{Askarian57}.  In the 1970's a comprehensive treatment of the effect was developed, including computer-based calculations of the acoustic pulse produced by particle showers, by both Askarian~\cite{Askarian79} and by John Learned~\cite{Learned79}.

Dedenko, Butkevich, \emph{et al.} performed calculations of the pressure pulse produced by neutrino interactions in the 1990's~\cite{Butkevich98}.  While the rough shape of the predicted pulses agreed, the normalization of the pressure pulse calculated by Learned, Askaryan, and Dedenko differed by an order of magnitude.  None of these early calculations included the Landau, Pomeranchuk, Migdal effect (see Section~\ref{lpmSection}).

We updated the treatment of Learned (in sea water) to include the LPM effect, in order to calculate the neutrino sensitivity of the SAUND experiment~\cite{Lehtinen02}, \cite{Vandenbroucke05}.  The same calculation framework used for SAUND was used for our studies of hybrid arrays in ice~\cite{Besson05}.  Niess and Bertin also calculated neutrino-induced pressure signals~\cite{Niess06} for both electromagnetic showers and for hadronic showers.  They included the LPM effect in the electromagnetic showers but not in the hadronic showers.

Most recently, the ACORNE group has done a careful study~\cite{Bevan07}, simulating hadronic showers by adopting the CORSIKA~\footnote{http://www-ik.fzk.de/corsika} air shower simulation package to work in water and ice (including the LPM effect).  They found that the radial distribution of energy deposition in the showers is peaked at smaller radii than for previous calculations, indicating that the signal content of neutrino-induced acoustic pulses is peaked at higher frequencies than previously calculated.  Calculations by the three most recent groups (\cite{Vandenbroucke05}, \cite{Niess06}, and \cite{Bevan07} all agree within a factor of two in pressure amplitude).
 
\subsection{Experiments in the laboratory}

The calculations done by Learned were verified in the laboratory by Sulak et al.~\cite{Sulak79} using a proton beam at Brookhaven National Laboratory.  This pioneering work verified many aspects of the thermoacoustic mechanism, and constrained contributions from other mechanisms (microbubble implosion and molecular dissociation) to be small.  Several aspects of the thermoacoustic mechanism were verified quantitatiely:

\begin{enumerate}
\item The frequency of the acoustic signal was inversely proportional to the beam diameter $c$, and roughly equal to $\frac{c}{2d}$ where $c$ is the sound speed.
\item The acoustic pressure amplitude increased linearly with the deposited shower energy.
\item The acoustic pressure amplitude scaled as $1/d^2$, i.e. proportional to the the shower energy deposition \emph{density} for fixed total shower energy deposition.
\item In tests comparing several different fluid target media, the acoustic pressure amplitude scaled with $\beta/C_P$ as expected, where $\beta$ is the thermal expansivity and $C_P$ is the specific heat capacity, over more than an order of magnitude in $\beta/C_P$.
\item In water, $\beta$ increases with temperature and is negative for temperatures between 0 and 4~$^{\circ}$C.  As expected, the acoustic signal amplitude increased with temperature and was inverted in polarity for temperatures near 0, crossing zero amplitude at a few degrees.  However, the exact zero-crossing temperature was 6~$^{\circ}$C.  This offset could indicate a small contribution from a secondary mechanism other than the thermoacoustic effect.
\end{enumerate}

Others (such as~\cite{DeBonis08} and~\cite{Simeone08}) have followed up on Sulak's work to further characterize the thermoacoustic effect using proton beams dumped in water.

The thermoacoustic effect was first demonstrated in ice using a proton beam at the University of Uppsala~\cite{Boeser05}.  Work is underway to experimentally verify the quantitative properties of the effect with a laser-induced signal in a large bubble-free block of ice at RWTH Aachen University~\cite{Karg09sensors}.

\subsection{Experiments in the field}

The DUMAND group did pioneering work on neutrino astronomy in ocean water in the 1970's and 1980's, mostly focused on the optical Cherenkov method but also including the acoustic method. In the last decade, the idea of acoustic neutrino detection has been taken up again and tested by several groups at several sites.  Recent summaries of this activity include~\cite{Vandenbroucke06Summary} and~\cite{Thompson09}.

Our group at Stanford was the first of these.  We read out an array of seven underwater microphones (hydrophones) on the sea floor in the Tongue of the Oceans, a deep ($\sim$1.5~km) ocean cul-de-sac in the Bahamas.  The array is the Atlantic Undersea Test and Evaluation Center (AUTEC), operated by the U. S. Navy.  Through an agreement with the Navy, our group operated the Study of Acoustic Underwater Neutrino Detection (SAUND) using the AUTEC hydrophones.  We installed a simple single-PC DAQ system to read out the seven hydrophones and searched for neutrino-induced signals with the array.  The sensitivity of the array was first estimated in~\cite{Lehtinen02}, and we presented the final analysis of the data, including the first neutrino flux limit produced with the acoustic method, in~\cite{Vandenbroucke05}.  We have now switched to a larger, upgraded array of 49 hydrophones and upgraded the DAQ system.  This is the SAUND-II project.  Data analysis is underway.  One interesting spin-off result from the SAUND-II project is that the data were used to verify for the first time~\cite{Kurahashi07} a general model of underwater noise produced by surface wind acting on the surface~\cite{Short05}.

The ANTARES (Astronomy with a Neutrino Telescope and Abyss environmental RESearch) optical Cherenkov neutrino array, located in the Mediterranean sea, also has an active acoustic neutrino detection research and development program, Antares Modules for Acoustic DEtection Under the Sea (AMADEUS).  AMADEUS consists of six clusters of six acoustic sensors each.  Within each cluster the sensors are arranged $\sim$~m from one another.  The distances between the clusters range from 15~m to more than 200~m~\cite{Lahmann09}.  The group has developed new source reconstruction techniques and applied them to the ANTARES data~\cite{Richardt09}, and has also characterized the noise environment and related it to wind patterns~\cite{Lahmann09}.

A group in the United Kingdom (Acoustic COsmic Ray Neutrino Experiment, ACORNE), has installed a data acquisition system at a military hydrophone array in Scotland.  The group has acquired a large data set and searched it for acoustic neutrino signals~\cite{Bevan09limit}.  They have also performed hydrophone array sensitivity studies~\cite{Perkin09}, studied the properties of hadronic shower deposition relevant to acoustic neutrino signal production~\cite{Bevan07}, and developed new mathematical methods for simulating the acoustic signal induced by thermal energy deposition~\cite{Bevan09simulation}.

Associated with the NEMO (NEutrino Mediterranean Observatory) optical Cherenkov neutrino telescope project, the O$\nu$DE (Ocean noise Detection Experiment) project is also studying acoustic neutrino detection in the Mediterranean Sea.  The detector is located 25~km off the shore of Sicily at 2~m depth.  It consists of four hydrophones, whose data are being used to characterize the noise environment and to study cetacean (sperm whale, in particular) activity.~\cite{Onde08}.


\chapter{Simulation of neutrino-induced signals and detector sensitivity}

\label{simulationChapter}

\noindent\emph{In this chapter we give an overview of the physics of hadronic showers and introduce a new parameterization of them, modified from a standard parameterization to include the LPM effect.  We then describe the software package that has been developed to simulate the acoustic signal as a function of sensor location, neutrino energy, and detector medium properties, and give results for the acoustic radiation pattern calculated in various media as a function of neutrino energy.  Finally we describe another software package that has been developed to determine the response (effective volume) of acoustic detector arrays as a function of neutrino energy.  We conclude with some notes on neutrino event reconstruction with acoustic signals.}

\section{Introduction}

In this chapter we describe several levels of simulation that we have performed.  First we describe a parameterization of hadronic shower energy deposition, which is necessary to determine the acoustic pulse produced by the hadronic shower.  Next we describe simulations of the acoustic pressure pulse as a function of time for arbitrary positions with respect to the neutrino-induced shower location and orientation, for neutrinos of arbitrary energy interacting in arbitrary media.  We then apply this simulation in particular to water, ice, and salt.

Next we developed a package to simulate detector responses and neutrino flux sensitivities for acoustic detector arrays intended to detect neutrino-induced acoustic pulses.  We developed a flexible framework to do this for arbitrary array designs in arbitrary media and have applied it to several example arrays.  In particular (assuming the acoustic properties of South Pole ice expected from theoretical predictions, which are different from those we have now measured) we determined that a large hybrid optical/radio/acoustic array centered on the IceCube array could detect $\sim$20 GZK events per year, with about half of them detected by more than one method simultaneously.


\section{Shower properties}

\subsection{Landau, Pomeranchuk, Migdal (LPM) effect}

\label{lpmSection}

In order to determine the acoustic signal induced by a particle shower, it is necessary to know the spatial distribution of thermal energy deposited by a particle shower in a dense (solid or liquid) medium.  The spatial distribution of the energy is as important as the total energy deposited, and this distribution determines the frequency spectrum of the acoustic signal as well as its radiation pattern.  We consider hadronic showers only.  This is because electromagnetic showers are elongated dramatically by the Landau, Pomeranchuk, Migdal (LPM) effect, sufficiently to reduce the deposited energy density and therefore the amplitude of the acoustic pulse.

The LPM effect is the following: ultra-high-energy interactions involve small transverse momentum transfer, occurring over a long time.  The interaction time is long enough for multiple scattering to occur.  This reduces the cross section of both bremsstrahlung and pair production interactions, effectively lengthening particle showers.  The effect is large for electromagnetic showers.  It is small but finite for hadronic showers, and occurs via $\pi^0$'s in sub-showers of the hadronic shower.  See~\cite{Klein99} for a review of the LPM effect.

The development of hadronic showers in solids and liquids is essentially the same as the development of hadronics showers in air, where they have been studied extensively by extensive air shower (EAS) arrays.  If column density (g/cm$^2$) is used instead of distance, the same treatment of hadronic showers can be used for air and for solids and liquids.

\subsection{Moli\`ere radius}
The Moli\`ere radius, $r_1$, is the radius of a cylinder containing on average 90\% of the energy deposited in an electromagnetic shower initiated by an electron or photon~\cite{Amsler08}.

\subsection{Radiation length}
The radiation length, $X_{rad}$, is the mean distance over which an electron's energy is reduced by one e-folding due to bremsstrahlung.  The radiation length is also 7/9 of the mean free path for pair production by a photon with sufficient energy for pair production.  See~\cite{Amsler08} for more details.


\subsection{Shower ``age''}

Showers of higher energy are longer and reach their maximum at greater depth in the medium than showers of lower energy.  The ``age'' parameter $s$ is a standard quantity introduced to quantify the longitudinal distance along a shower relative to its maximum (and therefore independent of energy), rather than relative to absolute distance in the medium.  The age is zero at the first shower interaction point, one at shower maximum, and two at the end of the shower (the point when the number of shower particles is less than one).  The age parameter was introduced for electromagnetic showers but can also be used for hadronic showers.~\cite{Sokolsky04}

\section{Parameterization of hadronic showers}

We use the hadronic shower parameterization (both longitudinal and radial) developed for the SAUND project and described in~\cite{Lehtinen02,Vandenbroucke05}.  The parameterization accounts for elongation due to the LPM effect.  Here are the full details of the parameterization, which was developed by Nikolai Lehtinen for the SAUND project.

The hadronic shower parameterization we use is the Nishimura-Kamata-Greisen (NKG) parameterization (presented e.g. in~\cite{Sokolsky04}), with the following features:

\begin{enumerate}

\item We normalize to the total shower energy (rather than total number of particles).

\item We use energy-dependent tail length and maximum shower depth parameterized from simulations that include the LPM effect for hadronic showers~\cite{Alvarez-Muniz98}.

\end{enumerate}

As presented in~\cite{Sokolsky04}, the particle density in a hadronic shower as a function of depth $X$ and radius $r$ can be approximated by the NKG parameterization,

\begin{equation}
D(X,r) = N(x) \rho(r),
\end{equation}

where we have decomposed the distribution into two independent functions, one giving the longitudinal distribution and one giving the radial distribution.  The two distributions are defined as follows (following the treatment of~\cite{Sokolsky04}):

\begin{equation}
N(X) = N_{max} \left(\frac {X-X_0}{X_{max}-X_0}\right)^{\frac{X_{max}-X_0}{\lambda}} e^{\frac{X_{max}-X}{\lambda}};
\end{equation}

\begin{equation}
\rho(r) = \frac{N_0}{r_1^2} f(s,r/r_1),
\end{equation}

\noindent where

\begin{equation}
f(s,r/r_1) = (\frac{r}{r_1})^{s-2}(1+\frac{r}{r_1})^{s-4.5} \frac{\Gamma(4.5-s)} {2 \pi \Gamma(s) \Gamma(4.5-2s)}.
\end{equation}

Here $X_0$ is the interaction depth, $N_{max}$ is the maximum particle density, $X_{max}$ is the depth at which it occurs, $\lambda = $~70~g/cm$^2$ is the tail length, $r_1$ is the Moli\`ere radius, and $s$ is the shower age.  This lateral distribution function is an analytical solution for electromagnetic showers, developed by Kamata and Nishimura~\cite{Kamata58}.  By choosing a \emph{constant} value $s =$~1.25 (the \emph{effective} age parameter), this function has been found experimentally to describe the average hadronic shower well.  Choosing a coordinate system where $X_0 =$~0 (such that $X$ is the longitudinal distance forward from the interaction point), and defining $t = X_{max} / \lambda$,

\begin{equation}
N(X) = N_{max} (\frac{X}{X_{max}})^t e^{t - X/\lambda}.
\end{equation}

We wish to renormalize such that the volume integral over the distribution is not the total number of particles but is unity:

\begin{equation}
\int D(X,r) 2 \pi r dr dX = 1.
\end{equation}

\noindent Then we can multiply by either total number of particles (or total shower energy) to get the particle density (or energy density).  The radial part is already normalized,

\begin{equation}
\int \rho(r) 2 \pi r dr = 1,
\end{equation}

\noindent if we choose

\begin{equation}
N_0 = 1.
\end{equation}

\noindent So it remains to normalize

\begin{equation}
\int N(X) dX = 1.
\end{equation}

\begin{figure}
\begin{center}
\subfigure[]{
\noindent\includegraphics[angle = 0, angle=90, width = 0.6\textwidth]{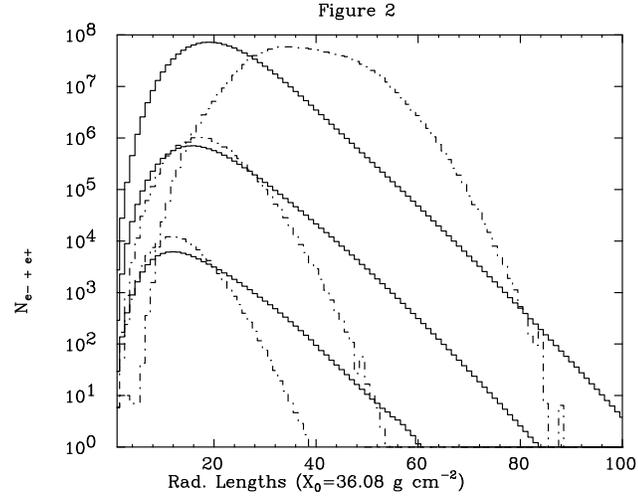}
\label{Alvarez-Muniz}
}
\subfigure[]{
\noindent\includegraphics[angle = 0, width = 0.6\textwidth]{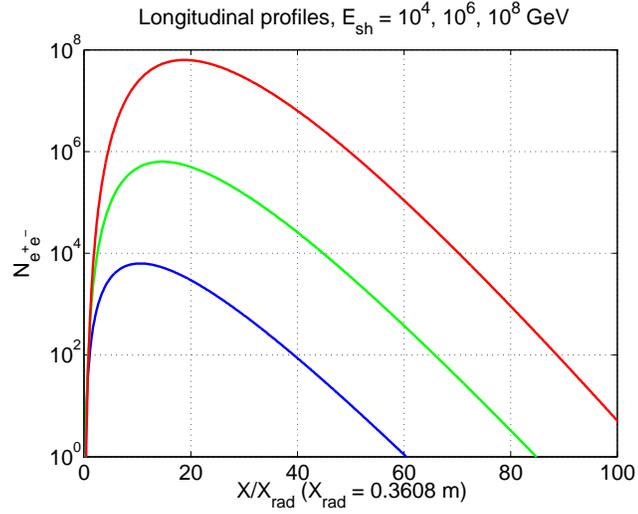}
\label{ourParameterization}
}
\caption[Longitudinal distribution of particles in hadronic showers]{Longitudinal distribution of particles in hadronic showers, for three representative shower energies.  \subref{Alvarez-Muniz} is taken from Figure~2 of~\cite{Alvarez-Muniz98}.  The solid curves, for hadronic showers, should be considered here; the dashed lines are for electromagnetic showers.  \subref{ourParameterization} is our parameterization of their result.
\label{longitudinal}
}
\label{plotLongitudinal_particles}
\end{center}
\end{figure}

This is achieved by choosing

\begin{equation}
N_{max} = \frac{t^{t-1}} {e^t \lambda \Gamma(t)}.
\label{normalization}
\end{equation}
Results from simulating hadronic showers including the LPM effect are presented in~\cite{Alvarez-Muniz98}, for shower energies $E_{sh} =$~10$^4$, 10$^6$, and 10$^8$~GeV.  For reference we reproduce Figure~2 of~\cite{Alvarez-Muniz98} here as Figure~\ref{longitudinal}\subref{Alvarez-Muniz}.  We parameterize their results as follows:

\begin{equation}
X_{max} = 0.9X_{rad} \ln \frac{E_{sh}} {E_{crit}};
\end{equation}

\begin{equation}
\lambda = \frac{1}{100} (130 - 5 \log \frac{E_{sh}}{10^4 \textrm{~GeV}})~\textrm{m}.
\end{equation}

\noindent To verify this parameterization, we can compare it directly to Figure~\ref{longitudinal}\subref{Alvarez-Muniz}.  First we normalize to total number of particles (rather than to unity as given by Equation~\ref{normalization}) by observing from Figure~\ref{longitudinal}\subref{Alvarez-Muniz} that

\begin{equation}
\log[N_{max}] \approx \log[\frac{E_{sh}}{\textrm{GeV}}] - 0.2.
\end{equation}

\noindent See Figure~\ref{longitudinal}\subref{ourParameterization} for our parameterization with this normalization, to be compared with Figure~\ref{longitudinal}\subref{Alvarez-Muniz}.

\begin{figure}[tbp]
\begin{center}
\includegraphics[angle = 0, width = 0.9\textwidth]{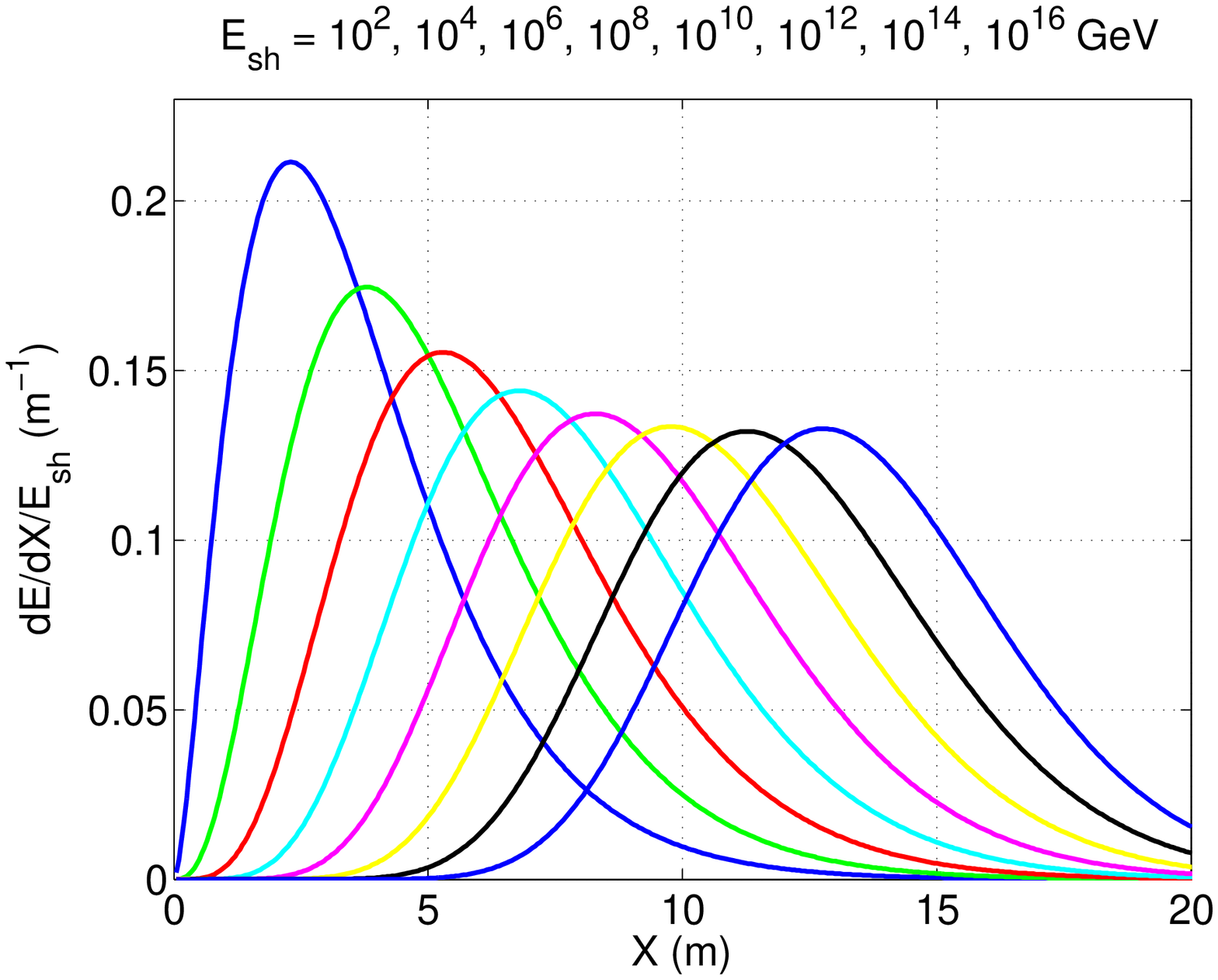}
\end{center}
\caption[Longitudinal distribution of energy in hadronic showers]{Longitudinal distribution of hadronic showers, in terms of energy density.  Note $\int [dE/dX] dX = E_{sh}$.  Shower energy is increasing from left to right (showers with higher energy are longer and have larger $X_{max}$.)}
\label{plotLongitudinal_energy}
\end{figure}

\begin{figure}[tbp]
\begin{center}
\includegraphics[angle = 0, width = 0.7\textwidth]{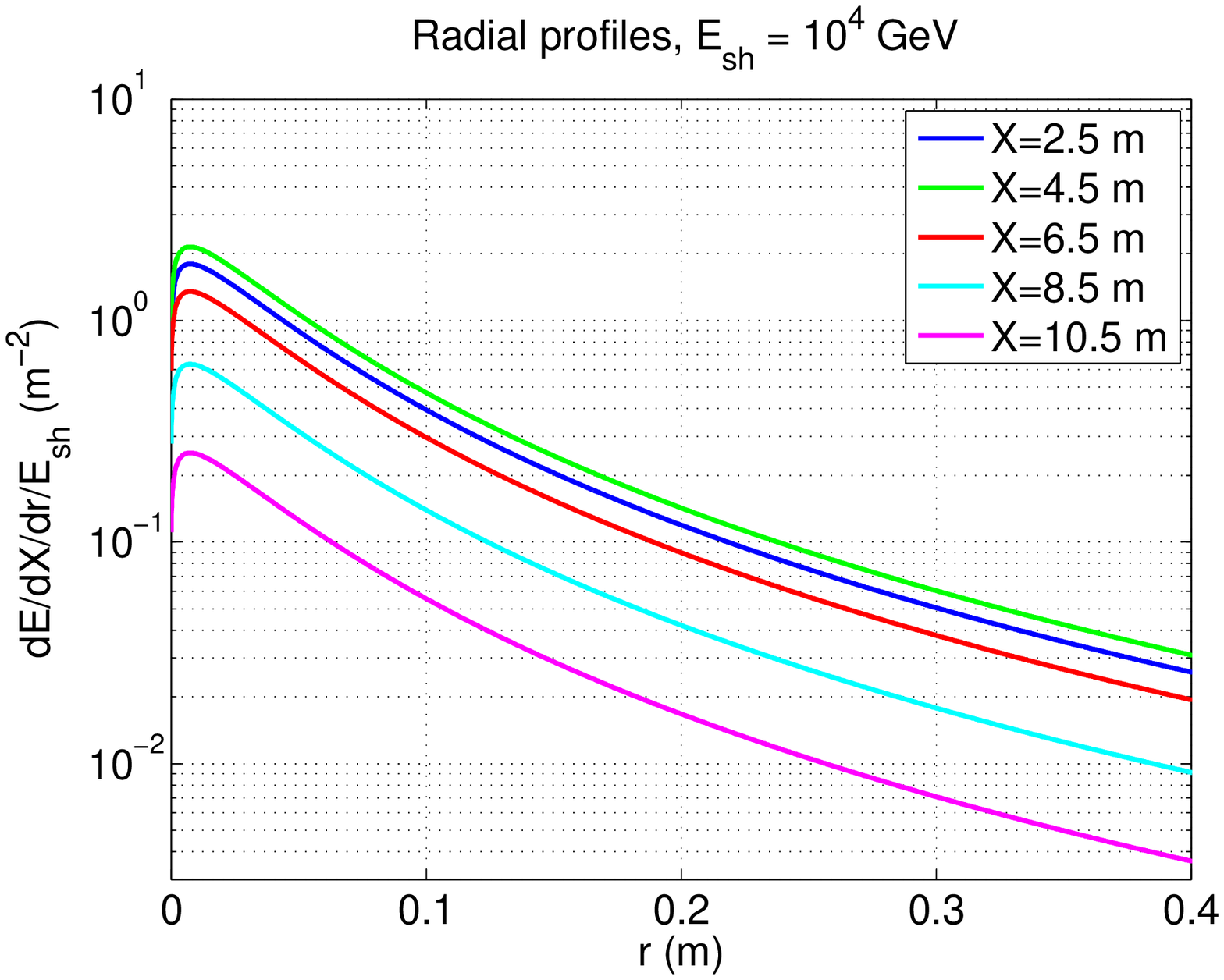}
\end{center}
\caption[Radial energy distribution for a 10$^4$~GeV hadronic shower]{Radial distribution for 10$^4$~GeV.  Note $\iint [dE/dX/dr]dXdr = E_{sh}$.  $X$ is longitudinal distance forward from interaction point.}
\label{plotRadial_log10Esh_04}
\end{figure}

\begin{figure}[tbp]
\begin{center}
\includegraphics[angle = 0, width = 0.7\textwidth]{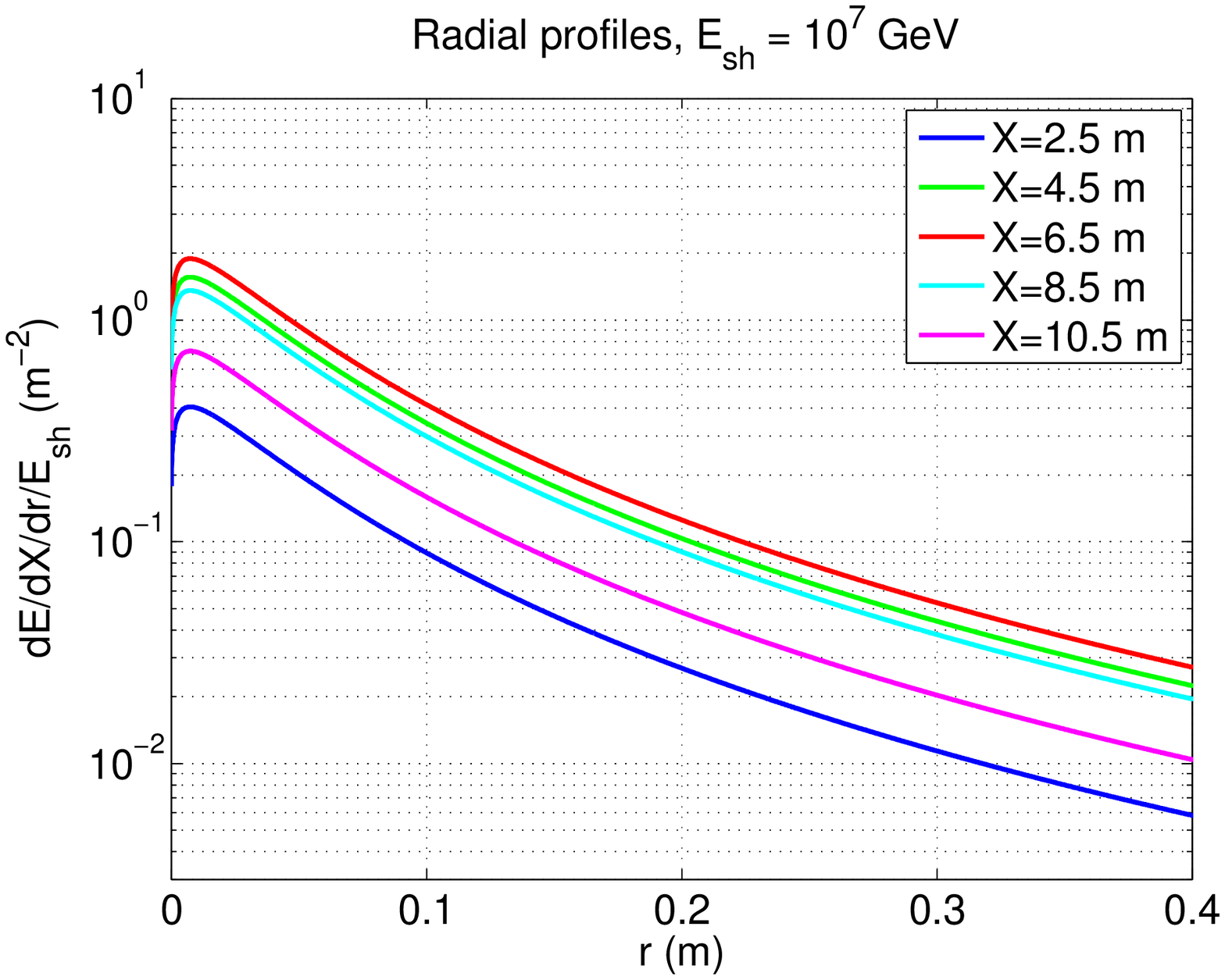}
\end{center}
\caption[Radial energy distribution for a 10$^7$~GeV hadronic shower]{Radial distribution for 10$^7$~GeV.  Note $\iint [dE/dX/dr]dXdr = E_{sh}$.  $X$ is longitudinal distance forward from interaction point.}
\label{plotRadial_log10Esh_07}
\end{figure}

\begin{figure}[tbp]
\begin{center}
\includegraphics[angle = 0, width = 0.7\textwidth]{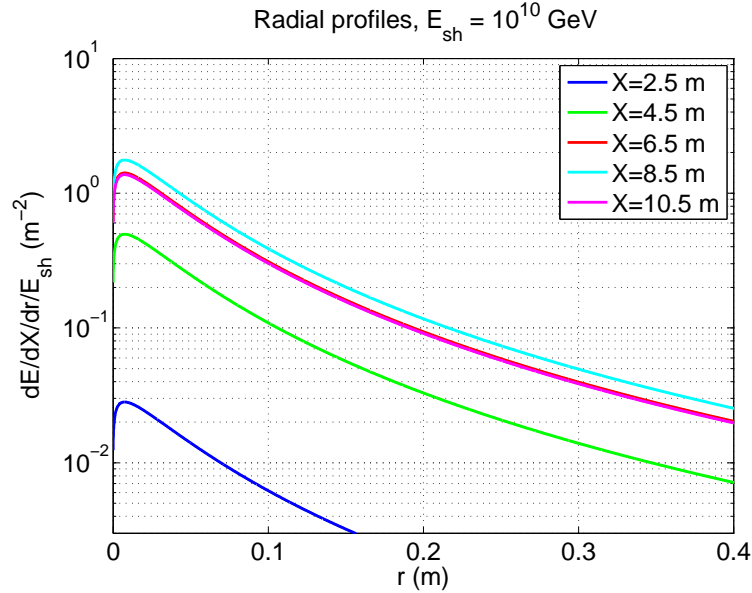}
\end{center}
\caption[Radial energy distribution for a 10$^{10}$~GeV hadronic shower]{Radial distribution for 10$^{10}$~GeV.  Note $\iint [dE/dX/dr]dXdr = E_{sh}$.  $X$ is longitudinal distance forward from interaction point.}
\label{plotRadial_log10Esh_10}
\end{figure}

Figure~\ref{plotLongitudinal_energy} gives the  the longitudinal distribution for a wider range of energies.  Figures~\ref{plotRadial_log10Esh_04}, \ref{plotRadial_log10Esh_07}, and \ref{plotRadial_log10Esh_10} give the radial distributions at 10$^4$, 10$^7$, and 10$^{10}$~GeV.  Integrating the radial distribution at a particular depth gives $dE/dX$ at that depth.

\section{Expected neutrino signals in water, ice, and salt}

\subsection{Computational method}
Given the spatial distribution of deposited heat in a homogeneous medium and the material properties of the medium, the thermoacoustic pressure pulse $P(t)$ can be calculated at an arbitrary position $r$ relative to the heat deposition.  N.~Lehtinen, J.~Vandenbroucke, and N.~Kurahashi wrote a computer program to do this calculation.  The program integrates over the energy deposition, including the contribution of each element of each spatial element to the acoustic signal, using a Green's function method.  The program was described in~\cite{Lehtinen02} and~\cite{Vandenbroucke05}.

Several energy distributions are supported, including those for a spherical Gaussian, an electromagnetic shower (including the LPM effect), a hadronic shower, and an electromagnetic discharge (such as that produced by a ``zapper'' calibration device developed for the SAUND project).  The material properties can be specified and in particular sea water, South Pole ice, and salt dome media are supported.

Each execution of the program calculates the pressure pulse at a grid of points, such that a radiation pattern contour can be determined with a single program execution.  Execution times are on the order of one pressure pulse (one point in the grid) per minute.

The material properties that we are using are given in Table~\ref{materialProperties}.  See~\cite{Vandenbroucke06Summary} for further comparison of the three media.

\begin{table}[tbp]
\centering
\caption[Material properties of water, ice, and salt relevant to acoustic neutrino detection]{Material properties used in the simulation of neutrino-induced pressure pulses in sea water, South Pole ice, and salt.}
\centering
\begin{tabular}{| c | c | c | c | c |}
\hline
\bf{Property} & \bf{Symbol (Units)} & \bf{Sea water} & \bf{South Pole ice} & \bf{Salt}\\
\hline
Temperature & $T$ ($^\circ$C) & 15 & -51 & 30 \\
\hline
Sound speed &  $v_L$ (m/s) & 1530 & 3920 & 4560 \\
\hline
Volume expansivity & $\beta$ (10$^{-5}$ K$^{-1}$) & 25.5 & 12.5 & 11.6 \\
\hline
Heat capacity & $C_P$ (J/kg/K) & 3900 & 1720 & 839 \\
\hline
Peak frequency & $f_p$ (kHz) & 7.7 & 20 & 42 \\
\hline
Gruneisen parameter & $\gamma = v_L^2 \beta / C_P$ & 0.153 & 1.12 & 2.87 \\
\hline
Radiation length & $X_{rad}$ (m) & 0.361 & 0.392 & 0.0997 \\
\hline
Moli\`ere radius & $r_M$ (m) & 0.0730 & 0.0794 & 0.0417 \\
\hline
Critical energy & $E_{crit}$ (GeV) & 0.0870 & 0.0870 & 0.0440 \\
\hline
\end{tabular} 
\label{materialProperties}
\end{table} 
\subsection{Acoustic radiation pattern in water, ice, and salt}

We used the software described above to calculate the acoustic radiation pattern for neutrinos of various energies in each of the media.  The software allows the user to choose a medium, a medium attenuation model, a signal processing method (raw or matched filter), and a threshold.  We assume that each neutrino of energy $E_{\nu}$ generates a hadronic shower of energy $E_{had} = y E_{\nu}$, where $y =$~0.2 for every interaction.  We give radiation patterns for multiple energies at a single threshold.  The contours can be rescaled to determine the radiation pattern at other thresholds.  For example, the contour for a 10$^{18}$~eV neutrino with a 10~mPa threshold is equivalent to the contour for a 10$^{19}$~eV neutrino with a 100~mPa threshold.  In other words, the neutrino energy threshold scales linearly with the ambient noise level.  This follows from the linear dependence of the acoustic pressure amplitude on the hadronic shower energy.

\begin{figure}[tbp]
\centering
\noindent\includegraphics[width=40pc]{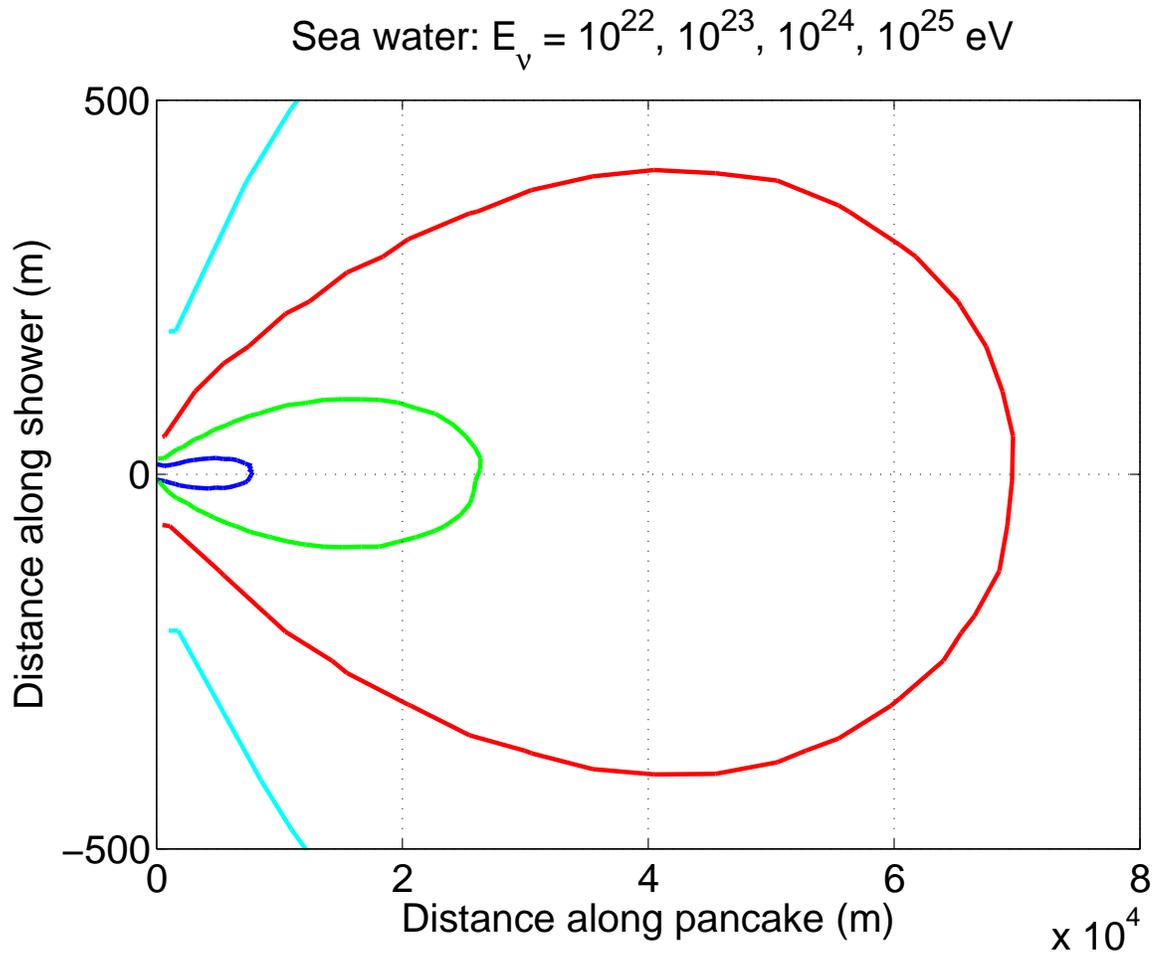}
\caption[Acoustic radiation pattern for a neutrino interacting in sea water]{Acoustic radiation pattern induced by a neutrino interacting in sea water, for various neutrino energies.  A sea water acoustic attenuation model is included.  A matched filter as described in~\cite{Vandenbroucke05} is used, with a threshold of 0.02.}
\label{plotWaterContours}
\end{figure}

Acoustic radiation contours for neutrino-induced signals in sea water are shown in Figure~\ref{plotWaterContours}.

\begin{figure}[tbp]
\centering
\noindent\includegraphics[width=30pc]{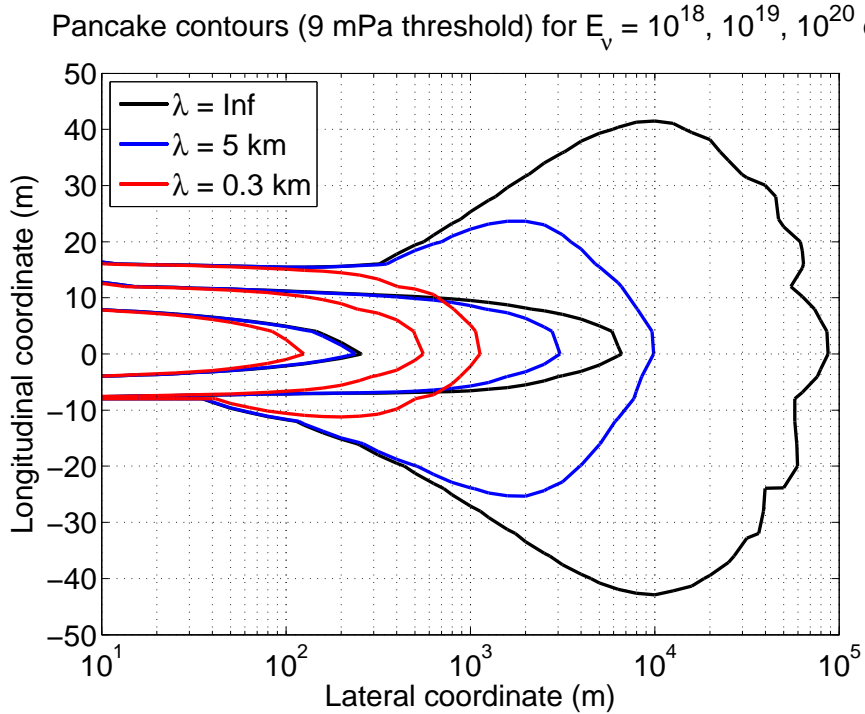}
\caption[Acoustic radiation pattern for a neutrino interacting in ice]{Acoustic radiation pattern induced by a neutrino interacting in ice, for various neutrino energies and acoustic attenuation lengths ($\lambda$) in ice.  A threshold of 9~mPa is used, with no matched filtering.}
\label{plotIceContours}
\end{figure}

Radiation contours for South Pole ice are shown in Figure~\ref{plotIceContours}.  The neutrino interacts at a lateral coordinate of zero and a longitudinal coordinate of +9~m (offset from zero in order that the shower max and acoustic radiation pattern are centered close to a longitudinal coordinate of zero).


\section{A large hybrid optical/radio/acoustic extension of IceCube}

We simulated a large hybrid optical/radio/acoustic extension of IceCube for GZK neutrino detection.  The encouraging results of this simulation motivated us to develop SPATS, in order to measure the acoustic properties of the ice and gain technical experience toward a larger detector array.  The simulations indicated that we could detect $\sim$10-20 GZK neutrinos per year, half of which were detected with both the radio and the acoustic method.  The results of this simulation were presented in~\cite{Besson05}, which is included here as Appendix~\ref{largeHybridAppendix}.

Note that these simulations used the theoretical estimates of acoustic attenuation in South Pole ice available at the time~\cite{Price06}.  Our new experimental measurement of the attenuation (Chapter~\ref{attenuationChapter}) shows that the true amount of attenuation is $\sim$30~times larger than the theoretical prediction used in the simulations.

\section{An intermediate-scale hybrid extension of IceCube}

Following the large hybrid detector simulation, we also simulated an intermediate hybrid detector which could be used to see a few GZK events and test the technology for the large array.  The results of this simulation are presented in~\cite{Besson08}.

\section{Acoustic and hybrid event reconstruction}

\subsection{Neutrino event reconstruction with only three hit sensors}

As described in Chapter~\ref{transientsChapter}, four or five hit receivers are generally necessary to reconstruct the location of a source.  Such algorithms will be necessary to reconstruct most events including background events which are expected to have roughly isotropic emission patterns for the most part.

However, in the case of neutrino event reconstruction, three hits on three strings is actually sufficient for good reconstruction if there are no noise hits: a plane fit through the hit sensors determines the plane of the very flat acoustic radiation ``pancake''.  The upward normal to the fitted plane then gives the neutrino direction with $\sim$1$^\circ$ precision.  If the signal is really neutrino induced, the mirror symmetry about the fitted plane is broken by the fact that upgoing neutrinos in this high energy range are absorbed by the Earth, so only downgoing neutrinos are detectable.  Therefore the upward normal to the fitted plane points to the direction from which the neutrino came.

Three hits on three strings is sufficient for neutrino direction determination with this method.  Vertex position and energy can also be determined in many cases, even with only three hits, as follows.  A change of coordinates to the pancake plane means two hits constrain the vertex to a hyperbola (not hyperboloid; it's a 2D problem now) and the third hit identifies one or two points on the first hyperbola.  In some cases a 4th hit is necessary to distinguish between two solutions to the intersection of two hyperbola.  But in many cases with three hits, one of the two solutions is unphysical so three hits are sufficient to determine neutrino direction as well as vertex position and shower energy.

This assumes there are no noise hits and that the three hits are already identified to be neutrino-induced.  Including 4 or more hits (with amplitude information) will usually be necessary to give some indication of the radiation pattern and thereby distinguish neutrino signals from background transients.


\subsection{Event reconstruction with shear waves}

In contrast to both radio and optical signals, two types of body waves are possible for acoustic waves in solids: pressure (longitudinal) and shear (transverse) waves.  There are also a variety of surface acoustic waves possible in solids, including Rayleigh waves.  In liquids, only longitudinal acoustic waves are supported.  But in a solid such as ice, a single acoustic emission event can produce both pressure (P) and shear (S) waves.  The S wave propagates at roughly half the speed of the P wave in many solid media including ice (see our experimental measurement of P and S wave speeds in South Pole ice, presented in Chapter~\ref{soundSpeedChapter}).  We have detected both P and S waves from both types of our emitters (frozen-in transmitters operated in ice, and retrievable pinger operated in water), as well as from ambient transient events.  The detection of shear waves from the source surrounded by water in particular was a surprise and is discussed in Section~\ref{shearSection}.

It is possible that neutrinos produce shear waves in addition to pressure waves.  While the pressure wave production has been verified and characterized in the laboratory by Sulak \emph{et al.}~\cite{Sulak79} as described in Chapter~\ref{introductionChapter}, little is known about shear wave production by the thermoacoustic effect.  It has been argued on theoretical grounds~\cite{Boeser06} that thermoacoustic shear wave production should be suppressed.  However, more laboratory work is necessary to resolve this question.

If particle showers do produce S waves as well as P waves, the S waves could contribute valuable information to the event reconstruction and background rejection potential of a detector array.  The distance to an acoustic source can be resolved with a \emph{single} module that detects both a P and an S wave.  If the acoustic emission is known to be isotropic, the energy of the event can then be determined from a single module.  In the case of pancake-shaped thermoacoustic radiation, more modules are likely necessary to determine the shower energy, although the sharpness of the radiation pattern could be quantified in a beaming factor and used to determine an order-of-magnitude estimate of the shower energy despite the severely anisotropic radiation pattern.  More generally, if P waves are detected on one or more modules, detection of additional S wave hits on one or more modules can be used to improve event reconstruction or reduce the number of hit modules necessary.

On the other hand, if particle showers are conclusively shown to \emph{not} produce shear waves, the presence of shear waves in detected events could be a valuable handle for distinguishing background signals from neutrino-induced signals.  We have already shown with SPATS that shear waves are present for at least one class of transient background events (see Chapter~\ref{transientsChapter}).

\subsection{Hybrid reconstruction}

If a future hybrid detector array detects signals from more than one of the possible (optical, radio, acoustic) methods, the information can be combined for improved event reconstruction.  The challenge is that the signals propagate with speeds that differ by five orders of magnitude.  For example, the time for a radio signal to propagate $\sim$1~km from a source to a receiver is much smaller than the time for an acoustic signal to propagate the same distance, and is in fact comparable to the time between individual samples in the acoustic signal.  The na\"ive conclusion is that the uncertainty in the acoustic signal arrival time is so large that it is impossible to combine the hits on the same footing with radio and optical hits, whose arrival time is determined much more precisely.

Nevertheless, the hit receiver information from the different methods can be put on the same footing, such that any hit on any receiver is as valuable as any other regardless of the signal type, by expressing the propagation equations in terms of distances rather than times.  The arrival times (and uncertainties of the arrival times) are simply scaled by the propagation speed.  When this is done, the uncertainty due to time-of-arrival resolution for the different methods is comparable.  The method was described and demonstrated with a Monte Carlo simulation in~\cite{Vandenbroucke06Hybrid}.



\chapter{Design of the South Pole Acoustic Test Setup}

\noindent\emph{In this chapter we give an overview of the development of SPATS and describe the array geometry and system design.  We give details about the design of individual components of SPATS and of the retrievable pinger that was developed and operated with SPATS.  For reference we include the surface layout of both the SPATS strings and the holes in which the pinger was operated.}

\label{designChapter}

\section{Timeline}

Following previous work on neutrino detector array simulation, sensor design, and laboratory tests, we conceived and designed the South Pole Acoustic Test Setup (SPATS) in 2005.  Construction, integration, and testing proceeded through 2006.  SPATS was initially conceived as three strings of acoustic sensors and transmitters, which were installed in three IceCube holes in January 2007.  Based on analysis of data from the first few months of SPATS in the three-string configuration, we decided to add a fourth string (with improvements in module design and layout relative to the first three strings).  The fourth string was constructed and tested in the remainder of 2007 and installed in December 2007.  In the same year (2007) we built a retrievable pinger, which we operated for the first time in December 2007 and January 2008.  Analysis of that data led to a significantly improved pinger, which we operated operated in December 2008 and January 2009.

\section{Array geometry} 

\begin{figure}[tbp]
\begin{center}
\includegraphics[angle = 0, width = 1\textwidth]{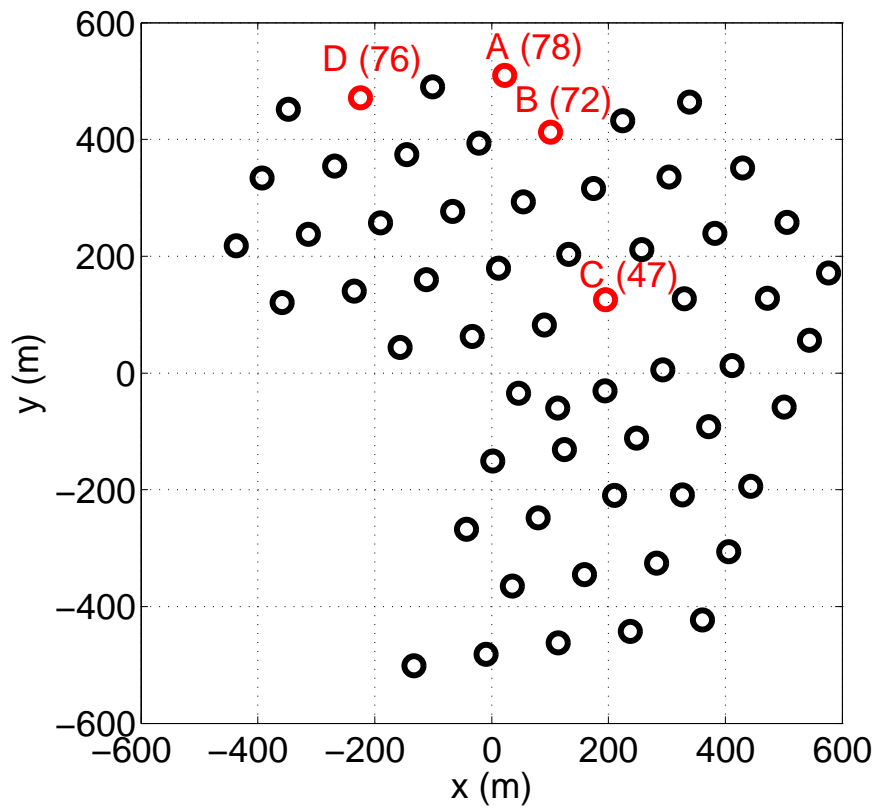}
\end{center}
\caption[Surface layout of IceCube and SPATS strings]{Surface layout of the 59 IceCube strings deployed by the end of January 2009.  The four holes with SPATS strings are labeled by SPATS string ID (letter) and IceCube string ID (number).}
\label{geometry_2009_no_pinger}
\end{figure}

The SPATS array consists of four strings, each deployed in an IceCube hole alongside an IceCube string (Figure~\ref{geometry_2009_no_pinger}).  Each string has seven acoustic stages, with each stage consisting of one transmitter module and one sensor module.  Strings A, B, and C were deployed in January 2007 and contain stages at depth 80, 100, 140, 190, 250, 320, and 400~m.  String D was deployed in December 2007 and contains stages at depth 140, 190, 250, 320, 400, 430, and 500~m.  The instrumentation at each depth is summarized in Table~\ref{depthsTable}.  The horizontal distances between strings are given in Table~\ref{baselines}. 

\begin{table}[tbp]
\centering
\caption[SPATS instrumentation on each string by depth]{Tabulation of which instrumentation is at which depth in each of the four SPATS strings.  ``HADES'' are the sensors of an alternative design.  ``L emitter'' is an alternative type of emitter (produced by the Lausanne group), connected to the same type of transmitter module (HV pulser) as other transmitters.}
\centering
\begin{tabular}{| c | c | c |}  
\hline
\bf{Depth (m)} & \bf{Strings A, B, and C} & \bf{String D} \\
\hline
80 & Stage 1 & - \\
\hline
100 & Stage 2 & - \\
\hline
140 & Stage 3 & Stage 1 \\
\hline
190 & Stage 4 & Stage 2 (inc. HADES 2A sensor) \\
\hline
250 & Stage 5 & Stage 3 \\
\hline
320 & Stage 6 & Stage 4 \\
\hline
400 & Stage 7 & Stage 5 \\
\hline
430 & - & Stage 6 (inc. HADES 2B sensor and L emitter) \\
\hline
500 & - & Stage 7 \\
\hline
\end{tabular} 
\label{depthsTable}
\end{table} 

\begin{table}[tbp]
\centering
\caption[Distances between SPATS strings]{Horizontal distances between SPATS strings.}
\centering
\begin{tabular}{| c | c |}  
\hline
\bf{Baseline} & \bf{Distance (m)} \\
\hline
AB & 124.8 \\
\hline
AC & 421.1 \\
\hline
AD & 249.2 \\
\hline
BC & 302.2 \\
\hline
BD & 330.3 \\
\hline
CD & 543.0 \\
\hline
\end{tabular} 
\label{baselines}
\end{table} 

\section{System overview}

\begin{figure}[tbp]
\begin{center}
\includegraphics[angle = 0, width = 1\textwidth]{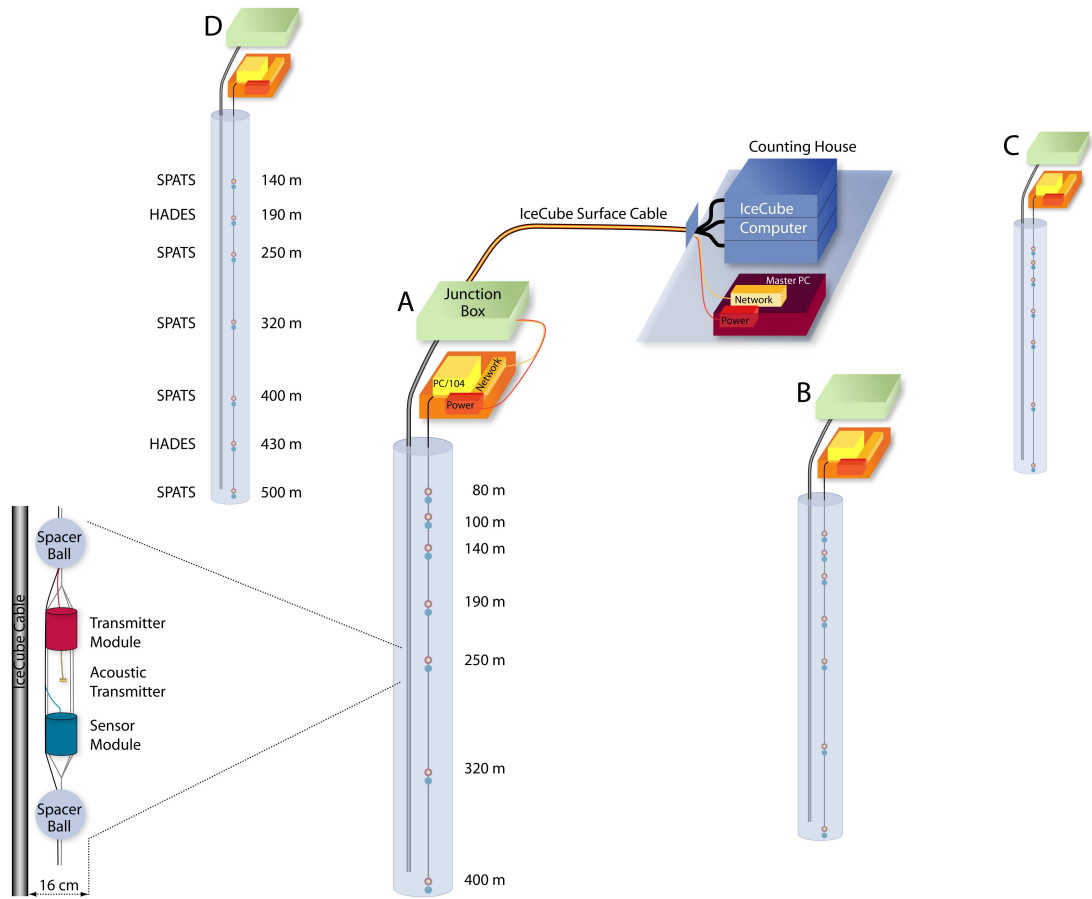}
\end{center}
\caption[SPATS array schematic]{Schematic of the SPATS system design.}
\label{spats_schematic_4_strings}
\end{figure}

A schematic of the SPATS array is given in Figure~\ref{spats_schematic_4_strings}.  Each of the SPATS strings was deployed in an IceCube hole, alongside the IceCube string, after the IceCube string was deployed and anchored.  Each of the strings consists of seven acoustic ``stages'' each at a different depth.  Each stage consists of a transmitter module and a sensor module.

The transmitters and sensors are connected to the surface via analog signals along copper wires.  At the surface of each string is an Acoustic Junction Box (AJB) inside of which is a waterproof compartment where the in-ice lines are connected to a rugged embedded computer (String PC).  The four String PC's are connected to a central Master PC, a rack-mounted server indoors in the IceCube Laboratory.  Power, communications, and timing are routed over surface cables from the Master PC to each of the String PC's.

The design of the SPATS system, and results from initial tests of SPATS hardware in the laboratory and in the field are described in detail in~\cite{Boeser06}.

\section{In-ice transmitters}	

SPATS includes 28 transmitter modules (seven per string) frozen into the ice between 80~m and 500~m depth.  Each module consists of a ring-shaped piezoelectric ceramic emitter and a high-voltage (HV) pulser module.  The emitter is in direct contact with the ice.  It is molded in epoxy for electrical insulation and is connected to the HV pulser via a short, stiff HV cable.  The emitter hangs directly beneath the HV module.  The HV module features a cylindrical steel pressure housing containing the HV pulser circuit.  The housing is penetrated on the bottom by the HV cable going to the emitter, and on the top by an 8-pin connector that mates to the cable going to the surface.

The 8 pins of the transmitter are described in Table~\ref{transmitterPinouts}.  The transmitter receives both +15~V and +12~V DC from the surface, as well as a steering voltage and a trigger signal.  The trigger is a digital signal.  The rising edge of the trigger signal initiates charging of an LC circuit, and the falling edge discharges it to generate an HV ($\sim$kV) pulse which is routed to the emitter.  The width of the trigger pulse determines the charge time, which influences the strength of the HV pulse and therefore of the acoustic pulse.  The amplitude can also be determined by a DC ``steering voltage''.  A periodic trigger signal can be used to pulse the transmitter repeatedly.

Each of the transmitter modules also features a temperature or pressure sensor.  The transmitter of the deepest stage (Stage 7) of each string has a pressure sensor built into a small port penetrating the pressure housing.  This was used to verify the depth of the string during deployment, determine the final installed depth of the string, and monitor the freeze-in process.  The other six stages of each string (Stages 1-6) have no pressure sensor and instead have a temperature sensor.  The temperature sensors were used to monitor the freeze-in process.  Both the pressure and temperature sensors are ``PT1000'' devices that output a current in the 4-20~mPa range.

A down-scaled version of the HV pulse that is routed down to the emitter is routed up to the surface along a dedicated wire.  The signal is digitized at the surface.  The shape and amplitude of the pulse can both be determined with good fidelity.  This is known as the ``HV read-back'' signal and is valuable for verifying transmitter performance as well for time-stamping the emission time of each transmitter pulse.

\begin{table}[tbp]
\centering
\caption[Transmitter pinouts]{Pinout of the SPATS transmitters.}
\centering
\begin{tabular}{| c | c | c |}  
\hline
\bf{Pin} & \bf{Name} & \bf{Function} \\
\hline
1 & +15 V & DC power for transmitter \\
\hline
2 & GND & ground for DC power \\
\hline
3 & TRG & trigger signal \\
\hline
4 & I$_{ret}$ & return (output) current from pressure/temperature sensor \\
\hline
5 & +12 V & power for pressure/temperature sensor (also necessary for HVRB) \\
\hline
6 & V$_{steer}$ & steering voltage, to control transmitter amplitude \\
\hline
7 & HVRB & down-scaled version of the HV pulse routed to the emitter \\
\hline
8 & HVRB GND & ground for the HVRB signal \\
\hline
\end{tabular} 
\label{transmitterPinouts}
\end{table} 

\section{In-ice sensors}

\subsection{``SPATS'' sensor design: First generation (Strings A, B, and C)}

\begin{table}[tbp]
\centering
\caption[Sensor pinouts]{Pinout of the SPATS sensors.}
\centering
\begin{tabular}{| c | c | c |}  
\hline
\bf{Pin} & \bf{Name} & \bf{Function} \\
\hline
1 & +15 V & DC power for sensor \\
\hline
2 & GND & ground for DC power \\
\hline
3 & CH1- (A-) & Negative side of Channel 1 differential output \\
\hline
4 & CH1+ (A+) & Postive side of Channel 1 differential output  \\
\hline
5 & CH2- (B-) & Negative side of Channel 2 differential output  \\
\hline
6 & CH2+ (B+) & Positive side of Channel 2 differential output  \\
\hline
7 & CH3- (C-) & Negative side of Channel 3 differential output \\
\hline
8 & CH3+ (C+) & Positive side of Channel 3 differential output \\
\hline
\end{tabular} 
\label{sensorPinouts}
\end{table}

Each of the first three strings features a sensor module in each stage.  Each sensor module consists of cylindrical steel pressure vessel housing three disk-shaped piezoelectric ceramic transducers.  The pressure cylinder is oriented with its axis vertical on the deployed strings.  The three transducers are in the equatorial plane of the module and are separated by 120$^\circ$ from one another to achieve good angular coverage by the module as a whole.  Each transducer is pressed against the inner surface of the pressure housing with a post and the contact force is adjusted with a screw.  Each transducer along with its amplifier constitutes a single sensor ``channel''.

The three channels of each module act independently, each with its own dedicated wires running to the surface.  The pinout of the SPATS sensor modules is shown in Table~\ref{sensorPinouts}.  Acoustic signals hit the outside of the pressure module and propagate along it.  For signals recorded by more than one channel, the $\sim$10~cm separation between channels results in resolvable time delays in the recorded signals, which can be used to determine the azimuthal direction to the source.  Alternatively if the source direction is known, the time delay can be used to determine the orientation of the sensor module.

\subsection{``SPATS'' sensor design: Second generation (String D)}

Based on experience with the first three strings, the design of the sensors was improved for String D.  The coupling of the three transducers to the steel housing is equalized better, and the electronic design was also improved.  Recordings with String D sensors have a signal-to-noise ratio that is superior to that of the first three strings.

\subsection{``HADES'' design}

In addition to five modules of the improved sensor design, two modules of an alternative design were included in String D.  The alternative design is known as ``HADES'' and is described in detail in~\cite{Semburg09}.  In contrast to the standard design, these sensors feature a single piezoelectric transducer (rather than three independent transducers), molded in epoxy and in direct contact with the ice.  The epoxy material was chosen to have acoustic impedance intermediate between ice and the piezoelectric, to facilitate coupling and reduce resonance.  The signal from the transducer is routed to amplifier electronics inside a steel pressure housing of the same type used for other SPATS sensor modules, from where the signal is routed to the surface in the same way as for the other sensors.  The locations of standard SPATS vs. HADES sensors on String D are listed in Table~\ref{depthsTable}.  In HADES modules, the single sensor is connected to channel 2, and channels 0 and 1 are not connected.

The SPATS and HADES designs are complementary, a feature which has been valuable for several analyses.  Relative to the standard SPATS sensor design, the HADES design features a flatter frequency response, a lower noise level, and a lower overall sensitivity.

\section{Cables}

\subsection{In-ice cables}
The stages are connected to the surface both mechanically and electrically by a cable bundle.  The bundle includes 14 cables each of the same design.  Each of the 14 cables runs from the surface to one transmitter or sensor module, such that the bundle thickness decreases with increasing depth.  Each cable is sufficiently strong to bear the load of its module individually.  The individual cables are not molded together but are wrapped together with a spiral of string.  Each cable in the bundle consists of 8 copper wires along with insulation, filler, and sheath.

\subsection{Surface cables}

The IceCube experiment features surface cables trenched $\sim$1~m beneath the snow surface.  There is one surface cable for each IceCube string, running from the string to the IceCube Laboratory (located in the center of the array) through a network of such trenches.  Each cable consists of an assembly of ``quads'', each quad containing four copper wires (two pairs).  We use two of these quads for each SPATS string.  One wire pair in each of our two quads carries DC power at +48 V.  The third pair is used for a DSL communications signal, and the fourth pair is used for an IRIG-B GPS timing signal.

\section{Surface hardware}

In addition to the hardware buried deep in the ice, surface hardware is necessary to control and read out the in-ice instrumentation.  We described the design of the SPATS data acquisition system (DAQ) in an article in Embedded Computing Design magazine~\cite{Vandenbroucke08Embedded}.

\subsection{Master PC}

A central ``Master PC'' is installed in the IceCube Laboratory, in the center of the IceCube array and several hundred meters across the ice surface from SPATS.  This server communicates with the computers installed at the surface of each SPATS string (the ``String PC's'').  The Master PC features a single GPS clock (which is connected to a GPS antenna on the roof of the ICL), as well as a SPATS Hub Service Board for each of the four String PC's.  Users can log on to the Master PC, and from there to each of the String PC's, to take data manually or to apply DAQ software upgrades.  Standard data taking occurs autonomously.  The Master PC is running a standard Linux operating system.

\subsection{SPATS Hub Service Board (SHSB)}

The Master PC includes one SHSB for each String PC.  The SHSB is a custom PCI board, designed by Kalle Sulanke at DESY Zeuthen.  The two surface cables arriving from each from each String PC connect to the SHSB.  A driver written for the SHSB allows the software commands on the Master PC to power each of the strings on/off, to monitor the DSL communications and and IRIG timing signal to each string, and to monitor the voltage and current of each of the power lines going to the strings.

\subsection{Acoustic Junction Boxes (AJB's)}

Near each IceCube hole is a $\sim$2 m deep, $\sim$4~m thick, and $\sim$10~m long trench that is dug for IceTop tanks and filled at the end of each construction season.  In addition to the IceTop tanks, there is a Surface Junction Box (SJB) where in-ice IceCube cables are mated to surface IceCube cables.

For the four SPATS strings, there is in addition an Acoustic Junction Box (AJB) located in this trench.  The in-ice SPATS cable bundle runs to this junction box.  The AJB contains DC-DC converters, a DSL modem, and a String PC.  A printed circuit board (PCB) inside each AJB routes the signals between the cables entering the AJB and the devices inside the AJB.

\subsection{String PC's}

Insisde of each AJB is a String PC.  This is a rugged embedded computer following the PC104 design (an industry standard for embedded computing).  The PC104 standard is a modular design allowing individual cards each serving a different purpose (e.g. CPU, ADC, DAC, GPS, or power supply) to be stacked together.  The boards communicate with one another via ISA (AT) and/or PCI buses that run vertically through the cards in the stack.

Our String PC is composed of modules produced by the Real Time Devices (RTD Embedded Technologies, Inc.) company.  Each string PC consits of 6 modules in a rugged aluminum enclosure to provide heat sinking, grounding, and additional protection inside the AJB.  The enclosure includes heat sinking fins for the CPU module, is splashproof, and has a lid on the top module such that the entire PC104 stack is enclosed (except for connectors in the IDAN enclosure that allow cables to be connected to the outside of the enclosure).

We chose the "extended range" (ER) modules, which are rated for temperatures between -40 $^\circ$C and +85 $^\circ$.  Here are the modules comprising each String PC

\subsubsection{CPU module}

There is one CPU module per String PC (RTD model IDAN-CML47786HX650ER-256D/D1GX).  This module features a 650 MHZ Celeron CPU board with 256 MB of memory.  The board also includes a 1 GB DiskOnChip flash disk module mounted directly on the board.  This module was intended to be used as the main disk for the system, but significant driver problems prevented us from running a stable operating system on it.  Instead we installed separate DiskOnModule flash disks (see below).  The The String PC's are running a minimal Linux operating system.

\subsubsection{Flash disks}

Instead of spinning disks which would have a high failure risk operating at the low temperature and humidity that the String PC's experience, we used robust low-temperature flash disks of the DiskOnModule model, made by PQI International.  We used the "wide temperature" version, rated for -40 $^\circ$C and +85 $^\circ$ (same as all other String PC components).  These modules have the distinct advantage that they are molded with a standard 40-pin IDE connector and integrated IDE controller, such that they look like a standard hard disk to the CPU.  We installed two modules per String PC (one master and one slave), each 1 GB.

All eight DiskOnModules are essentially clones of one another, containing the necessary operating system and data acquisition software.  The systems were designed such that a direct serial connection from the indoor Master PC could be used to obtain a String PC console remotely, in order to perform low level maintenance and troubleshooting including viewing the BIOS at system startup, in order to change from one DiskOnModule to another.  It is unclear if this emergency serial connection actually works, however, due to an issue in its wiring over the buried surface cables.  There have not been any serious problems in the String PC's to date and none of the envisioned emergency resources have been necessary.

\subsubsection{RAM disks}

While physical flash disks are used to store the system for each String PC, flash disks are known to survive a relatively small number of write cycles.  Therefore they are rarely written to, and in particular we do not write acquired data to them.  Instead we have a 100~MB RAM disk running on each String PC.  The RAM disk maps an area of RAM such that it looks to the operating system like a standard disk.  The String PC's have 256~MB of RAM, so $\sim$156~MB remain for use as standard RAM.  The String PC RAM is expected to survive many more write cycles than the flash disks, so this strategy was chosen to maximize the life of the String PC.  In standard operation, the RAM disk acts as a temporary buffer where data are written during a run.  One binary file is written per run.  As soon as a run is completed, the data are automatically transferred to the Master PC, compressed as they are transferred in order to save String PC - Master PC bandwidth and Master PC disk space.  The 100~MB RAM disk limits the total amount of data that can be acquired in any one run by any one string to 100~MB.  This is not a severe constraint in standard operation because the quota for for total SPATS data satellite transfer to the North is 150~MB per day.


\subsubsection{Fast analog input/output module}

There are three high-speed analog input/output modules per String PC (RTD model IDAN-SDM7540HR-8-68S).  These boards have both analog-to-digital (ADC) and digital-to-analog (DAC) channels.  Each module supports an ADC bandwidth of 1.25~megasamples per second, which can be divided arbitrarily among 16 single-ended or 8 differential channels, and can be configured as needed in software for each run (we use between 1 and 8 differential channels).  While this is the quoted bandwidth, in practice we have only achieved 200~kilosamples per second (see Chapter~\ref{performanceChapter}.  Digitized samples are stored in an 8 kilosample FIFO, which is read out by a driver.


A valuable feature of these fast ADC boards is the ``SyncBus'' which we have used to connect all three boards per stack together.  This allows a single clock signal to be distributed between the boards.  We use this to distribute a sample clock between the boards, such that multiple channels per string are sampled synchronously.  This is especially valuable because while each board has its own clock the clocks drift relative to one another (and relative to absolute time).  Using only one clock per String PC, and distributing it, simplifies this problem.

\subsubsection{Slow analog input/output board}

There is one low-speed analog input/output module per String PC (RTD model IDAN-6420HR-1-62S). These boards are similar to the fast boards but have a smaller sampling bandwidth and have a 1 kilosample FIFO instead of 8 kilosample.  They also do not have the SyncBus.

\subsubsection{Relay module}

There is one relay module (RTD model IDAN-DM6952HR-62D) per String PC.  This module features two boards, each with 8 electrical relays for a total of 16 relays.  One relay is used to control power to each of the seven transmitter modules independently, and one relay is used to control power to each sensor module independently.  One relay is also used for the pressure sensor in the bottom transmitter module of each string.  The final relay simultaneously controls power to the six temperature sensors in the remaining transmitter modules.  The ability to specifically power up/down only the components we need at a particular time enables us to save power.  It also gives us the capability of isolating and disabling particular channels in the event of severe damage such as power shorts which could otherwise threaten the entire string.



\subsection{DSL communictions}

A pair of ethernet extenders (Nexcomm NM220GKIT) is used to provide DSL communications between each string and the Master PC.  Four DSL boxes are connected to the Master PC, and one is in each of the AJB's connected to the String PC.  Each box converts an ethernet signal to a DSL signal that is routed over the surface cable.  The communication speed can be configured via DIP switches on the accessible (Master PC) side to be one of a set of speeds between 0.2 and 2.3~Mbps.  2.3~Mbps is currently used for all strings.  This bandwidth is available in each of the two directions.  We use encrypted communication protocols (SSH and SCP) for logging into the String PC's and transferring data from them, both in manual and autonomous data taking.  The encryption overhead results in an effective data transfer speed somewhat lower than the communication speed.  This speed limit is one of two bottlenecks in data acquisition and limits e.g. the total channel multiplicity and sampling frequency that can be used in pinger data taking, where the strategy is to acquire raw recordings for as long as possible from as many channels as possible simultaneously.

\subsection{GPS timing}

A Meinberg GPS clock (PCI board) is installed in the Master PC and connected to a GPS antenna on the roof of the IceCube Laboratory.  The GPS clock produces an IRIG-B timing signal which is distributed via the SHSB's over the surface cables to the String PC's.  The String PC's sample the IRIG-B signal synchronously with the transmitter and sensor data in order to determine the absolute time of each sample of each waveform with $\sim$10~$\mu$s precision.  IRIG-B is a standard digital timing signal that operates at 100 pulses per second.  The rising edge of each pulse is aligned to absolute GPS time.  The width of each ``high'' pulse encodes binary digits such that all 100 pulses of each one-second frame together specify the day of hear, hour, minute, and second at the first rising edge of the frame.




\section{Retrievable pinger}

A retrievable pinger was developed in 2007 and deployed in the 2007-2008 season.  The pinger, along with String D, was developed in order to measure the attenuation length after we were unable to do so with the first three strings.  The pinger was operated in each of six IceCube holes in the 2007-2008 season.  An upgraded version was operated in each of four IceCube holes in the 2008-2009 season.  The pinger was deployed in each water-filled hole just after the IceCube drill was removed, just before an IceCube string was deployed into it.  Details about the design of the retrievable pinger are given in Chapter~\ref{soundSpeedChapter} and also in ~\cite{Tosi09}.

In the 2007-2008 season, the horizontal distances between SPATS strings and holes in which the pinger were operated ranged between 124.3~m and 543.0~m.  In the 2008-2009 season, a much larger range of distances was achieved: 156.6~m to 1023.4~m.  The horizontal distances between SPATS strings and holes in which the pinger was operated are given in Table~\ref{pingerDistances}.

\subsection{Version 1 (2007-2008 season)}

\begin{figure}
\begin{center}
\subfigure[]{
\label{0708}
\noindent\includegraphics[width=20pc]{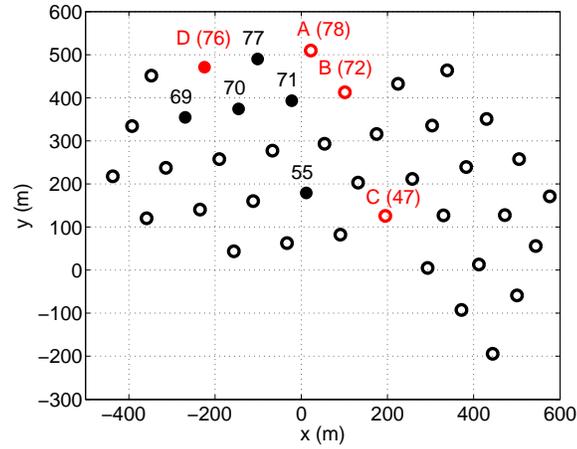}
}
\subfigure[]{
\label{0809}
\noindent\includegraphics[width=20pc]{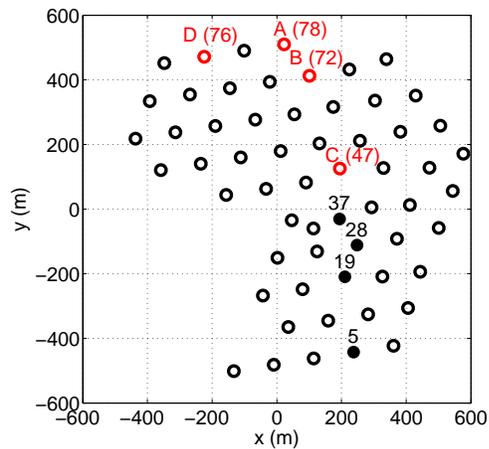}
}
\caption[Surface layout of pinger holes for both pinger deployment seasons]{Surface layout of pinger deployments.  Figure~\subref{0708} shows the 40 IceCube strings deployed by the end of January 2008, with the six holes pinged in the 2007-2008 season indicated by filled circles.  Note that the pinger was deployed and retrieved in Hole 76 just before deploying String D into it.  Figure~\subref{0809} shows the 59 IceCube strings deployed by the end of January 2009, with the four holes pinged in the 2008-2009 season indicated by filled circles.  In both plots, the numbers are IceCube hole ID's and the letters are SPATS string ID's.}
\label{pinger_geometry}
\end{center}
\end{figure}

The first version of the pinger was operated in six water-filled IceCube holes, as shown in Figure~\ref{pinger_geometry}.  The pinger was free to swing, twist, and bounce.  This resulted in significant pulse-to-pulse and run-to-run variation, and in significant shear wave production.

\subsection{Version 2 (2008-2009 season)}

The second version of the pinger was operated in four water-filled IceCube holes, as shown in Figure~\ref{pinger_geometry}.  For this version, centralizing spring ribs were added to the pinger assembly to keep it stable in the center of the IceCube hole during operation, in order to keep the angle of incidence with the hole wall constant and reduce pulse-to-pulse variations, run-to-run variations, and shear wave production.

\begin{table}[tbp]
\centering
\caption[Distances between pinger holes and SPATS strings]{Horizontal distances between pinger holes and SPATS strings.  The pinger was deployed in the first six holes in the 2007-2008 season (pinger Version 1), and in the last four holes in the 2008-2009 season (pinger Version 2).  Distances are in meters.  In the 2007-2008 season the horizontal distances ranged from 124.3~m to 543.0~m.  In the 2008-2009 season the horizontal distances ranged from 156.6~m to 1023.4~m.  In addition to the distances, the date that the pinger was operated in each hole is given.}
\centering
\begin{tabular}{| c | c | c | c | c | c |}  
\hline
\bf{Date} & \bf{Pinger hole} & \bf{String A} & \bf{String B} & \bf{String C} & \bf{String D} \\
\hline
December 15 2007 & 55 & 330.5 & 250.0 & 190.8 & 375.2 \\
\hline
December 18 2007 & 71 & 124.3 & 124.6 & 344.6 & 216.5 \\
\hline
December 21 2007 & 70 & 215.3 & 249.5 & 421.6 & 124.6 \\
\hline
December 24 2007 & 76 & 249.2 & 330.3 & 543.0 & (0) \\
\hline
December 28 2007 & 77 & 124.7 & 216.4 & 469.7 & 124.5 \\
\hline
January 2 2008 & 69 & 329.8 & 374.5 & 517.2 & 124.9 \\
\hline
December 12 2008 & 28 & 661.2 & 544.9 & 243.3 & 750.1 \\
\hline
December 16 2008 & 19 & 743.5 & 632.1 & 335.7 & 807.5 \\
\hline
December 29 2008 & 5 & 976.0 & 866.1 & 569.6 & 1023.4 \\
\hline
January 18 2009 & 37 & 567.2 & 453.4 & 156.5 & 653.4 \\
\hline
\end{tabular} 
\label{pingerDistances}
\end{table} 







\chapter{Performance of SPATS}

\noindent\emph{In this chapter we summarize the performance of the South Pole Acoustic Test Setup.  All transmitters and nearly all sensor channels are in good health.  The system as a whole continues to perform well, more than two years after being deployed and operated nearly continuously in harsh conditions.  In particular, the system has survived several unplanned power outages.  The String PC's recovered from power outages lasting up to $\sim$48~hours, enough time for the computers to equilibrate to the -51~$^{\circ}$C ambient temperture 2~m below the ice surface.  Our ADC boards have presented some technical challenges which we summarize.  The transmitters of the first three strings have some unexplained behavior, while the transmitters of the fourth string are behaving as expected.}

\label{performanceChapter}





\section{Transmitters}

\begin{figure}[tbp]
\centering
\noindent\includegraphics[width=20pc]{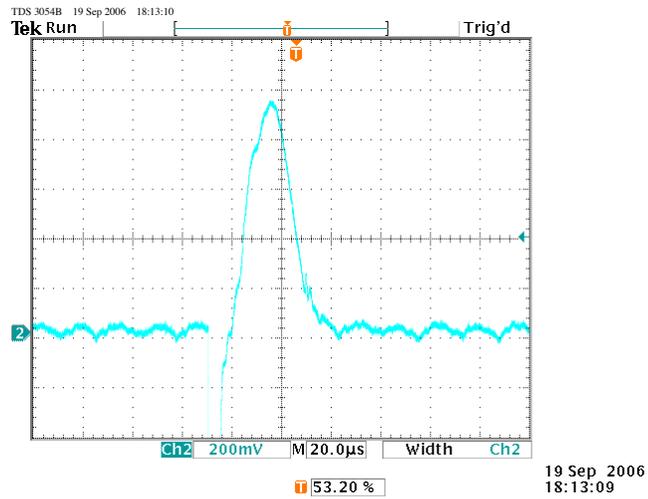}
\caption[HV read-back pulse recorded in laboratory with oscilloscope]{Example HV read-back pulse recorded with an oscilloscope in the laboratory prior to deployment, from a transmitter of the String A/B/C design.  Note the clean, unipolar shape.}
\label{TEK00000}
\end{figure}

\begin{figure}[tbp]
\centering
\noindent\includegraphics[width=25pc]{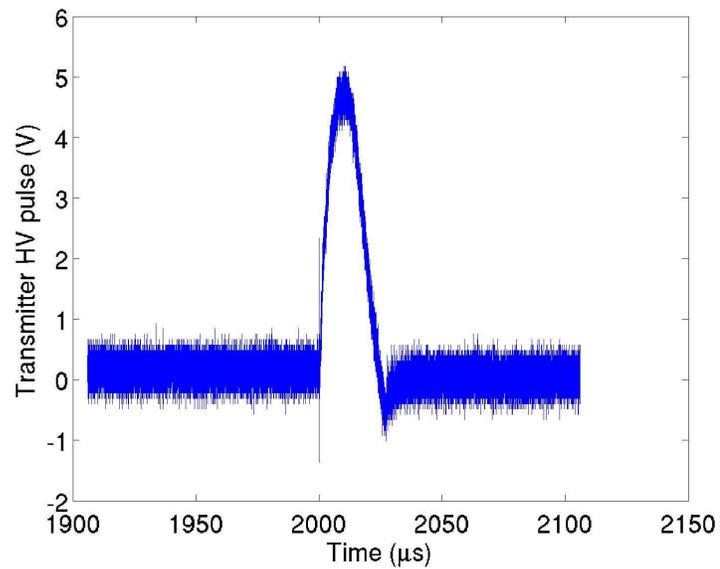}
\caption[HV read-back pulse recorded in laboratory with SPATS DAQ]{Example HV read-back pulse recorded with the SPATS DAQ in the laboratory prior to deployment, from a transmitter of the String A/B/C design.  Note the clean, unipolar shape.}
\label{hv}
\end{figure}

The transmitters of Strings A, B, and C behaved differently after deployment in the ice than prior to deployment in laboratory tests.  The electrical pulse shape generated by the HV pulser and discharged to the emitter was designed to be a simple unipolar pulse, and this shape was verified in the laboratory prior to deployment.  For an example HV read-back pulse recorded with an oscilloscope in the lab from a transmitter before deployment, see Figure~\ref{TEK00000}.  For an example recorded with the SPATS DAQ, see Figure~\ref{hv}.

The pulse shape can also be read out after deployment using the HV read-back feature.  Note that the HV read-back pulse is down-scaled by a factor of 500 in amplitude, relative to the HV pulse delivered to the piezoelectric ceramic, for both the String A/B/C and the String D design.  The transmitters of Strings A, B, and C exhibit complex multi-polar HV read-back pulse shapes after deployment in the ice, in contradiction to the clean unipolar pulses measured in the laboratory prior to deployment.

\begin{figure}
\begin{center}
\subfigure[AT2]{
\noindent\includegraphics[width=14pc]{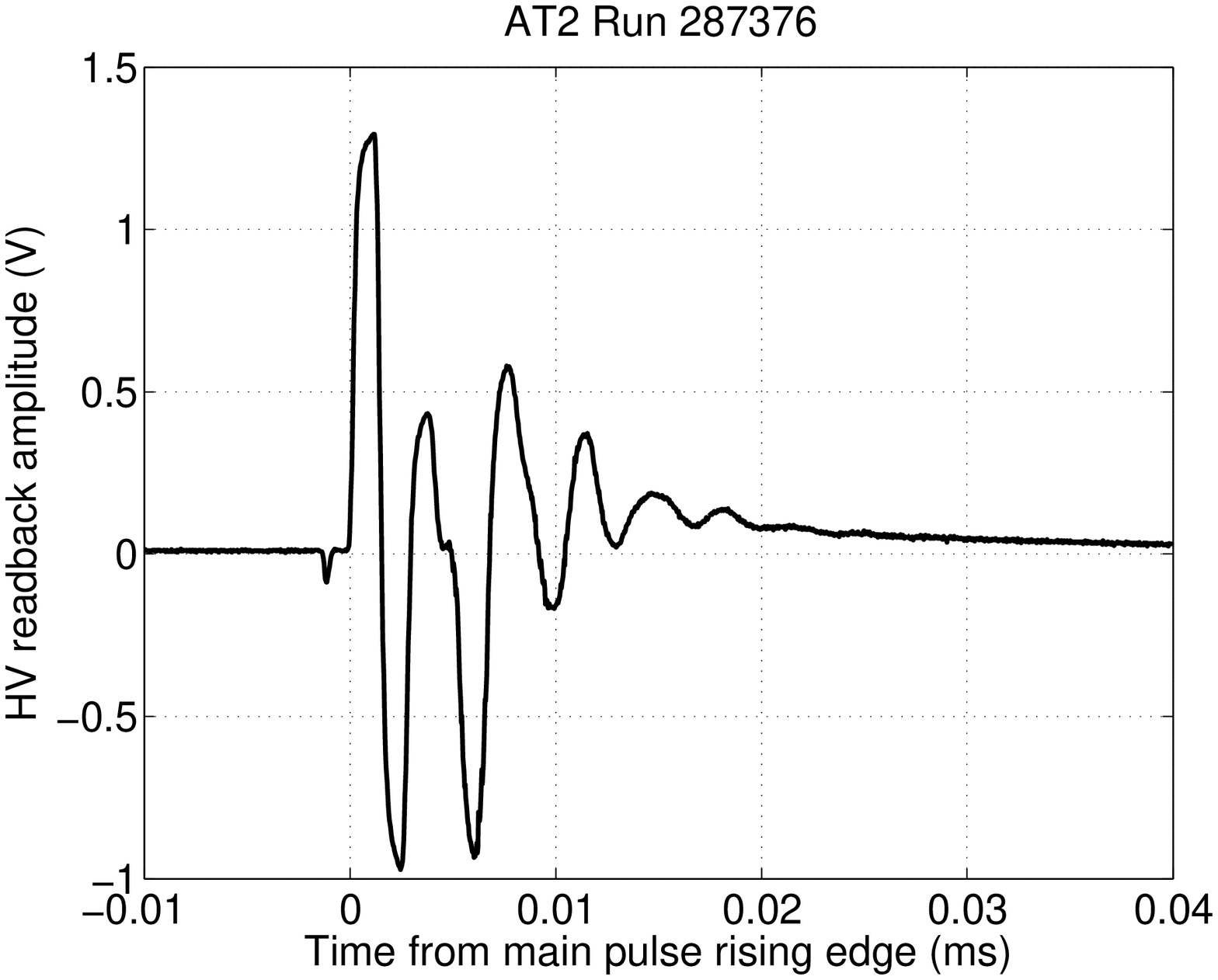}
}
\subfigure[AT3]{
\noindent\includegraphics[width=14pc]{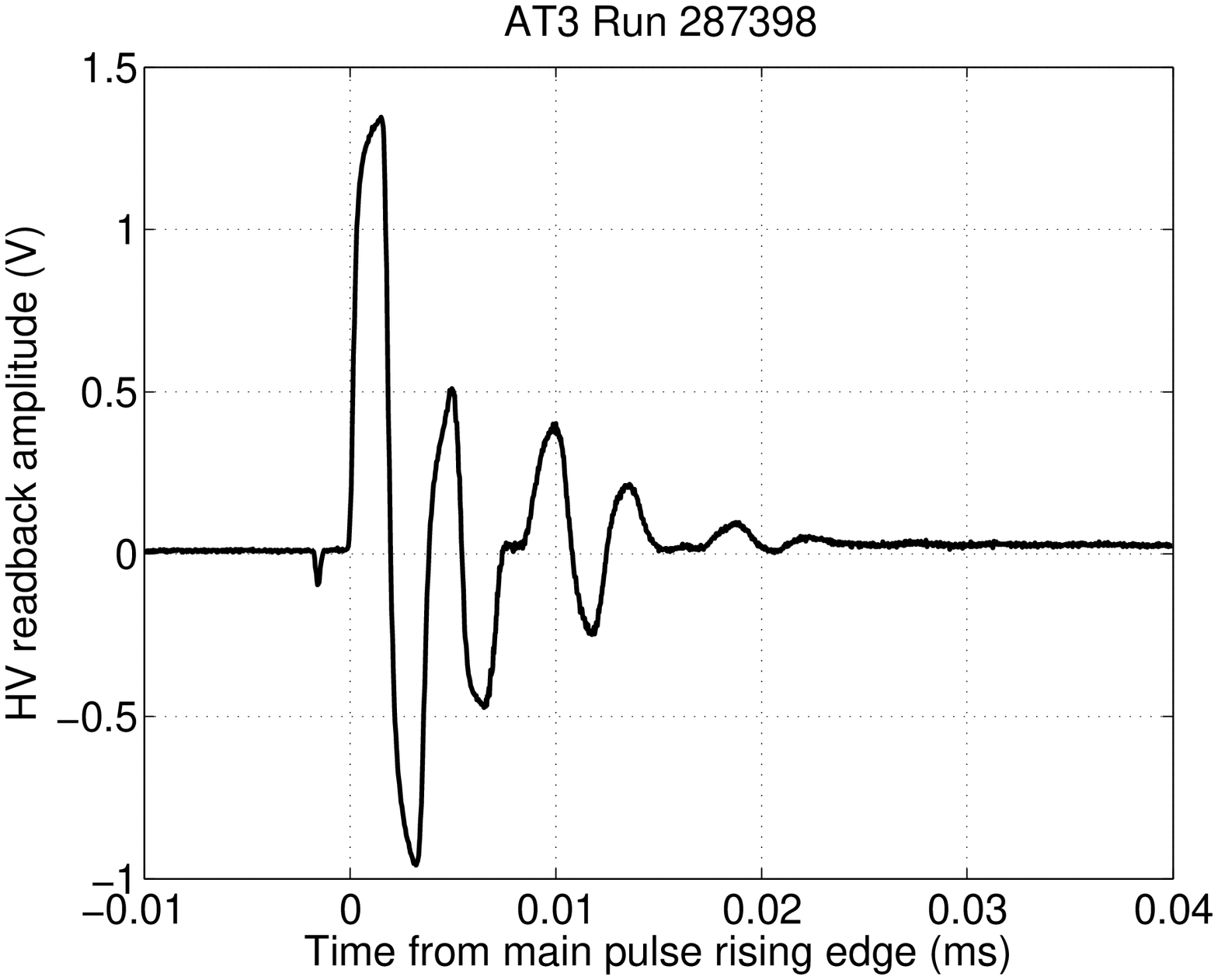}
}
\subfigure[AT4]{
\noindent\includegraphics[width=14pc]{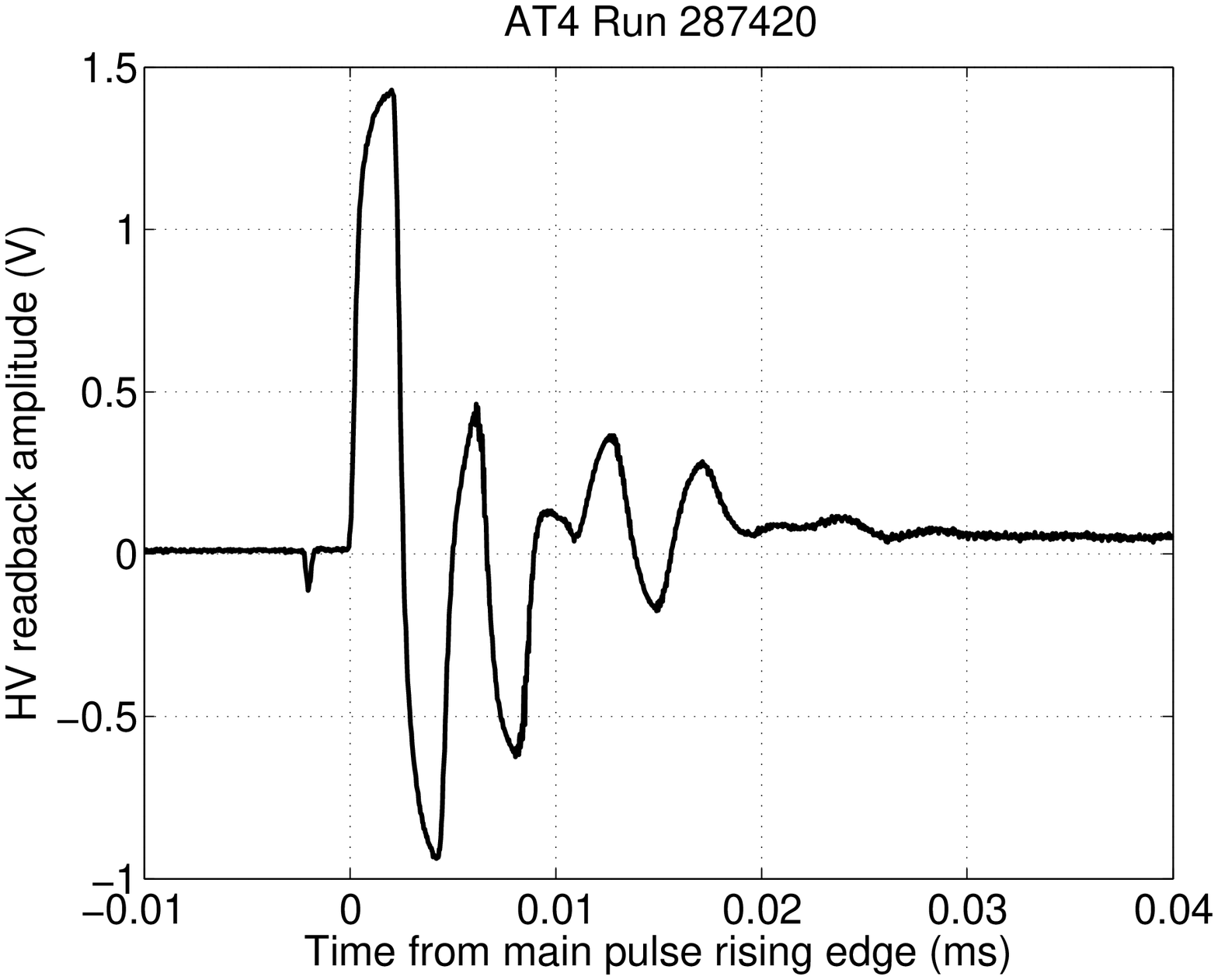}
}
\subfigure[AT5]{
\noindent\includegraphics[width=14pc]{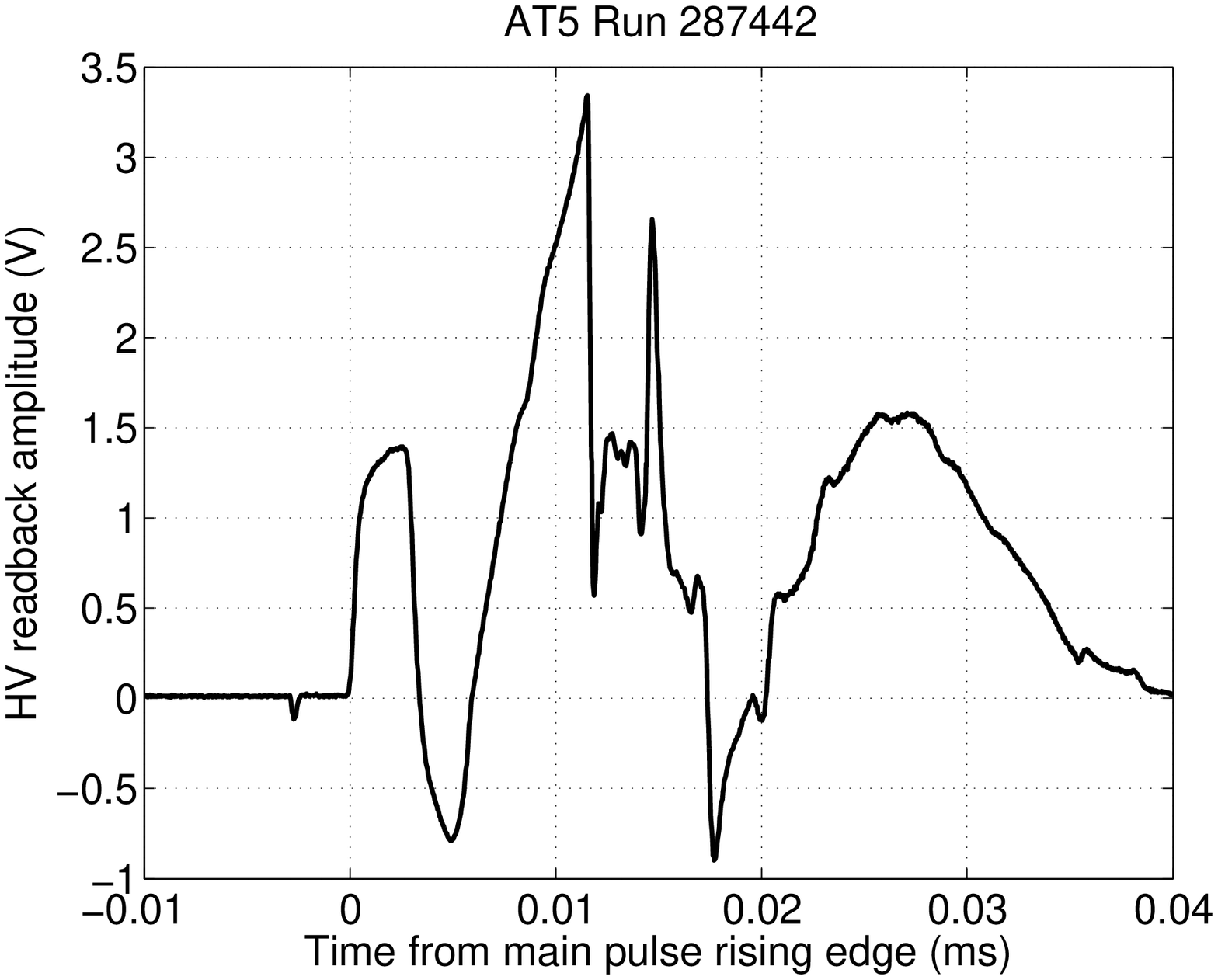}
}
\subfigure[AT6]{
\noindent\includegraphics[width=14pc]{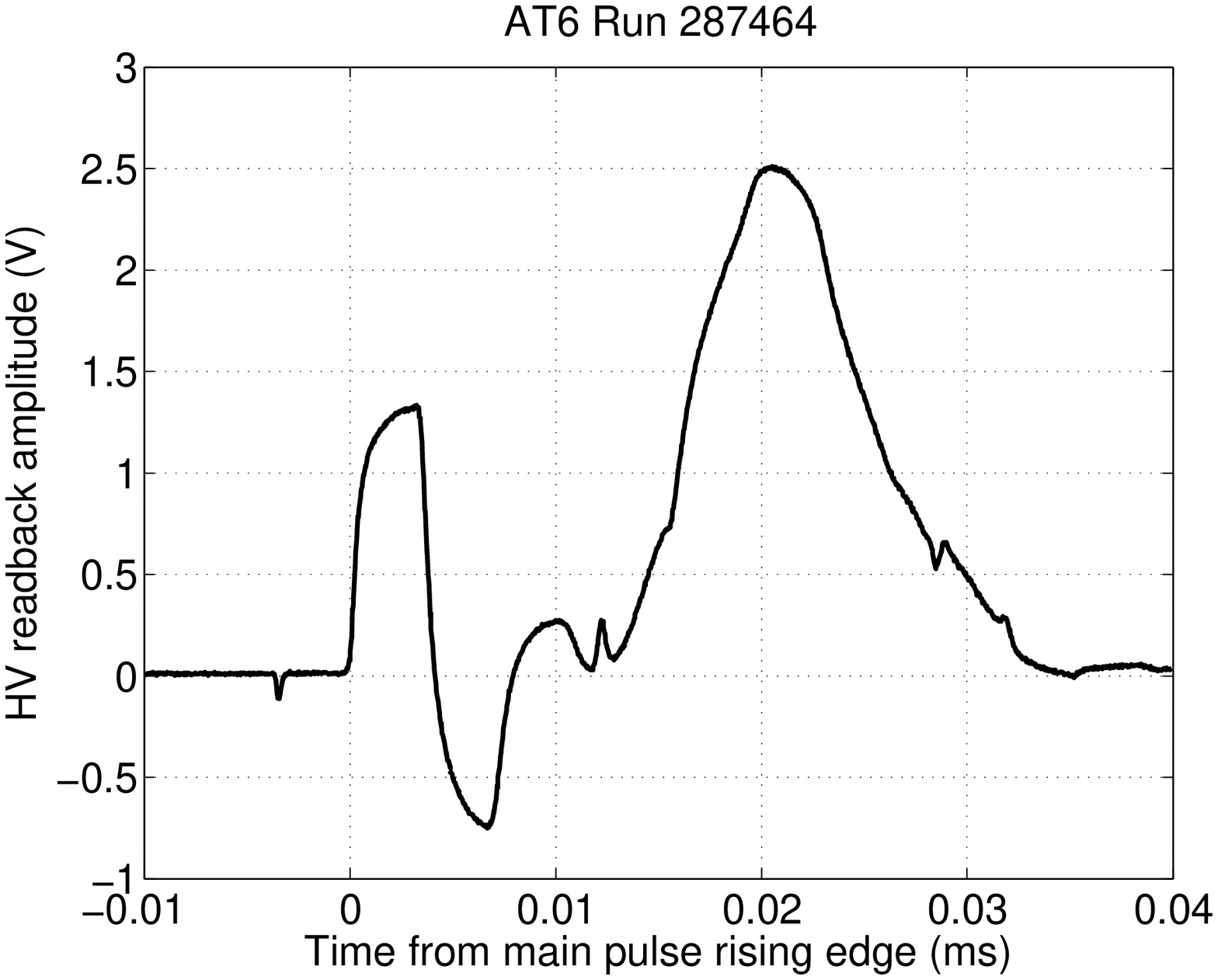}
}
\subfigure[AT7]{
\noindent\includegraphics[width=14pc]{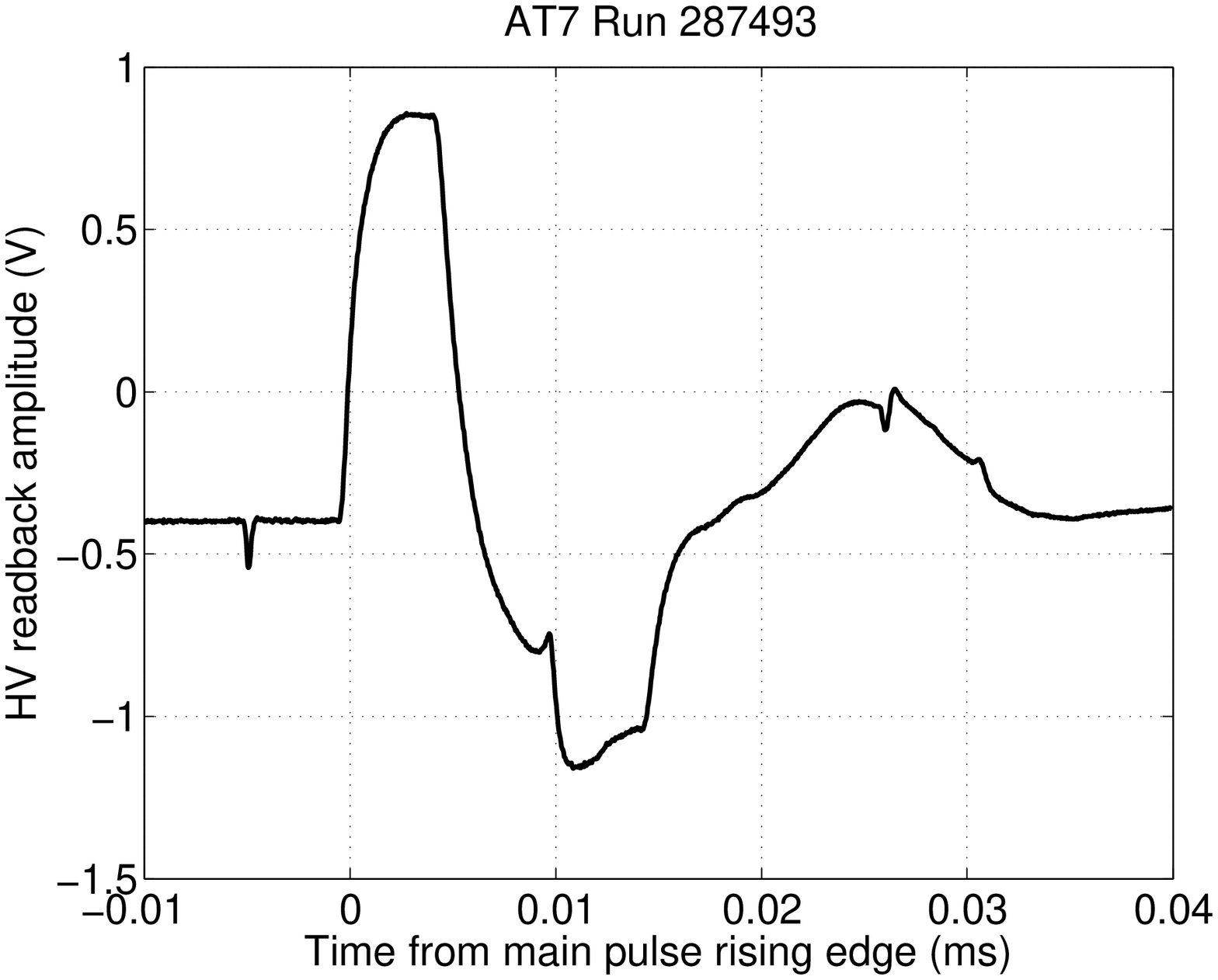}
}
\caption[String A transmitter performance]{HV read-back pulse shape for String A transmitters on Stages 2-7, as determined by overlaying $\sim$10$^3$ individual pulses with clock drift correction.  The shape was cleaner prior to deployment of the transmitters.}
\label{hvrbA}
\end{center}
\end{figure}

\begin{figure}
\begin{center}
\subfigure[BT2]{
\noindent\includegraphics[width=14pc]{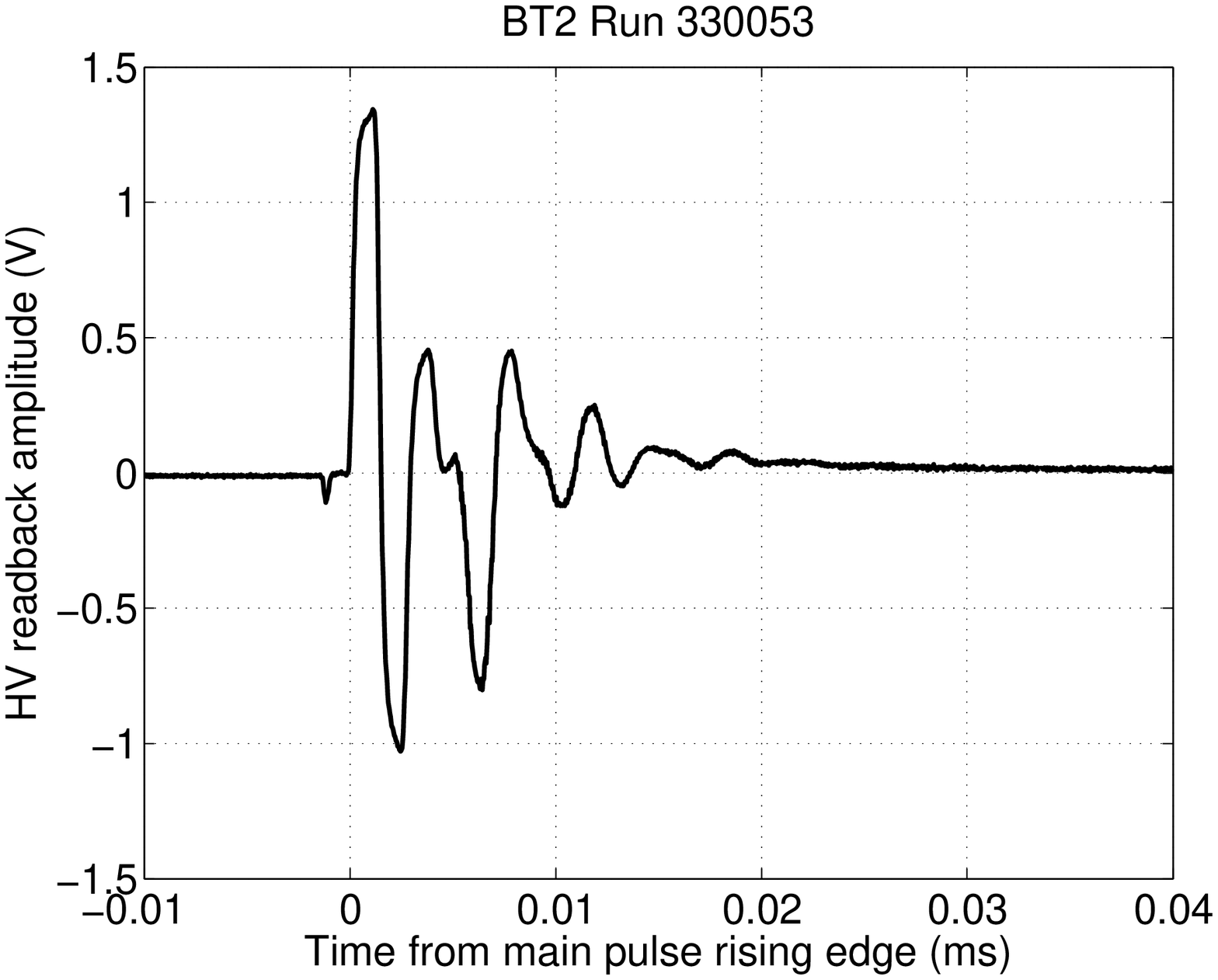}
}
\subfigure[BT3]{
\noindent\includegraphics[width=14pc]{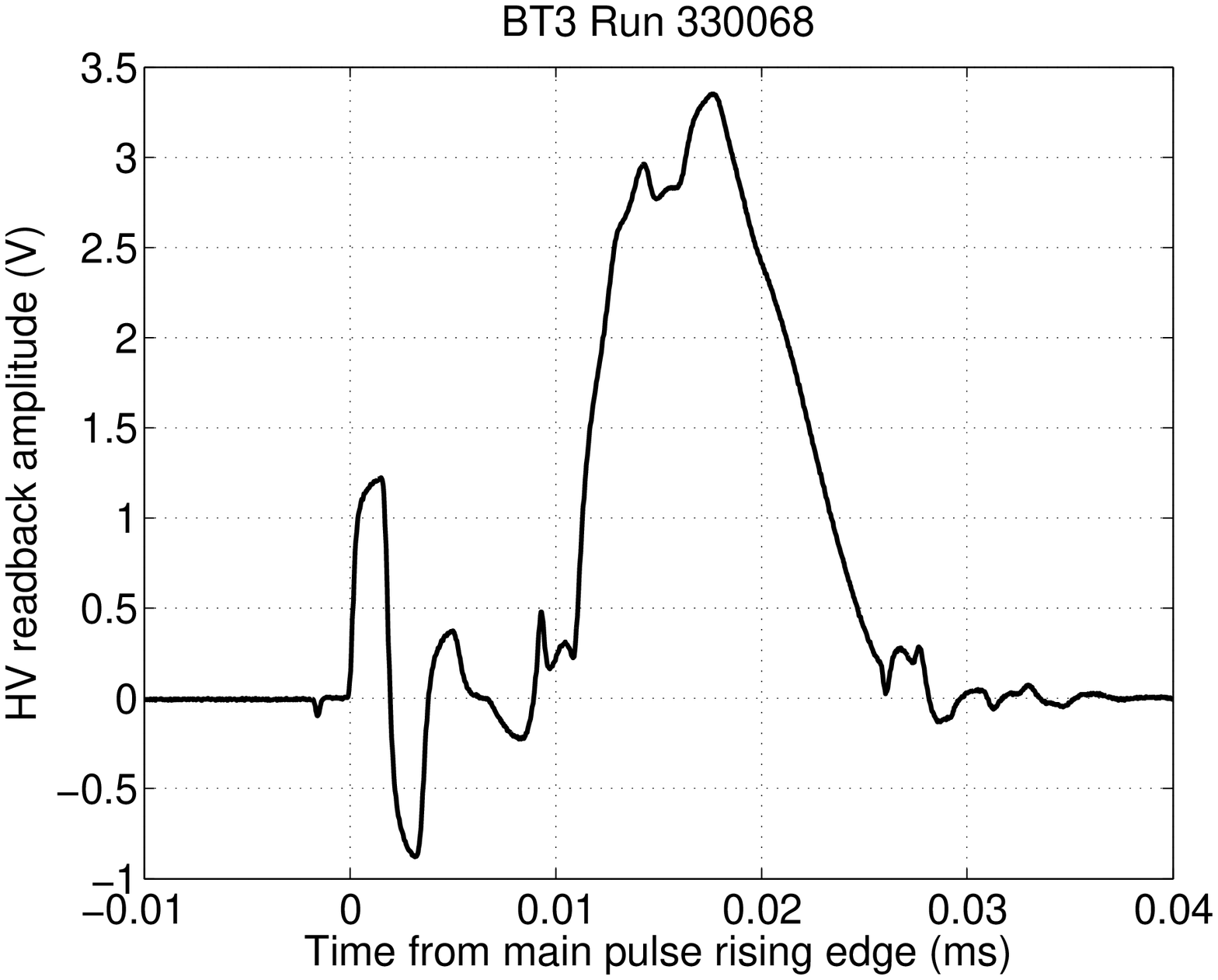}
}
\subfigure[BT4]{
\noindent\includegraphics[width=14pc]{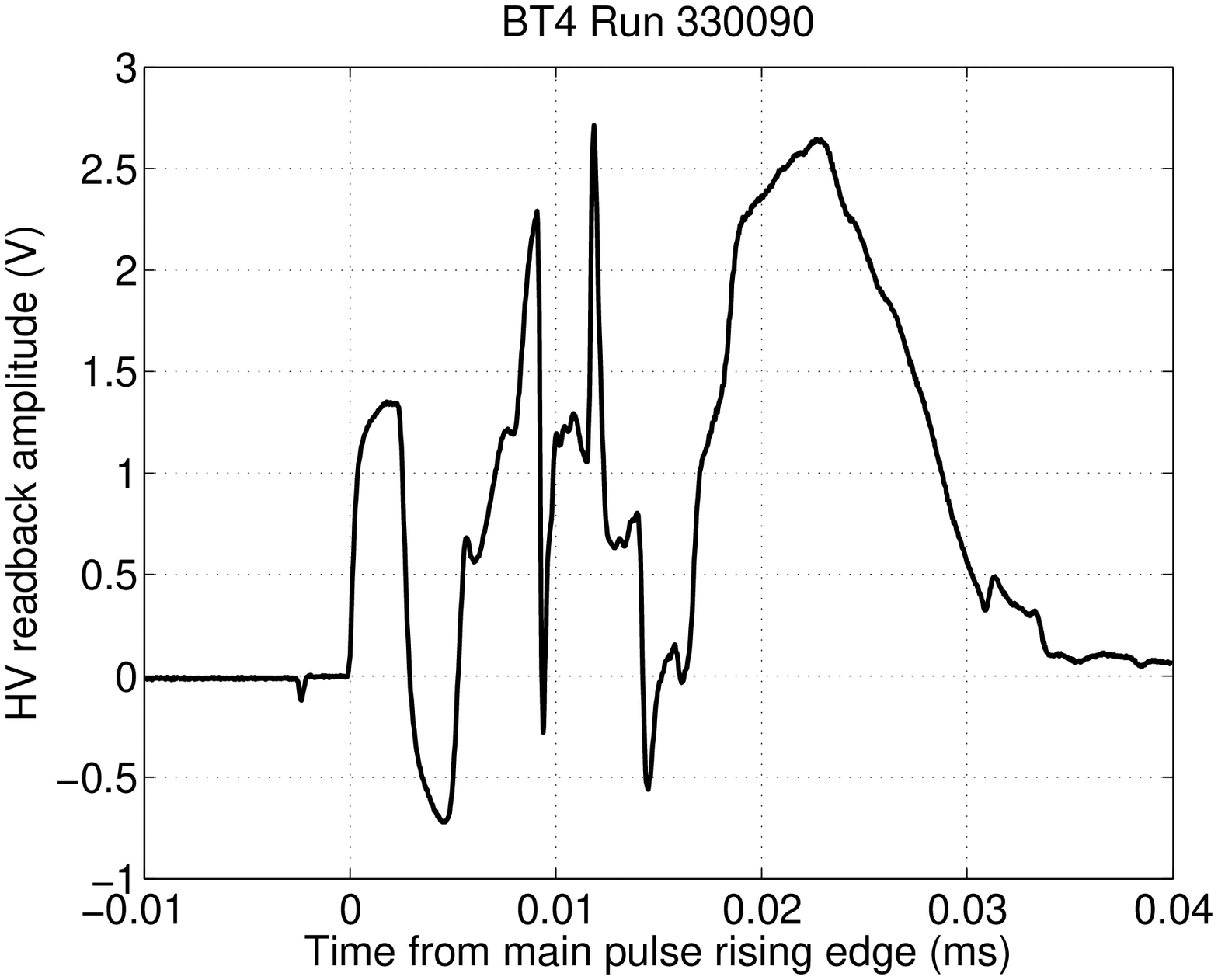}
}
\subfigure[BT5]{
\noindent\includegraphics[width=14pc]{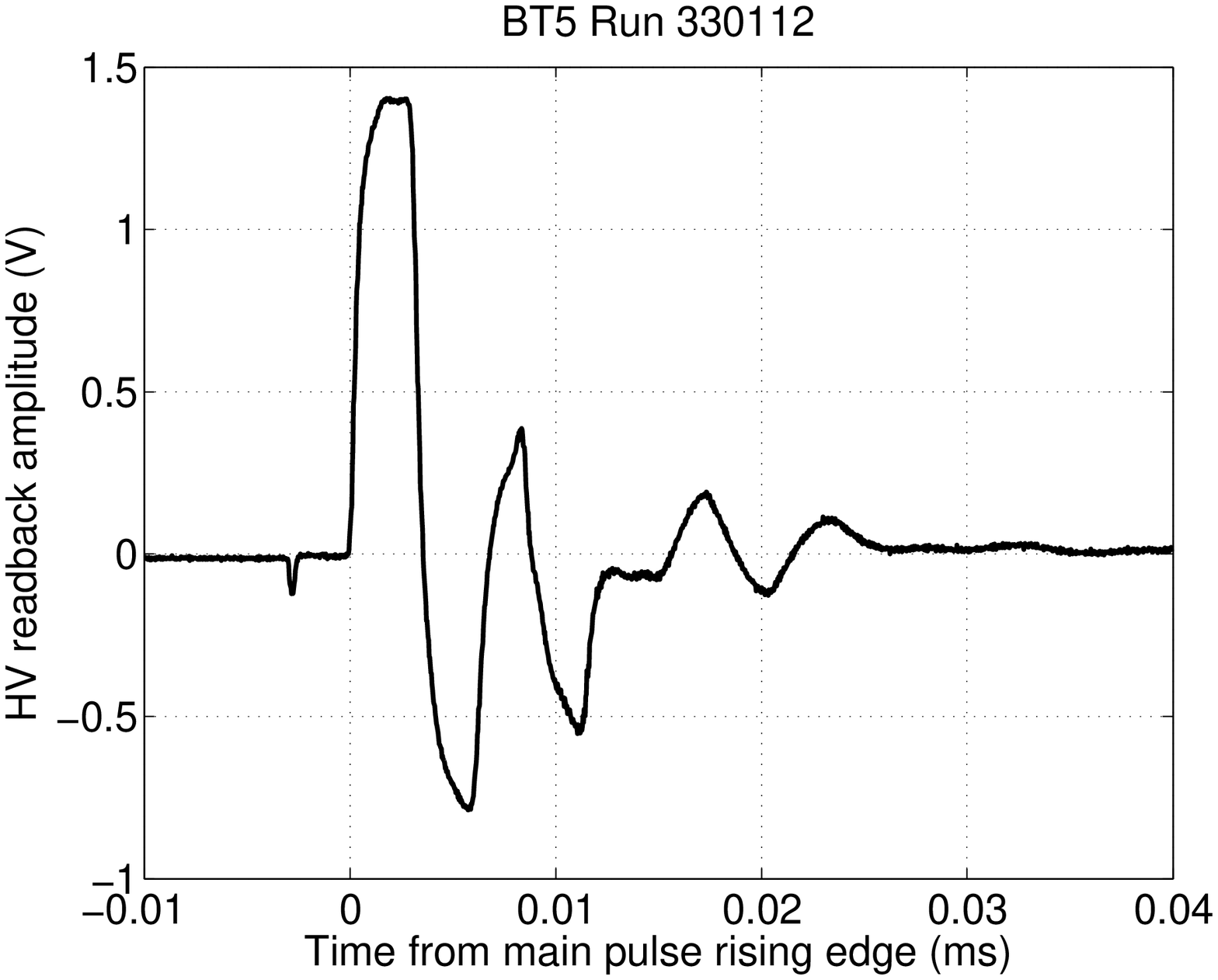}
}
\subfigure[BT6]{
\noindent\includegraphics[width=14pc]{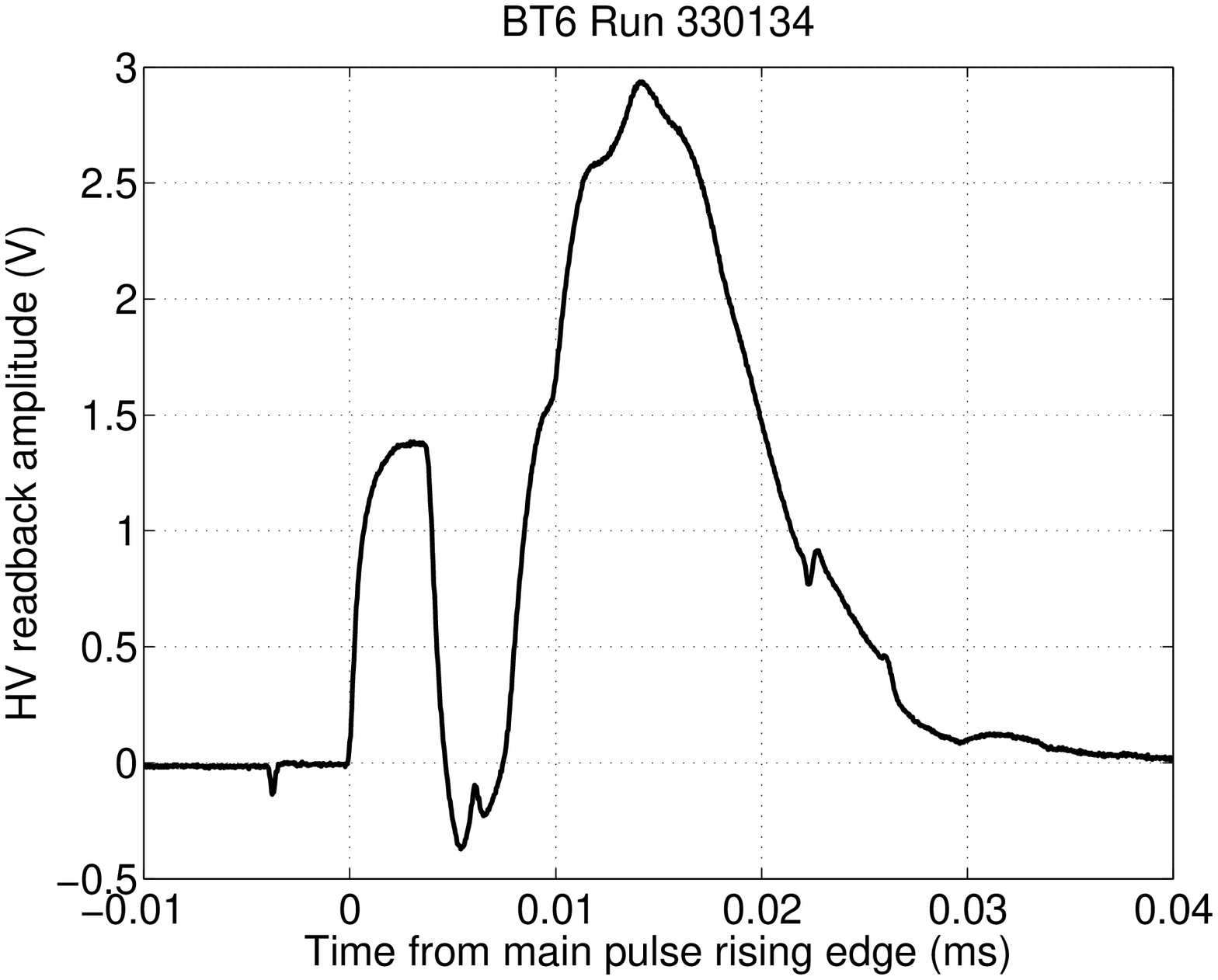}
}
\subfigure[BT7]{
\noindent\includegraphics[width=14pc]{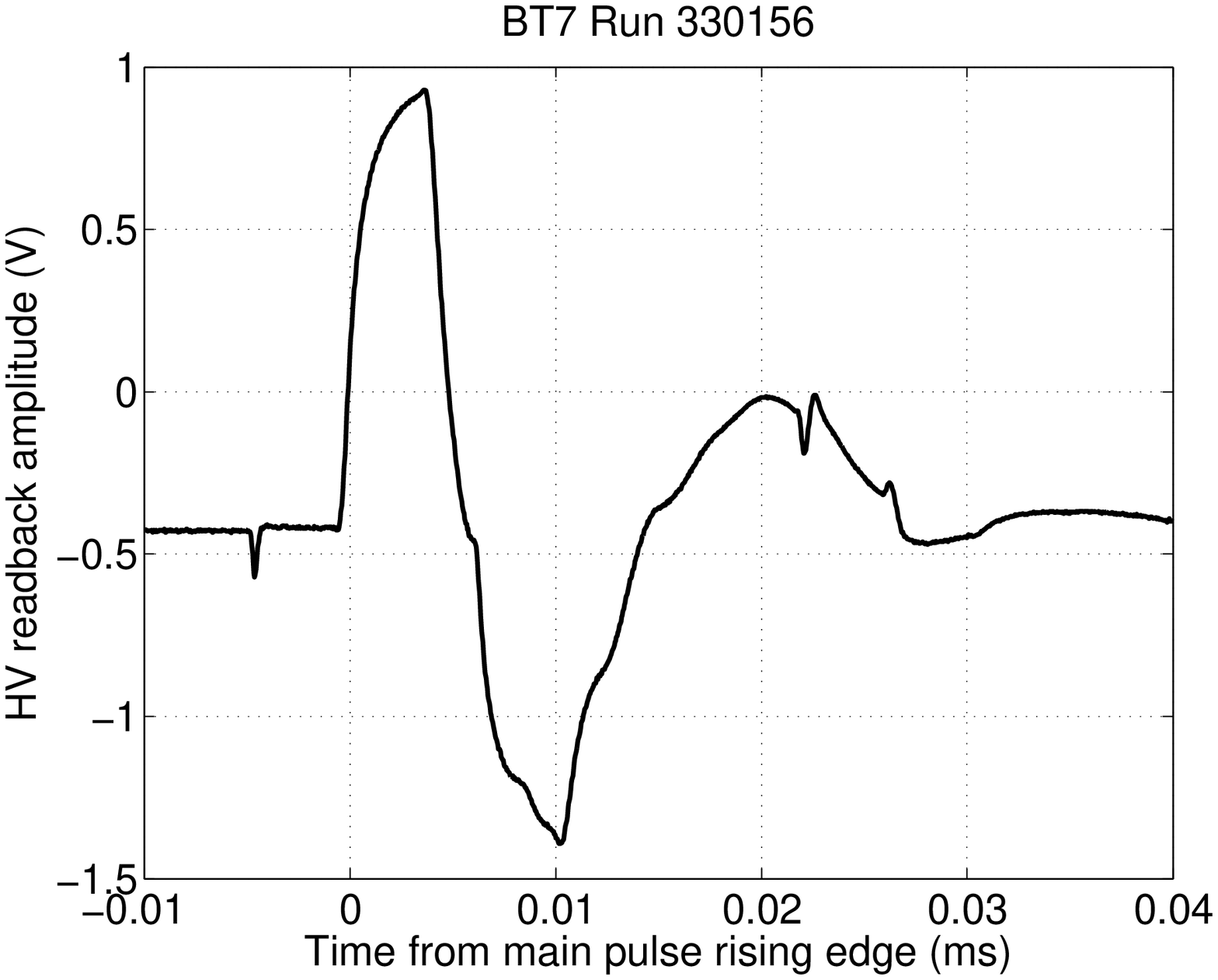}
}
\caption[String B transmitter performance]{HV read-back pulse shape for String B transmitters on Stages 2-7, as determined by overlaying $\sim$10$^3$ individual pulses with clock drift correction.  The shape was cleaner prior to deployment of the transmitters.}
\label{hvrbB}
\end{center}
\end{figure}

\begin{figure}
\begin{center}
\subfigure[CT2]{
\noindent\includegraphics[width=14pc]{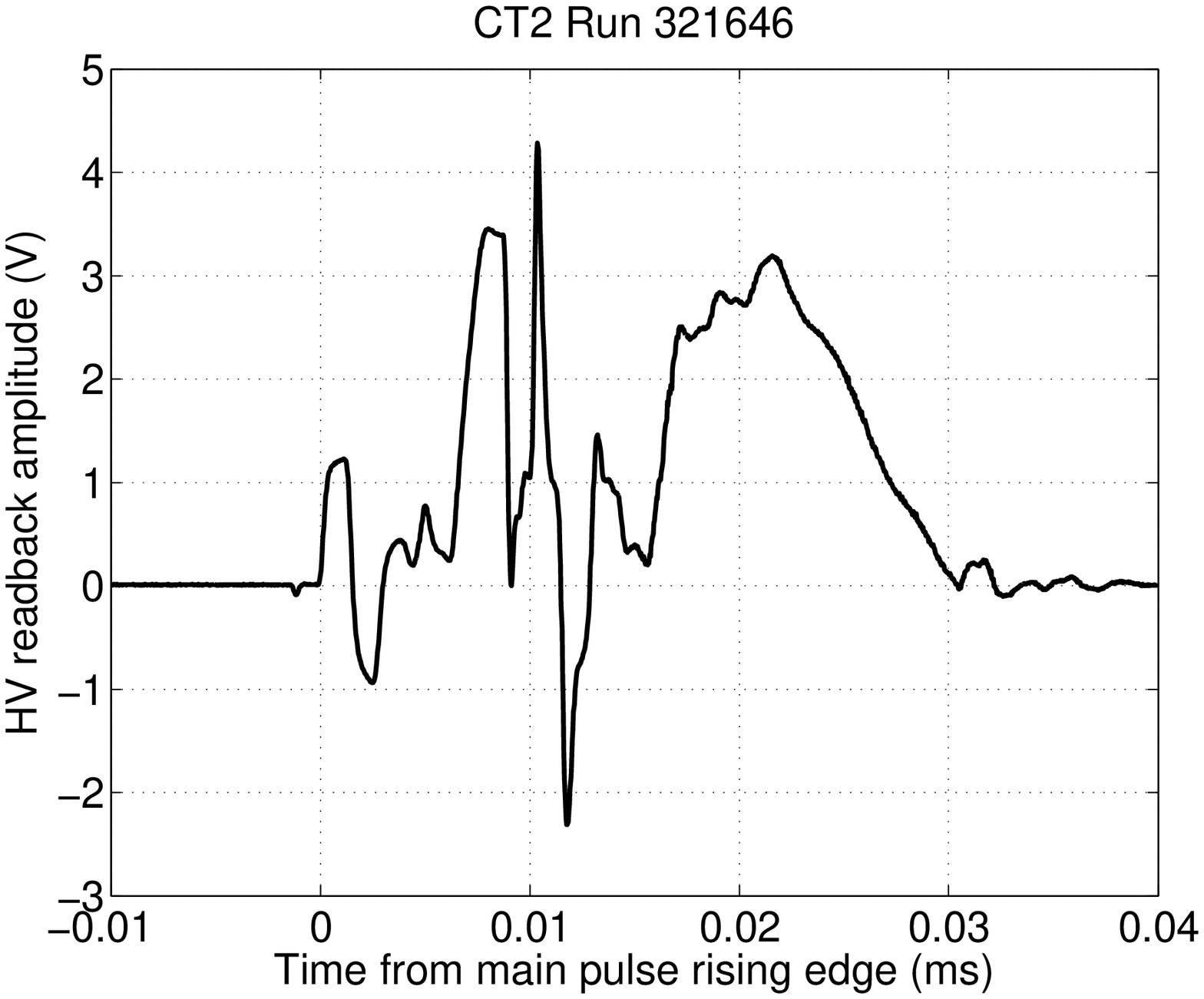}
}
\subfigure[CT3]{
\noindent\includegraphics[width=14pc]{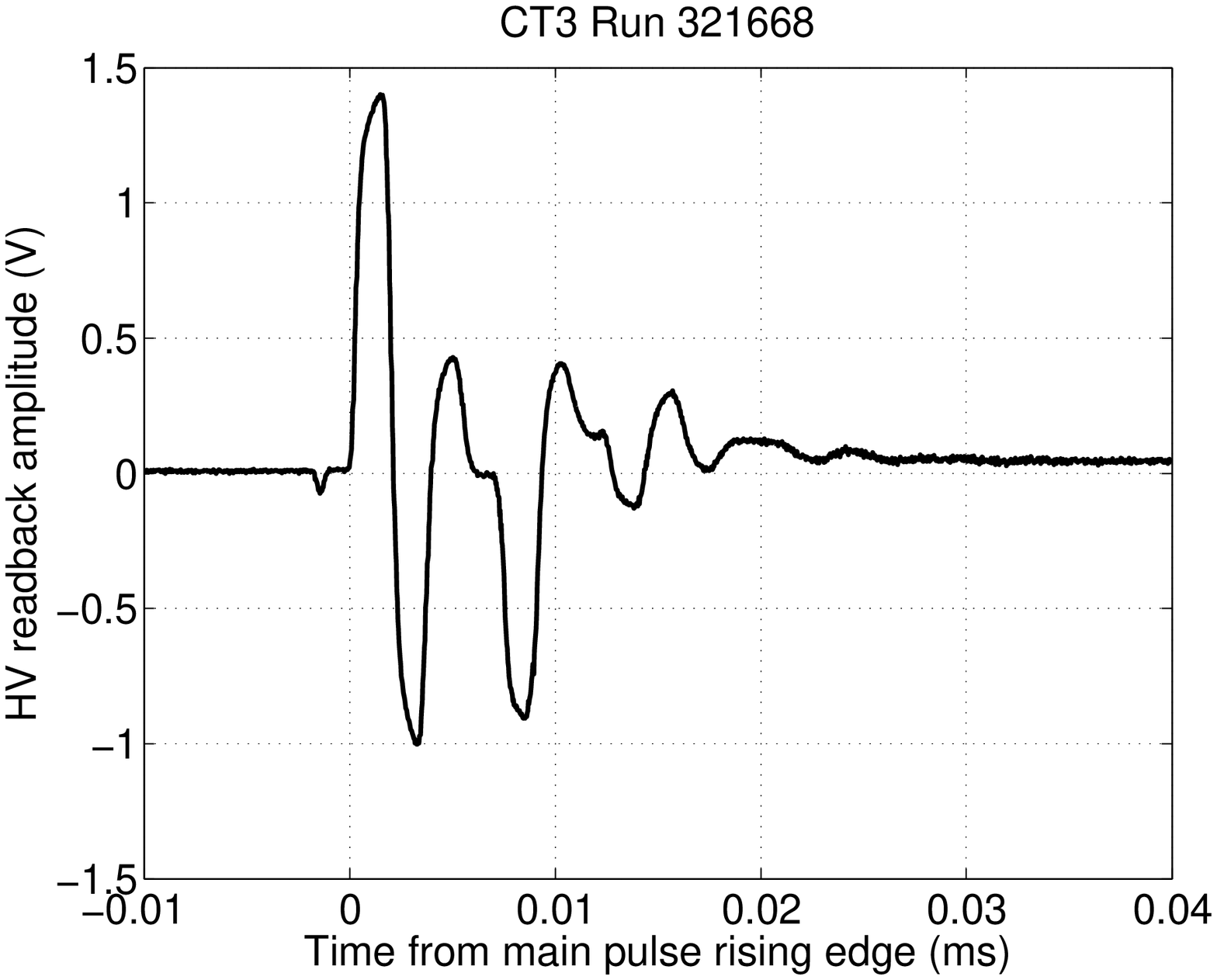}
}
\subfigure[CT4]{
\noindent\includegraphics[width=14pc]{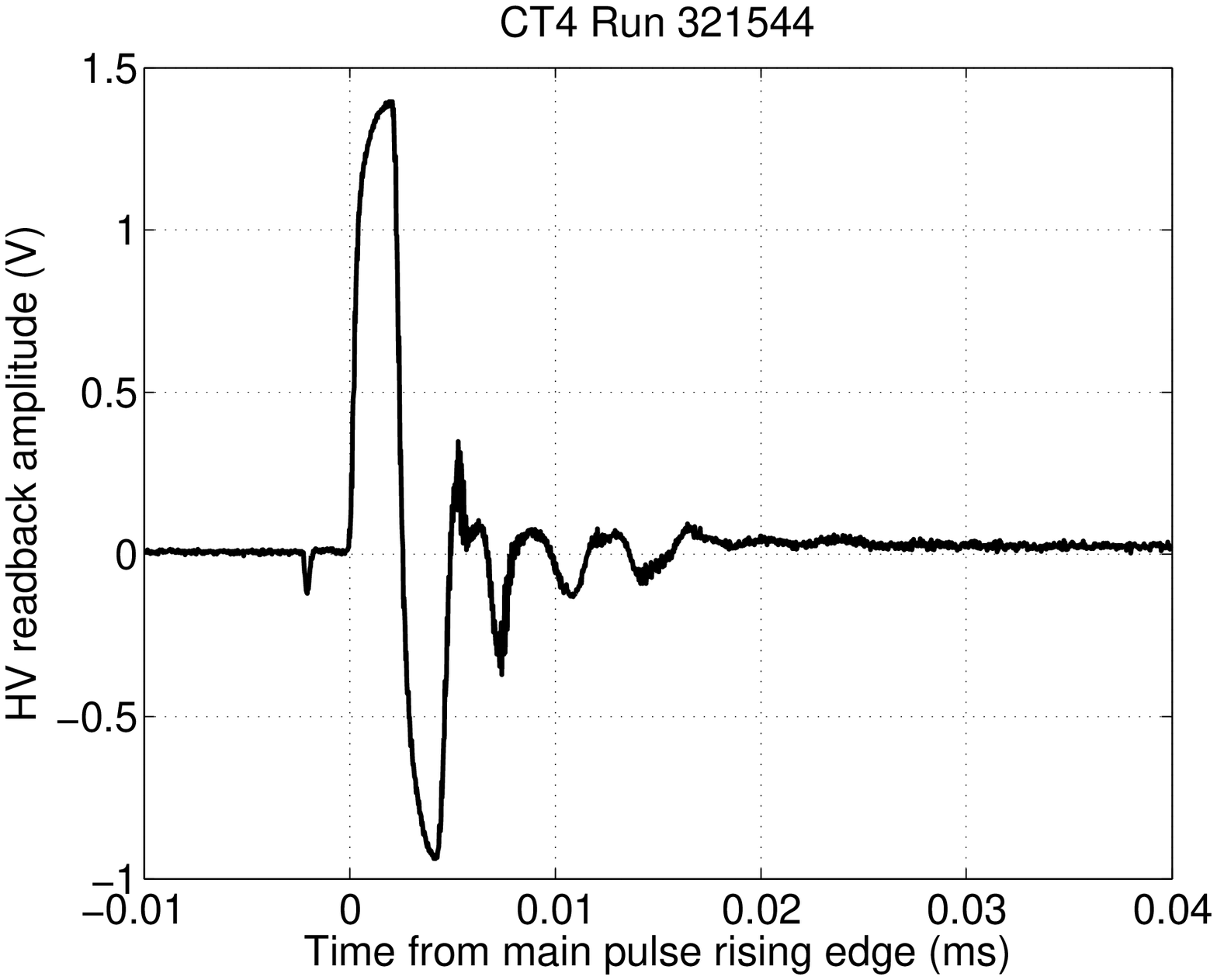}
}
\subfigure[CT5]{
\noindent\includegraphics[width=14pc]{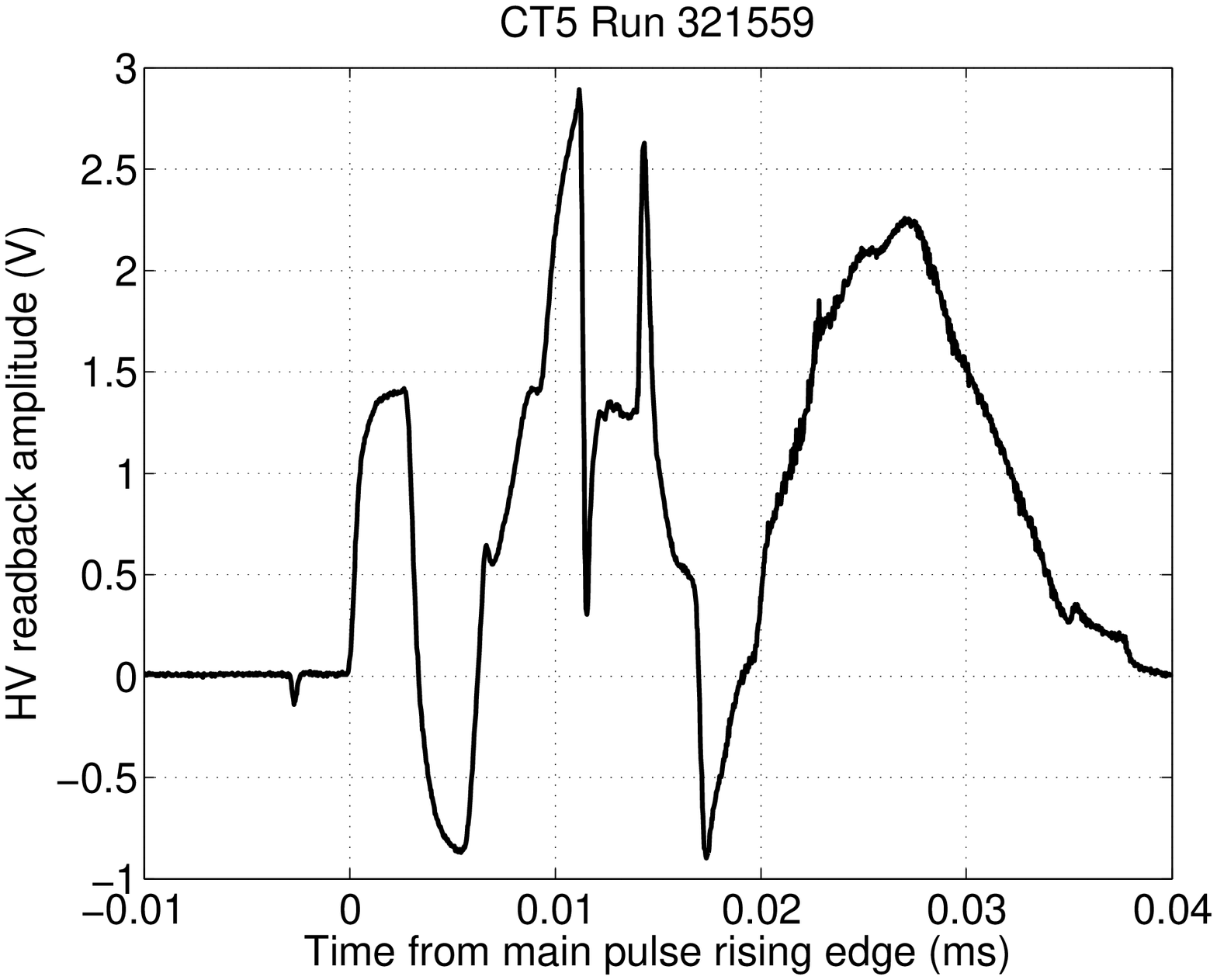}
}
\subfigure[CT6]{
\noindent\includegraphics[width=14pc]{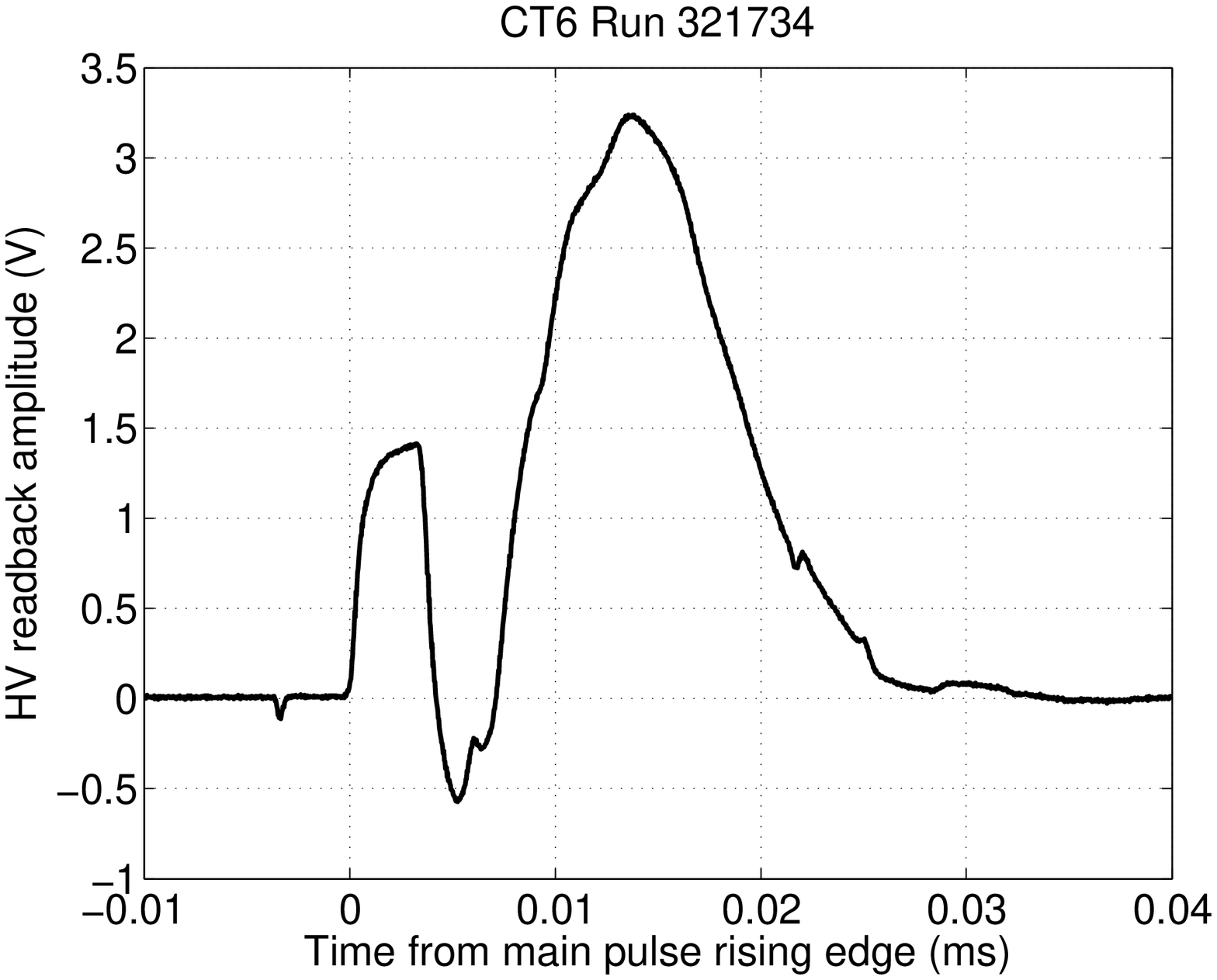}	
}
\subfigure[CT7]{
\noindent\includegraphics[width=14pc]{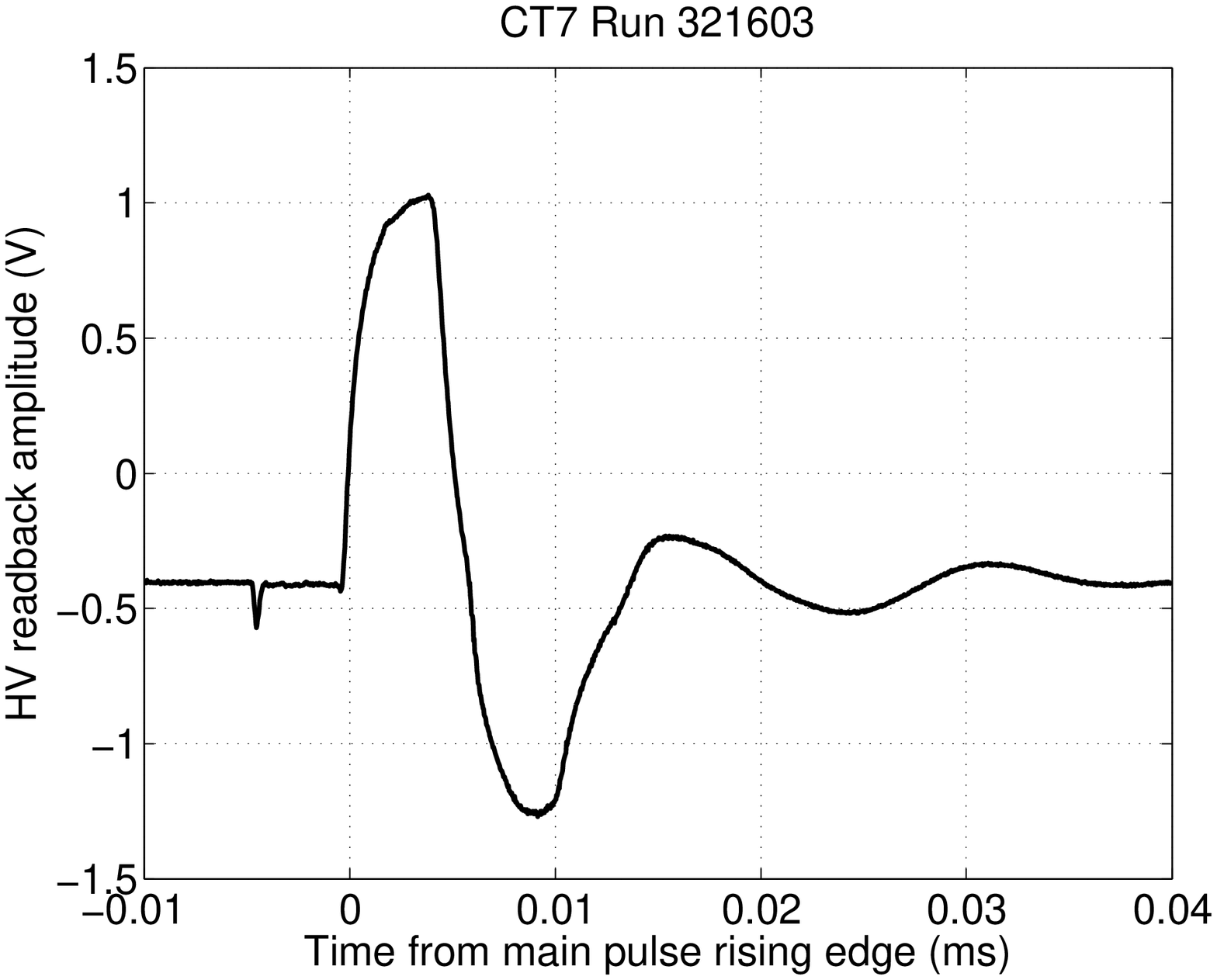}
}
\caption[String C transmitter performance]{HV read-back pulse shape for String C transmitters on Stages 2-7, as determined by overlaying $\sim$10$^3$ individual pulses with clock drift correction.  The shape was cleaner prior to deployment of the transmitters.}
\label{hvrbC}
\end{center}
\end{figure}

\begin{figure}
\begin{center}
\subfigure[DT2]{
\noindent\includegraphics[width=13pc]{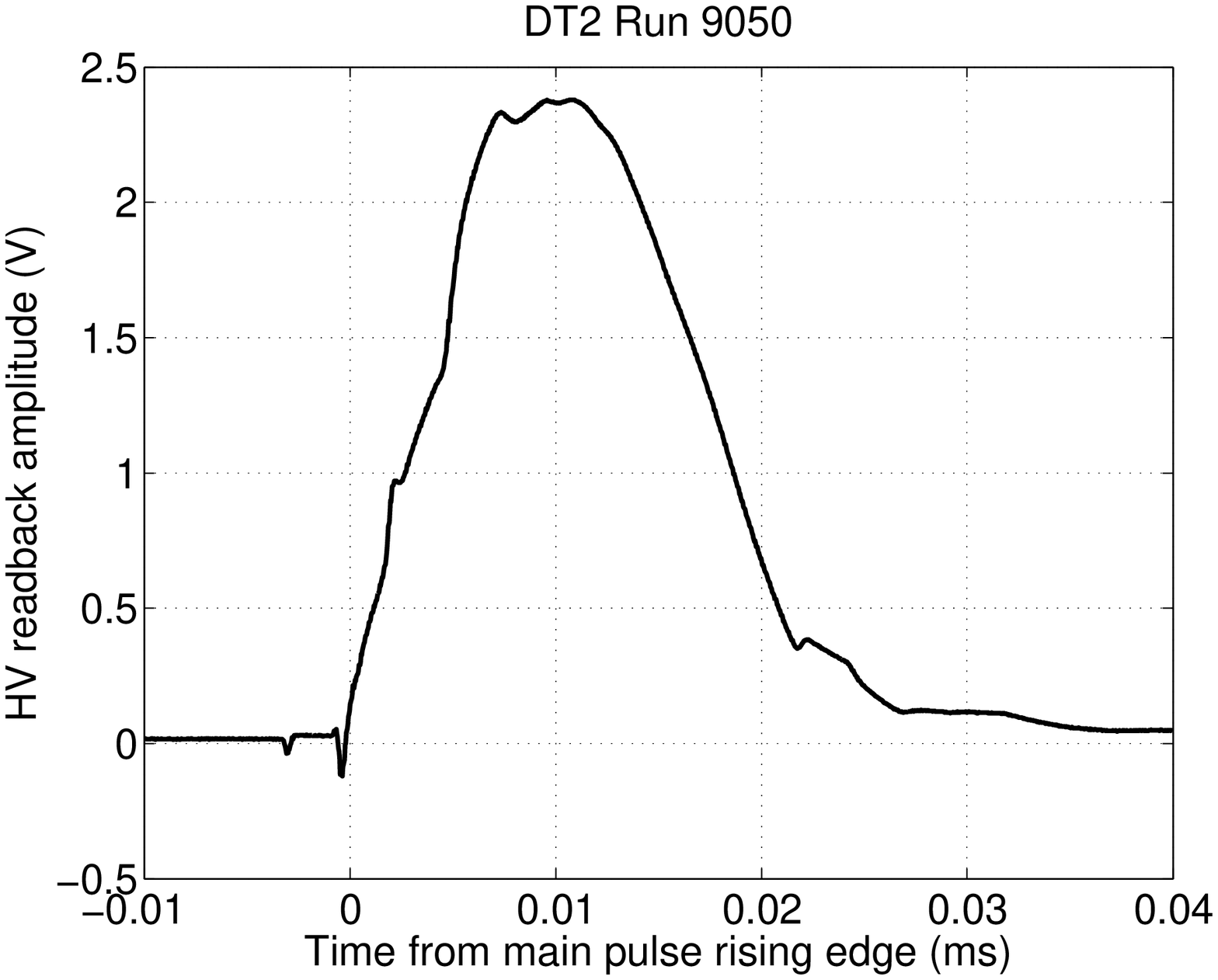}	
}
\subfigure[DT3]{
\noindent\includegraphics[width=13pc]{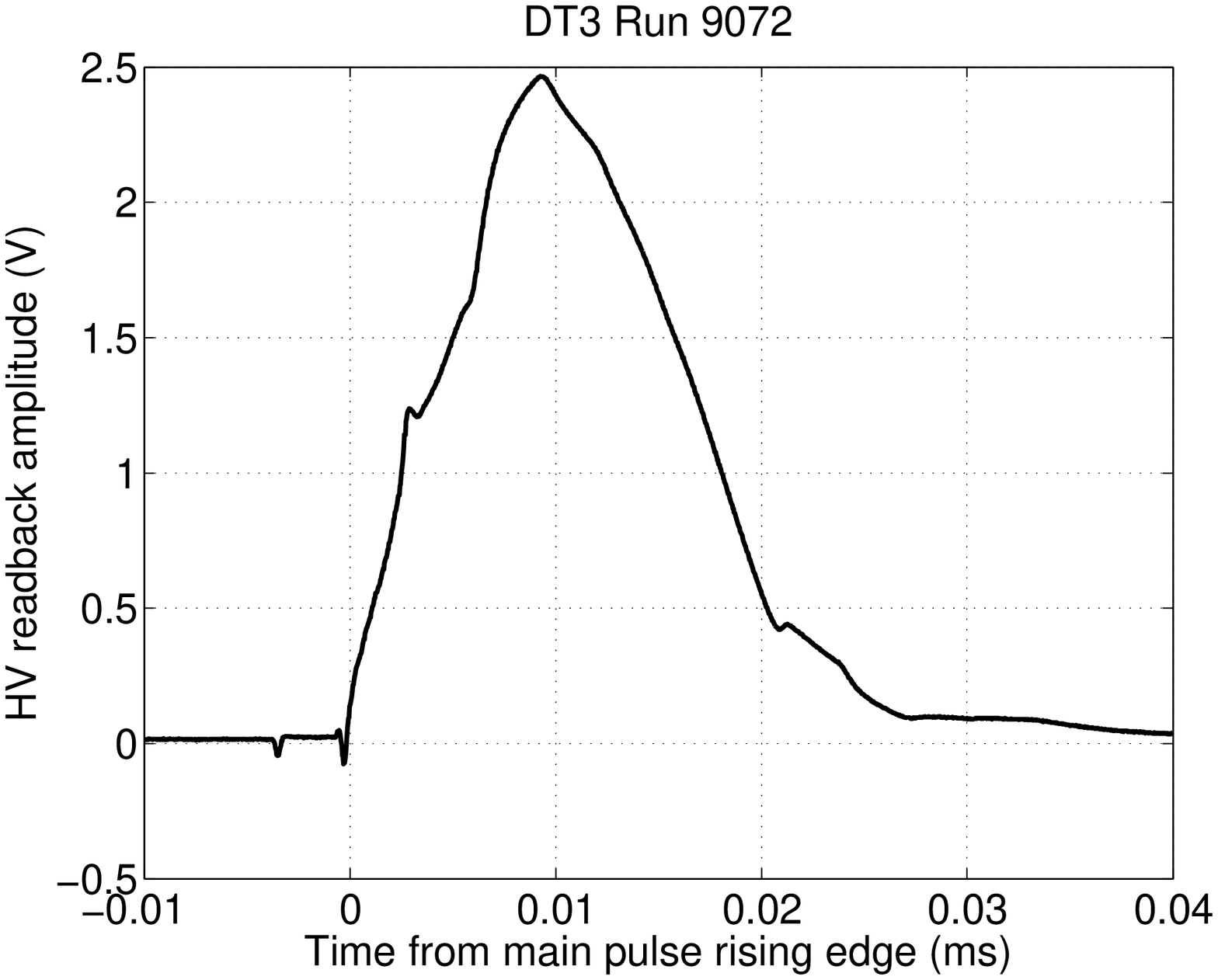}	
}
\subfigure[DT4]{
\noindent\includegraphics[width=13pc]{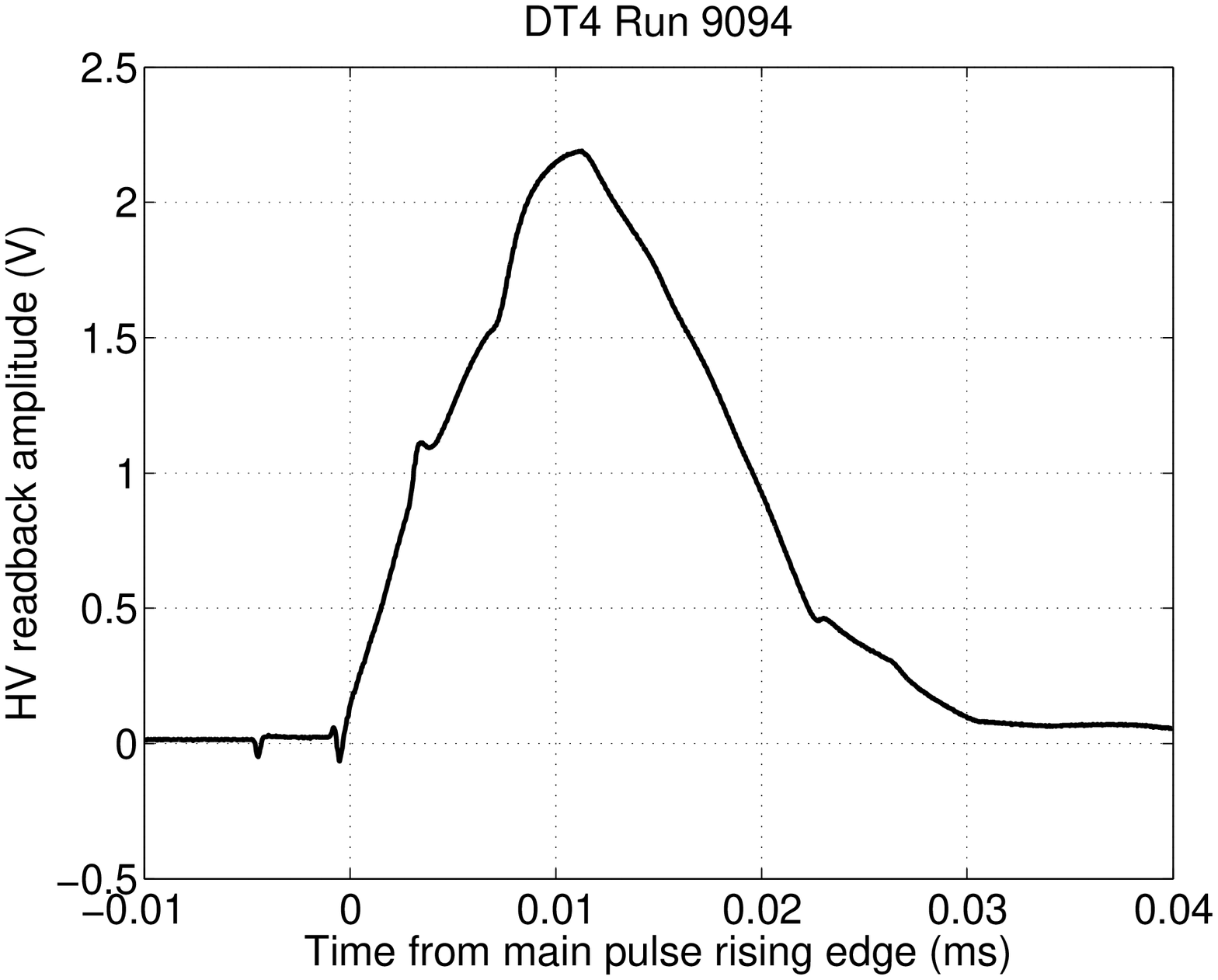}	
}
\subfigure[DT5]{
\noindent\includegraphics[width=13pc]{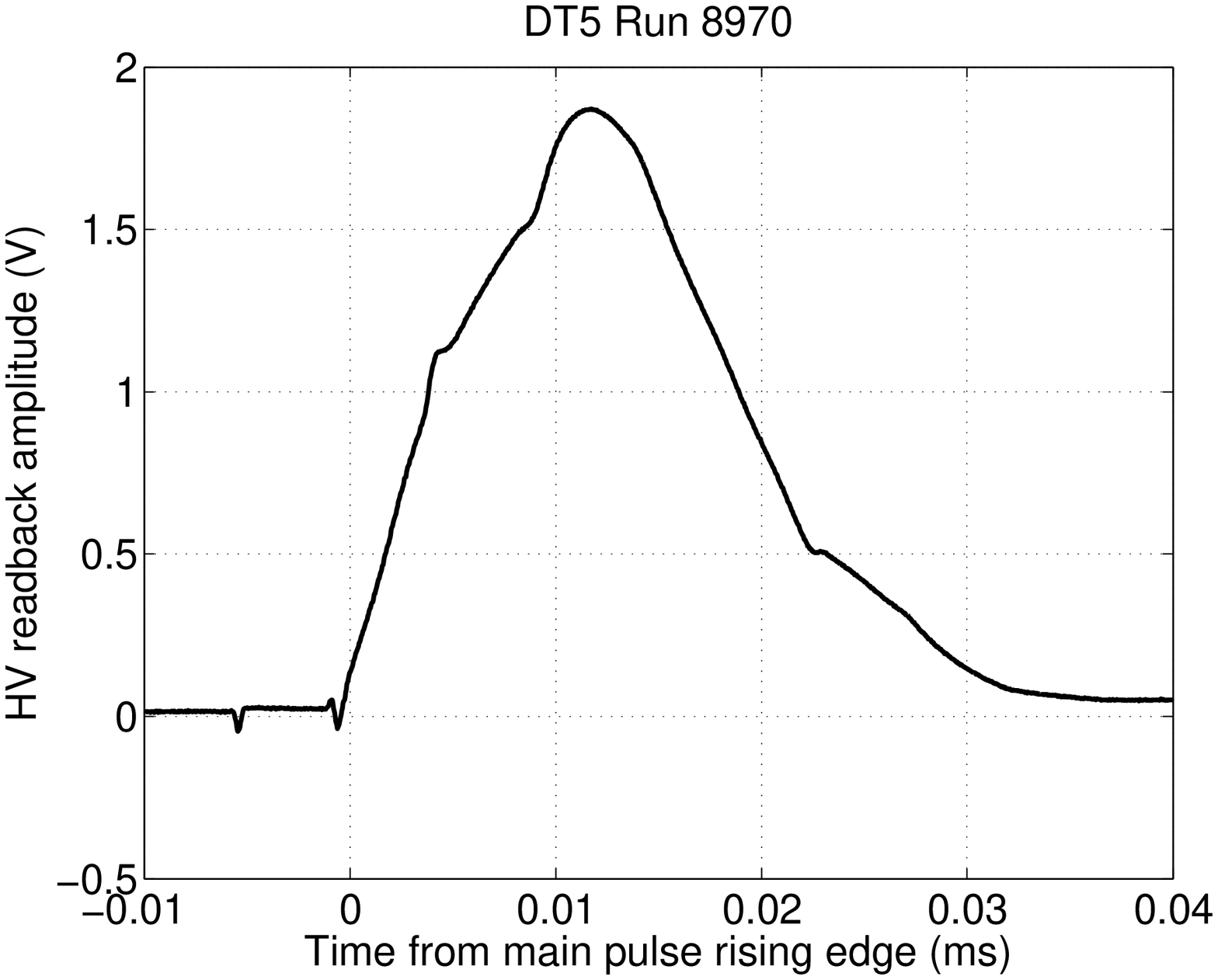}	
}
\subfigure[DT6]{
\noindent\includegraphics[width=13pc]{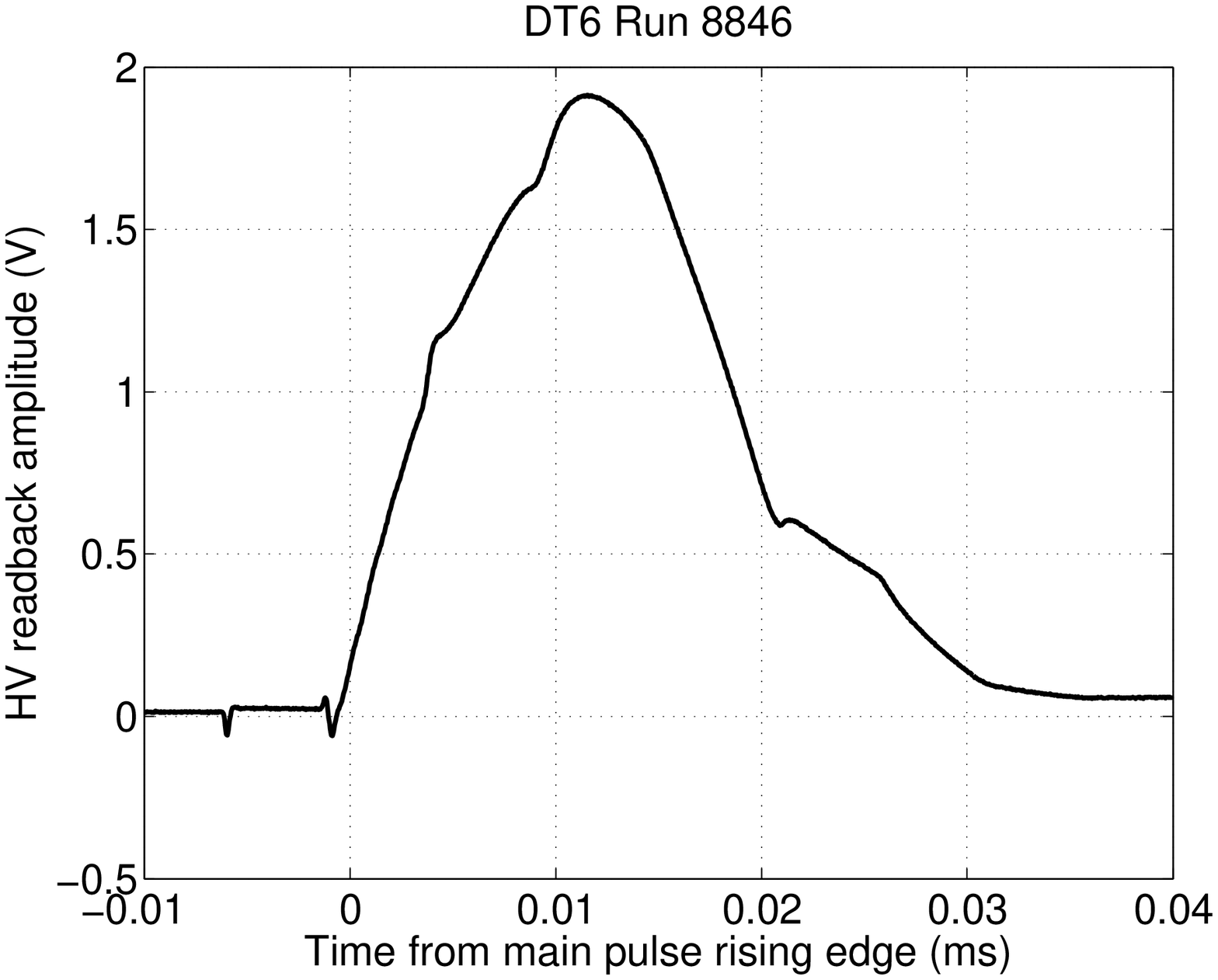}	
}
\subfigure[DT7]{
\noindent\includegraphics[width=13pc]{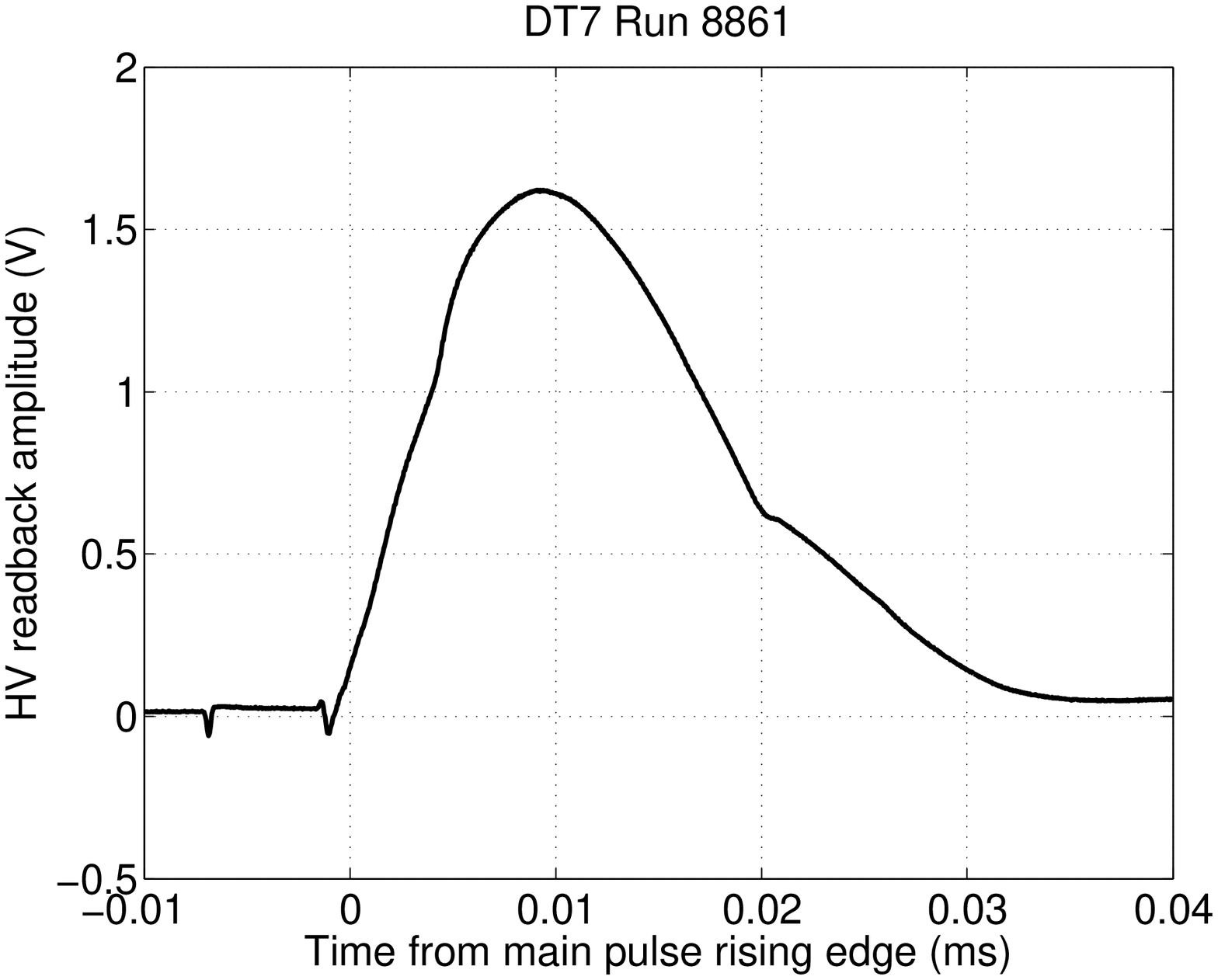}	
}
\caption[String D transmitter performance]{HV read-back pulse shapes for String D transmitters, as determined by overlaying many individual pulses with clock drift correction.  In comparison with the transmitters on Strings A, B, and C, the String D transmitters have an HV read-back pulse shape that is closer to the expectation from laboratory measurements and indicates healthy performance in comparison with the transmitters on Strings A, B, and C.}
\label{hvrbD}
\end{center}
\end{figure}

The HV read-back pulses are only sampled at 308~kHz (by the ``slow'' ADC board), too small a frequency to resolve details in the $\sim$10~$\mu$s HV read-back pulses.  However, with appropriate clock drift correction in the offline processing, many pulses can be overlaid to determine the pulse shape with very large effective sampling frequency.  This is a beneficial effect of the ADC clock drift.  If there were no clock drift (and the sampling frequency were a multiple of the transmitter pulse repetition rate), then corresponding samples of different pulses would overlap exactly.  The presence of clock drift, however, causes the sampling to scan (drift) over the pulse shape from one pulse to the next, resulting in an effective sampling frequency that is much larger than the nominal sampling frequency.

HV read-back pulse shapes determined by averaging many individual pulses with clock drift correction are shown for Strings A, B, C, and D in Figures~\ref{hvrbA}-\ref{hvrbD}.  These recordings are from inter-string runs taken on December 27, 2007.  The multiple oscillations and negative components of the waveforms on Strings A, B, and C are unexpected and not understood.  The pulses were expected to be similar to those recorded in the laboratory before deployment: simple unipolar, positive pulses.  One hypothesis is that the ground connection in the transmitters changed after deployment.

There are apparently multiple classes of strange performance of the String A/B/C transmitters, corresponding to multiple classes of pulse shapes in the HV readback recordings.  AT2, AT3, AT4, BT2, CT3, and CT4 have very similar shape and amplitude.

The trigger signal generated by the DAC's in the String PC's and used to fire the transmitters in the ice are square waves.  The rising edge initiates charging of the LC circuit in the HV pulser module, and the falling edge initiates discharge of this signal to the piezoelectric emitter.  A periodic square wave can be used to fire the transmitter cyclically, with the period of the wave determining the repetition rate of the transmitter.  For the HV readback pulses shown here, the repetition time was 55~ms.  The width of the trigger pulse (time between rising and falling edges) can be optimized to maximize the acoustic signal amplitude.  For Strings A, B, and C, the optimum charge time (trigger width) is 5~ms.  For String D, it is 2~ms.  For both types of transmitter, there is an electrical and acoustic pre-pulse emitted at the time of the rising edge, in addition to the main pulse at the time of the falling edge.

The pre-pulses on String D are small.  On Strings A, B, and C, the pre pulses are quite large, often larger than the main pulse.

For the HV read-back plots, drift correction was not performed in the standard way of dead-reckoning the absolute time of each sample using the IRIG GPS signal.  This is because the recordings are created by generating the transmitter pulses with a fast ADC board and recording the HV readback signals with the slow ADC board on the same String PC.  The two boards have different clocks with different drift rates.  Correcting for the drift using the dead-reckoning approach would require recording a signal from the fast board (to determine its drift rate) in addition to the slow board, which was not done when these data were acquired.  Instead of using that algorithm, we use an alternative drift correction algorithm, which minimizes the width of the pulse, where ``width'' is defined to be the time over a pre-defined threshold.  Minimizing this width with respect to the assumed clock drift achieves a very well aligned, coherent overlaid pulse.

The HV readback pulse shape is extremely stable, with negligible pulse-to-pulse or run-to-run variation in the pulse signal shape, and with negligible noise in the recordings.  To produce the plots shown we first determined the drift rate, then overlaid the samples of 1000 pulses according to the actual repetition period (determined from the drift rate).  Each set of 10 consecutive samples was then averaged together to produce a a waveform sampled with $\sim$100 times greater effective sampling frequency than that used by the ADC to sample the original series of pulses.

The transmitter of Stage 1 of each string is not included in the plots.  While we do have the capability to record these HV readback pulses, they use a different ADC configuration and the software to take these runs is not yet implemented.

For these recordings, the transmitter steering amplitude was 1200 DAC counts (5.86 V) for String D and 1536 DAC counts (7.5 V) for Strings A, B, and C.

Note that the curves resulting from overlying many curves have finite thickness, and some runs give thicker curves than others.  This could be due to residual clock drift that persists even after correcting for most of the drift.  However it is actually due to the ADC conversion jitter, which smears the curves by $\pm$~$\sim$1~ADC count vertically.  So the finite thickness is due to vertical smearing, not horizontal smearing.  The effect could be reduced by overlying fewer curves, or by averaging.  More pulses than necessary are used for the plots shown here.

The transmitters of String D were improved in design based on experience with transmitters on Strings A, B, and C.  The electrical design of the HV pulser was redesigned to be more robust with better signal to noise ratio.  The String D transmitters are performing better than those on the first three strings, as can be seen from the shape of the HV read-back pulses.  

The recorded inter-string transmitter signals from transmitters on the first three strings (A, B, and C) are weaker in South Pole ice than expected (from amplitude vs. distance measurements performed in water in Lake Tornetr\"ask in Abisko, Sweden), by an order of magnitude in absolute amplitude.  The explanation of this is unknown but may be related to the strange pulse shape in the transmitters from these strings.  The String D transmitters were not tested in a lake before deployment.

The cleanness of the String D HV read-back pulses, and their similarity to those from pre-deployment laboratory tests, indicate that the String D transmitters are performing better than those on the first three strings (A, B, and C).  Nevertheless, acoustic signals from the transmitters on the first three strings have clearly been detected by sensors on other strings hundreds of meters distant.  The pulse shape emitted into the ice is likely different from that expected from laboratory tests, but the signals can nevertheless be used both for sound speed and attenuation measurements if analysis strategies are used that do not depend on the shape or absolute normalization of the emitted transmitter signals.  This can be done by using a waveform energy method to integrate over the total energy in the pulse (regardless of shape) and by leaving the overall normalization of the signal as a free fit parameter.  Results using this strategy are presented in Chapter~\ref{attenuationChapter}.

On String D in contrast to Strings A, B, and C, the pulse shape is very similar from one stage to the next.  The amplitude decreases monotonically as the depth (cable length) increases.

In addition to the pulse shape on String D being healthier, the String D pulse shapes are very similar from one transmitter module to the next.  The amplitude of the pulses decreases with increasing module depth.  This is an expected effect due to voltage drop of the transmitter amplitude steering voltage in the 500~m long in-ice cable.  The HV read-back pulse itself is designed to not be affected by the long cable length (by using a differential signal), and therefore to represent the true HV pulse delivered to the piezoelectric emitter in the ice.  On the other hand, the DC steering voltage as well as the DC power voltage and ground level change with depth, and all of these affect the amplitude of the pulse generated by the HV pulser module.  Therefore for a fixed steering voltage at the surface, the actual amplitude delivered by the HV pulser module and therefore the acoustic amplitude emitted by the transmitter decreases with increasing module depth.  These two features (independence of HV read-back amplitude on cable length, and dependence of signal amplitude on cable length) are true for Strings A, B, and C as well as for String D.


\section{Sensors}

\subsection{Good and bad channels}

The sensors are performing well, with 74 of 80 (93\%) of deployed channels considered ``good''.  See Section~\ref{currentNoiseHistos} for a list of good/bad channels and discussion of their behavior.

\subsection{Azimuthal module orientation}

Each sensor module has three piezoelectric transducers inside, oriented 120$^\circ$ apart in azimuth.  The channels are labeled both with numbers and letters: A=2, B=1, C=0 (Note the counter-intuitive mapping).  Looking at the stages from above (with cable penetrator pointing up), the channels are ordered with A, B, C in counter-clockwisse order.  While each stage is fixed in orientation relative to the SPATS cable, the cable twists arbitrarily over its 400-500~m length and therefore the absolute orientation of the three channels inside each stage cannot be determined during deployment.

However, the absolute orientation of each channel can be determined using arrival time differences between the different channels.

This strategy was used successfully to determine the sensor module orientation in the Lake Tornetr\"ask test.  At this time the DAQ software was only capable of reading out two of the three channels simultaneously, and this was still sufficient to determine the rough orientation of the module.  Using all three channels, it should be possible to determine the absolute orientation of each module with $\sim$10$^\circ$ resolution.  This could be valuable for understanding the relative sensitivity of the different channels in each module, by understanding which channels are pointed toward the signal source and which have the module housing between the channel and the source.

In the lake test, a sensor module was fixed in one position, through a hole in the ice covering the lake.  A transmitter was then deployed 400 meters away from the sensor and pulsed repeatedly.  It was then moved to another location such that data were collected with the transmitter 400 m distant in each of the cardinal directions from the sensor.  A compass was not used to determine the cardinal directions exactly, but the four positions are labeled approximately according to the cardinal directions.  The software triggered the transmitter and recorded the sensors using the same ADC clock, such that absolute acoustic signal propagation time could be determined for each pulse without the need for any GPS timing.

\begin{figure}
\begin{center}
\subfigure["North"]{
\noindent\includegraphics[width=14pc]{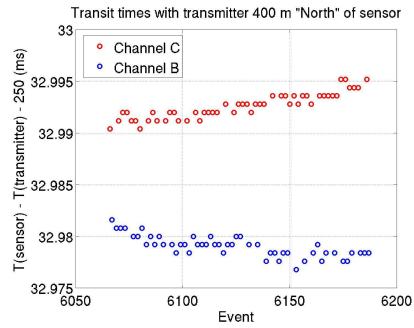}
}
\subfigure["East"]{
\noindent\includegraphics[width=14pc]{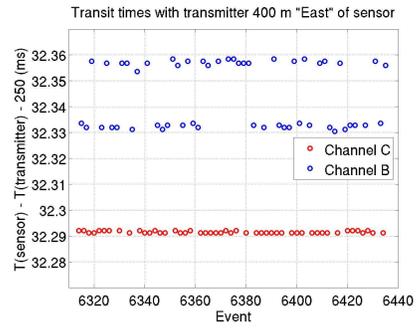}
}
\subfigure["South"]{
\noindent\includegraphics[width=14pc]{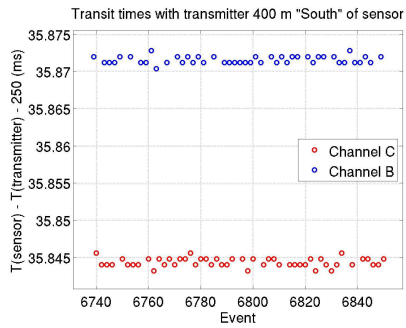}
}
\subfigure["West"]{
\noindent\includegraphics[width=14pc]{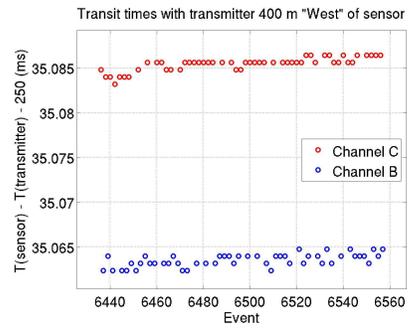}
}
\caption[Resolution of different arrival times for different channels in same sensor module]{Time of flight for each of two sensor channels in a sensor module in the lake test.  The transmitter was pulsed from four different directions at 400~m distance, while the sensor was held stationary.  The finite time between samples (0.8~$\mu$s) can be seen.  The oscillation period in the signal can also be seen in the ``East'' plot: for the Channel B events the algorithm sometimes identified the first cycle correctly and sometimes missed it and only found the second cycle.  The cycle oscillation time is $\sim$30~$\mu$s.  The data constrain the orientation of the sensor module.}
\label{abiskoOrientation}
\end{center}
\end{figure}

Results from the lake test are shown in Figure~\ref{abiskoOrientation}.  The path difference between the channels is $\sim$10~cm of steel.  The measured difference in arrival times between the two channels is in the range of $\sim$~10-50$\mu$s, depending on the direction to the transmitter.  The acoustic signal is expected to propagate through the steel cylinder at $\sim$6000~m/s, giving an expected time difference of $\sim$20~$\mu$s.  The time between ADC samples on each channel was 0.8$\mu$s.

The difference in arrival times constrains the sensor module orientation: Channel C is pointed toward the southeast, B toward the northwest, and A toward the northeast.  The transit times were varying over time while the transmitter was in the ``North'' position.  This could be explained if the cable was rotating clockwise (as seen from the top) during the measurement, such that Channel B rotated toward the transmitter and Channel C rotated away.  The total amount of rotation was less than 180$^\circ$ over the duration of the ``North'' recordings.


\begin{figure}[tbp]
\centering
\noindent\includegraphics[width=25pc]{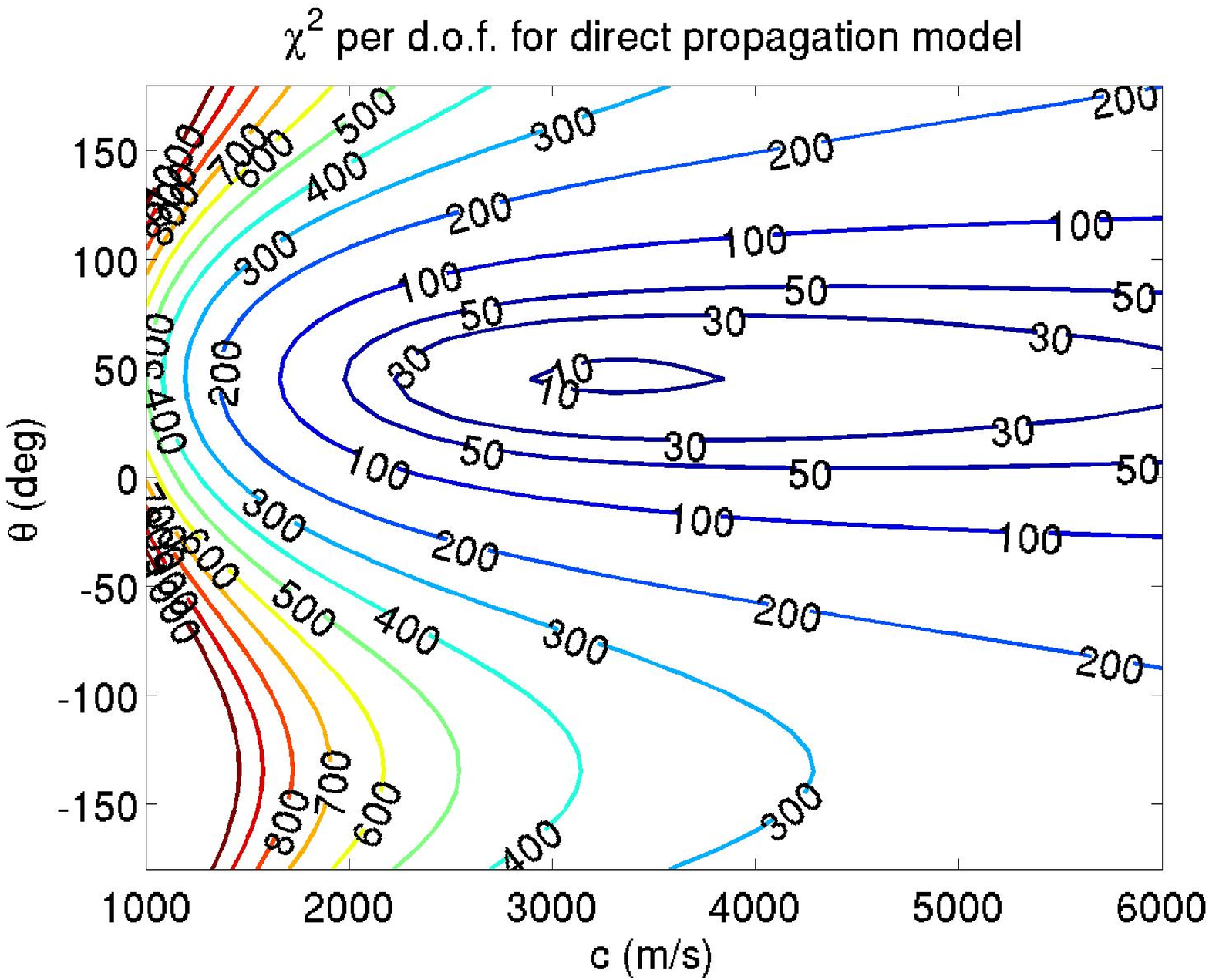}
\caption[Determination of sensor module orientation in lake test]{Reduced-$\chi^2$ contours as a function of azimuthal orientation and effective propagation speed between sensor channels.  This method has been demonstrated with two channels per module using lake test data.  It could be applied to South Pole data using all three channels per module to determine the orientation of every SPATS module in the ice.}
\label{abiskoOrientationContours}
\end{figure}

The orientation can be calculated quantitatively using a two-parameter fit: the azimuthal orientation of the module and the effective propagation speed through the pressure housing are free parameters.  The model uses the actual distance between channels and allows the propagation speed as well as the orientation to vary.  The chi-square statistic can be minimized to determine the actual azimuthal orientation of the module.  An example chi-square contour plot is shown in Figure~\ref{abiskoOrientationContours}.

\section{String PC performance}

The String PC's are performing well, even better than expected.  Despite problems getting a stable flash disk and a stable operating system in the laboratory, the String PC's have performed well over more than two years buried under 2~m of ice at the South Pole.  There have been several planned and unplanned power outages when one or more strings have been powered down for up to 48~hours.  During such periods the String PC's cool down and approach the ambient temperature in the ice, then warm up over several hours after being powered up again.

The String PC's have recovered from each of these outages without any problems.  In some cases the baseline noise level has changed very slightly on sensor channels after one of these power cycles.  This could be due to a change in the absolute calibration of the ADC's.  They have an auto-calibration feature, which we are currently not using, which could reduce this effect.  However it is barely noticeable already.


\section{ADC performance}

There have been several problems with both our fast and our slow ADC boards.  Some of these have been solved by improving the driver software provided by the ADC company.  Others have been mitigated by adapting our DAQ strategy.

\subsection{DMA timeouts, ADC hanging, and ADC sampling frequency limitations}

\label{sectionADCIssues}

\subsubsection{``Slow'' board: sampling frequency}

The ``slow'' ADC boards we have (one per String PC, Real Time Devices model DM6420HR) are specified to operate at up to 500~kHz total sampling frequency (integrated across channels).  Initially we could only operate it at $\sim$100~kHz before it hung.  After several iterations of upgrading the driver, we now operate it at 308~kHz without problems.  We cannot achieve the specified 500~kHz and the company has acknowledged that this is probably impossible, at least with a Linux operating system.  These boards are used primarily for recording the HV read-back signals.

\subsubsection{``Fast'' board: hanging, DMA timeouts, and sampling frequency}

The ``fast'' ADC boards we have (three per String PC, Real Time Devices model SDM7540HR) are specified to operate at a maximum sampling frequency (integrated across channels) of 1.25~MHz.  Again we had severe problems achieving reasonable sampling until we made significant modifications to the device driver.

The boards were chosen in order to use ADC boards 0, 1, and 2 to sample module channels 0, 1, and 2 respectively.  This would allow 1.25~MHz divided among 7 modules per string to achieve 179~kHz sampling frequency per channel and sample all 21 channels per string simultaneously.  Unfortunately we have only been able to achieve 200~kHz total sampling frequency per board, compared to the specified 1.25~MHz.  As for the slow board, the ADC company has confirmed that they do not know how to achieve the specified maximum throughput in Linux, despite their advertisements of full Linux support.

Increasing the sampling frequency (or the run duration) causes runs to hang and require being killed by an external signal (from the user or from a watchdog script).  These hanging events are caused by DMA time outs, and when they occur there is a kernel message that there was a DMA time out.  Currently we are able to acquire with one channel per board per string, in order to read out three channels per string, at 200~kHz each, for transient threshold-triggered runs of 45-minute duration.  This results in a few percent of runs failing with DMA time outs.  Increasing the sampling frequency, channel multiplicity, or run duration increases the failure rate.  As a workaround, we execute all runs inside a watchdog script that terminates hanging runs so that the next run can begin.

\subsection{Leading zeros}

With the ``fast'' board, even after the driver upgrades described above, the first read call of every run begins with a string of zero-valued samples returned by the driver.  There are typically on the order of a few thousand zeros at the beginning of every waveform due to this problem.  As a workaround, the software allows a user-configured ``dead time'' at the beginning of every run.  This duration of samples is acquired and discarded so that the actual waveform written to disk does not include any leading zeros.  Using a dead time of 100~ms is sufficient to remove all leading zeros.




\subsection{Clock drift}

\subsubsection{Description of the clock drift problem}

On each of the four strings there are four ADC boards, one ``slow'' and three ``fast'' models.  The fast boards are used to digitize the sensor waveforms.  The slow board is used to digitize the transmitter HV read-back signals, as well as for some transmitter control functions.  Each board has its own sample clock, which is used to drive both ADC and DAC operation.  The boards were discovered to drift over time.  That is, the sample time (number of samples actually processed) deviates from the true time (number of samples that should have been processed according to the absolute amount of time that has passed) by an amount that increases over time.  Put another way, the actual sampling frequency differs from the nominal sampling frequency.  The deviation is typically on the order of a few parts per million.  That is, the nominal time increases or decreases by a few microseconds per second, relative to the absolute time.   We define the ``drift rate'' as follows:

\begin{equation}
\textrm{drift rate } d = \frac{\textrm{actual sampling frequency}} {\textrm{nominal sampling frequency}} - 1.
\end{equation}

\noindent Typical drift rates are on the order of a few parts per million (positive or negative).  

Clock drift is most important when averaging pulses over a long duration in the time domain, and the severity of the problem increases with the recording duration.  The problem is severe for inter-string recordings of duration $\sim$100~s and moderate for retrievable pinger recordings of duration $\sim$20~s.

\subsubsection{Solution to the clock drift problem}

We have developed several algorithms to correct the clock drift problem.  The best one uses the IRIG GPS data stream, which is included in every sample of every waveform (as a single bit in the 16-bit word of each sample), to determine the absolute time of each sample directly.  The algorithm works directly on each waveform (run), without any other information necessary.  The algorithm works by first determining the rising edges in the IRIG data stream, which occur every 10 ms (absolute GPS time).  These are then used to determine the ``transfer function'' that gives absolute time as a function of the sample number in the recording.  The transfer function is automatically sampled every 10~ms by the IRIG rising edges.  This is frequently enough to determine the drift of the clock, but is much less frequent than typical sampling frequencies used for waveform recordings.  We then fit a polynomial to the function for smoothness (without the polynomial, the function is only defined every 10~ms and is discontinuous and introduces artifacts).  For most waveform durations (up to $\sim$100~ms), a linear fit is sufficient.  For longer durations, a second order fit is sometimes necessary (this means that we are including the ``drift of the drift'').  This transfer function is then used to determine the absolute time (measured since the waveform acquisition began) of each sample, as a function of its sample number.  After applying this clock drift correction algorithm, we know the time of each sample with $\sim$10~$\mu$s~accuracy.

\chapter{Measurement of sound speed vs. depth: pressure waves and shear waves}

\label{soundSpeedChapter}

\noindent\emph{Our measurement of the speed of both pressure and shear waves as a function of depth in South Pole ice is the first measurement we completed with SPATS, and has been the focus of much of my personal analysis effort.  I therefore include it as a stand-alone chapter, with all of my other analysis results described in the following chapter.  The measurement has implications for both neutrino astronomy and glaciology.  A modified version of this chapter has been submitted for publication and is available at~\cite{soundSpeed}.}

\paragraph{Abstract}
We have measured the speed of both pressure waves and shear waves as a function of depth between 80 and 500~m depth in South Pole ice with better than 1\% precision.  The measurements were made using the South Pole Acoustic Test Setup ({SPATS}), an array of transmitters and sensors deployed in the ice at South Pole Station in order to measure the acoustic properties relevant to acoustic detection of astrophysical neutrinos.  The transmitters and sensors use piezoelectric ceramics operating at $\sim$5-25~kHz.  Between 200~m and 500~m depth, the measured profile is consistent with zero variation of the sound speed with depth, resulting in zero refraction, for both pressure and shear waves.  We also performed a complementary study featuring an explosive signal propagating from 50 to 2250~m depth, from which we determined a value for the pressure wave speed consistent with that determined with the sensors operating at shallower depths and higher frequencies.  These results have encouraging implications for neutrino astronomy: The negligible refraction of acoustic waves deeper than 200~m indicates that good neutrino direction and energy reconstruction, as well as separation from background events, could be achieved.



%
%
%

%
%


%
%

%
 
%
%

\begin{figure}[tbp]
\centering
\noindent\includegraphics[width=30pc]{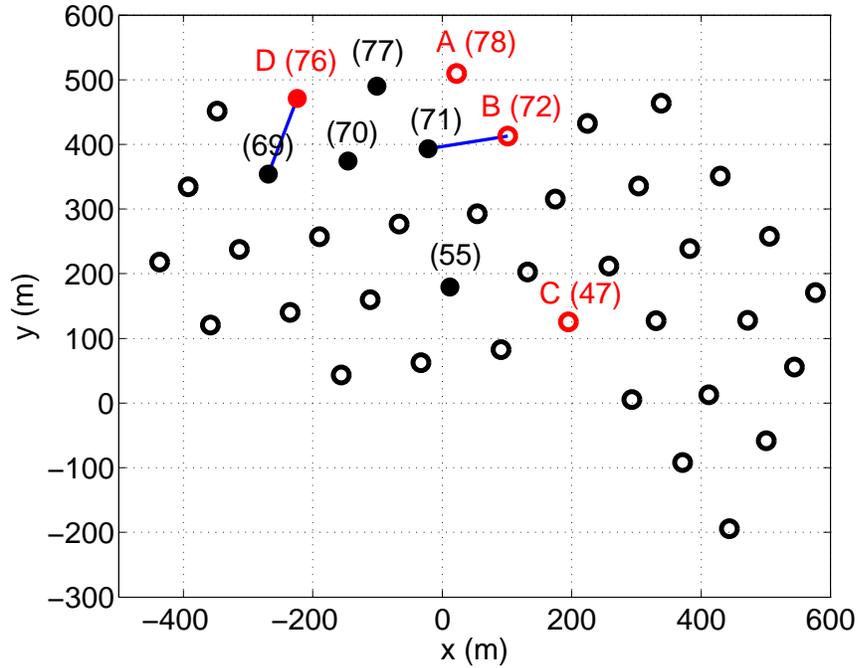}
\caption[Surface layout showing two baselines used in sound speed measurement]{Surface layout of the 40 strings constituting the IceCube array from February through November 2008.  The six holes in which the retrievable pinger was operated are indicated with filled circles.  The four holes with a SPATS string permanently deployed and frozen into the ice are indicated by SPATS ID letters.  IceCube hole ID numbers are given in parentheses.  The two baselines used in this analysis are indicated by line segments.}
\label{geometry}
\end{figure}

\section{Introduction}

The speed of sound in ice has been studied in theory, in the laboratory, and in the field.  In addition to pure interest in elastic materials physics, the measurement has applications to both geophysics~\cite{martinb:ice-density, Kohnen} and neutrino astronomy~\cite{Vandenbroucke08}.  At the South Pole, one measurement was made previously for pressure waves at seismic frequencies, for depths between 0~m and 186~m (i.e., in the layer of surface snow that is not yet fully densified to ice, known as ``firn''), using surface explosions~\cite{Weihaupt63}.

Beyond the South Pole, previous authors have also reported a variety of sound speed measurements in a wide range of conditions including laboratory and field measurements.  Field measurements have previously been made across the Antarctic and Greenland ice sheets and in temperate glaciers.  In principle the sound speed can vary from site to site due to differences in bubble concentration, temperature, and grain orientation.  The grain orientation as a function of position in a glacier is the ``fabric.''  The fabric can have a significant effect on the sound speed because the speed in monocrystalline ice varies by 7\% depending on the direction of propagation relative to the crystal axis~\cite{Price93}.  If the grain orientation is random, the sound speed is homogeneous and isotropic on macroscopic scales.  If there is non-random fabric, the sound speed can be inhomogeneous or anisotropic.

We report an \emph{in situ} measurement made using transmitters and sensors deployed between 80 and 500~m depth and operating in the audible to ultrasonic band.  In addition to making an independent measurement of pressure wave speed in the firn and extending Weihaupt's measurements from the firn deep into the bulk ice, we have measured for the first time the speed of shear waves in both the firn and bulk ice.  Previously, the best estimate of South Pole shear wave speed was a model based on the pressure wave speed and Poisson's ratio~\cite{martinb:ice-density}, and only applied in the firn where pressure wave speed measurements were available.

South Pole ice is uniquely suited as a medium for detection of high-energy (10$^{11}$-10$^{20}$~eV) neutrinos of astrophysical origin.  The interactions of these neutrinos in ice produce optical, radio, and acoustic radiation, each of which therefore provides a possible method of detecting the neutrinos.  The optical method is well suited for neutrinos of energy up to 10$^{17}$~eV, while the radio and acoustic methods are well suited for neutrinos of higher energy.  Deep ice at the South Pole has been shown to be extremely transparent in optical wavelengths~\cite{Ackermann06}.  The AMANDA ~\cite{Ackermann:2007km} and IceCube~\cite{Achterberg:2007bi} detectors have been developed to exploit this for optical neutrino detection.  Antarctic ice is even more transparent in radio wavelengths~\cite{Barwick05, Besson:2007ek}, and the Radio Ice Cherenkov Experiment (RICE)~\cite{Kravchenko06} was operated to search for radio signals from astrophysical neutrinos.  In situ measurements of the acoustic attenuation length in South Pole ice are in progress~\cite{Vandenbroucke08}.

To detect the ``cosmogenic'' neutrinos of energy $\sim$10$^{17-19}$~eV produced by ultra-high-energy cosmic rays interacting with the cosmic microwave background radiation, a detector with effective volume on the order of 100~km$^3$ is necessary.  While the optical method is well understood and calibrated with atmospheric neutrinos, it is prohibitively expensive to scale to such a size.  The acoustic and radio methods, on the other hand, can in principle be used to instrument a large volume sparsely and achieve good sensitivity per cost in this energy range.

The acoustic radiation is produced by the ``thermoacoustic'' mechanism: A neutrino interacts to produce a shower of particles, which locally heats the medium, causing it to expand and produce a bipolar shock wave.  The pulse width (peak frequency) and shape depend on the sensor location relative to the shower.  The acoustic source is simply the region over which significant heat is deposited by the shower: a filament with length of a few meters and diameter of a few centimeters.  The filament is aligned along the incident neutrino direction.  The acoustic radiation pattern is a wide, flat disk perpendicular to the filament and therefore perpendicular to the neutrino direction~\cite{Learned79}.  The peak frequency is $\sim$30~kHz at a distance of 1~km from the source, for points near the center of the radiation pattern~\cite{Bevan09simulation}.  

South Pole ice is predicted to be especially well suited for acoustic detection of extremely high-energy neutrinos~\cite{Price06}.  In comparison to ocean water, the signal amplitude is predicted to be larger in ice and the background noise has been determined by SPATS to be much more stable in ice~\cite{Karg09noise}.

As a solid, ice also has the unique advantage that it can support shear wave propagation.  If neutrinos produce shear waves in addition to pressure waves, a single acoustic sensor detecting both pulses could determine the distance to the interaction vertex as well as the particle shower energy.  Multiple sensors seeing some combination of pressure and shear waves could reconstruct the neutrino energy and direction better than if pressure waves alone were detected.  However while much theoretical and experimental work has been done on pressure waves generated by the thermoacoustic mechanism, little work has been done on shear waves.  It has been argued on theoretical grounds~\cite{Boeser06} that shear wave production by the thermoacoustic mechanism is suppressed, but other mechanisms could produce shear waves and in any case laboratory measurements are necessary.

The South Pole Acoustic Test Setup (SPATS)~\cite{Boeser08} was installed to measure the acoustic properties of South Pole ice relevant to neutrino astronomy, in particular the sound speed profile, the background noise (both the noise floor and the impulsive transients), and the attenuation length.  Here we focus on the first of these: the sound speed as a function of depth.  Sufficiently mapping this profile \emph{in situ} allows precise reconstruction of the location of transient acoustic sources in the ice, which has now been achieved with SPATS using the results presented here~\cite{Vandenbroucke09}.  In addition to the neutrino astronomy applications we consider here, transient source reconstruction could have interesting applications to geophysics.  These could include acoustic emission from cracking in the bulk ice as well as from stick-slip movement at the bedrock interface~\cite{Price06, Weiss97, Bindschadler03}.  Detecting, localizing, and characterizing either of these source classes would improve our understanding of glacial dynamics.  Moreover, precise knowledge of the the sound speed profile itself is necessary to measure ice thickness using reflection shooting~\cite{Kohnen}.

\section{Experimental method}

\subsection{Frozen-in sensors}

The IceCube array is currently under construction.  40 IceCube holes were drilled prior to February 2008 (Figure~\ref{geometry}).  An additional 19 holes were drilled between December 2008 and January 2009.  Each IceCube hole contains a string with 60 digital optical modules between 1450~m and 2450~m depth.  Each hole is drilled with hot water to produce a standing water column $\sim$60~cm in diameter and 2450~m in height.  The instrumentation is then installed in the hole and the water column re-freezes around it.

The SPATS array consists of 4 strings, each deployed alongside an IceCube string in an IceCube hole.  A schematic of the array is given in Figure~\ref{spats_schematic_4_strings}.  Each SPATS string contains 7 acoustic ``stages.''  Each stage comprises one transmitter module and one sensor module.  For the measurement presented here, only the sensor of each stage was used and a separate, retrievable pinger was used instead of the frozen-in transmitters.  On Strings A, B, and C the stages are at 80, 100, 140, 190, 250, 320, and 400~m depth.  On String D the stages are at 140, 190, 250, 320, 400, 430, and 500~m depth.  Each acoustic sensor module contains 3 piezoelectric sensor channels (each with its own pre-amplifier), with the exception of the modules at 190~m and 430~m depth on String D.  Each of these two modules contains a single sensor channel of an alternative design (``Hydrophone for Acoustic detection at South Pole'', HADES~\cite{Semburg09}).  The sensors are sensitive in the 5 to 100~kHz range.  The signal of the frozen-in transmitters (not used for this analysis) is broadband and peaked at $\sim$50~kHz.

The analog output of each channel is transmitted along copper cables to the surface, where it is digitized at 200 kilosamples per second by a rugged embedded computer (``String PC'') installed in a junction box buried 2~m beneath the snow surface.  Power, communications, and timing are distributed over surface cables several hundred meters long to each of the String PC's from an indoor server (``Master PC'') in the IceCube Laboratory.

In addition to digitizing the sensor waveforms, the String PC time stamps them.  Absolute time stamping is achieved with an IRIG-B signal routed to the String PC's from a GPS clock (Meinberg model \emph{GPS169PCI}) installed in the Master PC.  A single IRIG-B output from the clock is fanned out into four cables routed to the String PC's.  The GPS clock is specified to produce IRIG-B rising edges within $\pm$2~$\mu$s of absolute GPS time.  The delay introduced in the IRIG-B signals during propagation from Master PC to String PC is a few $\mu$s, negligible compared to other sources of timing uncertainty in this analysis.

To acquire the results presented here, each sensor channel was recorded continuously for 9~s in order to capture 9 consecutive pulses of the pinger which was operated at a 1~Hz repetition rate.  On each string the sensor channels were read out one by one and looped over sequentially.

\subsection{Retrievable pinger}

In addition to the permanently deployed array of sensors and transmitters, a retrievable pinger was operated in six water-filled IceCube holes, prior to IceCube string deployment in each hole, during the 2007-2008 season.\footnote{An upgraded version of the pinger was operated in four holes in the 2008-2009 season.  Here we focus on results from the 2007-2008 season.}  The pinger consisted of an isotropic piezoelectric emitter ball and a high voltage (HV) module.  The HV module consisted of a high voltage generator circuit contained in a steel pressure housing.  The emitter ball (model \emph{ITC-1001} from the International Transducer Company) produced a broadband pulse peaked in the $\sim$5-25~kHz range.  It was suspended $\sim$1.7~m below the HV module to reduce the effect of acoustic reflections off the steel housing.  The pinger was deployed on a steel-armored, four-conductor cable, which provided both the mechanical and electrical connection from the pinger to the surface.  It was lowered and raised from the surface with a winch.  The length of the cable was $\sim$2700~m, most of which remained spooled on the winch throughout the deployment.

On the surface, a GPS clock (Garmin model \emph{GPS 18 LVC}) was used to generate a 1 pulse per second (PPS) signal, with the rising edge of each pulse aligned to the start of each GPS second.  The PPS signal was routed over the armored cable to the pinger where it served as a trigger signal to pulse the pinger at 1~Hz with emission at known absolute times (modulo 1~s).  The rising edge of the PPS signal initiated charging of the HV pulser circuit, followed by discharge a time $t_e$ later, immediately resulting in acoustic emission.  This emission time delay introduced by the HV pulser was measured in the laboratory to be $t_e = $~1.9~$\pm$~0.05~ms over the range of temperatures in which the pinger operated ($-20$~$^\circ$C to $+20$~$^\circ$C).  The electrical pulse-to-pulse variation of the HV pulser module is negligible.

The electrical signal propagation speed through the $\sim$2700~m cable is 67\% of the speed of light in vacuum according to the manufacturer's specifications, resulting in a $\sim$13~$\mu$s predicted cable delay time.  This delay was verified in the laboratory to be on the order of 10~$\mu$s, negligible compared with other contributions to the timing uncertainty.  The GPS clock is specified to produce rising edges synchronized with absolute time within $\pm$1~$\mu$s.

Although the electrical pulse applied to the transmitter is unipolar, both the transmitter and the sensors ring.  This means that each pulse waveform contains many cycles (oscillations).  The rising edge of the first one is used to determine the acoustic signal propagation time.

The pinger pulsed at 1~Hz repetition rate while it was lowered from the surface to a maximum depth of 400-500~m and then raised back to the surface.  At each depth for which there was a frozen-in sensor on the recording SPATS strings, lowering was halted to keep the pinger stationary for 5 minutes.  This scheme guaranteed that every sensor channel of the SPATS array recorded one complete 9~s waveform while the pinger was stopped at each depth.

In addition to the expected pressure waves, shear waves were clearly detected for many pinger-sensor configurations.  Shear waves were previously detected from frozen-in SPATS transmitters, but it was a surprise to detect them from the pinger operating in water through which shear waves cannot propagate.  While the shear waves from the frozen-in transmitters were likely produced at the piezoelectric transducers themselves, the shear waves from the pinger in water were likely produced by mode conversion at the water/ice interface (hole wall).  Such mode conversion would be suppressed if the incident angle were normal, but if the pinger was not in the center of the hole the incident angle was oblique and shear wave production was favored.  Pinger pressure wave and shear wave identification and characterization are presented in detail in~\cite{Vandenbroucke08}

In the 2008-2009 season the pinger was operated with a mechanical centralizer to keep it close to the center of the hole, and the shear wave production was suppressed in that data compared to the 2007-2008 data analyzed here.  The transmission (for both pressure and shear waves) and reflection (for pressure waves only) coefficients in terms of the incident angle are given by the Zoeppritz equations~\cite{Aki02}.  However they are difficult to apply to this problem because the incident angle depends on the unknown lateral position of the pinger in the hole.

\subfiglabelskip=0pt
\begin{figure*}[tbp]	
\begin{center}
\subfigure[][]{
\label{waveform-a}
\noindent\includegraphics[width=19pc]{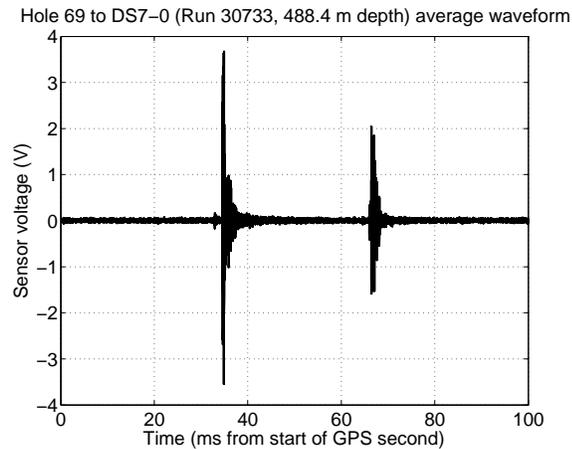}
}
\subfigure[][]{
\label{waveform-b}
\noindent\includegraphics[width=19pc]{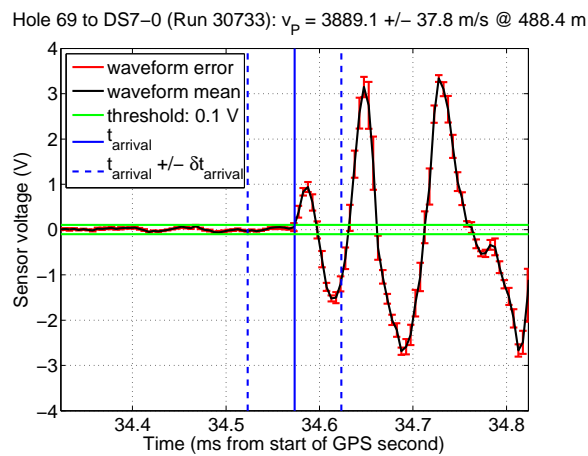}
}
\caption[Example pinger waveform used in sound speed measurement]{An example waveform recorded by a sensor.  \subref{waveform-a} shows the full average waveform resulting from averaging 9 pulses, accounting for clock drift.  Both the pressure pulse and the shear pulse are clearly visible above the noise.  The small pre-pulse before the main pressure pulse is an acoustic pulse initiated by the rising edge of the pinger trigger signal (the main pulse is discharged by the falling edge of the trigger signal, 1.9~ms later).  \subref{waveform-b} shows a close-up of the beginning of the main pressure wave.  For each sample in the mean waveform, the uncertainty is estimated to be $\pm$1 standard error of the mean of the 9 samples contributing to the average.  The threshold used to determine the signal start time is shown, as are the signal start time and uncertainty of the start time.}
\end{center}
\label{waveform}
\end{figure*}

\section{Data analysis}

\subsection{Geometry}
We analyzed two pinger-to-sensor hole combinations: Hole 69 to String D and Hole 71 to String B.  Two combinations were used both as a cross-check and to increase the number of depths included in the analysis.  The two hole combinations used for this analysis are nearest neighbors in the IceCube grid (125~m nominal spacing).  The horizontal distances between the holes (measured by a surveyor) are 124.9~m for Hole 69 to String D, and 124.6~m for Hole 71 to String B.

There are three contributions to the uncertainty in the horizontal separation between the pinger and sensor.  First, the center of each hole at the surface is determined by surveying to $\pm$0.1~m precision in each of the $x$ and $y$ coordinates.  Second, each IceCube hole has a radius of $\sim$0.3 m.  Assuming the pinger could be located laterally anywhere in the cylinder with equal probability, each of the $x$ and $y$ coordinates is within $\pm$0.17~m of the hole center at 68\% confidence.  Similar logic applies for the sensor.  Third, the drill head drifts laterally during drilling of each hole.  Using inclinometers located on the drill head, we estimate this drift to be $\pm$0.5~m in each coordinate.  This effect dominates the first two effects.  Therefore the uncertainty in the horizontal location of each of the pinger and sensor is $\pm$0.5~m and the uncertainty in the horizontal distance between the two is $\pm$0.5~m~$\times$~$\sqrt{2}$~=~$\pm$0.71~m.

The depth of each frozen-in SPATS sensor was verified during string deployment with a pressure sensor.  Each SPATS sensor depth is within $\pm$2~m of nominal.  The pinger depth was monitored in two ways: using pressure sensors and counting  the number of turns of the winch during lowering.  These two measurements were averaged together to determine the absolute pinger depth with $\pm$5~m uncertainty.  Due to a mistake made in converting winch turns to depth (later corrected and verified with the pressure sensors), the pinger was stopped at depths that are systematically shallower than the instrumented sensor depths, by an amount that increases with depth.  For the measurement with the sensor at 500~m depth, the pinger was at 477~m depth.  This has the effect that the relative uncertainty in the sound speed is $\pm$0.6\% for shallow depths (where it is dominated by the horizontal distance uncertainty) and increases to $\pm$1\% for deep depths (where the pinger depth uncertainty contributes nearly as much as the horizontal distance uncertainty).

While pressure and shear wave pulses were detected for many pinger-sensor combinations, for this analysis we selected only those with very high signal-to-noise ratio (SNR), sufficient to not only resolve the pulse but also to resolve its start time precisely.  For shear waves, only String D had sufficient SNR to identify the pulse start time precisely.  Within String D only the 5 non-HADES sensors had sufficient SNR, so there are 5 high-quality shear wave measurements.  For pressure waves, all 7 String D sensors, and 5 of 7 String B sensors, had runs with sufficient SNR.  This resulted in pressure wave measurements at 8 different depths, 4 of which have measurements with both strings.  For those depths with sound speed measured redundantly, the results agree well.

\subsection{Propagation time}
Each 9~s sensor waveform contains 9 pinger pulses, which were averaged together to increase the pulse SNR.  For each pulse sample both the mean and the standard deviation amplitude were determined.  This averaging procedure was designed to decrease the (incoherent) noise by a factor of 3 without affecting the signal amplitude.

While the pinger emission is driven by a clock which is continuously synchronized with GPS time, the sensor recording is driven by an analog-to-digital converter (ADC) clock which drifts by an amount on the order of 10 $\mu$s per second.  That is, the actual sampling frequency typically differs from the nominal sampling frequency by $\sim$10 parts per million.  Furthermore, the actual sampling frequency varies with time (the clock drift rate itself drifts).  This means that pulse averaging using the nominal time of each acquired sample results in large decoherence and a false average waveform.  This clock drift effect was removed by using the true absolute time of each sample as determined continuously from the IRIG-B GPS signal.  This is a 100 PPS digital signal that is sampled synchronously with the sensor voltage data.  Rising edges occur every 10~ms and pulse widths encode the absolute time.

After applying the clock drift correction algorithm, the absolute time of each sample of the waveform is known with a precision of $\pm$10~$\mu$s.  These absolute times were used in the pulse averaging: Absolute sample times were wrapped modulo the pulse repetition period (1 second), and were then sorted and binned to determine the average time and amplitude of each consecutive set of 9 samples.  Figure~\ref{waveform} shows a typical average waveform recorded by a sensor.  Although the signal start time is clear in this example, for other waveforms it was unclear if the algorithm selected the correct first signal oscillation or was wrong by $\sim$one oscillation period.  An uncertainty of $\pm$0.05~ms was therefore assigned to the start time for all waveforms.

For each averaged waveform, a bipolar discriminator was applied to determine the start time.  The noise level varied too much from channel to channel to use a fixed threshold, but for each channel the first cycle of the pinger signal was clearly visible above the noise.  Therefore a threshold was manually chosen for each channel.  The first threshold crossing was then verified by eye to be a good estimate of the signal start time for each channel.  The uncertainty on this arrival time determination is estimated to be $\pm$0.05~ms, corresponding to $\sim$1 signal oscillation period.

The uncertainty of the emission time is simply that of the HV pulser time delay, including variation with temperature: $\pm$0.05~ms.

\section{Results}

\subsection{Overview}

\begin{figure}[tbp]
\centering
\noindent\includegraphics[width=30pc]{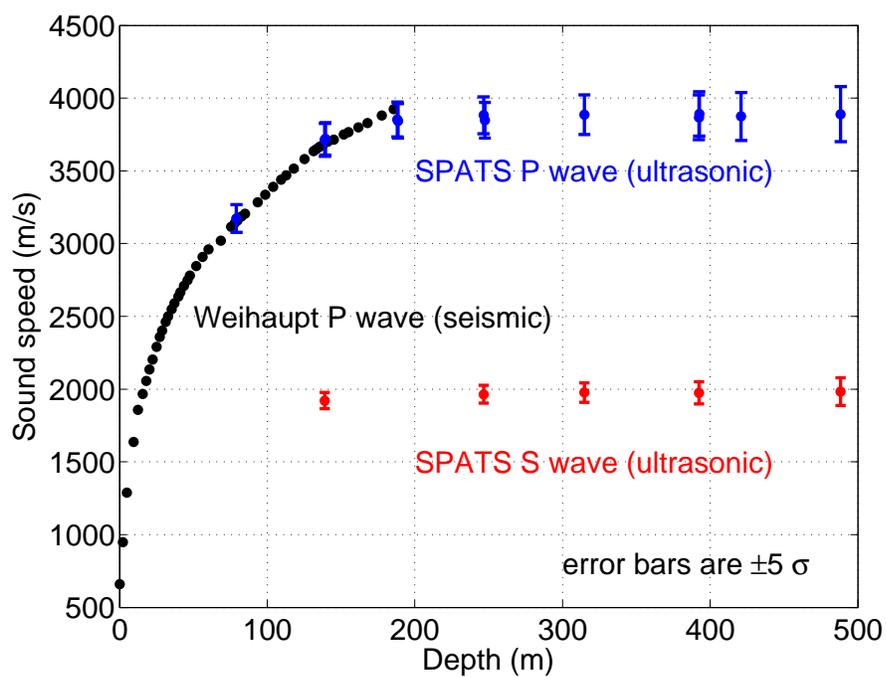}
\caption[Measurement of sound speed vs. depth for pressure and shear waves]{Measurements of sound speed for both pressure and shear waves at particular depths using the South Pole Acoustic Test Setup featuring transmitters and sensors at $\sim$5-25~kHz.  A previous measurement made at seismic (Hz) frequencies~\cite{Weihaupt63} is shown for comparison.  Note: the SPATS error bars are $\pm$5~$\sigma$ in order to be visible.  No uncertainty estimate is available for the Weihaupt result.}
\label{spats_and_weihaupt}
\end{figure}

Figure~\ref{spats_and_weihaupt} shows our measurement of the sound speed versus depth for both pressure and shear waves.  A previous measurement of pressure wave speed in firn~\cite{Weihaupt63} is shown for comparison.  Table~\ref{error_budget} shows the error budget for two example data points in the analysis.

\subsection{Pressure waves}

\begin{figure*}
\begin{center}
\subfigure[][]{
\label{fits-a}
\noindent\includegraphics[width=16pc]{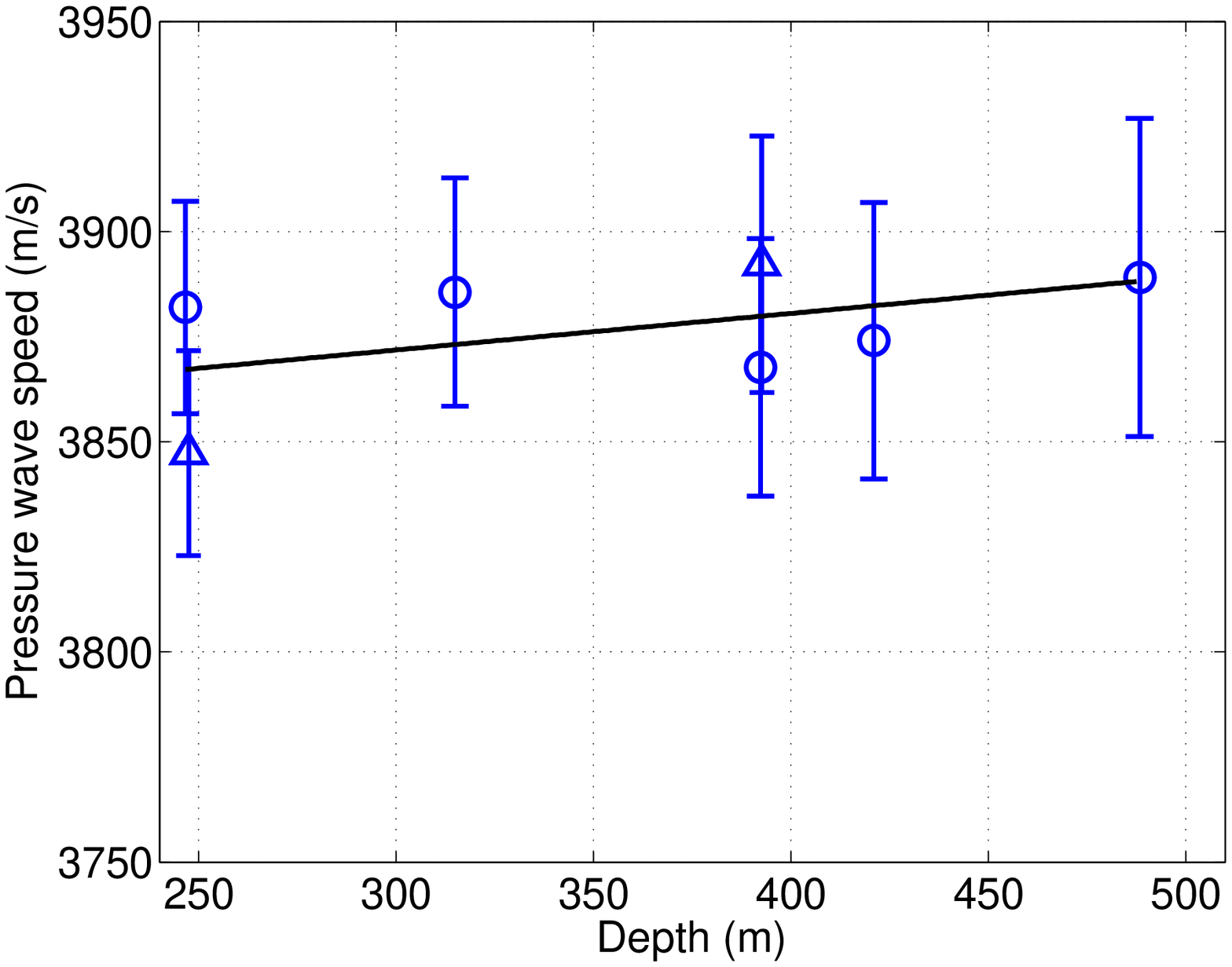}
}
\subfigure[][]{
\label{fits-b}
\noindent\includegraphics[width=16pc]{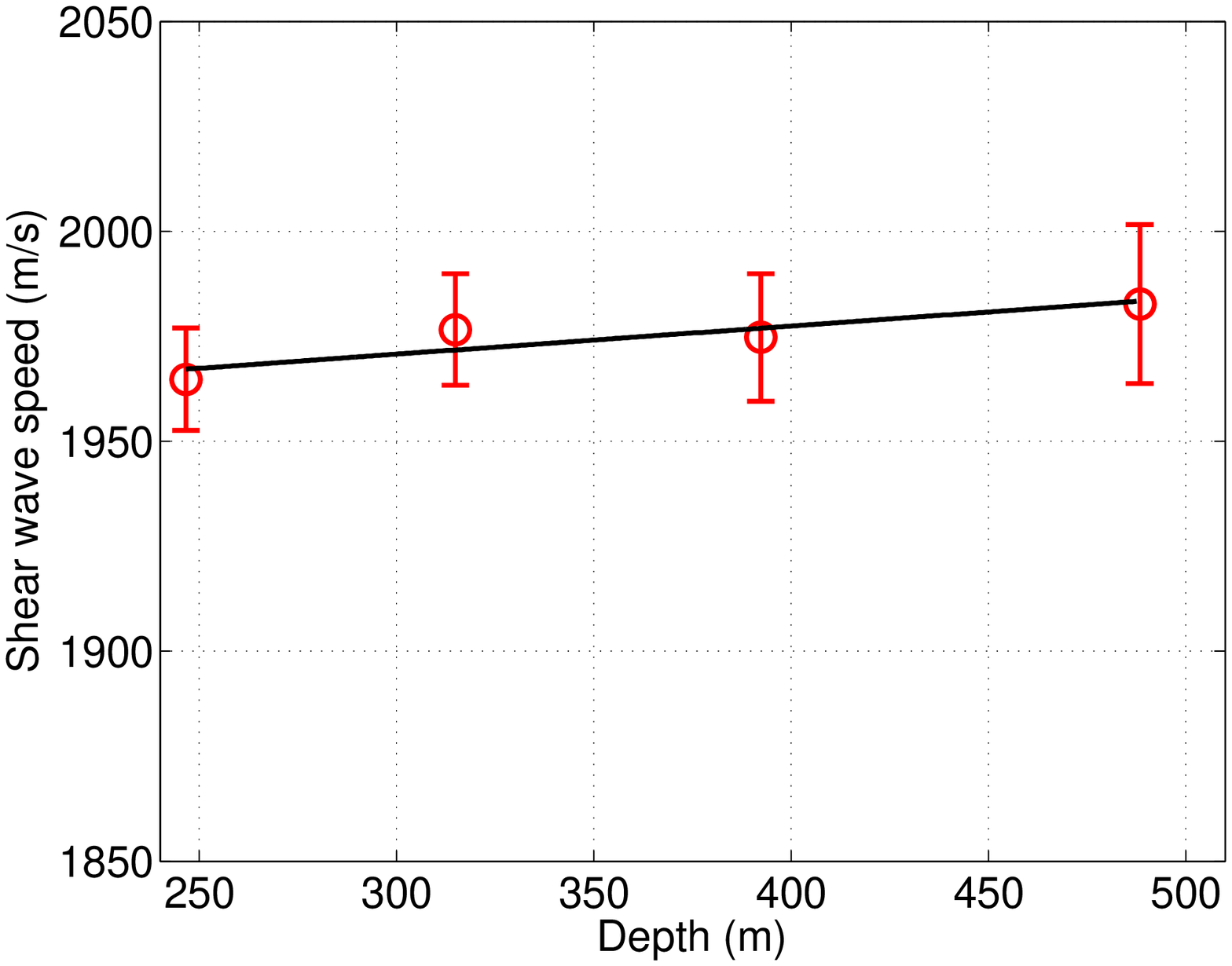}
}
\subfigure[][]{
\label{fits-c}
\noindent\includegraphics[width=16pc]{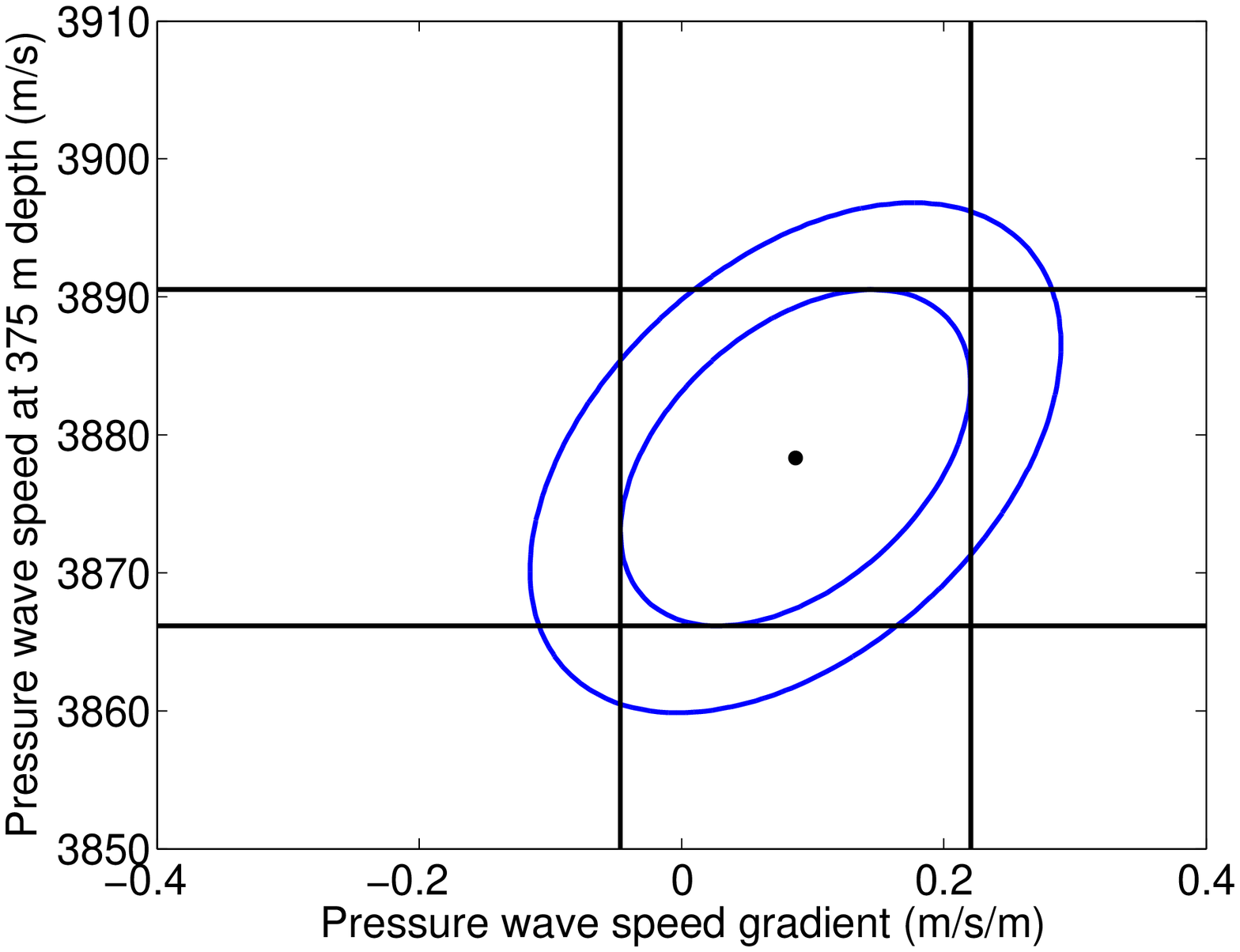}
}
\subfigure[][]{
\label{fits-d}
\noindent\includegraphics[width=16pc]{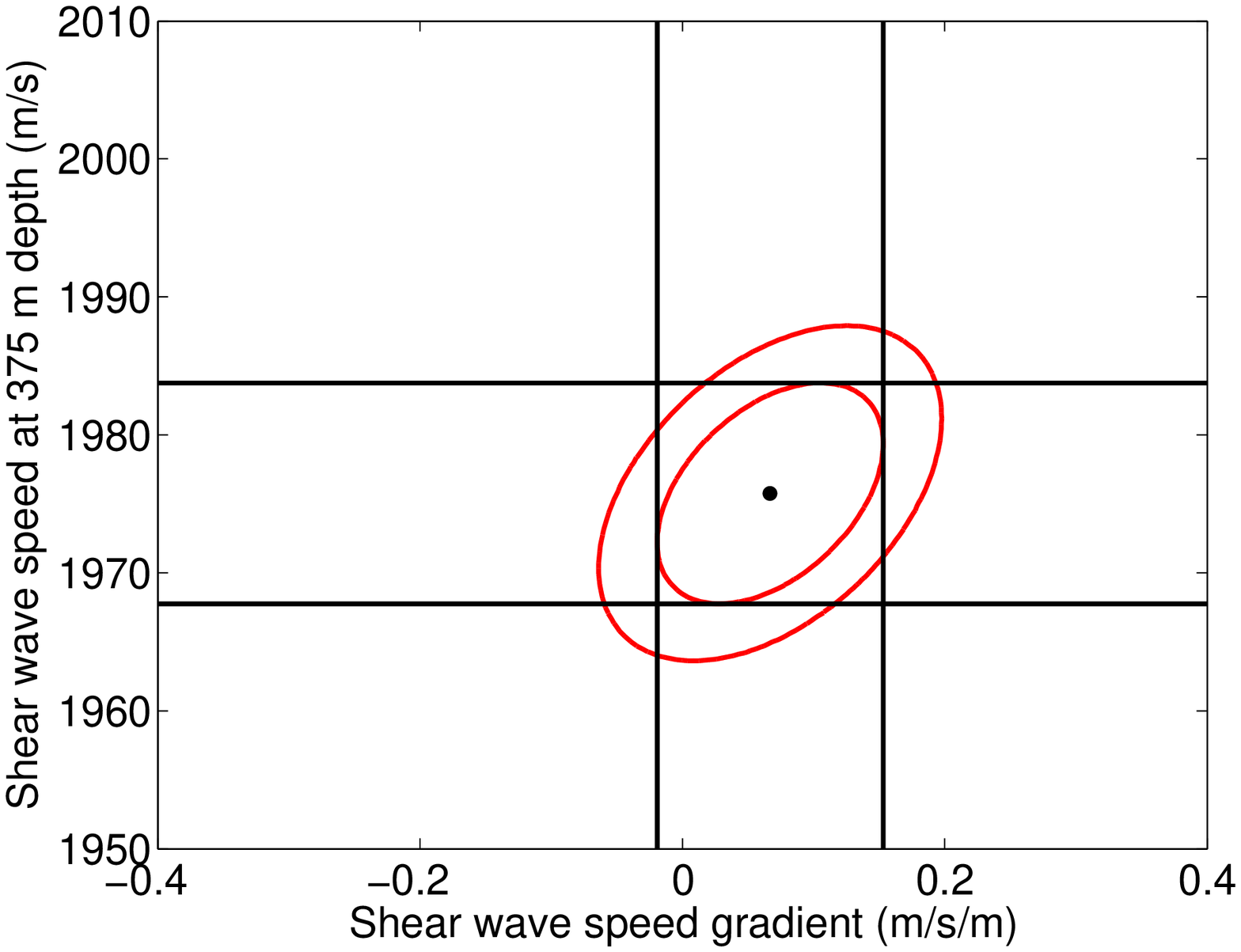}
}
\caption[Sound speed and sound speed gradient in deep ice, for P and S waves]{Pressure wave \subref{fits-a} and shear wave \subref{fits-b} speed vs. depth between 250 and 500~m depth.  Error bars are $\pm$1~$\sigma$.  Measurements made with String D (B) are shown as circles (triangles).  Confidence regions for a joint fit of sound speed and sound speed gradient are also shown, for both pressure waves \subref{fits-c} and shear waves \subref{fits-d}.  The dot gives the best fit ($\chi^2$ = 1.61 for 7-2=5 degrees of freedom for P waves and $\chi^2$ = 0.195 for 4-2=2 degrees of freedom for S waves).  The inner (outer) contour is drawn for $\Delta\chi^2$ = 1.00 (2.30).  The outer ellipse encloses the most likely 68\% of parameter space for the two parameters fit jointly.  The horizontal (vertical) lines give the one-sigma confidence region for the sound speed (sound speed gradient) fit individually.  Note that all errors are treated as uncorrelated.  It is possible that the systematic error contributions from the pinger and sensor positions are correlated between different measurements and that this is why the $\chi^2$ values are smaller than expected for uncorrelated errors.}
\end{center}
\label{fits}
\end{figure*}

Figure~\ref{fits}\subref{fits-a} shows a close up of the pressure wave speed versus depth in the deep, fully densified ice.  A linear fit was made to the data in the fully densified region from 250 to 500~m depth:
\begin{linenomath*}
\begin{equation}
v_P(z) = [z - (375\textrm{ m})] \times g_P + v_P(375\textrm{ m}),
\end{equation}
\end{linenomath*}
where $z$ is the depth (measured positive downward from the surface), $v_P(z)$ is the pressure wave speed at depth $z$, and $g_P$ is the pressure wave speed gradient in the 250-500~m depth range.  The parameterization was chosen such that the sound speed in the center of the fitted range is one of the parameters.  The best fit is:
\begin{linenomath*}
\begin{equation}
v_P(375\textrm{ m}) = (3878 \pm 12) \textrm{ m/s};
\end{equation}
\end{linenomath*}
\begin{linenomath*}
\begin{equation}
g_P = (0.087 \pm 0.13) \textrm{ m/s/m}.
\end{equation}
\end{linenomath*}

Figure~\ref{fits}\subref{fits-c} shows our constraints on the two-parameter fit (sound speed and sound speed gradient) describing the pressure wave propagation as a function of depth in the fully densified (bulk) ice.  The gradient is consistent with zero.

In the firn, our pressure speed results are consistent with the previous measurements by Weihaupt.


\subfiglabelskip=0pt

\subsection{Shear waves}

Figure~\ref{fits}\subref{fits-b} shows a close up of the shear wave speed versus depth in the deep, fully densified ice.  A linear fit was performed to the data in the fully densified region from 250 to 500~m depth:
\begin{linenomath*}
\begin{equation}
v_S(z) = [z - (375\textrm{ m})] \times g_S + v_S(375\textrm{ m}),
\end{equation}
\end{linenomath*}
where $v_S(z)$ is the shear wave speed at depth $z$, and $g_S$ is the shear wave speed gradient.  The best fit is:
\begin{linenomath*}
\begin{equation}
v_S(375\textrm{ m}) = (1975.8 \pm 8.0) \textrm{ m/s};
\end{equation}
\end{linenomath*}
\begin{equation}
g_S = (0.067 \pm 0.086) \textrm{ m/s/m}.
\end{equation}

Figure~\ref{fits}\subref{fits-d} shows our constraints on the two-parameter fit (sound speed and sound speed gradient) describing the shear wave propagation as a function of depth in the fully densified (bulk) ice.  The gradient is consistent with zero.

The shallowest depth for which we have a precise shear wave determination is 139~m depth.  At this depth the ice is still not fully densified.  As expected, the shear wave speed at this depth (1921~$\pm$~11~m/s) is slower than in the deep ice.
 
\section{Measurement with explosives}

In addition to the precision measurement using piezoelectric transmitters and sensors at $\sim$5-25~kHz, a complementary measurement was performed with explosives (seismic frequencies).  This measurement was performed in January 1999 as part of deployment of the AMANDA neutrino telescope.  Dynamite was attached to detonation cord and lowered to a depth $z_d = $~50~$\pm$~5~m in a mechanically drilled hole.  This hole was located $\sim$15~m horizontally from AMANDA Hole 13, which had an acoustic sensor (hydrophone) at depth $z_h = $~2250~$\pm$~10~m.  The detonation cord had an active core of PETN (pentaerythritol tetranitrate).  An electrical circuit near the blasting cap end of the cord triggered a digital oscilloscope to start recording the hydrophone signal.

A pulse was clearly visible above the noise, at an arrival time $t_a =$~566~$\pm$~5~ms after the trigger.  Assuming the detonation signal propagated\footnote{The detonation cord was ``Red Cord'' from Imperial Chemical Industries (ICI).  ICI was purchased by Orica Mining Services Worldwide, which now makes the same cord under the name ``Cordtex Pyrocord Detonating Cord.''  The detonation velocity we use is taken from the manufacturer's data sheet.} at $v_d =$~6750~$\pm$~250~m/s through the cord of length $L = $~52~$\pm$~3~m, the measured pressure wave speed is
\begin{linenomath*}
\begin{equation}
v_P = \frac{z_h-z_d}{t_a-L/v_d} = 3941 \pm 41 \textrm{ m/s.}
\end{equation}
\end{linenomath*}

The precision achieved in this measurement is $\pm$1.0\%.  It gives the pressure wave speed averaged over the depth profile from 50 to 2250~m depth.  It is consistent with the result obtained from the piezoelectric instrumentation, despite the significantly different frequency band and depth range.  This integral, vertical measurement is complementary to the differential, horizontal measurement.  It provides a valuable cross check and extends the range of measurement to nearly the entire thickness (2.8~km) of the South Pole ice.  The explosives measurement indicates that the pressure wave speed gradient is small not only in the 200-500~m depth range but also down to $\sim$2~km depth.

\begin{table}[tbp]	
\centering
\caption[Sound speed error budget]{Error budget for two example data points.  Each is for pinging from Hole 69 and receiving on String D.  While the error contribution from the horizontal distance is nearly the same for all data points, the contribution from the pinger depth increases with depth.  This is because the vertical distance between the pinger and sensor increases with depth, and the error contribution is proportional to this difference.} 
\centering      
\begin{tabular}{|  c  |  c  |  c  |}  
\hline                        
 & \bf{P wave} & \bf{S wave} \\
\hline
sensor depth (m) & 500 & 140 \\
\hline
pinger depth (m) & 477 & 138 \\
\hline
sound speed (m/s) & 3889 & 1921 \\
\hline
error due to horizontal distance (\%) & 0.54 & 0.57 \\
\hline
error due to pinger depth (\%) & 0.72 & 0.06 \\
\hline
error due to sensor depth (\%) & 0.29 & 0.02 \\
\hline
error due to emission time (\%) & 0.15 & 0.08 \\
\hline
error due to arrival time (\%) & 0.15 & 0.08 \\
\hline
\bf{total error} (\%) & \bf{0.97} & \bf{0.58} \\
\hline
\end{tabular} 
\label{error_budget}
\end{table} 

\section{Refraction}

\subsection{Calculated ray trajectories in firn and bulk ice}
\begin{figure}[tbp]
\centering
\noindent\includegraphics[width=30pc]{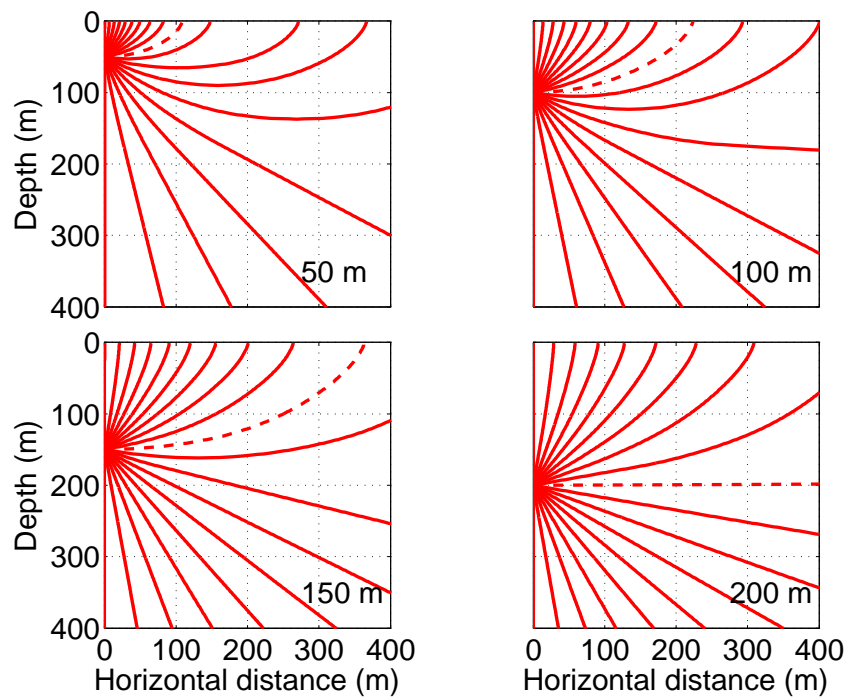}
\caption[Acoustic ray traces in South Pole ice]{Calculated pressure wave ray trajectories using the measured sound speed as a function of depth.  Refraction is significant in the firn (shallower than $\sim$174~m) and negligible below it.  Each panel shows rays emitted from a source at 50, 100, 150, or 200~m.  Rays are emitted every 10$^\circ$ from vertical upward to vertical downward.  The horizontally emitted ray is indicated by a dashed line.  The Weihaupt profile is used for depths between 0 and 174~m, and the SPATS linear best fit is used for depths between 174 and 500~m.  Although the two results agree within their error bars in the region from 174-186~m depth, the SPATS best fit predicts a sound speed slightly smaller than the Weihaupt results.  The two curves intersect at 174~m and the SPATS fit is chosen in the overlap region so that there is no kink in the velocity profile.}
\label{ray_traces}
\end{figure}

We have calculated the trajectory of individual acoustic rays to illustrate the degree of refraction for various source depths and emission directions.  Figure~\ref{ray_traces} shows example ray trajectories calculated for pressure waves.  The ray tracing was performed using an algorithm~\cite{Boyles84} that treats the ice as a layered medium, in each layer of which the sound speed gradient is constant.  Because the gradient is constant in each thin layer, the ray segment in each layer is an arc of a circle.  This algorithm gives a fast and accurate piecewise second order approximation to the true ray path and simultaneously calculates the integrated path length and travel time.  Note that in the presence of a vertical velocity gradient, even horizontally emitted rays are refracted toward the direction of decreasing sound speed.

\subsection{Radius of curvature in bulk ice}
Because the trajectory of a ray in a medium with constant sound speed gradient is a circle, a convenient way to quantify the amount of refraction is the radius of curvature:
\begin{linenomath*}
\begin{equation}
R = \frac{v}{|g|},
\end{equation}
\end{linenomath*}
where $v$ is the sound speed and $g$ is the gradient of the sound speed.  For pressure (shear) waves, our joint fit for the sound speed and sound speed gradient gives a best fit radius of curvature of $\sim$44~km ($\sim$29~km).  With a 44~km radius of curvature, a ray of length $L$=1~km (a possible propagation distance from source to sensor in a large neutrino detector) deflects by a small amount d with respect to straight-line propagation:
\begin{linenomath*}
\begin{equation}
d=\frac{L^2}{2R} = 11\textrm{ m}.
\end{equation}
\end{linenomath*}
This amount of deflection is smaller than the thickness of the radiation pattern induced by a neutrino.  Note that this is the deflection predicted using our best fit gradient.  Because our measurement of the gradient is consistent with zero, the radius of curvature is also consistent with infinity (zero deflection).

\section{Discussion}

\subsection{Comparison with previous results}

Kohnen~\cite{Kohnen} compiled sound speed measurements from Antarctica and Greenland.  After applying quality selection criteria to the existing measurements, he found a simple dependence of both pressure wave speed and shear wave speed on temperature: $v_p = -(2.30 \pm 0.17) T + 3795$ and $v_s = -(1.2 \pm 0.58) T + 1915$, where $v_p$ is the pressure wave speed in m/s, $v_s$ is the shear wave speed in m/s, and T is the temperature in~$^\circ$C. 

\begin{figure}[tbp]
\centering
\noindent\includegraphics[width=30pc]{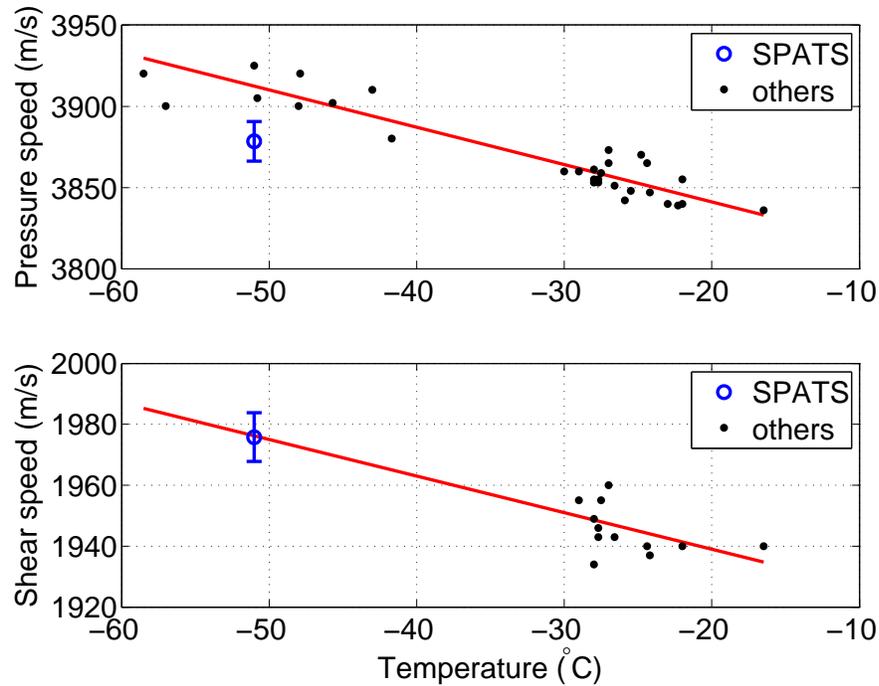}
\caption[Compilation of sound speed vs. temperature for P and S waves]{Compilation of sound speed vs. temperature in ice from different authors.  Only field (not laboratory) measurements are shown, and only measurements in the Greenland and Antarctic ice sheets (not temperate glaciers) are shown.  The non-SPATS data compilation is taken from~\cite{Kohnen}.  Previously the only shear wave measurements were between -15 and -30~$^\circ$C; SPATS has extended this to -51~$^\circ$C.  The lines give Kohnen's fits, without re-fitting to include SPATS.  The SPATS pressure wave result is slightly slower than previous results.  The shear wave result matches the low-temperature prediction of Kohnen (made by assuming Poisson's ratio is temperature independent, predicting a shear wave speed corresponding to each pressure wave speed, and fitting a straight line) very well.}
\label{speed_vs_temp}
\end{figure}

Figure~\ref{speed_vs_temp} shows the data points compiled by Kohnen along with the new SPATS measurement reported here.  Our pressure wave speed is slightly slower than the other measurements.  The other measurements do not include error estimates, so it is difficult to determine whether our result is consistent with them.  The other measurements were made with refraction shooting, in which rays are traced from a surface explosion to a surface sensor, and the maximum speed below the firn is deduced by unfolding the refraction through the firn.  Our \emph{in situ} measurement is less susceptible to systematic effects because it uses unrefracted rays between sources and sensors buried in the deep fully densified ice.

The SPATS shear wave measurement is the first below -30~$^\circ$C.  The shear wave fit by Kohnen was made using predictions at low temperature from the pressure speed and assuming temperature-independent Poisson's ratio.  Our measurement agrees well with his prediction.

A laboratory measurement of both pressure and shear wave speed in ice was reported recently~\cite{Vogt08}.  A degassing system was used to produce a $\sim$3~m$^3$ block of bubble-free ice in which the speeds were measured between 0 and -20~$^\circ$C.  The measured speeds were larger than predicted from the Kohnen fit by $\sim$50~m/s, perhaps due to the absence of bubbles or to grain orientation in the laboratory measurement.

The SPATS and Weihaupt results for pressure wave speed in South Pole firn are consistent in their region of overlap.  This is a valuable cross check because the two measurements use very different experimental methods and use signal frequencies that differ by 4 orders of magnitude.

While Weihaupt measured the pressure wave speed to a maximum of 186~m depth, all his measurements were in the firn ice (by necessity, because his measurement used waves that were refracted back to the surface).  We have confirmed Weihaupt's measurement in the firn and extended it into the fully densified bulk ice, to a maximum depth of 500~m.  Moreover, we have for the first time measured the shear wave speed in South Pole ice and have done so both in the firn and bulk ice, at depths from 140 to 500~m depth.

\subsection{Implications for neutrino astronomy and glaciology}

We have determined that the sound speed gradient in deep South Pole ice is consistent with zero, and therefore that the amount of refraction of acoustic waves is consistent with zero.  This is in contrast with most deep ocean sites, where refraction due to a vertical sound speed gradient is a significant challenge for acoustic neutrino detection~\cite{Vandenbroucke05}.

Optical photons are scattered in the ice such that typical photons detected in the AMANDA and IceCube arrays have scattered several times, losing much of their directionality.  Sophisticated algorithms have been developed to reconstruct the neutrino direction and energy, and the interaction location, in the presence of scattering~\cite{Ahrens04}.  Typically these algorithms fit the full scattered waveform shape and then use the rising edge to determine the arrival time of the ``direct'' unscattered photons.

Radio waves are refracted significantly in the firn and negligibly in the deep ice.  The RICE experiment spans both the firn and the bulk ice and therefore must account for refraction in its signal reconstruction, background rejection, and neutrino sensitivity determination.  We note that while radio waves are refracted downward in the firn, acoustic waves are refracted upward.  This means that while surface radio noise is waveguided down to a possible deep detector, surface acoustic noise is refracted back to the surface, such that the firn shields deep sensors from surface noise.  This expectation is confirmed by the observation that SPATS sensor ambient noise levels vary negligibly with time, despite the operation of large construction equipment directly above the array during each South Pole season~\cite{Karg09noise}.

We have shown that acoustic waves, similar to radio waves, propagate unrefracted in deep South Pole ice.  This means the location of an acoustic source can be reconstructed quickly and precisely using analytical methods.  Furthermore, our measurements imply that the acoustic radiation pattern (like the radio radiation pattern) is affected negligibly by refraction.  This unique pattern (a wide, flat ``pancake'') could be used as a signature to separate neutrino events from background events, which are likely to produce a spherical radiation pattern.  The radiation pattern could also be used (along with signal arrival time and amplitude) for neutrino event reconstruction.  For example, the neutrino arrival direction could be estimated by fitting a plane to the hit sensors; the upward normal points to the neutrino source~\cite{Vandenbroucke06Hybrid}.

We note that a similar array to RICE, deployed beneath the firn to avoid refraction, would benefit similarly to the acoustic method from preserved radiation pattern.  In fact, codeploying acoustic and radio arrays in the same volume of ice could allow the two arrays to operate in hybrid mode, detecting a significant fraction of events in coincidence~\cite{Besson05}.

If both a pressure and a shear wave pulse from a neutrino are detected by a single acoustic sensor, the time difference between them could be used to estimate the distance to the source and from this the neutrino energy, with a single sensor.  For distances less than $\sim$100~m, the precision of this reconstruction is dominated by the pulse arrival time resolution.  If the timing resolution is $\sim$0.1~ms, the distance resolution is $\sim$1~m, independent of distance within this ``near'' regime.  For distances larger than $\sim$100~m, the distance reconstruction precision is limited by the precision of our sound speed measurement.  Using the $\sim$1\% sound speed measurement presented here, the distance could be determined with $\sim$1\% precision (that is, the distance precision scales with distance in this ``far'' regime).

Now that the sound speed profile has been determined, it remains to determine the attenuation length and absolute noise level of South Pole ice, to determine its potential for acoustic neutrino detection.  Data taking is ongoing with the SPATS array to achieve this goal.  Furthermore, new data taken with longer transmitter-sensor baselines could provide sound speed measurements with several times greater precision than presented here.

In addition to its neutrino astronomy applications, this measurement has glaciology applications.  The horizontal and vertical measurements agree to better than 2\%, while the sound speed in monocrystalline ice is known to vary by 7\% with propagation direction relative to grain size.  This indicates there is not significant grain alignment in the South Pole fabric.  Furthermore, the sound speed profile can be used to track both artificially produced sounds (e.g. from explosives) and naturally occurring ones (e.g. seismic or acoustic emission events), both of which can contribute to our understanding of ice flow.  SPATS is now operating in a mode to trigger on ambient impulsive transients.  Such transients have been detected in coincidence between multiple sensors hundreds of meters apart.  The mechanism of these sounds is not yet understood but they could be due to stick-slip glacial flow at the bedrock interface or to (micro)cracks occurring in the bulk ice.  Using the sound speed results presented here, the locations of these acoustic sources can be localized precisely, information that could yield new insights into ice sheet dynamics.

\chapter{Gaussian noise}

\label{noiseChapter}

\noindent\emph{In this chapter we summarize the background noise detected in the SPATS sensor channels.  The noise is very well described by a normal distribution, and is very stable over time.  These are beneficial features for operating a transient acoustic detector.  We show the time evolution of the noise level in the sensors, both over a multi-year time period and during the initial freeze in process, and show the noise distribution on each individual channel.  For reference we tabulate the Gaussian noise parameters of each channels.}

\section{Gaussian noise distributions (as of May 2009)}

\label{currentNoiseHistos}

\label{noiseStatus}

\begin{figure}[tbp]
\begin{center}
\includegraphics[angle = 0, width = 1\textwidth]{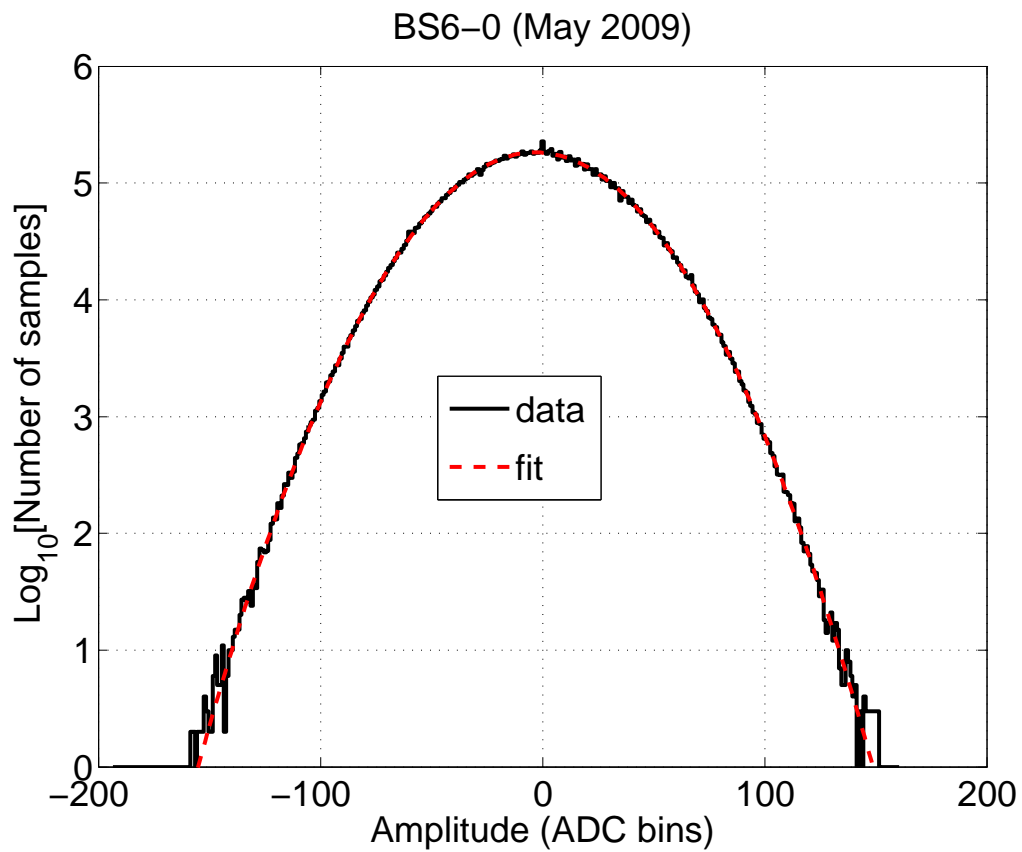}
\end{center}
\caption[Gaussian noise distribution for one example sensor channel]{Noise amplitude histogram for an example channel, BS6-0.  The histogram binning is exactly the same as the ADC binning to avoid artifacts.  The Gaussian fit is shown in addition to the histogram.  Like most channels, the noise on this channel is very well described by a normal distribution, over five orders of magnitude in amplitude counts.  Note the log scale.}
\label{gaussianHistogramBS6-0}
\end{figure}


\subfiglabelskip=0pt		
\begin{figure}
\begin{center}
\subfigure[AS1-0]{
\noindent\includegraphics[width=7pc]{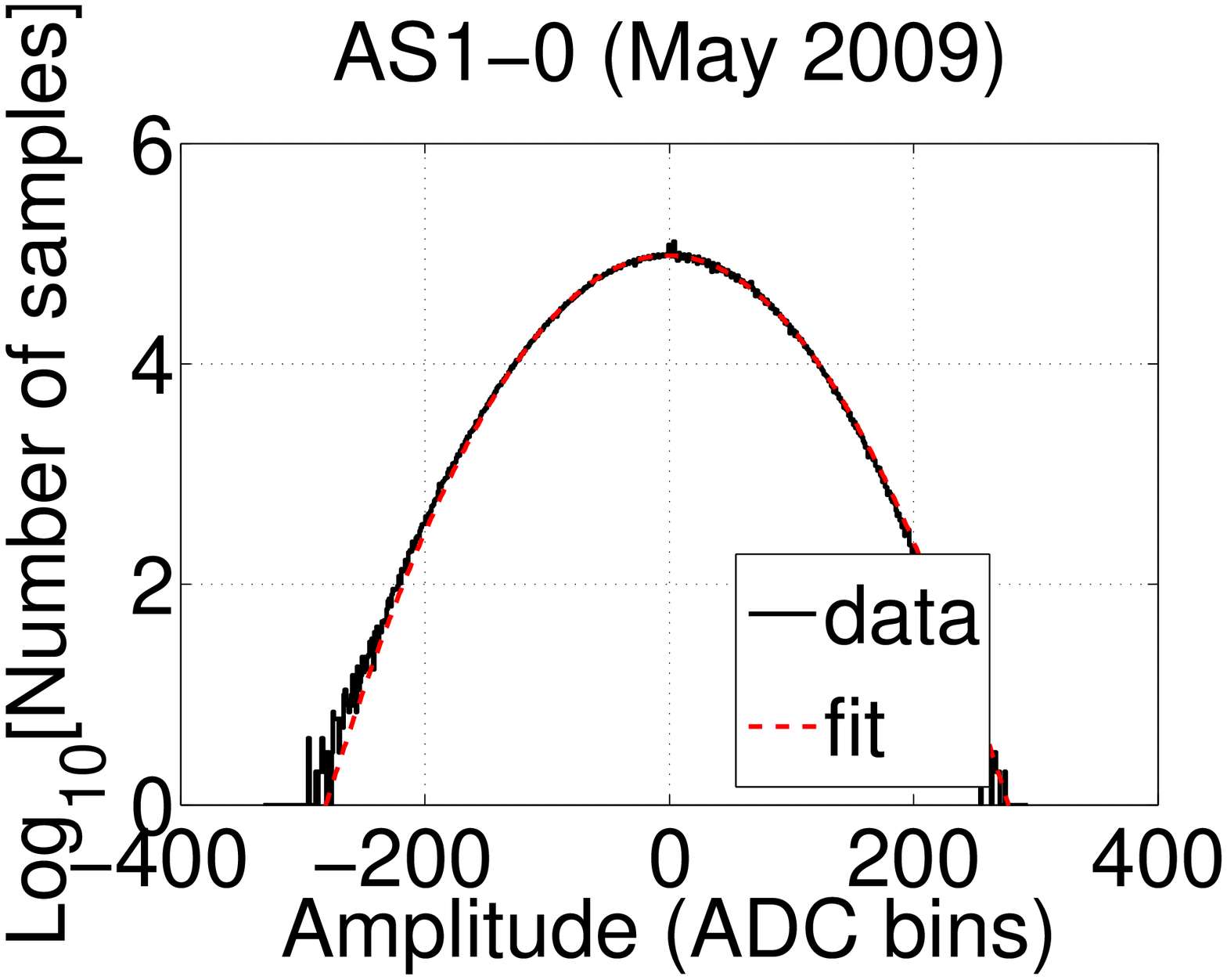}
}
\subfigure[AS1-1]{
\noindent\includegraphics[width=7pc]{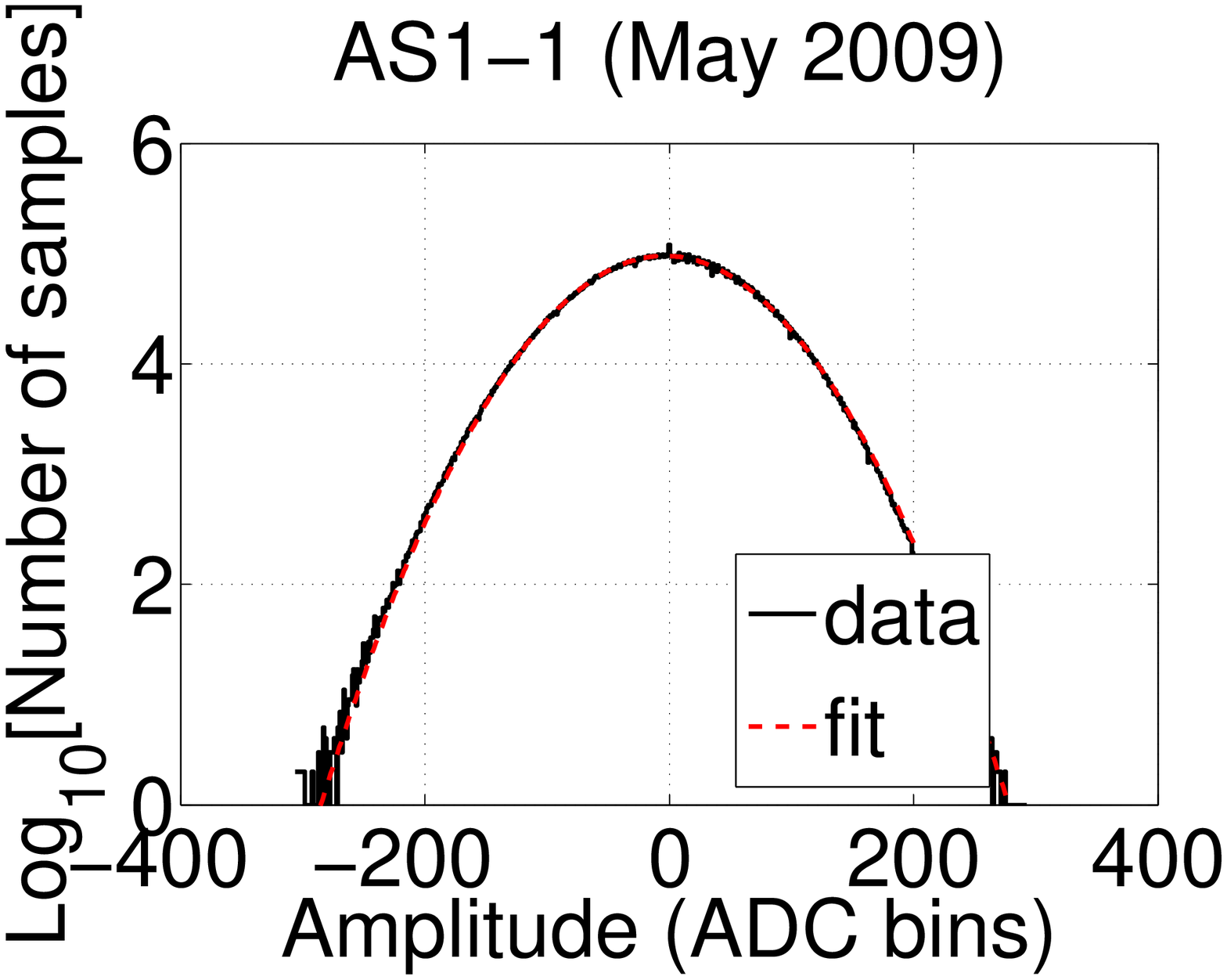}
}
\subfigure[AS1-2]{
\noindent\includegraphics[width=7pc]{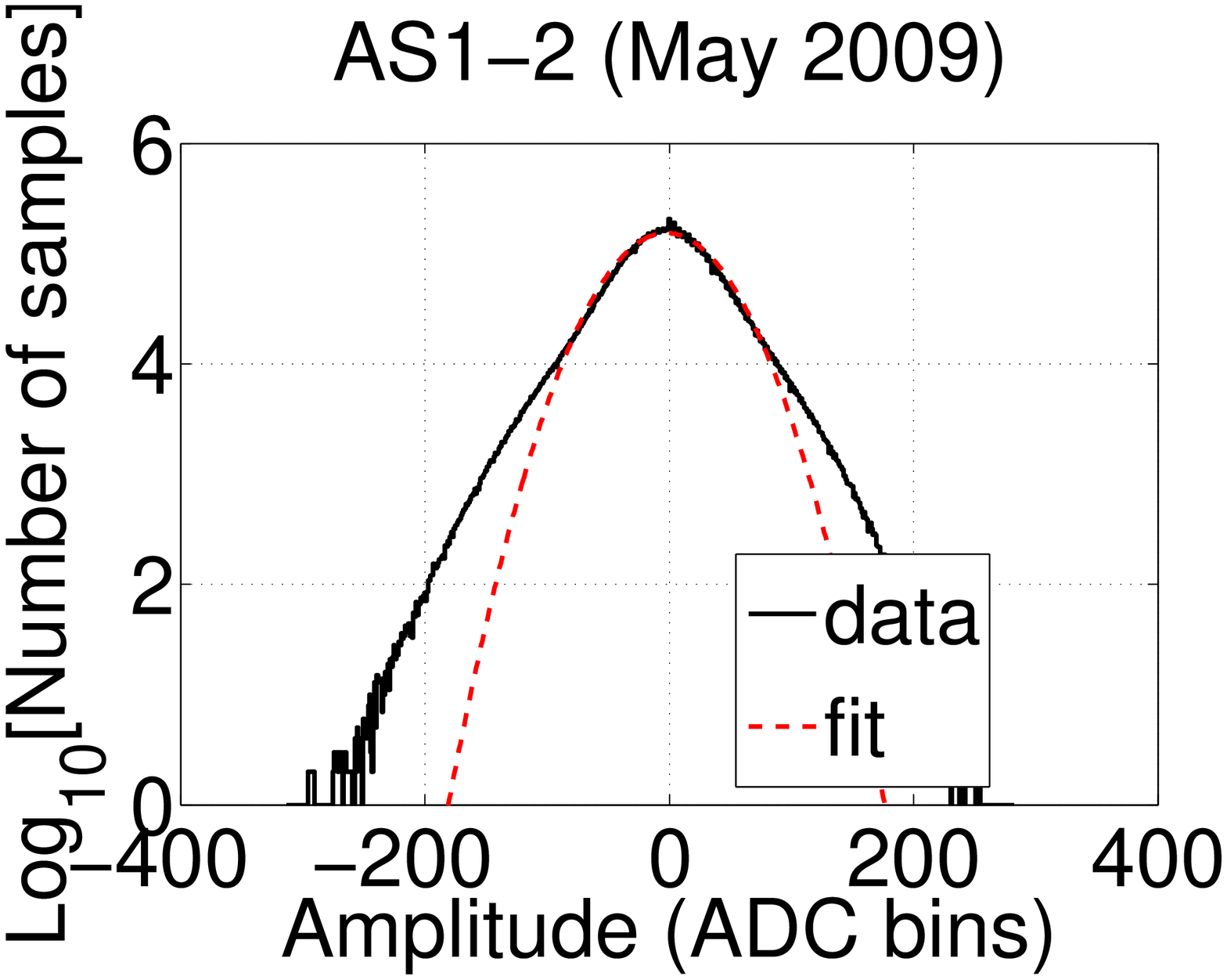}
}
\subfigure[AS2-0]{
\noindent\includegraphics[width=7pc]{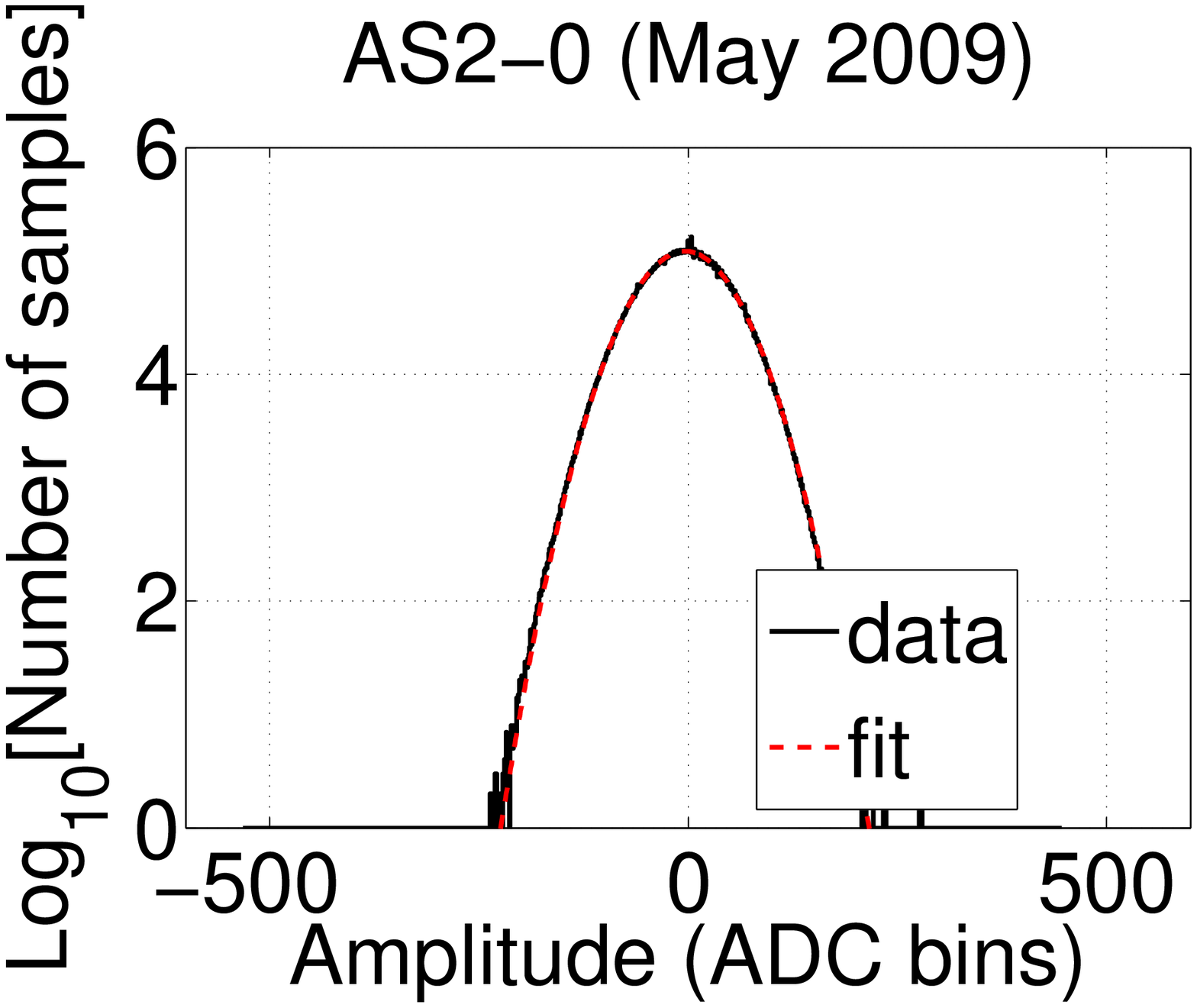}
}
\subfigure[AS2-1]{
\noindent\includegraphics[width=7pc]{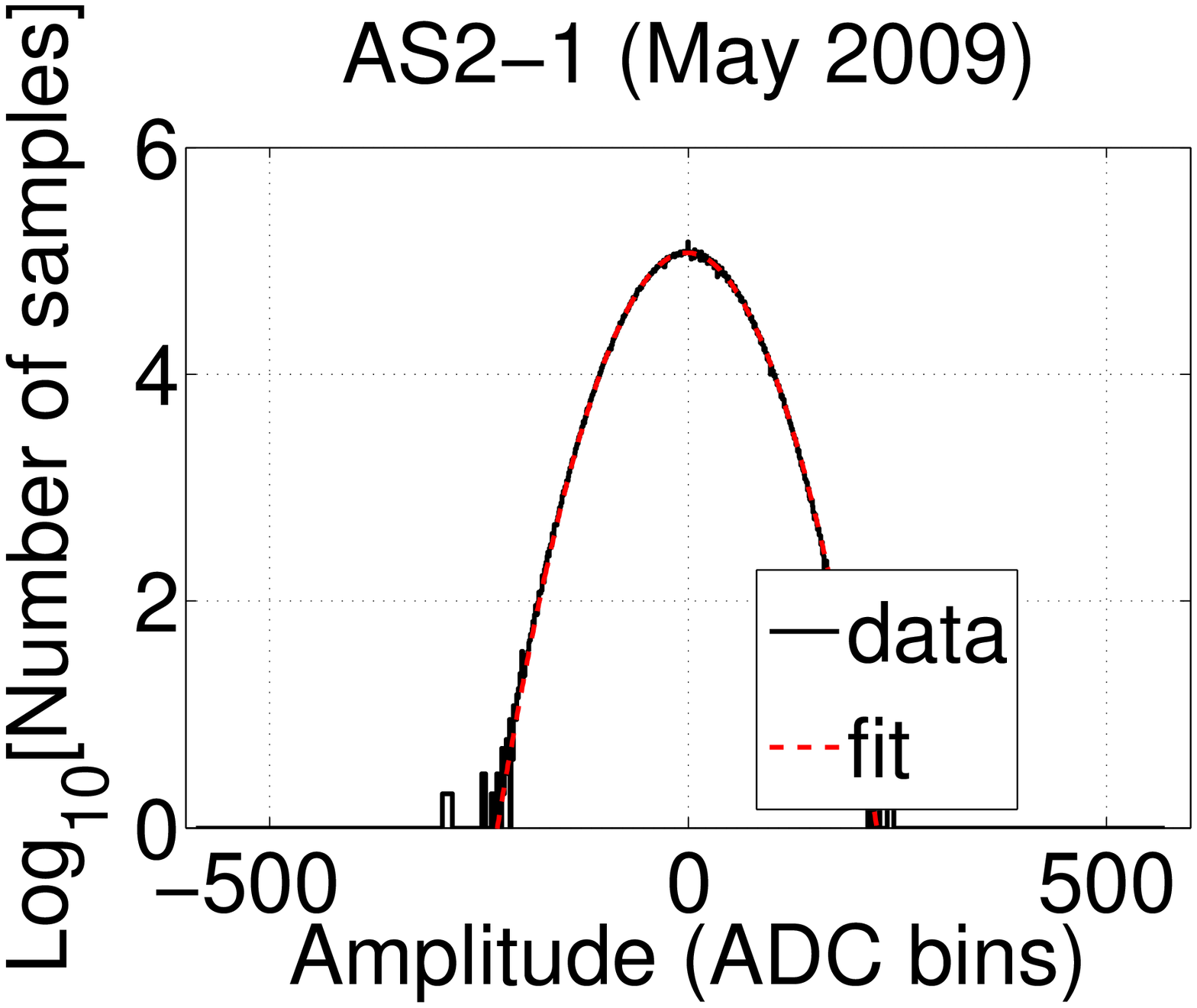}
}
\subfigure[AS2-2]{
\noindent\includegraphics[width=7pc]{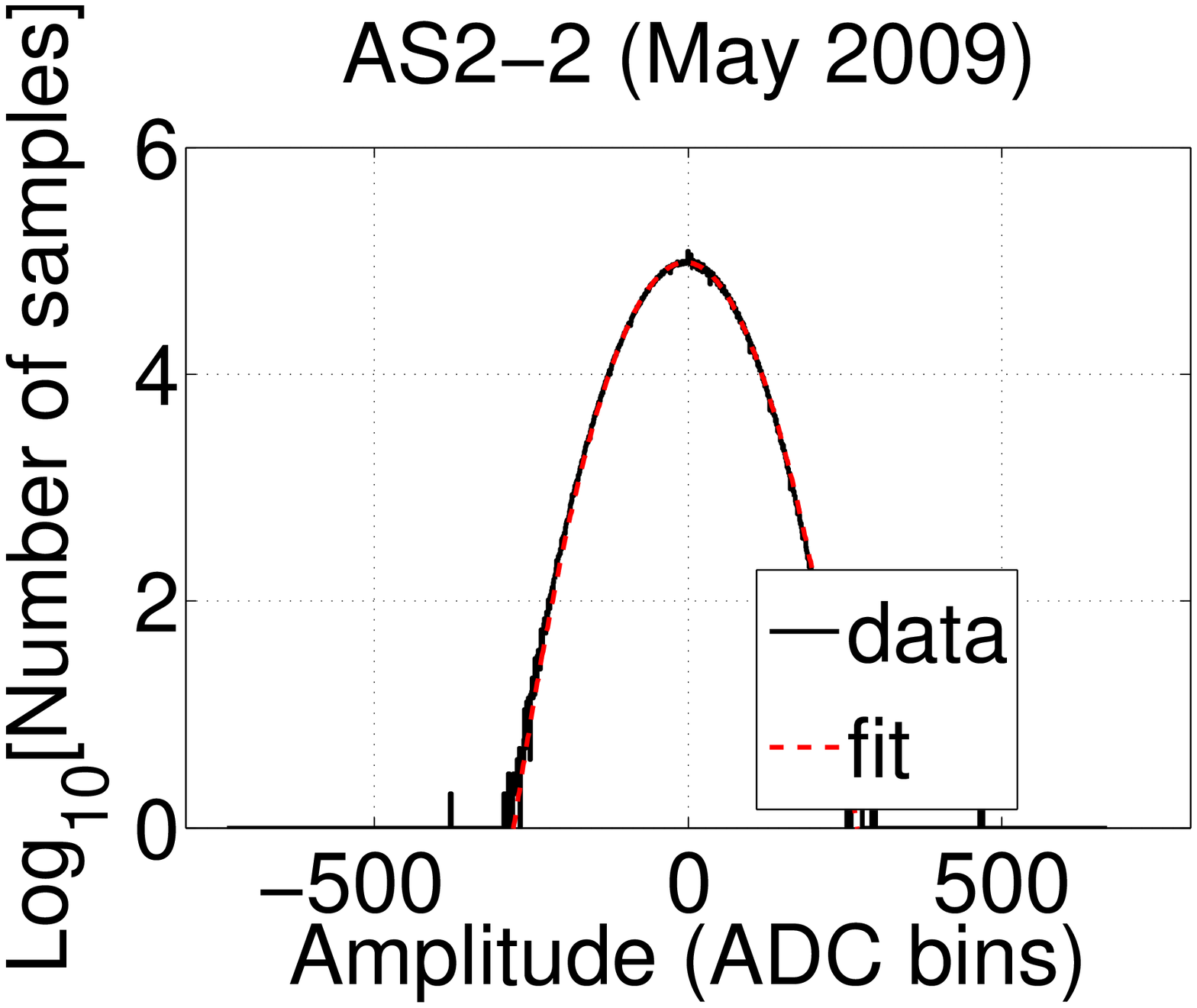}
}
\subfigure[AS3-0]{
\noindent\includegraphics[width=7pc]{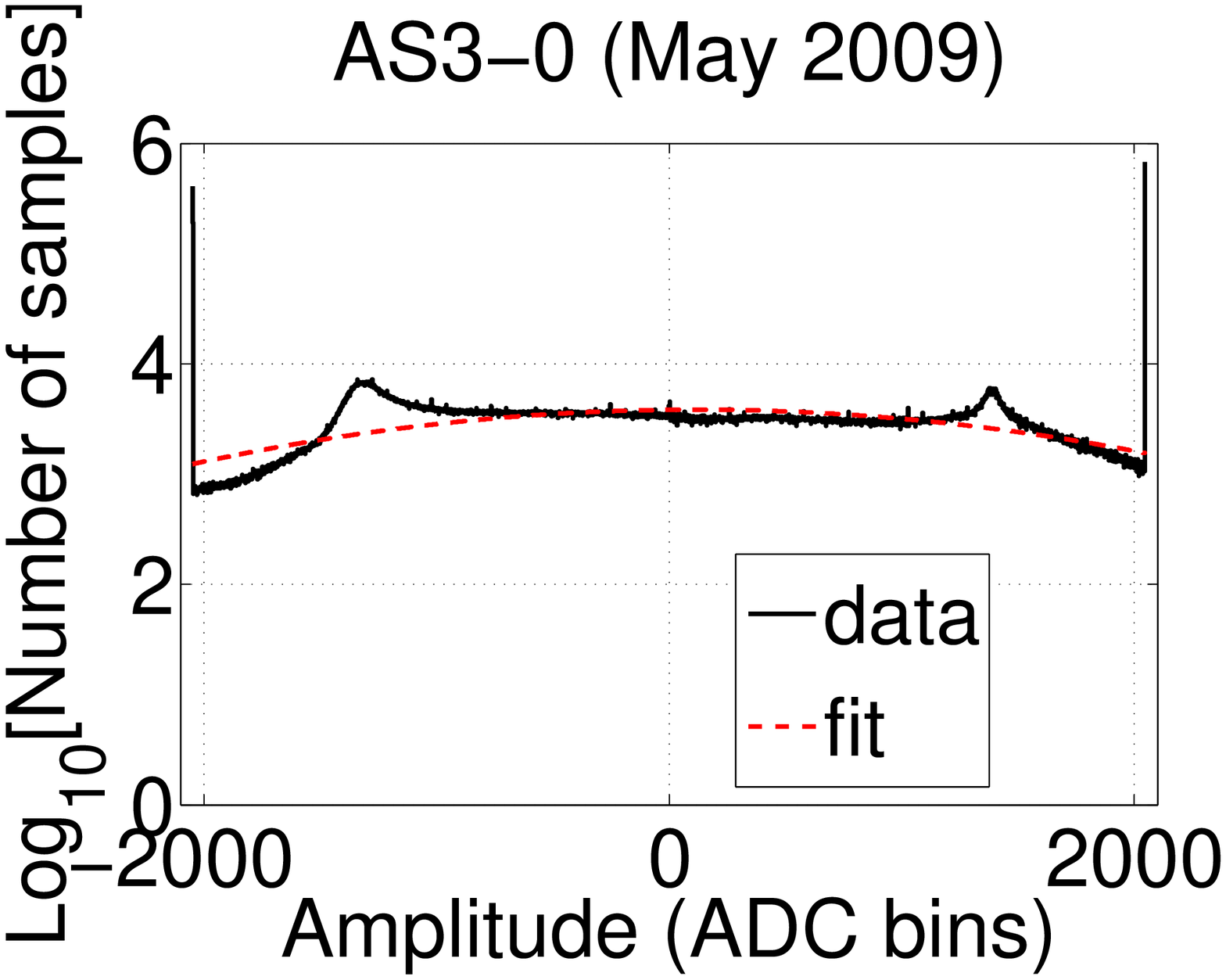}
}
\subfigure[AS3-1]{
\noindent\includegraphics[width=7pc]{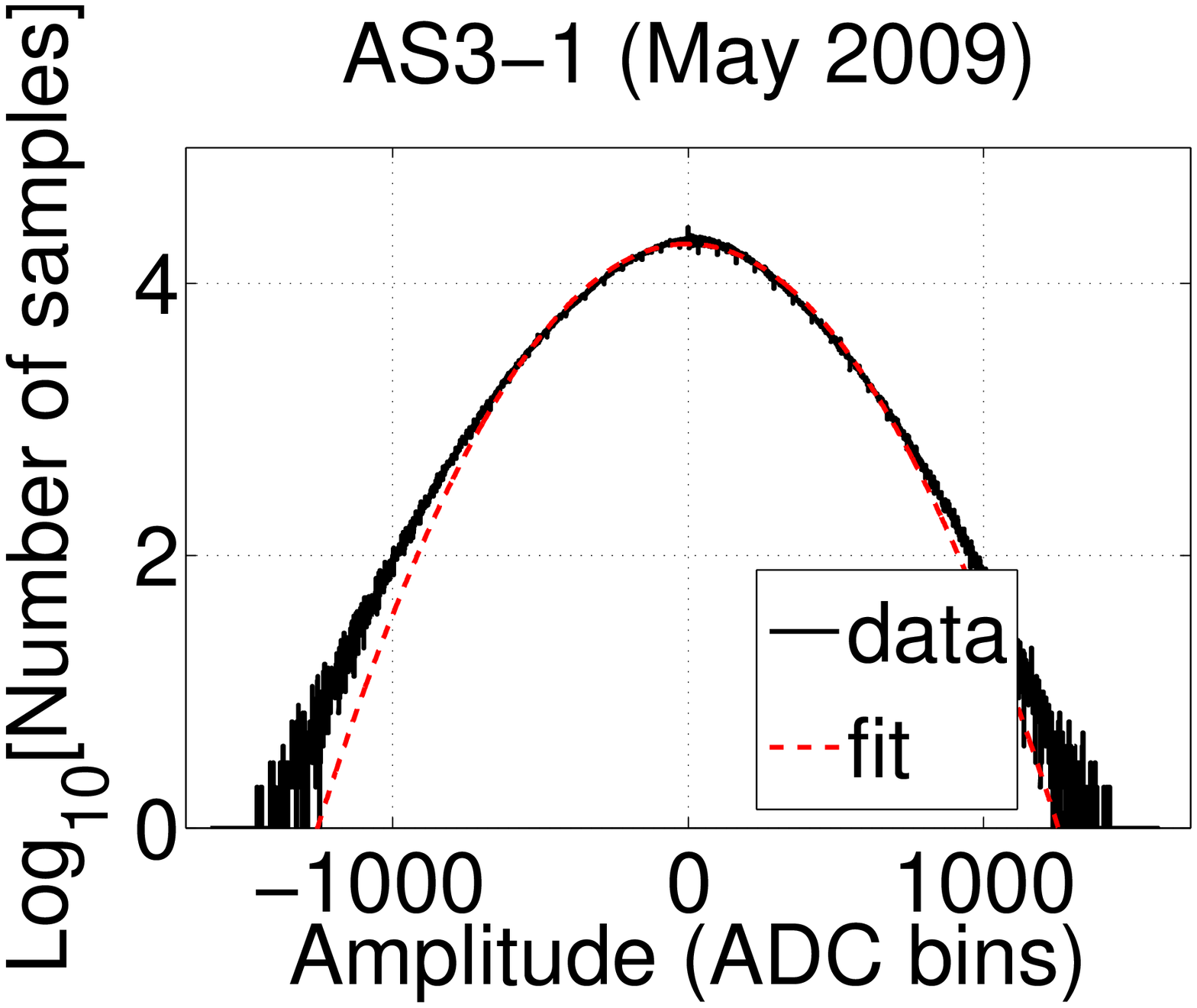}
}
\subfigure[AS3-2]{
\noindent\includegraphics[width=7pc]{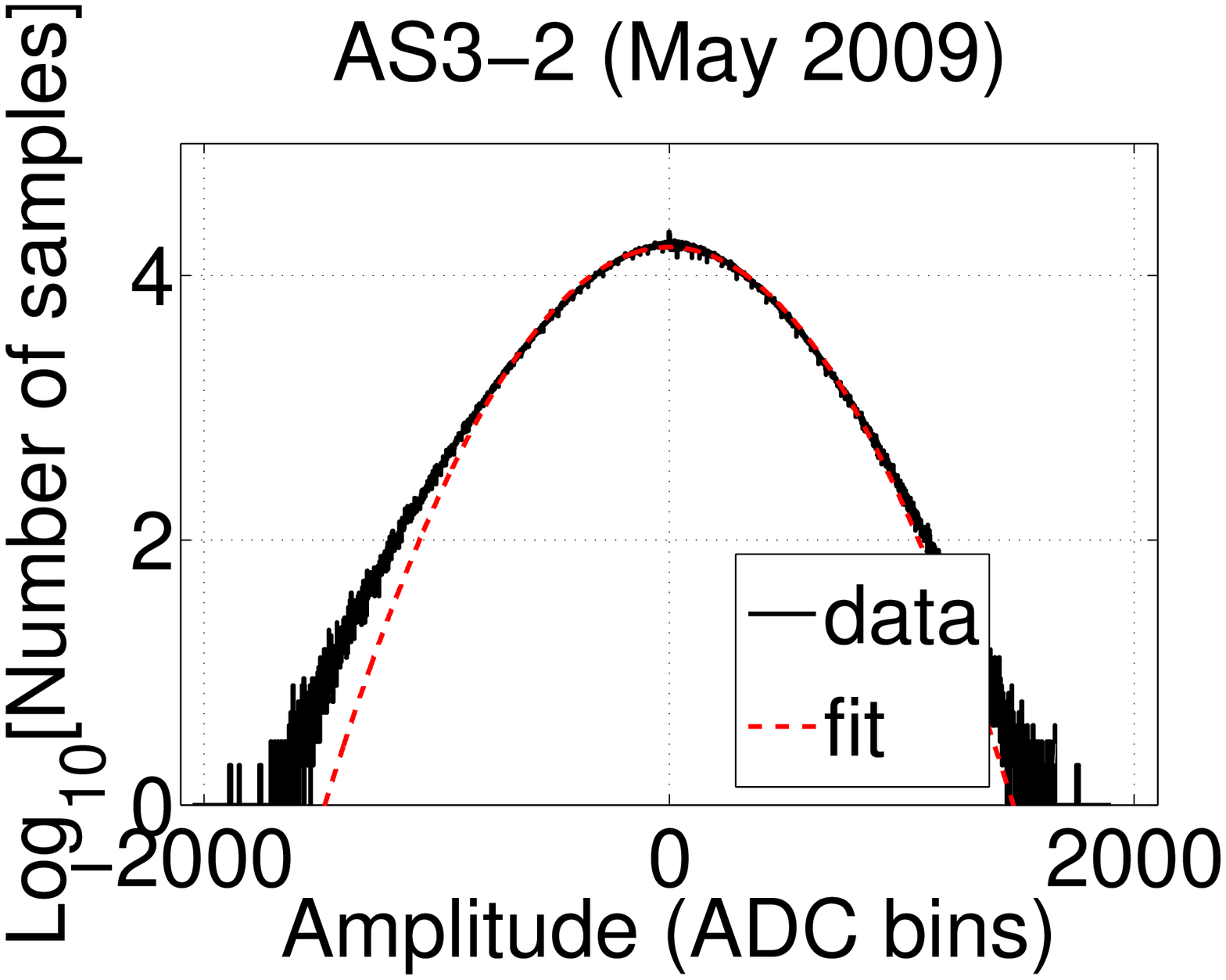}
}
\subfigure[AS4-0]{
\noindent\includegraphics[width=7pc]{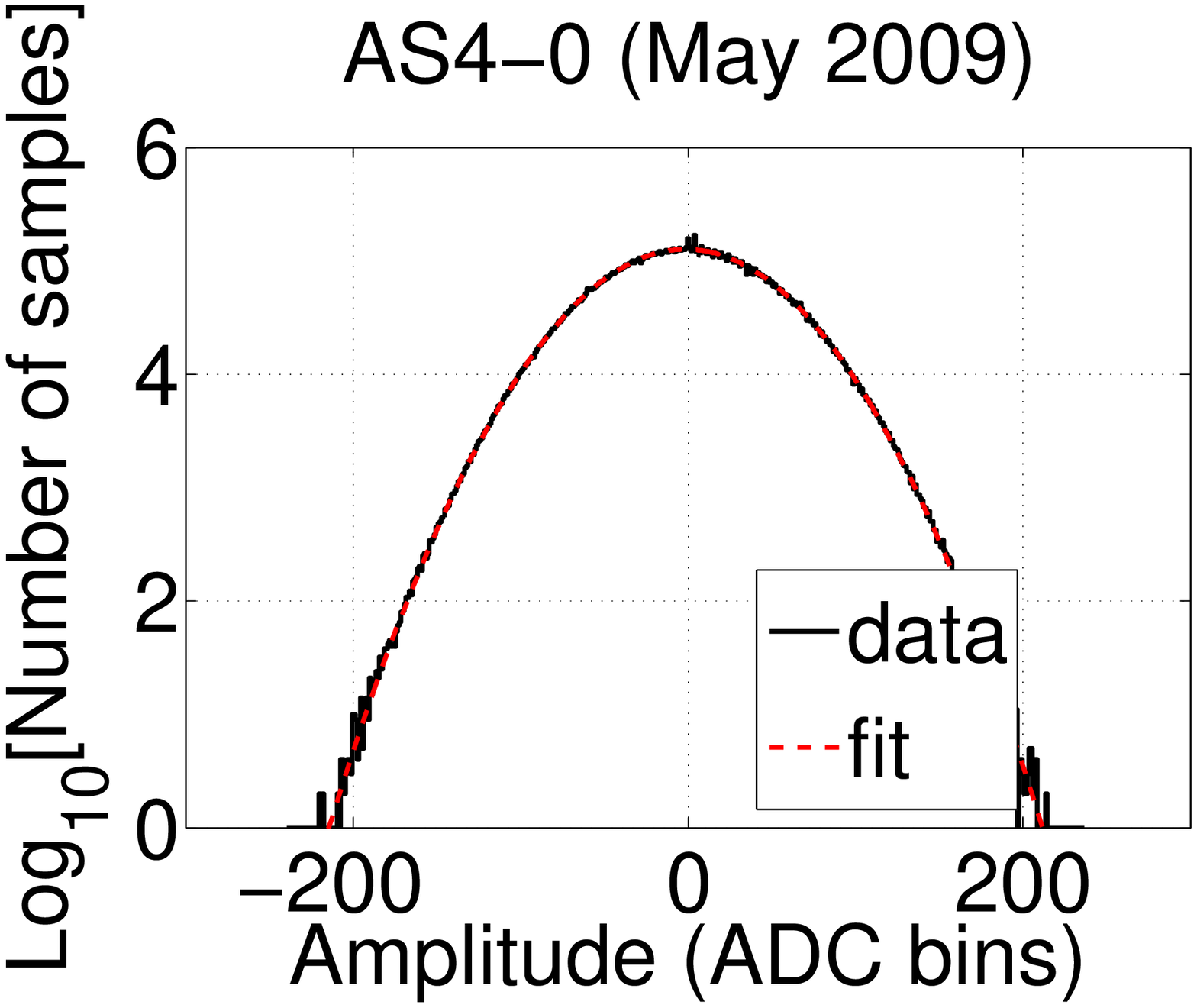}
}
\subfigure[AS4-1]{
\noindent\includegraphics[width=7pc]{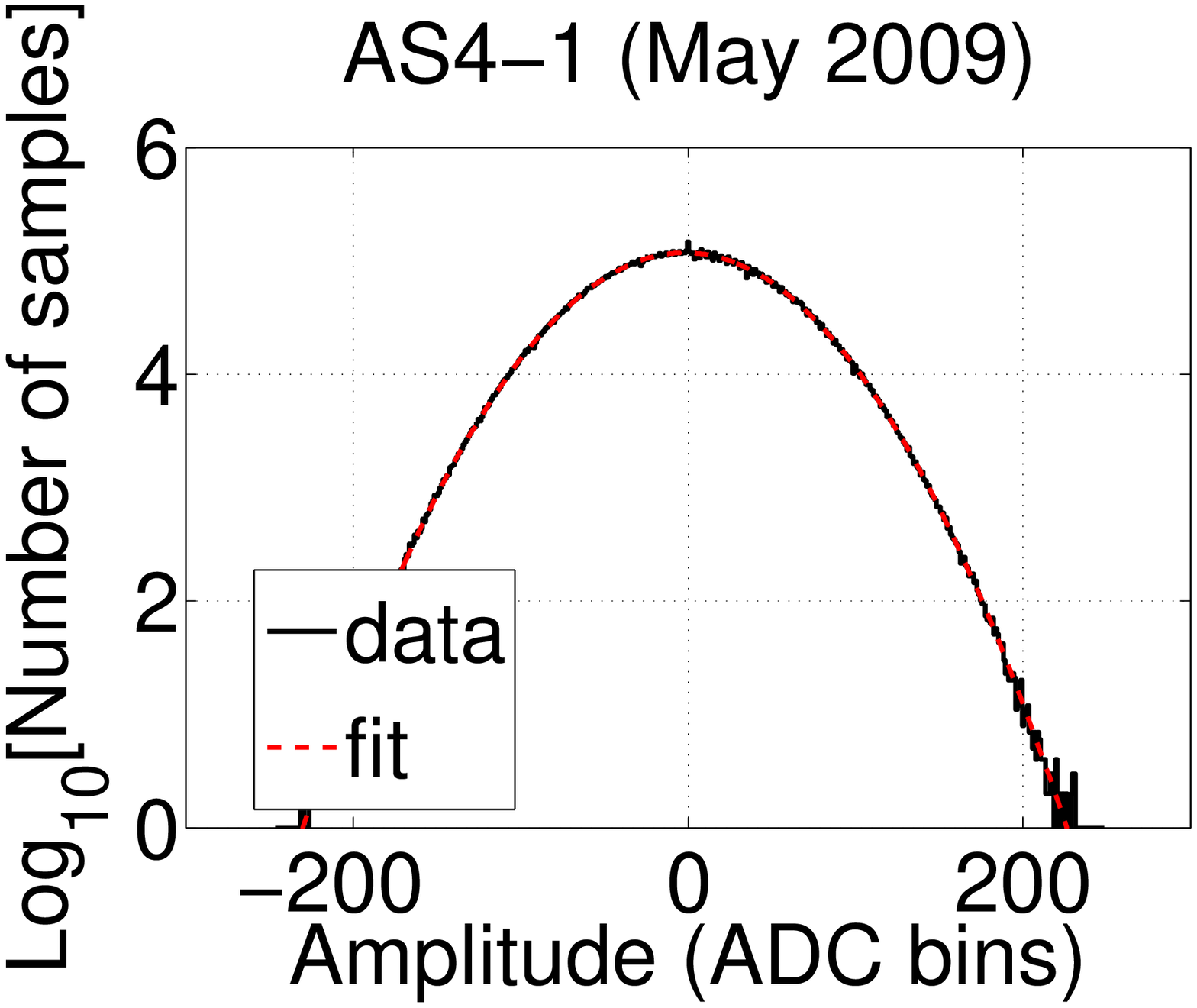}
}
\subfigure[AS4-2]{
\noindent\includegraphics[width=7pc]{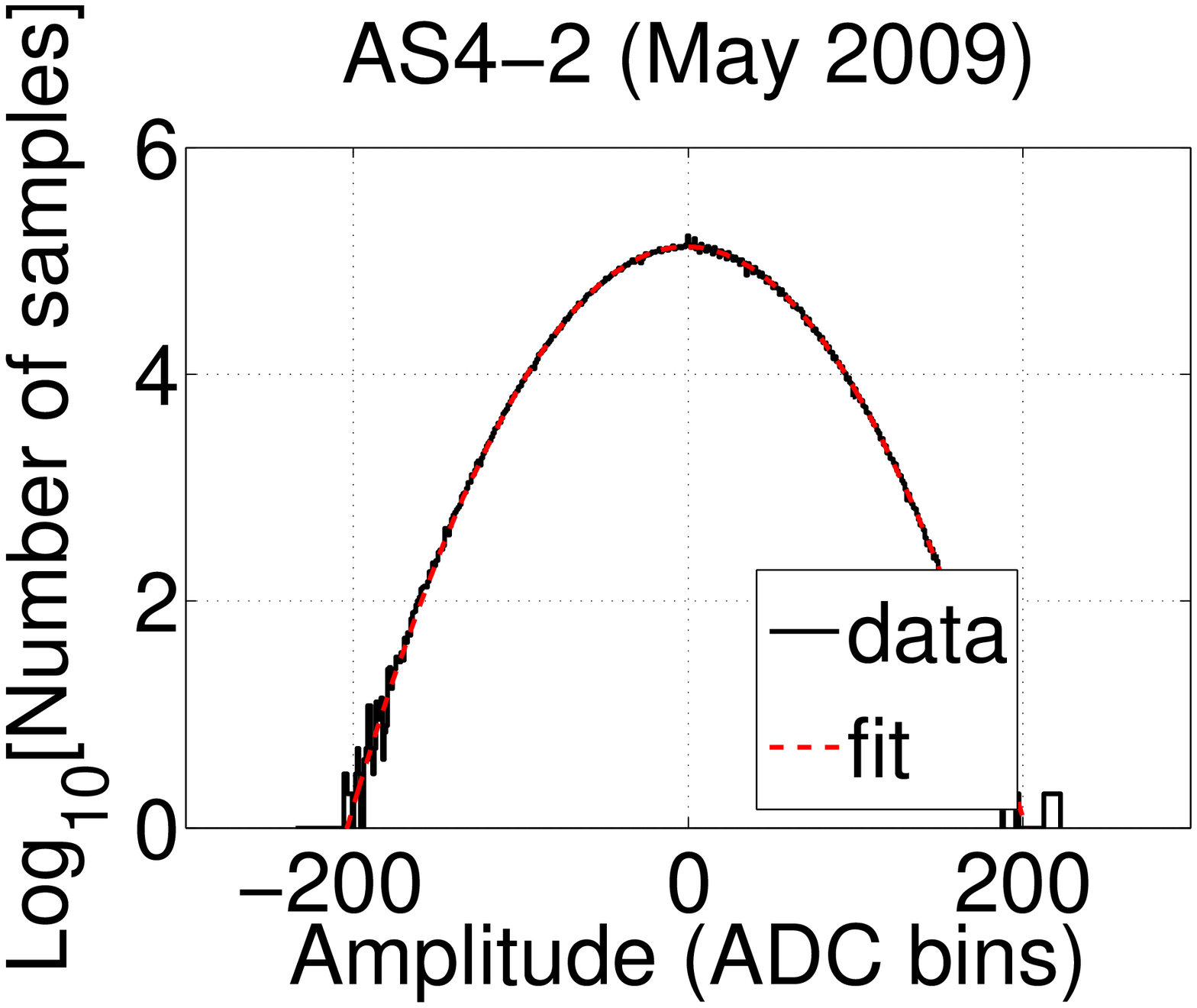}
}
\subfigure[AS5-0]{
\noindent\includegraphics[width=7pc]{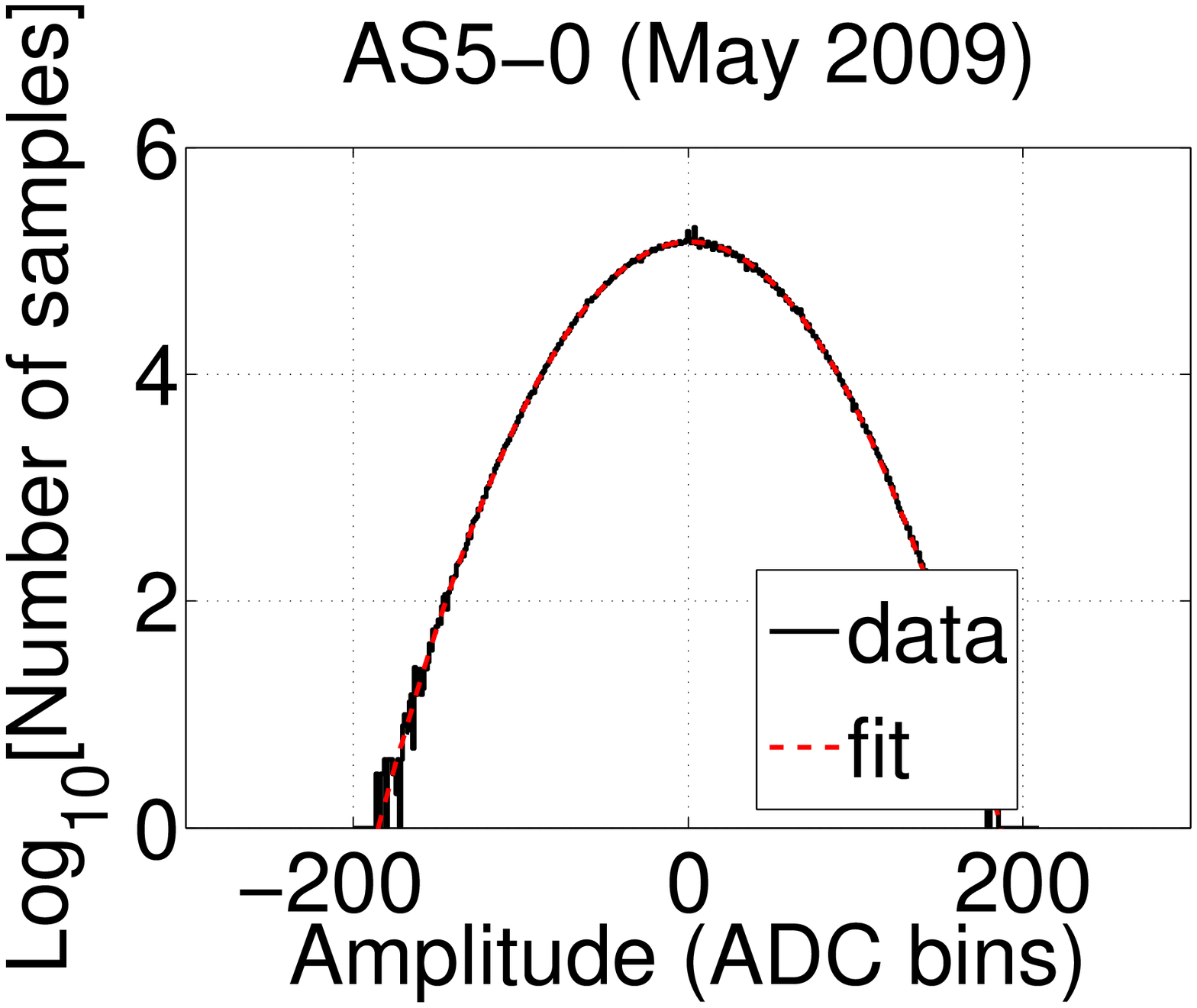}
}
\subfigure[AS5-1]{
\noindent\includegraphics[width=7pc]{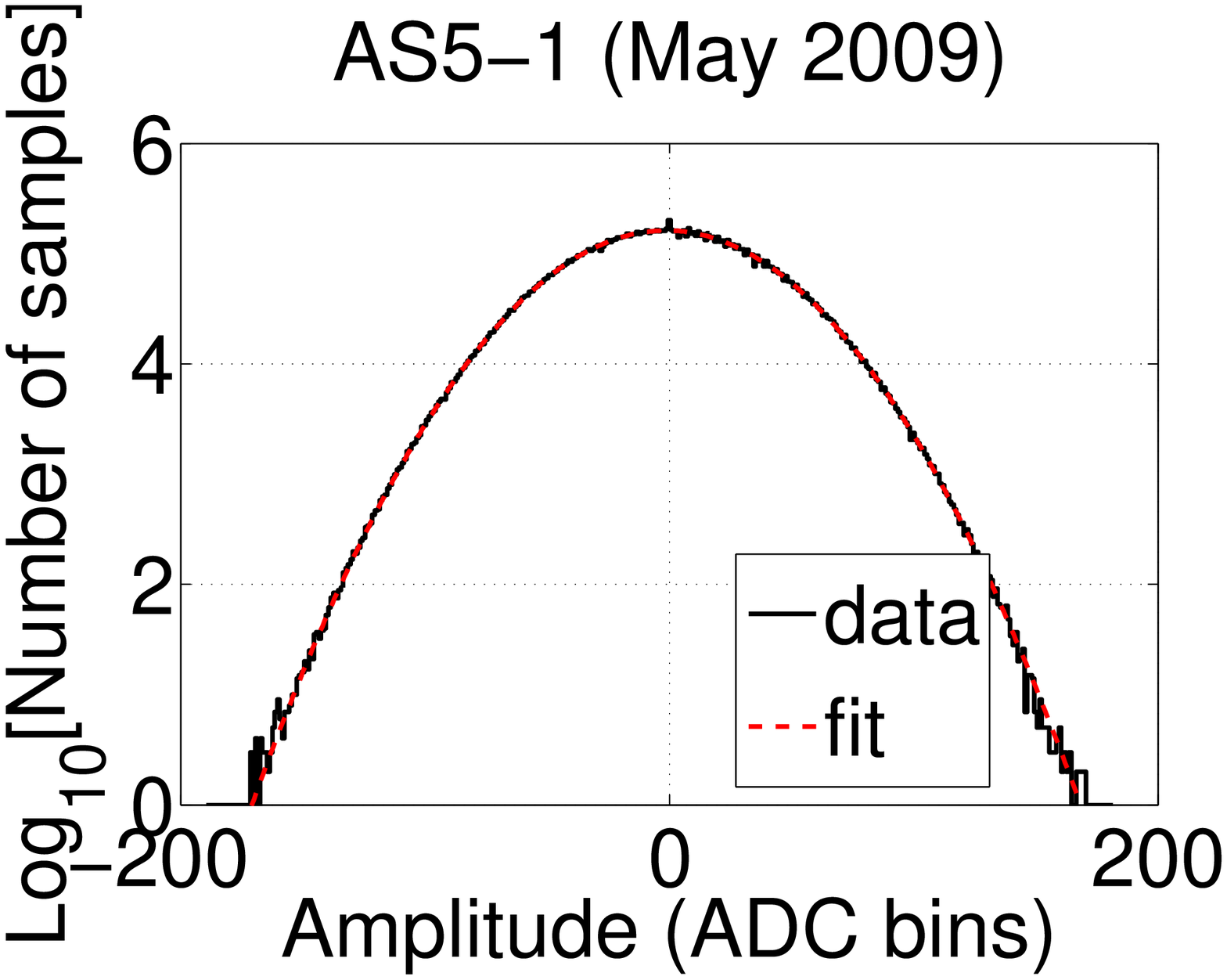}
}
\subfigure[AS5-2]{
\noindent\includegraphics[width=7pc]{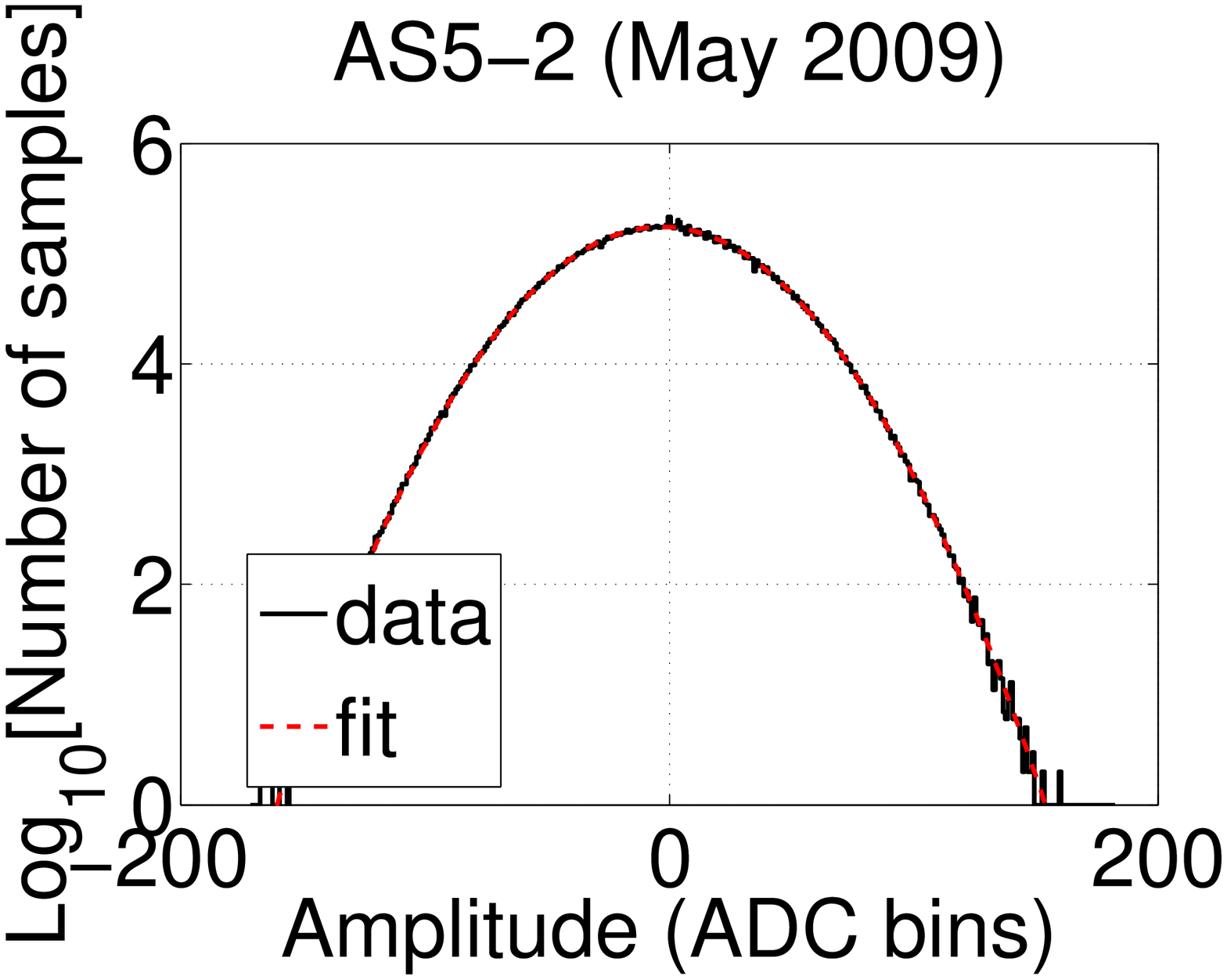}
}
\subfigure[AS6-0]{
\noindent\includegraphics[width=7pc]{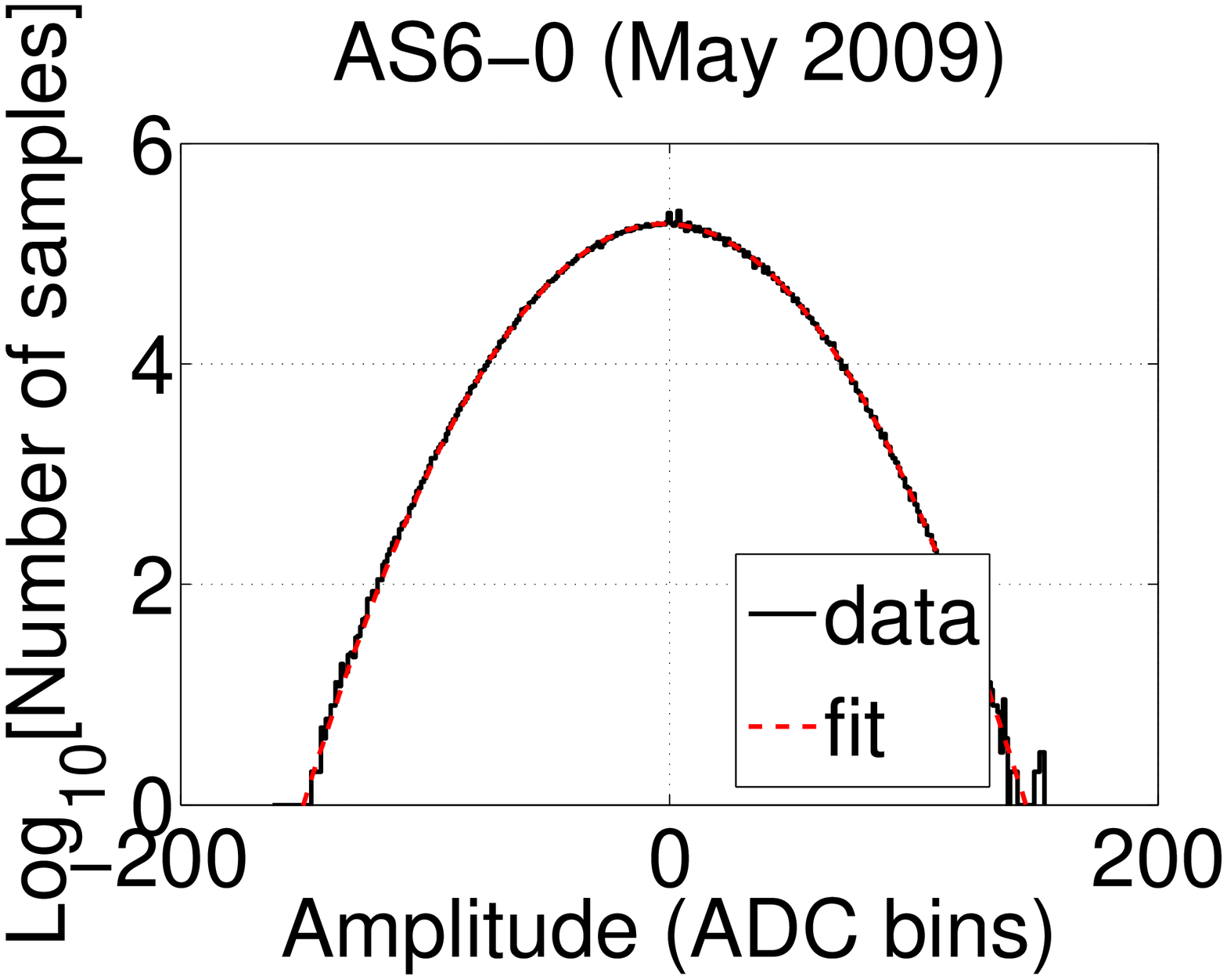}
}
\subfigure[AS6-1]{
\noindent\includegraphics[width=7pc]{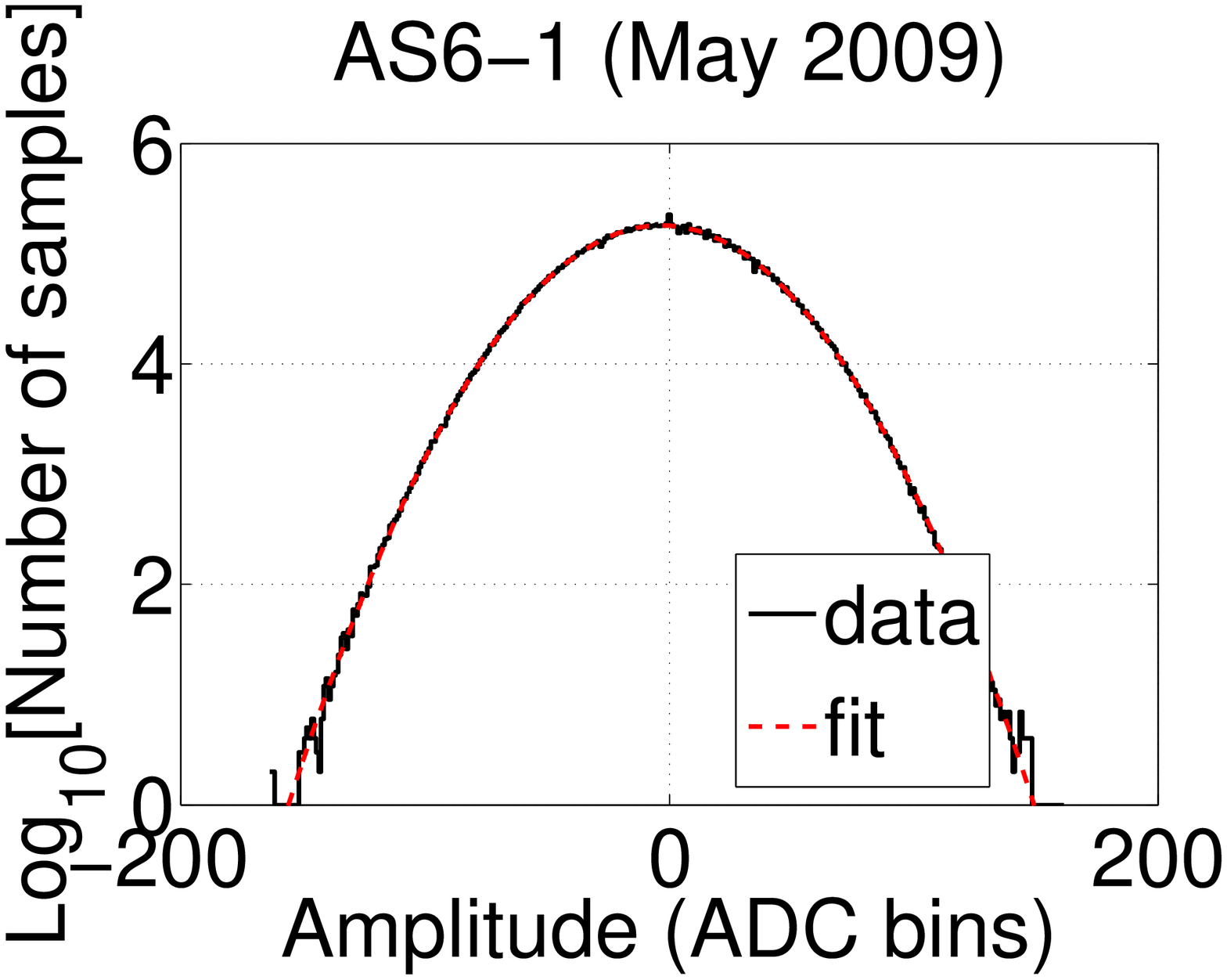}
}
\subfigure[AS6-2]{
\noindent\includegraphics[width=7pc]{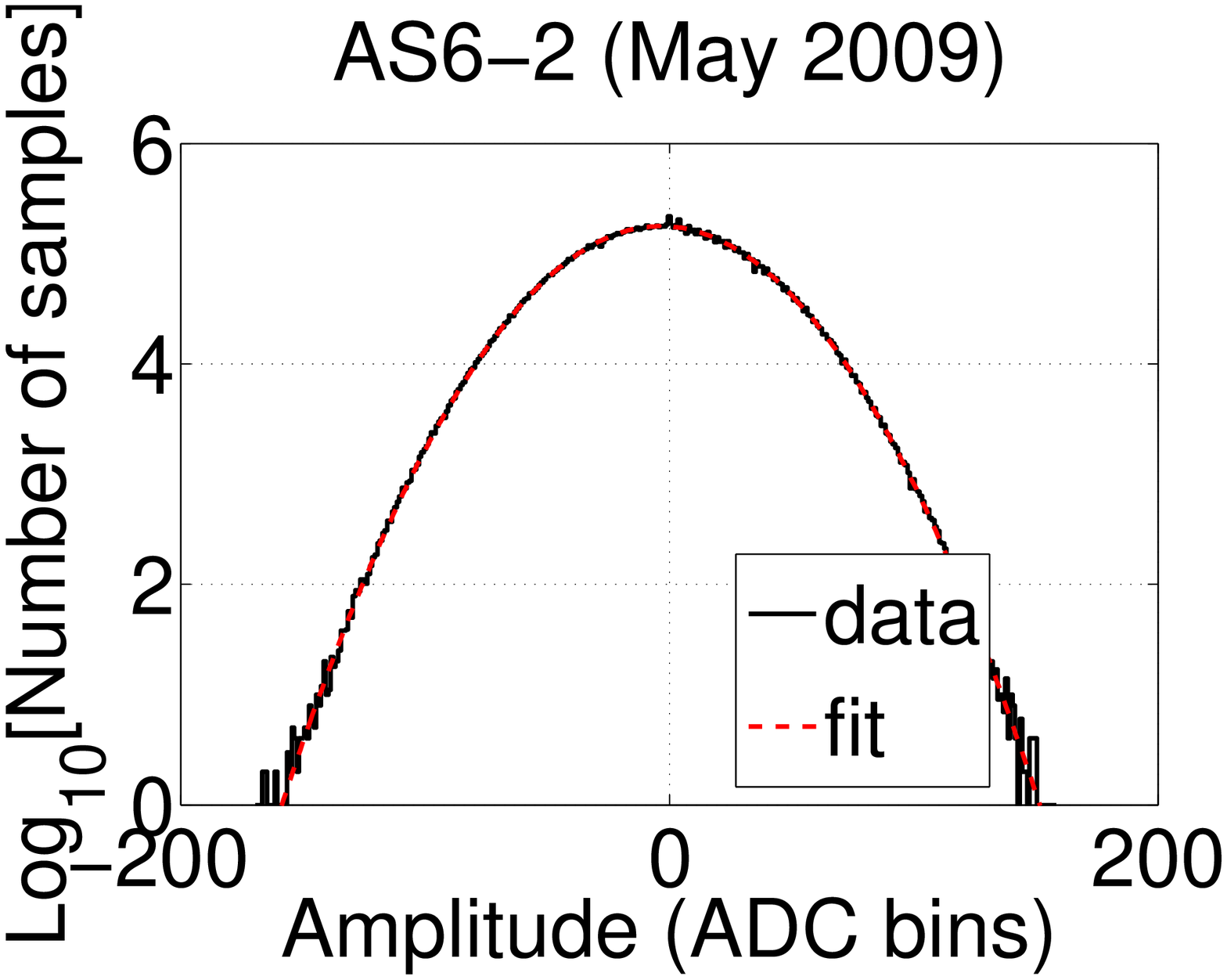}
}
\subfigure[AS7-0]{
\noindent\includegraphics[width=7pc]{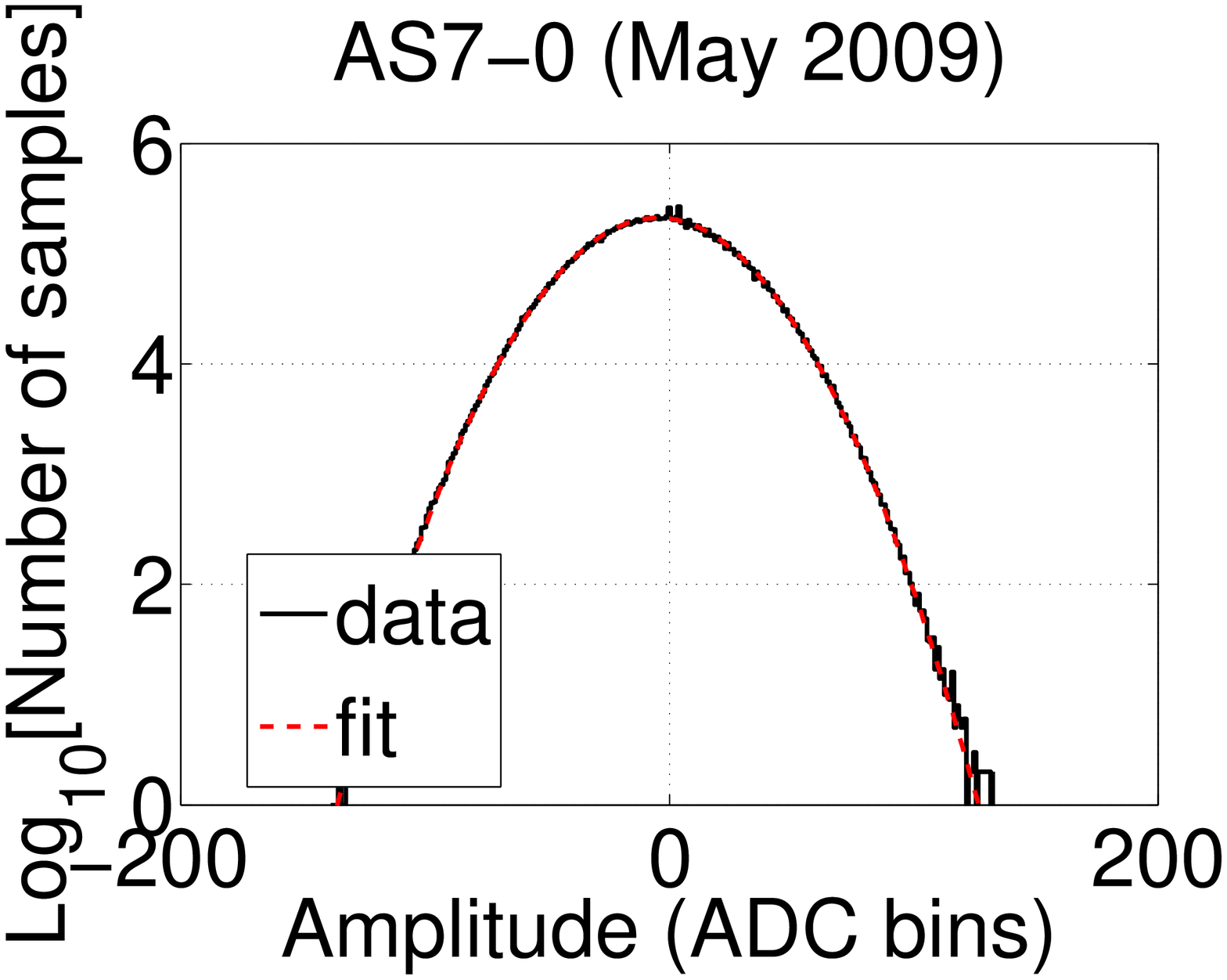}
}
\subfigure[AS7-1]{
\noindent\includegraphics[width=7pc]{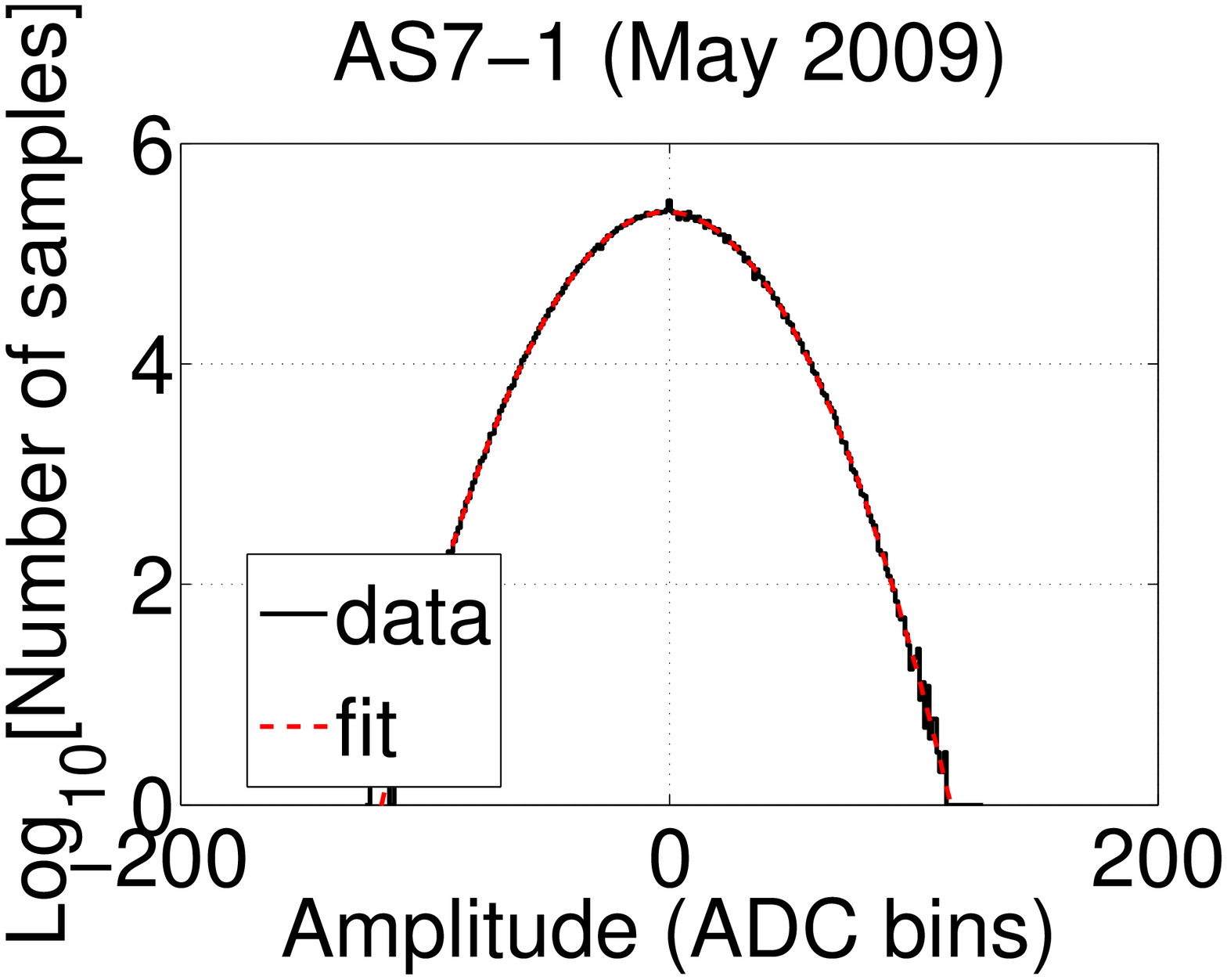}
}
\subfigure[AS7-2]{
\noindent\includegraphics[width=7pc]{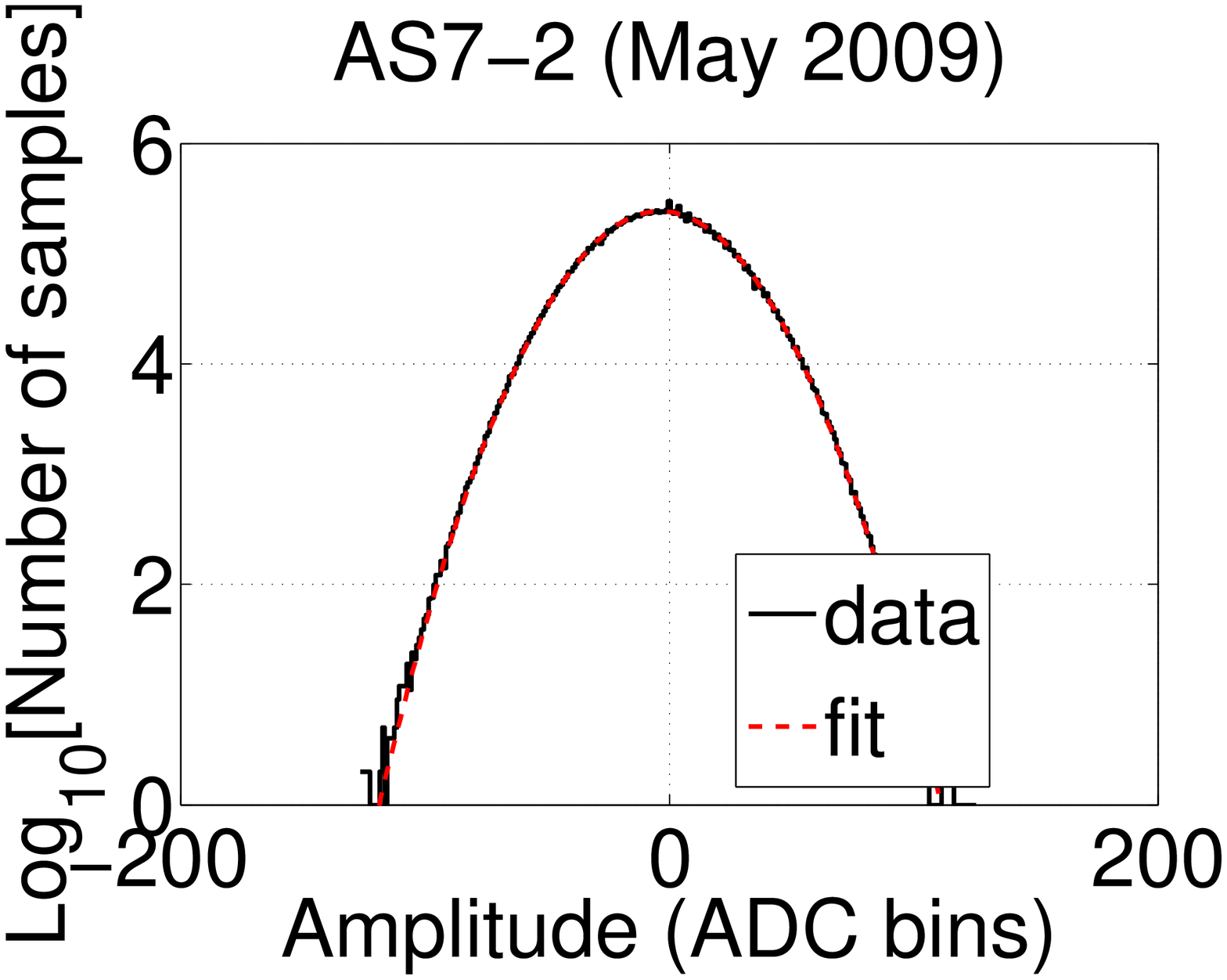}
}
\caption[Gaussian noise distributions for all String A channels]{Noise amplitude histogram (with Gaussian fit) for each channel of String A.}
\label{gaussianHistogramsA}
\end{center}
\end{figure}

\begin{figure}
\begin{center}
\subfigure[BS1-0]{
\noindent\includegraphics[width=7pc]{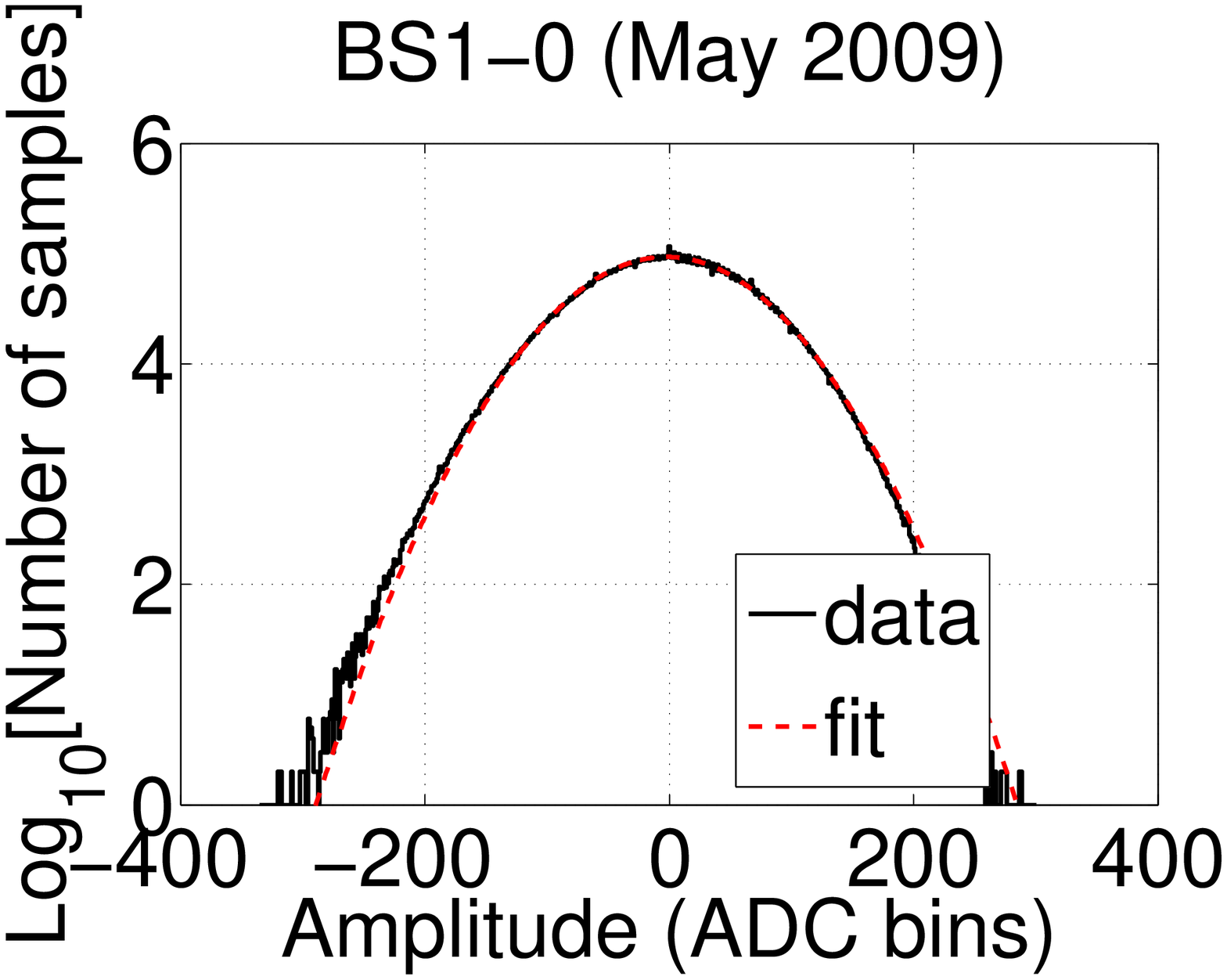}
}
\subfigure[BS1-1]{
\noindent\includegraphics[width=7pc]{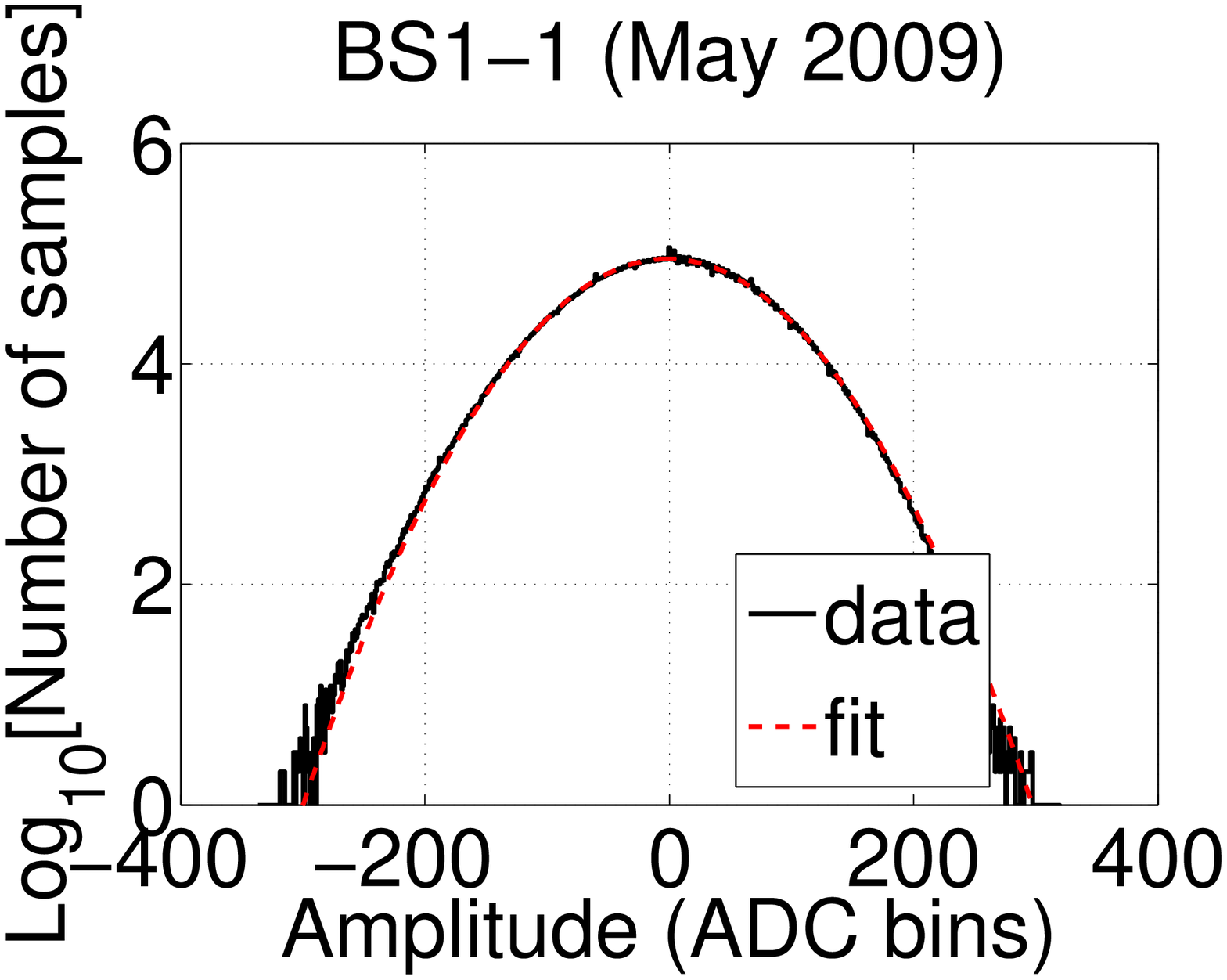}
}
\subfigure[BS1-2]{
\noindent\includegraphics[width=7pc]{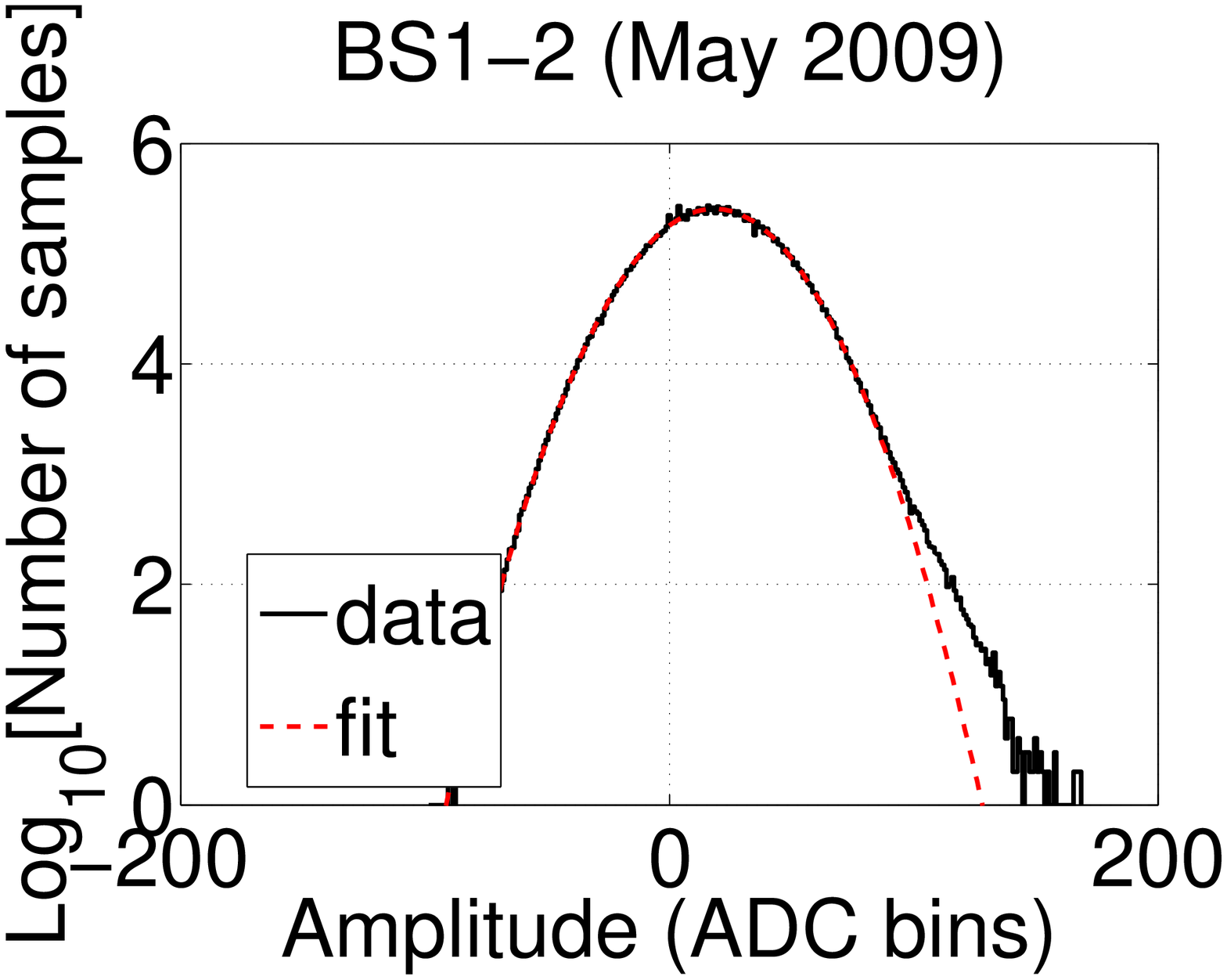}
}
\subfigure[BS2-0]{
\noindent\includegraphics[width=7pc]{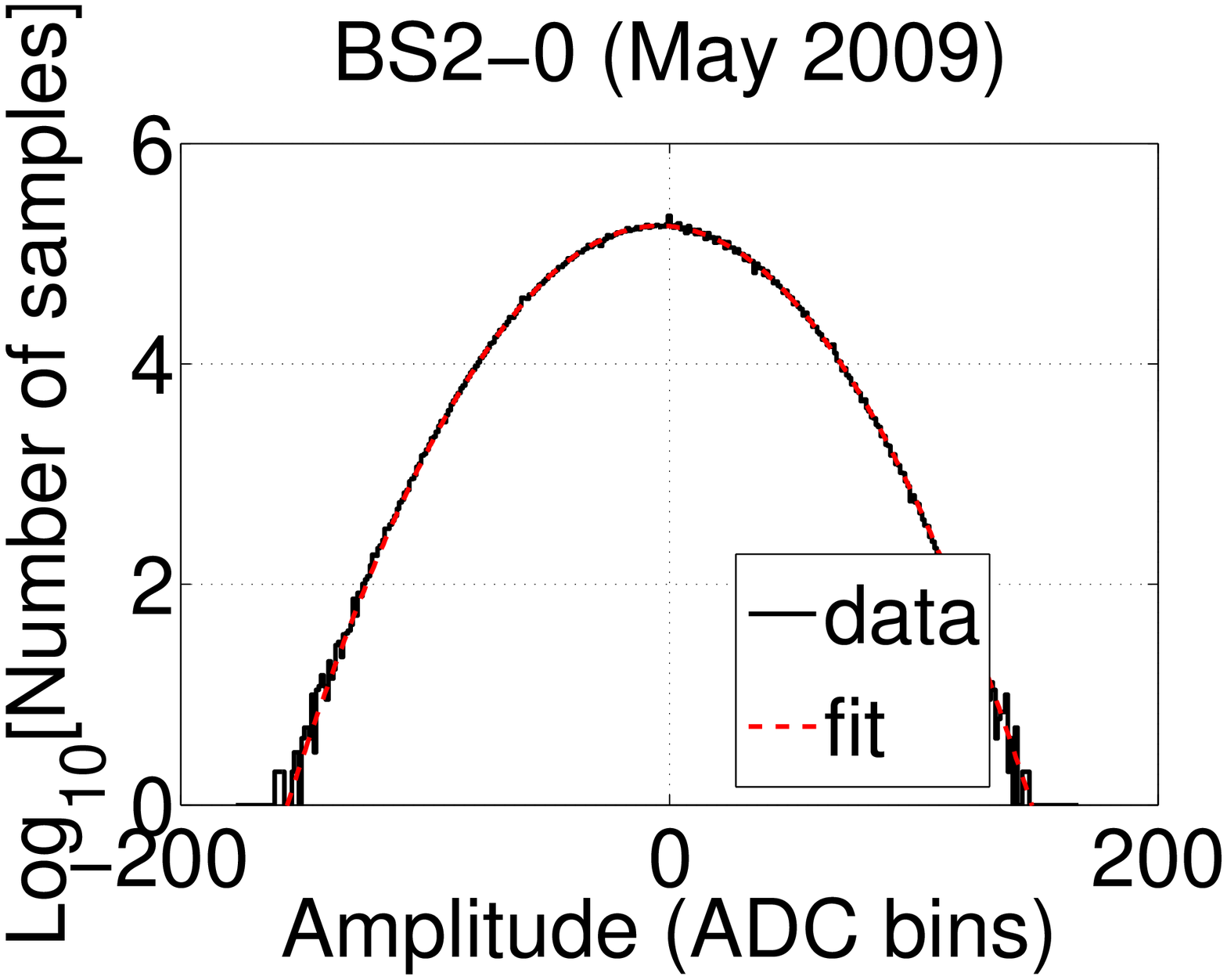}
}
\subfigure[BS2-1]{
\noindent\includegraphics[width=7pc]{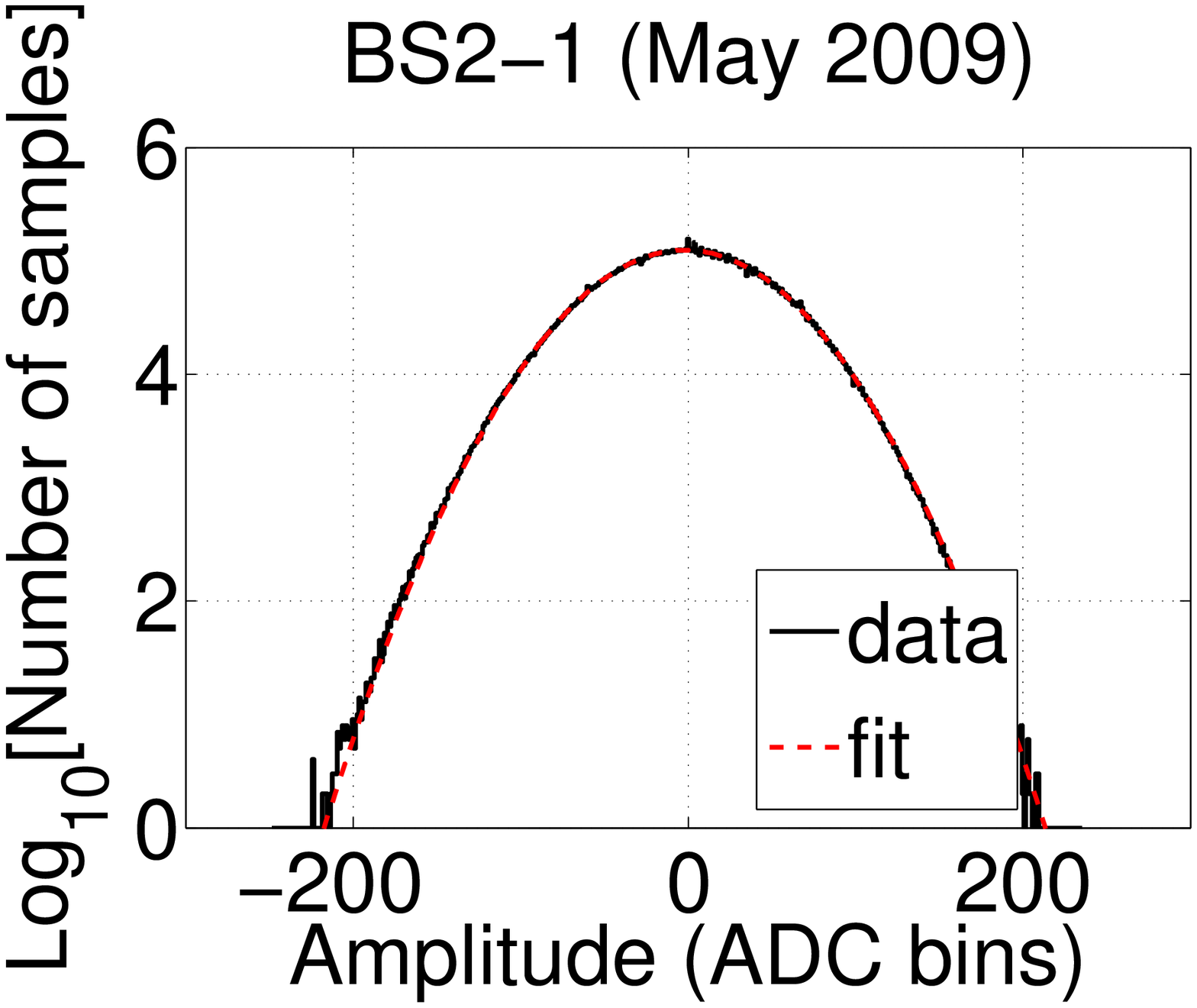}
}
\subfigure[BS2-2]{
\noindent\includegraphics[width=7pc]{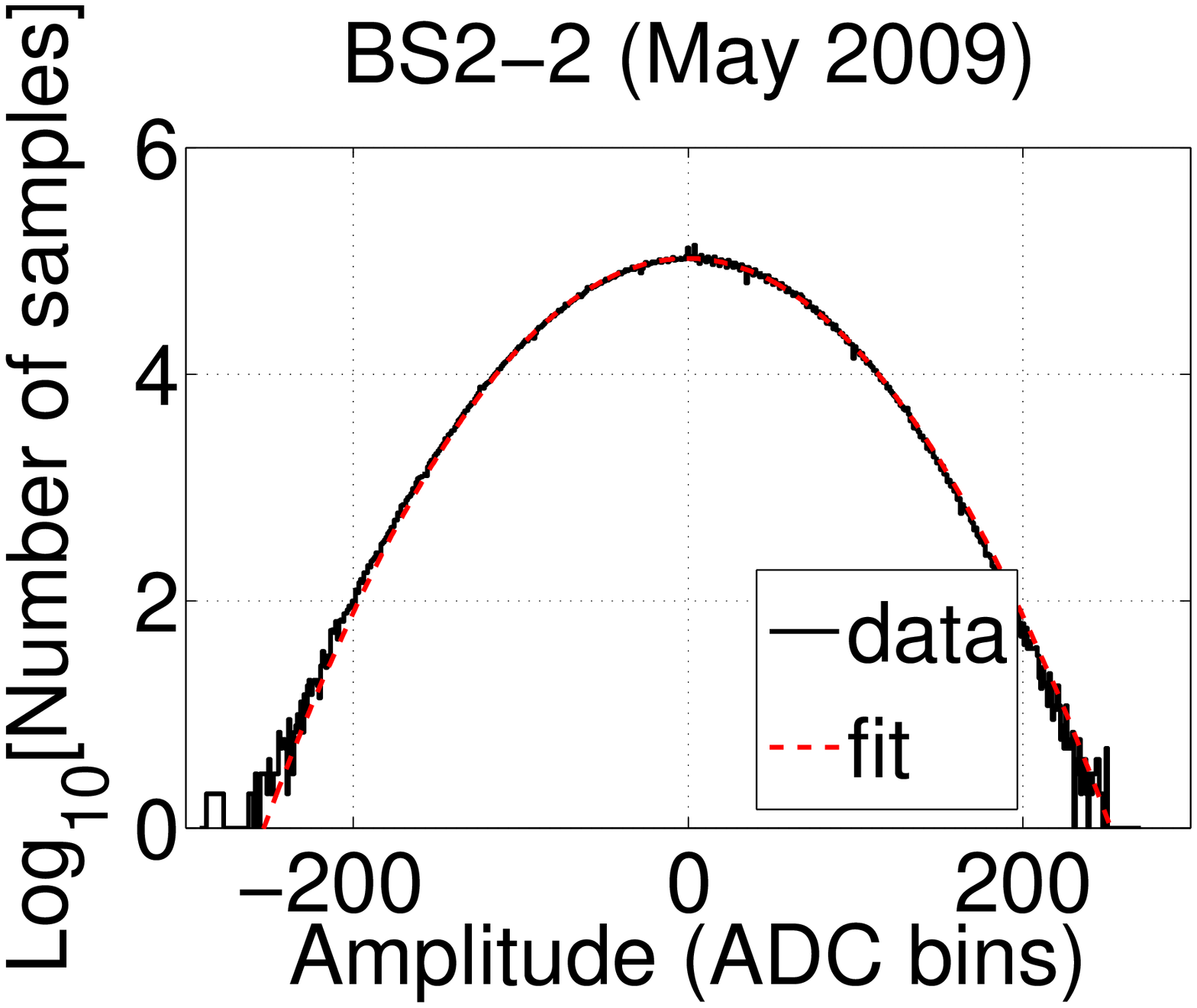}
}
\subfigure[BS3-0]{
\noindent\includegraphics[width=7pc]{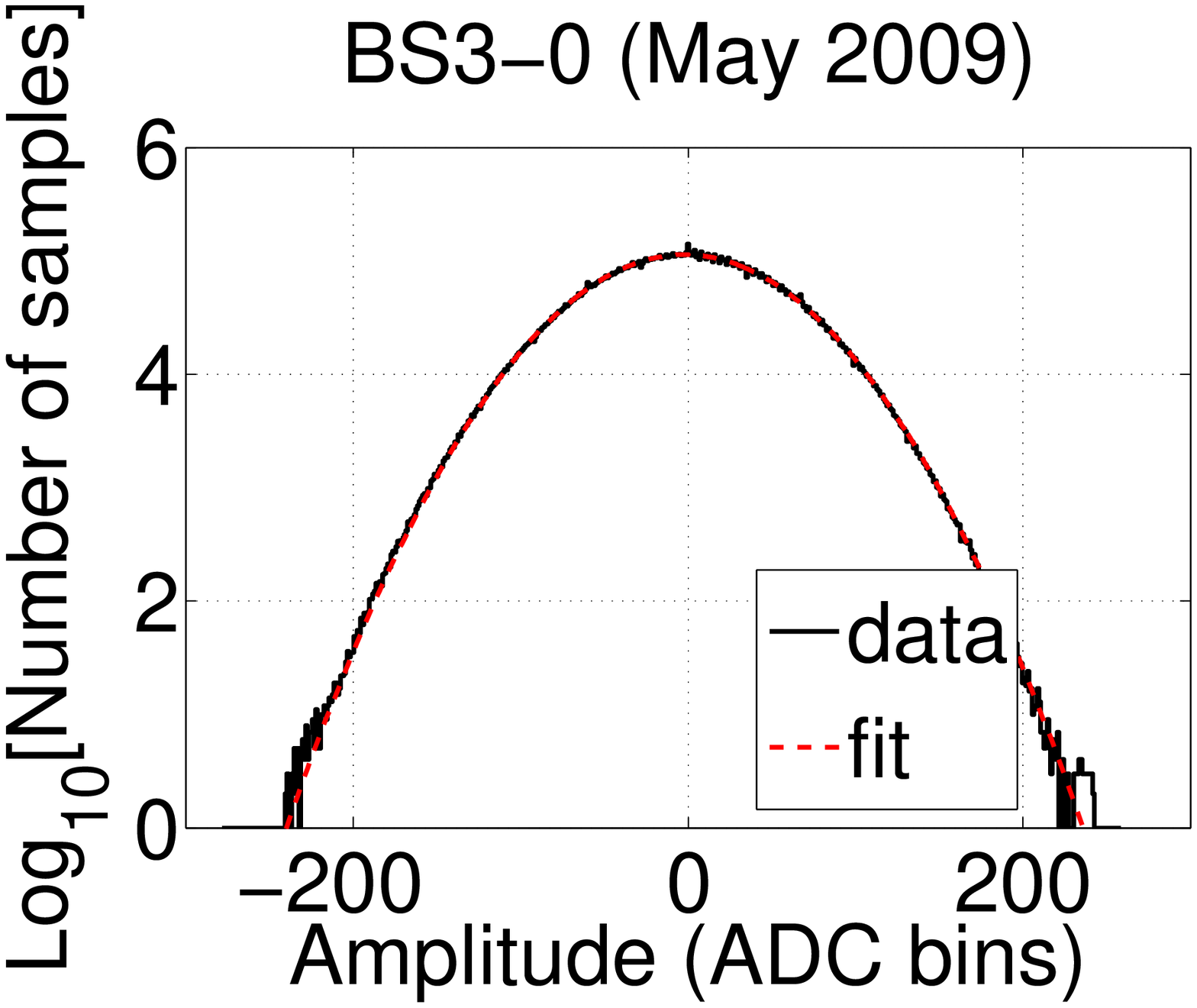}
}
\subfigure[BS3-1]{
\noindent\includegraphics[width=7pc]{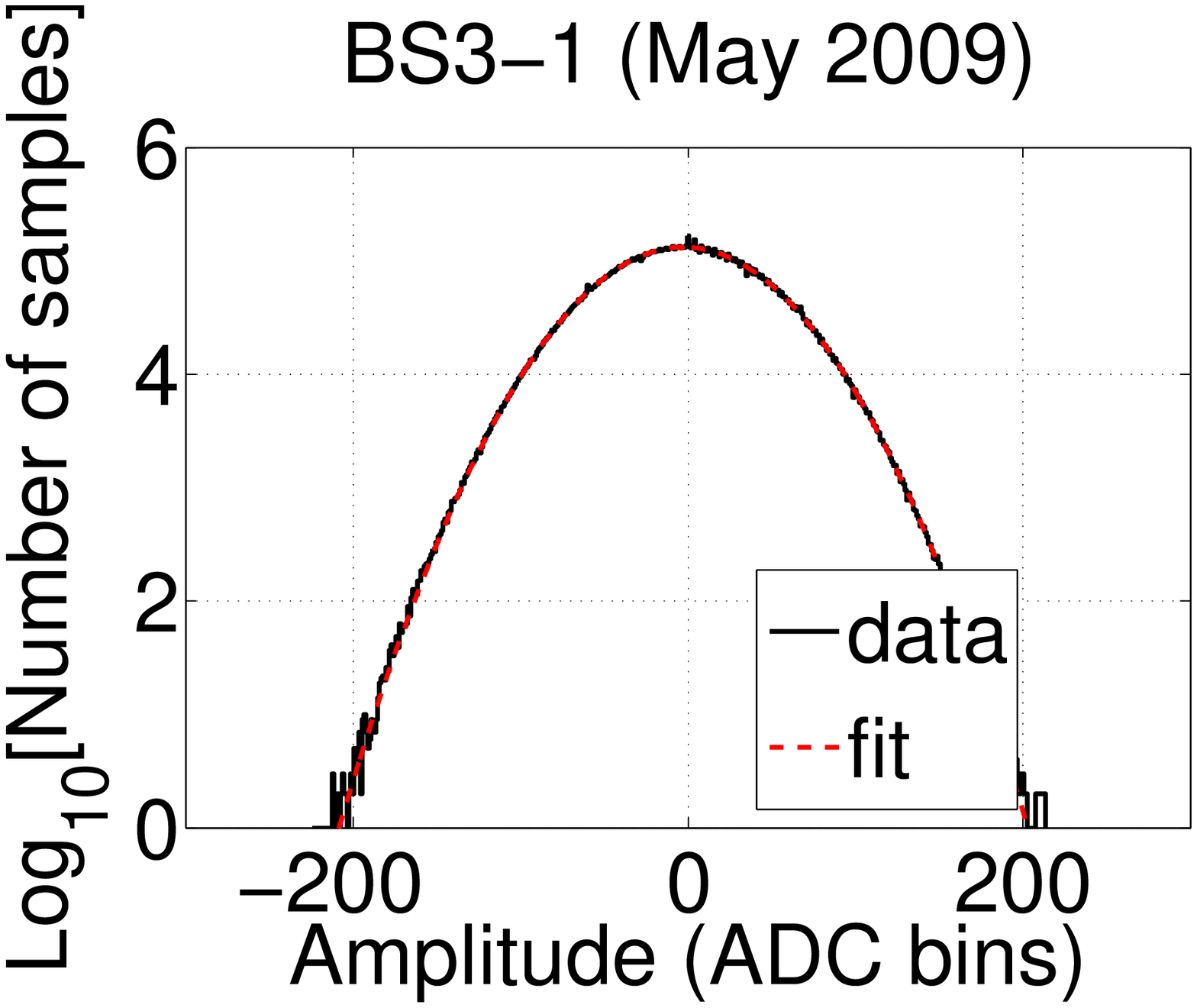}
}
\subfigure[BS3-2]{
\noindent\includegraphics[width=7pc]{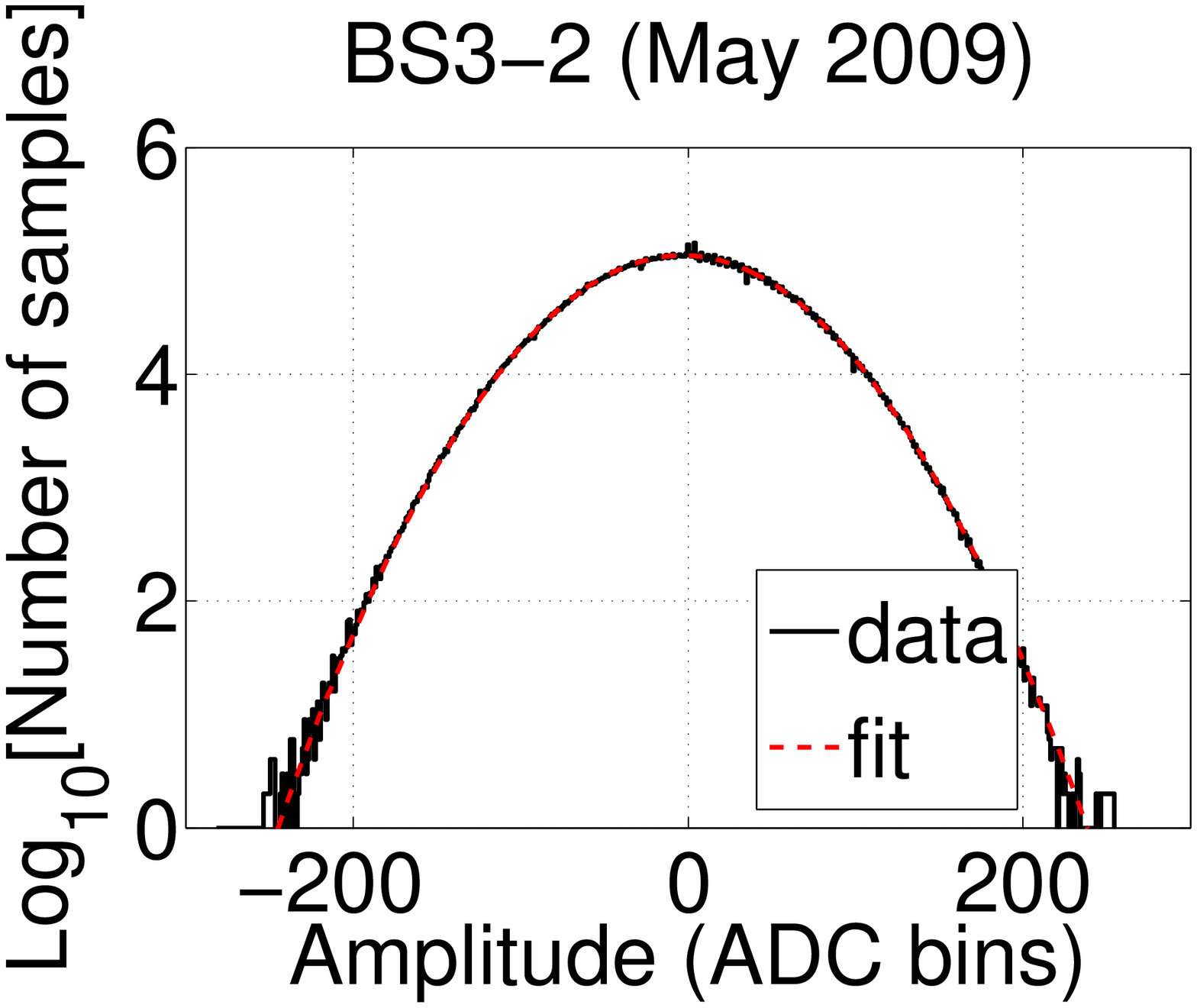}
}
\subfigure[BS4-0]{
\noindent\includegraphics[width=7pc]{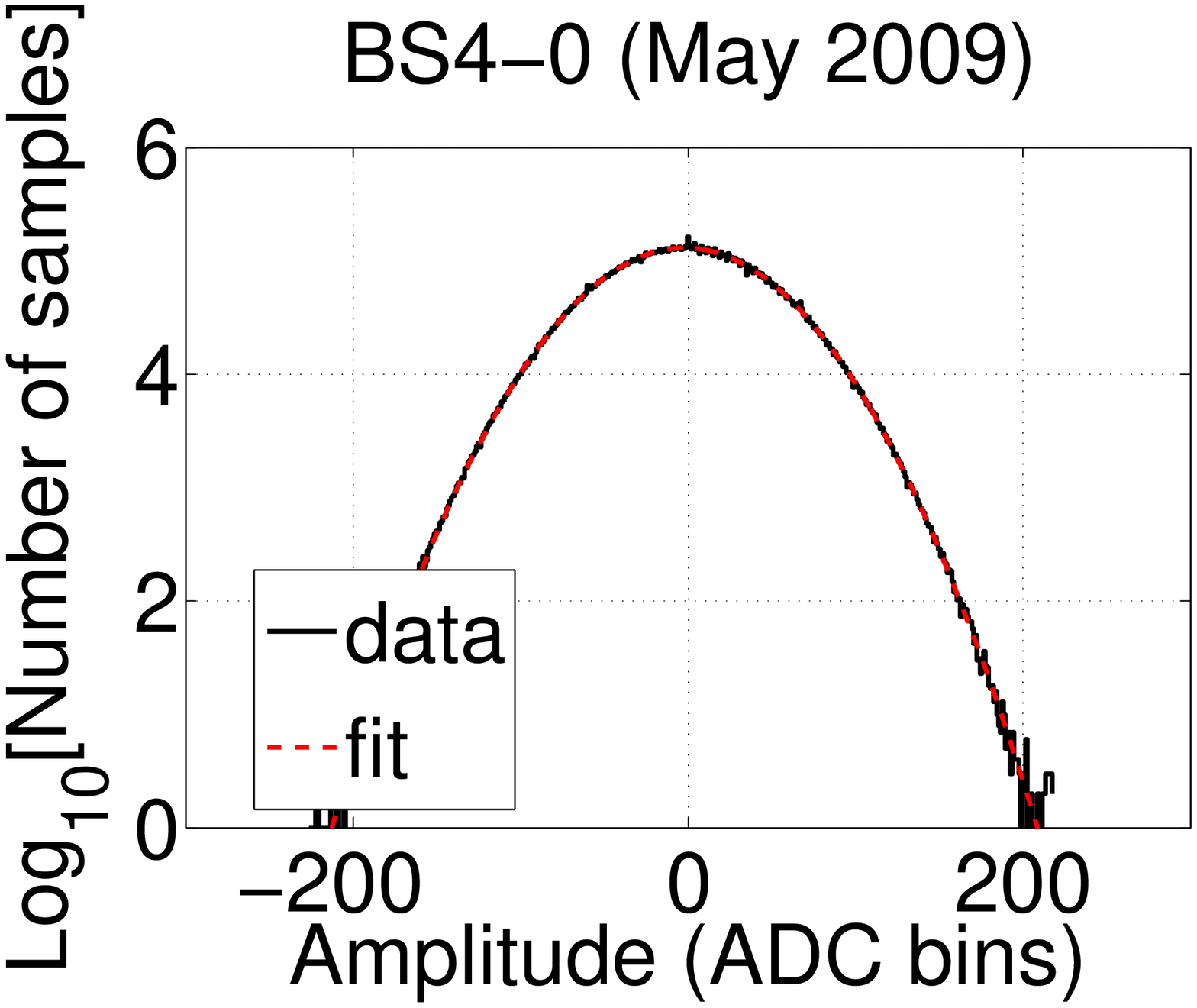}
}
\subfigure[BS4-1]{
\noindent\includegraphics[width=7pc]{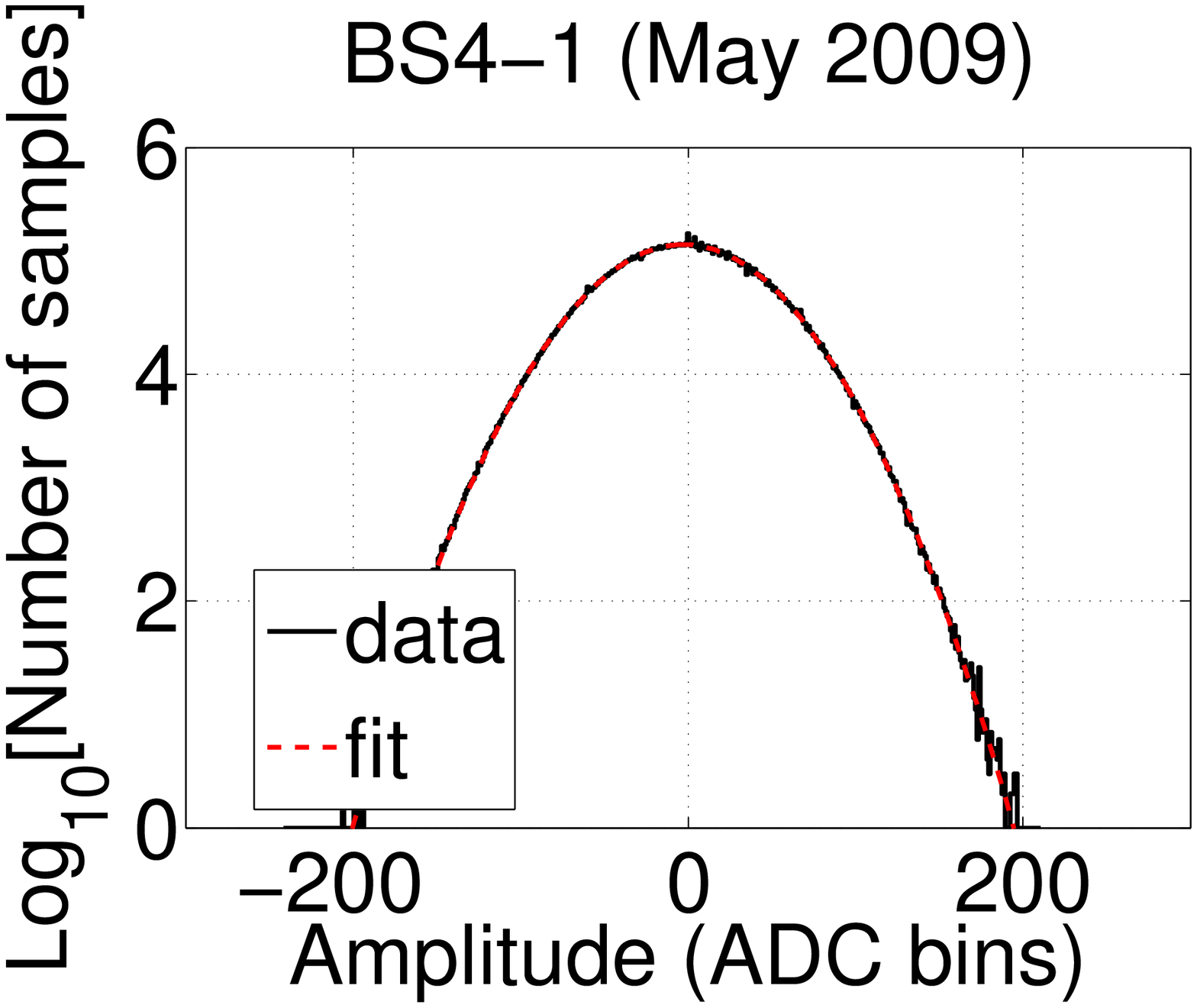}
}
\subfigure[BS4-2]{
\noindent\includegraphics[width=7pc]{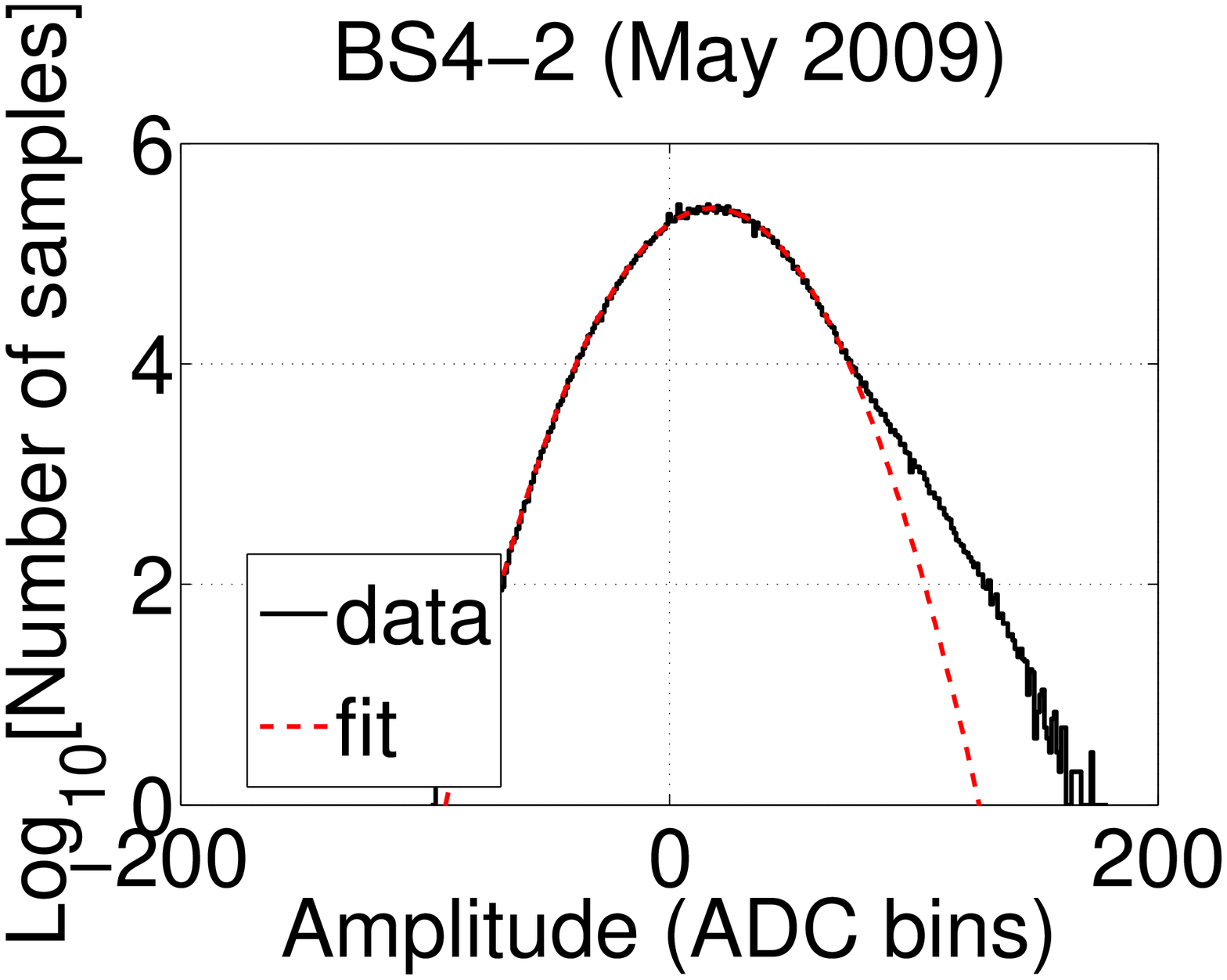}
}
\subfigure[BS5-0]{
\noindent\includegraphics[width=7pc]{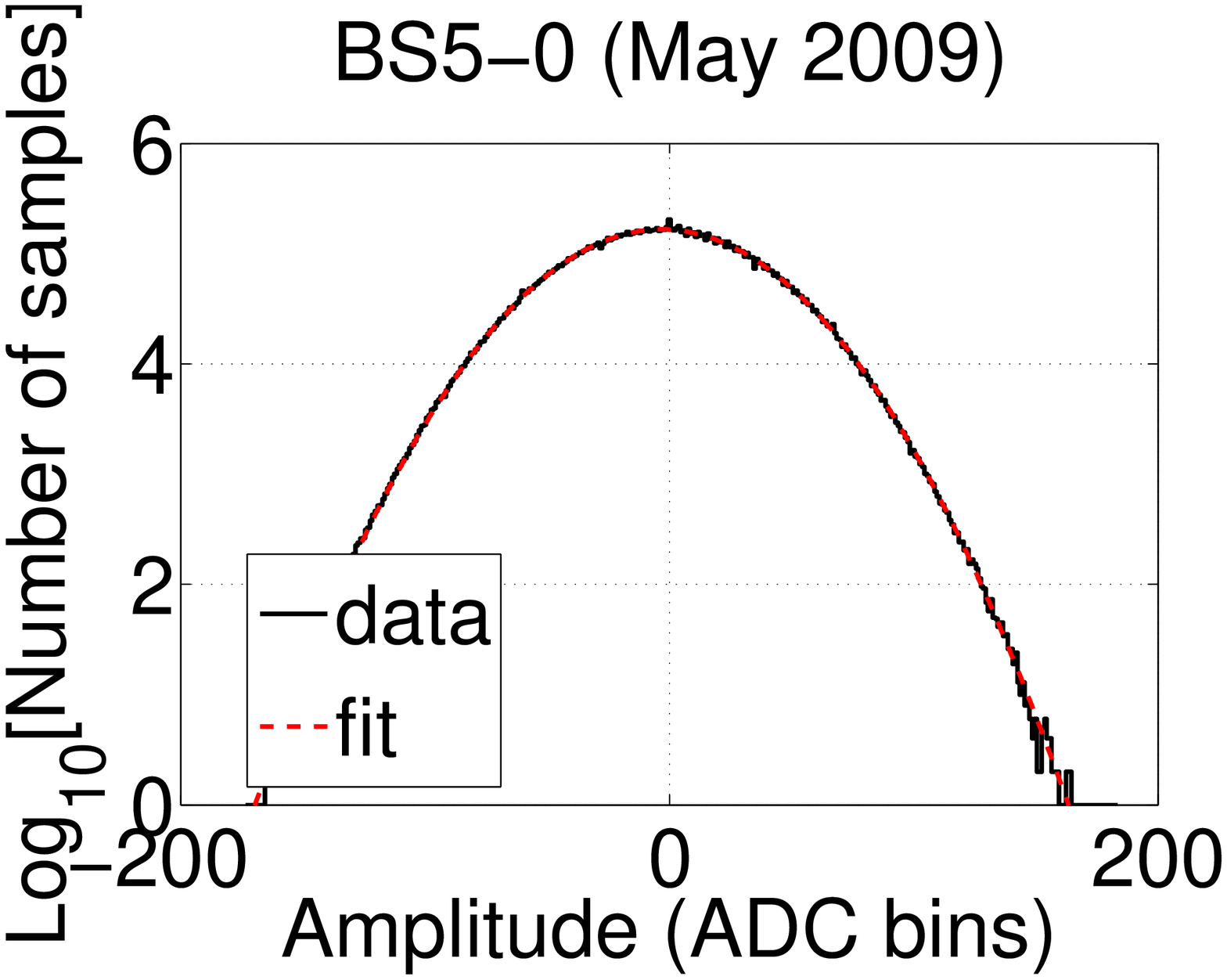}
}
\subfigure[BS5-1]{
\noindent\includegraphics[width=7pc]{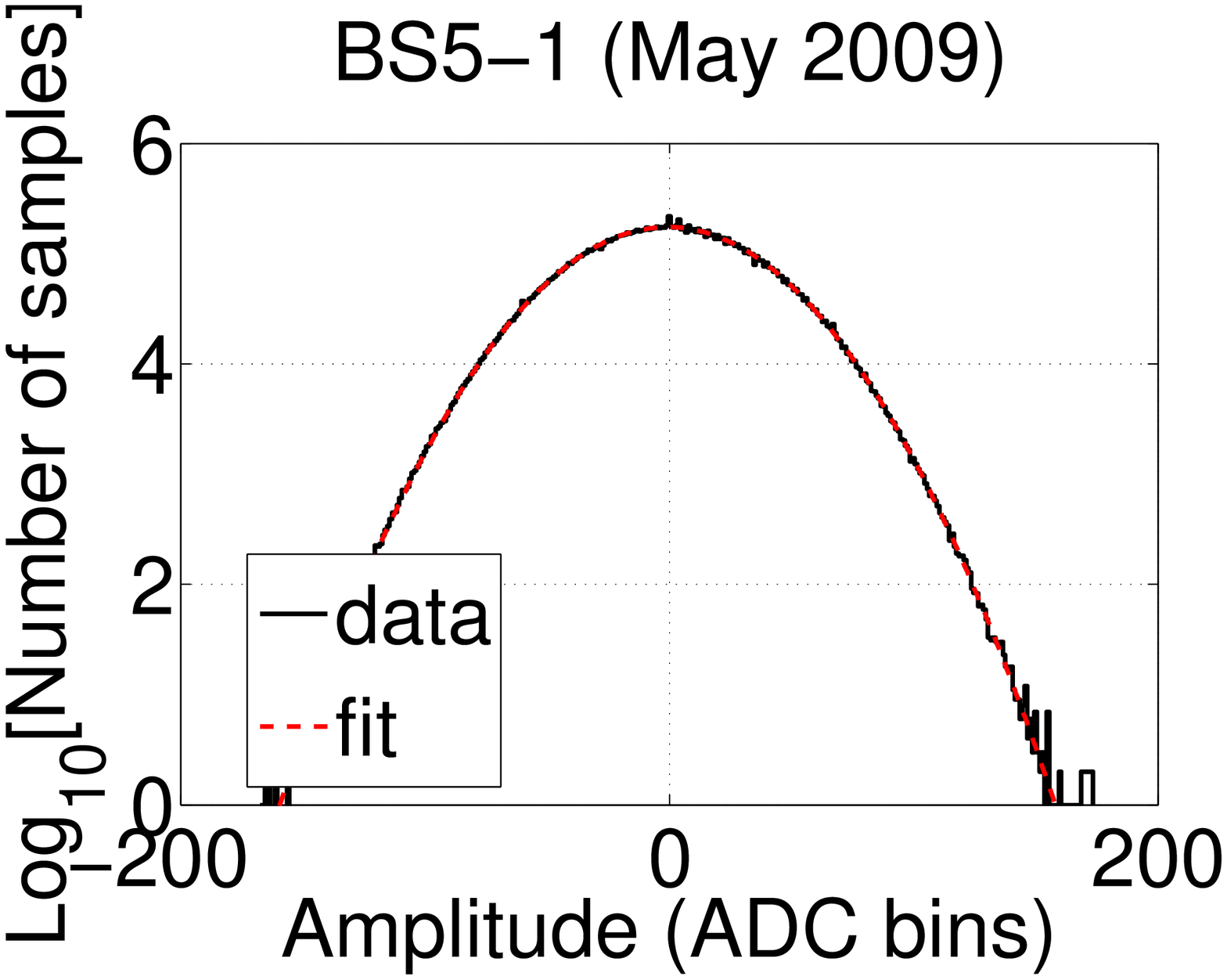}
}
\subfigure[BS5-2]{
\noindent\includegraphics[width=7pc]{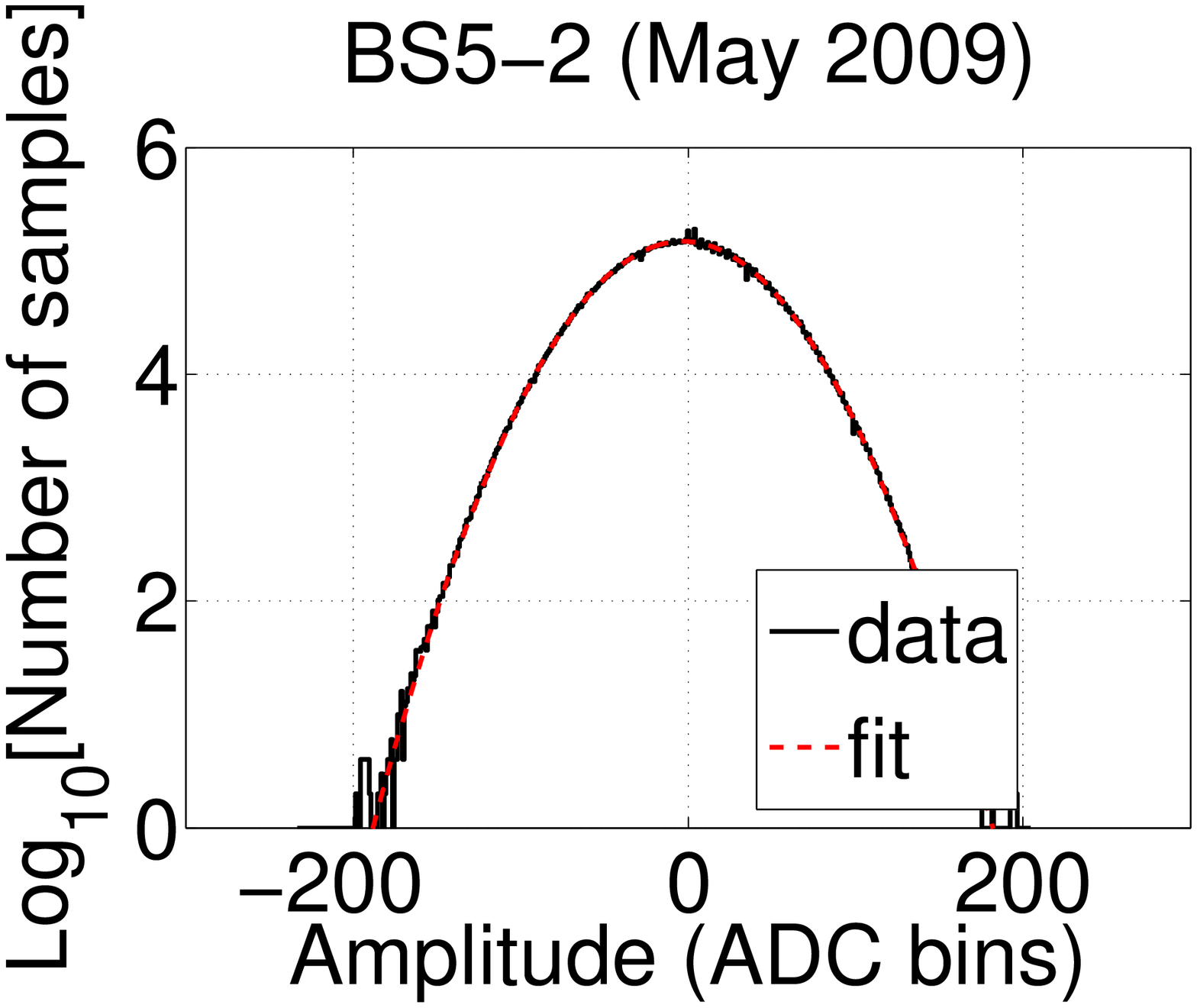}
}
\subfigure[BS6-0]{
\noindent\includegraphics[width=7pc]{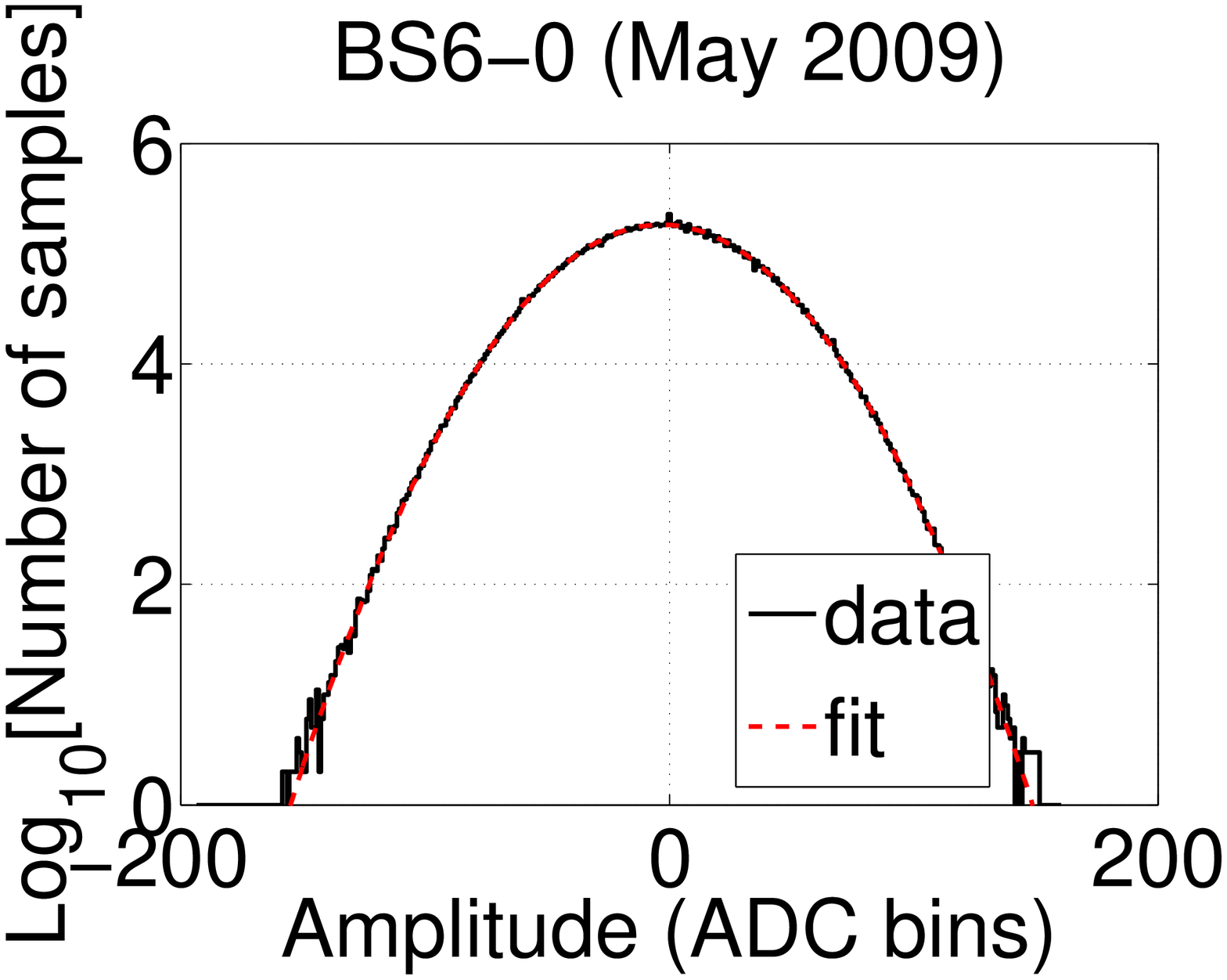}
}
\subfigure[BS6-1]{
\noindent\includegraphics[width=7pc]{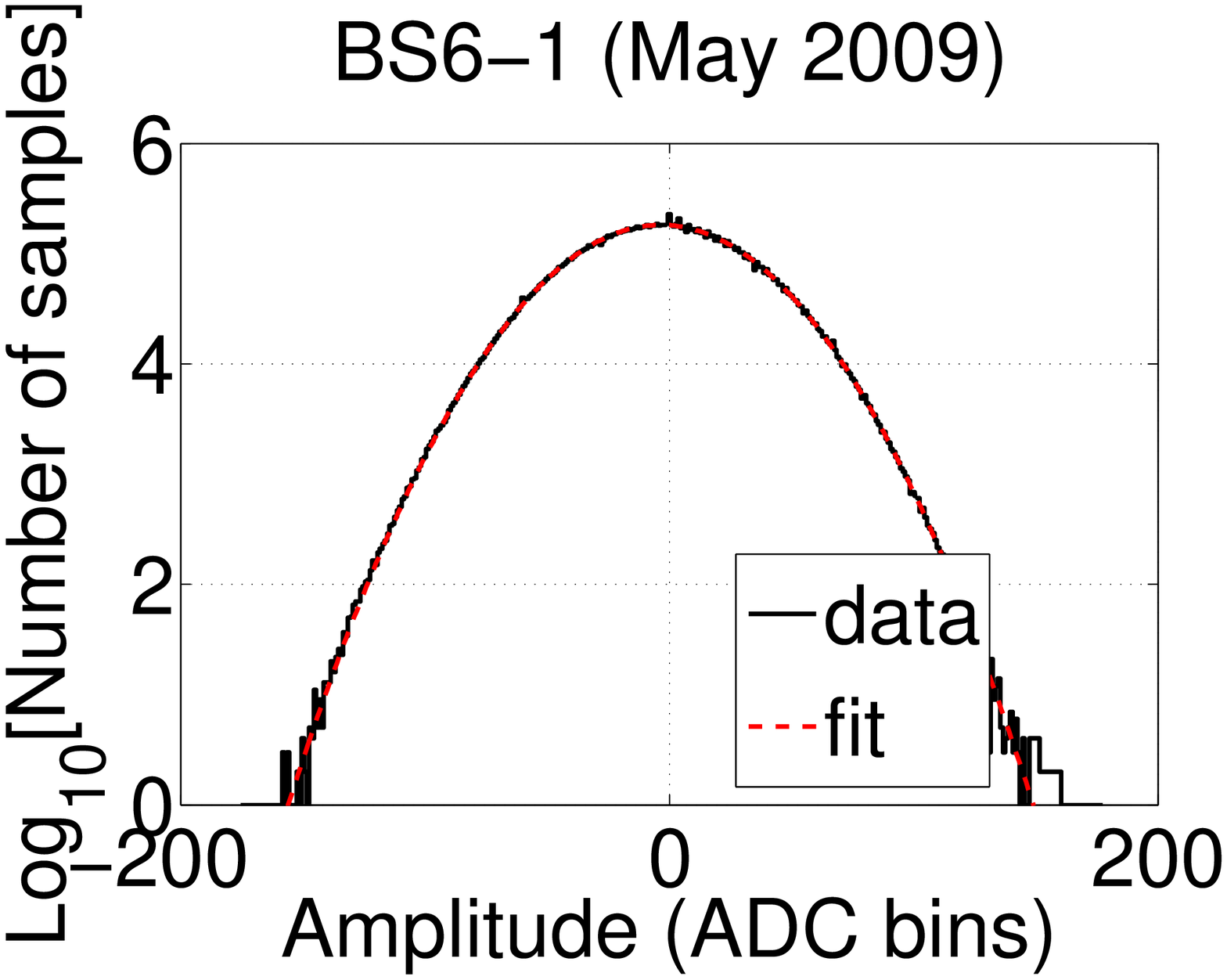}
}
\subfigure[BS6-2]{
\noindent\includegraphics[width=7pc]{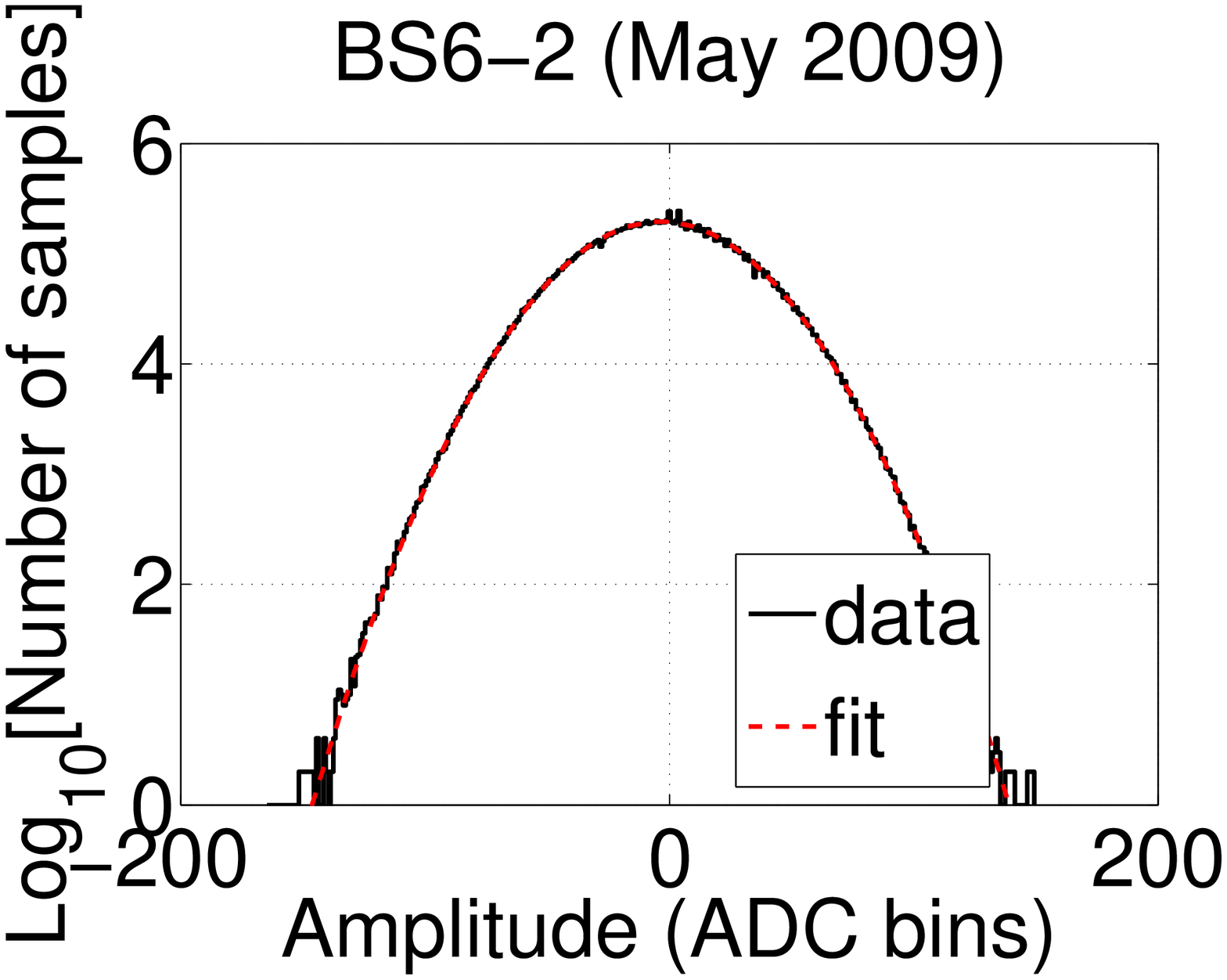}
}
\subfigure[BS7-0]{
\noindent\includegraphics[width=7pc]{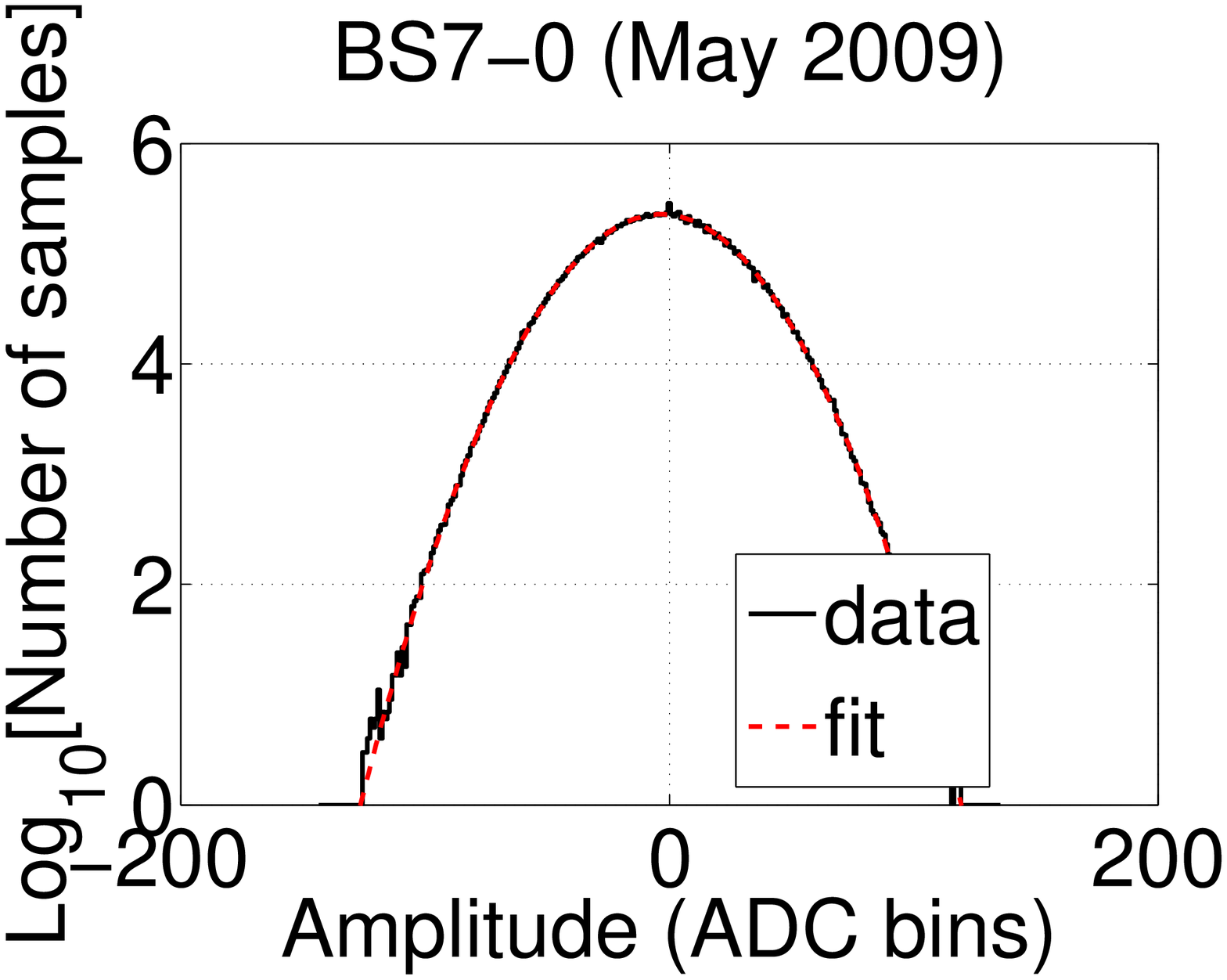}
}
\subfigure[BS7-1]{
\noindent\includegraphics[width=7pc]{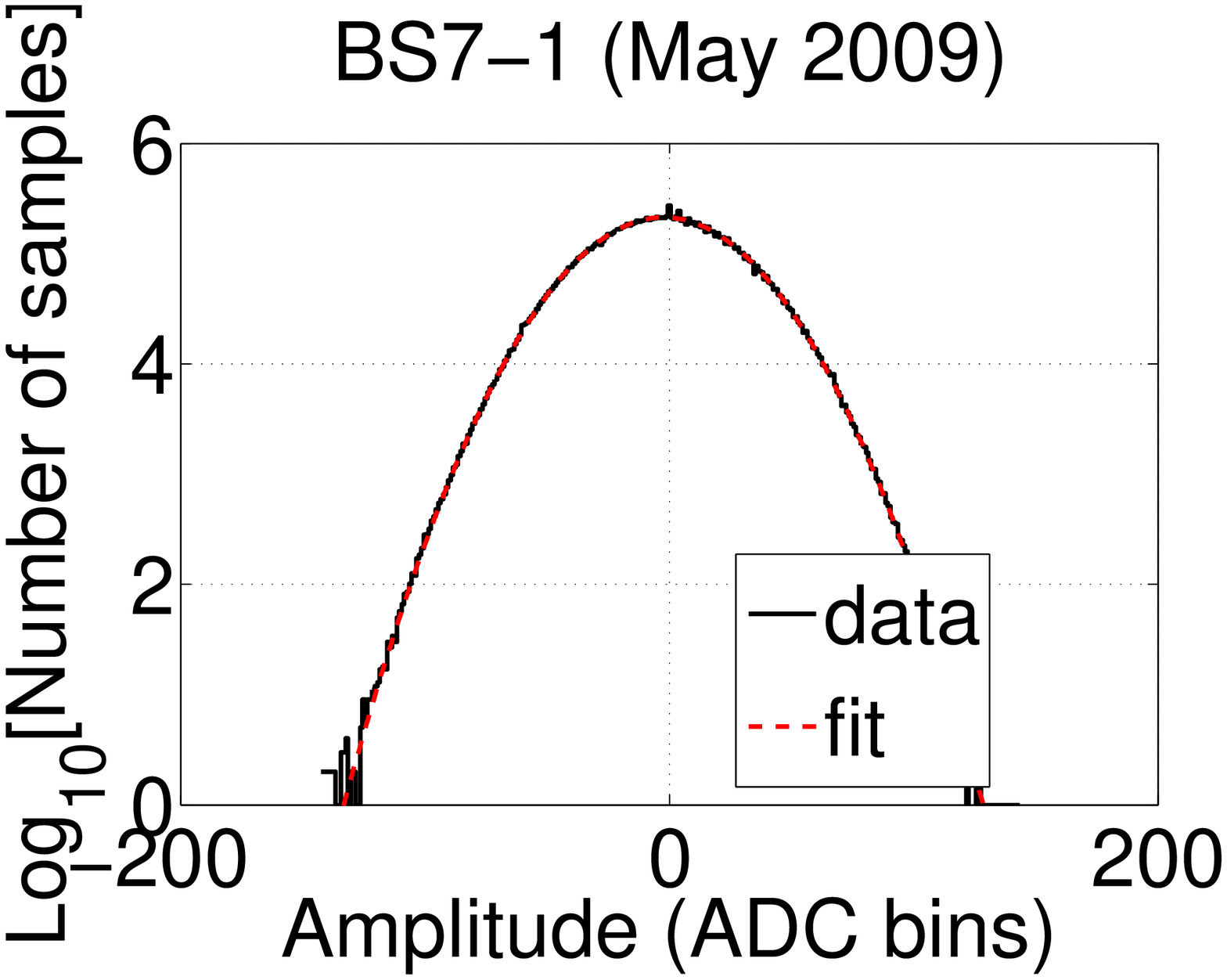}
}
\subfigure[BS7-2]{
\noindent\includegraphics[width=7pc]{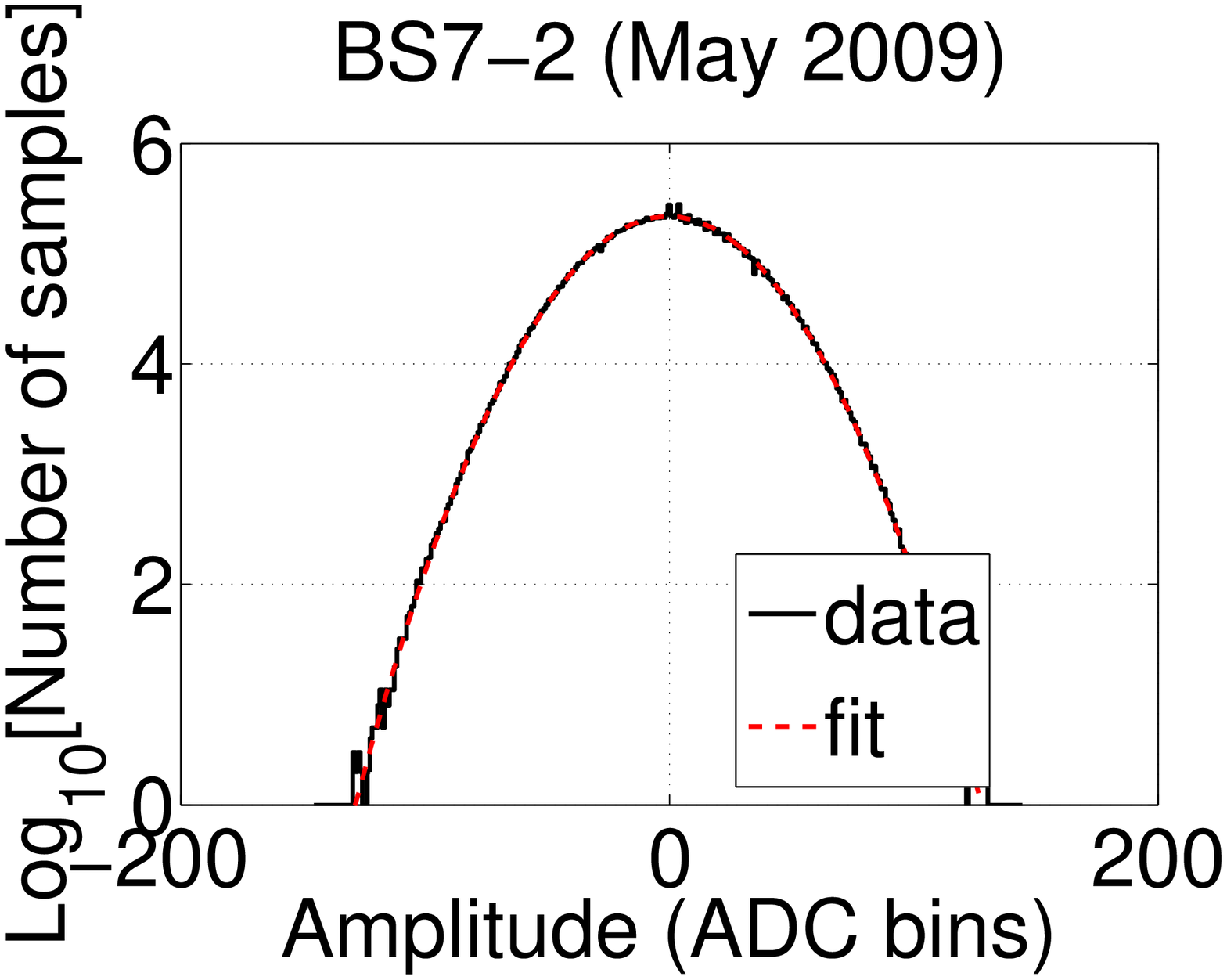}
}
\caption[Gaussian noise distributions for all String B channels]{Noise amplitude histogram (with Gaussian fit) for each channel of String B.}
\label{gaussianHistogramsB}
\end{center}
\end{figure}

\begin{figure}
\begin{center}
\subfigure[CS1-0]{
\noindent\includegraphics[width=7pc]{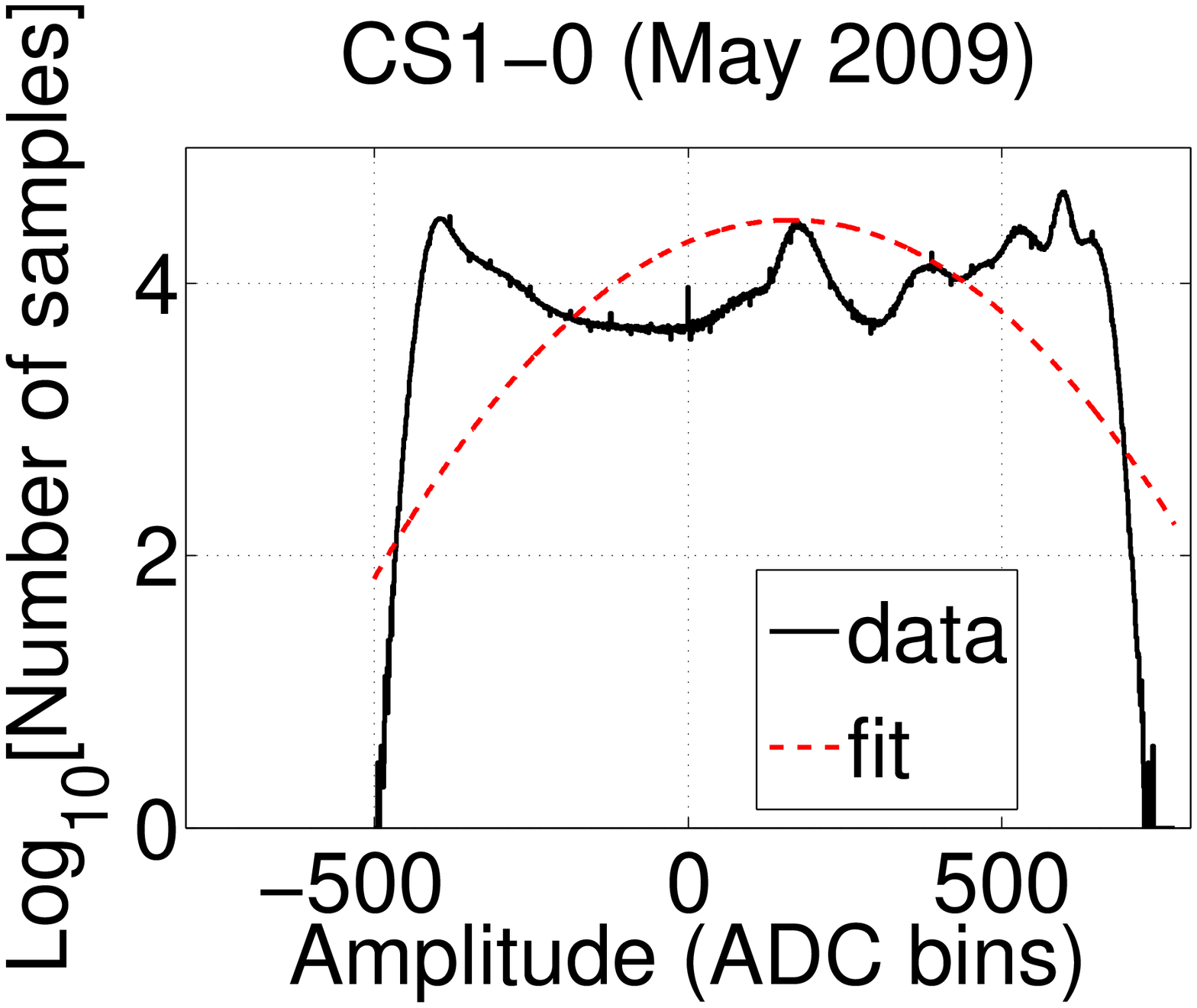}
}
\subfigure[CS1-1]{
\noindent\includegraphics[width=7pc]{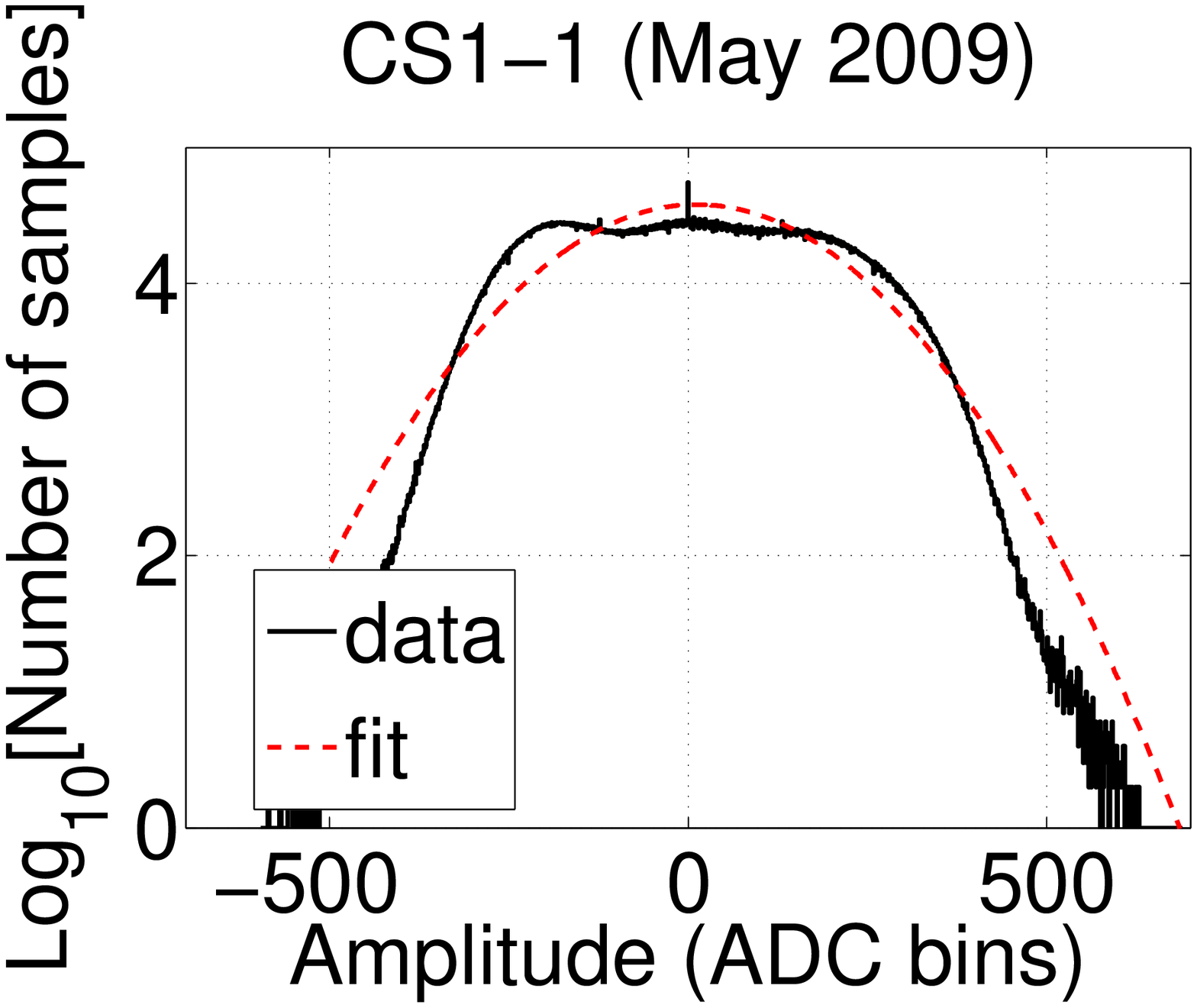}
}
\subfigure[CS1-2]{
\noindent\includegraphics[width=7pc]{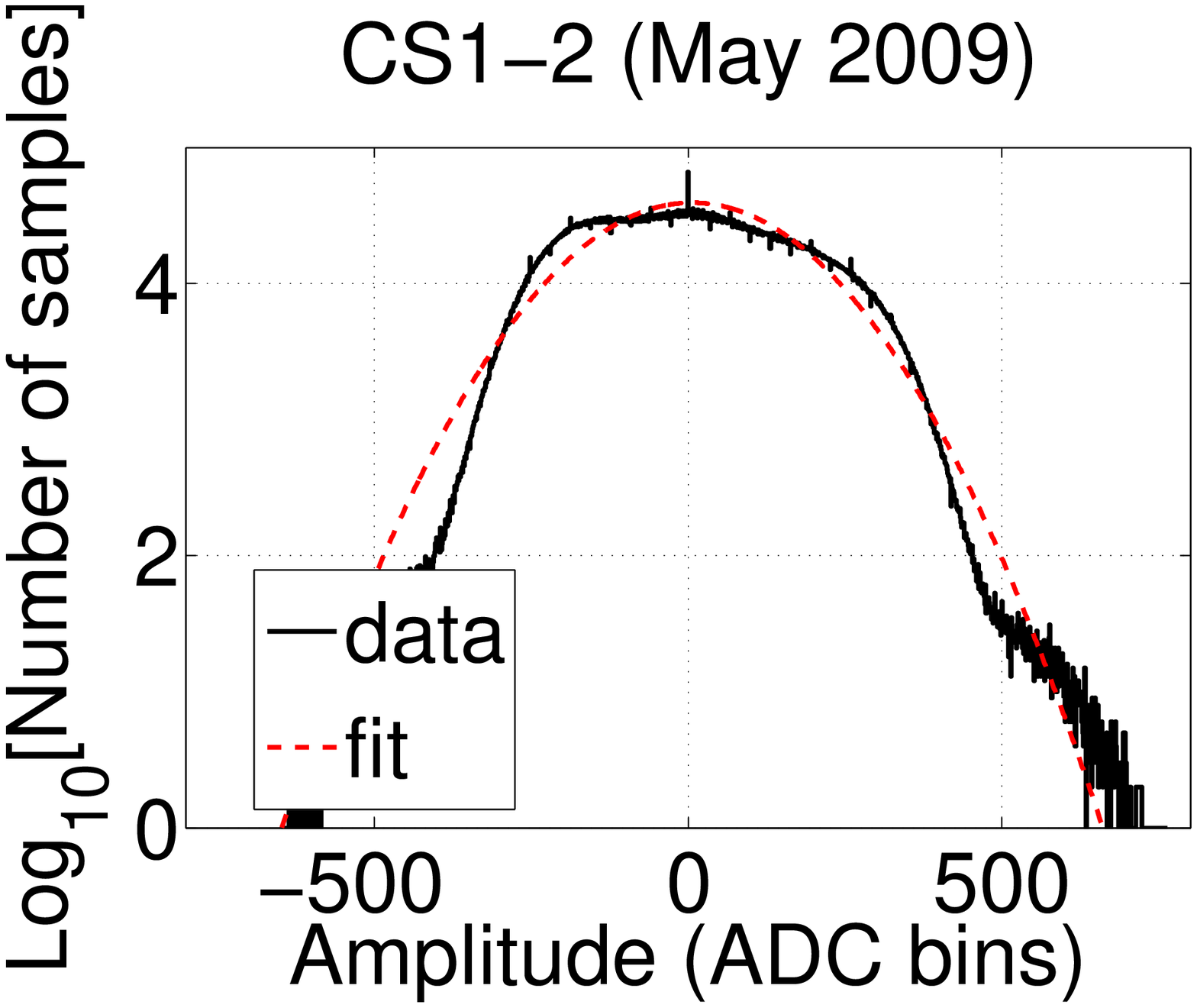}
}
\subfigure[CS2-0]{
\noindent\includegraphics[width=7pc]{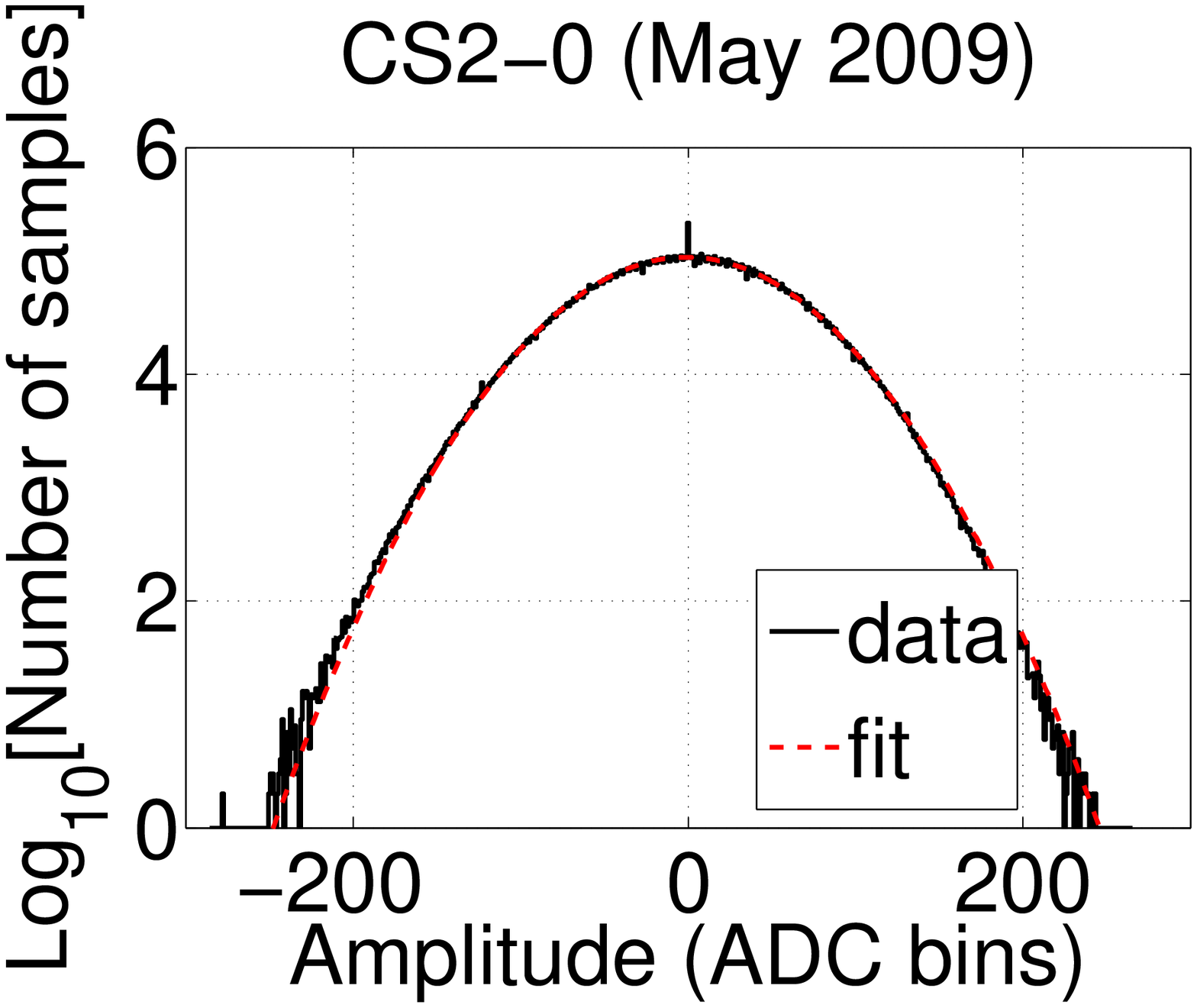}
}
\subfigure[CS2-1]{
\noindent\includegraphics[width=7pc]{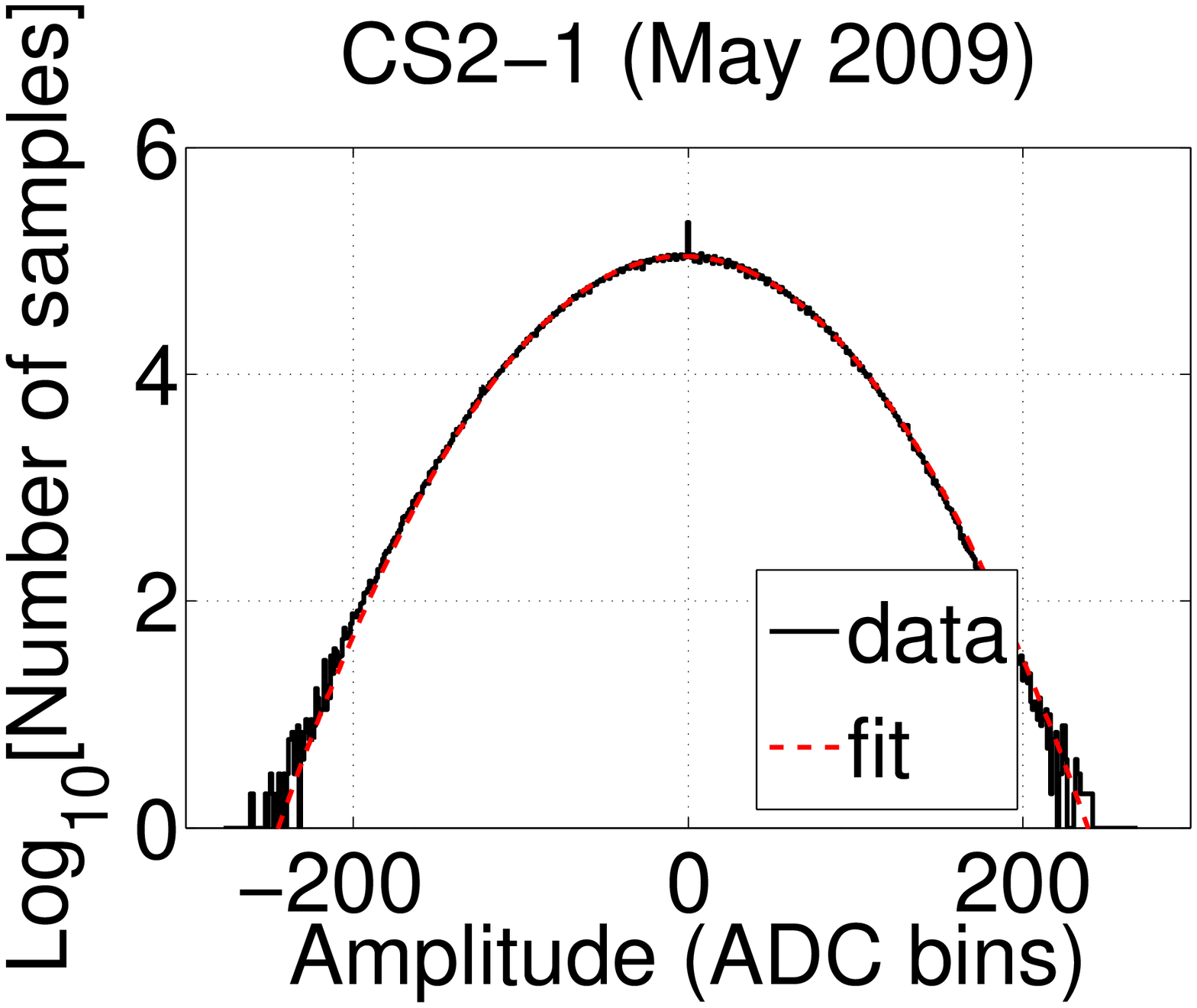}
}
\subfigure[CS2-2]{
\noindent\includegraphics[width=7pc]{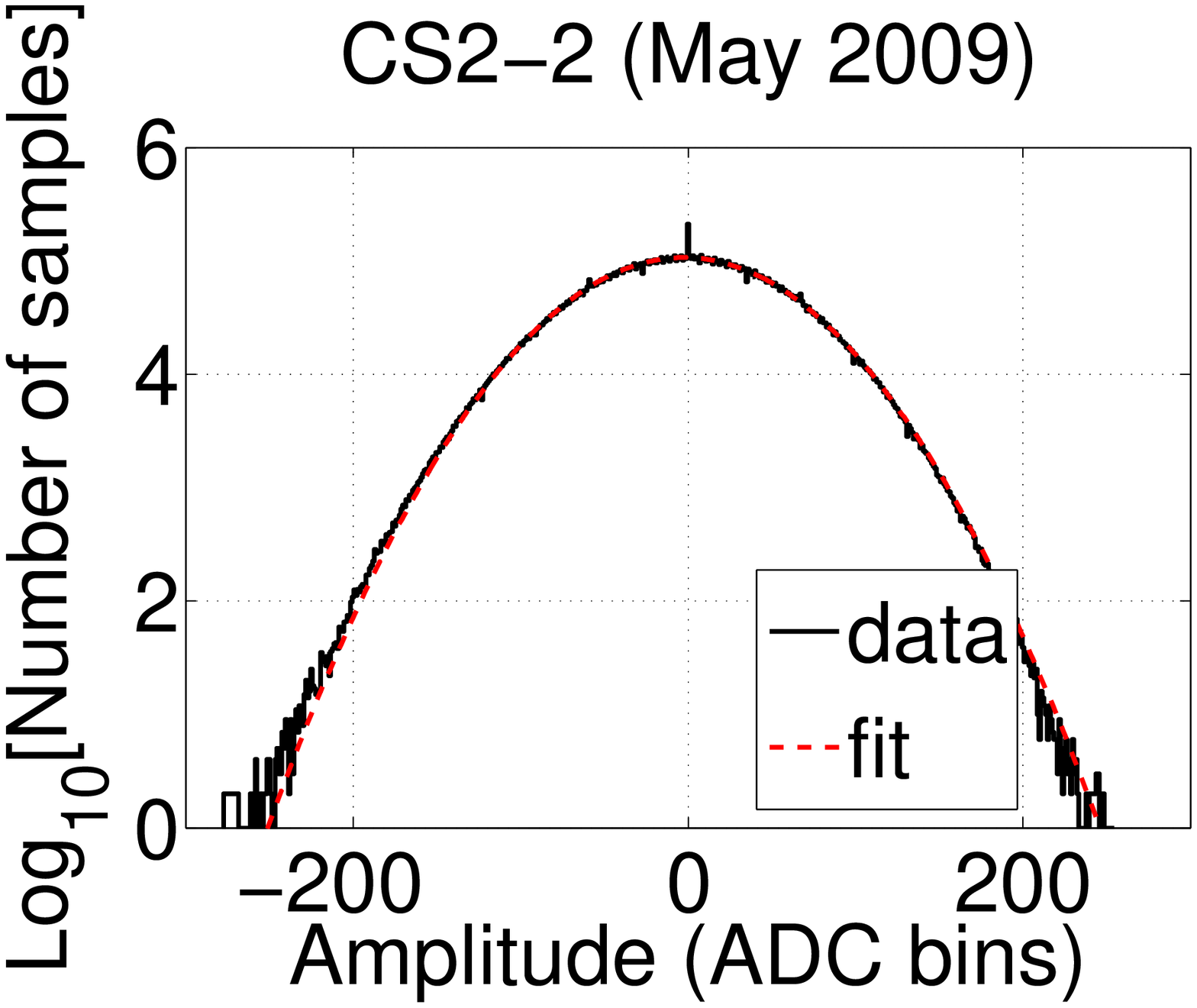}
}
\subfigure[CS3-0]{
\noindent\includegraphics[width=7pc]{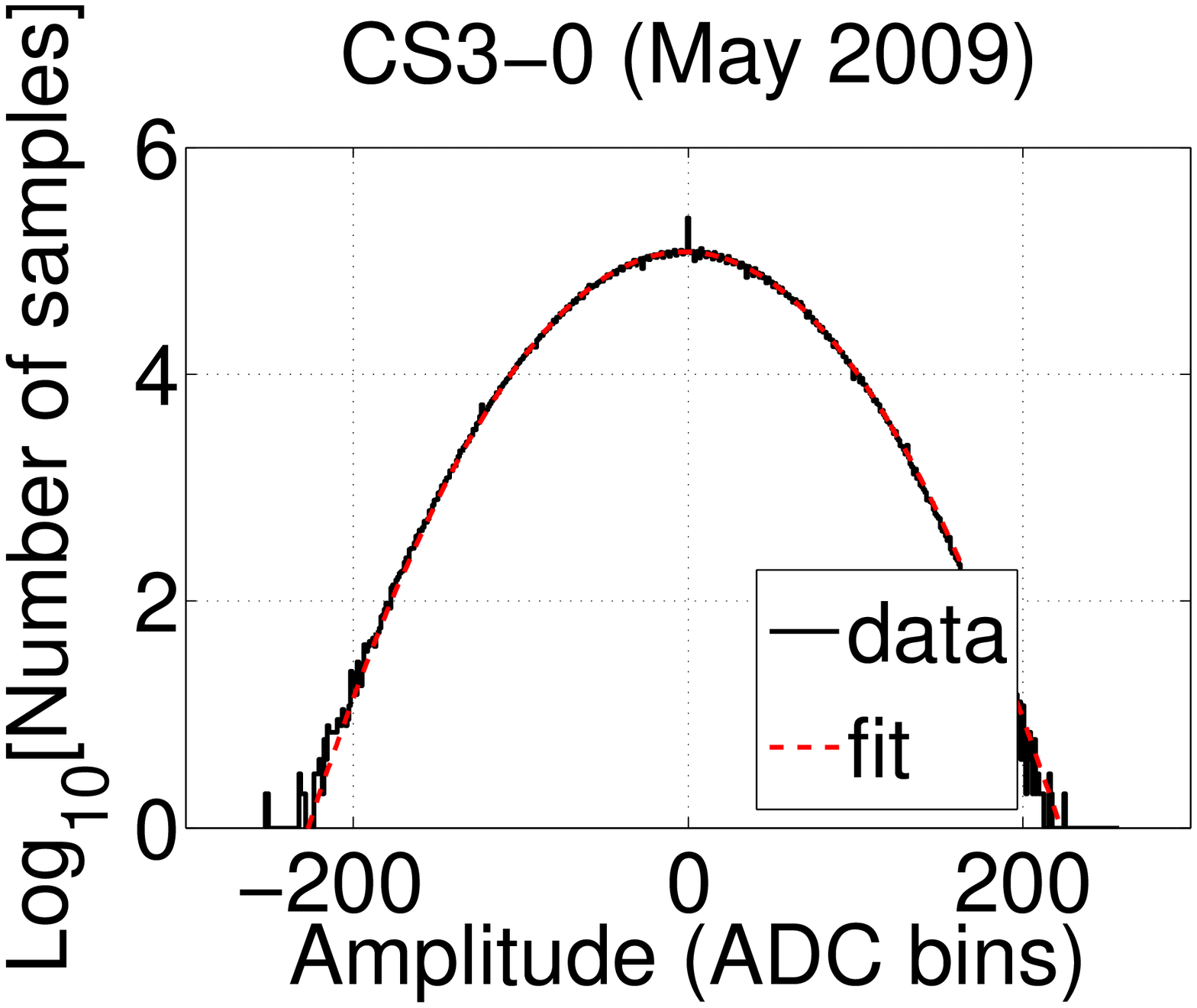}
}
\subfigure[CS3-1]{
\noindent\includegraphics[width=7pc]{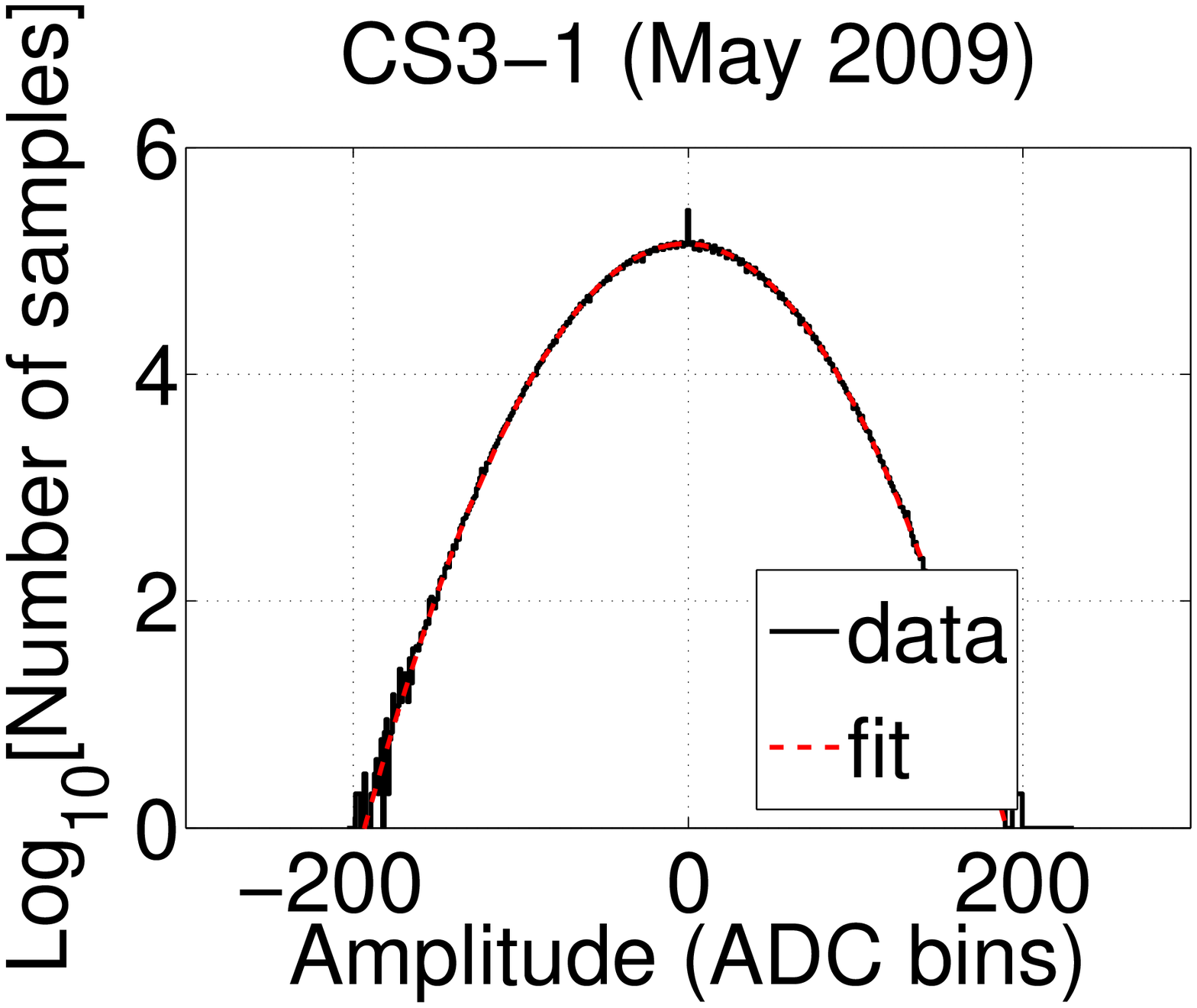}
}
\subfigure[CS3-2]{
\noindent\includegraphics[width=7pc]{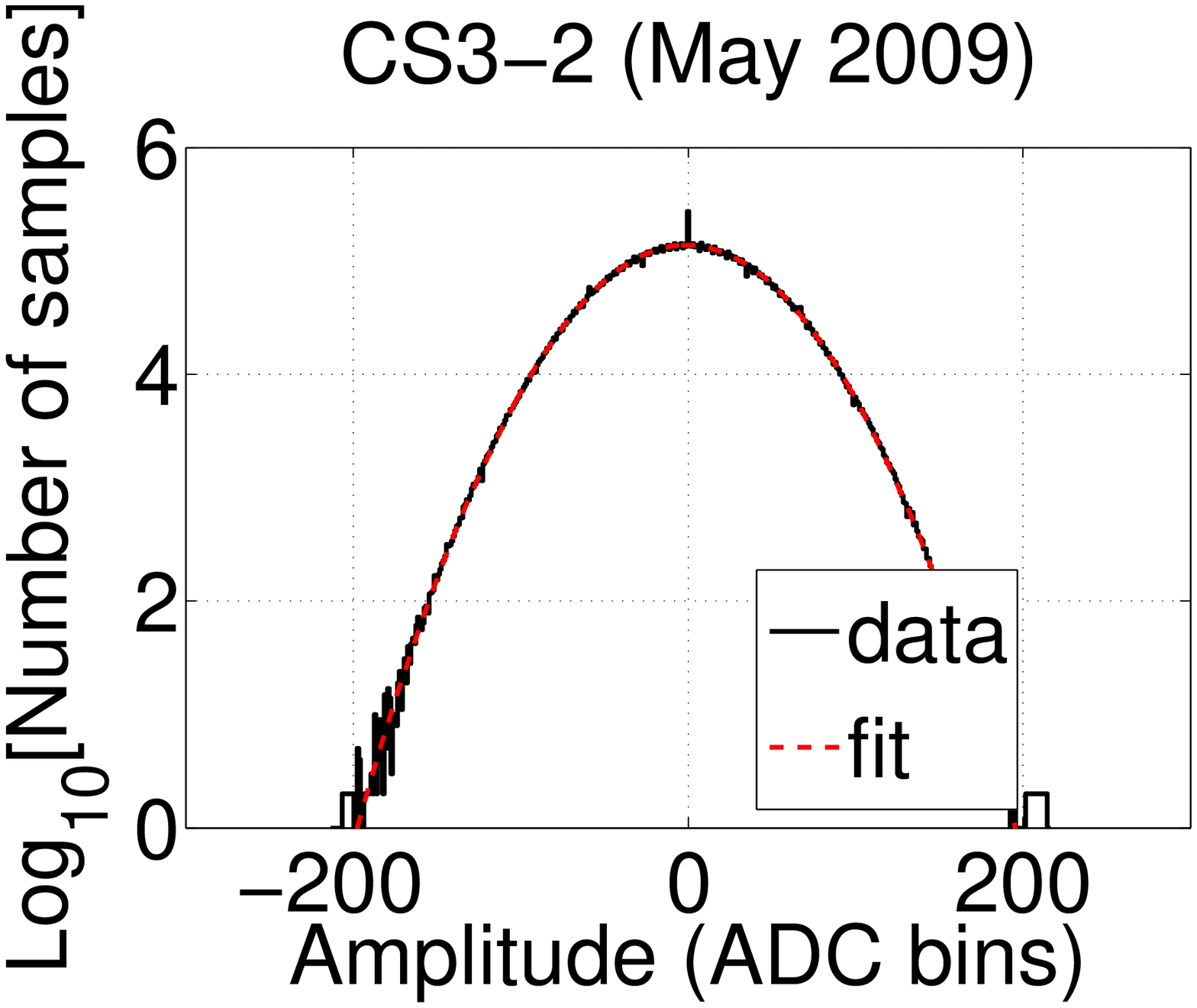}
}
\subfigure[CS4-0]{
\noindent\includegraphics[width=7pc]{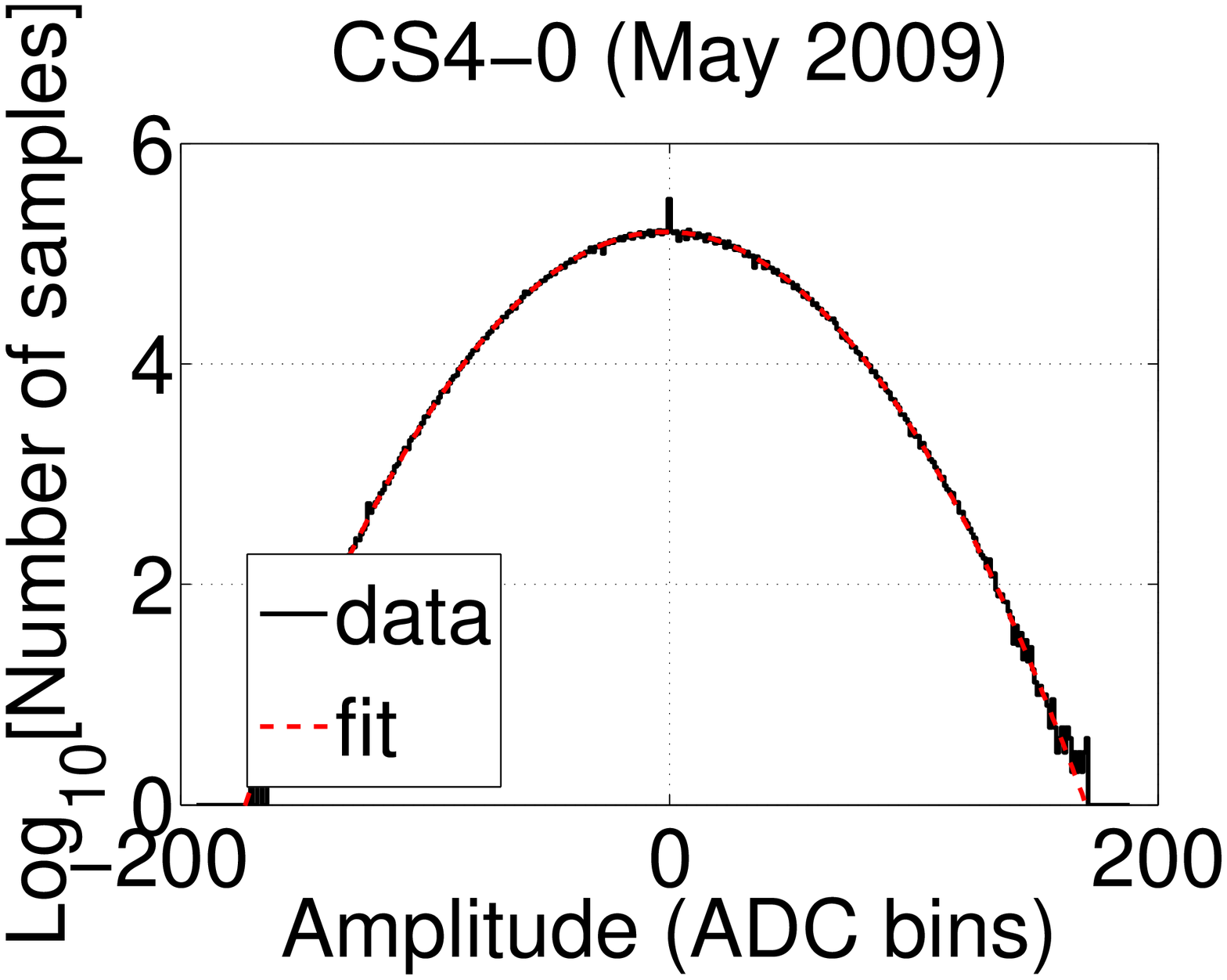}
}
\subfigure[CS4-1]{
\noindent\includegraphics[width=7pc]{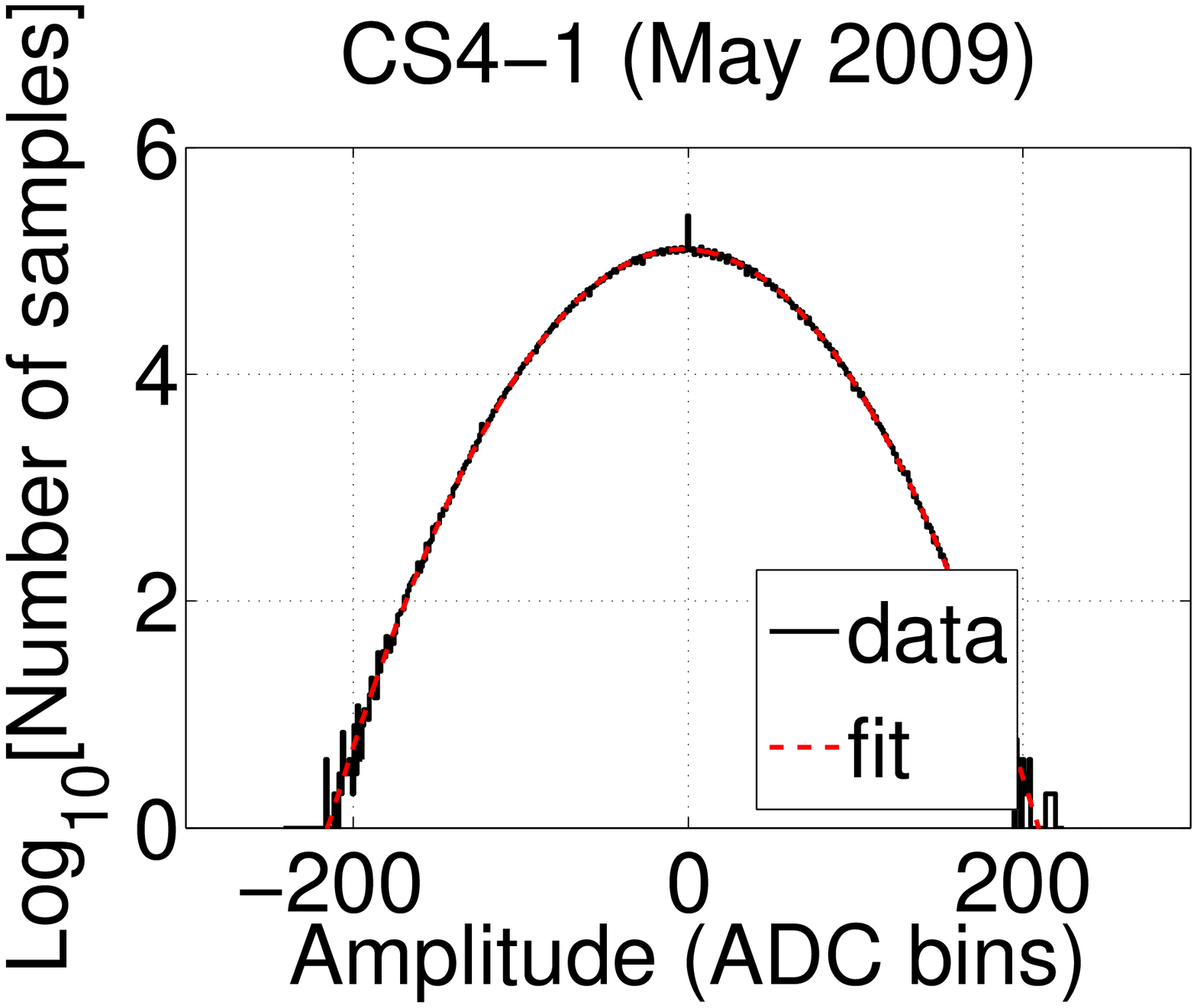}
}
\subfigure[CS4-2]{
\noindent\includegraphics[width=7pc]{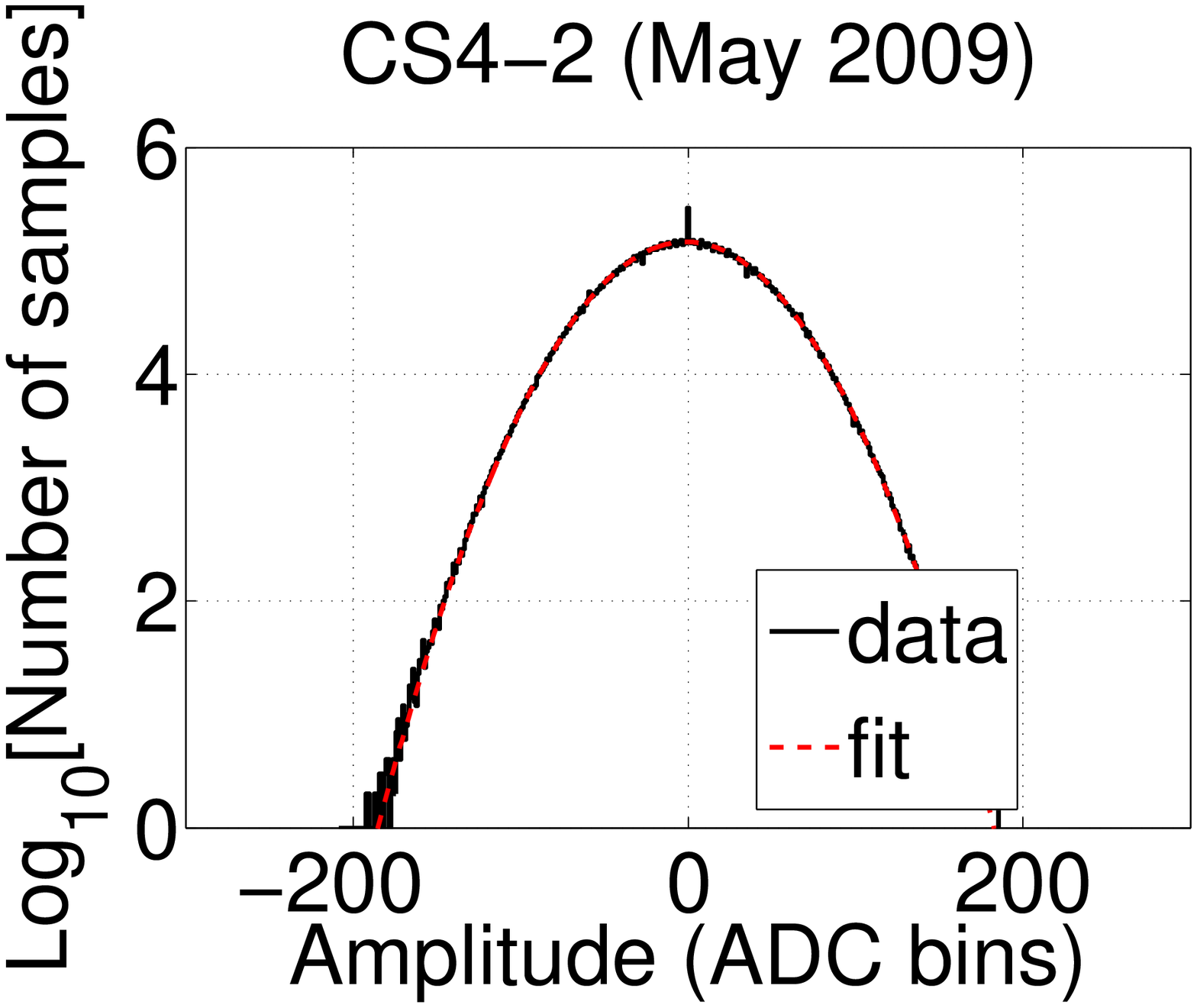}
}
\subfigure[CS5-0]{
\noindent\includegraphics[width=7pc]{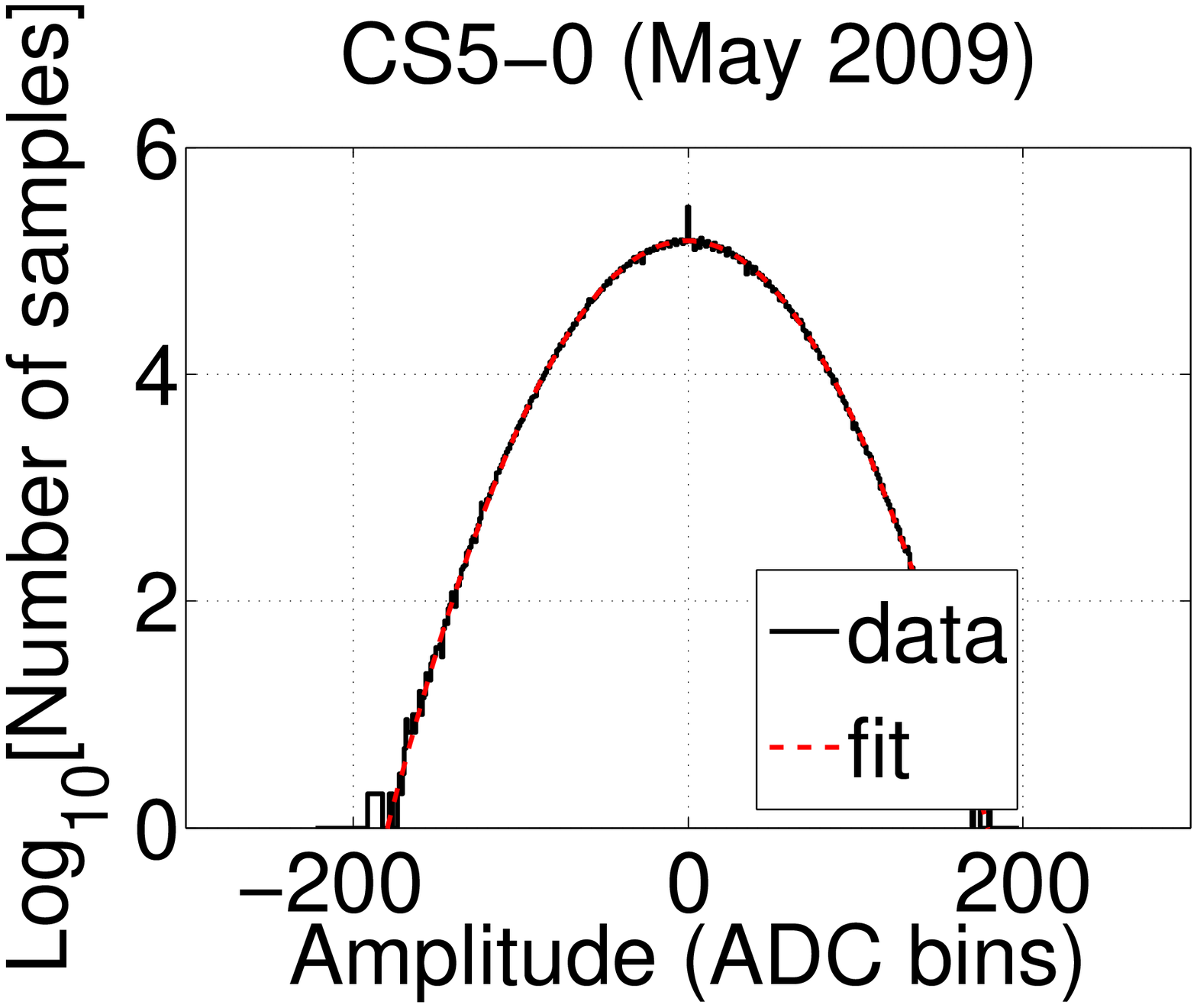}
}
\subfigure[CS5-1]{
\noindent\includegraphics[width=7pc]{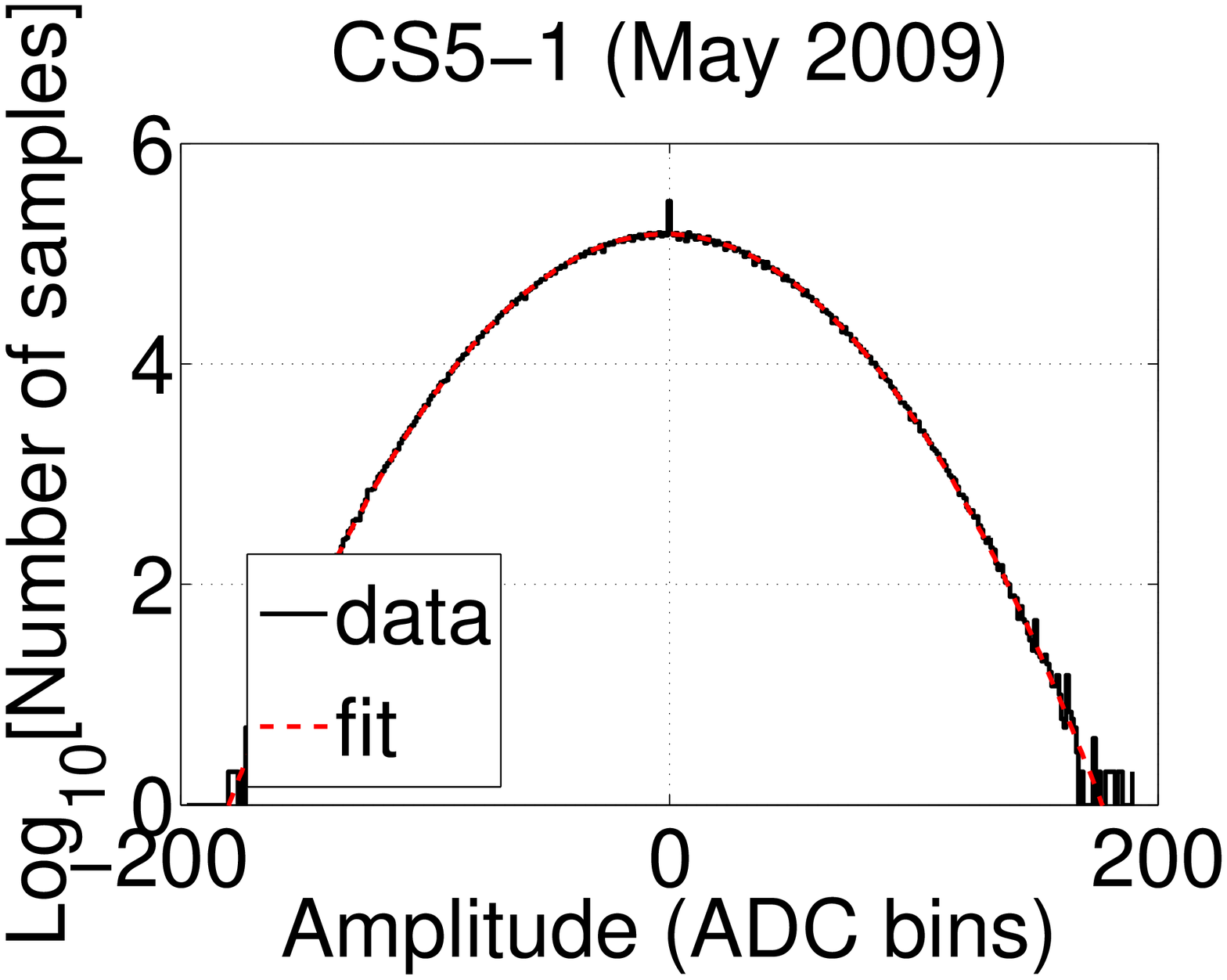}
}
\subfigure[CS5-2]{
\noindent\includegraphics[width=7pc]{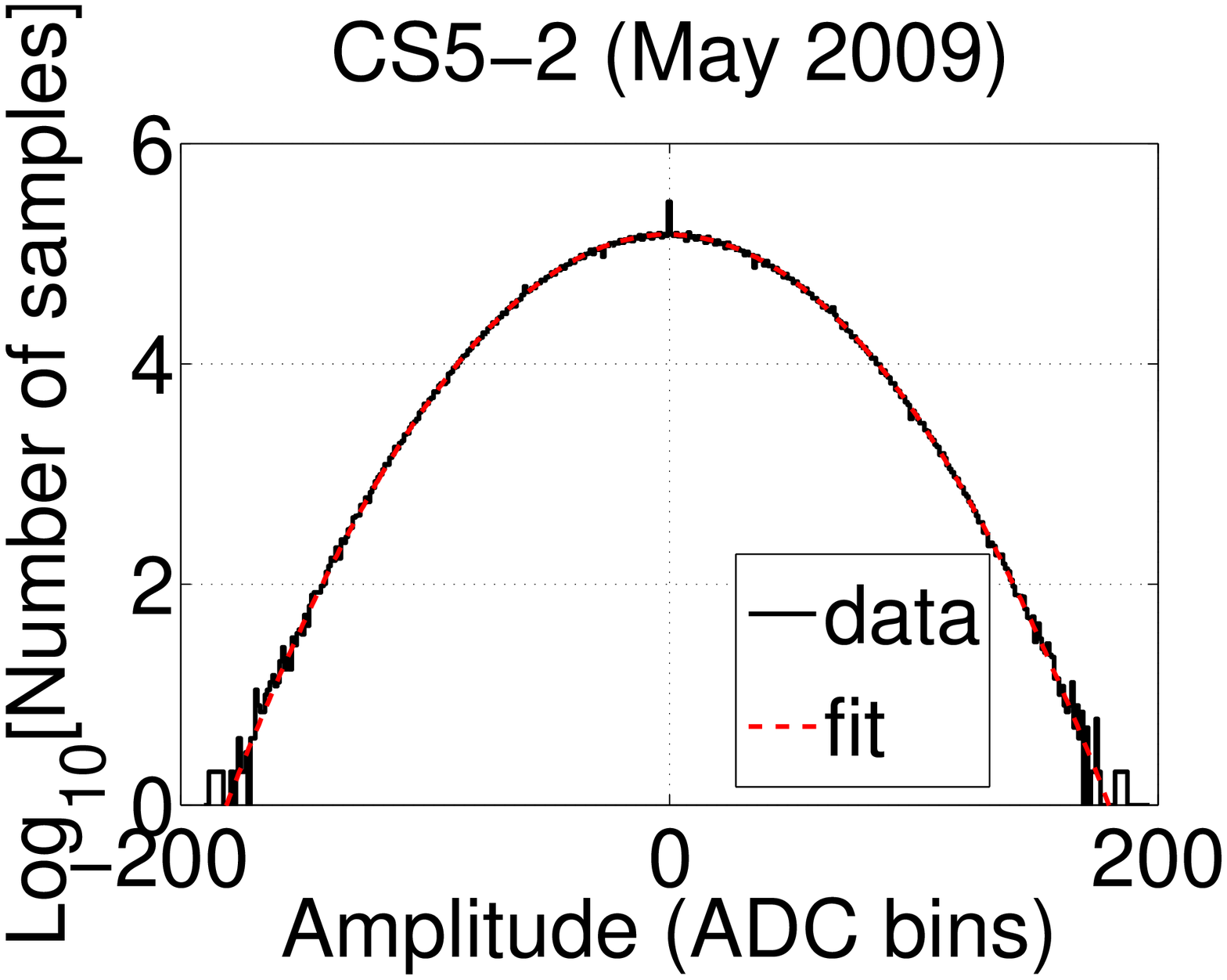}
}
\subfigure[CS6-0]{
\noindent\includegraphics[width=7pc]{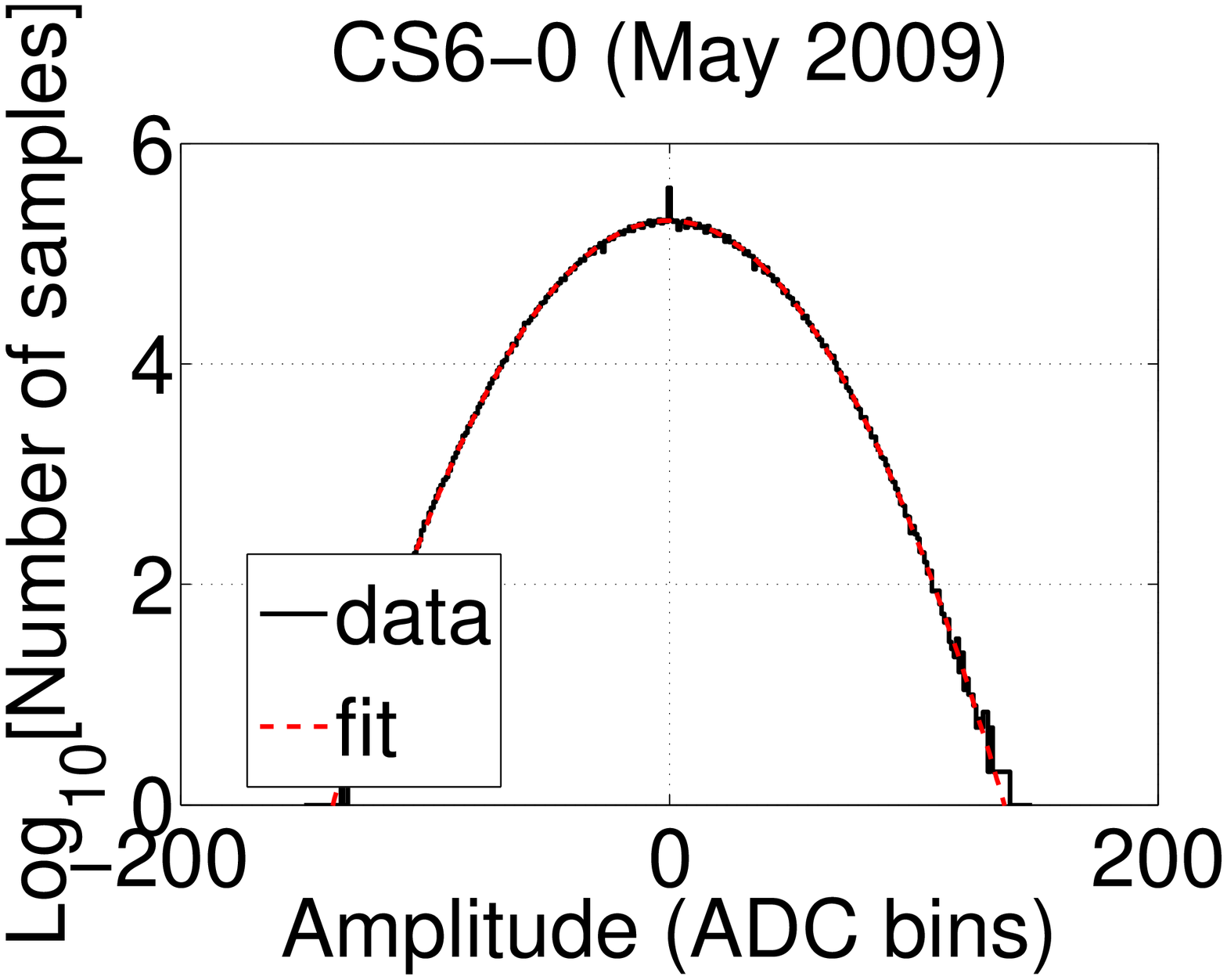}
}
\subfigure[CS6-1]{
\noindent\includegraphics[width=7pc]{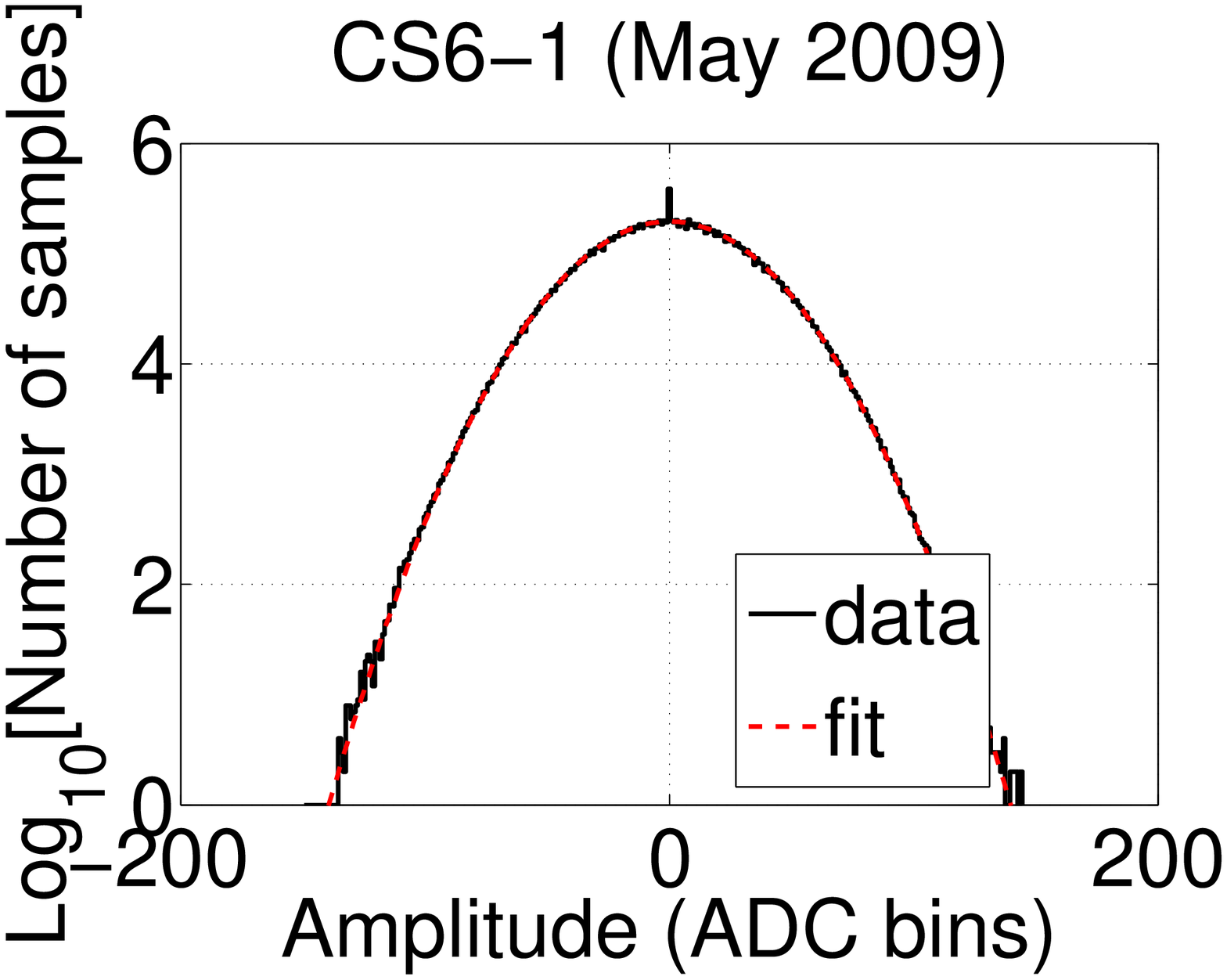}
}
\subfigure[CS6-2]{
\noindent\includegraphics[width=7pc]{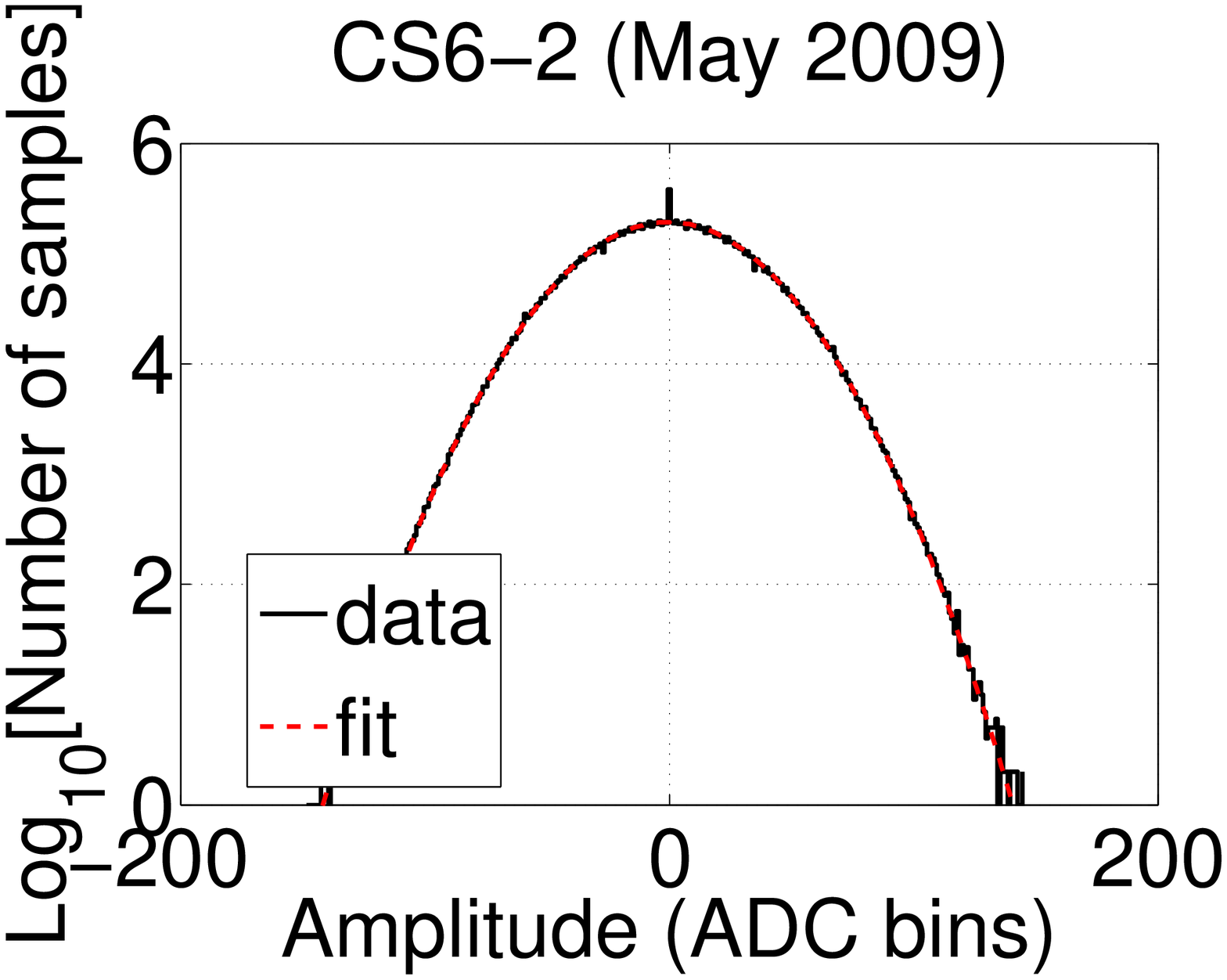}
}
\subfigure[CS7-0]{
\noindent\includegraphics[width=7pc]{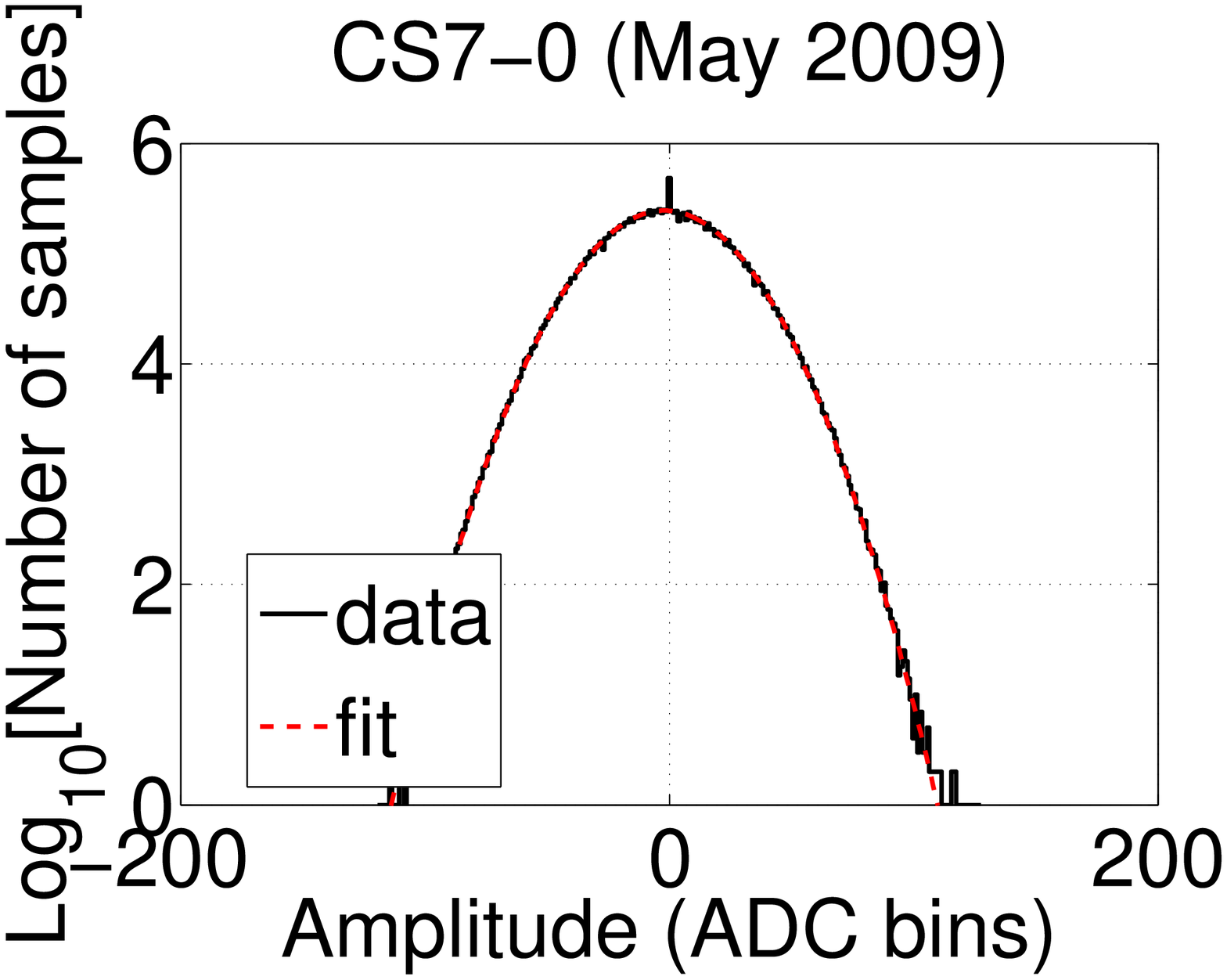}
}
\subfigure[CS7-1]{
\noindent\includegraphics[width=7pc]{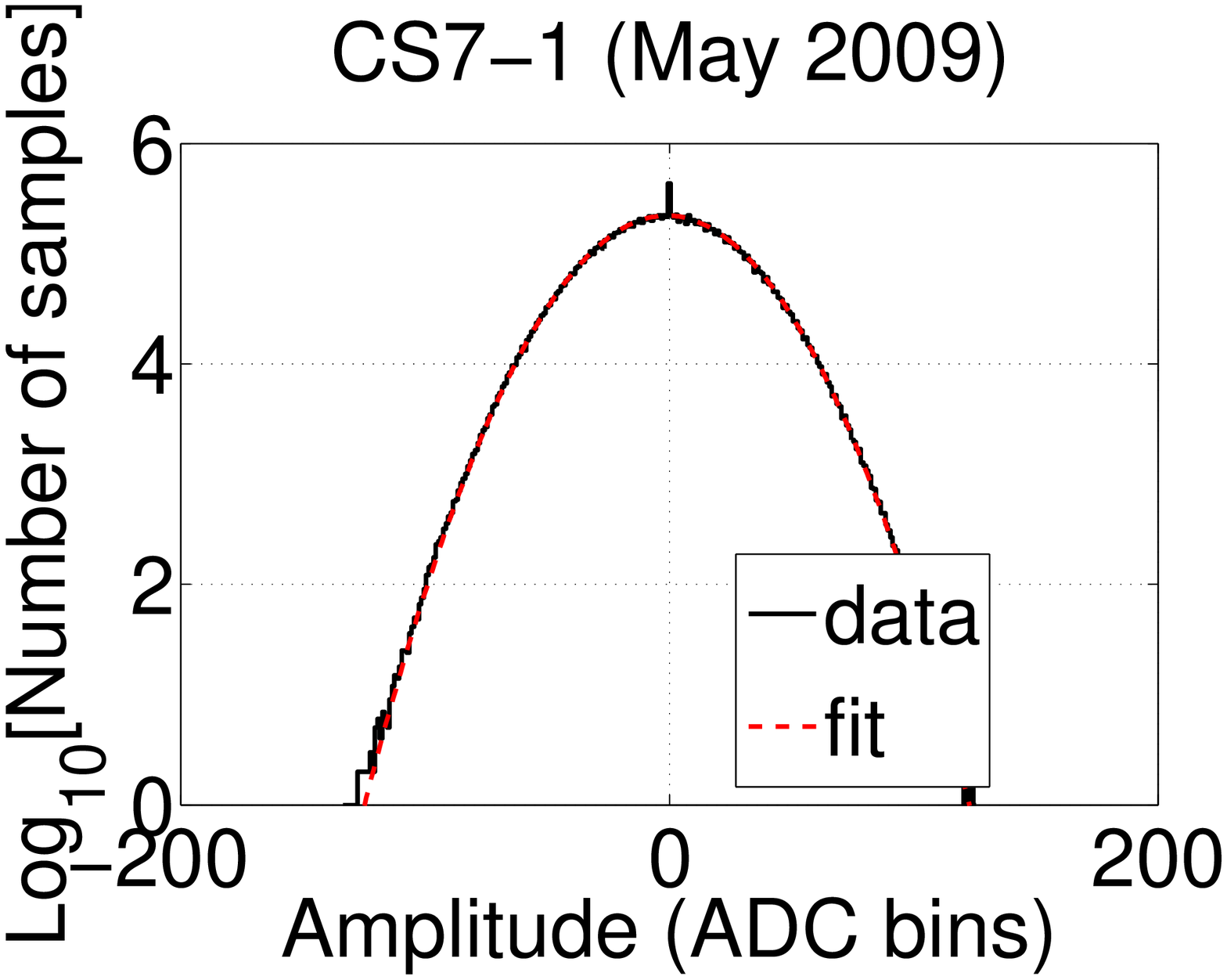}
}
\subfigure[CS7-2]{
\noindent\includegraphics[width=7pc]{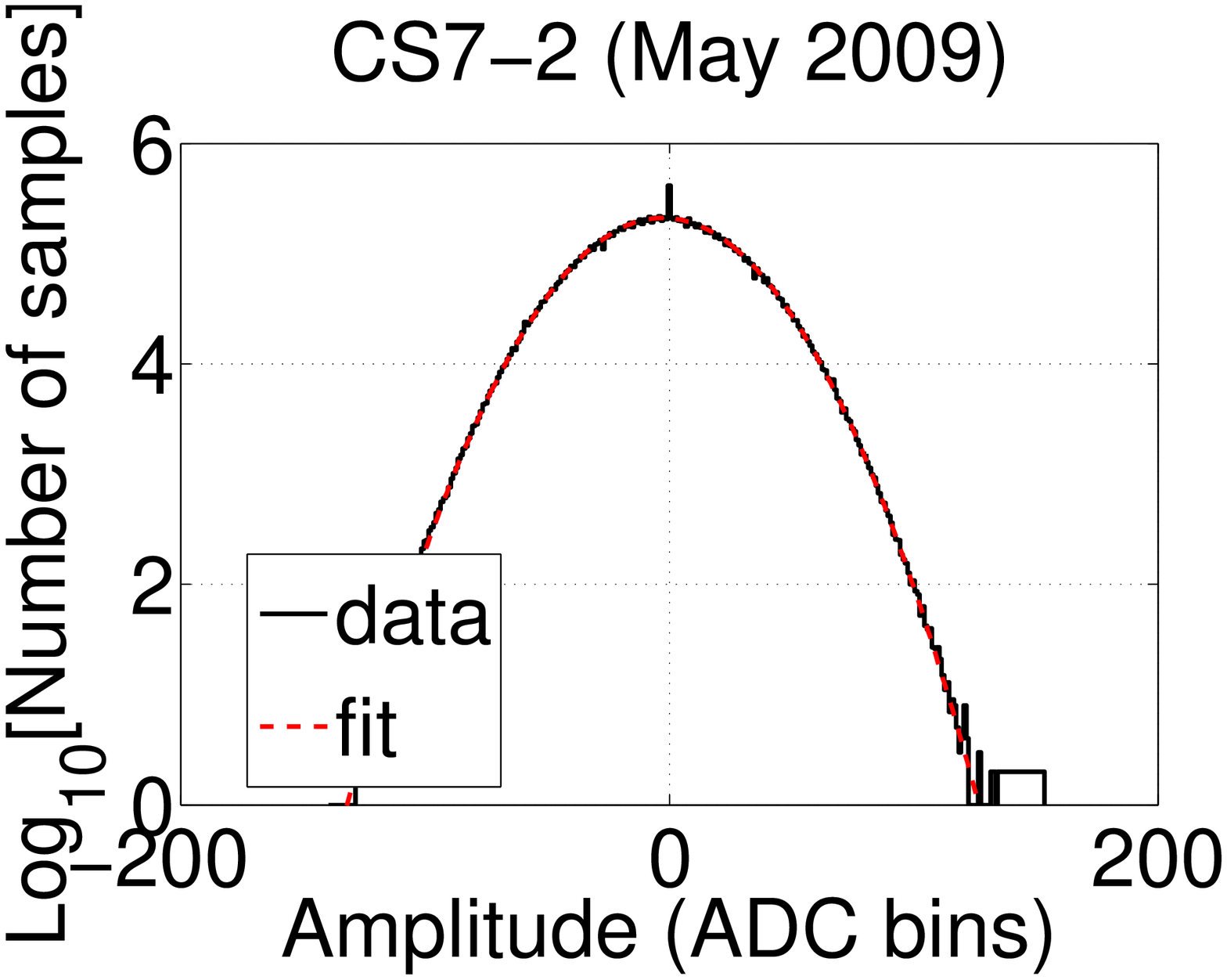}
}
\caption[Gaussian noise distributions for all String C channels]{Noise amplitude histogram (with Gaussian fit) for each channel of String C.}
\label{gaussianHistogramsC}
\end{center}
\end{figure}

\begin{figure}
\begin{center}
\subfigure[DS1-0]{
\noindent\includegraphics[width=7pc]{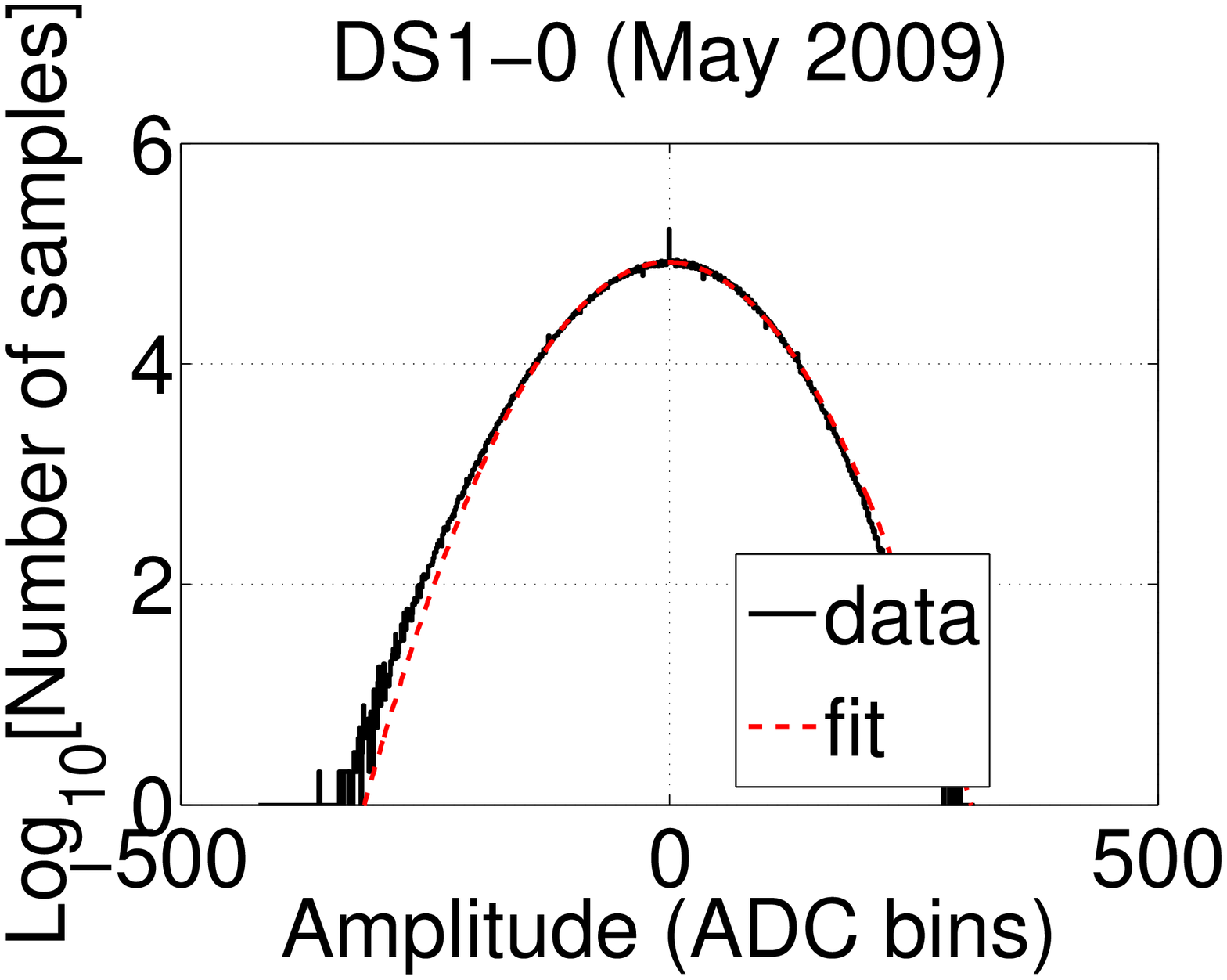}
}
\subfigure[DS1-1]{
\noindent\includegraphics[width=7pc]{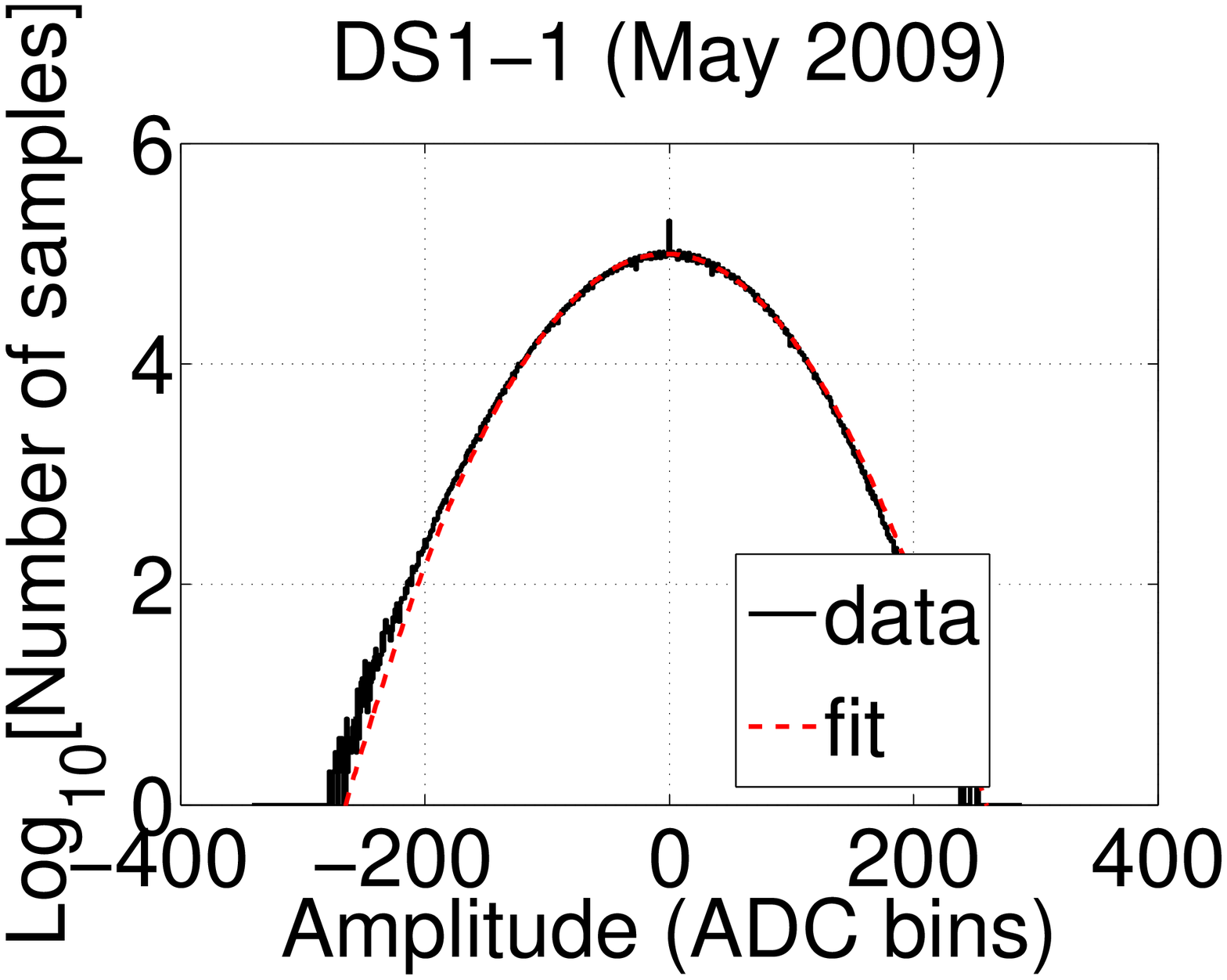}
}
\subfigure[DS1-2]{
\noindent\includegraphics[width=7pc]{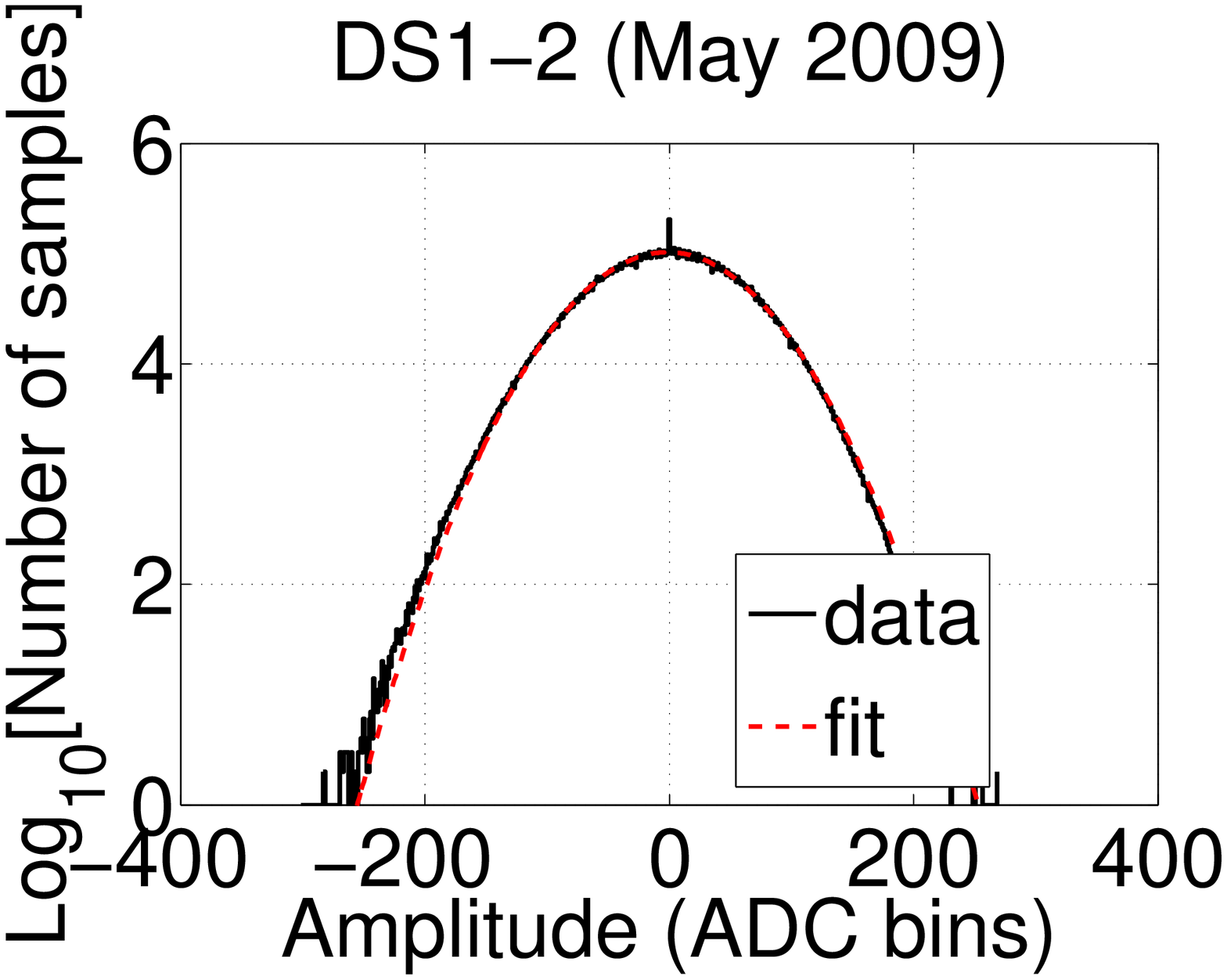}
}
\subfigure[DS2-2]{
\noindent\includegraphics[width=7pc]{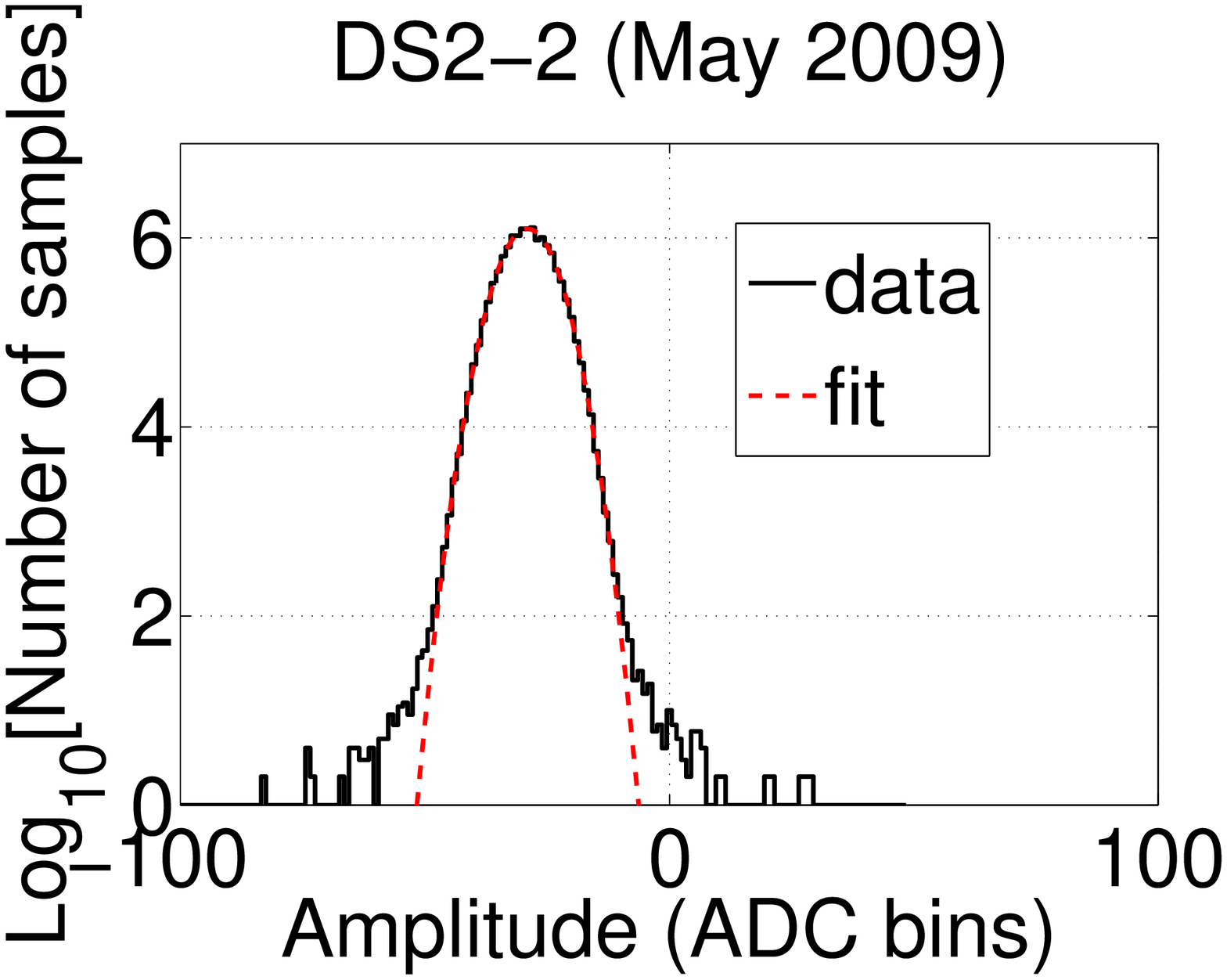}
}
\subfigure[DS3-0]{
\noindent\includegraphics[width=7pc]{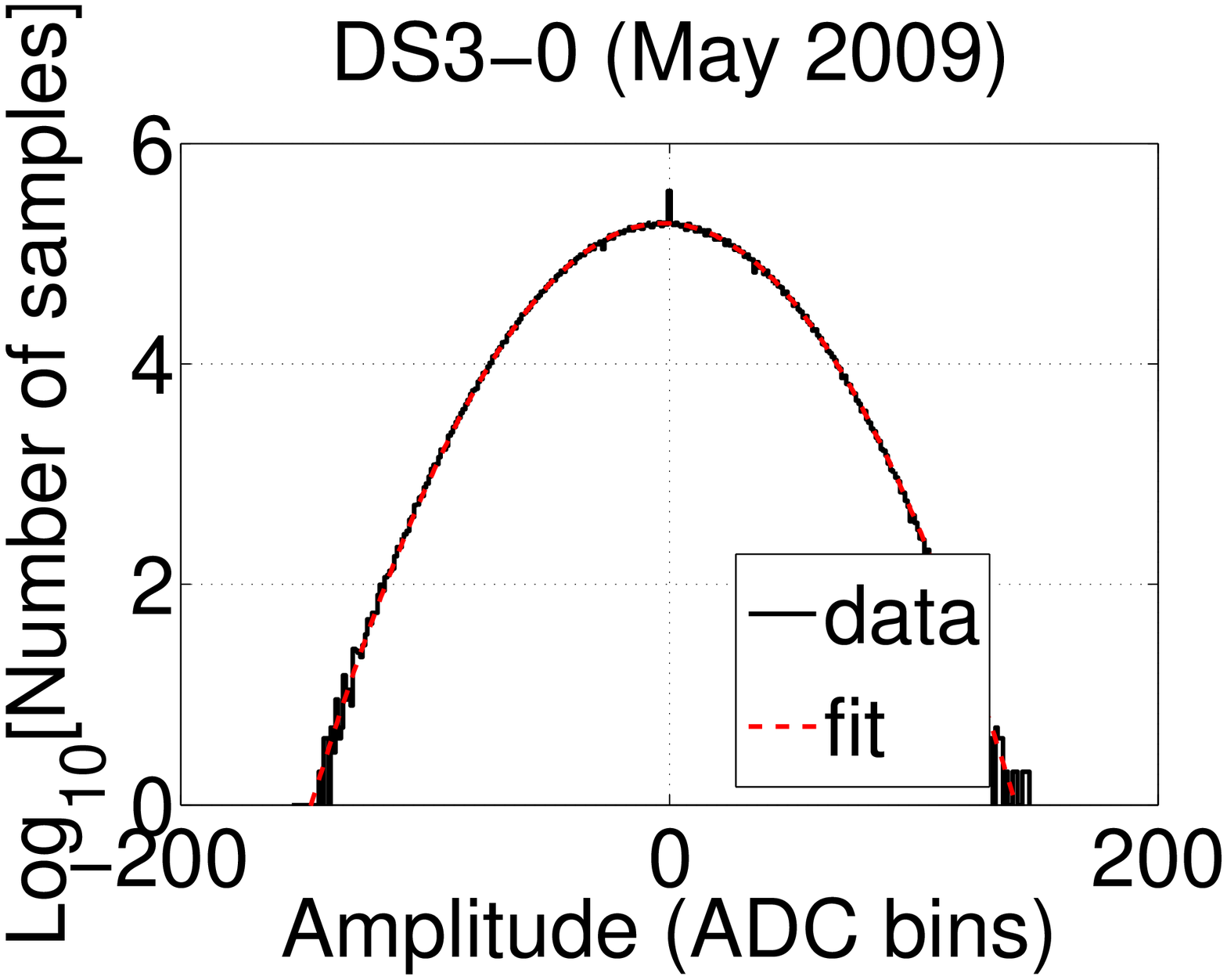}
}
\subfigure[DS3-1]{
\noindent\includegraphics[width=7pc]{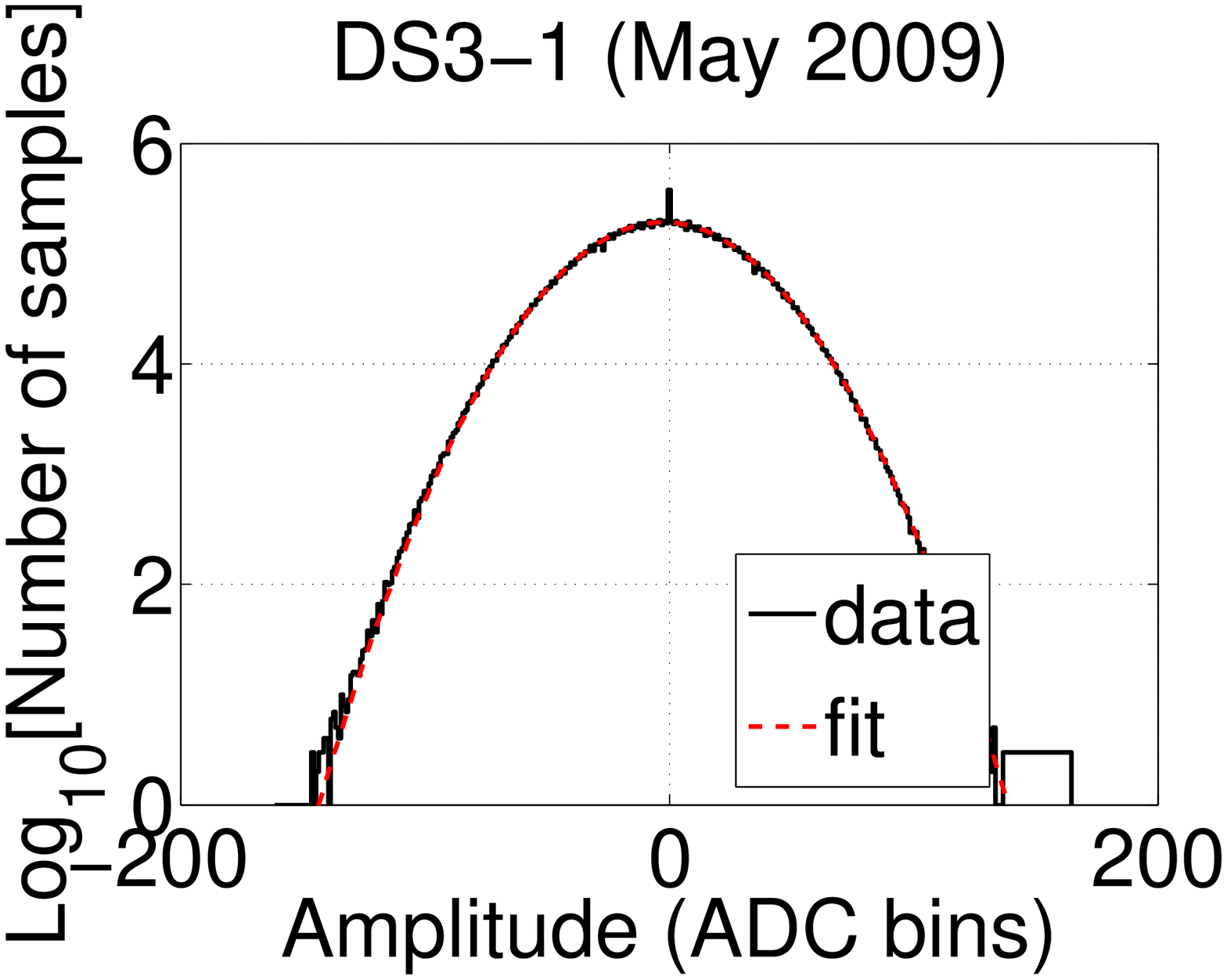}
}
\subfigure[DS3-2]{
\noindent\includegraphics[width=7pc]{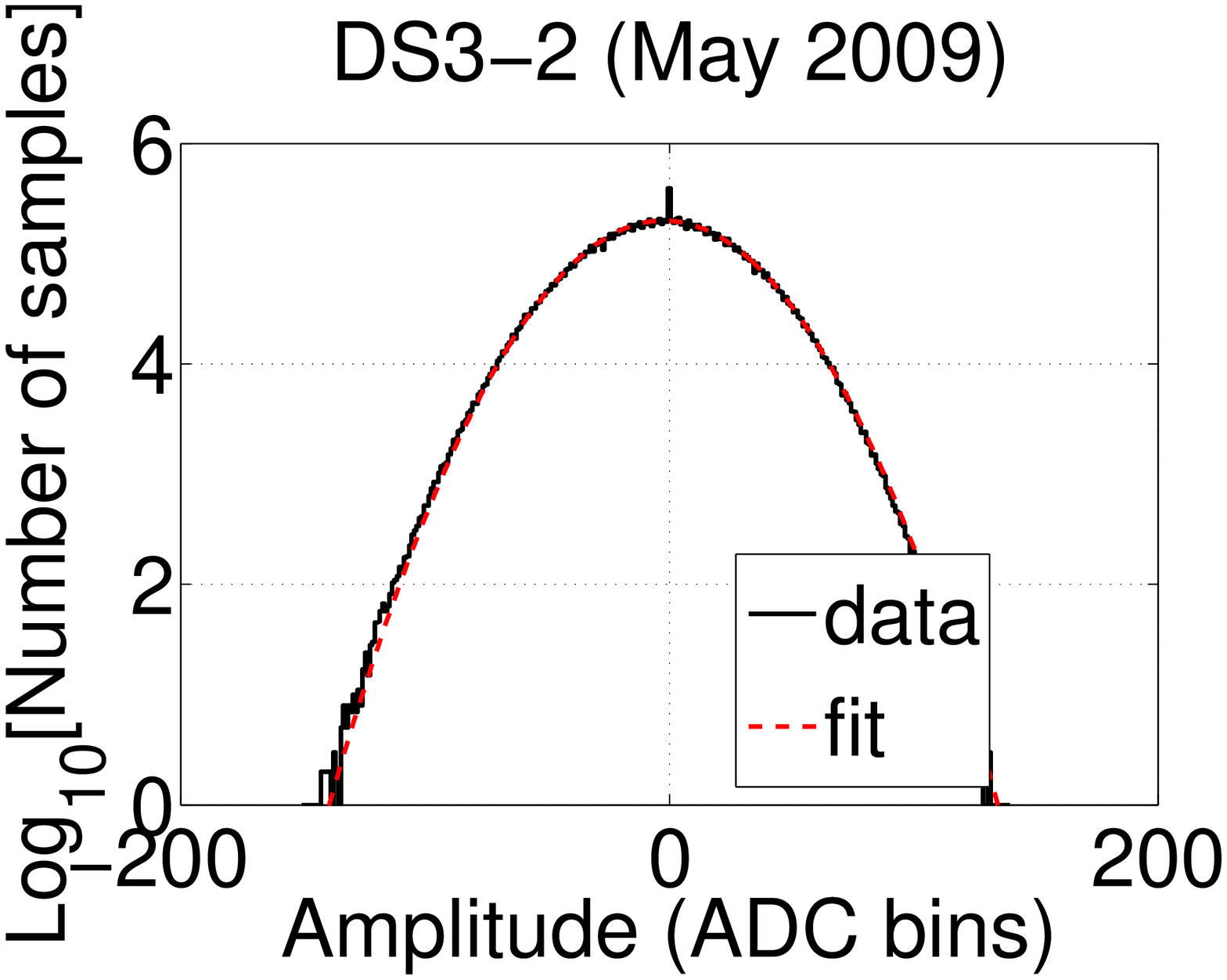}
}
\subfigure[DS4-0]{
\noindent\includegraphics[width=7pc]{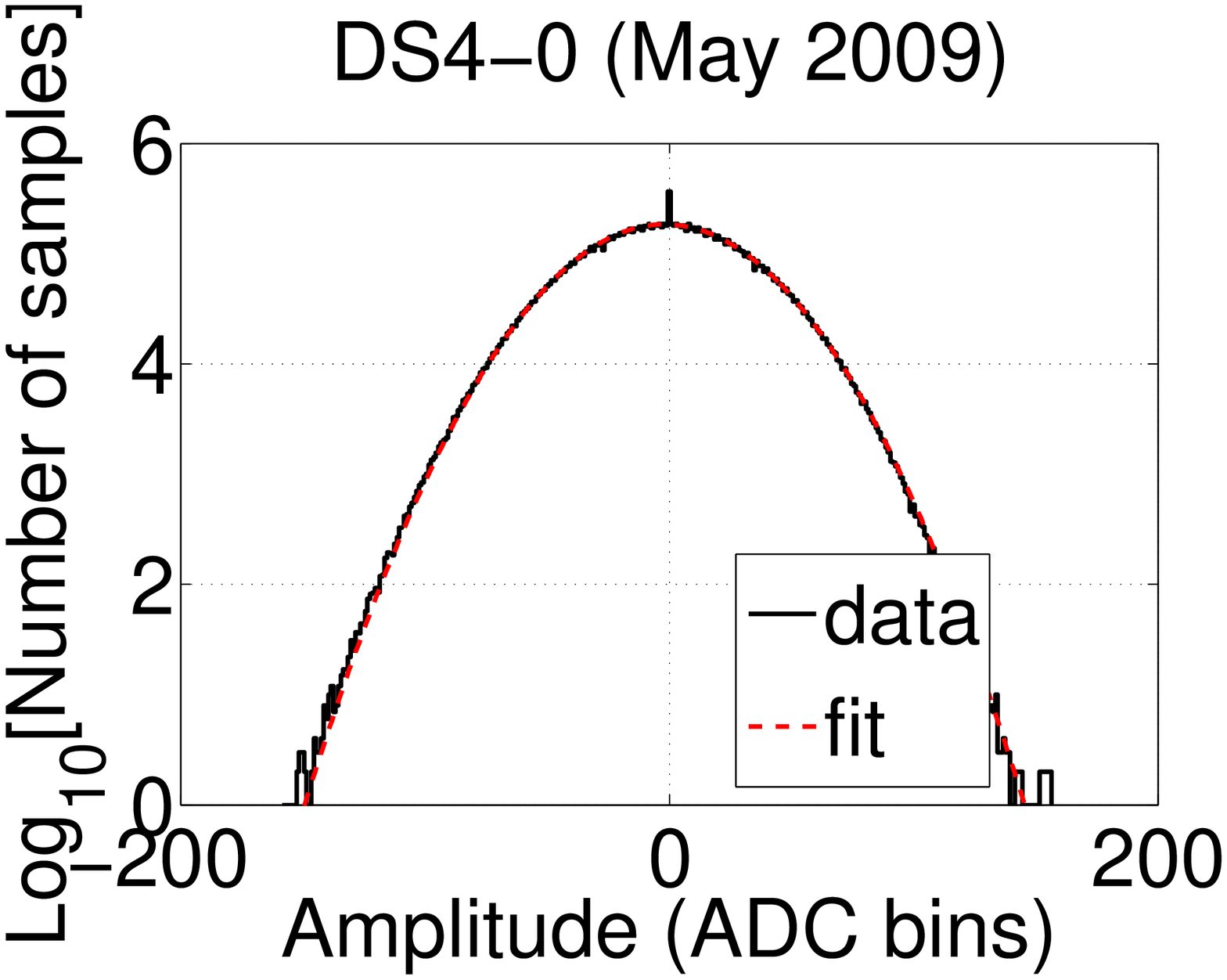}
}
\subfigure[DS4-1]{
\noindent\includegraphics[width=7pc]{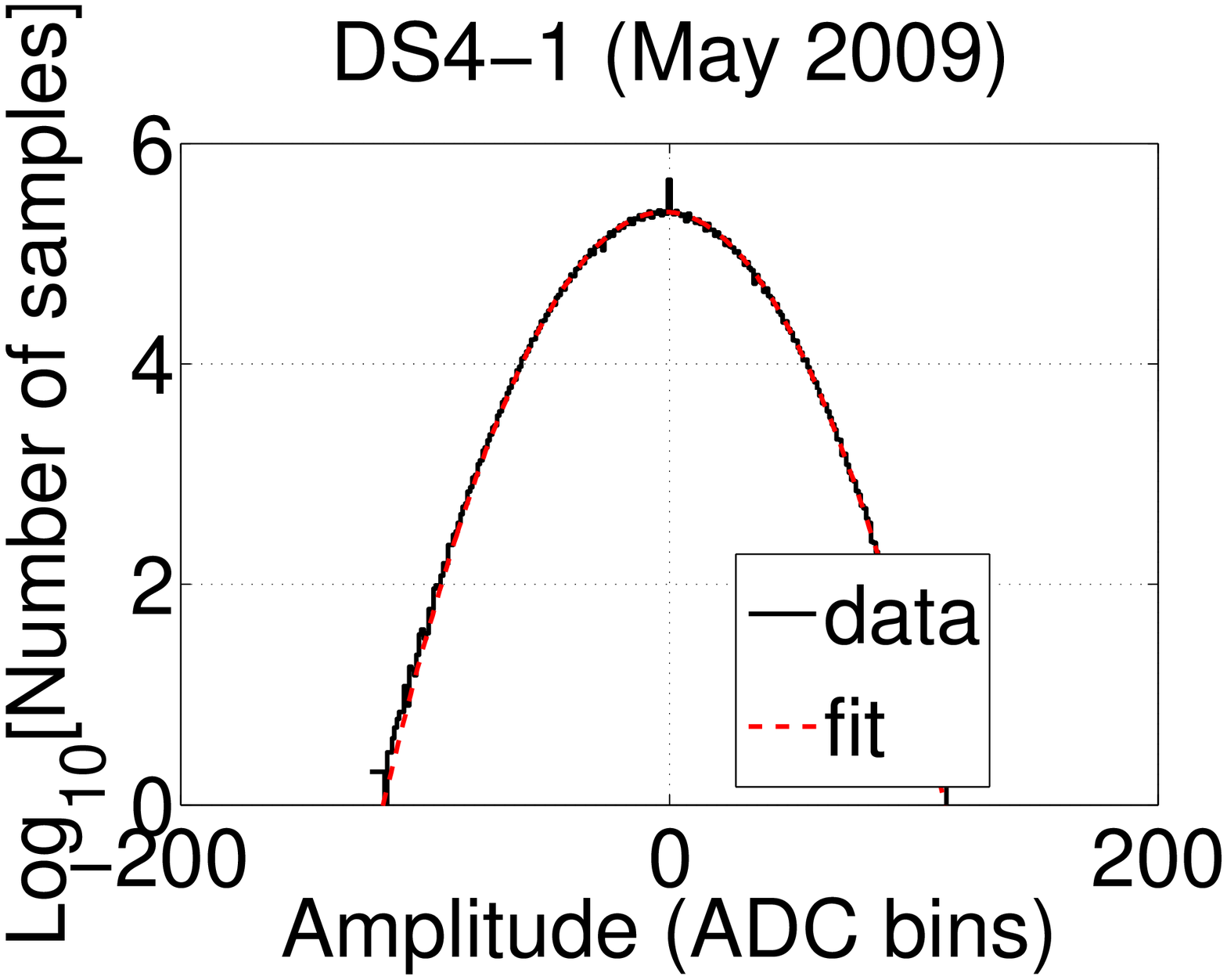}
}
\subfigure[DS4-2]{
\noindent\includegraphics[width=7pc]{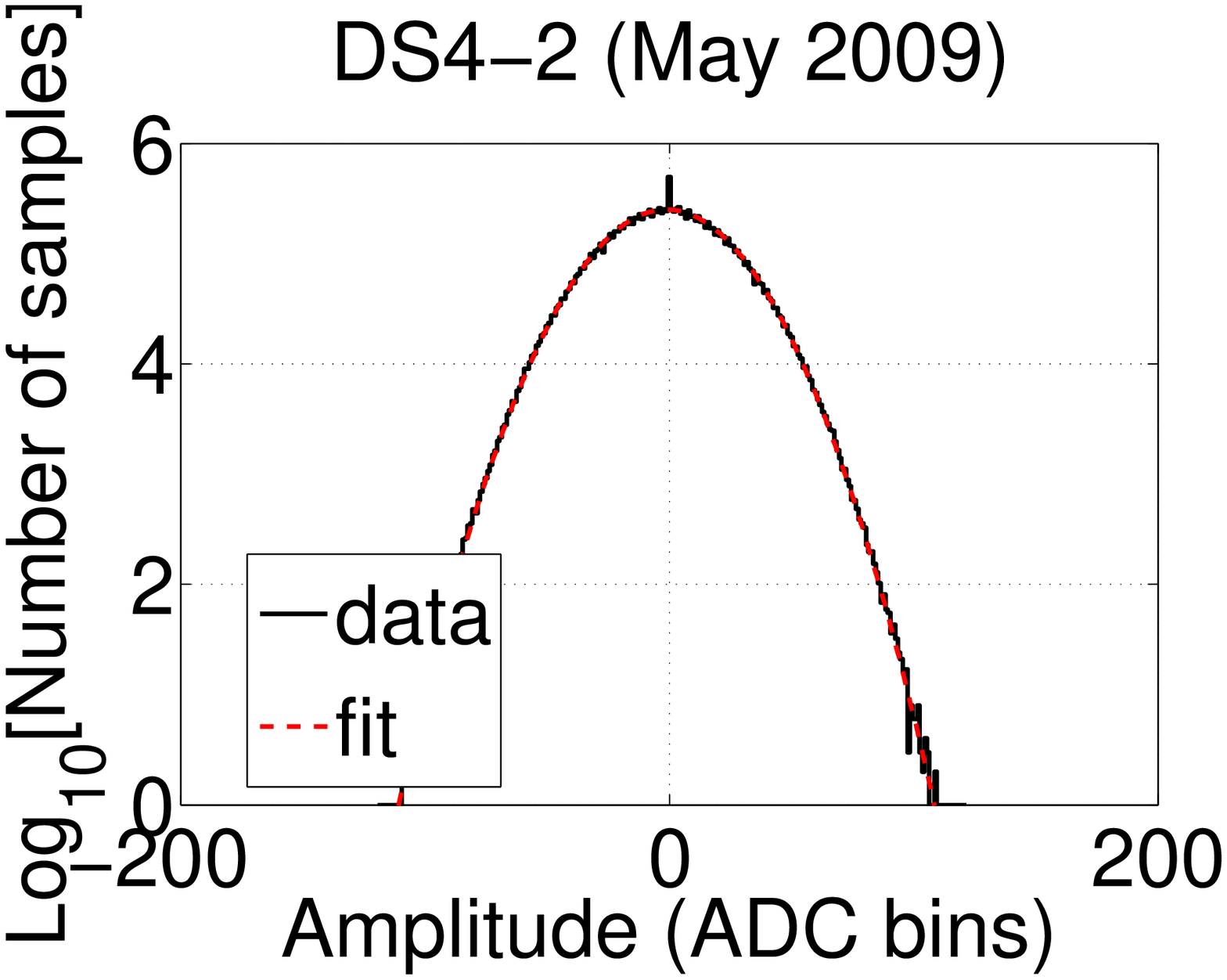}
}
\subfigure[DS5-0]{
\noindent\includegraphics[width=7pc]{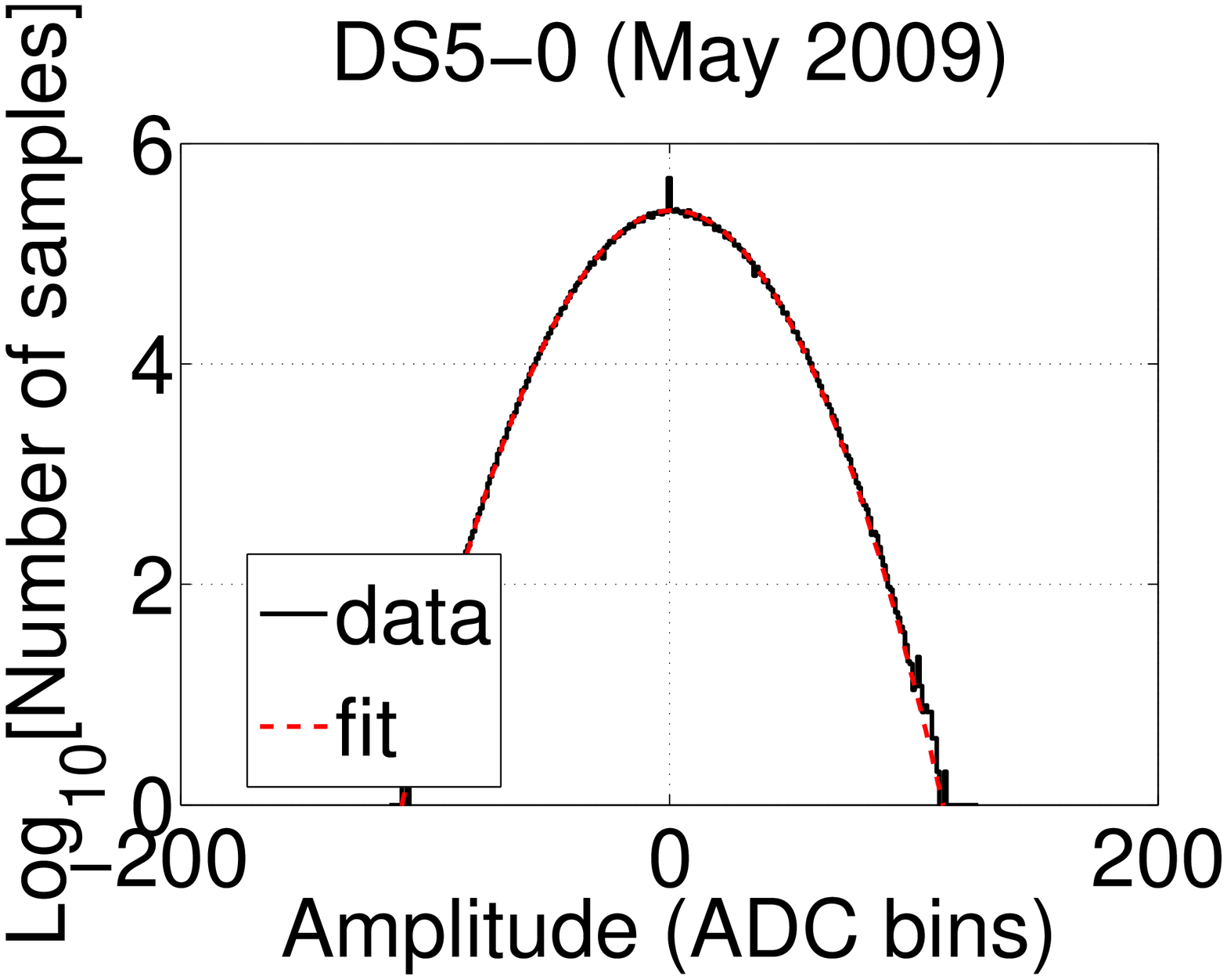}
}
\subfigure[DS5-1]{
\noindent\includegraphics[width=7pc]{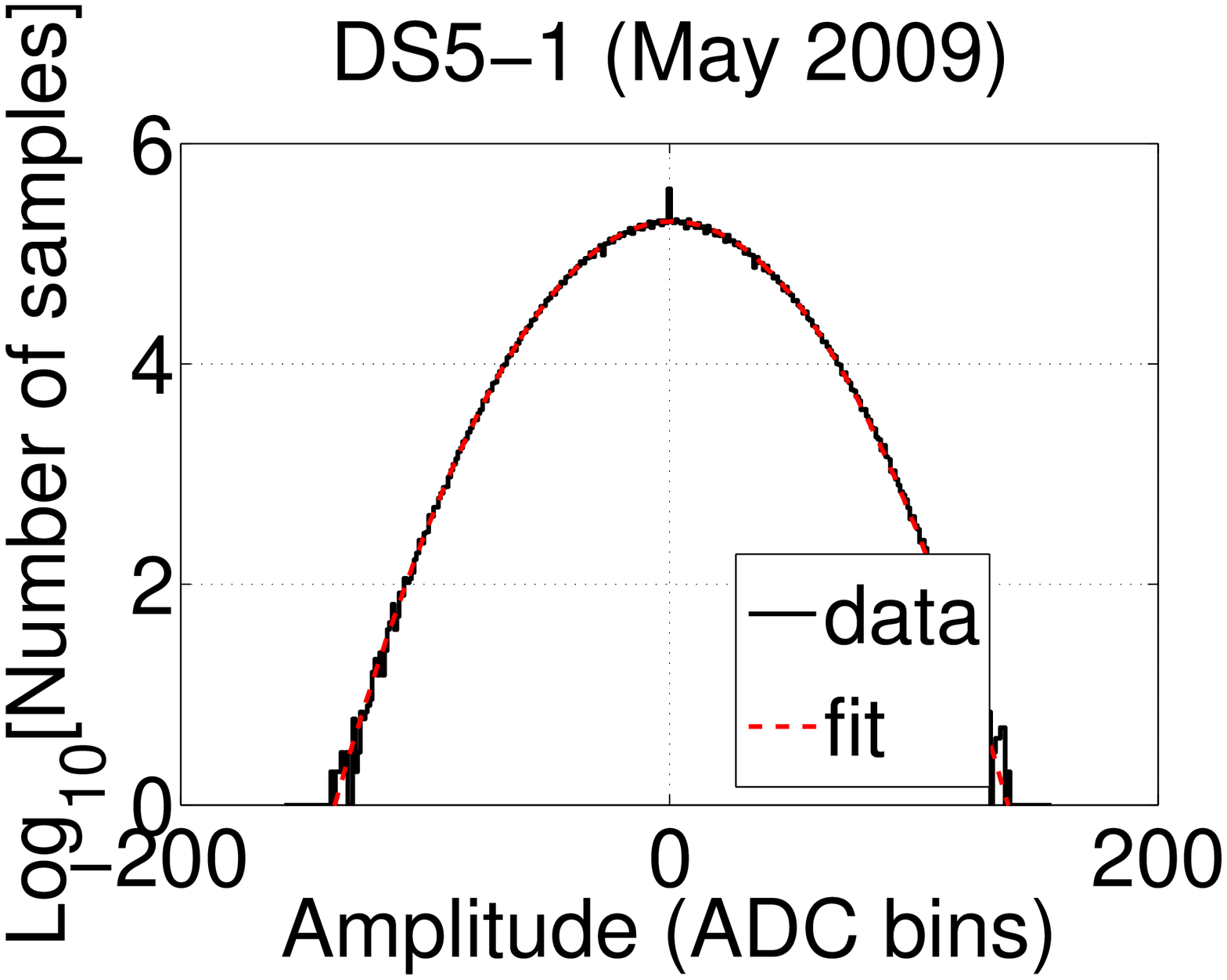}
}
\subfigure[DS5-2]{
\noindent\includegraphics[width=7pc]{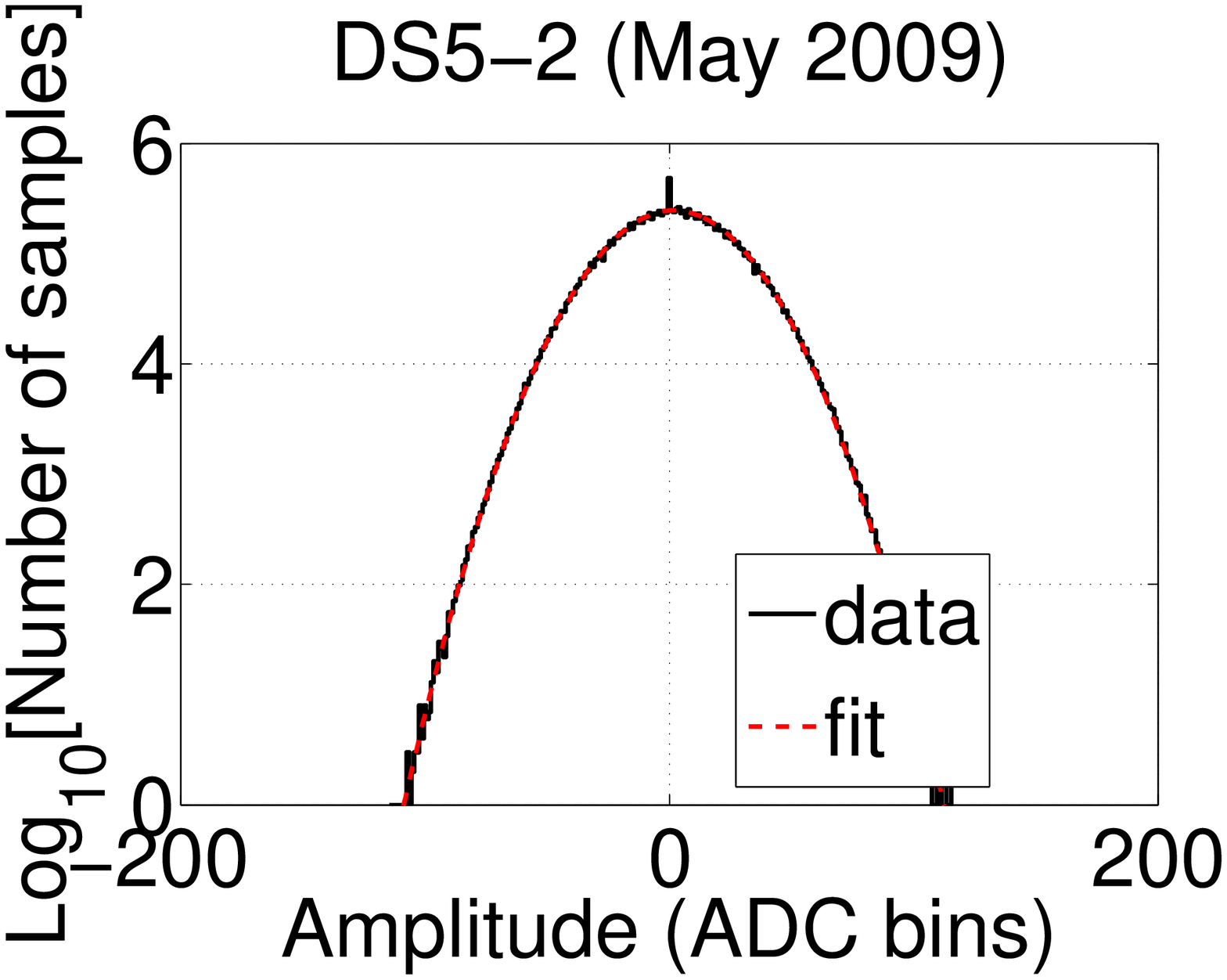}
}
\subfigure[DS6-2]{
\noindent\includegraphics[width=7pc]{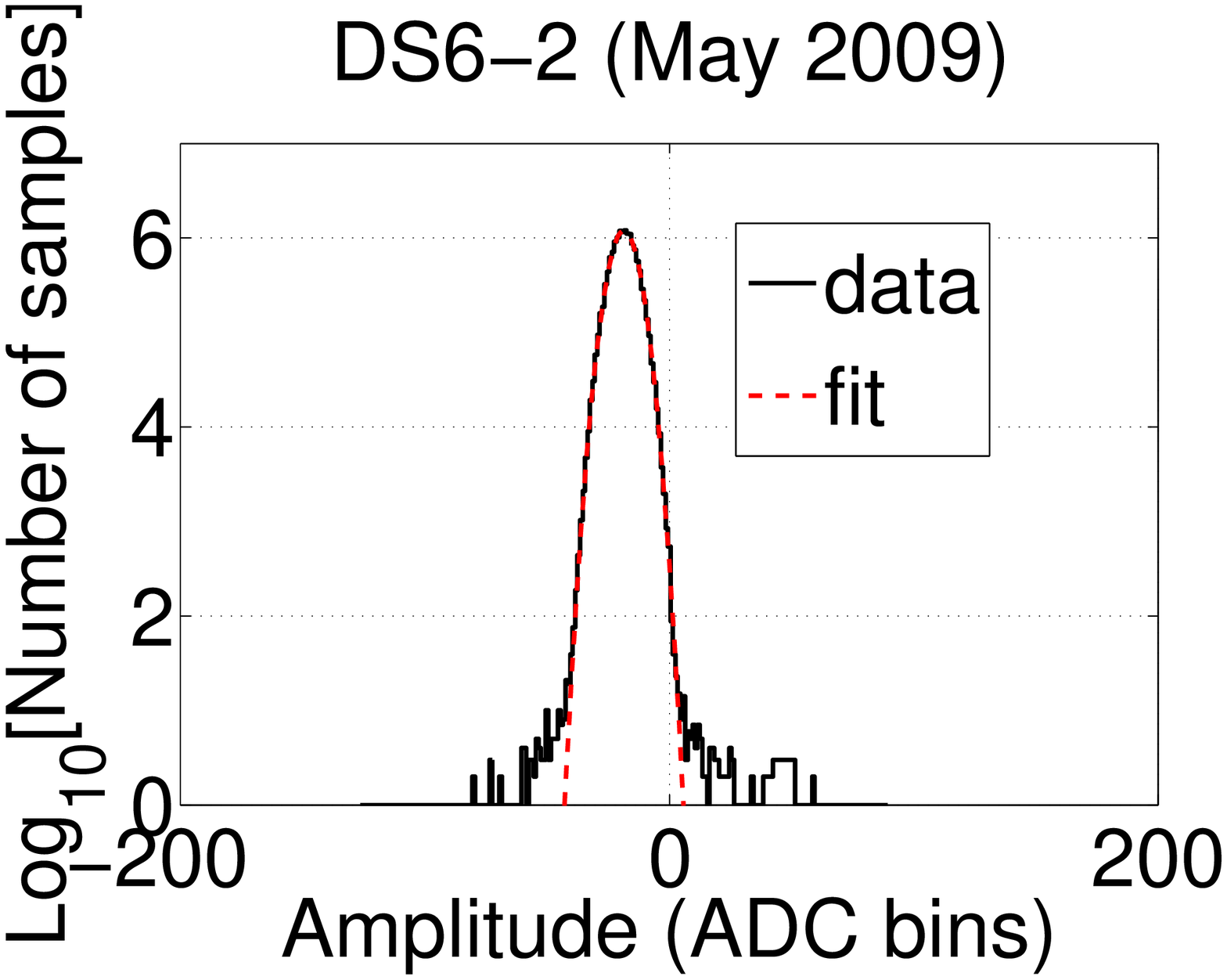}
}
\subfigure[DS7-0]{
\noindent\includegraphics[width=7pc]{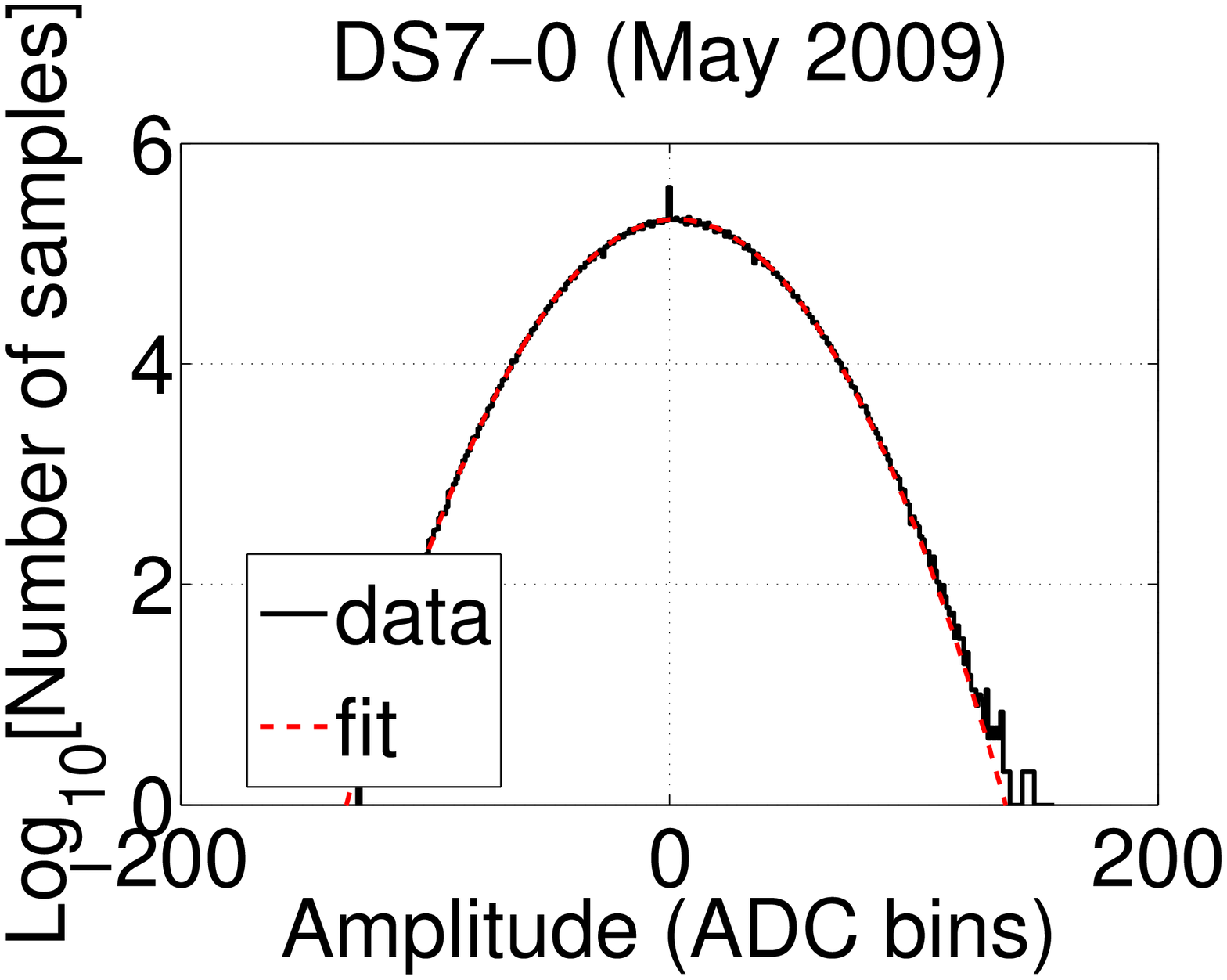}
}
\subfigure[DS7-1]{
\noindent\includegraphics[width=7pc]{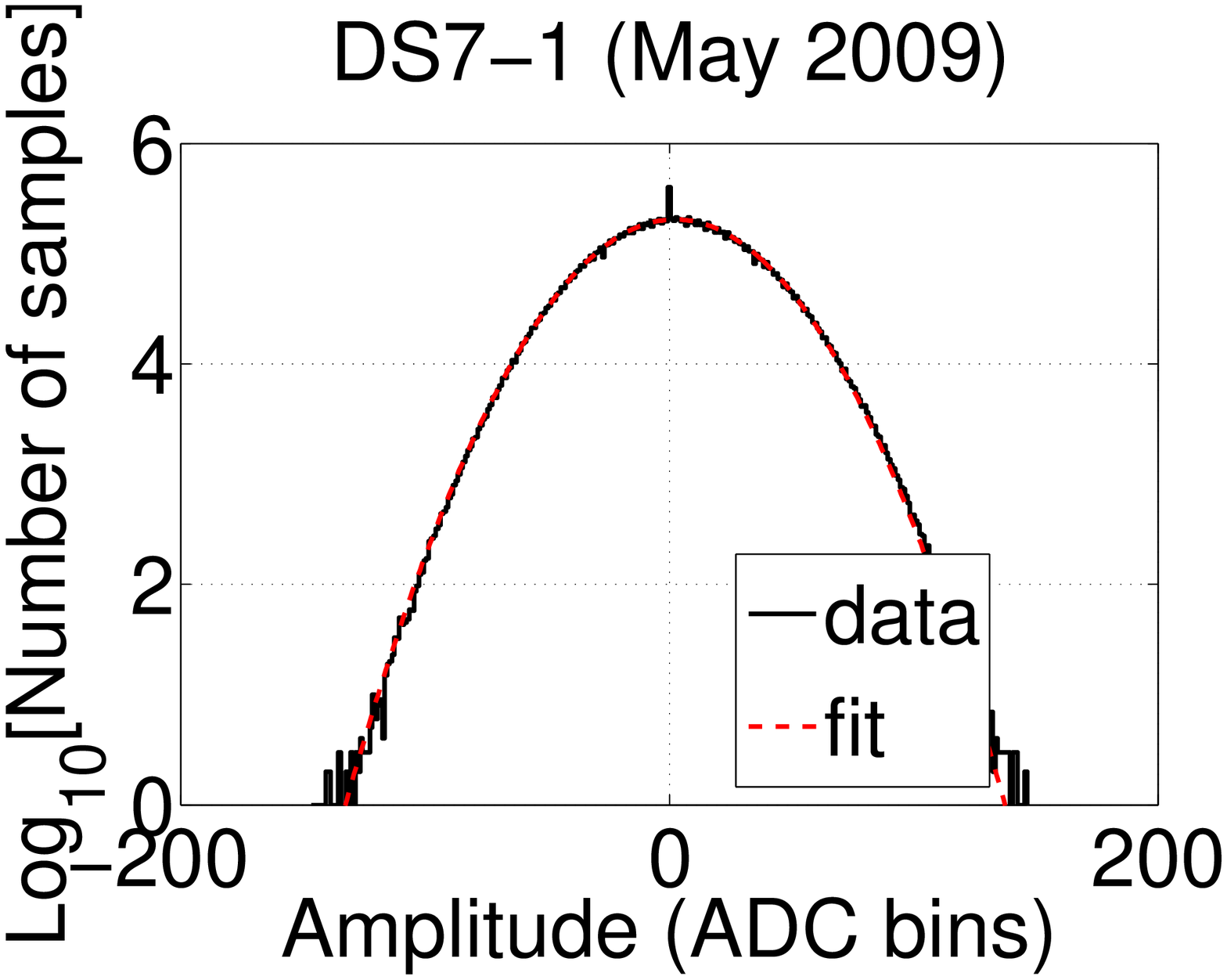}
}
\subfigure[DS7-2]{
\noindent\includegraphics[width=7pc]{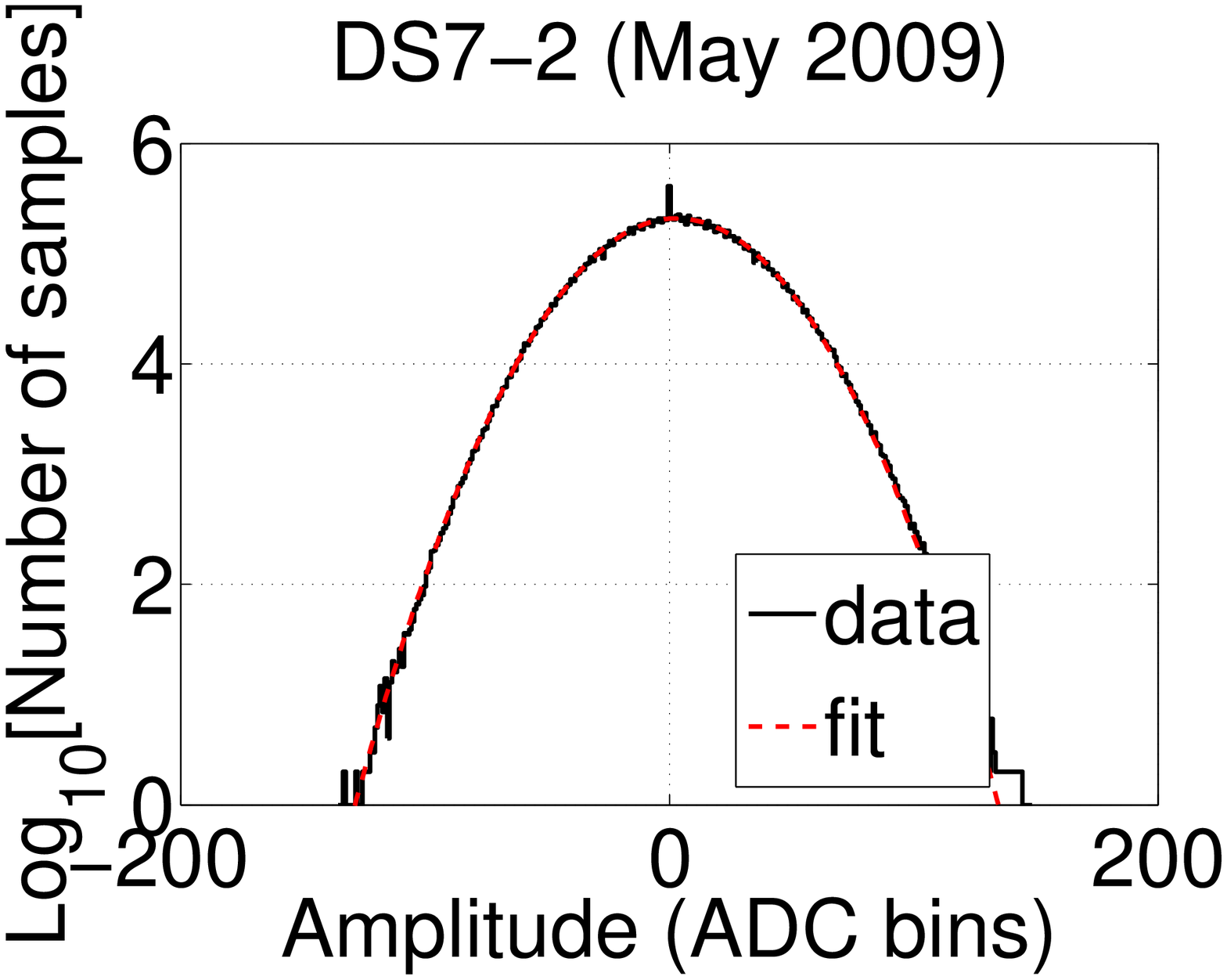}
}
\caption[Gaussian noise distributions for all String D channels]{Noise amplitude histogram (with Gaussian fit) for each channel of String D.  Note there are only 17 channels on String D, not 21, because there is only one channel on each of the HADES stages (DS2 and DS6).}
\label{gaussianHistogramsD}
\end{center}
\end{figure}

Figure~\ref{gaussianHistogramBS6-0} shows a typical noise amplitude histogram.  The data used are the ``monitor-noise'' stream, a data stream in which every channel is read out for 0.1~s once per hour, in forced mode.  All noise samples from one month on one channel were combined to make the amplitude histogram.  The distribution is very well described by a Gaussian.

Figures~\ref{gaussianHistogramsA}-\ref{gaussianHistogramsD} show noise amplitude histograms for all channels of all strings.  On Strings A, B, C, and D, there are 21+21+21+17 = 80 channels in all.  As of May 2009, every channel is well described by a Gaussian, with a few exceptions worth noting:

\begin{enumerate}

\item AS1-2 has fat tails on both sides.
\item AS3-0 is in saturation.  Note the spikes at amplitude = $\pm$2047 (the two extremes of the ADC output).  The small peaks at $\pm$~$\sim$1300 ADC bins are interesting.
\item AS3-1 and AS3-2 are Gaussian but with very large $\sigma$, an order of magnitude larger than on normal channels.
\item BS4-2 has a fat tail, interestingly on only one side (the positive side).
\item All 3 channels of CS1 have strange distributions.  CS1-0 has a particular non-Gaussian shape.  CS1-1 and CS1-2 also have a non-Gaussian shape, similar to one another but different from CS1-0.
\item The two HADES channels (DS2-2 and DS6-2) have excesses at high positive and negative amplitude.  The excesses have a different shape than the fat tails in those SPATS channels that have them.  The fat tails are smooth deviations from the Gaussian fits and appear to be the normal behavior of the HADES channels.

\end{enumerate}

\begin{table}[]
\centering
\caption[Dead channels]{Dead channels.  The channels can be determined either by the noise amplitude distribution shape, or by the amplitude of the fit to the Gaussian noise ($\sigma$).  Both methods result in the same six channels being classified dead.}
\begin{tabular}{| c | c |}	
\hline
\bf{Sensor module} & \bf {Dead channels} \\
\hline
\hline
AS3 & 0, 1, 2 \\
\hline
CS1 & 0, 1, 2 \\
\hline
\end{tabular} 
\label{deadChannels}
\end{table} 

Six channels are considered ``dead'' and are excluded from most analyses.  These channels are listed in Table~\ref{deadChannels}.

The parameters of the Gaussian fits to the amplitude distributions are shown for all channels in Table~\ref{gaussianFits}.  These are the parameters of the fits shown in Figures~\ref{gaussianHistogramsA}-\ref{gaussianHistogramsD}, using all data from May 2009.  For each channel, both $\mu$ and $\sigma$ are given in ADC bins.  Note that in the HADES modules only one channel is connected, and the other two are not connected (NC).  ``Dead'' channels are indicated in bold face.  The set of six dead channels (as determined primarily from noise amplitude histogram \emph{shape}) is exactly the same as the set of channels with Gaussian noise \emph{amplitude} ($\sigma$) greater than 100 ADC counts, so this could have been used as the criterion to determine dead channels.

\begin{table}[]
\centering
\caption[Gaussian noise fit parameters for all channels]{DC offset and standard deviation ($\mu$, $\sigma$) from Gaussian noise fits.}
\begin{tabular}{| c | c | c | c |}  
\hline
\bf{Sensor module} & \bf{Channel 0} & \bf{Channel 1} & \bf{Channel 2}\\
\hline
\hline
AS1 & (-1.70, 58.32) & (-3.64, 58.81) & (-2.23, 36.55) \\ 
\hline
AS2 & (-3.66, 45.45) & (-1.07, 46.90) & (-4.13, 57.11) \\ 
\hline
AS3 & \bf{(114.84, 1435.20)} & \bf{(-0.51, 282.00)} & \bf{(0.16, 335.96)} \\ 
\hline
AS4 & (-1.42, 43.99) & (-1.95, 47.19) & (-0.95, 41.74) \\ 
\hline
AS5 & (1.02, 38.13) & (-1.24, 34.63) & (-3.31, 32.03) \\ 
\hline
AS6 & (-1.89, 30.03) & (-3.24, 31.02) & (-3.46, 31.56) \\ 
\hline
AS7 & (-4.63, 26.51) & (-1.23, 23.43) & (-3.72, 23.11) \\ 
\hline
\hline
BS1 & (-2.24, 60.02) & (-1.14, 62.54) & (18.43, 22.01) \\
\hline
BS2 & (-4.03, 30.98) & (-1.88, 44.48) & (-0.44, 52.61) \\
\hline
BS3 & (-1.97, 49.31) & (-2.73, 42.38) & (-3.28, 50.12) \\
\hline
BS4 & (-2.07, 43.45) & (-2.87, 40.59) & (17.69, 21.88) \\
\hline
BS5 & (-3.08, 33.96) & (-0.69, 32.32) & (-3.13, 37.91) \\
\hline
BS6 & (-3.20, 30.86) & (-3.47, 30.99) & (-3.48, 28.91) \\
\hline
BS7 & (-3.45, 24.76) & (-2.35, 26.43) & (-0.32, 25.86) \\
\hline
\hline
CS1 & \bf{(163.56, 190.47)} & \bf{(11.97, 146.54)} & \bf{(6.68, 142.16)} \\
\hline
CS2 & (-0.99, 51.28) & (-2.97, 50.18) & (-2.53, 51.66) \\
\hline
CS3 & (-2.16, 46.45) & (-1.50, 39.36) & (-1.03, 40.42) \\
\hline
CS4 & (-1.49, 35.18) & (-2.94, 43.81) & (-1.21, 37.71) \\
\hline
CS5 & (-0.27, 36.70) & (-1.52, 36.61) & (-0.61, 37.01) \\
\hline
CS6 & (-0.26, 27.82) & (0.20, 28.27) & (-0.49, 28.63) \\
\hline
CS7 & (-1.97, 22.45) & (-0.93, 24.96) & (-2.53, 26.12) \\
\hline
\hline
DS1 & (-1.49, 65.22) & (-2.63, 54.67) & (-1.35, 52.90) \\
\hline
DS2 & NC & NC & (-28.97, 4.28) \\
\hline
DS3 & (-2.33, 29.27) & (-2.38, 28.60) & (-2.28, 27.72) \\
\hline
DS4 & (-1.85, 29.93) & (-2.04, 23.12) & (-1.13, 22.00) \\
\hline
DS5 & (1.34, 22.23) & (0.92, 27.92) & (1.86, 22.19) \\
\hline
DS6 & NC & NC & (-18.54, 4.60) \\
\hline
DS7 & (2.81, 27.28) & (2.47, 27.30) & (3.03, 26.61) \\
\hline
\end{tabular} 
\label{gaussianFits}
\end{table} 

\section{Time evolution of Gaussian noise}

\subsection{Initial noise evolution during freeze-in period}

\begin{figure}
\begin{center}
\subfigure[ DS1]{
\noindent\includegraphics[width=14pc]{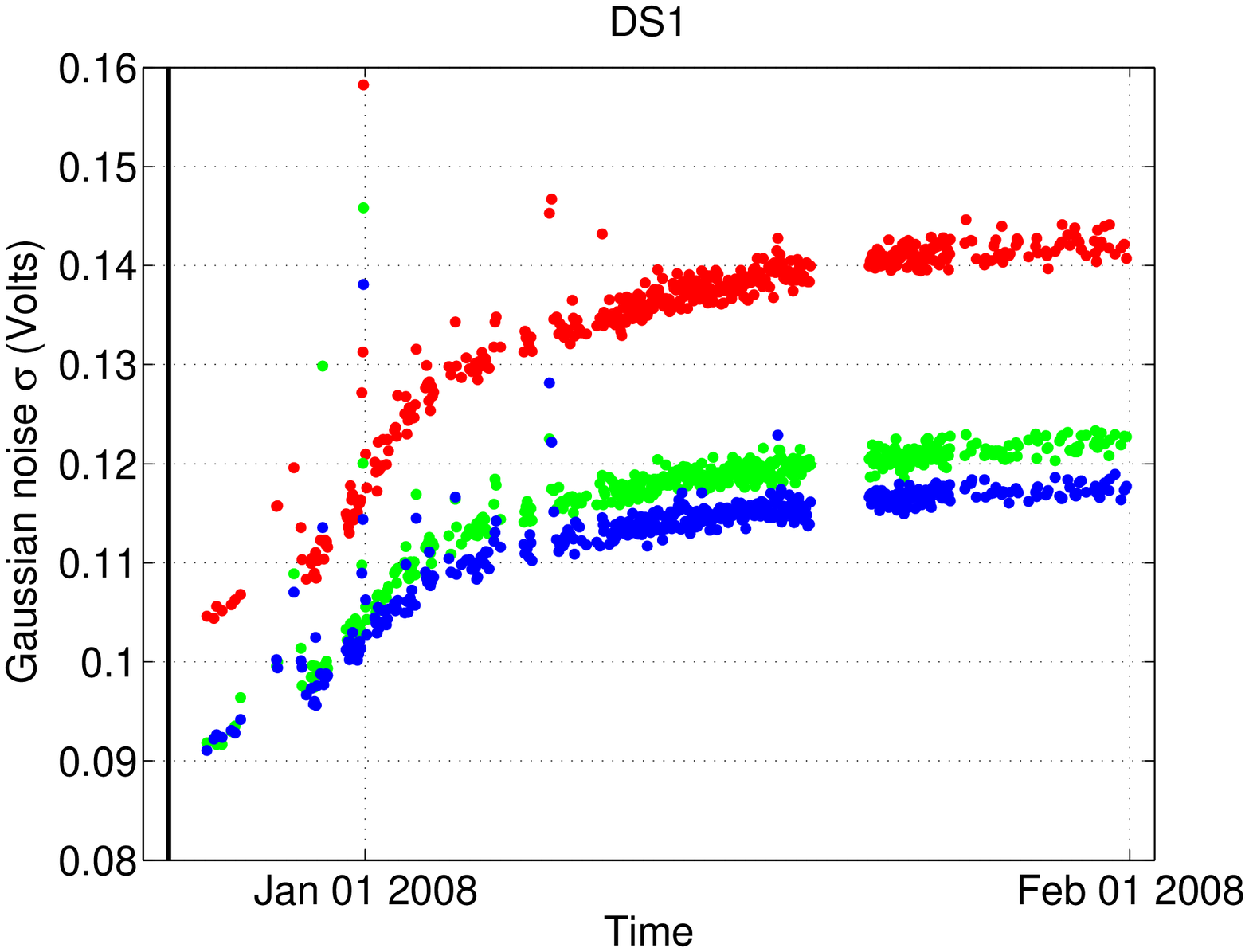}
}
\subfigure[ DS3]{
\noindent\includegraphics[width=14pc]{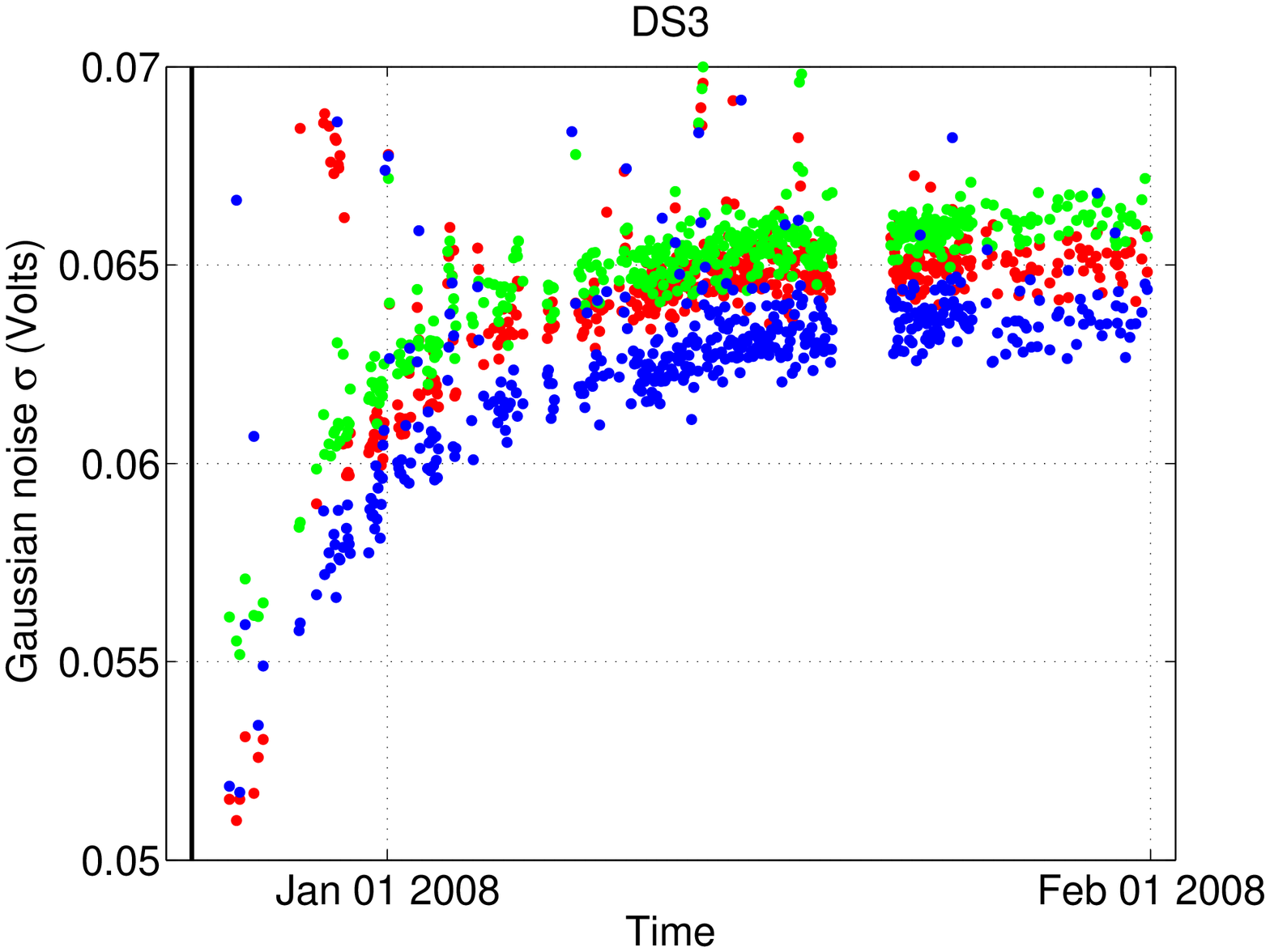}
}
\subfigure[ DS4]{
\noindent\includegraphics[width=14pc]{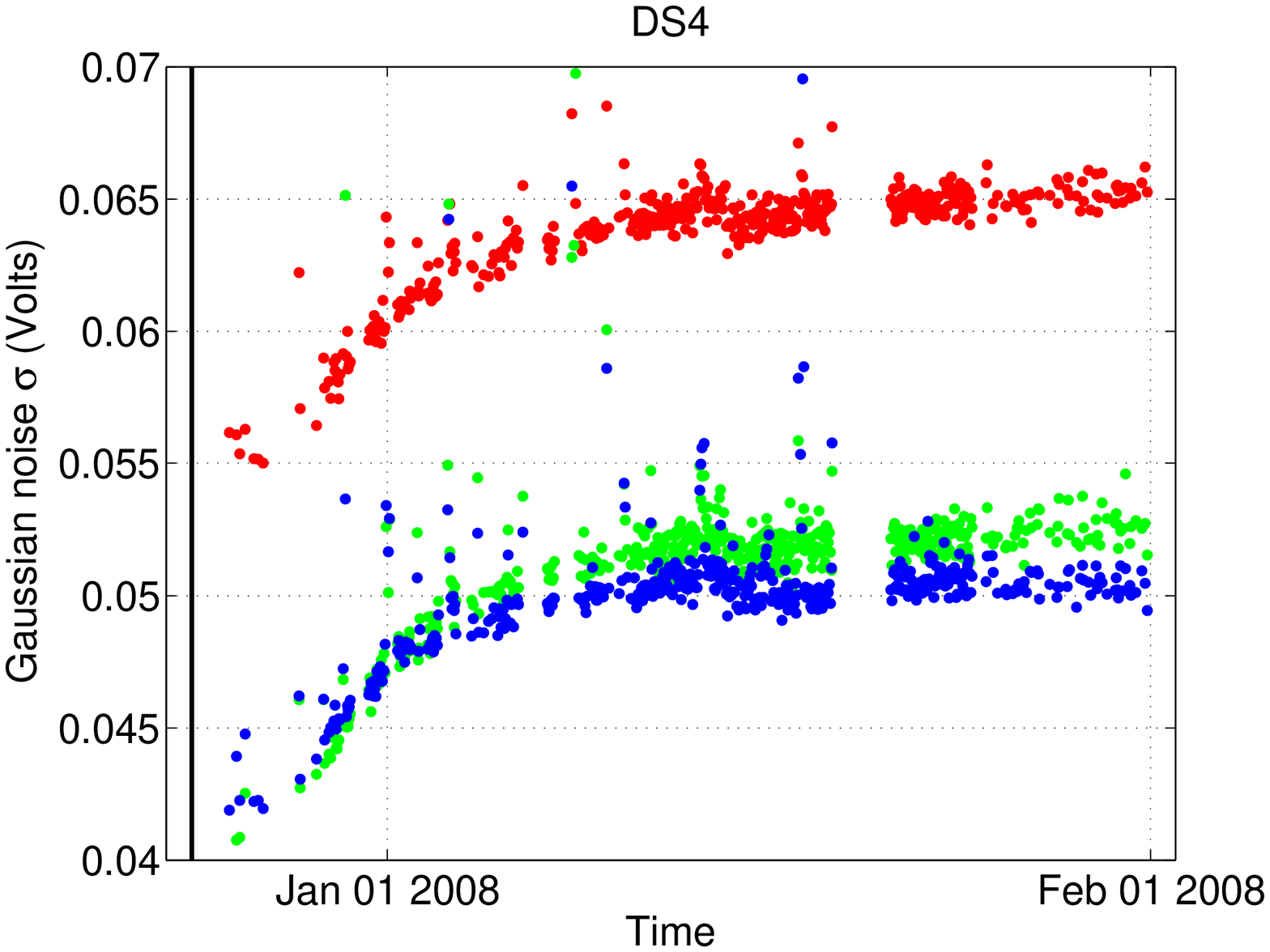}
}
\subfigure[ DS5]{
\noindent\includegraphics[width=14pc]{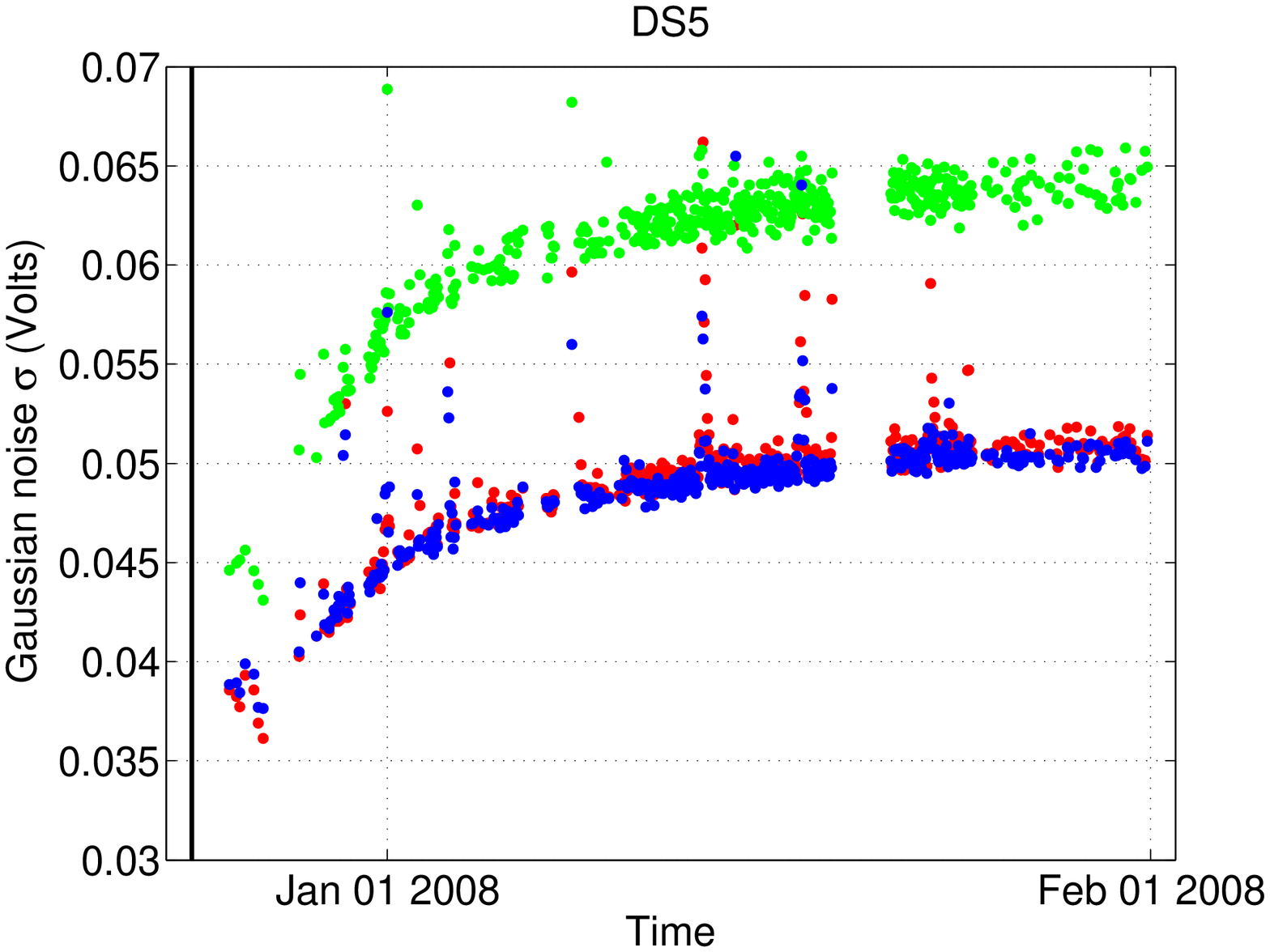}
}
\subfigure[ DS7]{
\noindent\includegraphics[width=14pc]{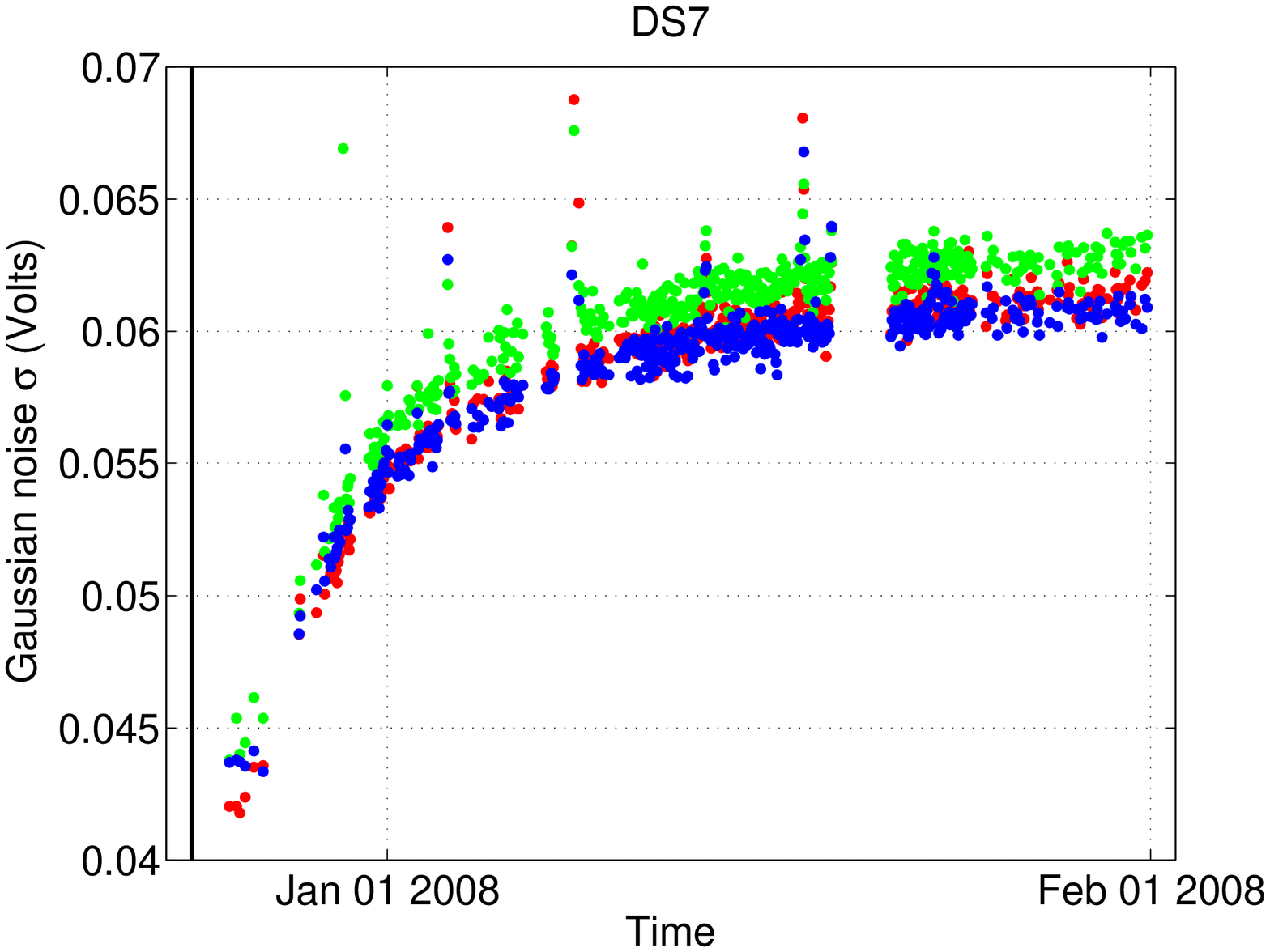}
}
\caption[Noise evolution during sensor feeze-in]{Gaussian noise amplitude ($\sigma$) for each SPATS channel on String D, during the first few weeks after String D was deployed, during which the string was freezing into place.  One plot is shown for each module, with all three channels of the module included.  Channel 0, 1, 2 is colored red, green, blue respectively.    The spikes visible on top of the smoothly increasing baseline are due to drilling of individual IceCube holes.}
\label{freezeNoise}
\end{center}
\end{figure}

The Gaussian noise on each channel increased steadily after freeze-in for several weeks and then stabilized.  We believe the noise increase during the freezing process is due to the acoustic coupling of the modules to the bulk ice improving as the module freezes to the ice and as its temperature drops to that of the surrounding ice.  Figure~\ref{freezeNoise} shows noise vs. time for all String D channels during the first few weeks after deployment.  HADES channels were operating with a gain that was too low for the data to be useful, and are not shown.  The time that we completed String D deployment (01:14 UTC on December 24, 2007) is shown with a vertical line.  These plots show one point per run, with one run taken every hour.  From each run we extract a continuous recording of 0.1~s duration, sampled at 200~kHz sampling frequency.  At least 0.1~s of data are recorded and discarded before selecting the 0.1~s recording to use, in order to avoid artifacts from the ``leading zeros'' ADC problem.  A histogram is built from the 20,000 sample amplitudes and a Gaussian curve is fit to the histogram as shown in Section~\ref{currentNoiseHistos}.

Similar behavior was observed on Strings A, B, and C.  The noise level of each channel increased steadily over several weeks during the freeze-in process and then reached a stable value for the next 1.5 (for String D) to 2.5 (for Strings A, B, and C) years until the present.

From freeze-in until now (a period of 1.5-2.5 years), the noise level on every channel has been very stable.  There is no evidence of noise variation occurring during short-term weather changes (storms) on the surface or during the intense periods of human activity on the surface (using heavy construction equipment) extending from October to February during IceCube construction in the austral summers.

There is one notable source of time variation in the Gaussian noise, which is the IceCube hot water drill.  During construction seasons, we hear the drill go down and up again in each hole that it drills.  A spike is resolved by each sensor channel as it passes the depth of the channel on the way down, and another spike is resolved on the way up.  The narrowness of this spike is interesting and may indicate directional acoustic emission from the IceCube drill.  Another possible explanation is layering of the bulk ice acoustic properties.  But this is difficult to reconcile with the fact that we have detected acoustic signals from a wide range of zenith angles from both the retrievable pinger and from frozen-in SPATS sensors.

We do not hear any surface noise and indeed the only clear source of noise variation we have seen is due to a source (the IceCube drill) below the surface, in the deep ice.  This is in distinct contrast to ocean noise which varies strongly on multiple time scales and is known to be correlated with surface wind.  This could be confirmation of our expectation that the waveguide effect in the firn reflects surface noise back to the surface and shields the deep ice from surface noise.

\subsection{Multi-year noise evolution}

\begin{figure}
\begin{center}
\subfigure[ AS1]{
\noindent\includegraphics[width=11pc]{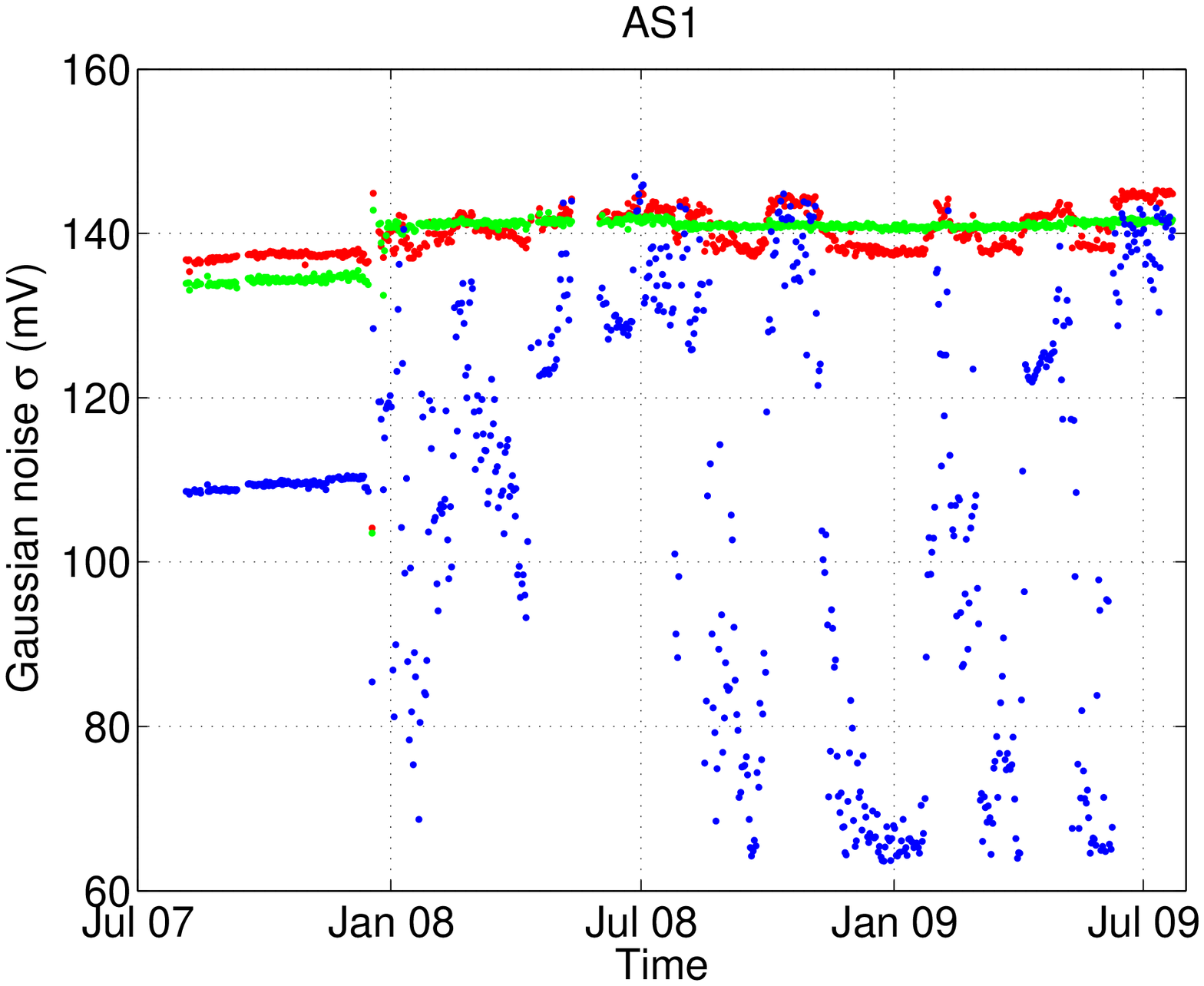}
}
\subfigure[ AS2]{
\noindent\includegraphics[width=11pc]{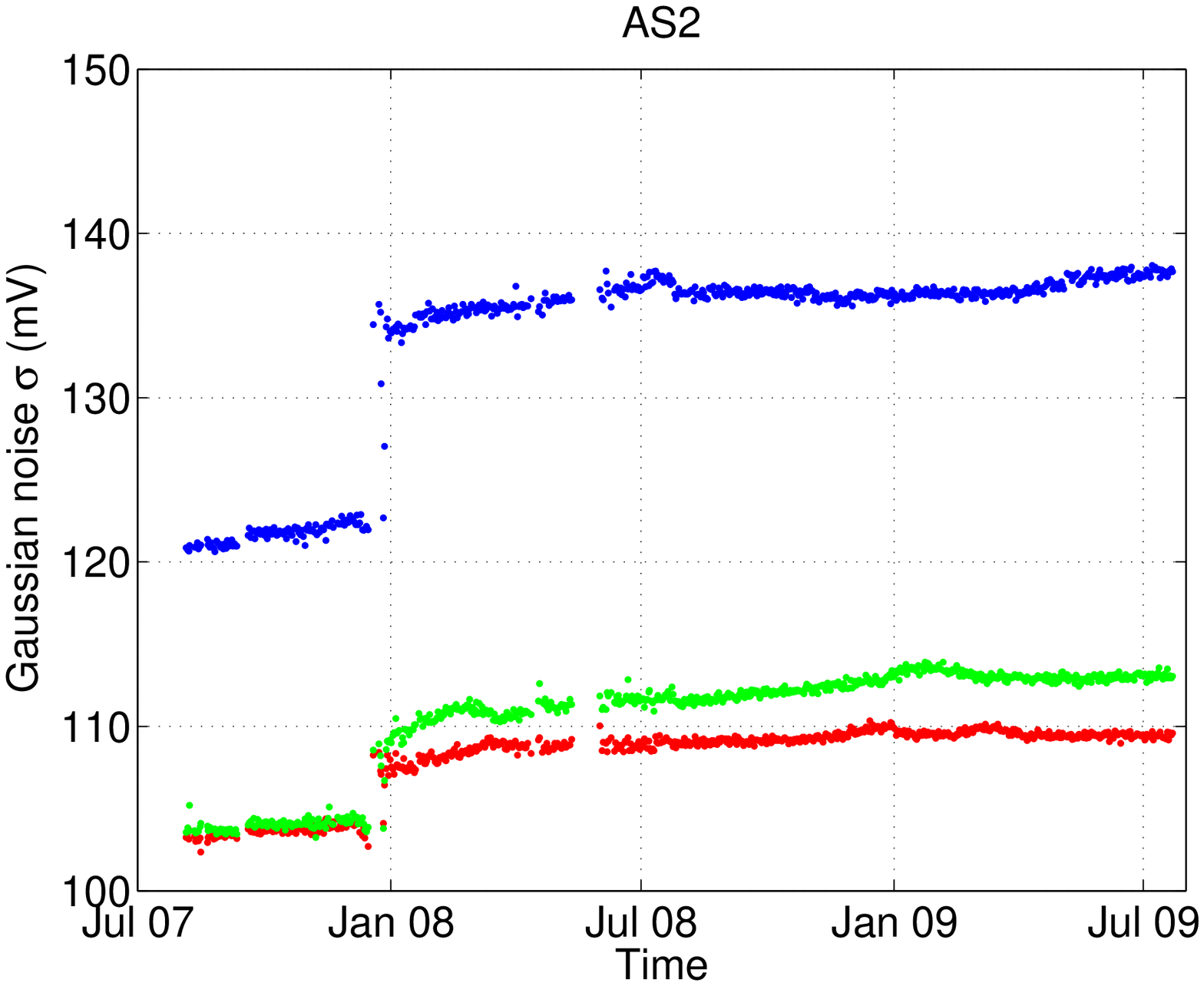}
}
\subfigure[ AS3]{
\noindent\includegraphics[width=11pc]{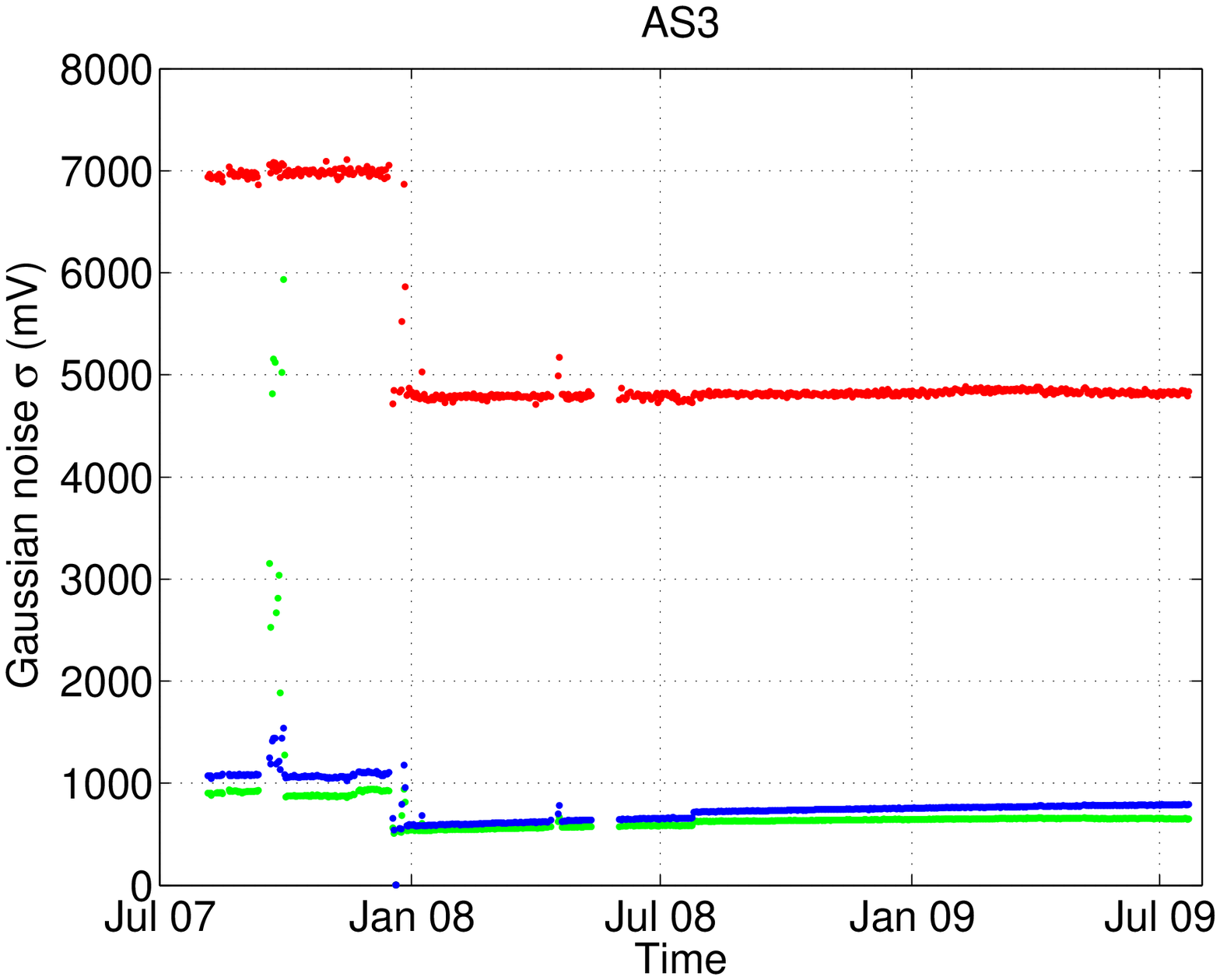}
}
\subfigure[ AS4]{
\noindent\includegraphics[width=11pc]{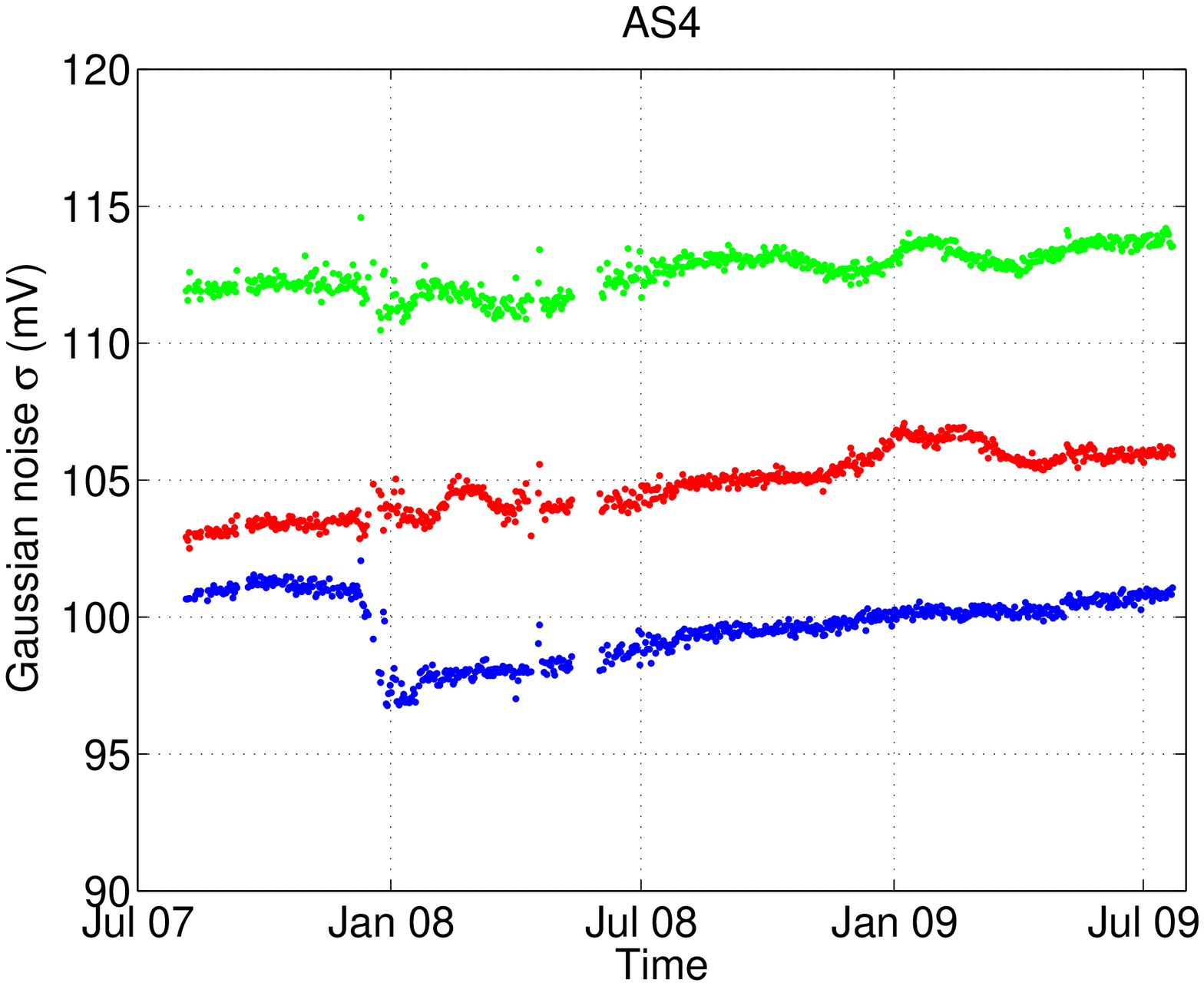}
}
\subfigure[ AS5]{
\noindent\includegraphics[width=11pc]{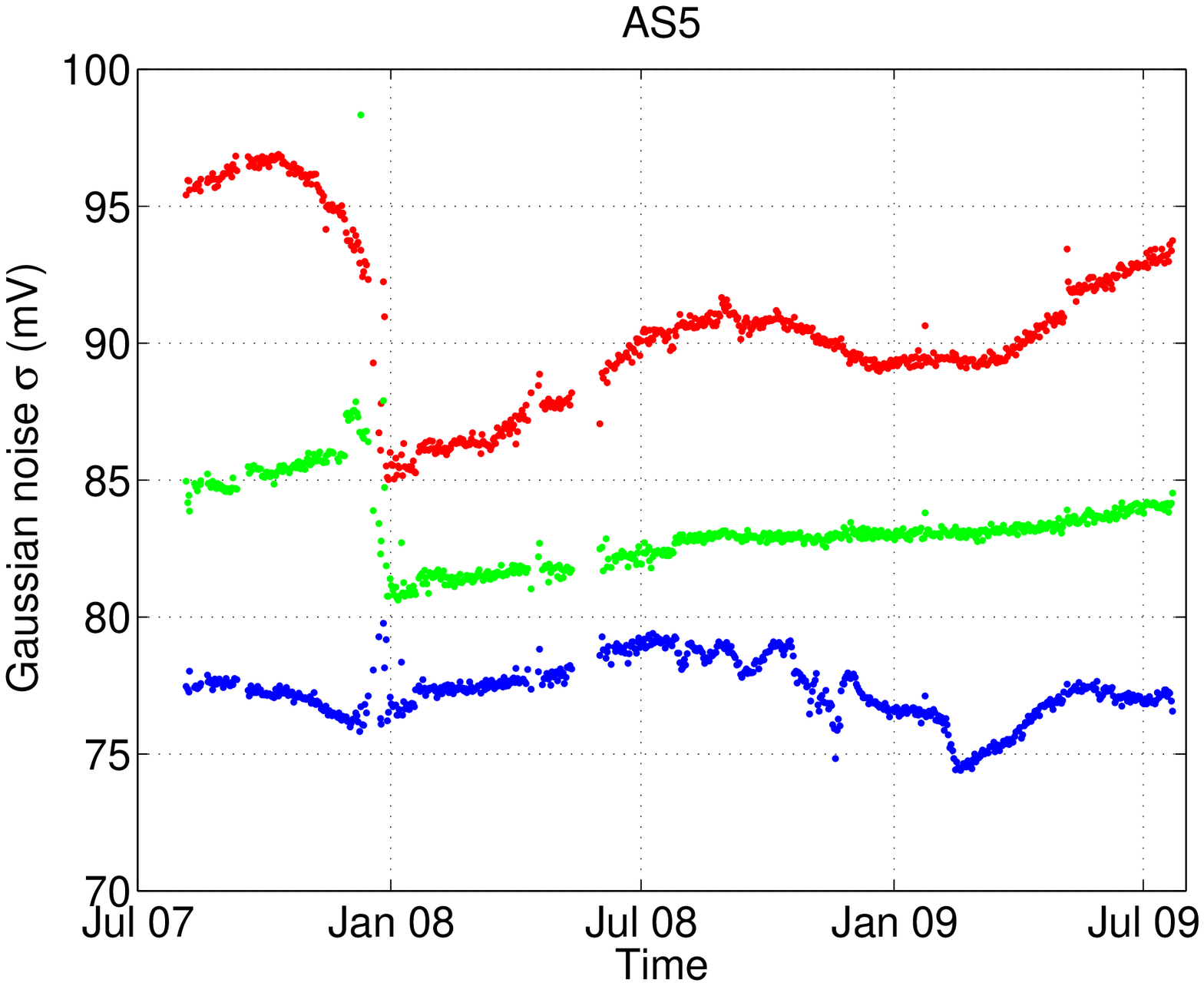}
}
\subfigure[ AS6]{
\noindent\includegraphics[width=11pc]{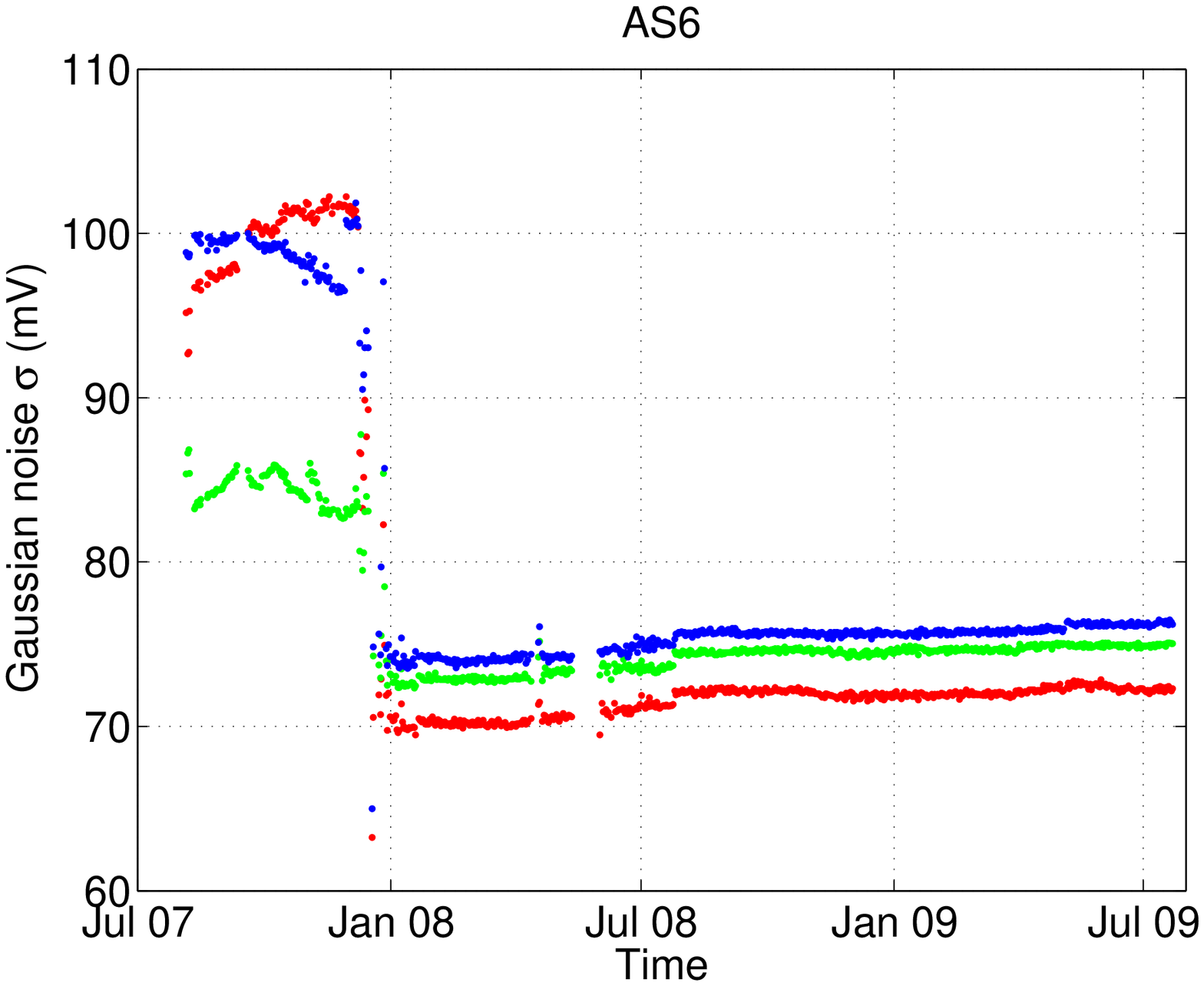}
}
\subfigure[ AS7]{
\noindent\includegraphics[width=11pc]{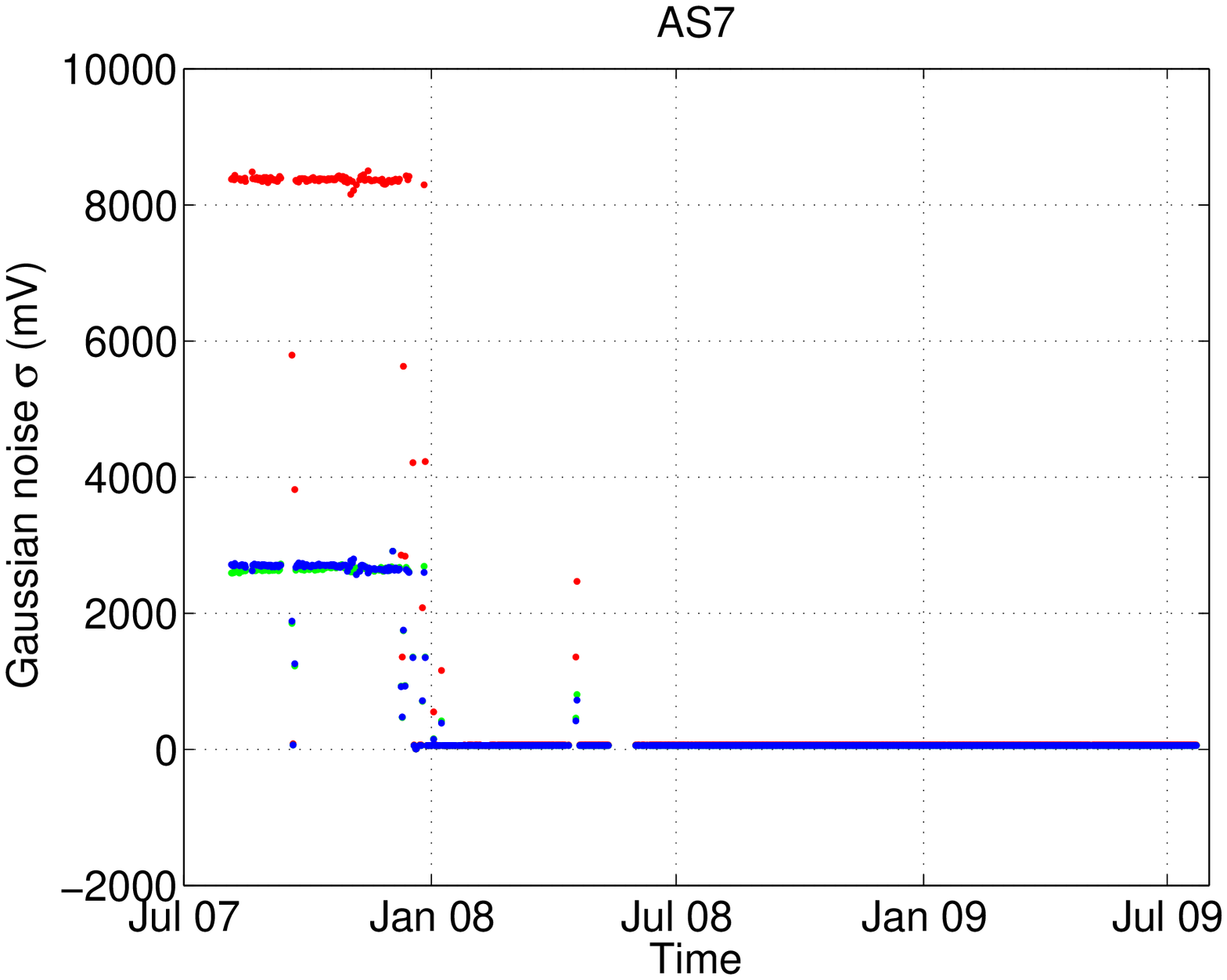}
}
\caption[Multi-year noise evolution on String A]{Long-term evolution of noise level of each channel on String A.  One plot is shown for each module, and each plot shows all three channels colored red, green, and blue for channel 0, 1, and 2 respectively.}
\label{noiseEvolutionA}
\end{center}
\end{figure}

\begin{figure}
\begin{center}
\subfigure[ BS1]{
\noindent\includegraphics[width=11pc]{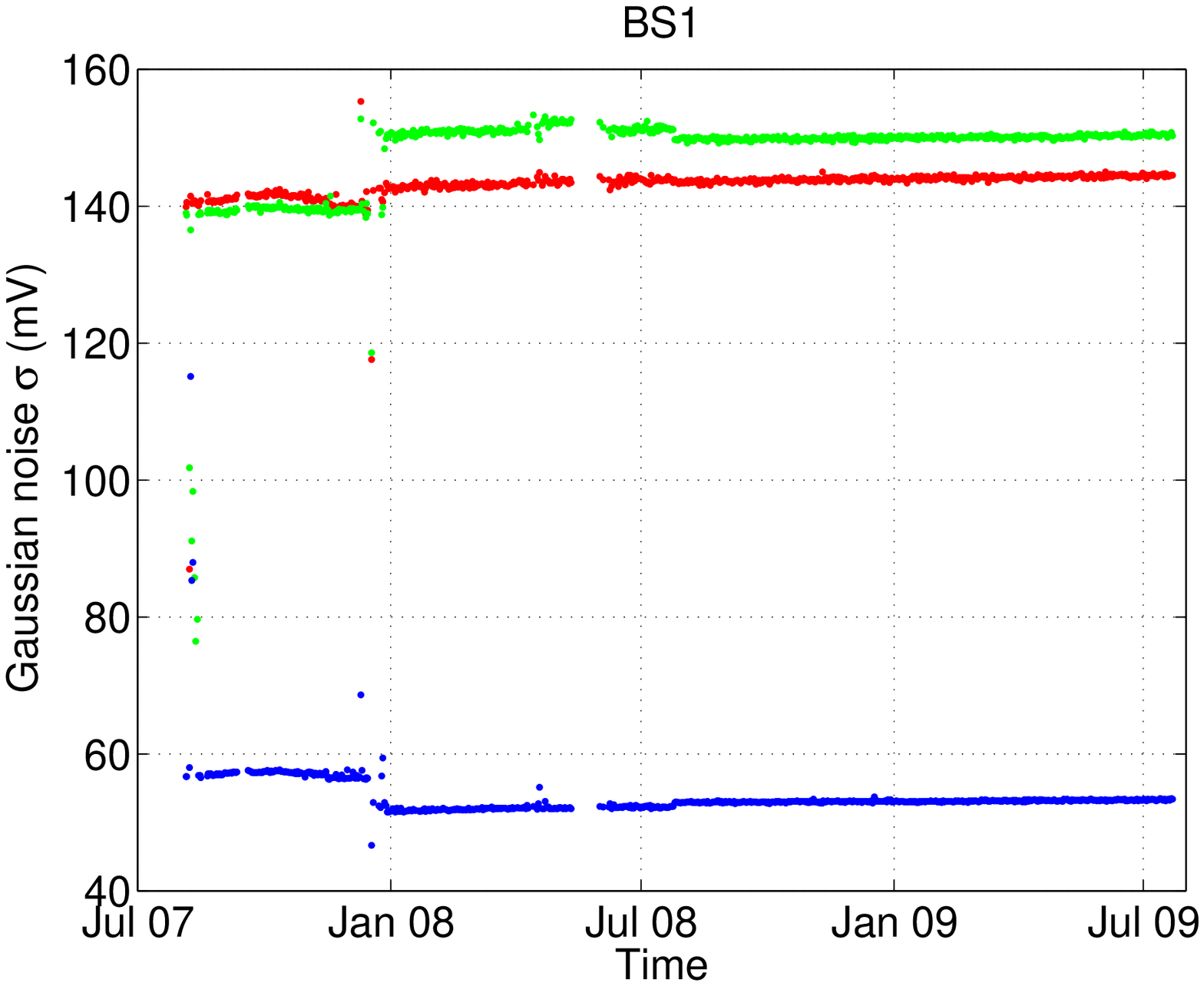}
}
\subfigure[ BS2]{
\noindent\includegraphics[width=11pc]{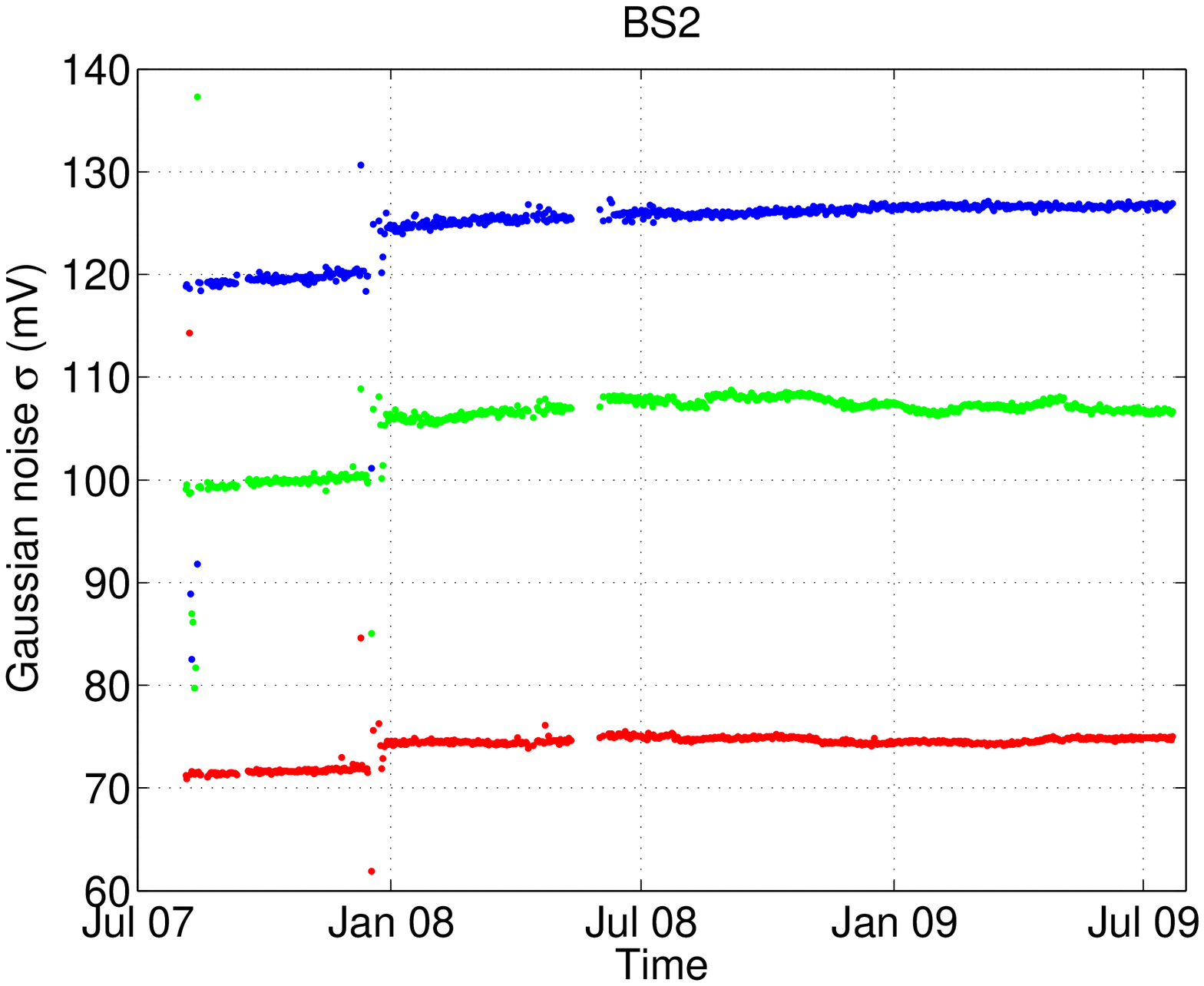}
}
\subfigure[ BS3]{
\noindent\includegraphics[width=11pc]{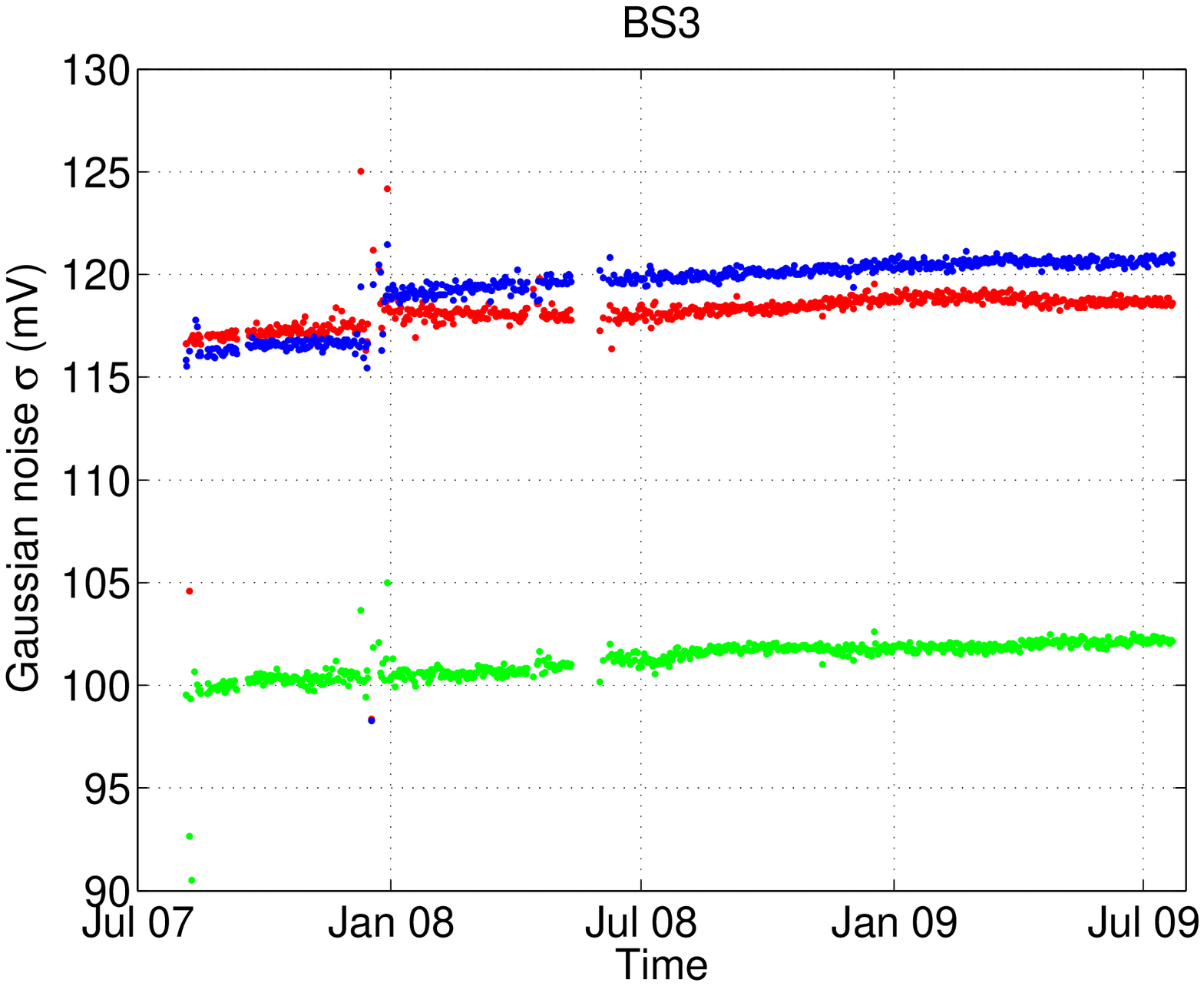}
}
\subfigure[ BS4]{
\noindent\includegraphics[width=11pc]{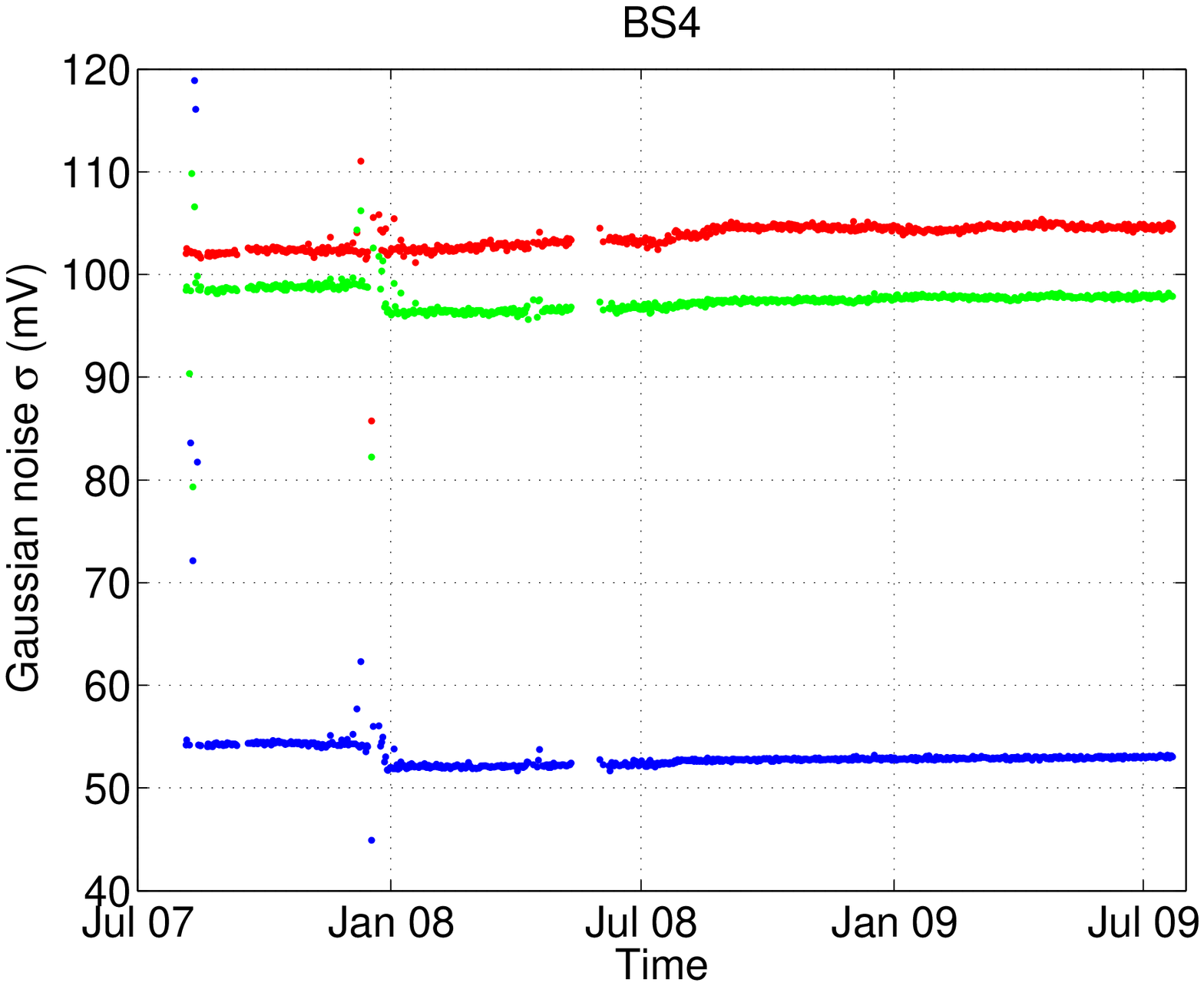}
}
\subfigure[ BS5]{
\noindent\includegraphics[width=11pc]{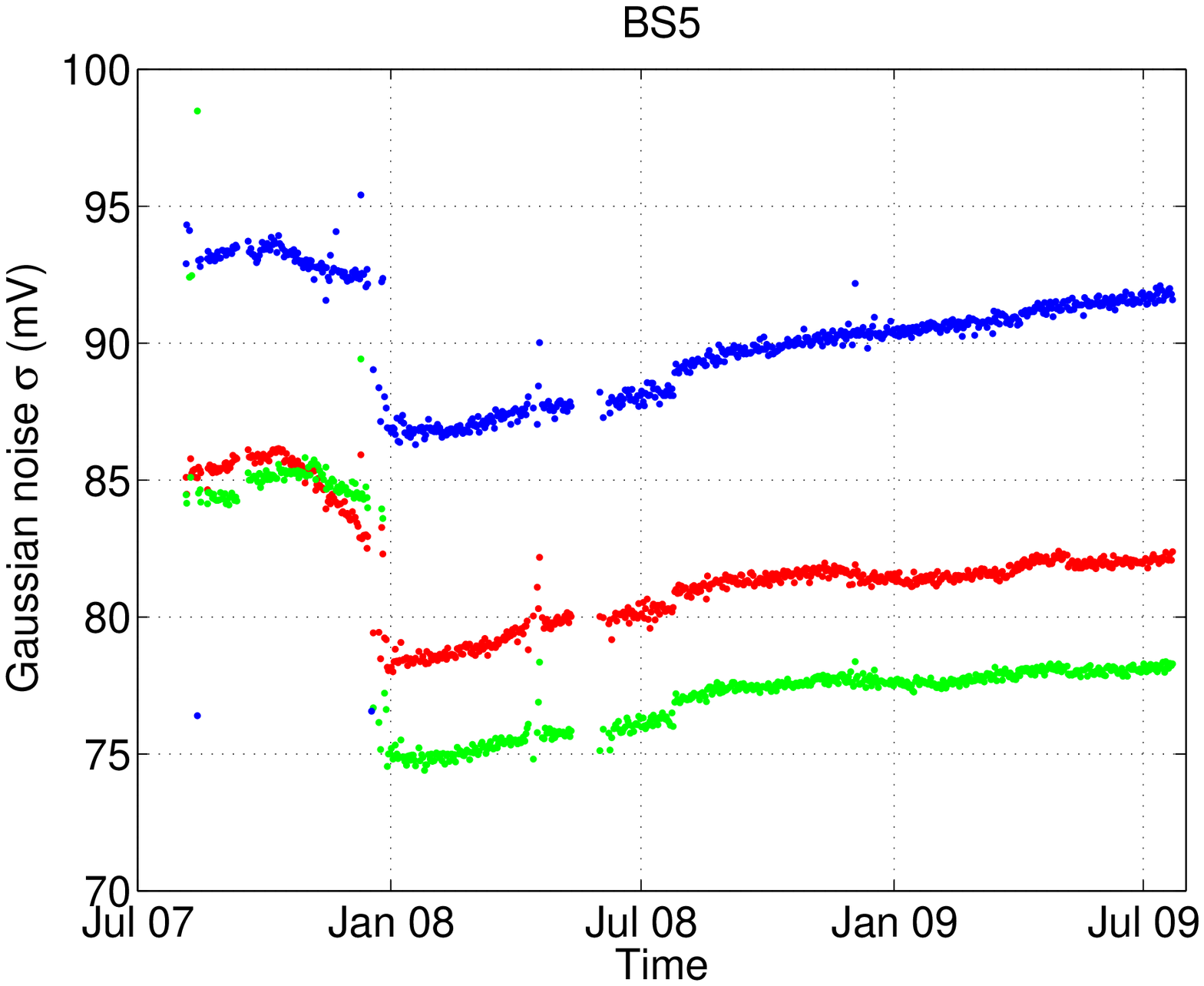}
}
\subfigure[ BS6]{
\noindent\includegraphics[width=11pc]{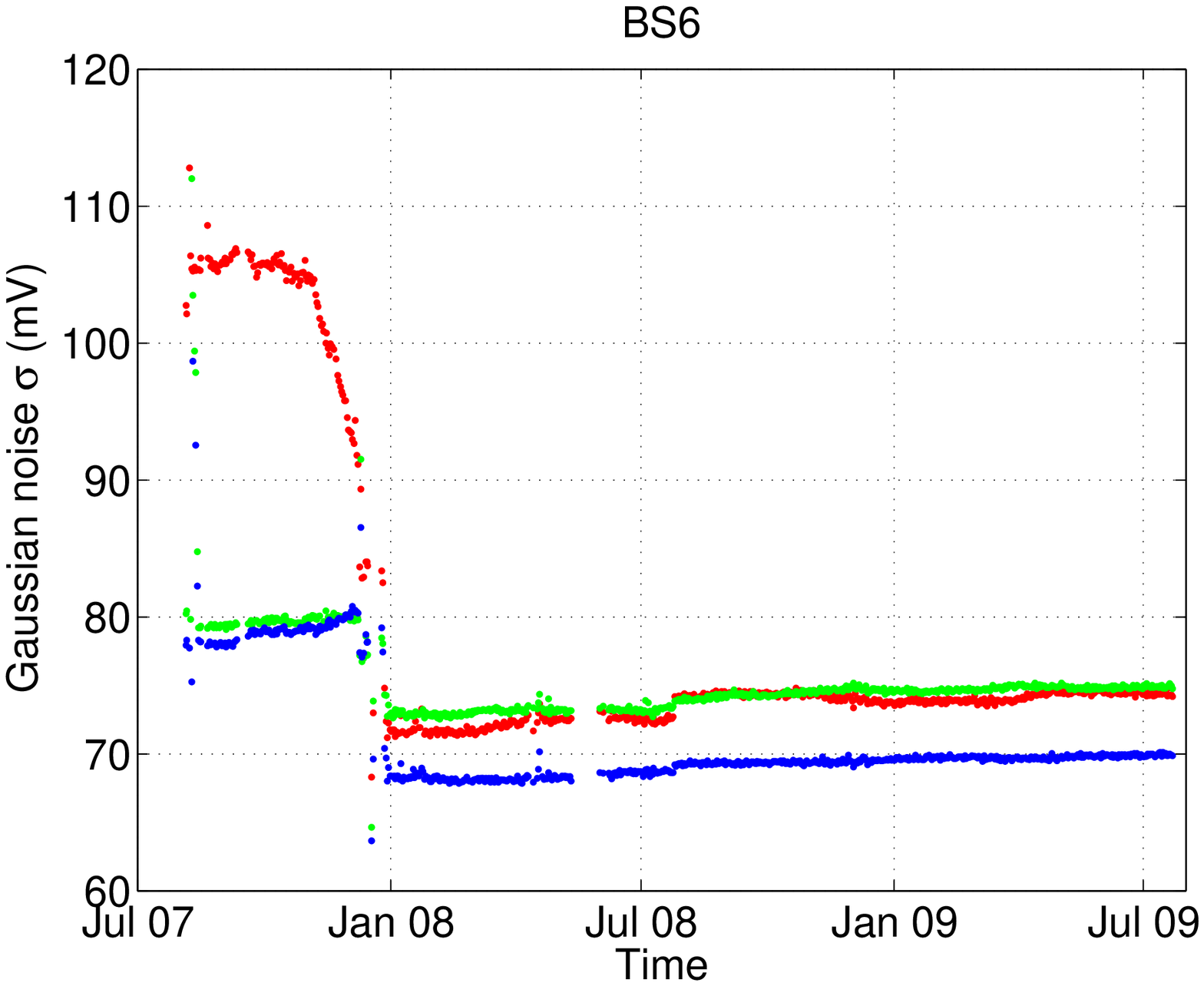}
}
\subfigure[ BS7]{
\noindent\includegraphics[width=11pc]{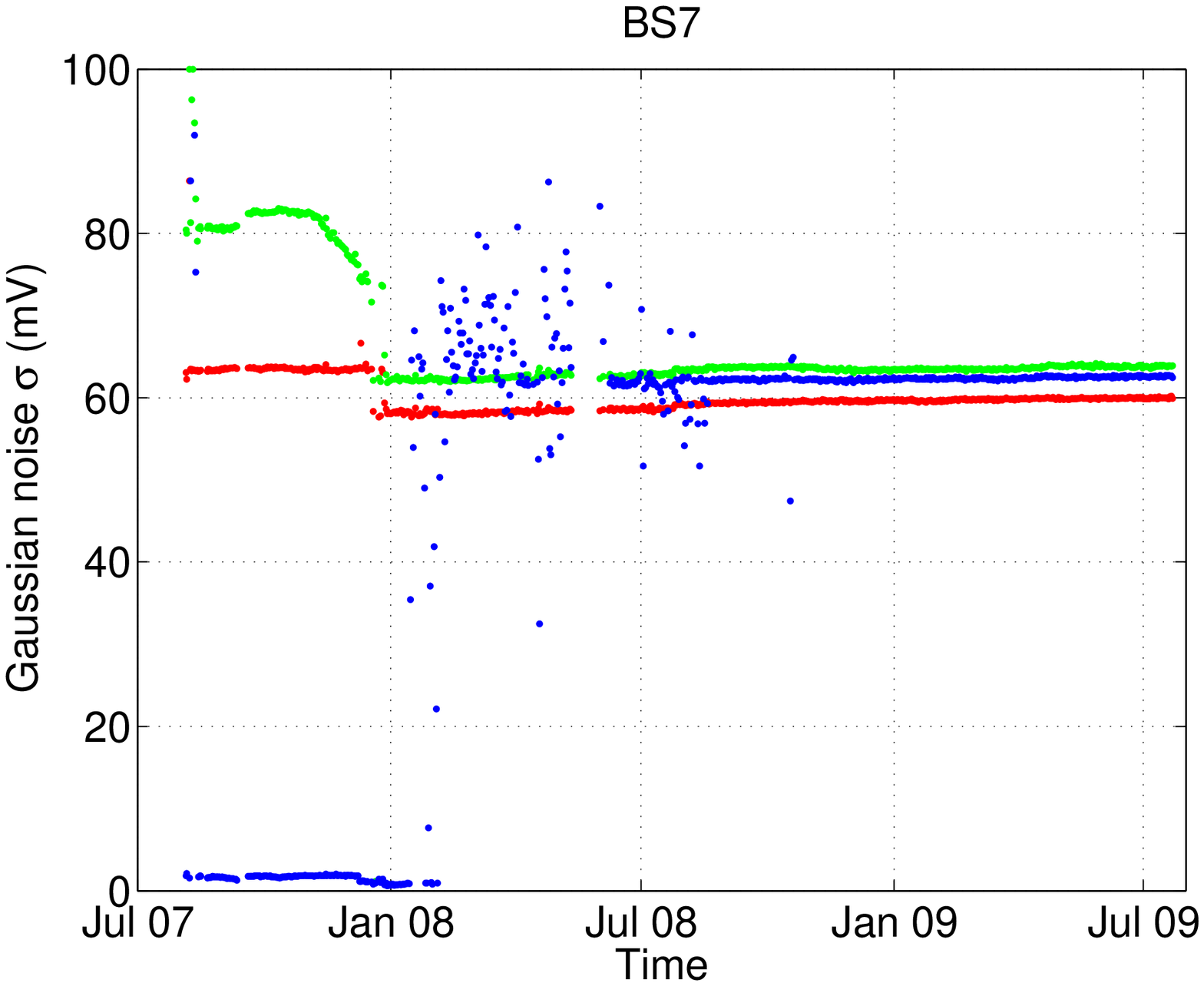}
}
\caption[Multi-year noise evolution on String B]{Long-term evolution of noise level of each channel on String B.  One plot is shown for each module, and each plot shows all three channels colored red, green, and blue for channel 0, 1, and 2 respectively.}
\label{noiseEvolutionB}
\end{center}
\end{figure}

\begin{figure}
\begin{center}
\subfigure[ CS1]{
\noindent\includegraphics[width=11pc]{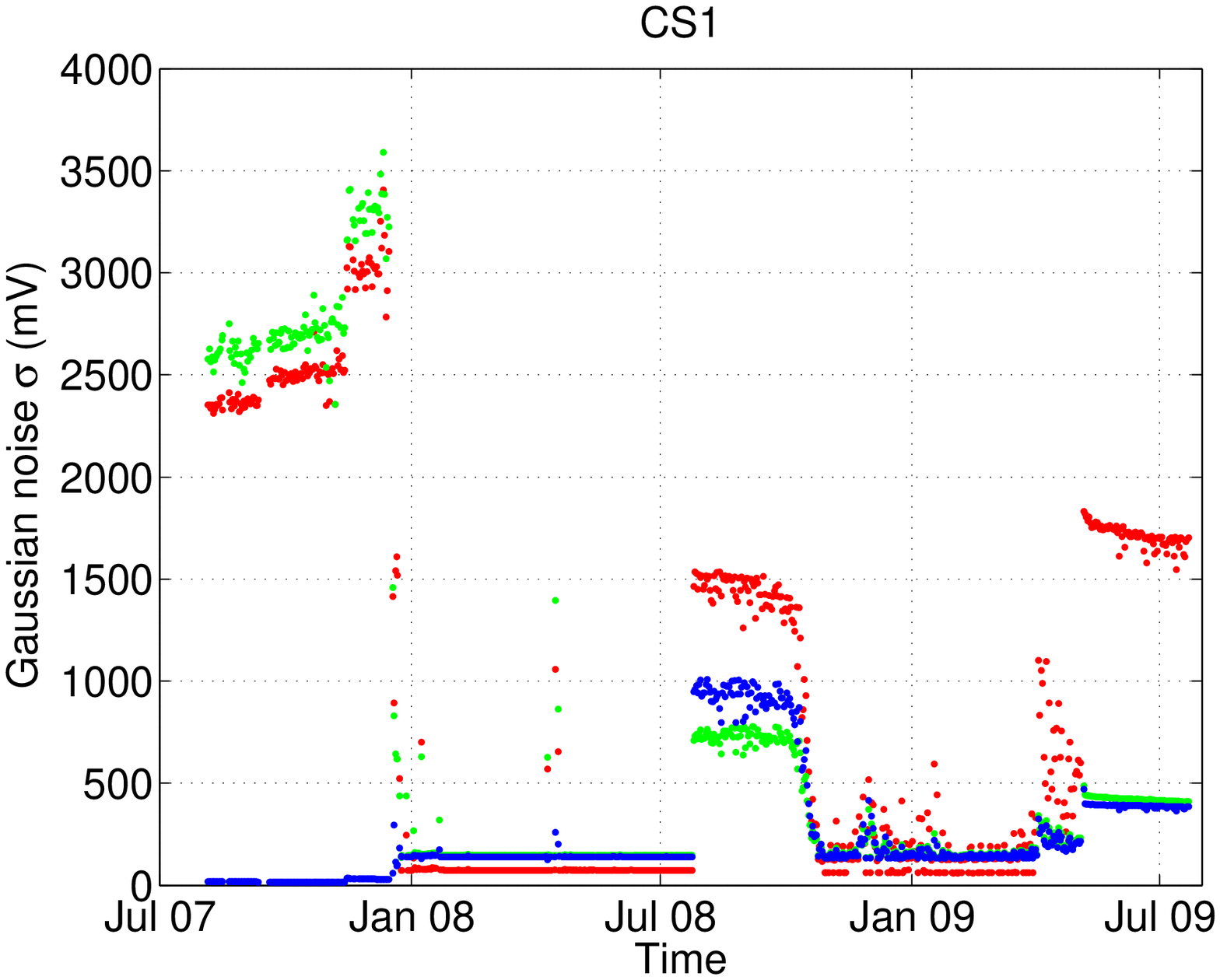}
}
\subfigure[ CS2]{
\noindent\includegraphics[width=11pc]{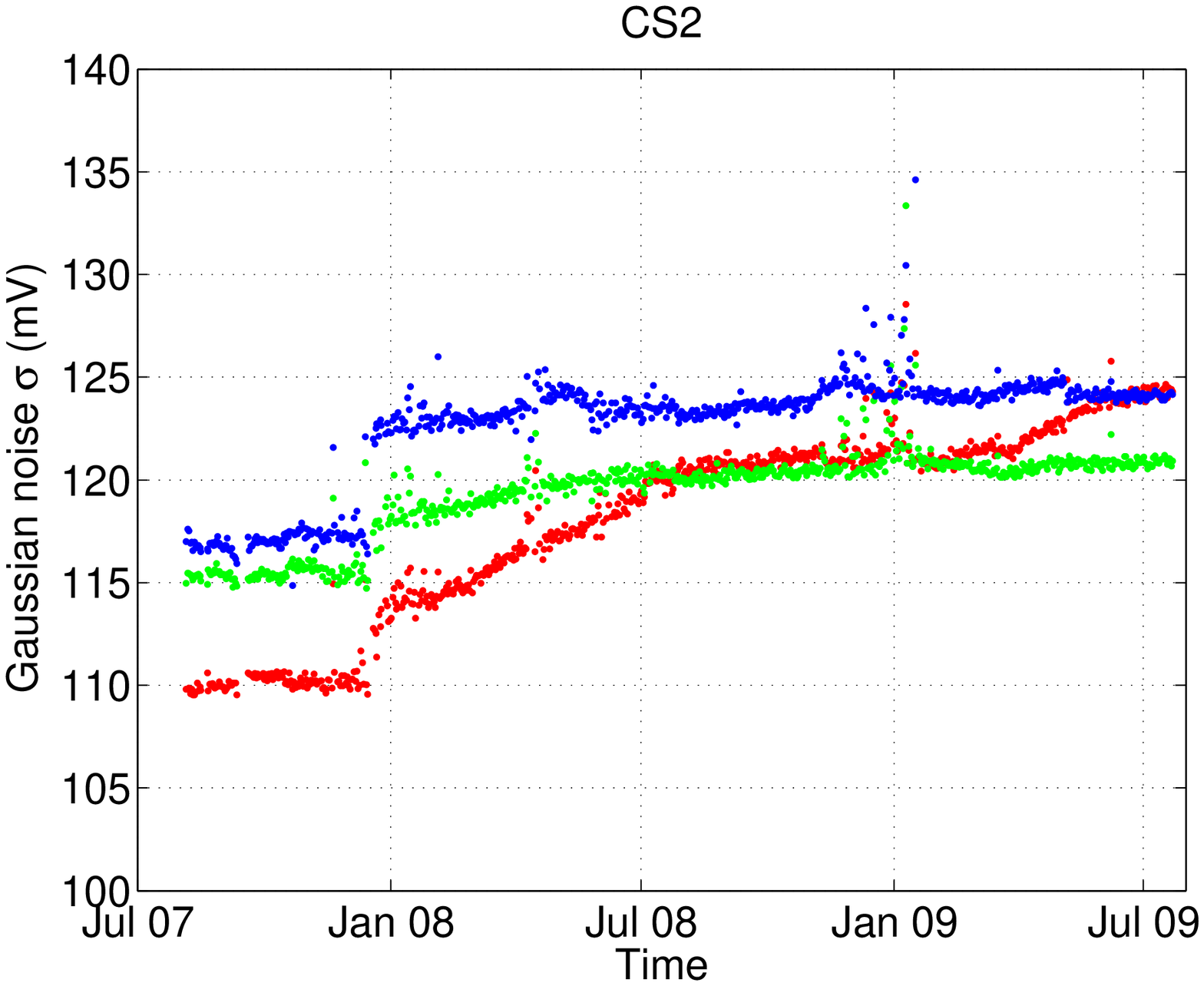}
}
\subfigure[ CS3]{
\noindent\includegraphics[width=11pc]{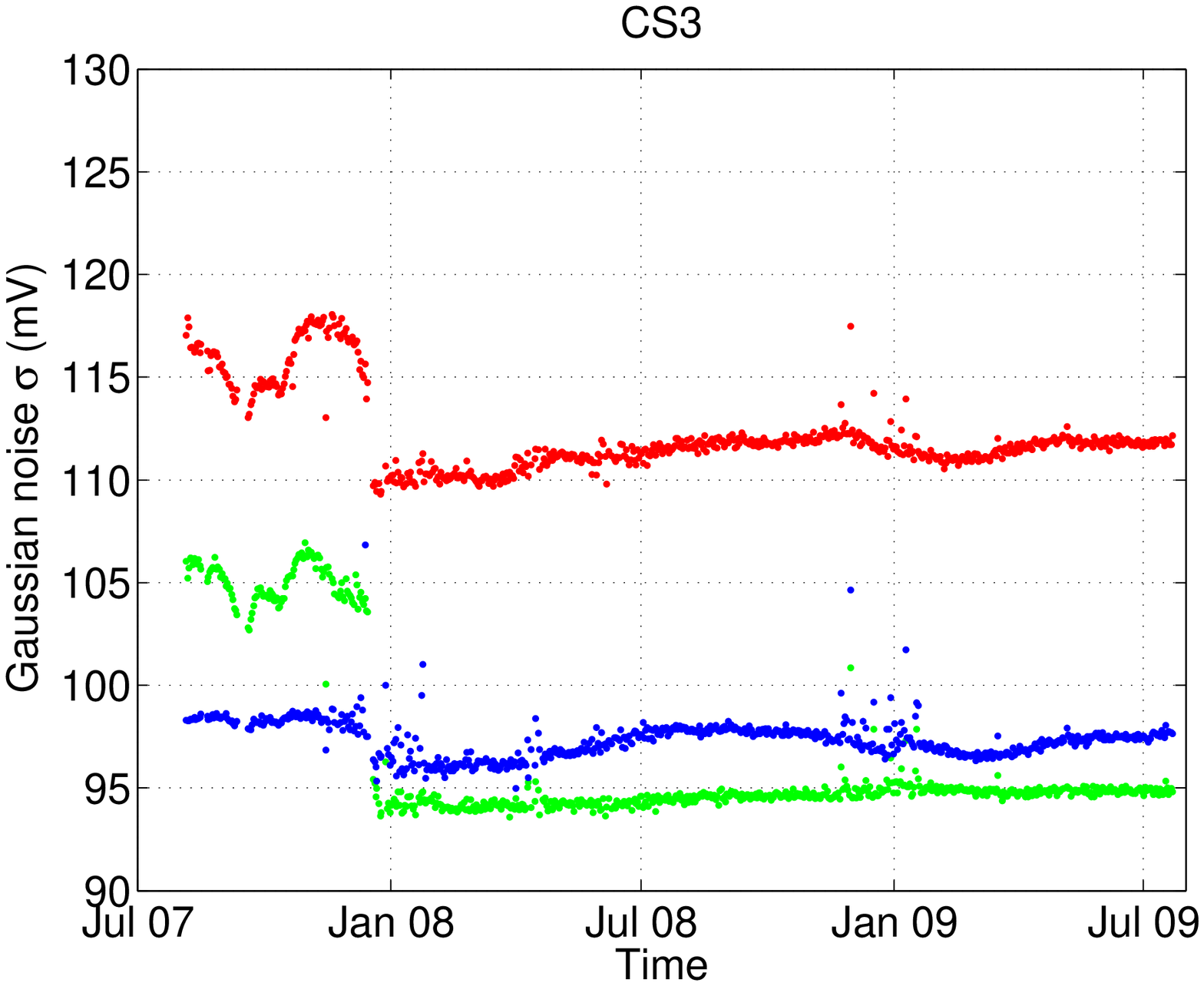}
}
\subfigure[ CS4]{
\noindent\includegraphics[width=11pc]{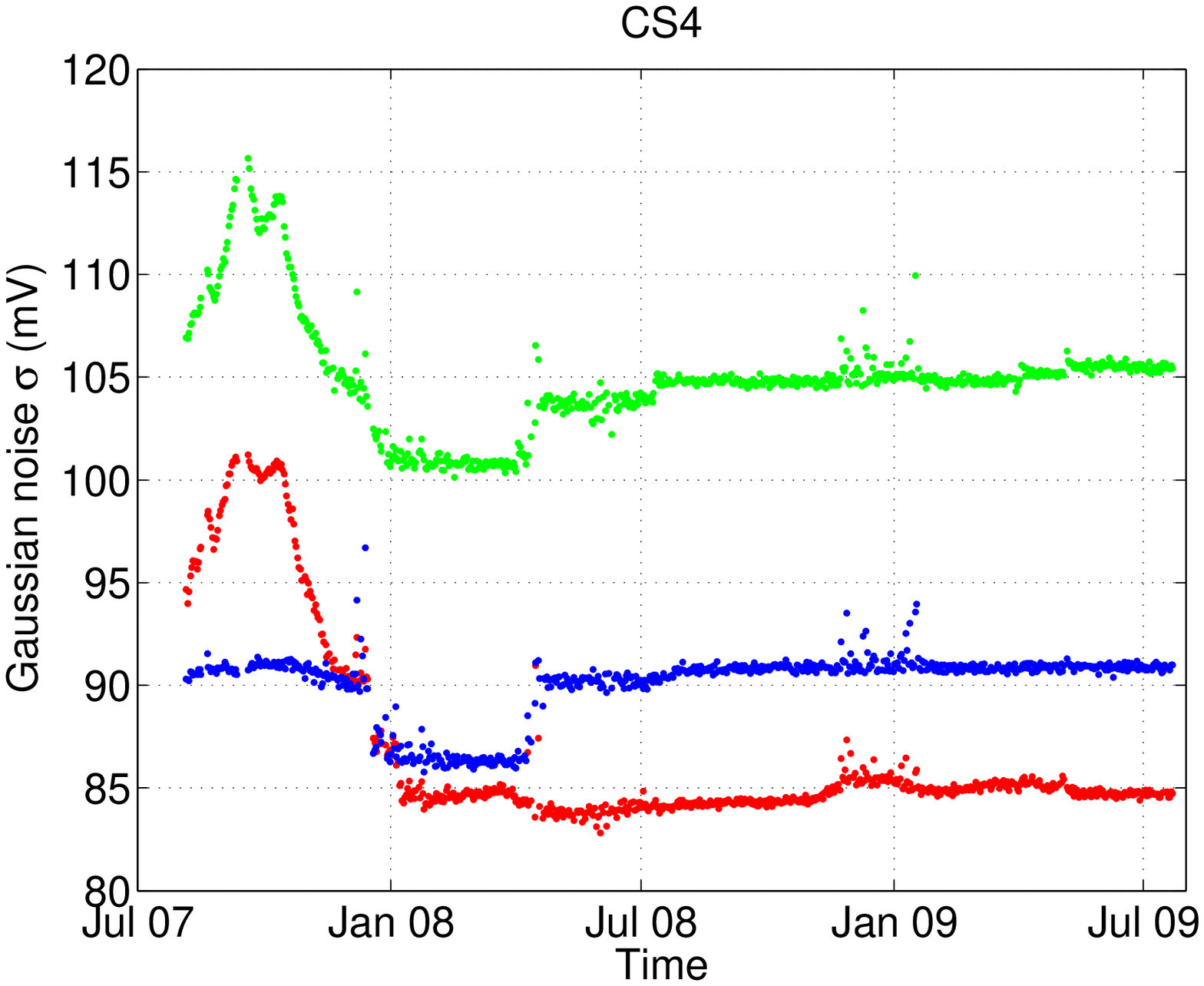}
}
\subfigure[ CS5]{
\noindent\includegraphics[width=11pc]{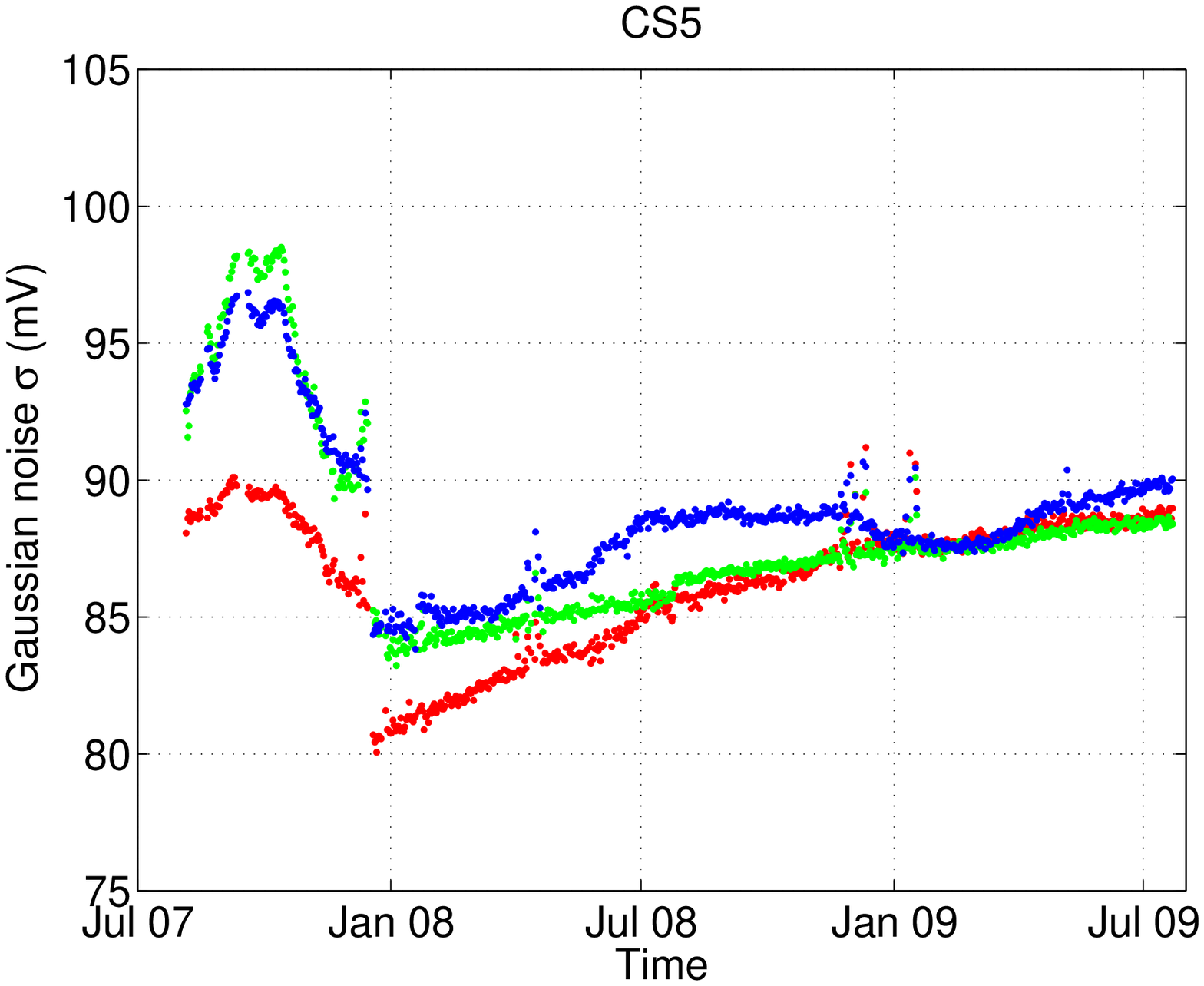}
}
\subfigure[ CS6]{
\noindent\includegraphics[width=11pc]{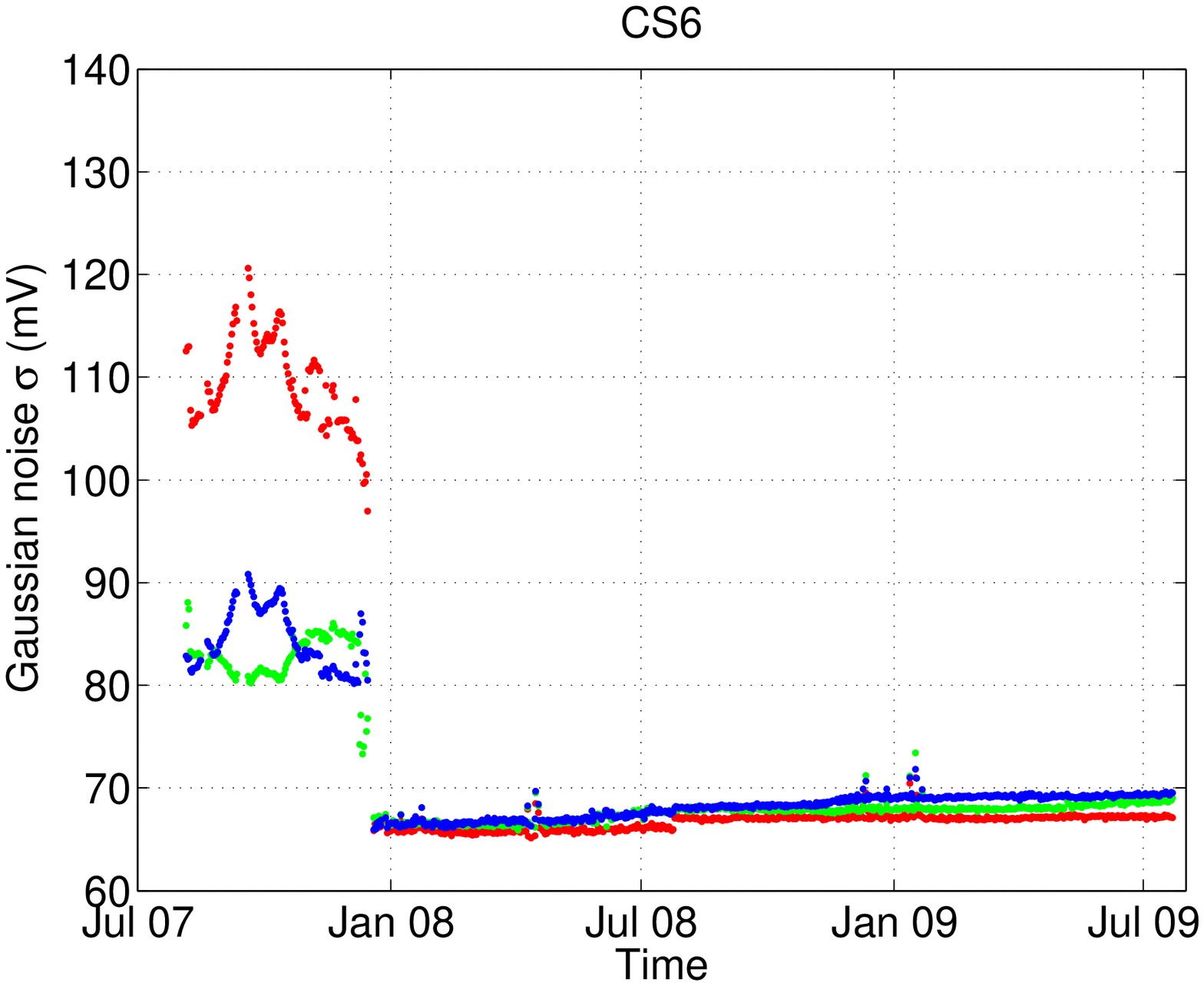}
}
\subfigure[ CS7]{
\noindent\includegraphics[width=11pc]{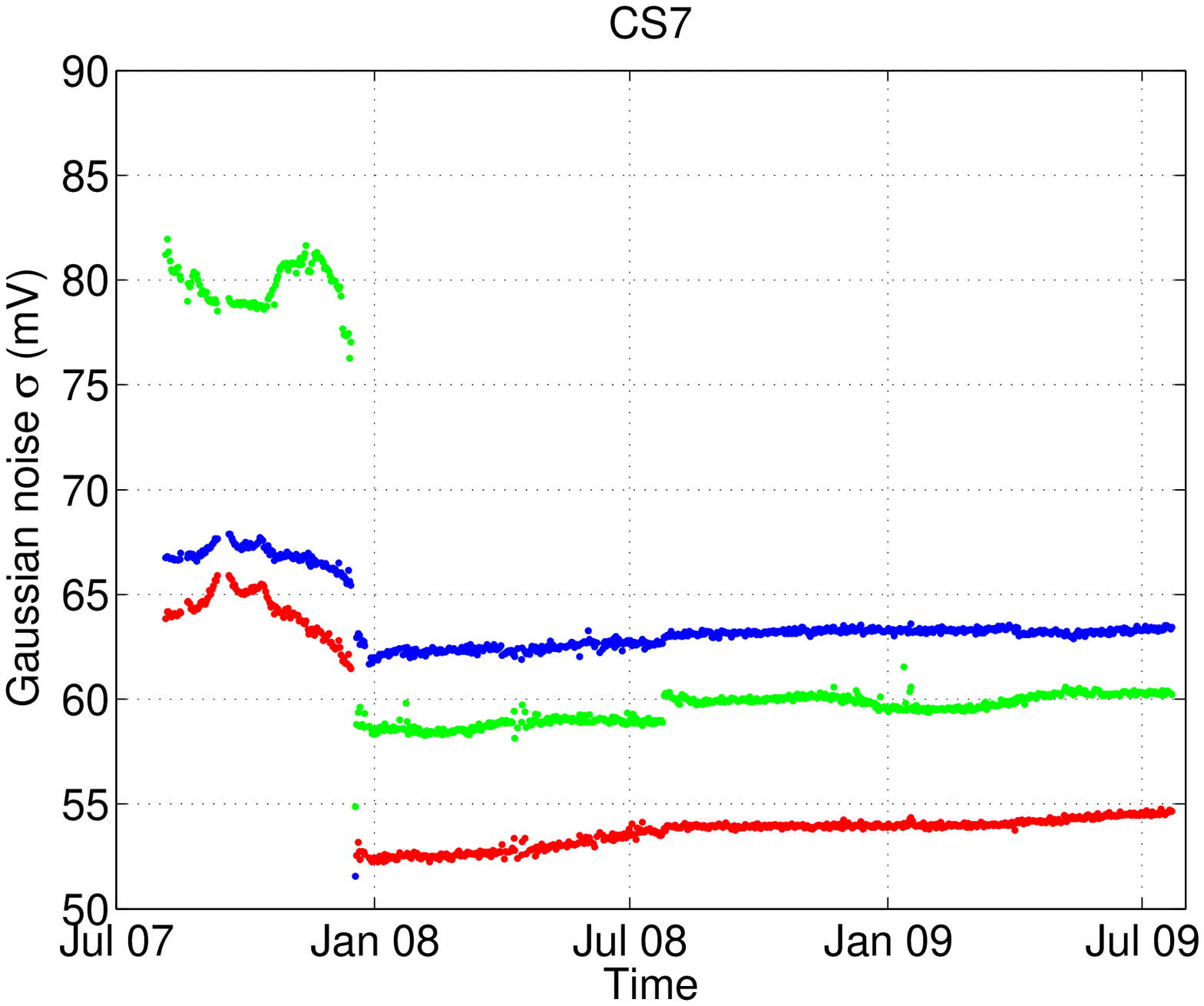}
}
\caption[Multi-year noise evolution on String C]{Long-term evolution of noise level of each channel on String C.  One plot is shown for each module, and each plot shows all three channels colored red, green, and blue for channel 0, 1, and 2 respectively.}
\label{noiseEvolutionC}
\end{center}
\end{figure}

\begin{figure}
\begin{center}
\subfigure[ DS1]{
\noindent\includegraphics[width=11pc]{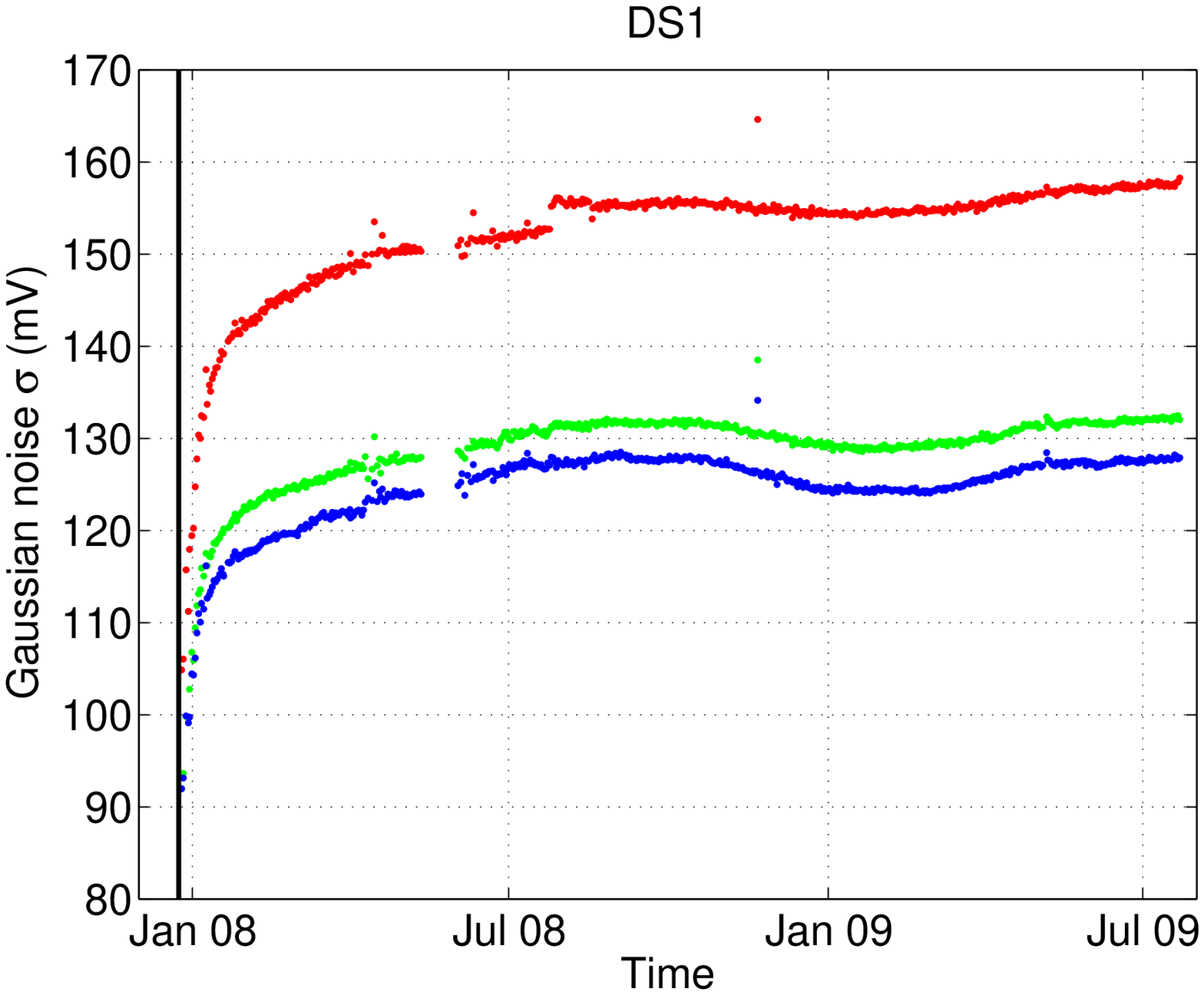}
}
\subfigure[ DS2]{
\noindent\includegraphics[width=11pc]{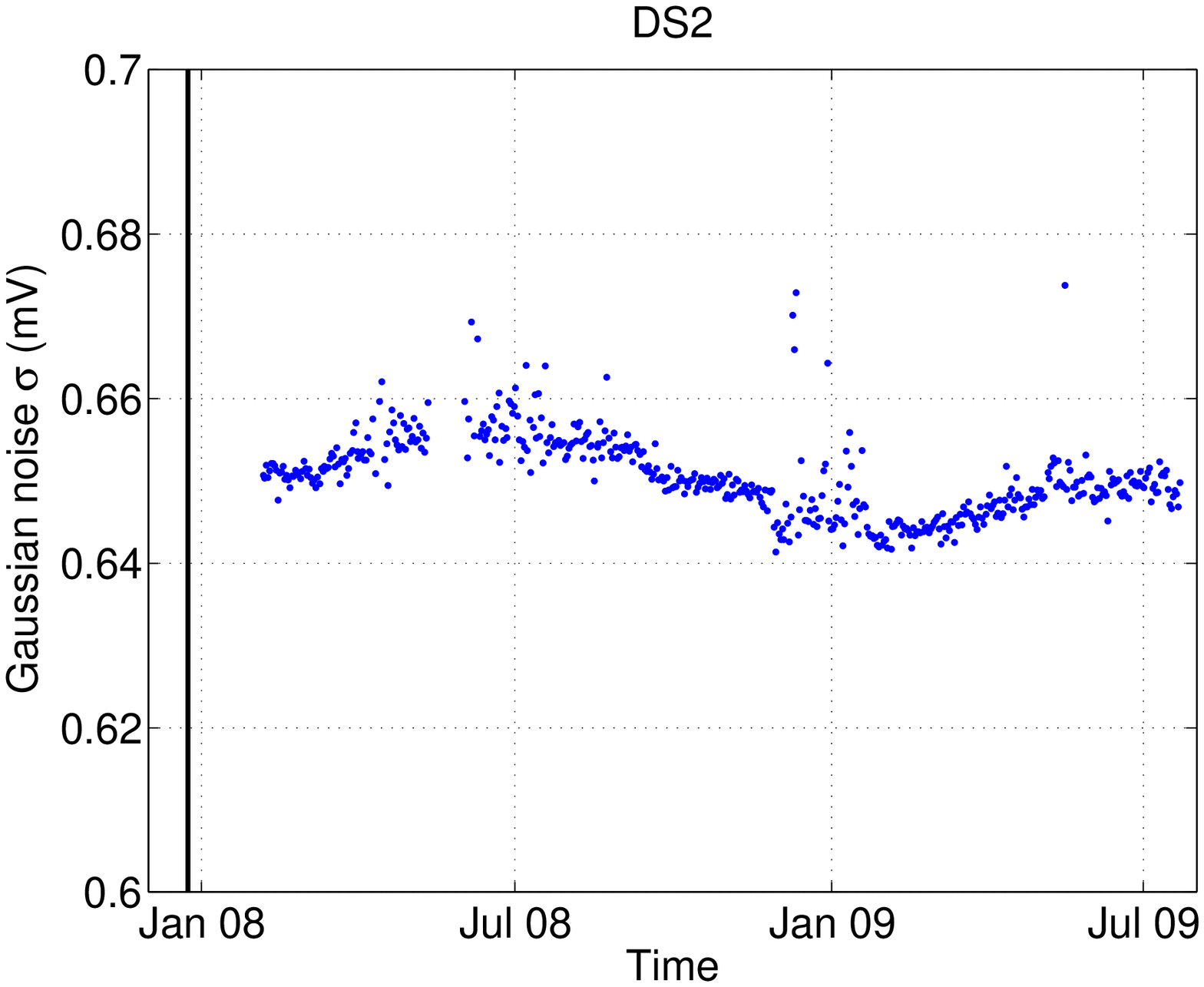}
}
\subfigure[ DS3]{
\noindent\includegraphics[width=11pc]{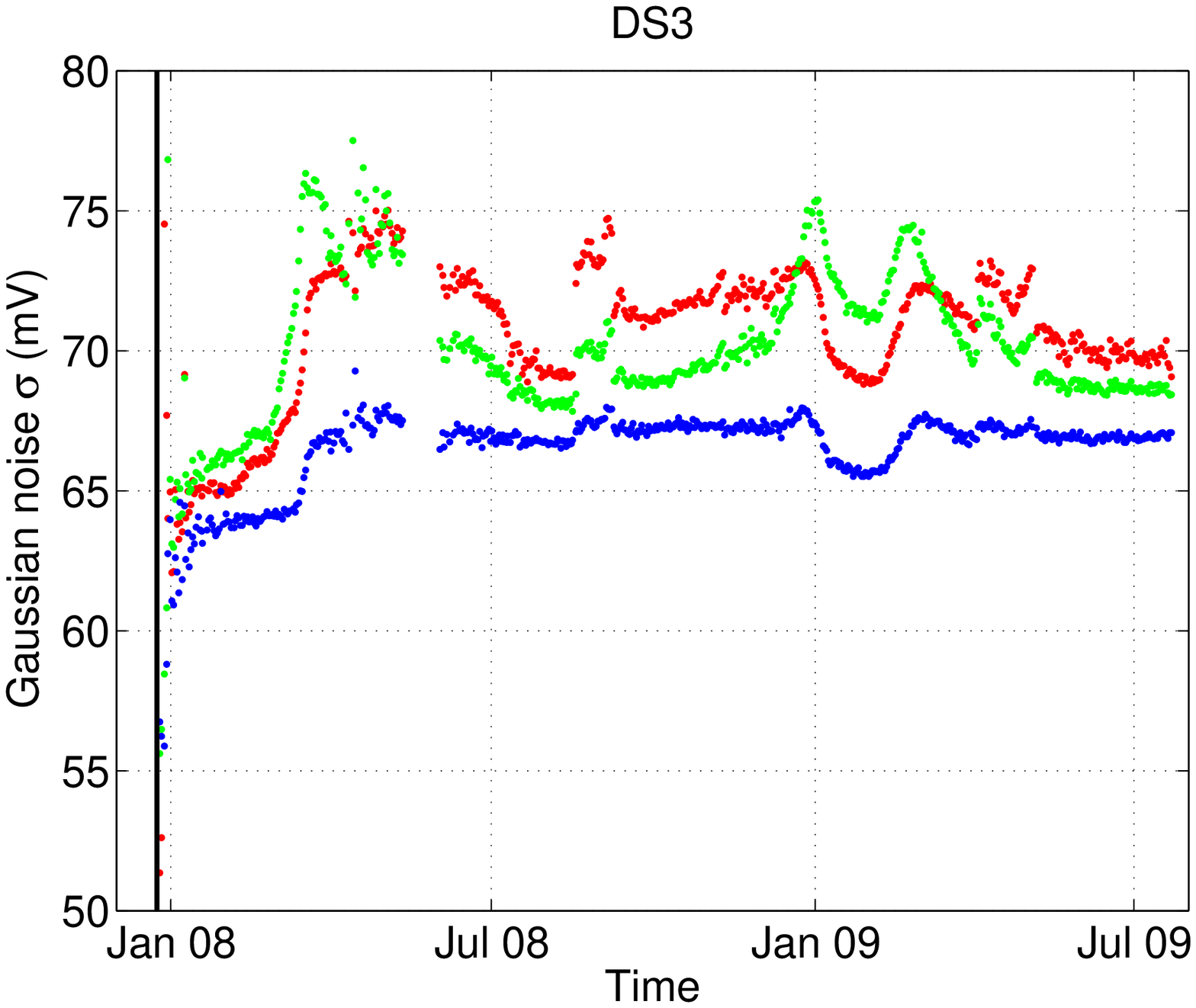}
}
\subfigure[ DS4]{
\noindent\includegraphics[width=11pc]{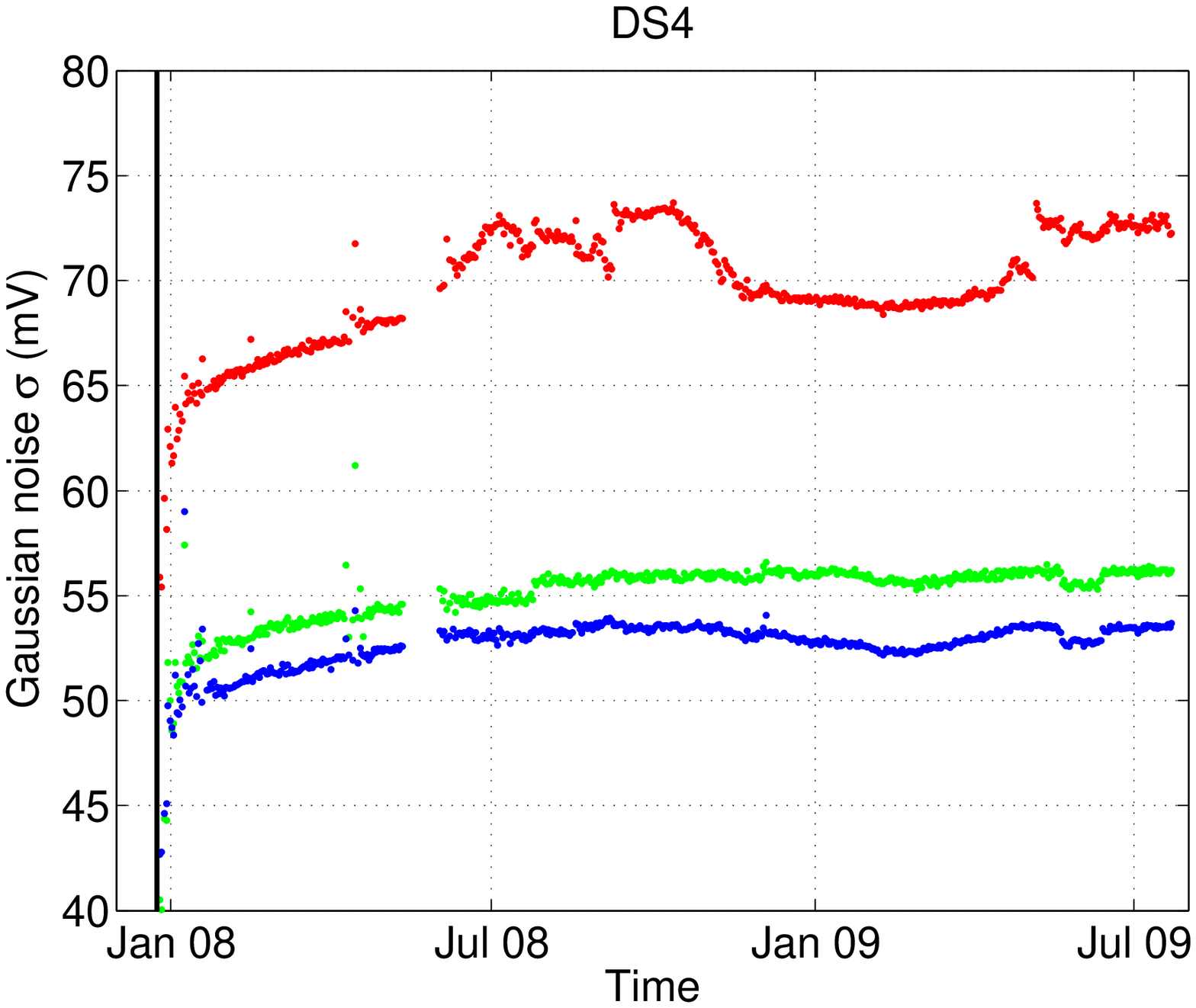}
}
\subfigure[ DS5]{
\noindent\includegraphics[width=11pc]{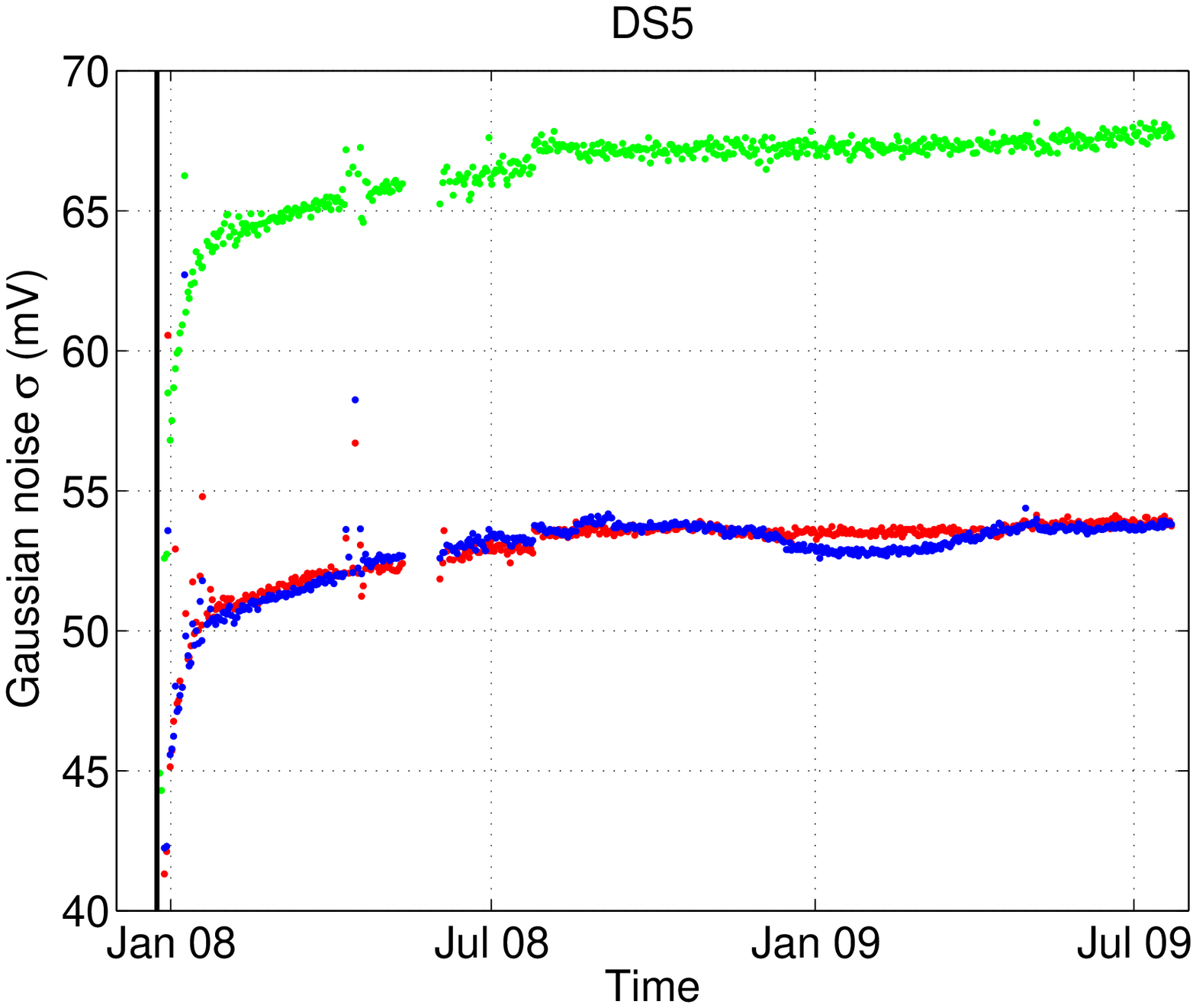}
}
\subfigure[ DS6]{
\noindent\includegraphics[width=11pc]{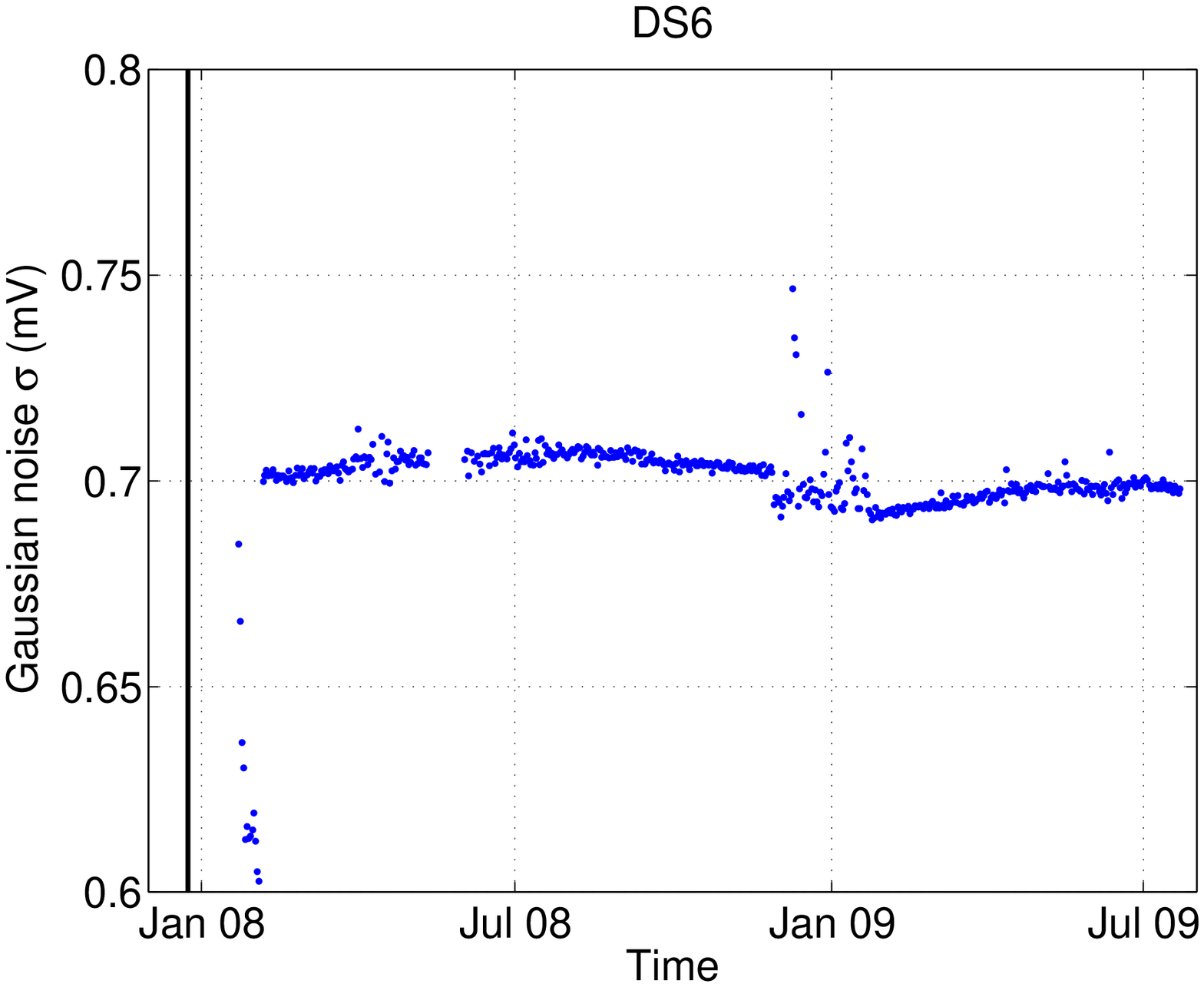}
}
\subfigure[ DS7]{
\noindent\includegraphics[width=11pc]{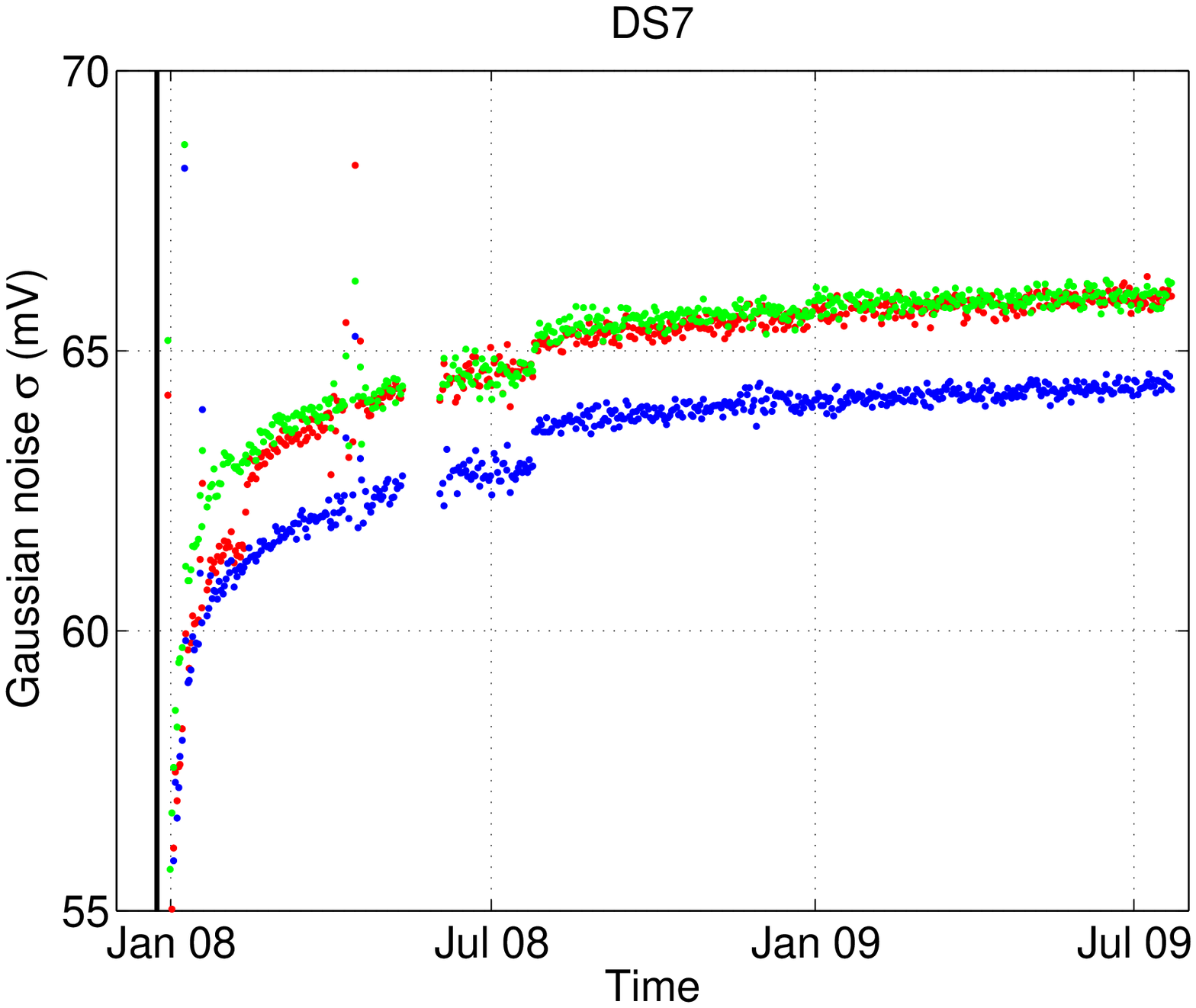}
}
\caption[Multi-year noise evolution on String D]{Long-term evolution of noise level of each channel on String D.  One plot is shown for each module, and each plot shows all three channels colored red, green, and blue for channel 0, 1, and 2 respectively (except for HADES modules which only have channel 2 connected).  String D deployment is indicated with a solid vertical line.}
\label{noiseEvolutionD}
\end{center}
\end{figure}

Figures~\ref{noiseEvolutionA}-\ref{noiseEvolutionD} show the long-term evolution of the noise level on all SPATS strings, from deployment to the present.  These plots were made using the Gaussian fits for each hourly run.  Instead of including every run individually, the runs of each day were averaged together to produce a single data point for each day.  The noise level is generally very stable.  Exceptions to this include:

\begin{enumerate}

\item The noise levels slowly increase during the freezing period, asymptotically approaching the stable frozen-in noise level.  The time to achieve equilibrium apparently increases with depth, reaching 6-12 months on the deepest modules.  See for example the deepest String D module, where the equilibration time was $\sim$12 months.
\item In String D, which has sensors with the best signal-to-noise ratio, there is a hint of annual (seasonal) modulation in the noise level.  Particularly in the channels with the cleanest noise-vs-time behavior, there is apparently a minimum during the South Pole summer and a maximum during the South Pole winter.  It is not yet clear how significant this modulation is, or if it is a real effect in the ambient acoustic noise or an instrumental artifact such as variation in the surface ADC electronics.  It will be interesting to watch this trend over another annual cycle to see if it persists.
\item Noise from drilling of individual IceCube holes is heard during the December-January annual IceCube construction season.
\item On Strings A, B, and C, the noise behavior was more stable after January 2008 than before.  In January 2008 our sensor operation protocol was changed: prior to this time we powered off the sensor modules between runs; from this time to the present we leave the sensor modules powered on at all times.
\item There are discontinuities when the strings are powered off, or data taking is stopped, for short periods (happens only a few times per year).  This is not surprising because we have seen that sensors require some equilibration time to reach a stable noise level after being powered down, probably due to electronics self-heating equilibrating.  What is surprising is that sometimes the new steady-state noise level after power cycling is different from that prior to power cycling.
\item Dead/bad channels generally fluctuate, sometimes dramatically, and often have very high output voltages (see discussion of dead/bad channels in Section~\ref{noiseStatus}).

\end{enumerate}







\chapter{Transient events}

\label{transientsChapter}

\noindent\emph{In this chapter we present the acquisition and analysis of transient events with SPATS.  Over a year of high quality transients data have been acquired.  We have detected both pressure and shear waves from individual transient events.  Our offline coincidence determination and event reconstruction algorithms are described, and results of applying them to the data are presented.  We clearly hear acoustic emission from recently drilled IceCube holes as well as from most of the large, shallow water wells drilled for use by both the IceCube drill and, remarkably, by the AMANDA drill.  The AMANDA well was last heated nine years ago.  We estimate the improvement that could be achieved by upgrading from offline to online coincidence and conclude that it is small.}

\section{Interlude: shear waves in SPATS}

\label{shearSection}

An interesting effect of acoustic studies in ice compared to water (where most previous acoustic neutrino detection studies have been performed) is that ice can support shear waves in addition to pressure waves, while water can only support pressure waves.  It is unclear whether particle showers produce any shear waves via the thermoacoustic or other mechanisms, and laboratory work is underway to determine this.

A surprising discovery that we made in the SPATS data is that both our frozen-in transmitters and our retrievable pinger (operated in water-filled IceCube holes) produce shear waves detected in SPATS sensors.  The shear wave characteristics are described in detail in~\cite{Vandenbroucke08}.  It is not very surprising that the frozen-in piezoelectric emitters produce a component of shear in addition to pressure waves, via motion of the emitter relative to the ice.  What is surprising is the detection of shear waves from the pinger operating in water.

Shear waves cannot propagate through fluids, so they must be produced by mode conversion from the pressure waves.  This could occur during propagation through the bulk ice, at grain boundaries, with small shear waves being produced throughout propagation of the primary pressure pulse.  However these shear waves would be incoherent, occur at various times relative to the primary wave, and would likely be small amplitude and possibly buried in the noise.

Because single large-amplitude shear waves were detected in the pinger data, with arrival times scaling with distance in a way that is consistent with emission near the same source as the pressure waves, the likely explanation for shear wave production in the pinger data is mode conversion at the water-ice interface (the hole wall).  Such mode conversion is favored when a pressure wave impinges on an interface at glancing incidence, and is suppressed under normal incidence.  The acoustic analog of the Fresnel equations, giving the transmitted and reflected amplitudes for both pressure and shear waves as a function of incidence angle and medium properties on both sides of the interface, are the Zoeppritz equations.  They are described in~\cite{Aki02}.

Given the geometry of the pinger, the hole wall, and the sensor, there are two ways of achieving glancing incidence and favoring shear wave production.  In the horizontal plane (if the sensor and pinger are at the same depth), this can be achieved if the pinger is not centered in the hole, and/or if the hole is not exactly cylindrical.  In the vertical plane, this can be achieved if the sensor is at a different depth than the pinger.  By this mechanism shear waves can be produced even if the hole is cylindrical and the pinger is perfectly centered.

In the 2007-2008 season, the pinger had no centralizer and is believed to have bounced/swung/twisted in the hole.  This hypothesis is supported by the fact that the shear waves were significantly suppressed in the 2008-2009 pinger data, for which the pinger was centralized in the hole with brass spring ribs.  Furthermore, in the 2007-2008 data there was significant pulse-to-pulse variation among the nine pulses required in each nine-second run.  The pressure and shear wave amplitudes both varied from pulse to pulse.  Moreover, the amplitudes were anti-correlated, indicating that the total amount of energy was roughly conserved but the partitioning between transmitted pressure energy and transmitted shear energy varied from pulse to pulse as the pinger swung in the hole and the angle of incidence changed.  This picture is consistent with that expected from the Zoeppritz equations.

Shear waves have now been detected from in-ice transmitters, from the retrievable pinger, and from ambient transients.  While we have heard pressure waves at up to $\sim$1~km distance, we have not yet heard shear waves beyond 200~m distance.  This could be due to lower intrinsic amplitude in the shear waves, or it could be due to greater attenuation of the shear waves during propagation.  It would be interesting to measure the shear wave attenuation in addition to the pressure wave attenuation that we have measured.  Unfortunately we have been unable to do this so far because none of our shear wave data sets have a sufficiently long baseline over which to distinguish attenuation from $1/r$ divergence.

\section{Transients data taking}

Over the summer of 2008, a series of DAQ upgrades enabled SPATS to operate as a 4-string, 12-channel, high-duty transients detector.  Every hour, a 45-minute run is executed simultaneously on each of the 4 strings.  The run starts on the hour and finishes 45 minutes after the hour.  Three channels per string are read out, each at 200~kHz.  The final 15 minutes of every hour are used for other types of data taking, including acquisition of raw noise recordings.

The strings run independently but simultaneously.  Because of problems in the ADC driver, some runs hang and need to be terminated.  This happens to a few percent of runs.

\begin{table}[tbp]
\centering
\caption[Time period for each transients configuration]{Time period of operation for each of the eight transients DAQ channel configurations used to date.  All times are UTC.}
\centering
\begin{tabular}{| c | c | c |}	
\hline
\bf{Configuration} & \bf{Beginning time} & \bf{Ending time} \\
\hline
1 & Aug 30 2008 06:00 & Oct 13 2008 18:45 \\
\hline
2 & Oct 13 2008 19:00 & Oct 23 2008 20:45 \\
\hline
3 & Oct 23 2008 21:00 & Dec 3 2008 19:45 \\
\hline
4 & Dec 3 2008 20:00 & Jan 16 2009 13:45 \\
\hline
5 & Jan 16 2009 14:00 & Jan 20 2009 12:45 \\
\hline
6 & Jan 20 2009 13:00 & Feb 10 2009 02:45 \\
\hline
7 & Feb 10 2009 19:00 & Feb 20 2009 10:45 \\
\hline
8 & Feb 20 2009 11:00 & (ongoing) \\
\hline
\end{tabular}
\label{channelConfigurationTimes}
\end{table}

The first high-quality runs with all 4 strings participating began on August 30, 2008, at 06:00 UTC.  Since then we have changed the configuration of channels and thresholds that we are using several times.  A total of eight different configurations have been used.  The configuration run periods are tabulated in Table~\ref{channelConfigurationTimes}.  The details of two example configurations (the first configuration and the current configuration) are listed in Table~\ref{channel_configurations}.

\begin{table}[tbp]
\centering
\caption[Example channel configurations used for transient data acquisition]{Two example channel configurations used for transient data acquisition. Configuration 1 is the first used for high quality transients data acquisition, and Configuration 8 is the stable configuration currently in use.  Channels that were read out in one configuration and not the other are shown in bold.  The depth of each channel is shown in parentheses.  The low and high threshold of each channel (in ADC counts) is shown in square brackets.
\label{channel_configurations}
}
\centering
\begin{tabular}{| c | c |}  
\hline
\bf{Configuration 1} & \bf{Configuration 8} \\
\hline
Beginning Aug 30 2008, 06:00 UTC & Beginning Feb 20 2009, 11:00 UTC \\
\hline
Ending Oct 13 2008, 18:45 & (ongoing) \\
\hline
\textbf{AS4-2 (190 m) [-212, 210]} & AS5-1 (250 m) [-178, 176] \\
AS5-1 (250 m) [-177, 175] & AS6-0 (320 m) [-154, 150] \\
AS6-0 (320 m) [-154, 150] & \textbf{AS7-2 (400 m) [-124, 116]} \\
\hline
BS5-1 (250 m) [-163, 163] & BS5-1 (250 m) [-164, 162] \\
BS6-0 (320 m) [-159, 153] & BS6-0 (320 m) [-160, 154] \\
\textbf{BS6-2 (320 m) [-149, 143]} & \textbf{BS7-2 (400 m) [-134, 134]} \\
\hline
\textbf{CS4-2 (190 m) [-195, 193]} & CS5-1 (250 m) [-187, 183] \\
CS5-1 (250 m) [-186, 184] & CS6-0 (320 m) [-145, 143] \\
CS6-0 (320 m) [-143, 143] & \textbf{CS7-2 (400 m) [-139, 133]} \\
\hline
\textbf{DS2-2 (190 m)} [-31, 3] & DS3-1 (250 m) [-162, 156] \\
DS3-1 (250 m) [-160, 156] & DS4-0 (320 m) [-151, 147] \\
DS4-0 (320 m) [-150, 146] & \textbf{DS5-2 (400 m) [-112, 114]} \\
\hline
\end{tabular} 
\end{table} 

During the transient runs, each channel is triggered using a bipolar discriminator implemented in software.  Both a high and low threshold, in ADC counts, is specified for each channel.  If that channel's output voltage becomes greater than or equal to the high threshold, or less than or equal to the low threshold, the channel is triggered.  When a channel is triggered, the 1001 samples centered on the first triggering sample are written to disk.  At 200~kHz sampling frequency, each captured waveform is of duration 5.005~ms.  As in all runs, the waveform is time-stamped with the IRIG GPS signal online, and the absolute time of the first sample of the waveform is written in the data stream.  The three channels per string are triggered independently.

When a channel triggers, it is examined for triggers again starting with the first sample after the captured waveform (501 samples after the triggering sample of the captured waveform).  This choice ensures that many consecutive samples exceeding threshold do not cause individual triggers, while also ensuring that any samples meeting the trigger condition are written to disk in at least one event (and sometimes in two events).  If a waveform is over threshold for a long time, it can result in many consecutive individual triggers.

\section{Trigger rates}

\subsection{Threshold choice}

As described earlier, the background noise on each channel is well described by a normal distribution with mean $\mu$ and standard deviation $\sigma$ (both measured in ADC counts).  We have determined $\mu$ and $\sigma$ for each channel and used it to choose high and low discriminator thresholds for that channel, corresponding to $n \sigma$ from the mean.  That is:

\begin{equation}
T_{low} = round(\mu - n \sigma)
\end{equation}
\begin{equation}
T_{high} = round(\mu + n \sigma),
\end{equation}

\noindent where $T_{low}$ is the low threshold and $T_{high}$ is the high threshold, both in ADC counts.  The \emph{round} function is used because although $\mu$ and $\sigma$ can take any real value, $T_{low}$ and $T_{high}$ must be integers.  This choice of thresholds allows us to directly choose the expected trigger rate on each channel, such that the trigger rate is stable on each channel, the rate is consistent from channel to channel, and the number of high triggers is roughly equal to the number of low triggers.

We chose thresholds in order to trigger several times per minute per channel, or 100-200 triggers per channel per 45-minute run.  This rate was chosen to achieve $\sim$100~MB of transients data per day in total for four strings, which fits safely in the daily SPATS bandwidth quota of 150~MB.  In order to achieve this trigger rate, we chose $n =$~5.2~$\sigma$ for the SPATS channels and $n =$~7.0$\sigma$ for the HADES channels.  In the following sections we calculate the expected trigger rate and compare it with the actual rate.

\subsection{Expected trigger rate from Gaussian noise}

\label{rateSection}

In the configuration we have used since SPATS was installed, the DAQ operates on each sensor independently.  If any sensor exceeds threshold, a waveform surrounding the triggering sample is captured and written to disk.  Analysis of coincidence between multiple sensor channels, and using such coincidence to perform source position and emission time reconstruction, is then performed offline.

With such a trigger we can calculate the expected trigger rate, under the hypothesis of Gaussian-only noise, as follows.

The probability of choosing a value $x$ from a normal distribution ($\mu$, $\sigma$) that is less than $x_0$ is given by the cumulative distribution function for the normal distribution,

\begin{equation}
P(x<x_0) = \frac{1}{2}[1 + \textrm{erf}[\frac{x_0-\mu}{\sigma \sqrt{2}}]],
\end{equation}

\noindent where \emph{erf} is the error function.  This function approaches 1 as $x_0$ approaches $\infty$, and it approaches 0 as $x_0$ approaches $-\infty$.  From this we can determine the probability that $x$ is greater than $x_0$:

\begin{equation}
P(x>x_0) = 1-P(x<x_0) = \frac{1}{2}[1 - \textrm{erf}[\frac{x_0-\mu}{\sigma \sqrt{2}}]].
\end{equation}

Next we can substitute our high threshold, $T_{high}$, to determine the probability that any individual sample of the Gaussian noise is greater than it:

\begin{equation}
P(x>\mu + n \sigma) = \frac{1}{2}[1 - \textrm{erf}[\frac{n}{\sqrt{2}}]],
\end{equation}

\noindent and similarly we can calculate the probability that any sample is less than the low threshold:

\begin{equation}
P(x<\mu - n \sigma) = \frac{1}{2}[1 + \textrm{erf}[{-n}{\sqrt{2}}]] = \frac{1}{2}[1 - \textrm{erf}[{n}{\sqrt{2}}]] ,
\end{equation}

\noindent where we have used the fact that the error function is odd.  The result is that the probability of triggering high is equal to the probability of triggering low, as expected from the symmetry of the Gaussian distribution and our choice of thresholds centered on the DC offset.

The probability of any one sample triggering is the sum of the probability of triggering low and the probability of triggering high:

\begin{equation}\noindent
P = 1 - \textrm{erf}[\frac{n}{\sqrt{2}}].
\end{equation}

Finally, the trigger rate $R$ is given by scaling this by the sampling frequency $f$:

\begin{equation}
R = f P = f(1 - \textrm{erf}[\frac{n}{\sqrt{2}}]).
\end{equation}

For our choice $n =$~5.2$\sigma$ used for SPATS channels, the probability of any one sample triggering is $p =$~1.99~x~10$^{-7}$.  The single-channel trigger rate is 0.0399~Hz, or 2.39 events per minute.  In each 45-minute run, we expect 108 events per channel, or $\sim$1290 events for all 12 channels summed over 4 strings.  At 2.089~kB per event, the expected data rate is 64.7~MB per day from transient runs, summed over all four strings.

\begin{figure}
\begin{center}
\subfigure[]{
\noindent\includegraphics[width=17pc]{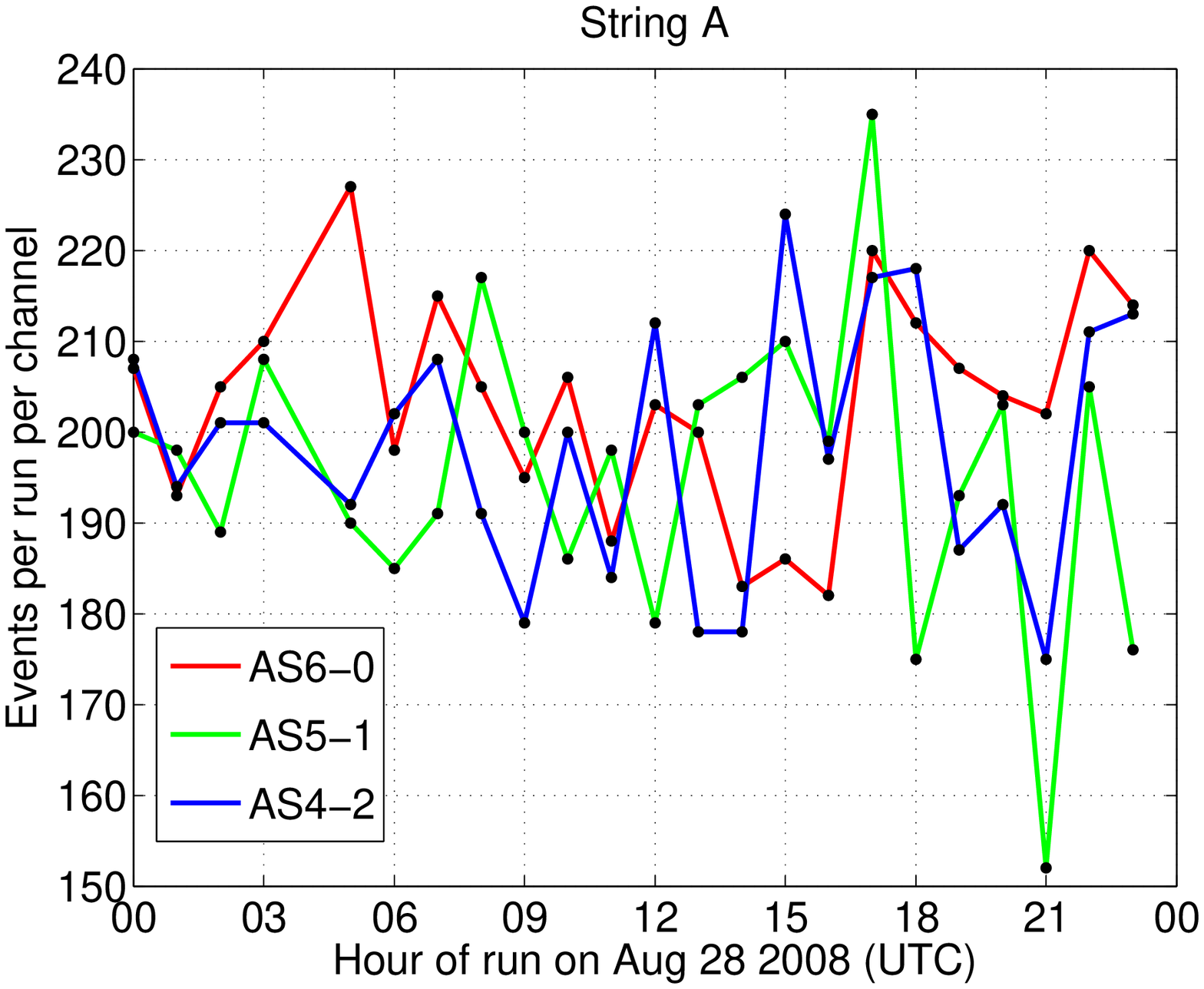}
}
\subfigure[]{
\noindent\includegraphics[width=17pc]{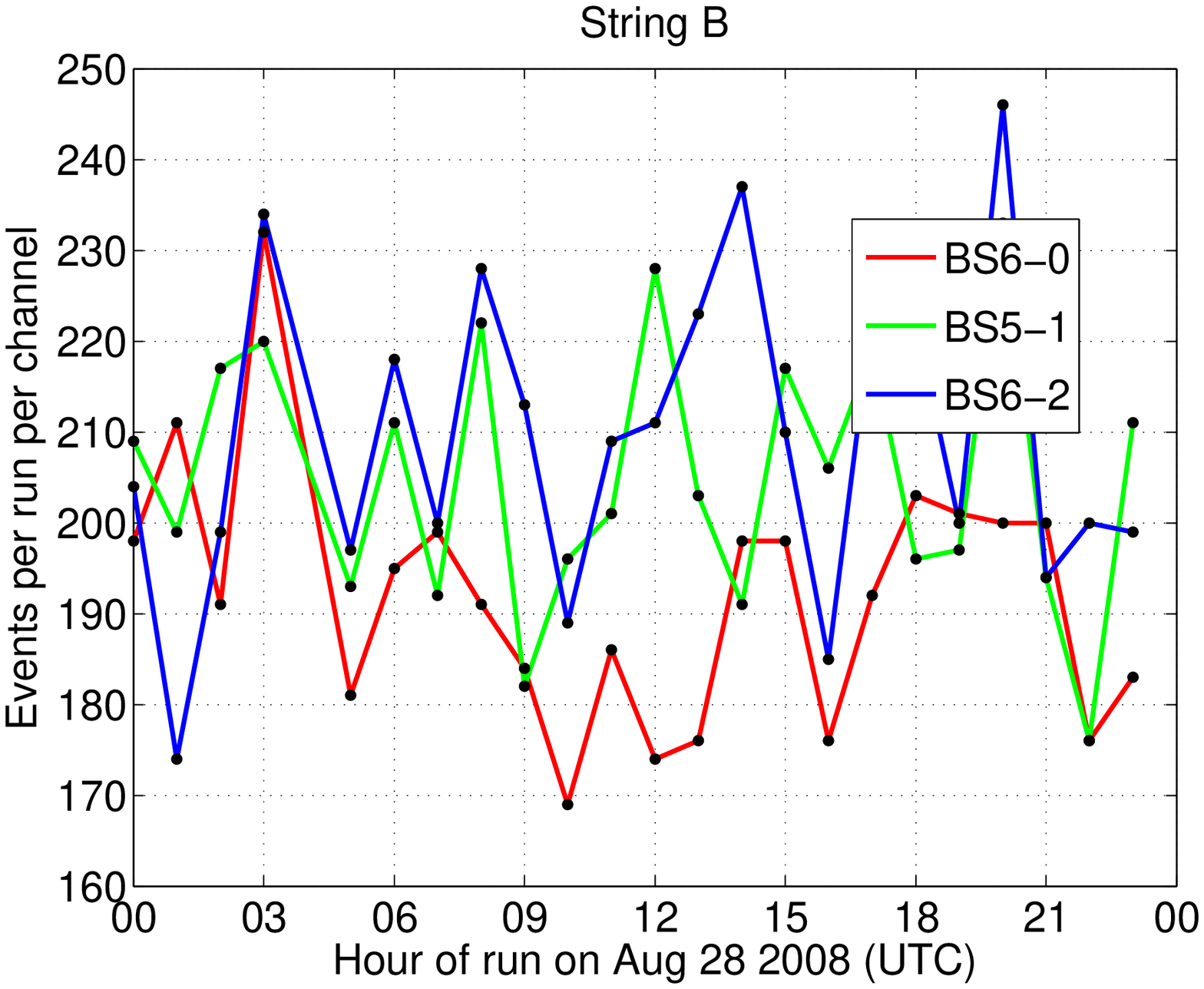}
}
\subfigure[]{
\noindent\includegraphics[width=17pc]{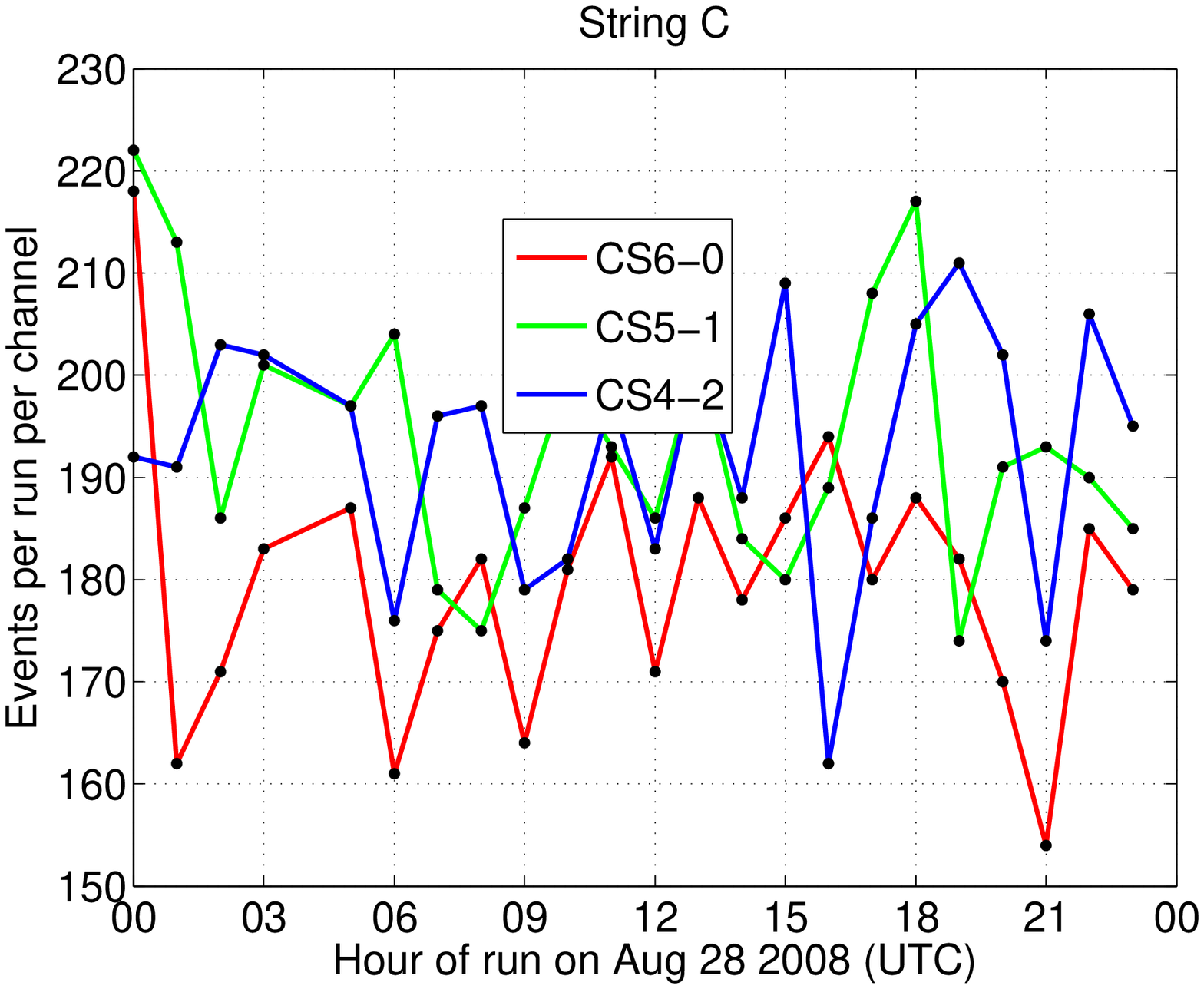}
}
\subfigure[]{
\noindent\includegraphics[width=17pc]{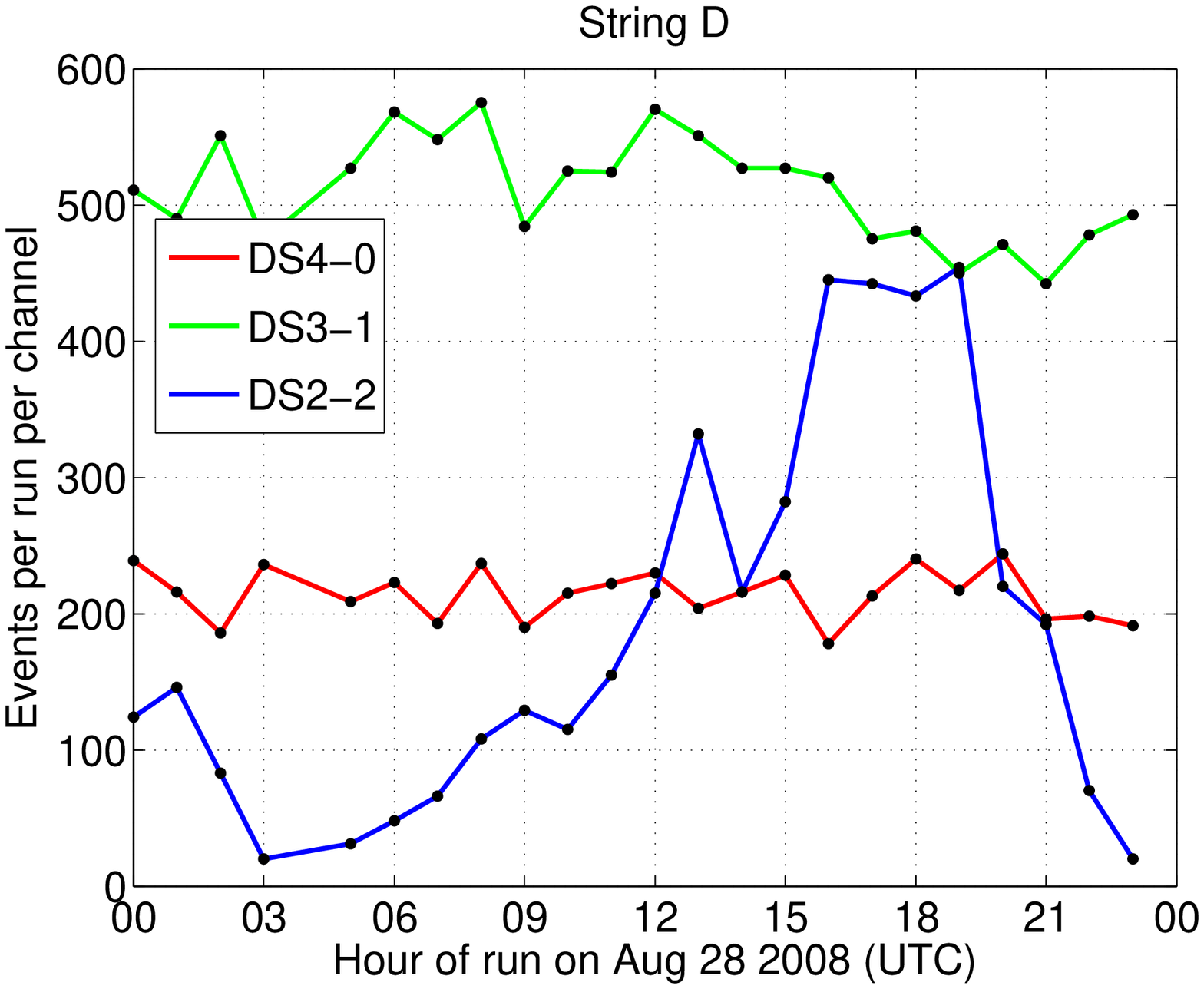}
}
\caption[Example transient trigger rates from August 2008]{Number of triggered events per run vs. start time of run, for each channel of each string on August 28, 2008.  Each transients run is 45 minutes long.  The channel and threshold configuration used for these runs was Configuration 1, details of which are given in Table~\ref{channel_configurations}.}

\label{transientRates_080828}
\end{center}
\end{figure}

\begin{figure}
\begin{center}
\subfigure[]{
\noindent\includegraphics[width=17pc]{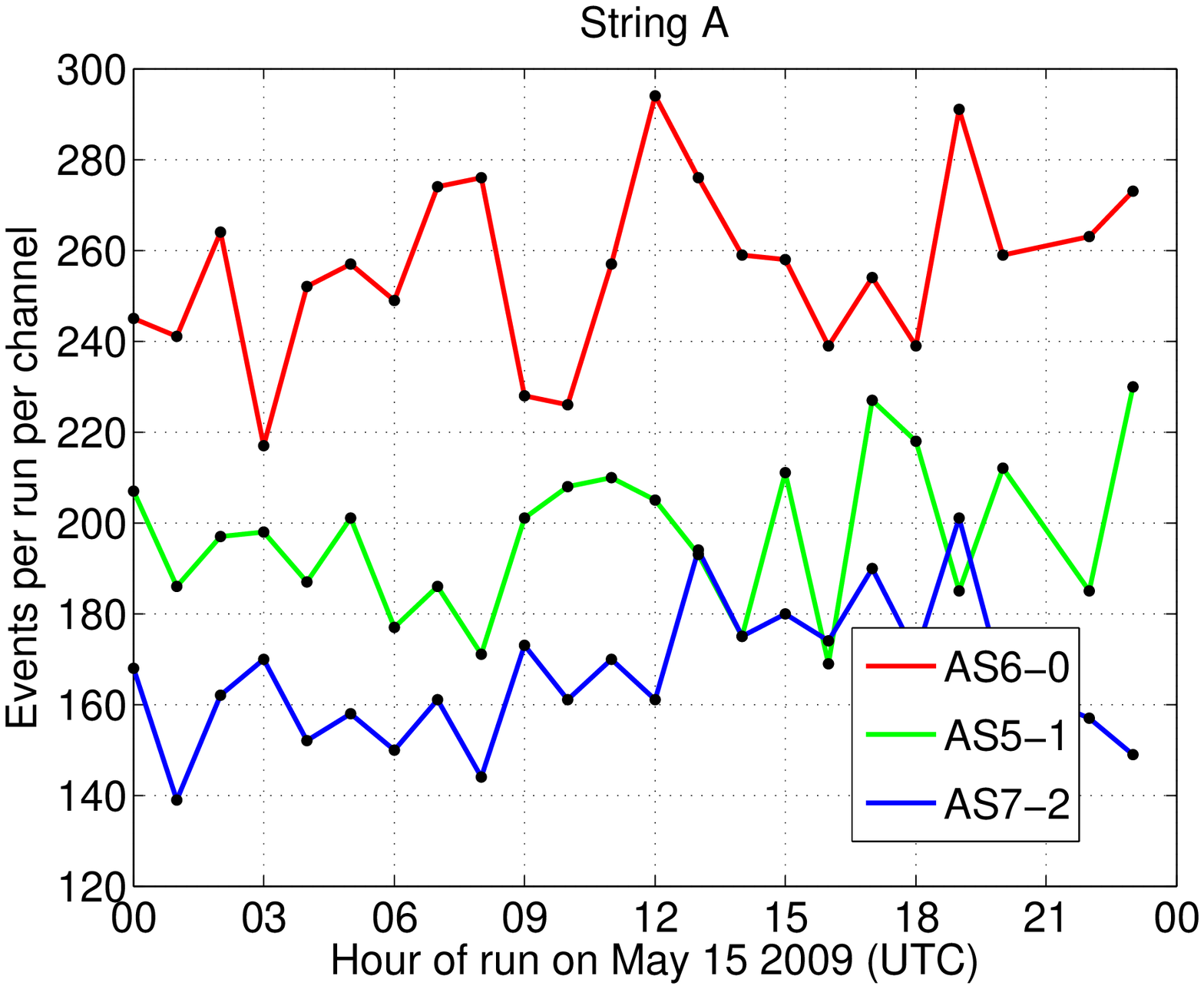}
}
\subfigure[]{
\noindent\includegraphics[width=17pc]{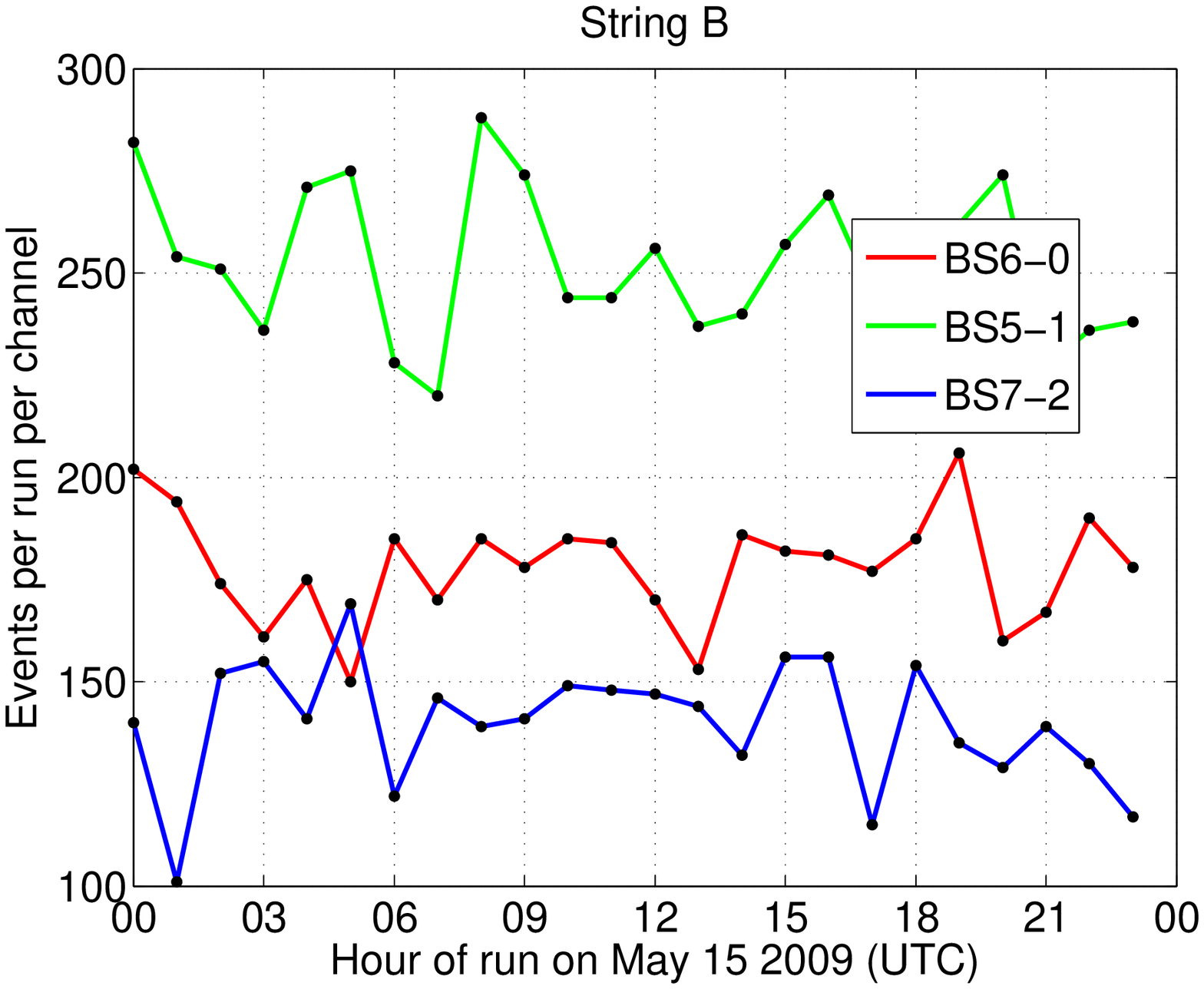}
}
\subfigure[]{
\noindent\includegraphics[width=17pc]{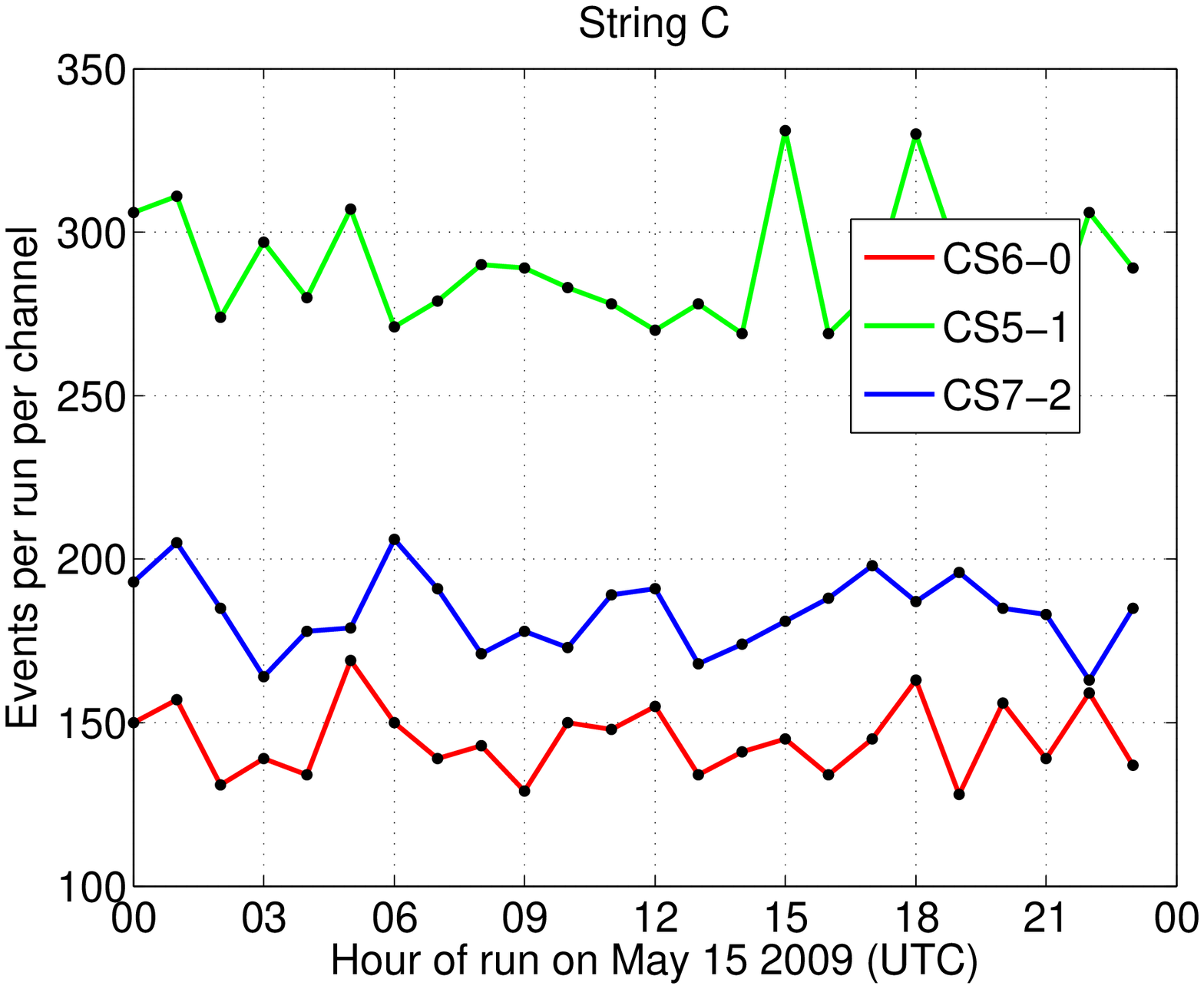}
}
\subfigure[]{
\noindent\includegraphics[width=17pc]{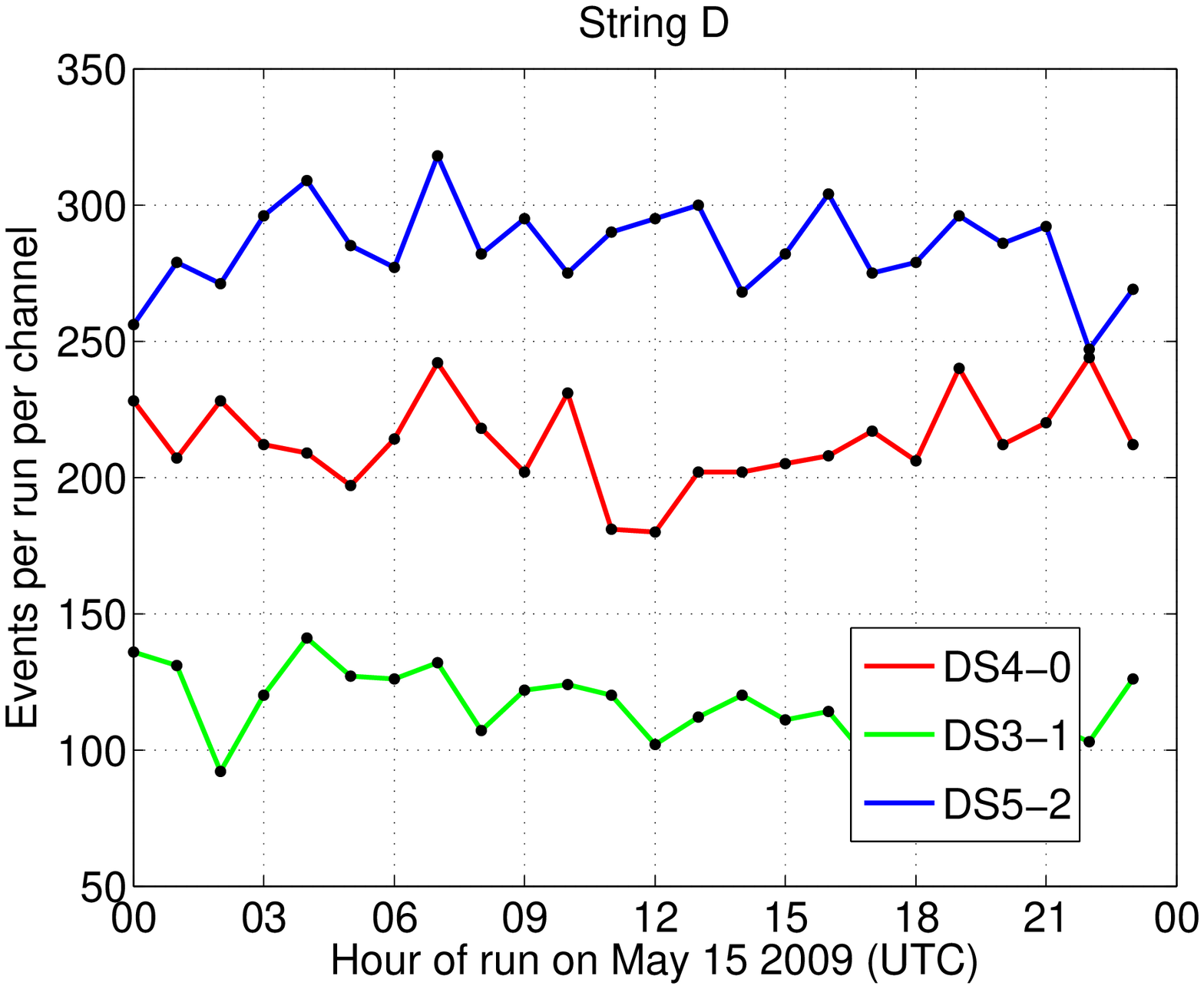}
}
\caption[Example transient trigger rates from May 2009]{Number of triggered events per run vs. start time of run, for each channel of each string on May 15, 2009.  Each transients run is 45 minutes long.  The channel and threshold configuration used for these runs was Configuration 8, details of which are given in Table~\ref{channel_configurations}.}
\label{transientRates_090515}
\end{center}
\end{figure}

\subsection{Actual trigger rate}

Figure~\ref{transientRates_080828} shows the number of triggers per run vs. time of day for all channels on August 28, 2009, and Figure~\ref{transientRates_090515} shows the same plot for May 15, 2009.  Several points are worth noting.  First, the rate is very stable on each channel.  This is a significant benefit of the Gaussianity of the noise distribution on each channel.  It is a very different situation from hydrophone arrays in water, for example, where the noise level varies on multiple time scales and so do the trigger rates, unless sophisticated adaptive threshold algorithms are used~\cite{Vandenbroucke05, Kurahashi07}.  Second, the rate is consistent from channel to channel.  This is the benefit of choosing channel-dependent thresholds of $\pm n \sigma$ relative to the channel-dependent mean.  Third, the trigger rate is comparable to that expected from Gaussian statistics.

The actual trigger rate is $\sim$200 triggers per channel per run, compared with the expected rate of $\sim$100 triggers per channel per run from Gaussian statistics as estimated above.  The excess raises our typical daily data rate for transients data from $\sim$65~MB per day to $\sim$130~MB per day, still safely within our daily $\sim$150~MB quota.  There are several possible contributions to the discrepancy between the actual trigger rate and that expected from the Gaussian noise estimate:
\begin{enumerate}
\item Non-Gaussianity of the background noise amplitude distribution.
\item Variation in the Gaussian $\mu$ and $\sigma$ with respect to the values assumed in choosing the trigger thresholds.
\item Non-Gaussian impulsive transients (actual transient ``signals'' above the noise).
\end{enumerate}

Finally, while the trigger rate on each channel is stable over a long time scale, there are some clear differences between the rates in August 2008 and May 2009.  In particular, in August 2008 the rates were more consistent from channel to channel, and in May 2009 there is more channel-to-channel variation in the trigger rates.  This is likely because the Gaussian noise $\mu$ and $\sigma$ have drifted slightly over many months, while we have not changed the trigger thresholds over the same time period.

The hypothesis that the observed slow drift in trigger rates is due to slow drift in Gaussian noise parameters can be tested with a calculation.  As an example we consider channel CS5-1, for which the mean trigger rate (in events per run) increased from 194 on August 28, 2008 to 288 on May 15, 2009.  During this time period, the Gaussian noise of the channel did in fact drift slightly higher, such that the actual (low, high) thresholds (measured in number of $\sigma$ from $\mu$) were (-5.20, +5.23) on August 28 but were (-5.13, 5.10) on May 15.  The thresholds were originally chosen to be at (-5.2, +5.2).  The expected number of triggers per run for pure Gaussian noise is 108 events per run for a threshold of (-5.2, +5.2) and 183 events per run for a threshold of (-5.1, +5.1).  For both time periods, the difference between the actual and the expected Gaussian-only trigger rate is likely due to genuine non-Gaussian impulsive signal events above the noise.  The conclusion is that for this example time period, the trigger rate increased roughly by the proportion expected, in the same time period that the Gaussian noise increased while the absolute thresholds remained fixed.

If precisely stable event rates are necessary, the thresholds can be updated to match drifting Gaussian noise parameters every $\sim$month.  However, the drift is small and there are also benefits of having absolute voltage thresholds that are constant over multiple months.

\section{Noise hit cleaning with sample multiplicity cut}

Before applying a coincidence requirement, we introduced a ``sample multiplicity'' cut.  The purpose of this cut is to reduce the effect of noise hits.  Most of our triggers are ``Gaussian tail'' events in which a single sample fluctuated high in the Gaussian tails.  These events compose the Gaussian-only expected rate described above.  For real acoustic transients, we expect a coherent signal with a clear waveform shape different from the Gaussian noise recordings in which each sample amplitude is independent of the others.  Except for events very close to threshold where we likely cannot distinguish them from noise in any case, the real acoustic transients should have more than one sample meeting the bipolar discriminator trigger requirement.  Therefore for each event we count the number of samples passing the trigger (the ``sample multiplicity'') and require it to be greater than one.

This cut reduces the rate of Gaussian noise hits by several orders of magnitude, while presumably having a small effect on real acoustic transients.  The Gaussian noise rejection factor is estimated as follows.  Whenever there is a trigger, the DAQ records 1001 samples from the triggering channel, centered on the first sample sample to meet the trigger requirement.  First we consider an arbitrary recording of 1001 samples (not necessarily one that passed the trigger).  Given an arbitrary recording of $n =$~1001 samples, we assume that if the recording is a Gaussian noise capture then all the samples are independent.  Each sample has a probability $p =$~1.99~x~10$^{-7}$ of meeting the trigger requirement, as shown in section~\ref{rateSection}.  The probability that exactly $k$ of the $n$ samples match the trigger requirement is given by the binomial distribution:

\begin{equation}
P(k) = {n \choose k} p^k (1-p)^{n-k}.
\label{binomialDistribution}
\end{equation}

\begin{figure}[tbp]
\begin{center}
\includegraphics[width= 0.5\textwidth]{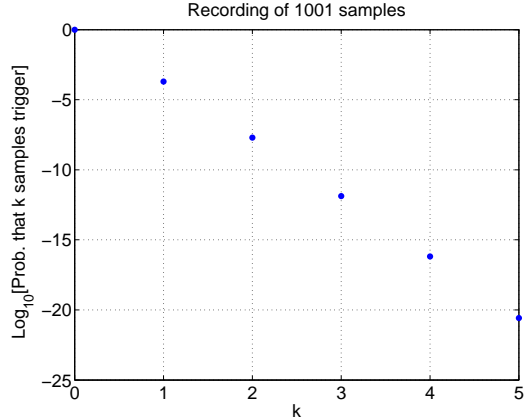}
\end{center}
\caption[Probability that $k$ of $n$ waveform samples meet trigger condition]{For an arbitrary recording of $n = $~1001 independent noise samples, the probability that exactly $k$ of them pass the trigger condition is given by the binomial distribution.  We have assumed the probability that any particular sample triggers is $p =$~1.99~x~10$^{-7}$, as is expected under our trigger conditions.}
\label{triggerMultiplicityStats}
\end{figure}

This equation is plotted in Figure~\ref{triggerMultiplicityStats}.  Note that increasing $k$ by 1 decreases the probability of occurrence dramatically.  In the general case,

\begin{equation}
\frac{P(k+1)}{P(k)} = \frac{p}{1-p} \frac{n-k}{k+1}.
\end{equation}

\noindent For our particular values of $p$ and $n$, increasing $k$ by 1 decreases the probability of occurrence by four orders of magnitude.

Two particular probabilities are worth calculating for what follows (we are assuming $p <<$~1).

\begin{equation}
P(0) = 1-np;
\end{equation}

\begin{equation}
P(1) = (np)[1-p(n-1)].
\end{equation}

We are interested in the probability that two samples trigger in a single capture, given that one sample has triggered.  First recall Bayes' Theorem:

\begin{equation}
P(A|B) = P(B|A) \frac{P(A)}{P(B)}.
\end{equation}

\noindent Using this we can determine the probability that 2 or more samples pass the trigger, given that 1 or more samples pass the trigger:

\begin{equation}
P(k \ge 2 | k \ge 1) = P(k \ge 1 | k \ge 2) \frac{P(k \ge 2)}{P(k \ge 1)} = \frac{1 - P(0) - P(1)}{P(0)} = p(n-1).
\end{equation}

We can cross-check this with another method.  Given that the center sample in the captured waveform of $n =$~1001 samples passed the trigger, what is the chance that at least one of the remaining $n-1$ samples passed the trigger?  It is the complement of the probability that zero samples passed the trigger.  Then using Equation~\ref{binomialDistribution}, it is

\begin{equation}
1- {n-1\choose 0}p^0(1-p)^{n-1-0} = 1-(1-p)^{n-1} = 1-(1-p(n-1)) = p(n-1),
\end{equation}

\noindent which is the same answer we got using the first method.

For our values of $p$ and $n$, this gives a probability of 1.99~x~10$^{-4}$.  This is the fraction of triggered events that have more than 1 sample passing the trigger, under the hypothesis that all samples are independently selected from the same Gaussian amplitude distribution.  That is, requiring at least two samples to pass the trigger condition for each triggered waveform reduces the rate of Gaussian tail events by a factor of $\sim$2~x~10$^{-4}$.

\section{Multi-channel coincidence}

Processing was performed offline to determine multi-channel coincidence clusters.  For each hour, it was first required that runs from all 4 strings were present.  A small fraction of hours are excluded by this requirement because some runs are halted prematurely due to a hanging ADC driver.  When this happens the runs are terminated by the DAQ and deleted.

The DAQ executes one transients run per string per hour.  Each run produces a single file.  The maximum file size is $\sim$34~MB (compressed) and occurs when the trigger rate is so high that it fills the 100~MB RAM disk with the file (before compression).  A trigger rate orders of magnitude larger than typical is necessary to cause such a saturated run.  When these runs occur the data are retained, but the run terminates as soon as the disk fills and therefore the run has an abnormally short duration.  To avoid these saturated runs, only runs of size $<$~10~MB were included.  This requirement also improves offline processing because large runs take longer to process.

Only hours with all four single-string runs present were considered for further analysis.  For each hour, the runs from the four strings are merged and sorted to produce a time-ordered list of single-channel triggers.  Next the sample multiplicity cut described above is applied.  Finally, a coincidence window is applied to determine ``clusters'' of single-channel triggers that could be causally related.  A coincidence window of length 200~ms is used.  The maximum distance between sensors in the 4-string SPATS array is 686~m (between CS1 and DS7).  At 3878~m/s, these sensors are separated by 177~ms.  The extra 13\% buffer achieved by using 200~ms allows for refraction and finite time resolution.

For each time-ordered list of single-channel triggers, one per hour, we loop over events in the list to determine coincidence clusters.  For each event $i$, we determine if any other events occurred within 200~ms after event $i$.  If any did, we add them to a ``cluster'' and search for any other events occurring within 200~ms of those events.  The cluster is extended until there are no more events within 200~ms after the last event in the cluster.  Each single-channel event belongs to either one cluster or no cluster.  This algorithm ensures that any two events within 200~ms are members of the same cluster.  The duration of the cluster can in principle be much longer than ~200 ms if the trigger rate is high, with the first and last events in the cluster separated by much more than 200~ms.  But this happens rarely due to our very low trigger rate ($\sim$0.07~events per channel per second without the sample multiplicity cut, and much lower with it).

Next, we restrict the cluster by including only the first trigger on each sensor \emph{module}: we only include the first hit at a given channel, and if two different channels of the same module are hit, we only include the first hit.  All other hits are ignored.  This solves three problems: multiple triggering on a single long pulse arriving at a single channel; shear waves causing two pulses arriving at a single channel; and degeneracies in the reconstruction algorithm due to including two hits at two channels of the same module, which occur at nearly the same time and the same location with respect to other modules and therefore do not contribute independently to vertex reconstruction due to the small distance ($\sim$10~cm) separating channels of the same module.

For event reconstruction, we require that at least 3 strings were hit.  This is because while 3 or 4 strings allows good vertex reconstruction, the 3D vertex cannot be determined if only 1 or 2 strings are hit (no matter how many modules on them are hit).  If 1 string is hit, the source can only be constrained to a circle around the string.  If only 2 strings are hit, the source can only be constrained to two points, mirror images of one another about the plane defined by the two strings.

\begin{table}[tbp]
\centering
\caption[Parameters used in transients analysis]{Parameters used in transients analysis.  The analysis includes two steps: multi-module coincidence determination followed by source location and emission time reconstruction.}
\centering
\begin{tabular}{| c | c |}  
\hline
\bf{Parameter} & \bf{Value} \\
\hline
Maximum run file size & 10~MB \\
\hline
Minimum number of strings with good runs present & 4 \\
\hline
Minimum sample multiplicity & 2 \\
\hline
Minimum string multiplicity & 3 \\
\hline
Minimum module multiplicity & 5 \\
\hline
Coincidence time window & 200 ms \\
\hline
Sound speed & 3878~m/s \\
\hline
Maximum $\tau$ (see Equation~\ref{tauEquation}) & 10~ms \\
\hline
\end{tabular} 
\label{transientsCuts}
\end{table} 

Finally, we require that at least 5 modules are hit in the cluster in order to reconstruct it.  The cuts used in transients analysis are summarized in Table~\ref{transientsCuts}.

\section{Reconstruction of source location and time}

Consider a signal emitted by a source at location $\bf{r_0}$, at emission time $t_0$.  Assume the signal propagation speed is $c$ and is homogeneous and isotropic throughout the propagation medium.  The signal is detected by $n$ receivers.  The signal arrives at receiver $i$ with location $\bf{r_i}$ at time $t_i$, $i =$~1...n.  The challenge is to solve for ($t_0$, $\bf{r_0}$), given ($t_i$, $\bf{r_i}$).

This is a common problem with applications including: localization of emergency distress calls using cell tower signals; localization of animals from their vocalizations for field biology; localization of gamma-ray bursts using the interplanetary network, and localization of seismic emissions using distributed seismometers.  A class of solutions uses the differences in arrival times at the various receivers, and is known as the ``time difference of arrival'' (TDOA) technique.  The value of time differences of arrival follows from the fact that, while the propagation time from the source to the receivers is not directly known, the differences between them are.

In three dimensions, any pair of receivers constrains the source to a hyperboloid, symmetric about the axis between the two receivers.  Adding a third receiver adds a second hyperboloid, whose intersection with the first hyperboloid is generally a curve.  Adding a fourth receiver (third hyperboloid) generally results in two point-like solutions.  A fifth receiver (fourth hyperboloid) is generally necessary to distinguish between the two solutions, as emphasized in~\cite{Spiesberger01}.  However in some cases there is only one solution, and in many cases where there are two solutions one is unphysical, for example lying far outside the instrumented volume.

If the propagation speed is homogeneous and isotropic, the problem can be solved analytically as described in~\cite{Spiesberger90}.  The solution is determined by first writing down the propagation equation, $d_i = c \Delta t_i$, for each receiver.  The system of $n$ equations is linearized to a system of $n-1$ equations by subtracting off the equation for the first hit receiver.  The system can then be written as a matrix equation and solved robustly using singular value decomposition (SVD).  One benefit of this approach is that the $n$ receivers are all treated on the same footing, even if there are much more than 5 of them.  The emission time is solved simultaneously with the vertex position and both are determined with a direct analytical equation.  No iterating over subsets of the $n$ receivers, as is done by other algorithms, is necessary.

This method can be further improved by subtracting the average equation, rather than the first equation, from the system, as demonstrated in~\cite{Wahlberg01}.  This method retains $n$ equations rather than $n-1$ and can find unique solutions in many (but not all) cases where $n= $~4.

With SPATS we are currently reading out sensors between 250~m and 400~m depth, where we have measured the sound speed to be highly independent of depth.  The analytical algorithms therefore perform well, as long as the source is also deeper than $\sim$200~m.  For shallower sources, refraction in the firn (upper $\sim$200~m of ice, which is still converting from snow and is less dense than the fully dense ice below) should be included.  Unlike the analytical, invertible problem described above, this is an inverse problem that must be solved numerically.  An algorithm that solves it was demonstrated in ~\cite{Vandenbroucke05} and should work well for SPATS.  To begin with, however, we use the simpler and faster algorithm that requires at least 5 receivers and is described in~\cite{Spiesberger90}.  We have implemented this algorithm in MATLAB, and for events with 12 receivers hit it can reconstruct 3,000 events per second on a 1.8~GHz Dual Core AMD Opteron 265.

While the sample multiplicity cut described above rejects many of the noise hits, some noise hits remain and cause false vertex localization.  Refraction also causes systematic mis-reconstruction, an effect that is worst for sources in the firn and is worse for shallow events than deep events within the firn.  Because of these two effects, we introduced a final quality parameter defined as follows:

\begin{equation}
\tau = \sqrt{ \frac{1}{n} \sum_{i=1}^n (t_i - t_i^{expected})^2},
\label{tauEquation}
\end{equation}

\noindent where $t_i$ is the actual experimental arrival time at sensor module $i$, and $t_i^{expected}$ is the expected arrival time under the hypothesis that the acoustic signal was emitted at the time and location given by the vertex reconstruction algorithm, and propagated at constant $c$ to the receiver.  This quality parameter $\tau$ is simply the root-mean-square of the time residuals of the actual arrival times with respect to the expected arrival times under the hypothesis of homogeneous $c =$~3878~m/s.  For an event to be considered well-reconstructed, we require $\tau <$~10~ms.  This is a rather conservative cut designed to retain most events that have been affected by refraction, while rejecting events whose expected times dramatically mismatch the actual times due to the presence of noise hits.

\section{Results of transient source reconstruction}

We ran the coincidence determination and event reconstruction algorithm descried above on the full set of transients data collected to date (August 2008 - September 2009).  Basic statistics of the data set are described in Table~\ref{transientsStats}.  During the South Pole summer season (Nov-Feb), the live time for transients acquisition is reduced due to other types of data taking such as pinger runs.  However after this time period the duty cycle is stable at $\sim$70\%.  In standard data taking, 25\% of the live time is used for other types of data taking such as raw noise recordings.  The remaining 5\% corresponds to time periods in which transient acquisition was running but all four strings did not contribute complete transient runs, due to the ADC hanging problem (see Section~\ref{sectionADCIssues}).

\begin{table}[tbp]
\centering
\caption[Transient event statistics]{Summary of transient event statistics.  ``Runs'' includes all single-string runs for the month, including those without 4 strings running at once and those with file size $>$10~MB.  ``Good hrs'' is the total number of hours for which all four strings are present and have file size $<$10~MB.  ``Live fraction'' is the total fraction of time in the month for which ``good'' transient runs were occurring, defined to be (45 minutes live per hour) / (60 minutes) x (number of complete hours) / (number of hours in the month).  The number of vertices after applying the $\tau$ cut is given, followed in parentheses by the number of vertices before applying the $\tau$ cut.  Stable data taking began at the end of August 2008, so that month is incomplete.  The run period considered in this analysis is 00:00 UTC on August 30, 2008 through 23:45 UTC on September 30, 2009.  The rate of transient events is gradually decreasing, likely due to the Rod wells becoming quieter as their freeze-in proceeds.  The total number of good hours analyzed is 8,013, corresponding to 250.4 days of live time given that there are 45 minutes of live time per good hour.}
\centering
\begin{tabular}{| c  | c  | c | c  | c  | c  |}  
\hline
\bf{Month} & \bf{Runs} & \bf{Good hrs} & \bf{Live fraction} & \bf{Vertices} & \bf{Vertices/hr} \\
\hline
Aug 08 & 143 & 24 & - & 30 (43) & 1.25 \\
\hline
Sep 08 & 2,670 & 623 & 65\% & 1028 (1333) & 1.65 \\
\hline
Oct 08 & 2,825 & 655 & 66\% & 1158 (1445) & 4.54 \\
\hline
Nov 08 & 2,354 & 484 & 50\% & 754 (927) & 1.56 \\
\hline
Dec 08 & 2,244 & 411 & 41\% & 459 (618) & 1.12 \\
\hline
Jan 09 & 2,712 & 571 & 58\% & 722 (940) & 1.26 \\
\hline
Feb 09 & 2,626 & 566 & 63\% & 260 (380) & 0.46 \\
\hline
Mar 09 & 2,814 & 661 & 67\% & 349 (485) & 0.53 \\
\hline
Apr 09 & 2,749 & 644 & 67\% & 361 (476) & 0.56 \\
\hline
May 09 & 2,781 & 648 & 65\% & 292 (363) & 0.45 \\
\hline
Jun 09 & 2,837 & 677 & 71\% & 53 (53) & 0.08 \\
\hline
Jul 09 & 2,928 & 696 & 70\% & 71 (71) & 0.10 \\
\hline
Aug 09 & 2,920 & 691 & 70\% & 79 (81) & 0.11 \\
\hline
Sep 09 & 2,817 & 662 & 69\% & 62 (62) & 0.09 \\
\hline
\bf{Total} & \bf{35,420} & \bf{8,013} & - & \bf{5,678} & - \\
\hline
\end{tabular} 
\label{transientsStats}
\end{table} 

\begin{figure}[tbp]
\begin{center}
\includegraphics[angle = 0, width = 1\textwidth]{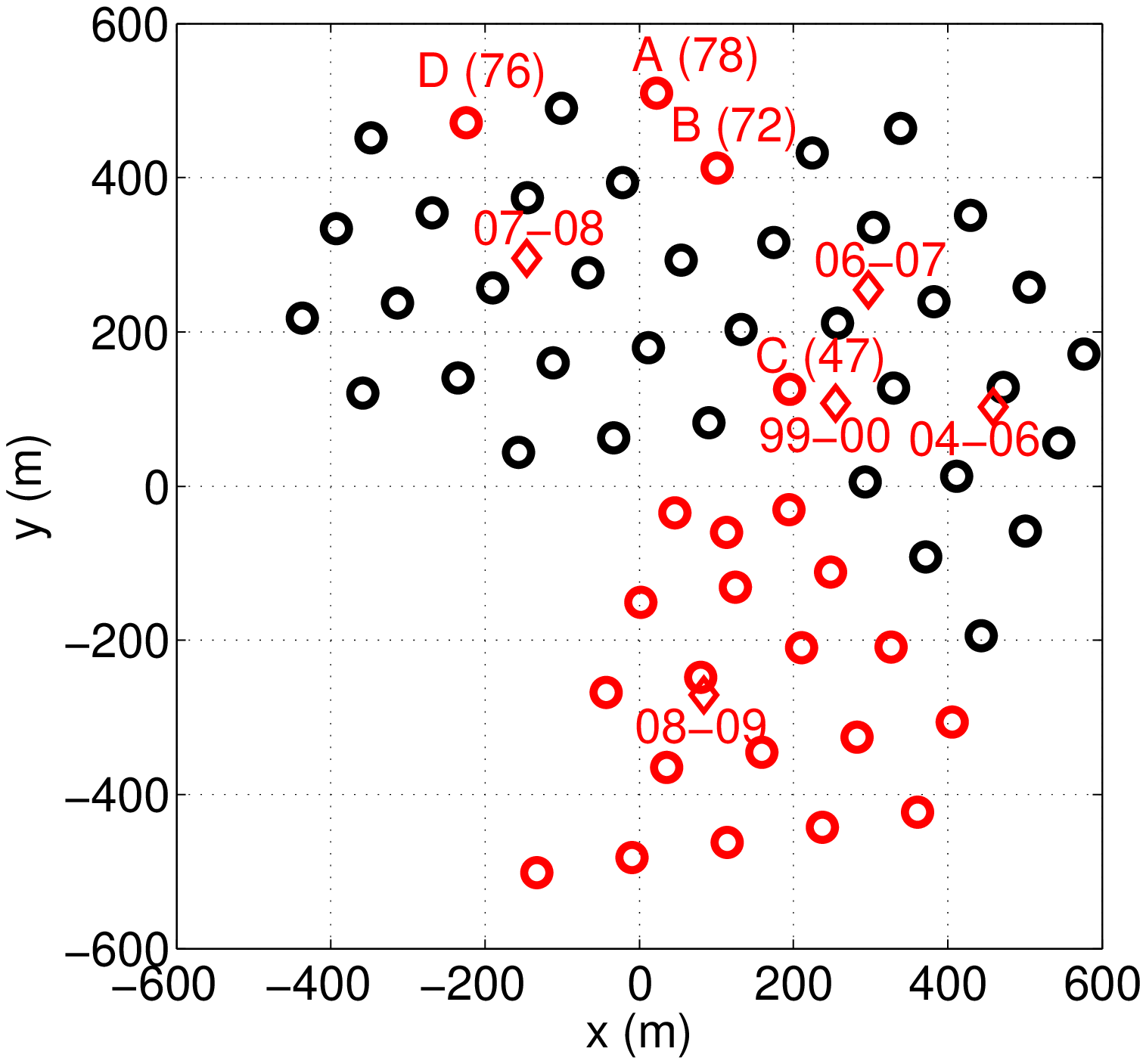}
\end{center}
\caption[Location of Icecube and AMANDA Rodriguez wells]{The 59-string IceCube configuration, in operation in 2009.  The four IceCube and one AMANDA Rodriguez well locations are identified by diamonds and labeled by the drill season(s) in which they were used.  The 19 strings that were drilled during the 2008-2009 season, many of which have been detected by SPATS to emit acoustic signals, are colored red.  While the IceCube hole locations are known with better than 1~m precision, the uncertainty of the Rod well locations is $\sim$10~m.}
\label{geometry_2009}
\end{figure}

Figure~\ref{geometry_2009} shows the layout of IceCube and SPATS during 2009.  All 59 IceCube strings deployed and operating during 2009 are shown.  The 19 that were deployed in the 2008-2009 season are colored red.  Finally, the five Rodriguez, or ``Rod'' wells, are marked with diamonds and labeled by season.  Four of the Rod wells were drilled for IceCube, and one was drilled for AMANDA.

\begin{figure}
\begin{center}
\subfigure[Aug 2008]{
\noindent\includegraphics[width=17pc]{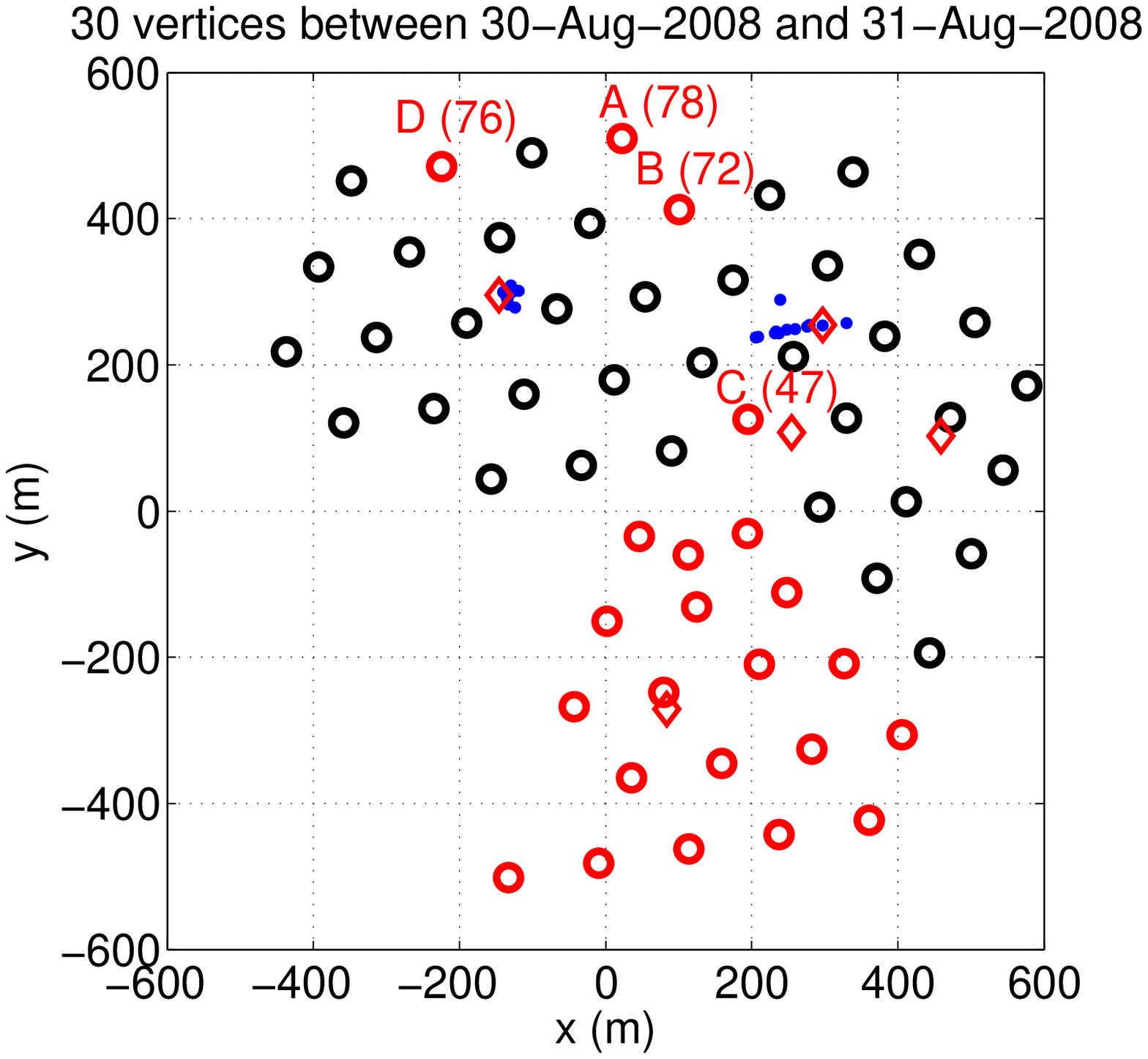}
}
\subfigure[Sep 2008]{
\noindent\includegraphics[width=17pc]{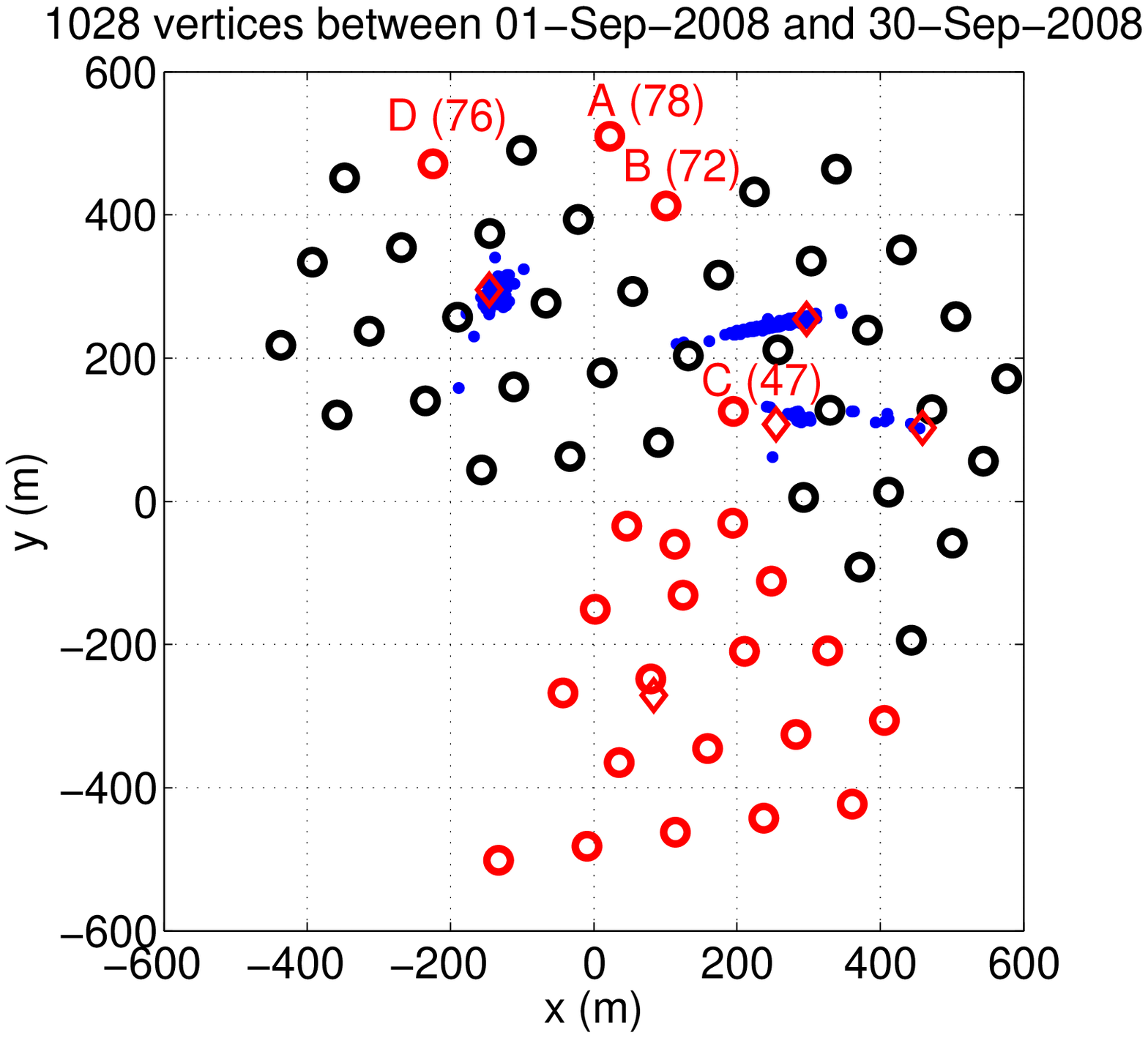}
}
\subfigure[Oct 2008]{
\noindent\includegraphics[width=17pc]{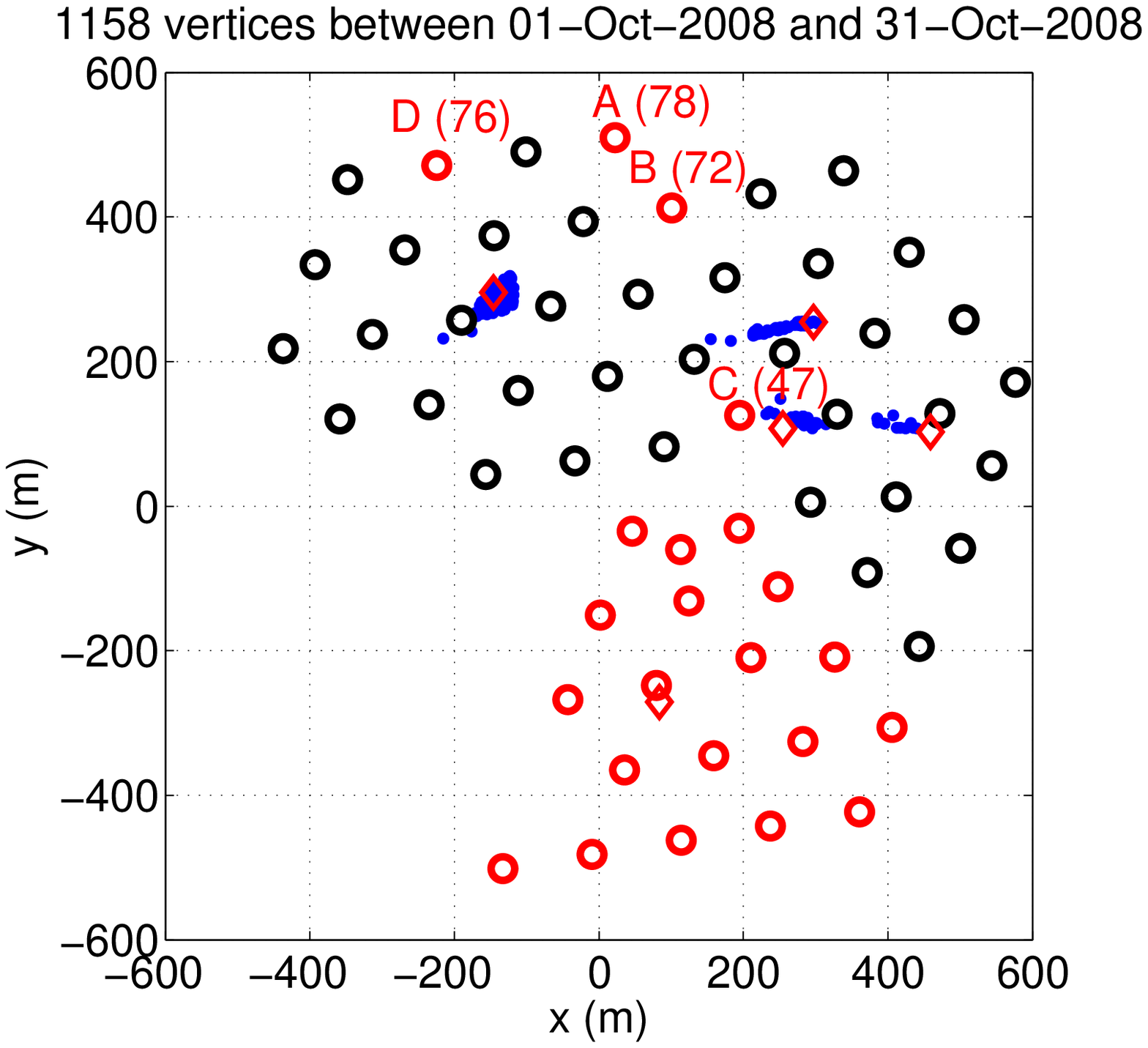}
}
\subfigure[Nov 2008]{
\noindent\includegraphics[width=17pc]{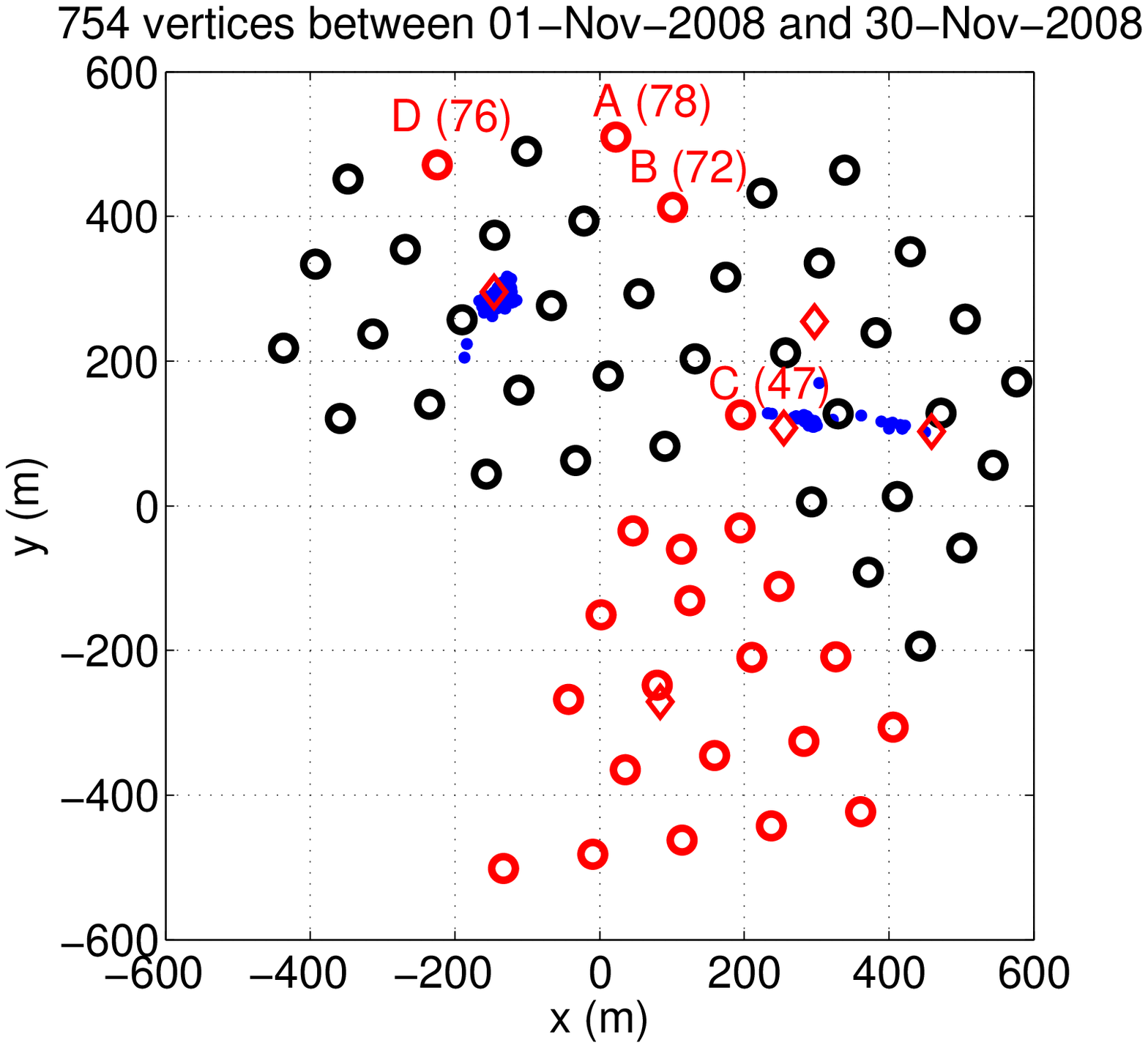}
}
\caption[Transient vertex locations for Aug-Nov 2008]{Map of reconstructed vertex locations for transient events in August-November 2008.}
\label{transientsXY1}
\end{center}
\end{figure}

\begin{figure}
\begin{center}
\subfigure[Dec 2008]{
\noindent\includegraphics[width=17pc]{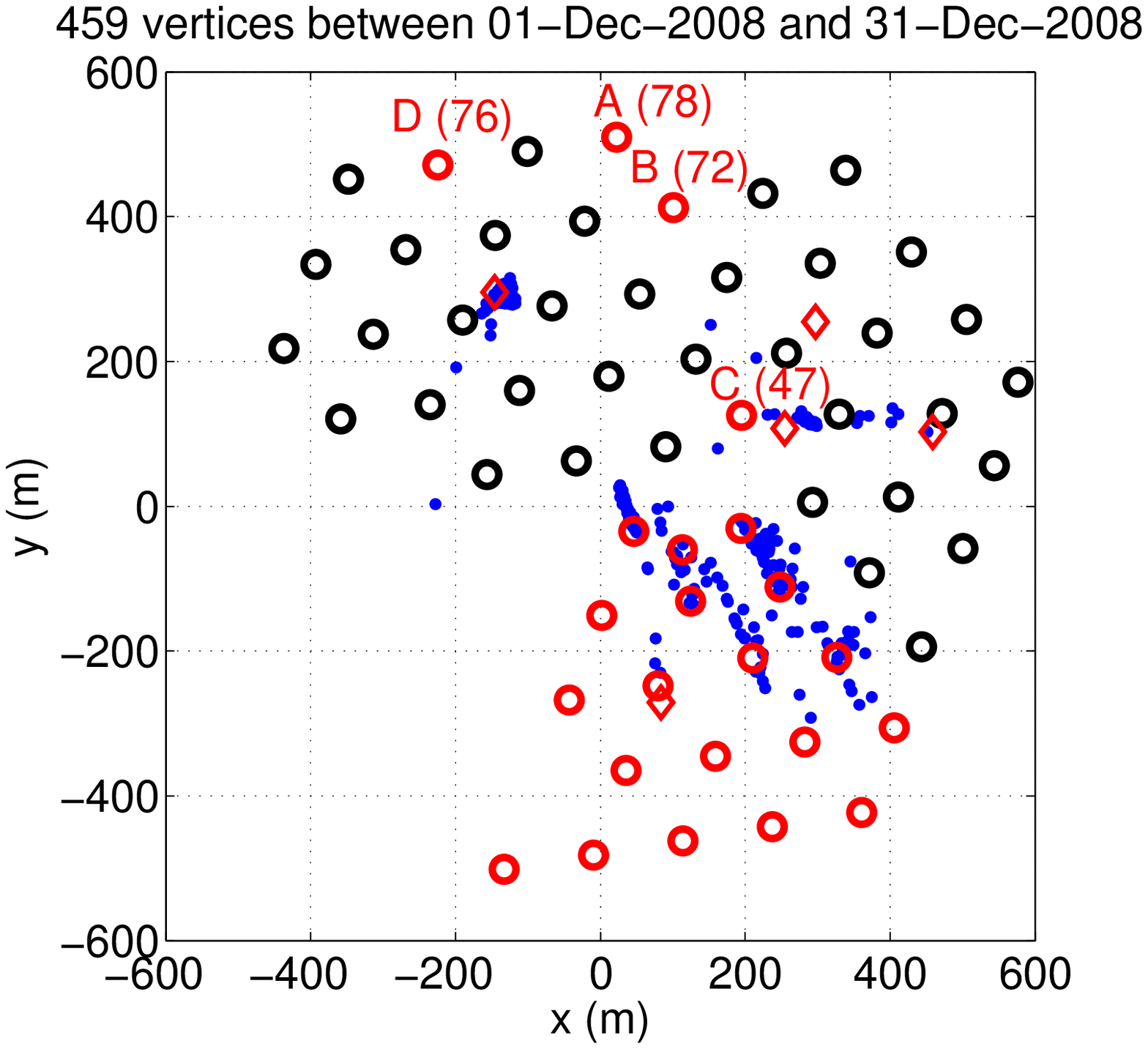}
}
\subfigure[Jan 2009]{
\noindent\includegraphics[width=17pc]{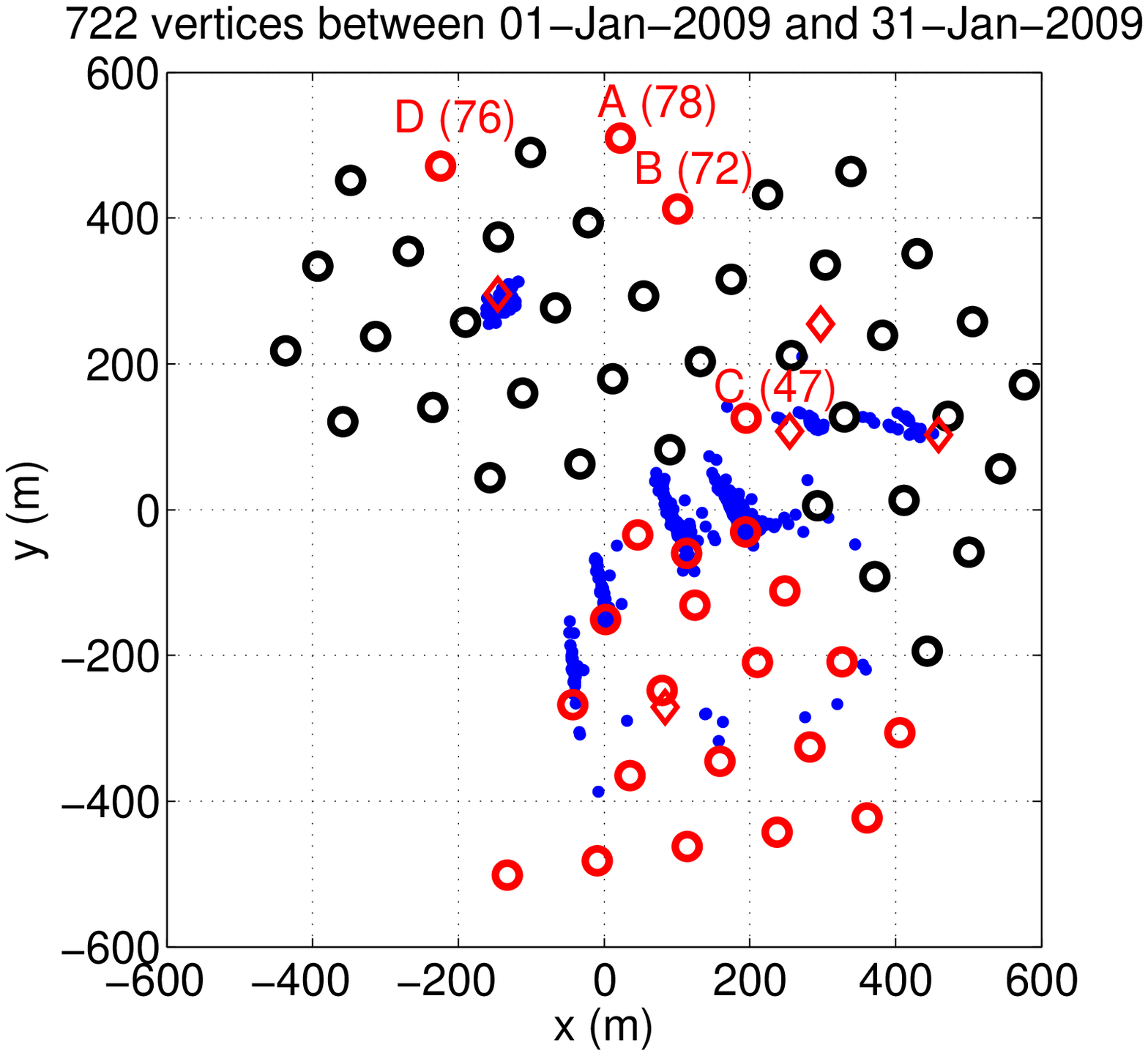}
}
\subfigure[Feb 2009]{
\noindent\includegraphics[width=17pc]{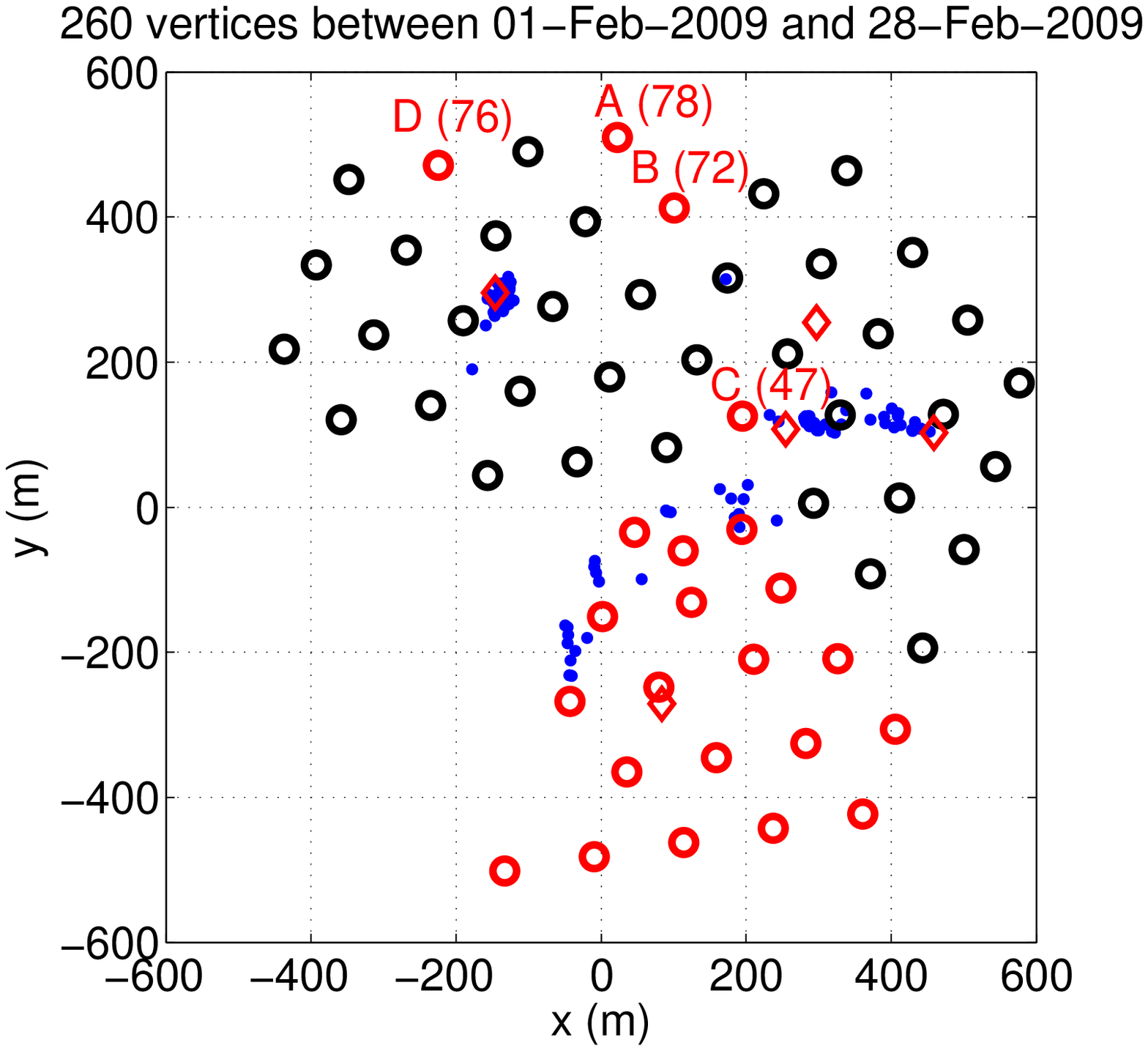}
}
\subfigure[Mar 2009]{
\noindent\includegraphics[width=17pc]{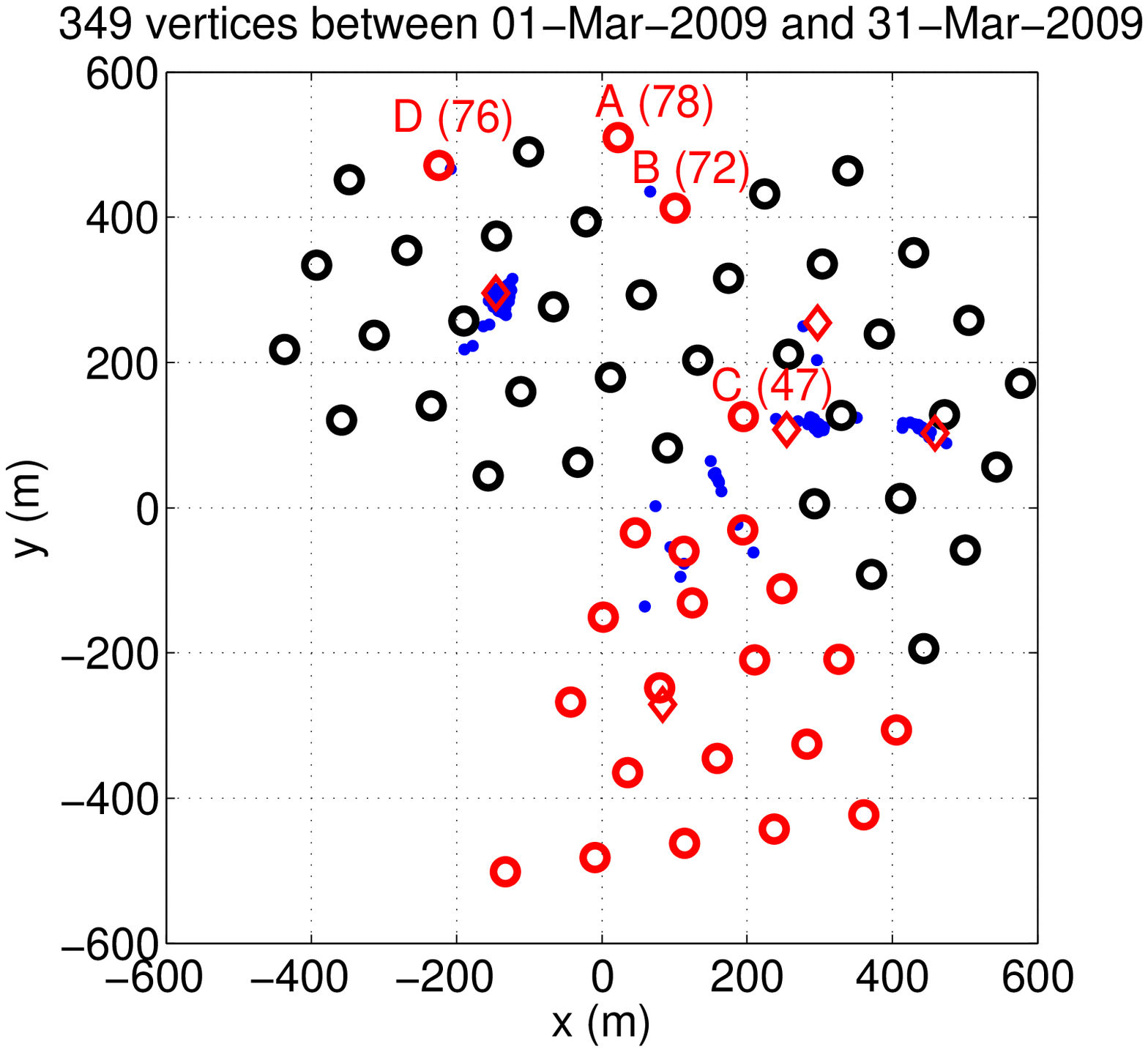}
}
\caption[Transient vertex locations for Dec 2008 - Mar 2009]{Map of reconstructed vertex locations for transient events in December 2008 - March 2009.}
\label{transientsXY2}
\end{center}
\end{figure}

\begin{figure}
\begin{center}
\subfigure[Apr 2008]{
\noindent\includegraphics[width=17pc]{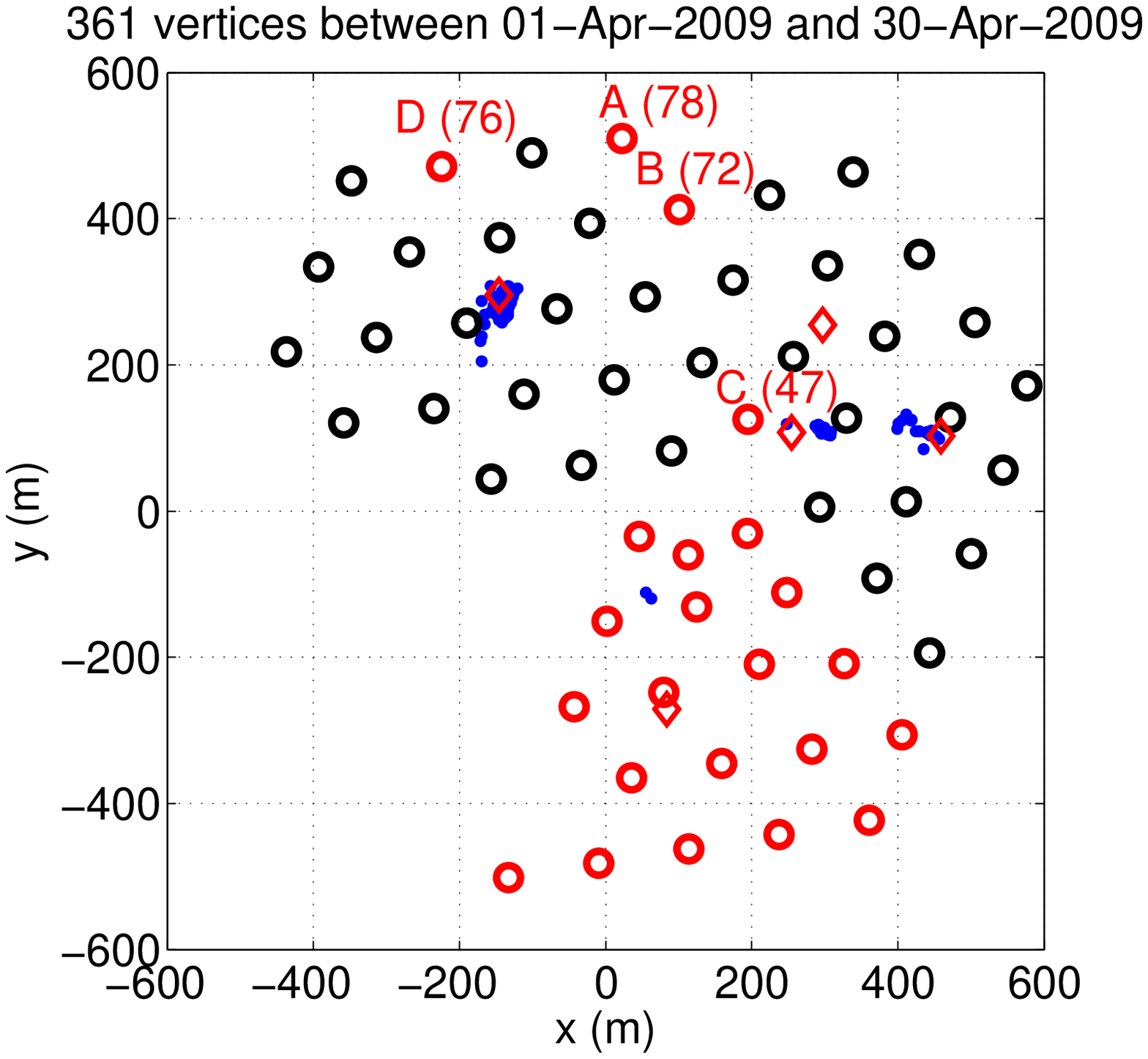}
}
\subfigure[May 2009]{
\noindent\includegraphics[width=17pc]{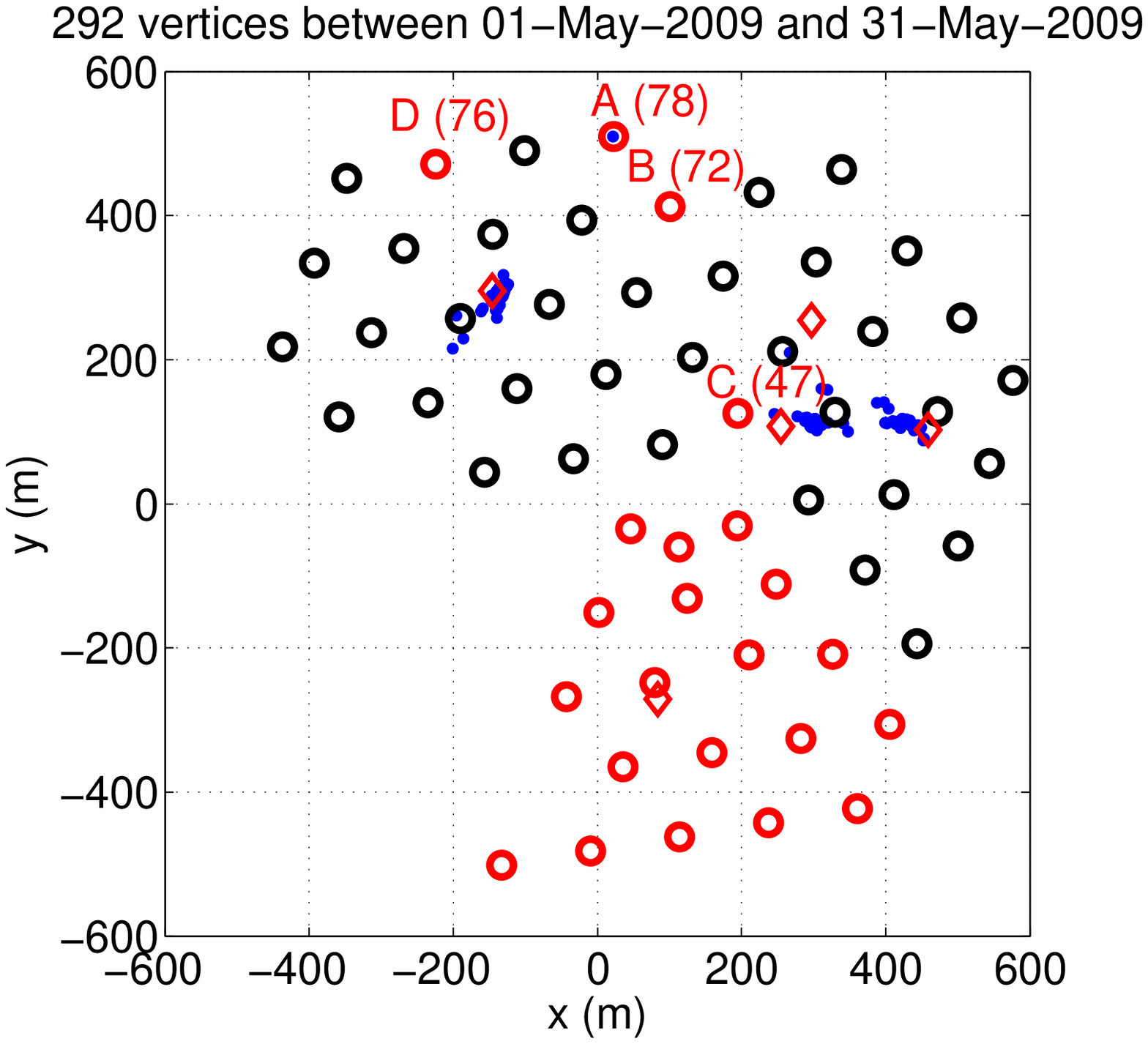}
}
\subfigure[Jun 2009]{
\noindent\includegraphics[width=17pc]{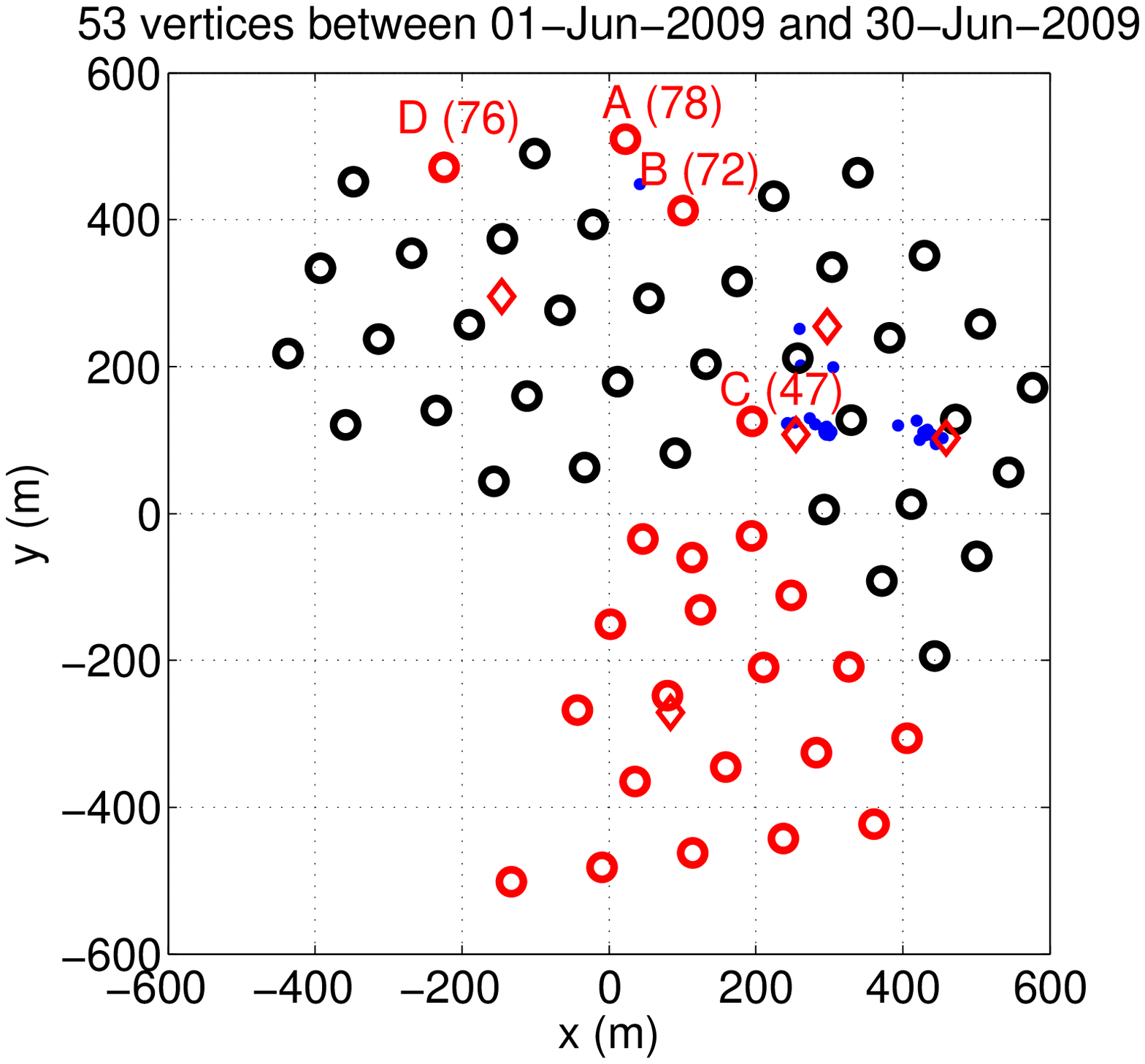}
}
\subfigure[Jul 2009]{
\noindent\includegraphics[width=17pc]{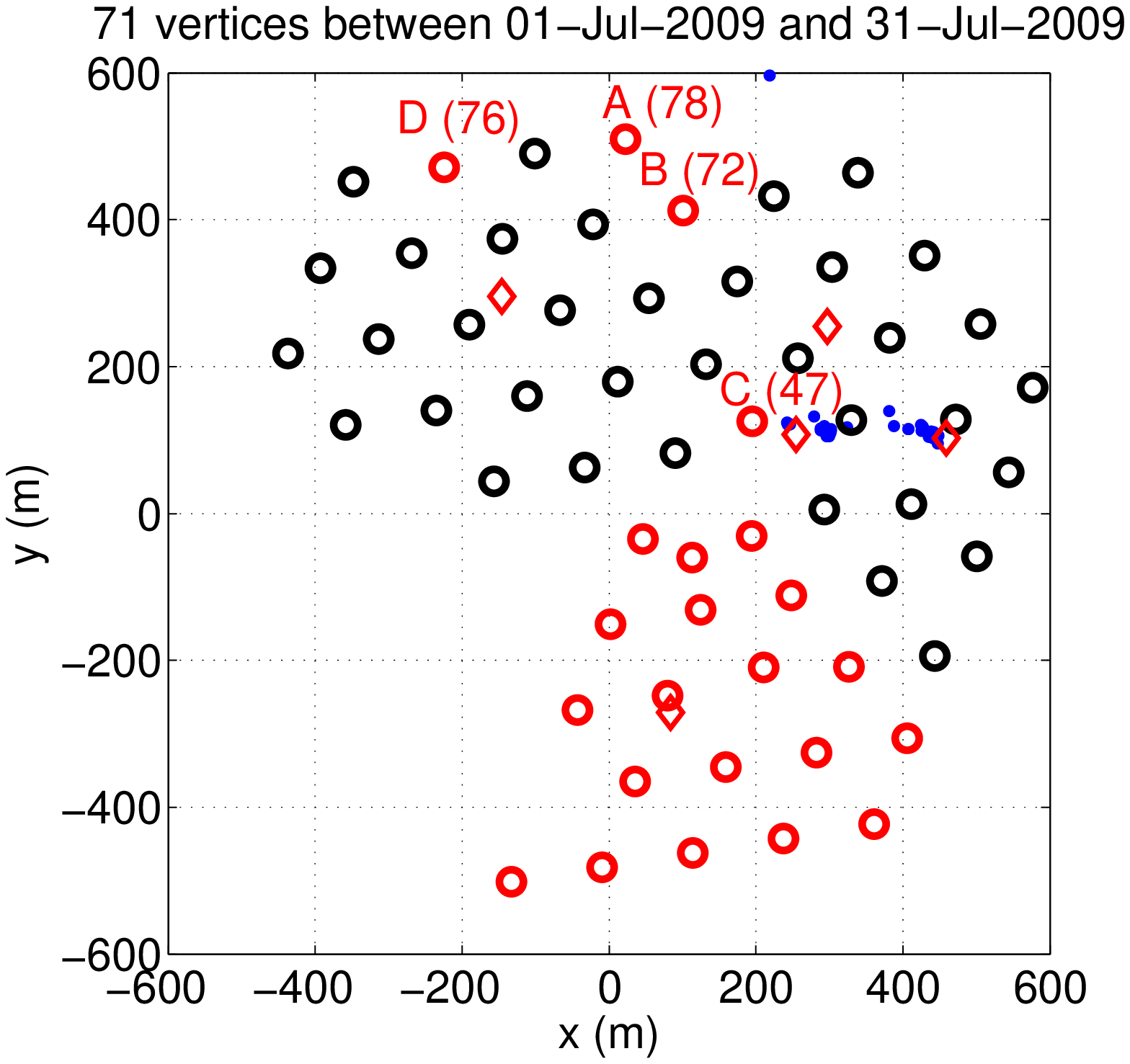}
}
\caption[Transient vertex locations for Apr-Jul 2009]{Map of reconstructed vertex locations for transient events in April-July 2009.}
\label{transientsXY3}
\end{center}
\end{figure}

\begin{figure}
\begin{center}
\subfigure[Aug 2009]{
\noindent\includegraphics[width=17pc]{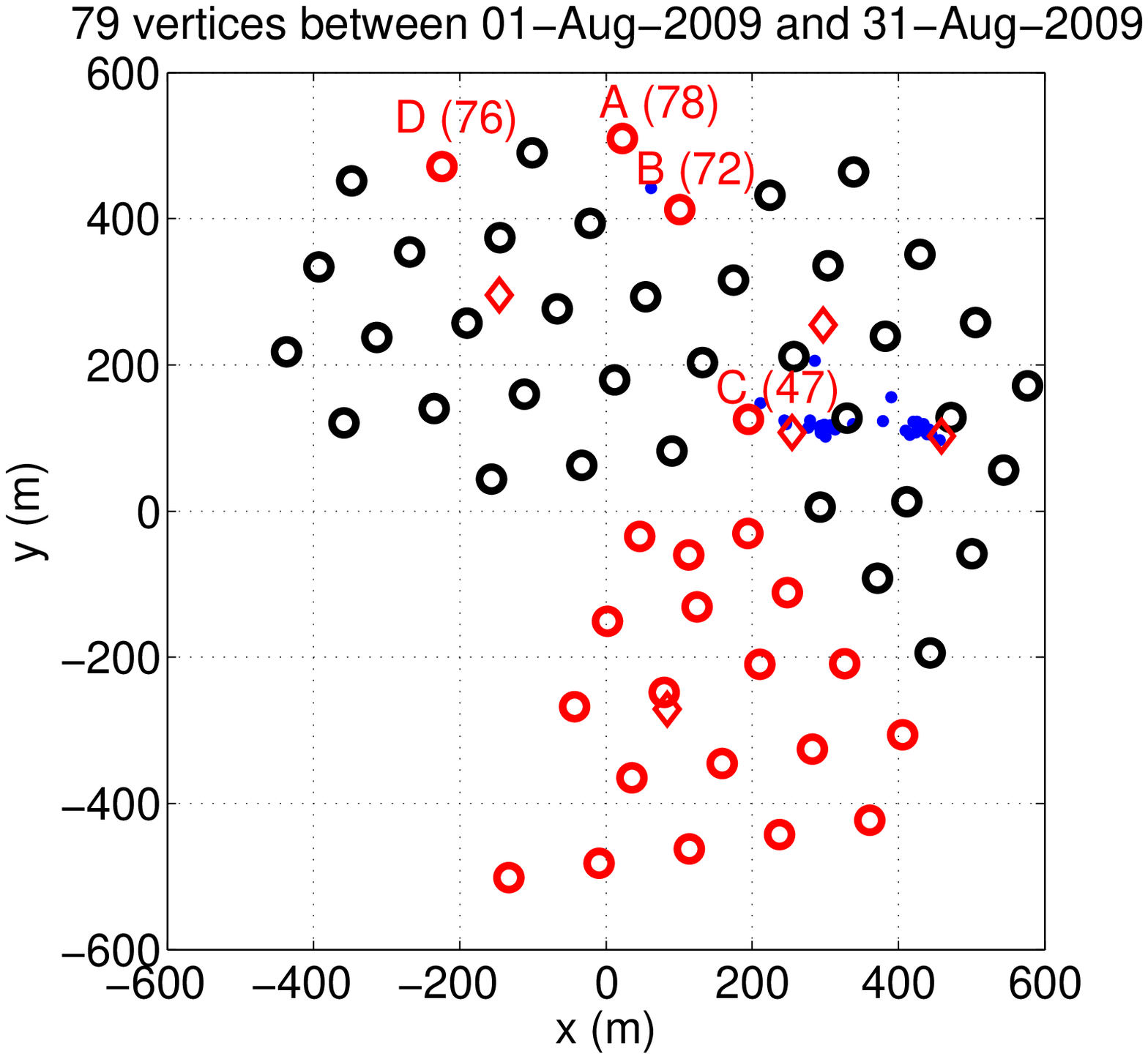}
}
\subfigure[Sep 2009]{
\noindent\includegraphics[width=17pc]{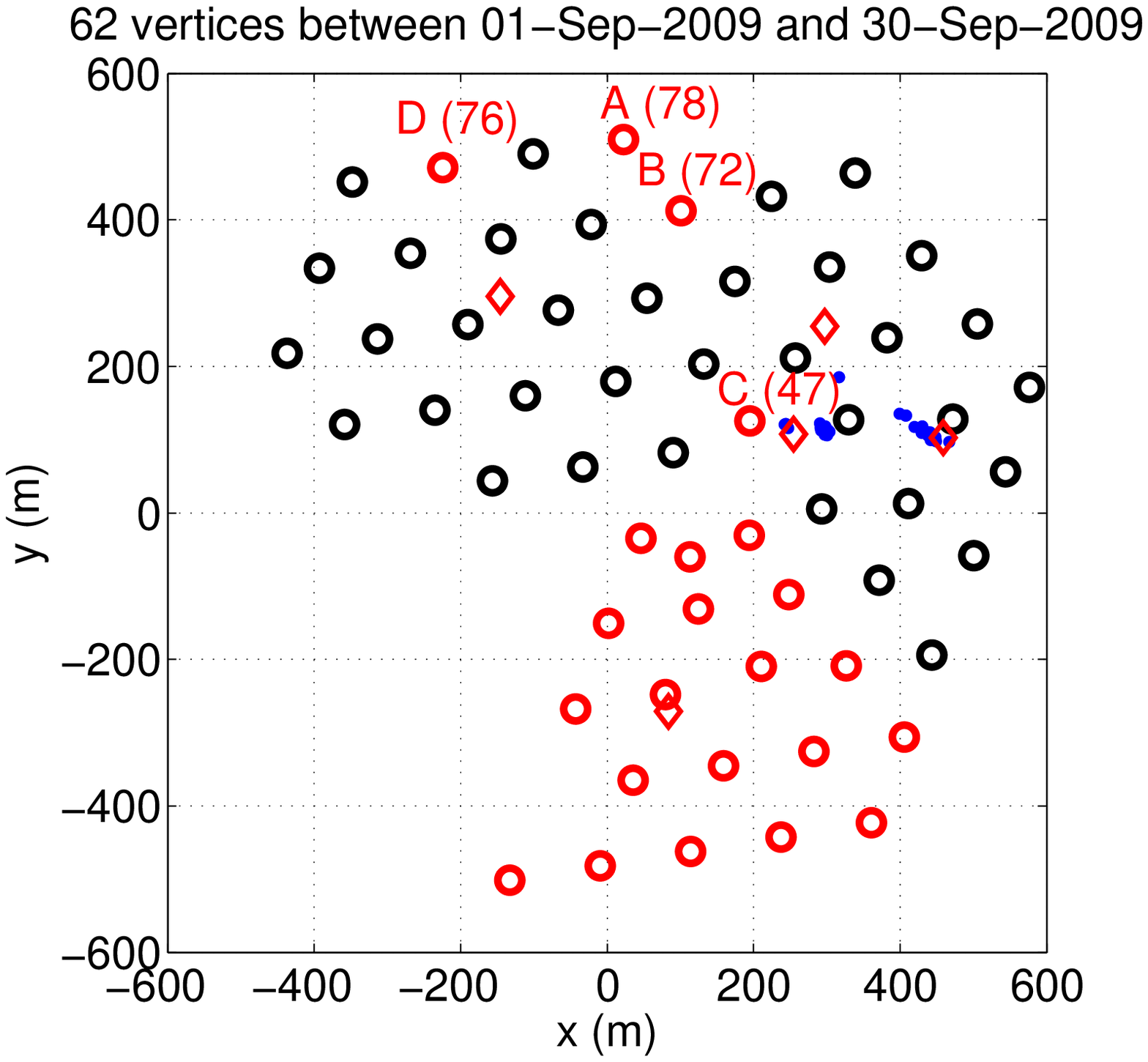}
}
\caption[Transient vertex locations for Aug-Sep 2009]{Map of reconstructed vertex locations for transient events in August-September 2009.}
\label{transientsXY4}
\end{center}
\end{figure}

\begin{figure}[tbp]
\begin{center}
\includegraphics[angle = 0, width = 1\textwidth]{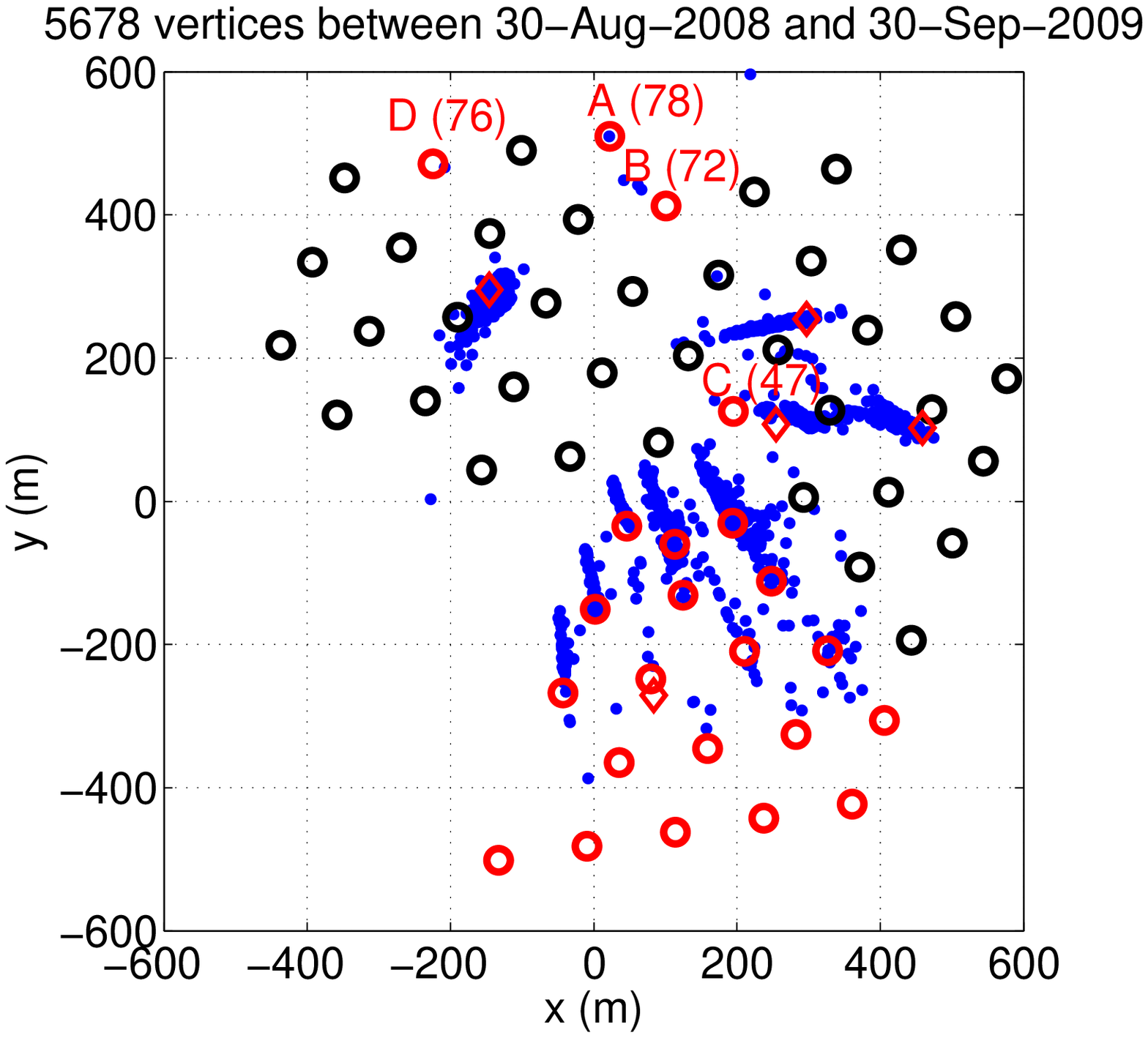}
\end{center}
\caption[Transient vertex locations for entire 14-month run period]{All 5,678 vertices reconstructed over the $\sim$14-month run period.  The five Rod wells are indicated by diamonds, and the 19 young (drilled in the 2008-2009 season)IceCube holes are indicated by red circles.  The Rod wells and young IceCube holes are the sources contributing most of the reconstructed transient events.}
\label{transientsXYAll}
\end{figure}

Figures~\ref{transientsXY1}-\ref{transientsXY4} show the $(x, y)$ distribution of reconstructed transient source locations for each of the 14 months of data taking.  Figure~\ref{transientsXYAll} shows the $(x, y)$ distribution of all transient vertex reconstructions for the entire 14-month run period.  We clearly detect acoustic emission from both AMANDA and IceCube Rod wells, as well as from the IceCube holes themselves.

\subsection{Detection of acoustic emission from AMANDA and IceCube Rodriguez wells}

A Rodriguez (``Rod'') well is typically drilled at the beginning of each IceCube construction season (in November).  A narrow ($\sim$0.6~m diameter) hole is drilled to a depth of $\sim$45~m (beyond the minimum depth necessary for water to pool; above this depth the water leaks out into the firn)\footnote{Thanks to Jeff Cherwinka and Jonas Kalin for providing the Rod well size and shape estimates given in this section.}.  A heater is lowered to the bottom of the well and used to create a cavern of diameter $\sim$15~m, i.e. a bulb at the bottom of the long neck.  This well is used as a source for the heated and circulated water used by the IceCube hot water drill system.  Another well of the same type, located close to the new South Pole Station, is used as the water source for all other purposes at the station.

As the season progresses, the well sinks deeper due to the ice below the well melting as a result of continuous heat input from the submerged heater.  By the end of the season the well is typically $\sim$60-80~m deep.  At this time the well consists of a long, narrow neck capping a large bulb.  For example, the 2007-2008 well shape is estimated to be as follows: neck of diameter $\sim$0.6~m down to a depth of $\sim$45~m, at which point it gradually widens, reaching $\sim$16~m diameter at its deepest point ($\sim$61.5~m).  For each IceCube hole, $\sim$20,000 gallons of water are extracted from the Rod well.  At the end of each season, the bulb has a layer of water at the bottom, with a cavern of air on the top of the bulb and air filling the neck to the surface.

As can be seen in Figures~\ref{transientsXY1}-\ref{transientsXY3}, we clearly detect the IceCube Rod wells as point-like acoustic emitters.  A surprising fact is that we still hear Rod wells several years after they were last heated.  Because the Rod wells feature large volumes of water, they likely take a long time (months or even years) to freeze.  For comparison, the IceTop tanks (water tanks comprising a surface air shower detector component of IceCube) of size $\sim$1~m take several months to freeze.  Moreover, even after the water in the Rod wells has frozen, we expect there to be a significant over-density that could take a long time to relieve and could cause acoustic emission from cracking events.

Rod wells were also used for the AMANDA drill.  The last AMANDA Rod well was last heated during the 1999-2000 season.  This well is located in the middle of the AMANDA array, near the 2004-2006 IceCube Rod well.  Although these two sources are near one another, two individual clusters of acoustic emission vertices can be resolved in the SPATS data.  It is remarkable that not only the oldest IceCube Rod well, last heated in 2006, but also the last AMANDA Rod well, last heated in 2000, is still producing significant acoustic emission after nine years.

While we have detected four of the five Rod wells, some contribute significantly higher event rates than others.  This is likely due to a combination of geometry (distance from SPATS strings) and volume of water left to freeze in the wells.  The properties of each well are as follows:

\begin{enumerate}

\item 1999-2000 (AMANDA) well: The amount of water left in the well at the end of the drilling season was medium to large.
\item 2004-2006 well: A single well was used for both the 2004-2005 and 2005-2006 seasons.  This well is likely deeper than others due to this re-use.  It is expected to be very deep but not very wide, with $\sim$50,000 gallons of water remaining at the end of use.
\item 2006-2007 well: Estimated to have been left with a moderate amount of water, maybe similar to the 2004-2006 well.  This well was relatively deep and narrow.
\item 2007-2008 well: This well was left with a very large amount of water, $\sim$120,000 gallons or more.  This is consistent with the fact that we observe many more acoustic emission events from this well than from other wells.  The well diameter was $\sim$15~$\pm$~2~m.  Six days prior to the end of drilling, the well depth was 67~m and the water surface was at 61.5~m depth.
\item 2008-2009 well: This well was left with very little water, $\sim$10,000-15,000 gallons.  This is consistent with the fact that we have not observed any acoustic events from this well, although the non-observation of acoustic emission could also be due to the large distance of this well from SPATS sensors.  At the end of the season, the diameter of the Rod well bulb (as a function of depth) ranged between 6 and 17~m, with a mean of $\sim$12~m.  The final depth was 75.5~m and the final water surface was at 73~m depth.
\end{enumerate}

\subsection{Detection of acoustic emission from IceCube holes}

In addition to the Rod wells, we clearly detected acoustic emission from many (more than half) of the IceCube holes that were drilled in the 2008-2009 IceCube construction season.  Unlike the Rod wells, which have emitted over the many months we have been running our transients data acquisition, emission from each IceCube hole started typically five days after drilling of the hole, and decreased after $\sim$two weeks.  This indicates that, like the Rod wells, we hear the re-freeze process in the IceCube holes.  The IceCube holes are much narrower than the bulbs in the IceCube holes so it is not surprising that their emission occurs over a shorter time period.

For the most part we hear transient events from the IceCube holes during their re-freeze phase, and do not hear transient emission from them while they are being drilled.  Instead, we hear the IceCube drill itself, as it goes down and up in each hole, as a Gaussian noise level increase, but generally we do not hear associated transient events.  An exception to this is that for a single hole we have seen several transient events whose vertex reconstruction corresponds in space and time with the IceCube drill.

\subsection{``Smearing'' reconstruction artifact due to not including refraction in reconstruction}

As can be seen in the transient vertex distributions, each source is smeared along a line relative to the point-like (in $(x, y)$) source expected.  This is an artifact of not accounting for refraction through the firn in the source reconstruction.  Although the sensors being read out are deeper than the firn, much of the detected emission originates in the firn.  In particular, the Rod wells are only $\sim$70~m deep, where the sound speed is significantly smaller than in the deep bulk ice.  A more accurate reconstruction algorithm would account for refraction through the firn and is expected to remove the smearing artifact.

Our hypothesis that the smearing is an artifact due to not including refraction has been confirmed by Jens Berdermann (at DESY Zeuthen), who simulated vertical lines of sources (point-like in $(x, y)$ but with a range of $z$ values) with and without including refraction to determine ``true'' Monte Carlo propagation times.  In each case an algorithm assuming zero refraction was used to reconstruct the sources.  In the case with no refraction included in the true times, point-like $(x, y)$ reconstructed source distributions were achieved.  But in the case where refraction was included in the calculation of the true travel times but was not included in the reconstruction algorithm, the smearing effect seen in real South Pole data was reproduced.

This smearing effect is therefore expected to be removed by applying a reconstruction algorithm that accounts for refraction.  Such an algorithm was successfully applied in analysis of SAUND data~\cite{Vandenbroucke05} and should be applied to SPATS data.  Similar to the South Pole data, we observed strange distributions in the SAUND data when refraction was not included, whereas the distributions were reasonable when refraction was included.

\section{Detection of shear waves from transient events}

We have detected shear waves from some of the background transient events.  A search was performed by first reconstructing the vertices as described above (using the first hit per module).  Because the coincidence algorithm automatically extends the time window as long as there are still hits within 200~ms of the last hit, shear waves are included in the coincidence cluster determination as long as the source is within a few hundred m of the sensor.  Because we only include the first hit per module, any detected shear waves are neglected in the currently used reconstruction algorithm.  But we can search for shear wave hits after reconstruction by plotting arrival time vs. distance from reconstructed vertex, for all hits in a coincidence cluster.

\begin{figure}[tbp]
\begin{center}
\includegraphics[angle = 0, width = 0.6\textwidth]{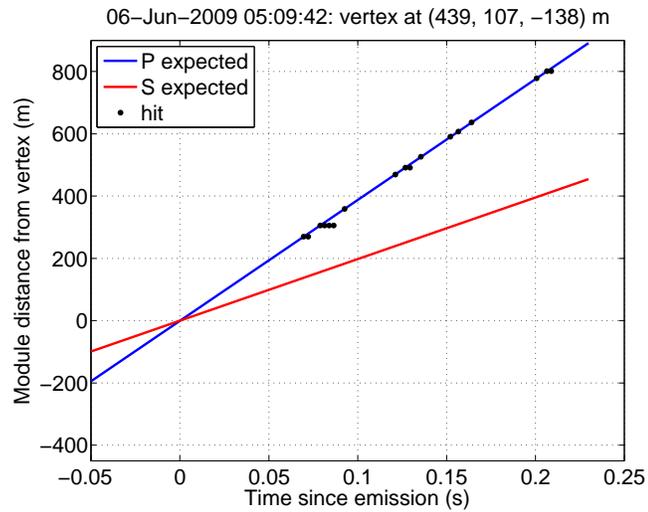}
\end{center}
\caption[Transient event without shear pulse detected on any channel]{Distance to reconstructed vertex, versus time since reconstructed emission, for hits of a single coincidence cluster.  This event has no shear wave pulse detected.}
\label{shear0}
\end{figure}

\begin{figure}[tbp]
\begin{center}
\includegraphics[angle = 0, width = 0.6\textwidth]{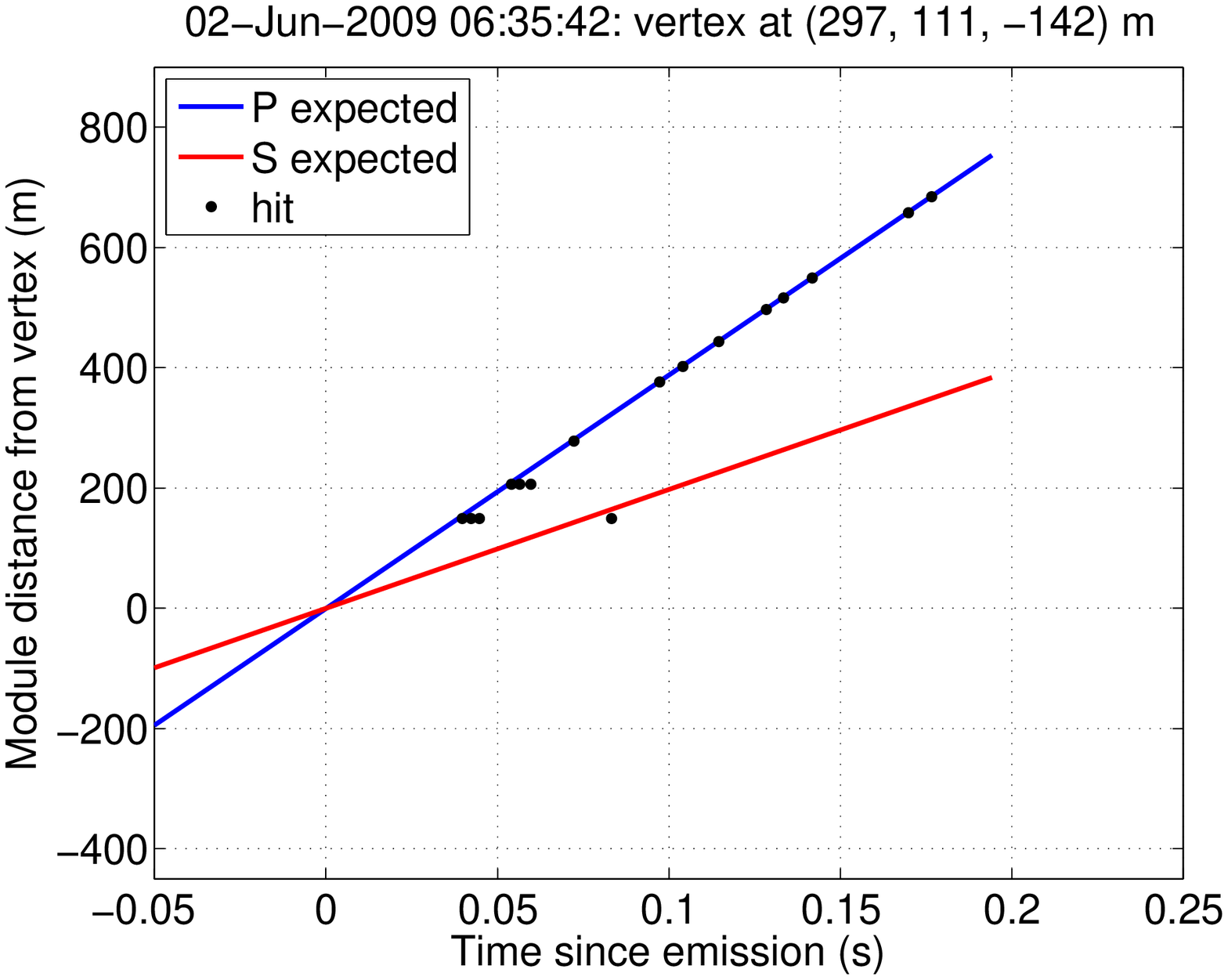}
\end{center}
\caption[Transient event with shear pulse detected on one channel]{Distance to reconstructed vertex, versus time since reconstructed emission, for hits of a single coincidence cluster.  This event has a shear pulse detected on one channel.}
\label{shear1}
\end{figure}

\begin{figure}[tbp]
\begin{center}
\includegraphics[angle = 0, width = 0.6\textwidth]{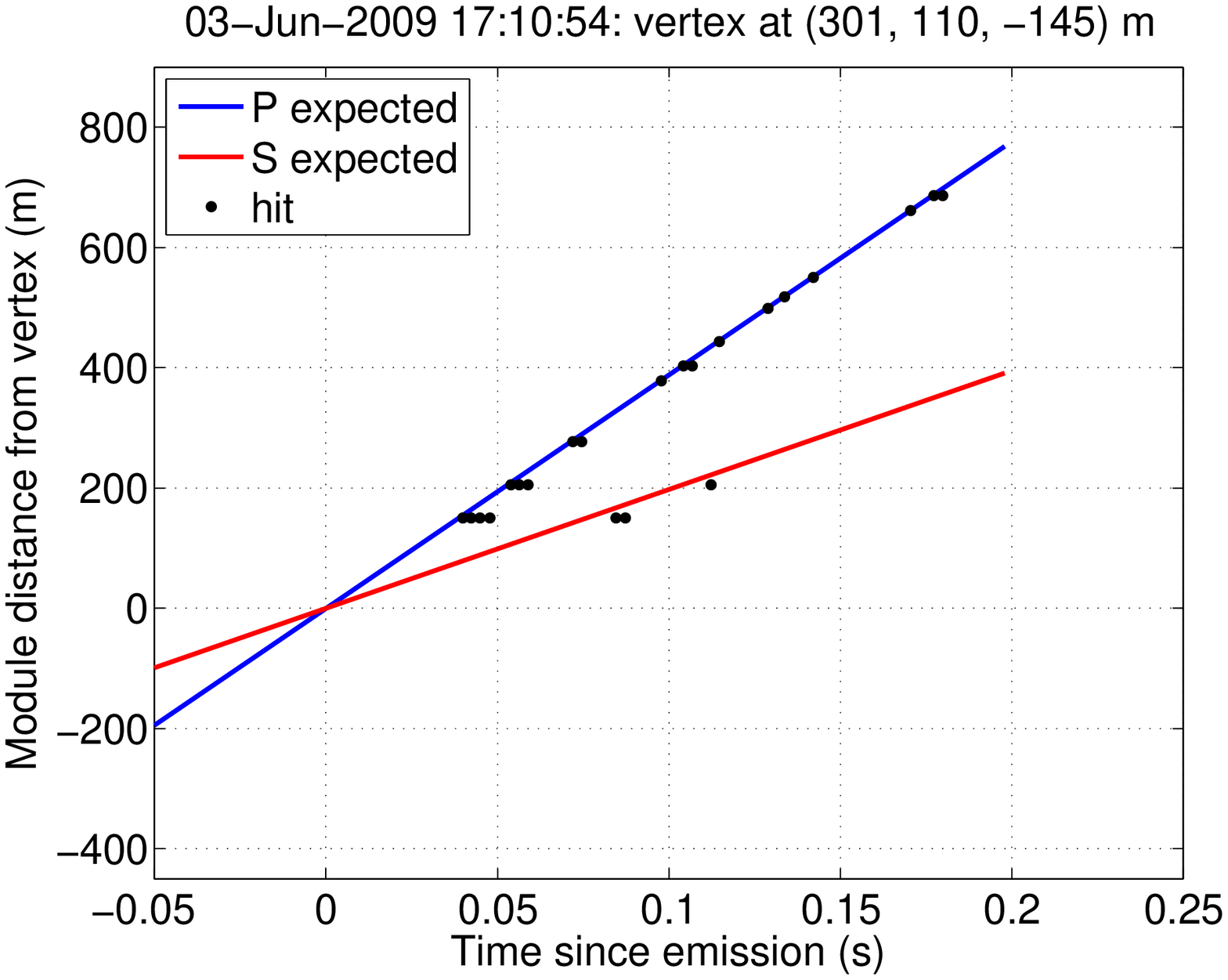}
\end{center}
\caption[Transient event with shear pulse detected on two channels]{Distance to reconstructed vertex, versus time since reconstructed emission, for hits of a single coincidence cluster.  This event has a shear pulse detected on two channels.}
\label{shear2}
\end{figure}

\begin{figure}[tbp]
\begin{center}
\includegraphics[angle = 0, width = 0.6\textwidth]{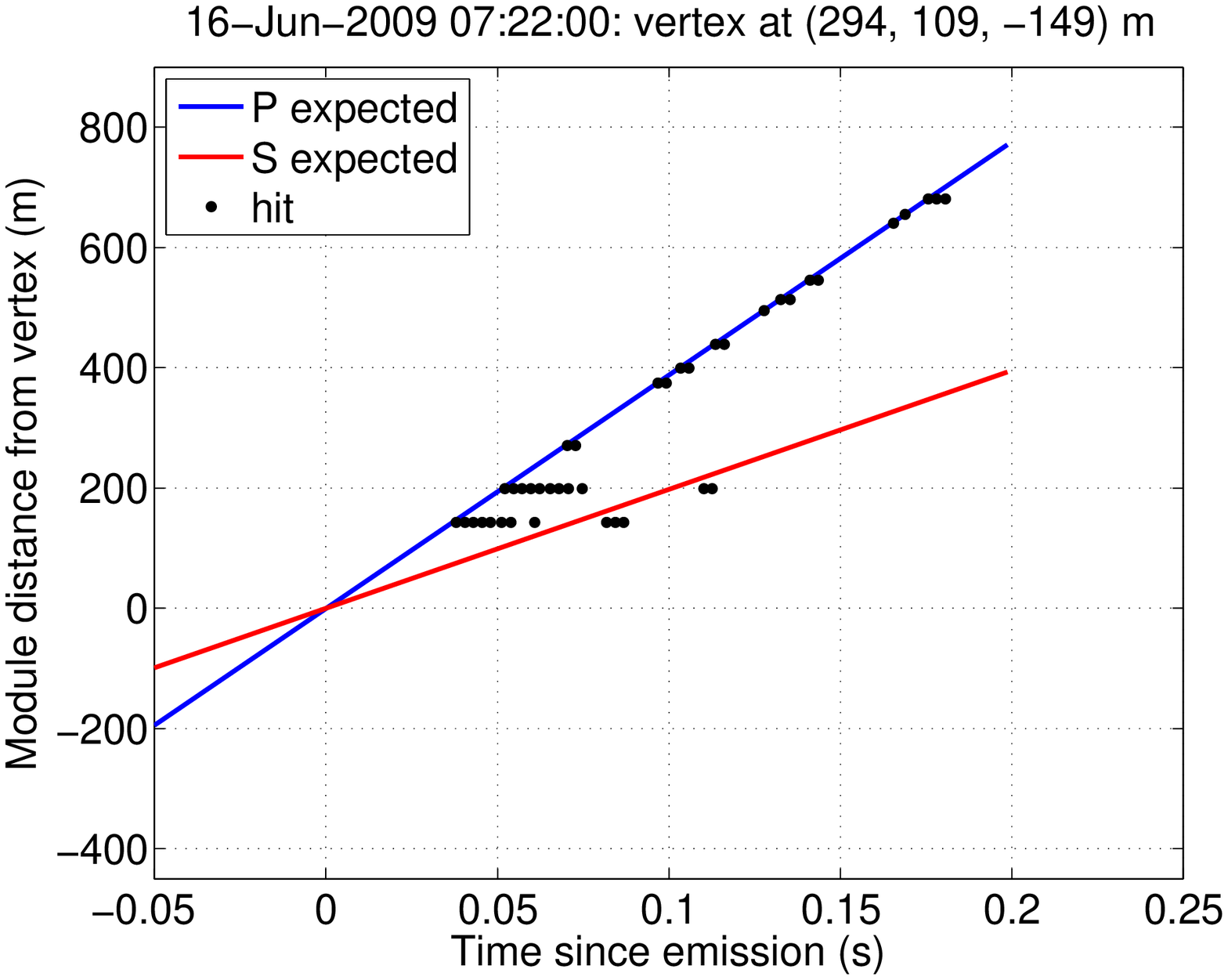}
\end{center}
\caption[Transient event with ambiguous shear pulse detection]{Distance to reconstructed vertex, versus time since reconstructed emission, for hits of a single coincidence cluster.  This is an example of ambiguous shear pulse determination due to the long duration of the pressure wave emission.}
\label{shearAmbiguous}
\end{figure}

Examples of such plots are shown in Figures~\ref{shear0}-\ref{shearAmbiguous}.  In each plot, pressure waves are expected to lie on the steeper line, and shear wave pulses are expected to lie on the shallower line.  For this study we focus on coincidence events from June 2009 with at least 11 channels hit (out of 12 being read out).  Most such events come from the region of the 2004-2006 Rod well.  The figures demonstrate that the detected events lie in four classes: shear detected on 0, 1, or 2 channels, or ambiguous shear detection due to long-duration pressure emission which extends beyond the expected start time for shear emission.

Shear wave hits can be used to improve source position and time reconstruction.  They can also be used to improve event characterization for distinguishing neutrino signals from background signals.  For example, if the thermoacoustic effect can be proven to not produce shear waves, then the presence of shear emission in an event identifies it as background.

\section{Expected threshold improvement with online coincidence}

Many detector arrays operate with multi-channel coincidence required online.  For example, both the SAUND and the HESS experiments have performed upgrades from offline to online coincidence.  The rate of mutli-channel coincidence is lower than the rate of single-channel events, so a lower threshold can be achieved for a fixed data rate.  Another benefit of online coincidence is that the whole array (including non-triggering channels) can be read out when a trigger occurs.  If the rate of noise triggers in the coincidence time window is small, then requiring multiple channels to be hit in this time window can provide a significant threshold improvement.  However, if the rate of noise triggers in the coincidence window is large, then requiring coincidence does not provide a significant benefit.

Such an upgrade, from offline to online coincidence, could be applied to the SPATS DAQ.  However, we show here that such an upgrade would only achieve a small threshold improvement (15\%).

We assume that such an online coincidence trigger would require 3 strings to be hit in a time window of 200 ms.  We also assume that whenever a trigger occurs, we read out all 12 channels currently being read out for transient data acquisition.  We also assume that the same number of samples (1001) would be captured per channel, and that the amount of disk space required per event per channel would be the same as it is currently for offline coincidence.

If a single channel on string $S$ exceeds threshold (call this hit $h$), we require that at least one channel on each of at least two other strings (the two strings must be other than $S$) also trigger within 200 ms after hit $h$.  Let $c$ be the probability that any particular channel triggers in a 200 ms time window.  For a sampling frequency $f$ and time window $t$, there are $N = f t$ samples in the time window.  In our case, $f =$~200~kHz and $t =$~200~ms, so $N =$~4 x 10$^4$.  Assume that the probability of any particular sample triggering is $p$.  In Section~\ref{rateSection} we calculated $p$ as a function of the trigger threshold for Gaussian noise.  The probability of a particular channel triggering in a time window $t$ is then 

\begin{equation}
c = 1-(1-p)^N.
\end{equation}

For $s =$~1.99e-7 and $N =$~4 x 10$^4$, $c =$~7.94 x 10$^{-3}$.  This is the probability that a particular channel triggers in our 200~ms time window.  Now let $P$ be the probability that at least one channel out of three on a string trigger in the time window.  It is given by

\begin{equation}
P = 1-(1-c)^3.
\end{equation}.

For our current settings, this is $P =$~2.36 x 10$^{-2}$.  This is the probability that a particular string is hit in our time window.  Now we require that at least two strings other than $S$ are hit.  This is given by

\begin{equation}
G = 3 P^2 - 2 P^3
\end{equation}

For our numbers, this gives $G =$~1.65 x 10$^{-3}$.  Any time a particular channel exceeds threshold, $G$ is the probability that we meet the global online coincidence trigger requirement.  This whole calculation assumes Gaussian noise triggers dominate.

Now assume we read out all 12 channels any time the global trigger is satisfied.  This means that for the current DAQ settings, switching from offline to online coincidence would lower the data rate by a factor of 12$G =$~1.98 x 10$^{-2}$.  Alternatively, for a fixed data rate, we can use this factor of $\sim$50 to lower our threshold.  This would raise the single-channel trigger rate by a factor of 50 while maintaining the same total data rate.  Switching from a threshold of 5.2$\sigma$ to 4.42$\sigma$ would give the desired factor of 50 increase in single-channel trigger rate.  This is an improvement of $1-4.42/5.2 = 15\%$.
\chapter{Measurement of acoustic attenuation length}

\label{attenuationChapter}

\noindent\emph{In this chapter we present our measurement of the acoustic attenuation length.  We summarize results obtained from another analysis using pinger data, then give details of an analysis using inter-string data.  The conclusion from these two analyses, as well as others, is that the attenuation is more than an order of magnitude larger than expected from theoretical estimates.}

\section{Inventory of data available for attenuation measurement}

\subsection{Inter-string}

This was the methodology for which SPATS was originally designed.  SPATS frozen-in transmitters are recorded with frozen-in sensors on different strings.  With the first three strings, the available baselines were insufficient to overcome the large systematic uncertainties, which motivated construction and deployment of String D.  With String D, along with improvements in data taking and processing (especially clock drift correction), inter-string signals can be detected between many transmitter-sensor combinations, and an attenuation measurement can be made.  Data selection cuts are described below, but to summarize: considering only horizontal propagation paths, high-quality inter-string recordings are available at depths of 100~m, 140~m, 190~m, 250~m, 320~m, 400~m.  Unfortunately no data are available for the 80~m level, because it is controlled by a different ADC board from the others, for which we have not (yet) developed the necessary DAQ software.  We describe the inter-string attenuation analysis in detail below.

\subsection{Pinger}

In the 2007-2008 pinger data taking campaign, data were taken for all 9 instrumented SPATS depths.  Unfortunately the pinger was swinging/bouncing/twisting significantly, and recordings by the same channel of the pinger operating in different IceCube holes produced irreproducible waveforms.  This was improved significantly with the addition of centralizers for the 2008-2009 pinger data taking.  The 2008-2009 pinger data have allowed our best determination of the attenuation length, and the data are much better for this than the 2007-2008 data.  However for this data taking campaign we focused on our best five instrumented depths: 190~m, 250~m, 320~m, 400~m, and 500~m.  Therefore we only have good information about the depth dependence of the attenuation in this depth range.  It may be possible to extract a rough estimate from the 2007-2008 data, and a pinger campaign is planned for the 2009-2010 season to pin down the depth and frequency dependence of the attenuation.  The pinger attenuation results are described in detail in~\cite{Tosi09} and~\cite{attenuationPaper}.

\subsection{Gaussian and transient noise sources}

The IceCube drill noise is heard as an increase in the Gaussian noise level as the drill approaches a given SPATS sensor channel in depth.  The noise then decreases as the drill proceeds deeper.  The drill is heard again, and recedes again, as the drill passes the depth of the channel again on the way back up.  Assuming the emitted drill noise is isotropic, these data can be used to estimate the attenuation length.

Impulsive transients can also be used to estimate the attenuation length.  For example, if a single transient event is observed by several sensors separated by hundreds of meters, the transient source location can be reconstructed and studying the amplitude vs. distance can then be used to constrain the attenuation length.  An alternative method is to study the distribution of sources in space: for example if the distribution of background events were homogeneous, and a cutoff in the observed rate were observed as a function of distance from the SPATS sensors, this could also be used to constrain the attenuation length.

Both of these analyses are in progress.

\section{Pinger attenuation measurement}

We have used several methods to estimate the amount by which ice attenuates acoustic signals.  We analyzed each of the following types of data to estimate the attenuation length: Gaussian noise from the IceCube drill, inter-string data, pinger data, transients data.  I focus on the results from the inter-string analysis, which has been my personal analysis focus.

Parallel analyses were performed with the retrievable pinger, and are described in~\cite{attenuationPaper,Tosi09}.  The most successful pinger analysis used the data from the 4 pinger deployments of the 2008-2009 season.  The 6 pinger deployments of the 2007-2008 season were not as useful.  The 2008-2009 data were of significantly higher quality for two reasons: (1) the repetition rate was increased from 1~Hz to 10~Hz (except for one of the 2008-2009 pinger deployments, for which the repetition rate was accidentally set to 8~Hz), which increased the signal-to-noise ratio by $\sqrt{10}$ after waveform averaging; and (2) centralizing ribs were added to the pinger assembly to prevent the swinging/bouncing/twisting motion that caused significant pulse-to-pulse variation in the 2007-2008 data.

The acoustic attenuation length was measured with better than 30\% precision using the following method, details of which are given in~\cite{Tosi09}.  For each sensor channel, pinger runs recorded at the same level in all 4 different pinger deployment holes were considered.  For each, the pressure wave energy and effective amplitude were determined.  Then the attenuation length $\lambda$, as well as an overall normalization factor $b$, was fit for each sensor channel independently.  The resulting independent estimates of $\lambda$ were consistent with one another, and systematic effects were dramatically reduced because only one sensor channel was used for each fit and because the 4 pinger holes lay roughly along a single line of sight as seen from the SPATS strings, such that azimuthal effects in both the sensors and pinger were reduced.  The $b$ quantity fit for each channel constitutes an \emph{in situ} estimate of its relative sensitivity.

\section{Inter-string attenuation measurement}

\subsection{Data set}

Several different inter-string data sets were acquired with different parameters.  Several iterations in data taking and analysis were taken over a two-year period as we improved both data taking and analysis techniques.  Here I focus on the most recent (and most useful) data set.

For this most recent data set, two data acquisition software upgrades were performed.  First, the software was upgraded so that only one clock per string was used for all data taking, including both transmitter pulsing and sensor recording.  Although each ADC/DAC board has its own clock, a ``SyncBus'' connection allows the clock of any board to drive acquisition on another board or boards on the same String PC.  We took advantage of this to use the first board on each string to control all boards on the string, so that acquisition among the different boards (corresponding to different sensor channels) was synchronous and only one clock's drift had to be monitored on each string.

The second upgrade was to record one channel of the sensor that is on the same stage as the firing transmitter.  In addition to providing a useful ``intra-stage'' recording of the transmitter signal at very close ($<$1~m) range, this includes a recording of the IRIG-B GPS signal in order to monitor the clock drift on the transmitting string.

Data were recorded for each transmitter recorded by each sensor in the SPATS array.  In each run, one transmitter was pulsed with a 25~Hz repetition period for 40~s.  All inter-string transmitter-sensor combinations were collected over a two-day period (April 1-2, 2009).  The transmitters are known to require several seconds to reach steady pulse-to-pulse performance, likely due to self-heating of the electronics during initial operation followed by temperature equilibration.  The pulse amplitude is initially somewhat larger than in the steady state and then decays to the steady-state amplitude with a time constant of $\sim$2~s.  To be sure we are recording in the steady state, we start the sensor recording 11~s after the transmitter begins pulsing.  In each run, all three channels of a single sensor module are recorded continuously for 20~s, enough time to record 500 transmitter pulses.  The three channels are sampled synchronously, at 200 kilosamples per second on each channel.  

\subsection{Selection of transmitter-sensor combinations}

We designed SPATS with many transmitters and sensors, for two purposes: reliability and redundancy.  We did not know how much of the instrumentation would perform sufficiently well to be used for different analyses, particularly for the attenuation analysis which we knew would be most important.  Furthermore, we did not know which systematic effects would be important.  At first we attempted to combine all possible inter-string combinations from a transmitter on one string to a sensor on another string.  Results of this first analysis are reported in~\cite{Boeser07}.  However, this analysis method suffers from the following effect: Given an arbitrary transmitter on one string and sensor on a different string, those for which the transmitter and sensor are at different depths (``diagonal'' paths) tend to be longer than those for which the transmitter and sensor are at the same depth (``horizontal'' paths).  Stated another way, for inter-string combinations in the SPATS array geometry, both the transmitter emission angle and the sensor arrival angle are correlated with the transmitter-sensor distance.  The transmitters are known to emit more in their equatorial plane that at diagonal angles, and the sensors are known to be more sensitive in the horizontal plane than for diagonal arrival angles.  Taken together, these facts mean that the signal recorded for diagonal paths (which tend to be longer than horizontal paths) are systematically smaller than for horizontal paths, due to the angular response of both transmitters and sensors.  Because of the systematic correlation between path length and emission or arrival angle, measuring amplitude vs. distance is degenerate with measuring amplitude vs. zenith angle at sensor or transmitter.  Instead of mapping out the attenuation of the length, we are mapping out the angular response of our sensors and transmitters if we include diagonal as well as horizontal paths.

Because of this effect, we chose not to include diagonal transmitter-sensor paths and instead focused on single-depth, horizontal paths.  

This analysis uses a single transmitter recorded by all sensor channels at the same depth as the transmitter.  A single transmitter is used because the different transmitters are known to have different inherent transmittivity and perhaps different coupling to the ice with respect to one another.  For each transmitter we perform a complete attenuation analysis using all sensor channels as the same depth as the transmitter.  We then repeat this analysis for each transmitter (allowing transmittivity to be a free overall normalization constant in the fits, different from transmitter module to module) and finally combine all attenuation results from all transmitters.

\subfiglabelskip=0pt		
\begin{figure}
\begin{center}
\subfigure[][]{
\label{driftA}
\noindent\includegraphics[width=16pc]{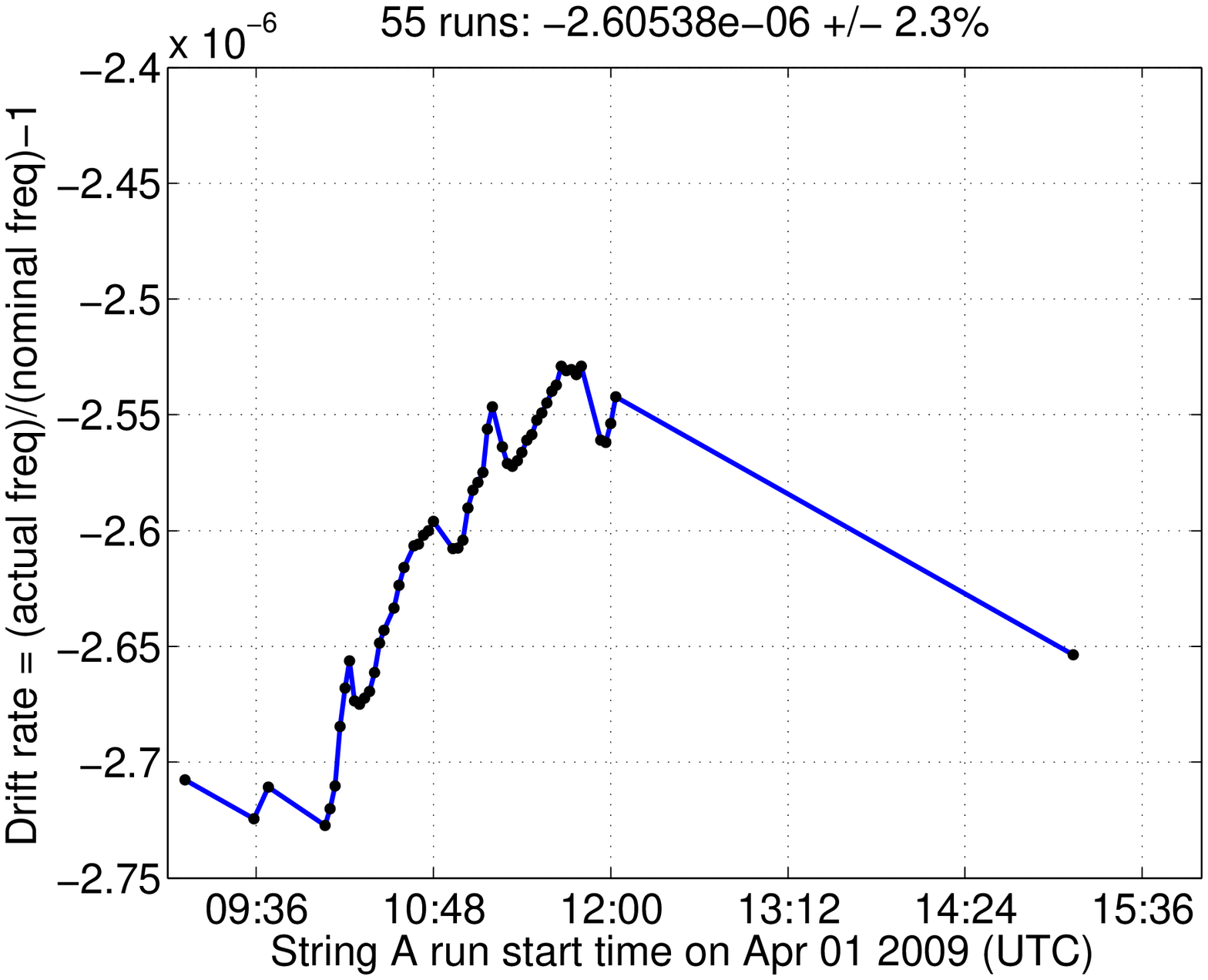}
}
\subfigure[][]{
\label{driftB}
\noindent\includegraphics[width=16pc]{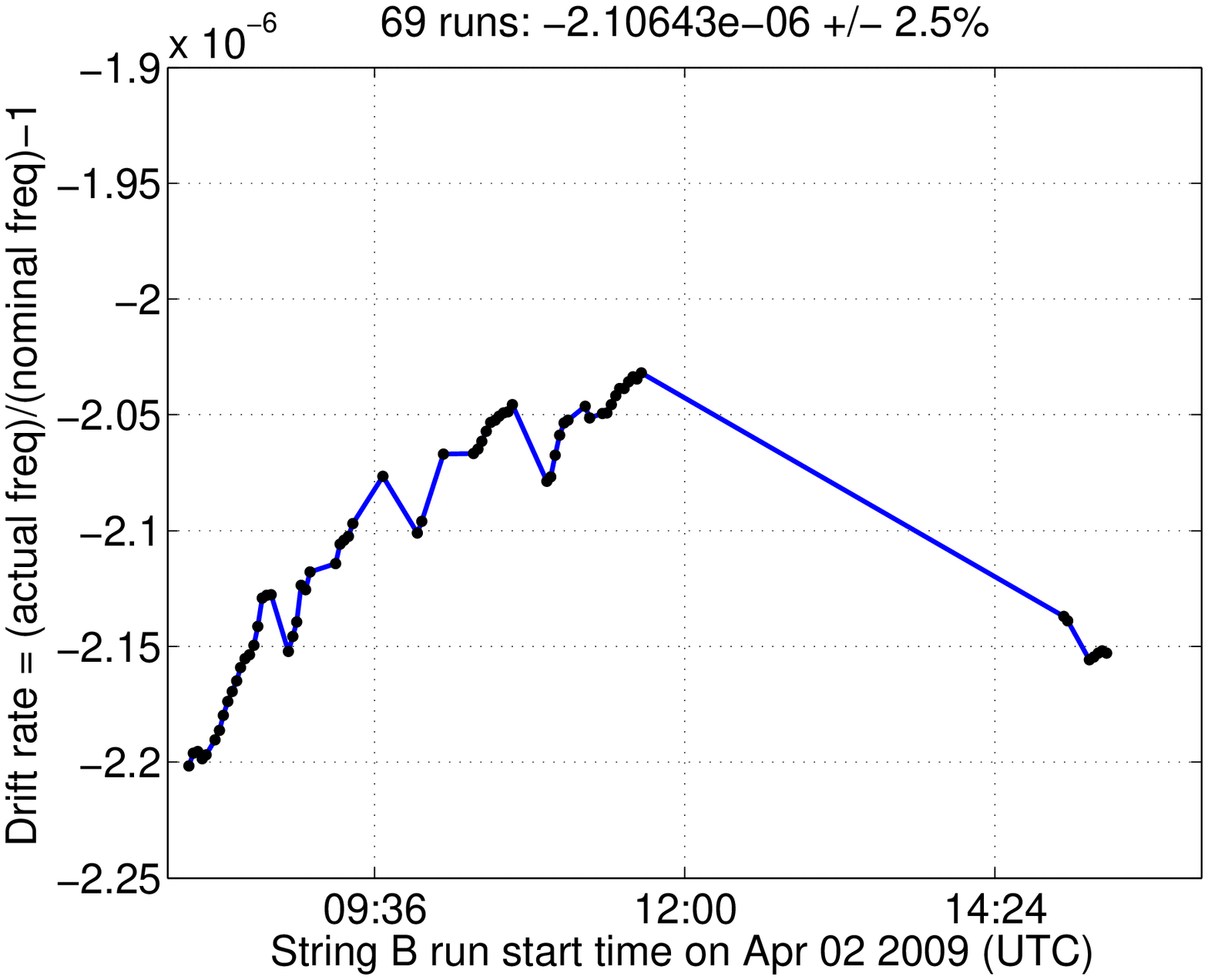}
}
\subfigure[][]{
\hspace{8pt}
\label{driftC}
\noindent\includegraphics[width=16pc]{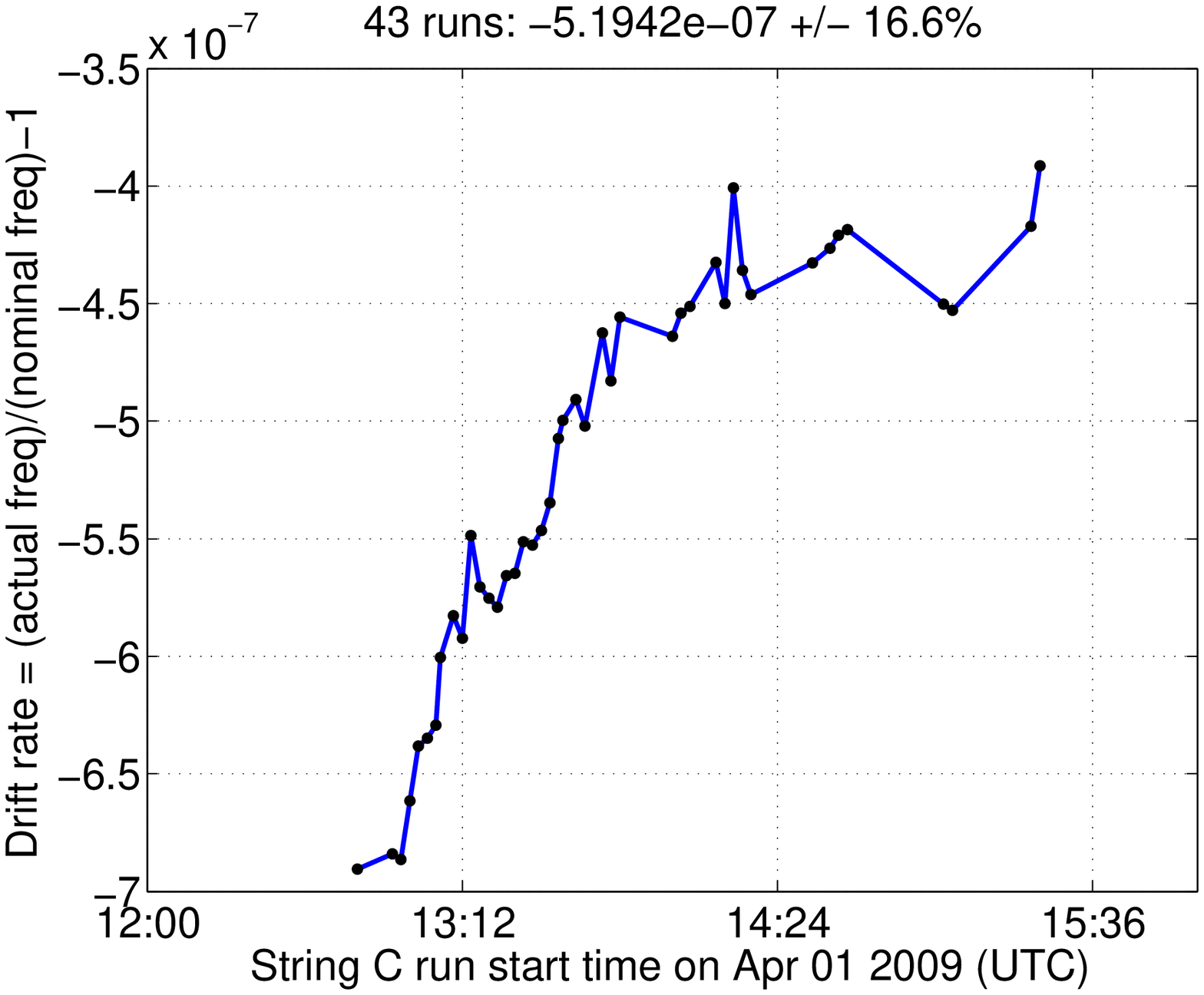}
}
\subfigure[][]{
\label{driftD}
\noindent\includegraphics[width=16pc]{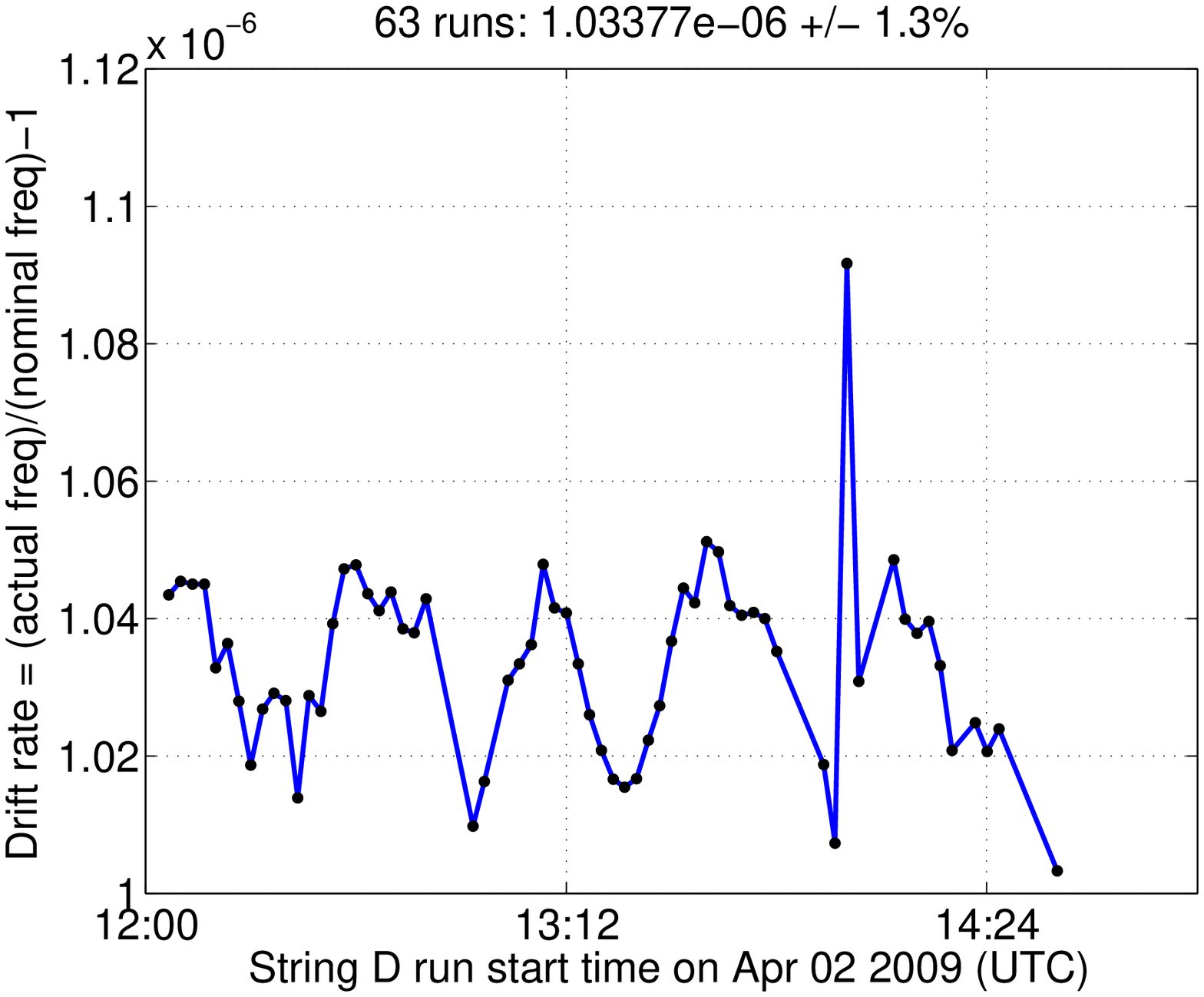}
}
\caption[Clock drift rate for each string during inter-string data acquisition]{Drift rate vs. time for each of the 4 clocks (one per string) in the inter-string data-taking campaign of April 1-2, 2009 (UTC).  The drift rate was determined every time the transmitter was run.  Over the two-day data taking period, the drift rate on each string was stable at the few-percent level, with the exception of String C.  On String C the drift rate was small, so the absolute variation of the rate was small even though the relative variation was larger than a few percent (16.6\%).  For each string we take the mean of the drift rate measurements as an estimate of the drift rate for the entire data-taking period and use it for the inter-string attenuation analysis.}
\label{transmitterDrift}
\end{center}
\end{figure}

\subsection{Sensor and transmitter clock drift correction}

For each sensor channel, the 500 consecutive pulses are averaged together to improve the signal-to-noise ratio.  Each string uses a single clock to drive both its analog-to-digital converters (ADC's) and its digital-to-analog converters (DAC's).  The ADC's are used to record the sensor waveforms, and the DAC's are used to pulse the transmitters.  The clocks drift at a rate that is typically several parts per million, or tens of $\mu$s over the course of the 20~s waveform recording.  This cumulative amount of drift is on the order of one signal oscillation period and therefore can cause severe decoherence in the pulse averaging if the nominal rather than true sampling frequency is used.  This clock drift effect is corrected on the sensor side by using the IRIG-B GPS signal (which is recorded synchronously with each sensor channel recording) to determine the actual sampling frequency at the time of the recording.  This actual sampling frequency (which differs from the nominal sampling frequency by several parts per million) is used in the pulse averaging.

We also must correct for the clock drift on the transmitting string.  The drift causes the actual transmitter repetition rate to be different from the nominal repetition rate by a few parts per million.  Similar to the effect on the sensing string, this effect would cause decoherence by an amount on the order of a signal oscillation period, resulting in destructive interference in the pulse averaging, if not accounted for.  The actual repetition rate was measured for each run using IRIG-B GPS signals recorded on the transmitting string while it was transmitting, by using the sensor recorded at the same stage as the transmitter and applying the same drift determination algorithm as used for the sensing string.  The true repetition rate varied from string to string but was constant on each string at the few-percent level over the course of the two-day data taking campaign, as shown in Figure~\ref{transmitterDrift}.  Therefore the mean value of the transmitter repetition rate for each string was used in the pulse averaging.

\subsection{Waveform averaging and signal pulse stability}

\subfiglabelskip=0pt		
\begin{figure}
\begin{center}
\subfigure[][]{
\label{40ms}
\noindent\includegraphics[width=17pc]{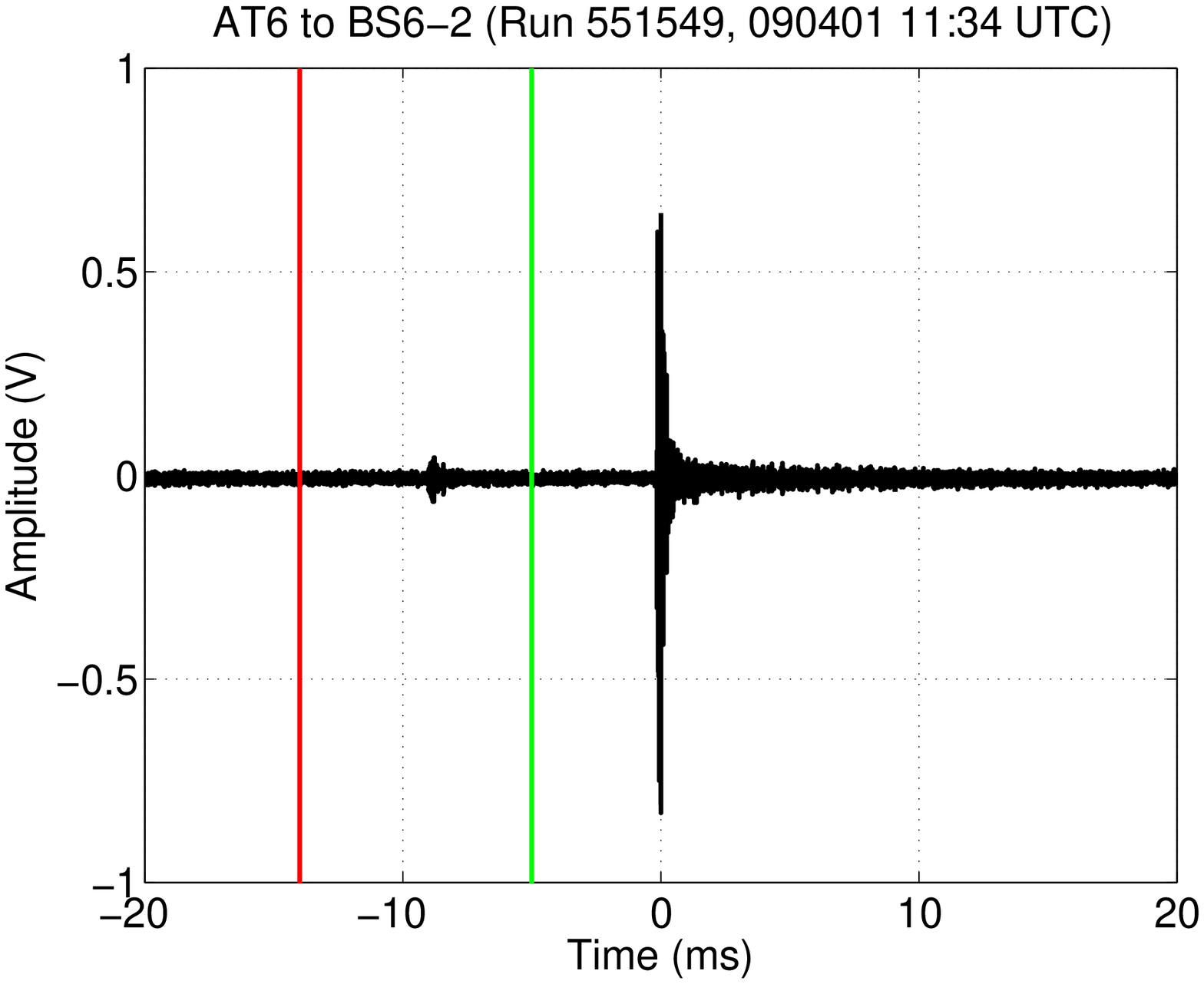}
}
\subfigure[][]{
\label{4ms}
\noindent\includegraphics[width=17pc]{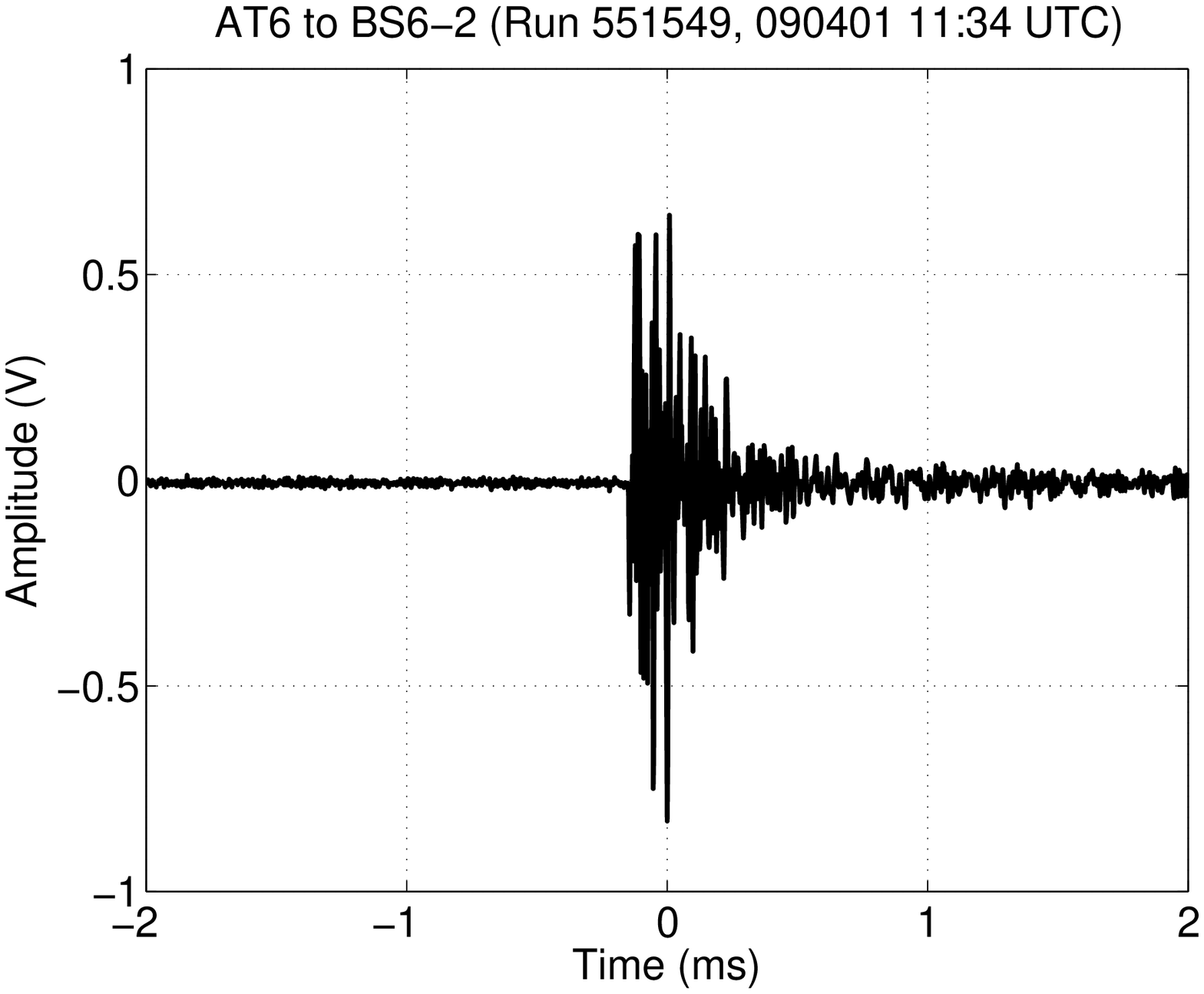}
}
\subfigure[][]{
\hspace{8pt}
\label{400us}
\noindent\includegraphics[width=17pc]{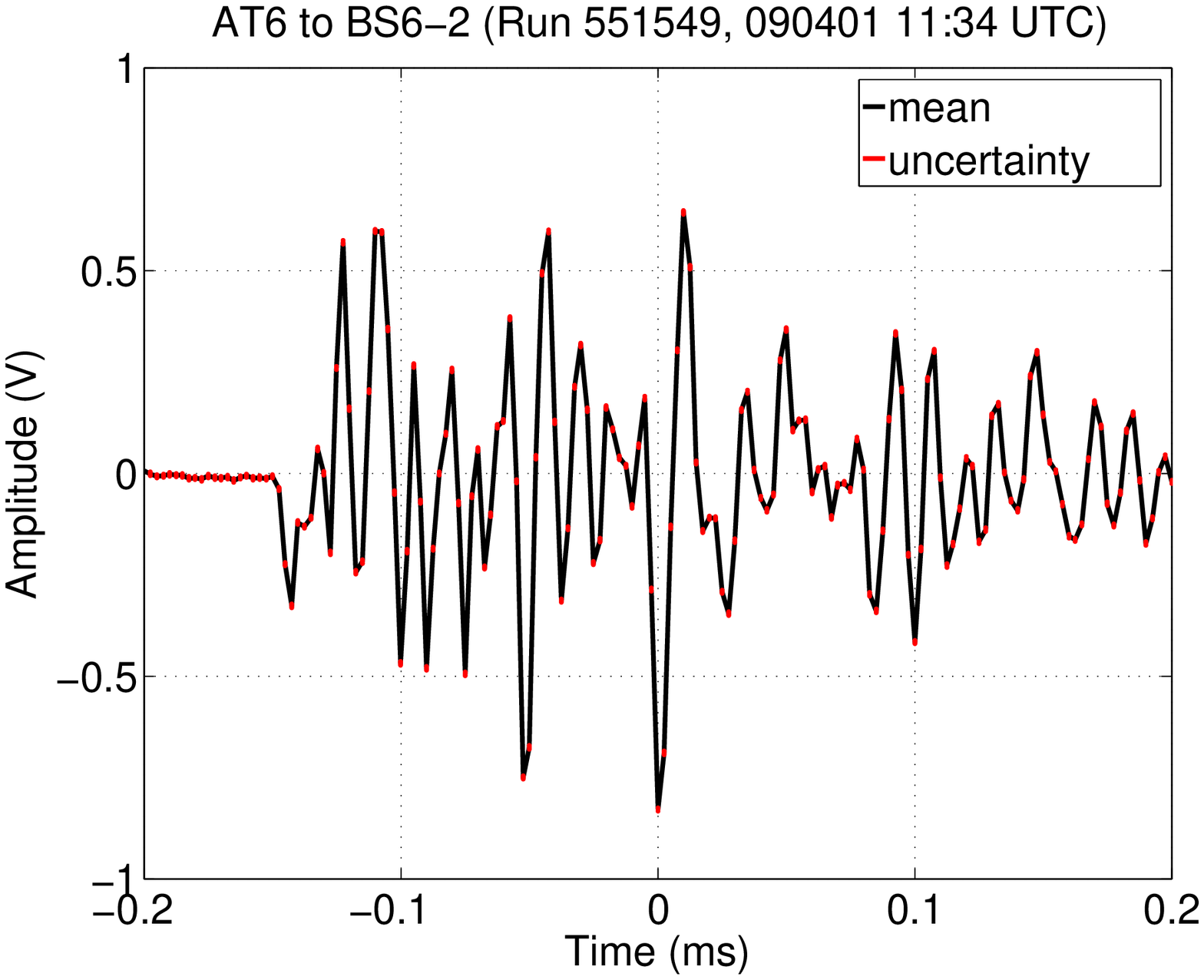}
}
\caption[Example inter-string waveform]{Example averaged inter-string waveform, for AT6 transmitting and BS6-2 (at a distance of 125~m) recording.  The time axis is centered with the maximum-amplitude sample at $t=0$, and time is wrapped modulo the repetition rate (roughly 40~ms; the precise actual period accounting for clock drift is used).  The pulse at $t=0$ is the pressure wave, and the smaller pulse is the shear wave.  In this example a shear wave is visible at a time of 31.0~ms after the pressure wave (equivalent to 9.0~ms before the pressure wave, modulo 40~ms), as expected.  The time window used for integration of the pressure wave signal is indicated with vertical lines.  It begins 5~ms before the maximum-amplitude sample and ends 26~ms after the maximum-amplitude sample.  Figures~\subref{40ms},~\subref{4ms}, and~\subref{400us} show the same waveform with time windows of 40~ms, 4~ms, and 400~$\mu$s respectively.  Only Figure~\subref{400us} shows error bars on each sample, visible as very small red vertical lines.}
\label{waveform_example}
\end{center}
\end{figure}

\subfiglabelskip=0pt		
\begin{figure}
\begin{center}
\subfigure[][]{
\label{runA}
\noindent\includegraphics[width=16pc]{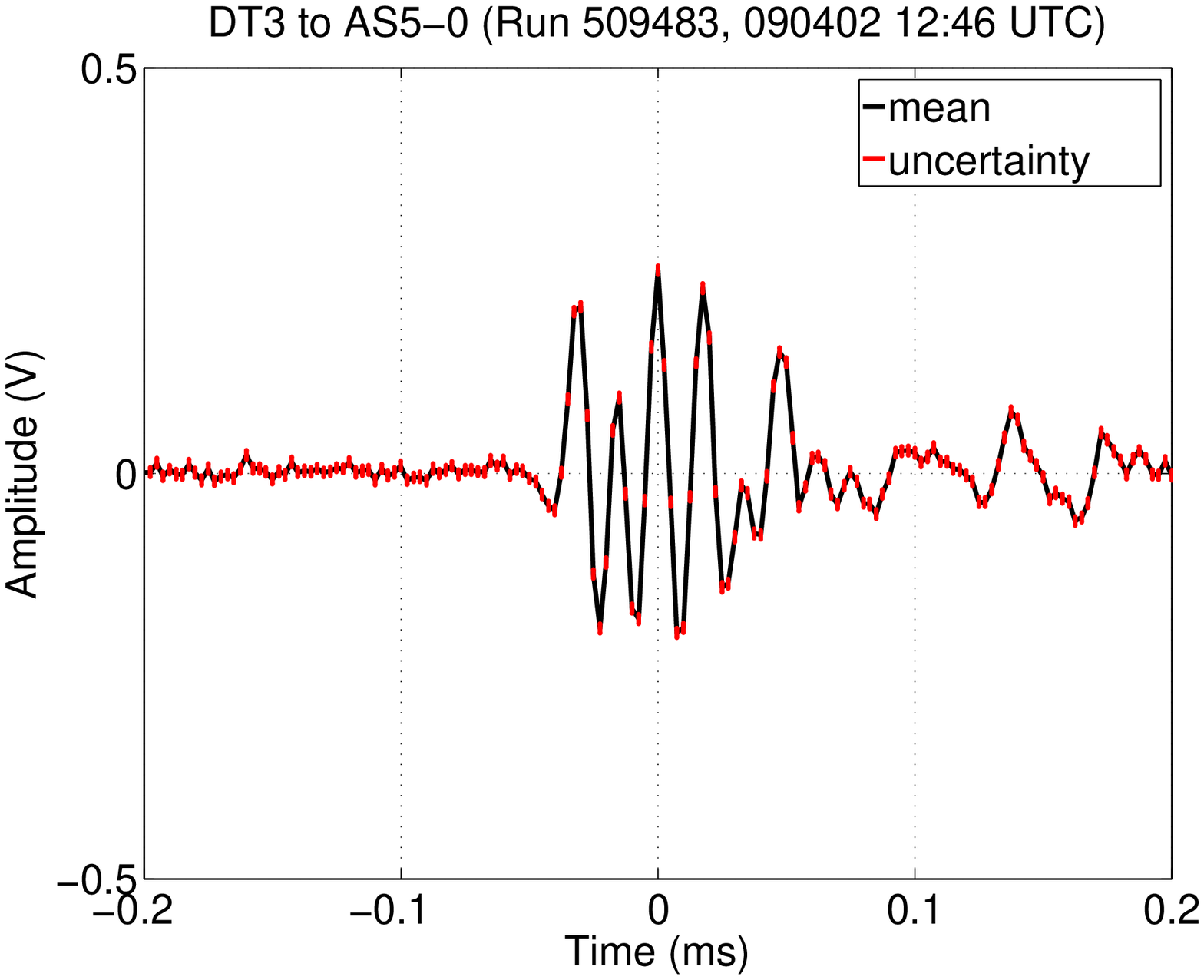}
}
\subfigure[][]{
\label{runB}
\noindent\includegraphics[width=16pc]{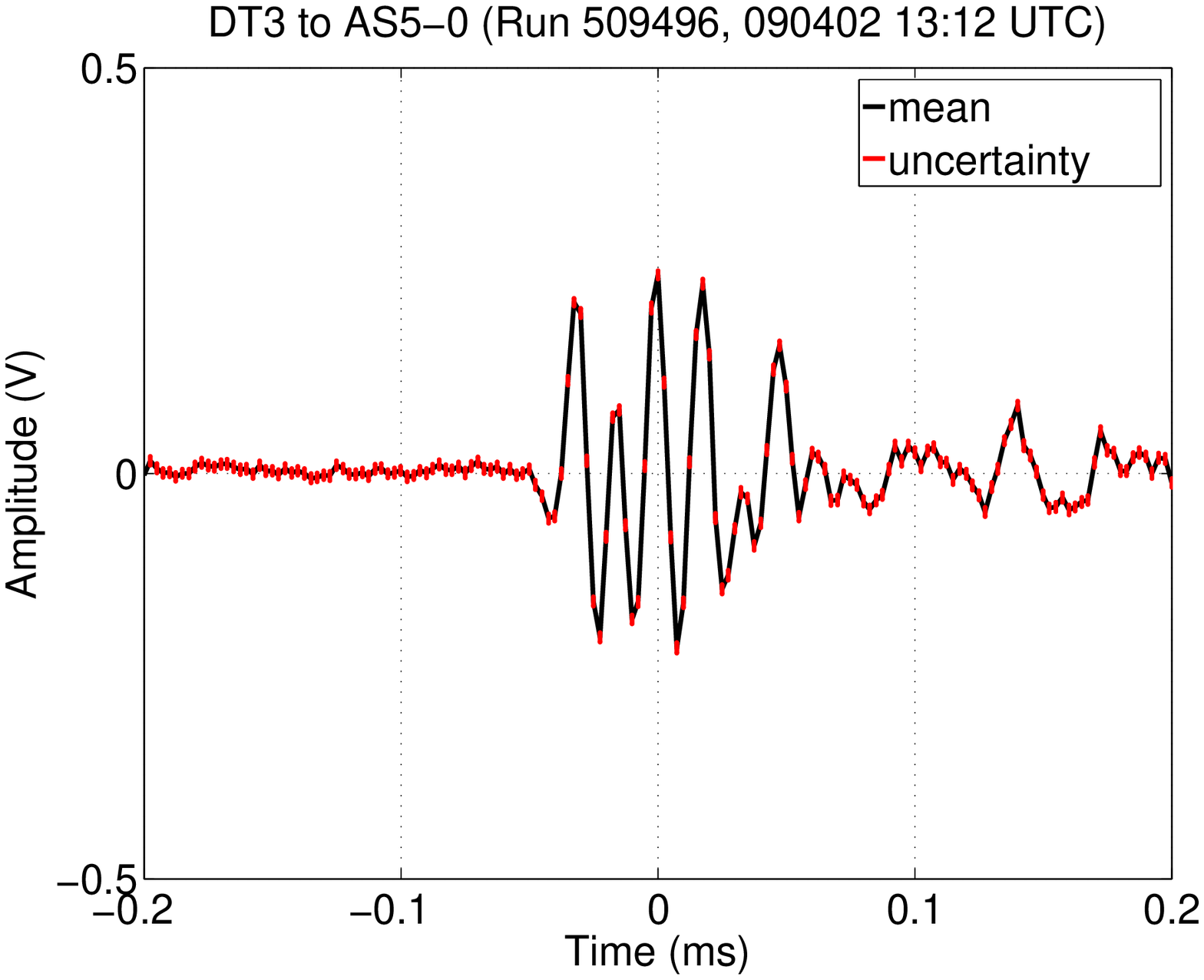}
}
\caption[Reproducibility of inter-string pulses and runs]{Demonstration of the reproducibility of pulses and runs.  Two mean waveforms are shown for two different runs with DT3 transmitting and AS5-0 recording.  Each mean waveform is determined from 500 individual pulses.  The two waveforms have very similar shape and amplitude, and the variations between them are consistent with the statistical error bars shown as vertical red lines.  This verifies the reproducibility of transmitter emission, sensor recording, clock drift correction, and waveform averaging.}
\label{waveform_reproducibility}
\end{center}
\end{figure}

\subfiglabelskip=0pt		
\begin{figure}
\begin{center}
\subfigure[][]{
\label{AS6-0}
\noindent\includegraphics[width=11pc]{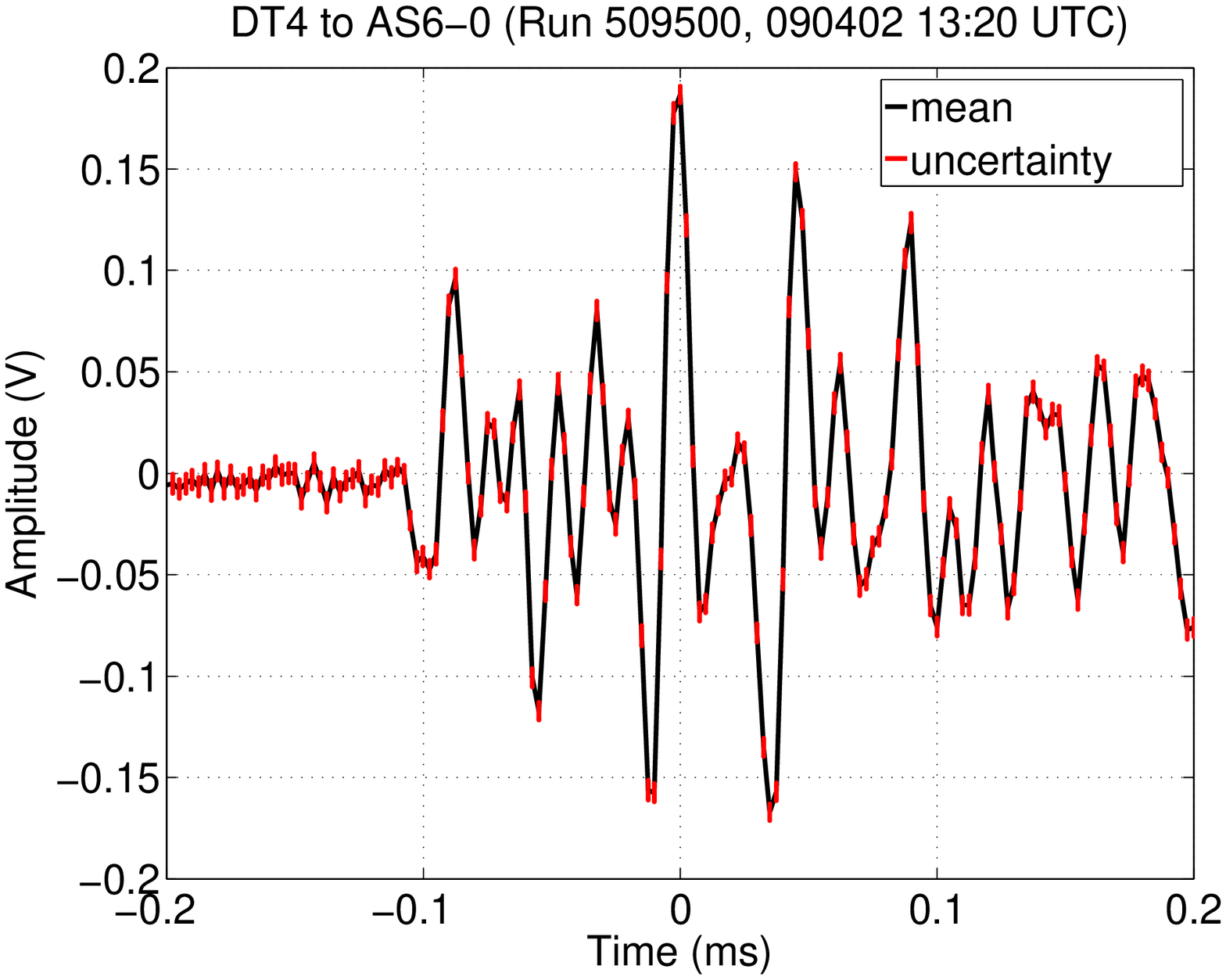}
}
\subfigure[][]{
\label{AS6-1}
\noindent\includegraphics[width=11pc]{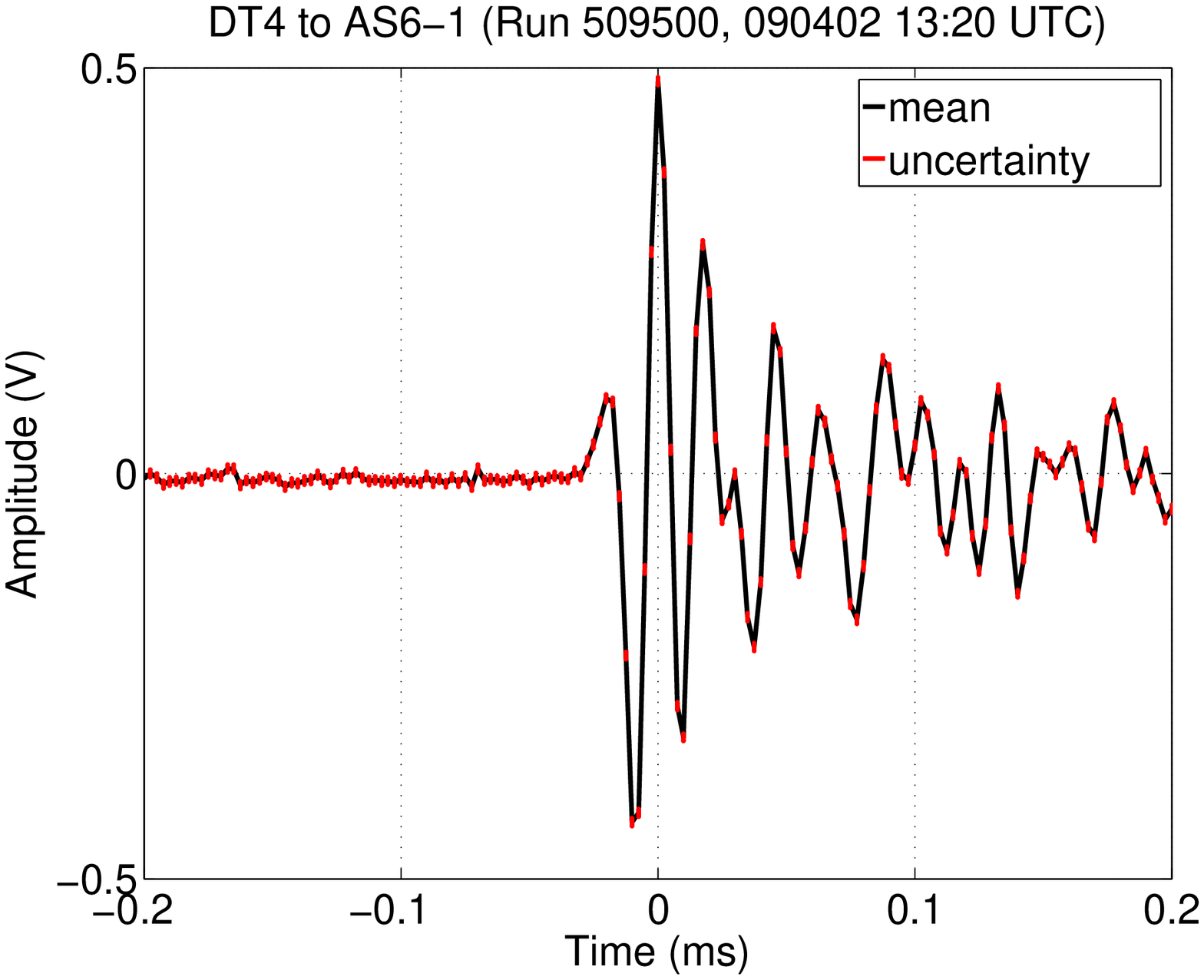}
}
\subfigure[][]{
\label{AS6-2}
\noindent\includegraphics[width=11pc]{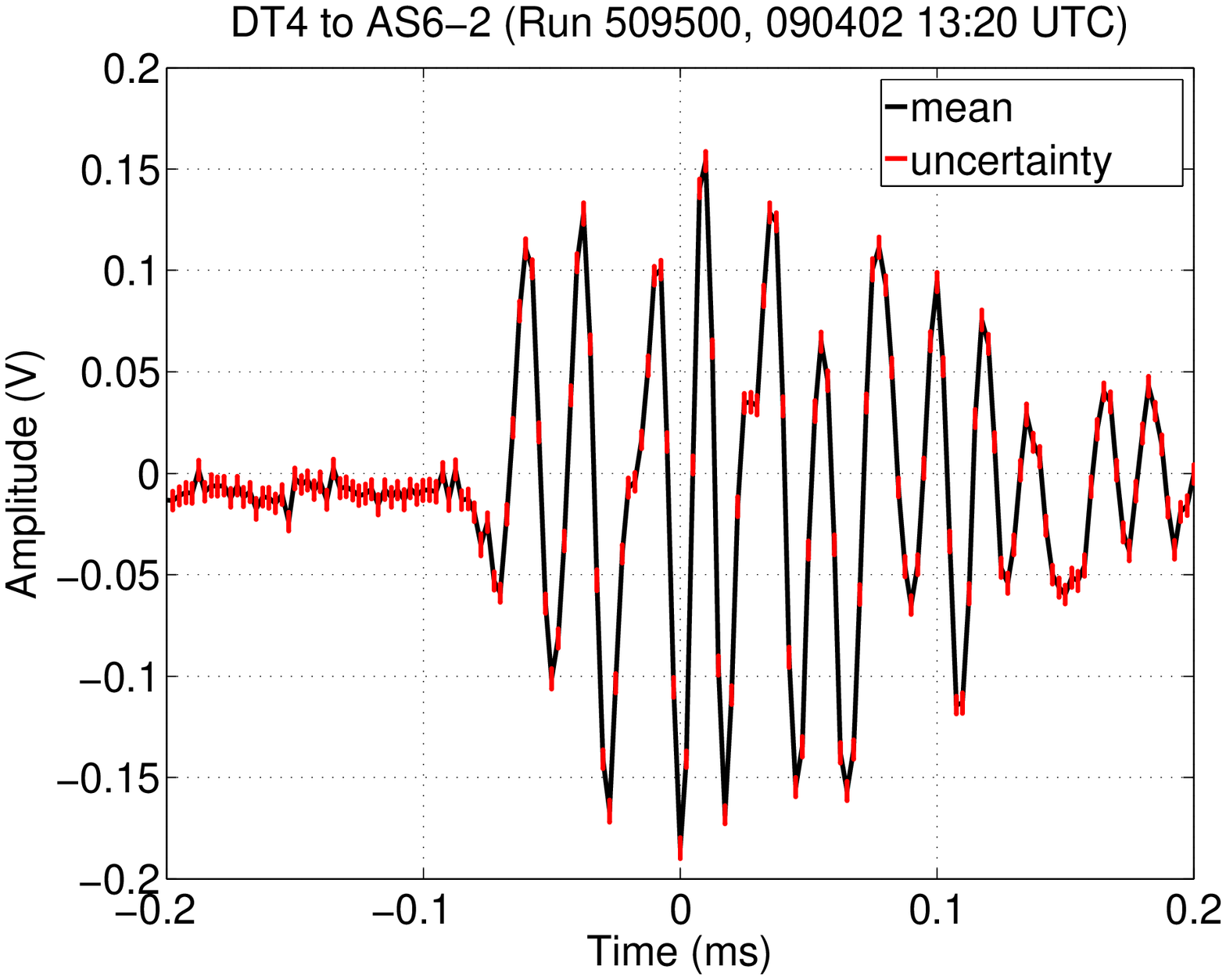}
}
\subfigure[][]{
\label{BS6-0}
\noindent\includegraphics[width=11pc]{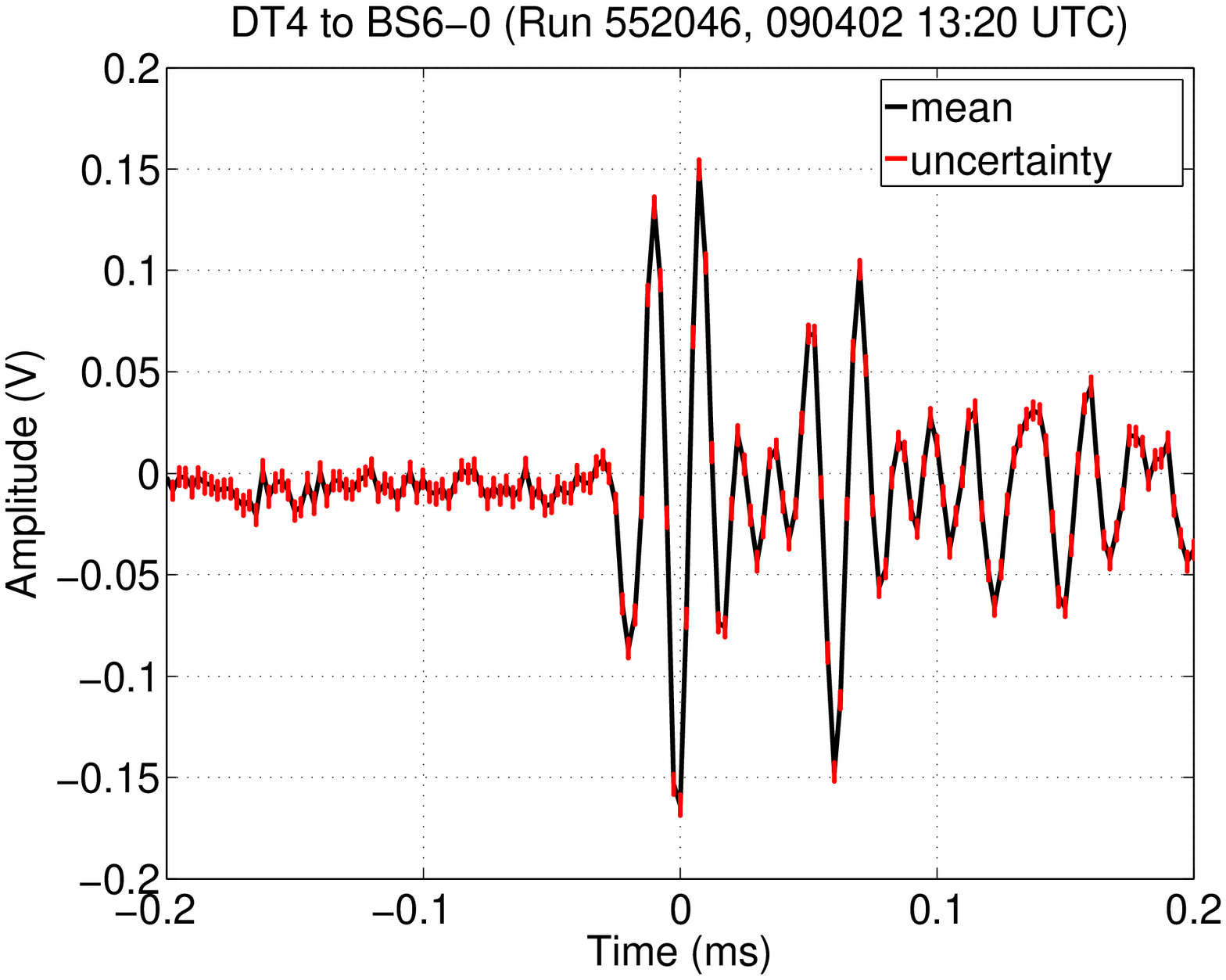}
}
\subfigure[][]{
\label{BS6-1}
\noindent\includegraphics[width=11pc]{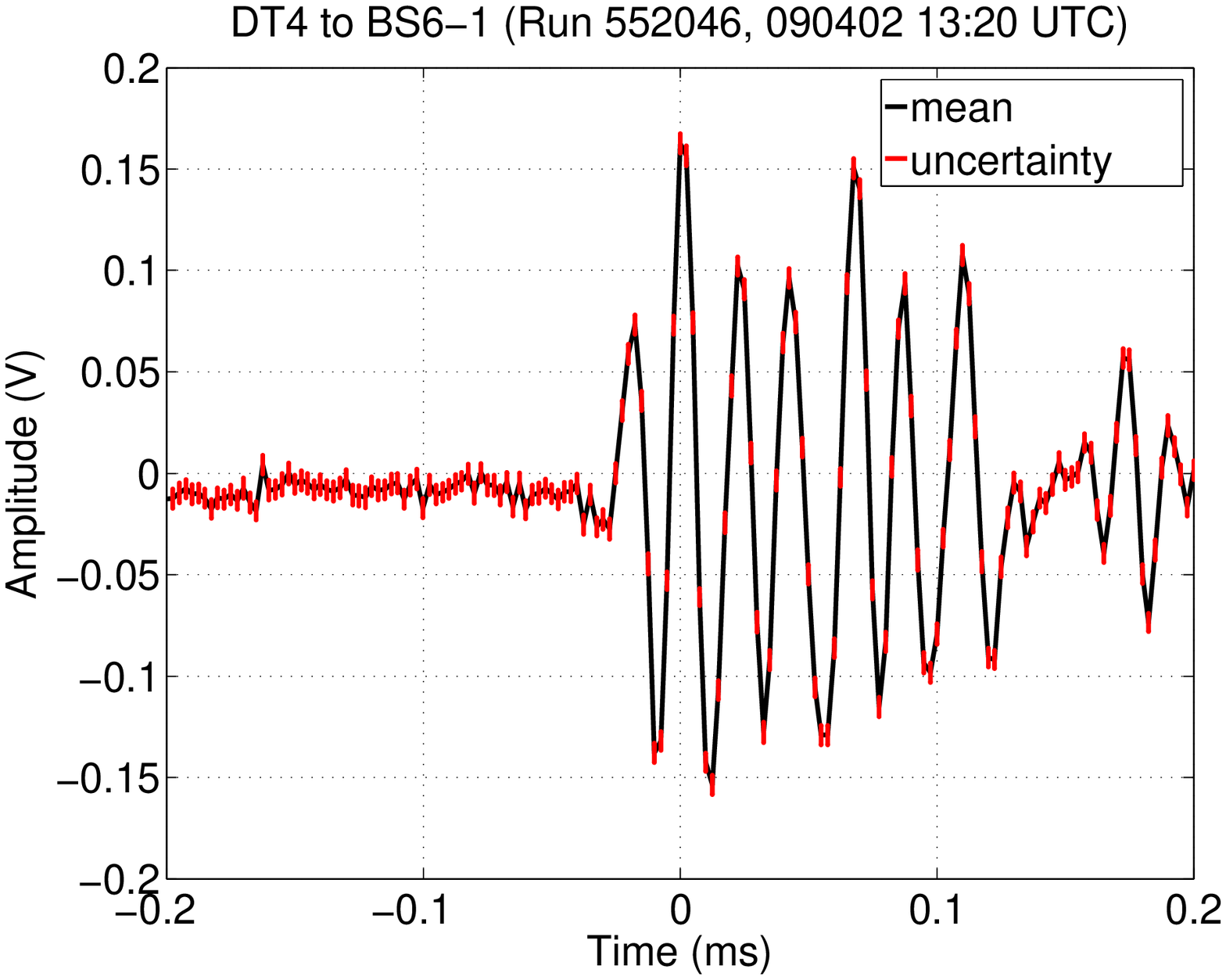}
}
\subfigure[][]{
\label{BS6-2}
\noindent\includegraphics[width=11pc]{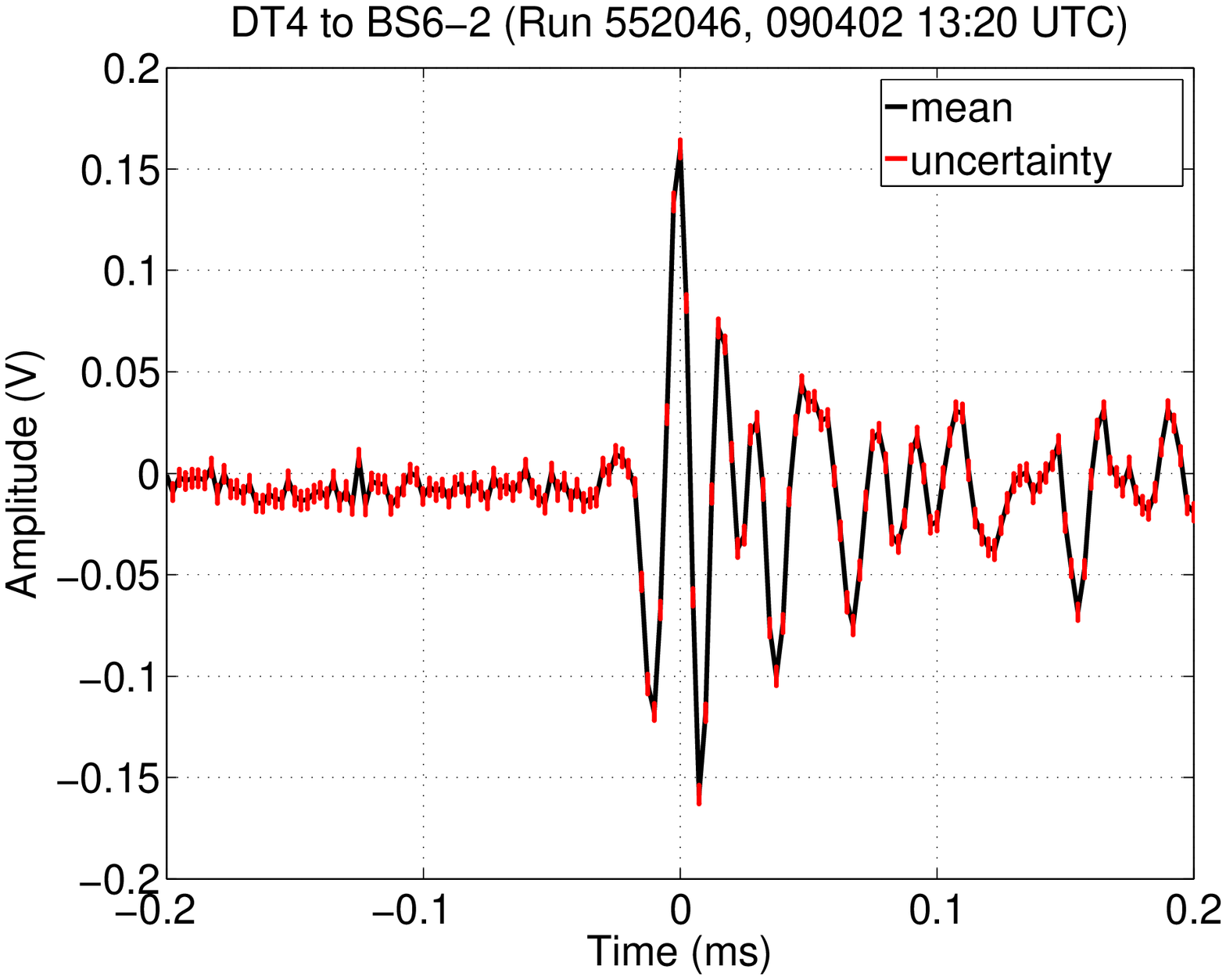}
}
\subfigure[][]{
\label{CS6-0}
\noindent\includegraphics[width=11pc]{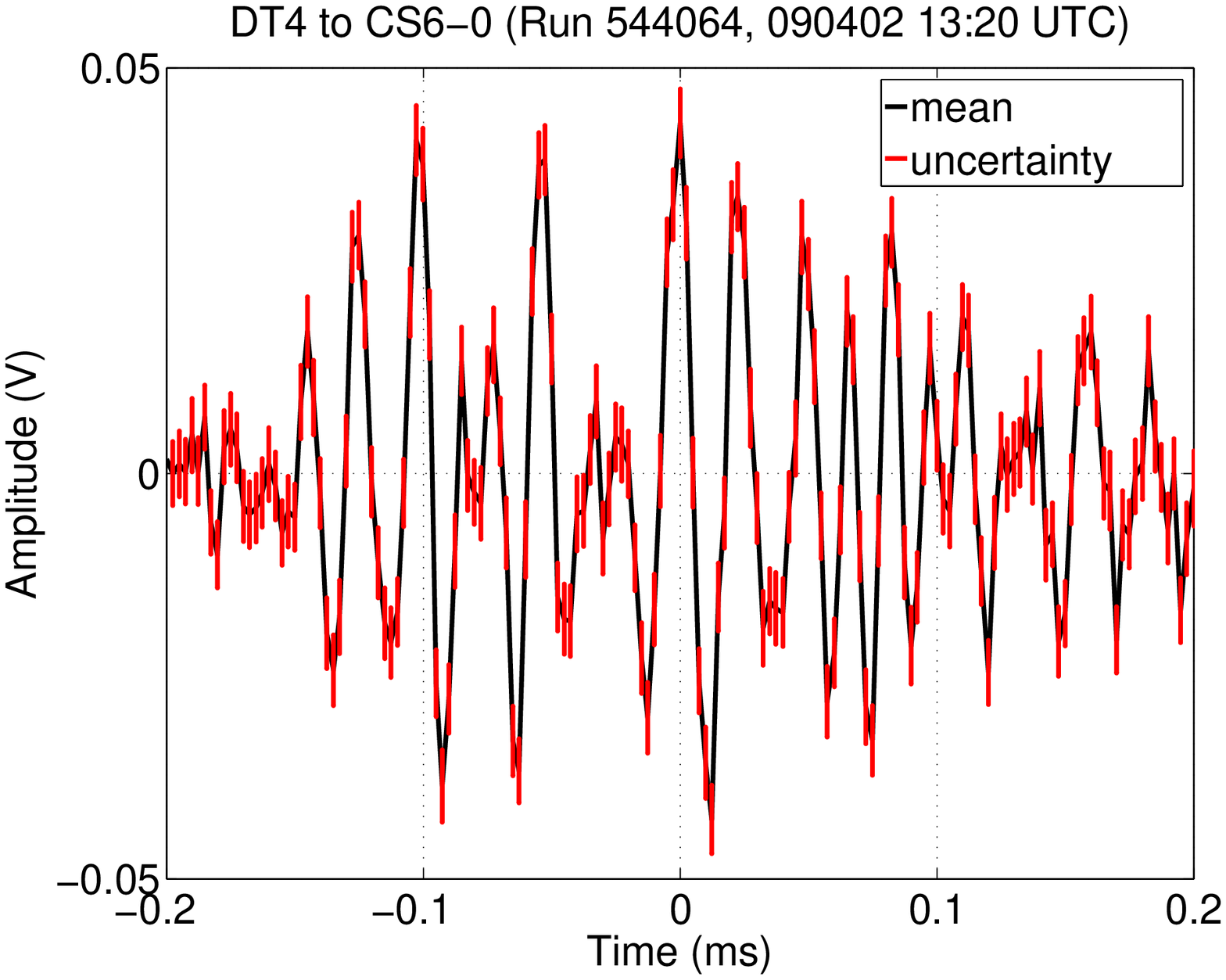}
}
\subfigure[][]{
\label{CS6-1}
\noindent\includegraphics[width=11pc]{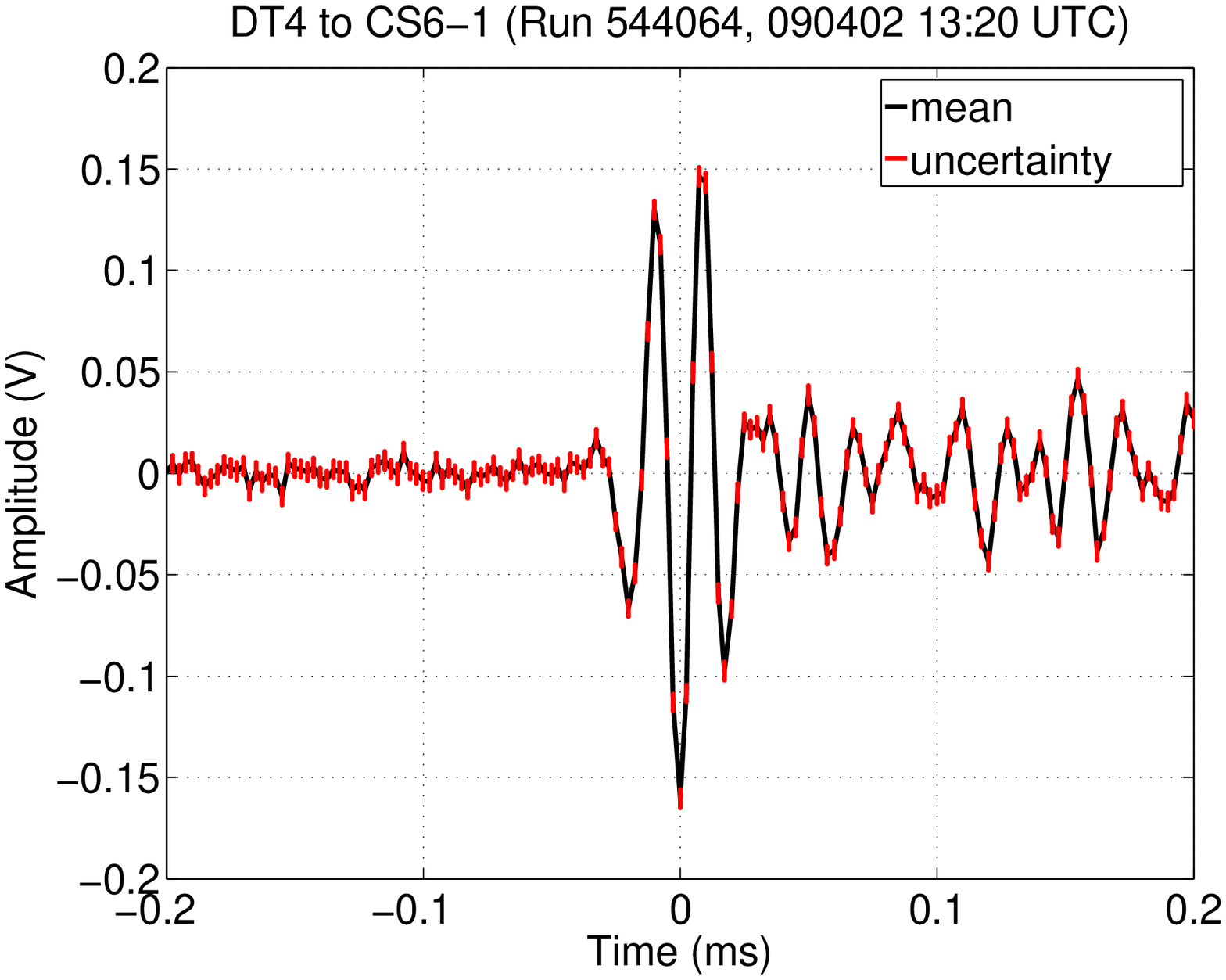}
}
\subfigure[][]{
\label{CS6-2}
\noindent\includegraphics[width=11pc]{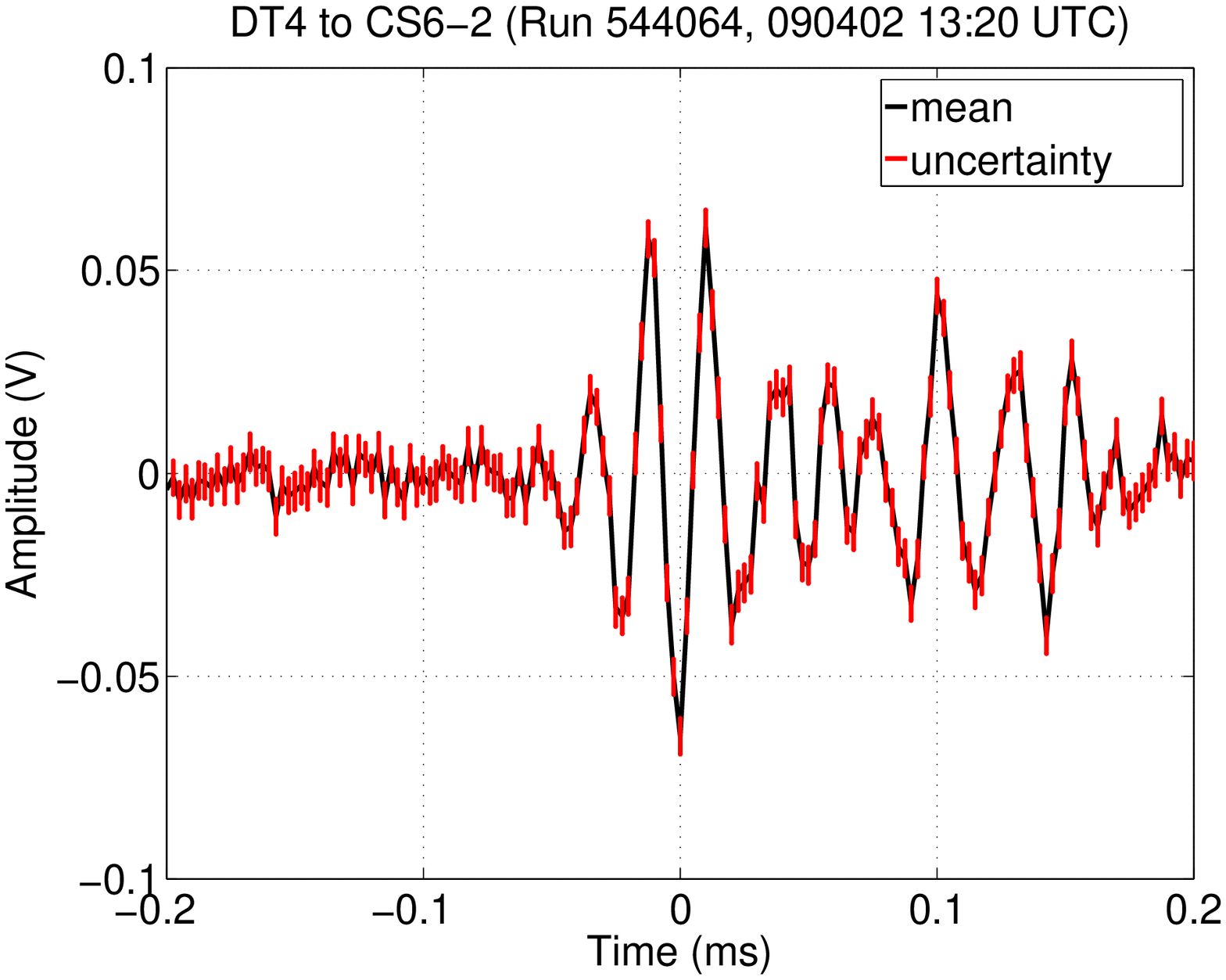}
}
\caption[Waveforms for all sensor channels recording a transmitter at the same depth]{One transmitter (DT4, at 320~m depth) recorded by all nine sensor channels at the same depth (3 per module on each of 3 strings).  Each waveform is centered on the maximum-amplitude sample.  The top row is String A; the middle row is String B; and the bottom row is String C.  The channels are ordered 0, 1, 2 from left to right.  Each waveform shows a well determined signal shape well above noise.  The shape is, however, significantly different from channel to channel.  This is due to both the azimuthal variation of the transmitter output, and the differing characteristic (resonant) response of each sensor channel.}
\label{all_channels_one_transmitter}
\end{center}
\end{figure}

To average the pulses, the actual (after drift correction) times of all samples in the recording were wrapped in time modulo the repetition period.  Each set of 250 consecutive samples (one-half the number of pulses) was then grouped together to form one bin.  In each bin the mean and standard deviation amplitude were then determined.  This algorithm results in a mean waveform with an effective sampling frequency twice as large as that used online (because the number of points per bin is half of the total number of pulses, so the time between bins is half of the online sampling time).  In addition to the mean, the uncertainty of each sample amplitude is estimated by this algorithm and is simply the standard error of the mean (standard deviation divided by $\sqrt{250}$) for each bin.  The standard error of the mean is used because the noise is stable and Gaussian, and therefore the mean of the Gaussian can be determined with a precision greater than the width of the Gaussian by $\sqrt{250}$.

An example average waveform is shown in Figure~\ref{waveform_example}.  The errors shown are statistical only and are the standard error of the mean from the waveform averaging procedure.  The small statistical errors indicate that the clock drift correction and waveform averaging algorithms perform well.  The errors include three components: pulse-to-pulse variation in the signal, pulse-to-pulse variation in the noise, and residual clock drift.  The pulse-to-pulse variation in the signal waveform amplitude and shape is evidently very small.  Moreover, the pulse shape and amplitude are reproducible from run to run, as shown in Figure~\ref{waveform_reproducibility}.  The statistical uncertainty on each waveform's shape and amplitude (including all three components) is small.  That is, the voltage waveform is well-determined.  This is the waveform as output by the sensor.  What is more uncertain is the original acoustic pressure waveform (both the shape and the amplitude), before it is convolved with the sensor response which has both frequency and angular (both polar and azimuthal) dependence.  This is the dominant systematic uncertainty and is addressed below.

The ratio between pressure and shear wave amplitude varies with transmitter-sensor configuration, but the shear waves are always lower amplitude than the pressure waves.

Although the pulses are very stable from run to run and pulse to pulse on a single sensor channel, the shape and amplitude differ significantly from channel to channel, as shown in Figure~\ref{all_channels_one_transmitter}.

\subsection{Noise stability}

\begin{figure}[tbp]
\begin{center}
\includegraphics[angle = 0, width = 0.7\textwidth]{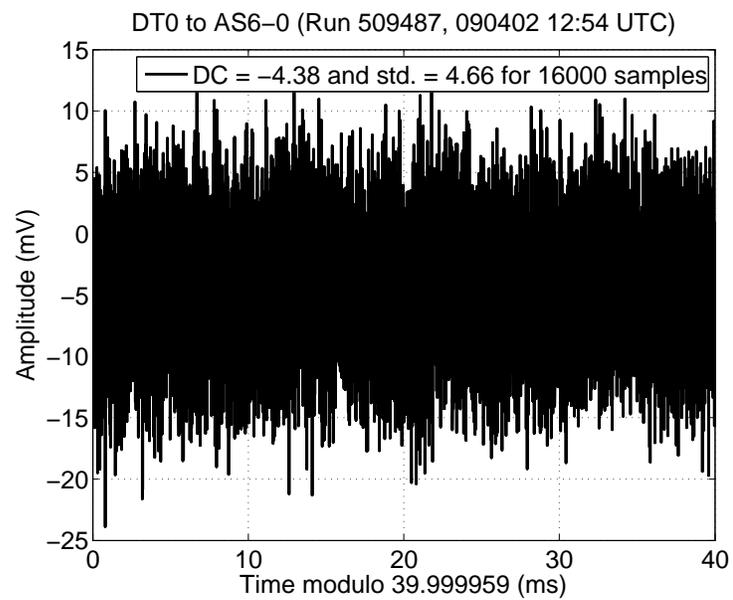}
\end{center}
\caption[Noise recording for inter-string noise subtraction]{Example ``T0'' (noise) recording, which has been processed using drift correction and waveform averaging in the same way as for waveforms with a real transmitter pulsing.  The mean and standard deviation for the run are shown.  Noise recordings like this were used to subtract the noise in the inter-string attenuation analysis}
\label{DT0_AS6-0_090402_125412_run509487}
\end{figure}

The acoustic noise conditions at South Pole have been determined by SPATS to be very stable~\cite{Karg09noise}.  In particular we checked that they were very stable during the two-day inter-string data taking campaign.  Several raw noise runs were recorded on each channel, interspersed among the transmitter recordings, by recording with the same sampling frequency and duration but with no transmitter pulsing.  The noise runs were processed with the same waveform averaging algorithm as the transmitter runs.  An example average noise recording is shown in Figure~\ref{DT0_AS6-0_090402_125412_run509487}.  For each noise run, after waveform averaging the mean amplitude (DC offset, $\mu$) and the standard deviation of the noise samples ($\sigma$) was calculated.  Each of $\mu$ and $\sigma$, on each sensor channel, was stable at the few-percent level during the two-day period.

\subsection{Windowing}

For each average waveform, in order to measure the signal amplitude or energy content we need to apply a time window in order to select the pressure wave signal and separate it from the shear wave signal.  The signal content often extends well beyond the maximum-amplitude sample, so it is necessary to include as large a window as possible while avoiding shear wave contamination.  For the inter-string data set there are three possible windowing algorithms:

\begin{enumerate}
\item Full window (no windowing).  This is the easiest choice but it suffers from shear wave contamination.
\item Dead-reckoning using emission and arrival times.  This is in principle the best algorithm, but it requires a difficult sequence of steps to determine the emission time and from that to calculate the expected arrival times of both pressure and shear waves, when their separation, the repetition rate, and the times of flight are all on the order of tens of ms and the delay from the beginning of the transmitter firing to the beginning of the sensor recording is on the order of 10 s.  While the accounting is straightforward for the pinger which has emission at a known time driven directly by a GPS clock, it is difficult for the inter-string data where the transmitter emission time is asynchronous and can occur at any time relative to GPS absolute time.  While possible in principle, this is a difficult and unnecessary algorithm.
\item Fixed-width windowing based on peak amplitude.  This is the algorithm we use for this analysis.  From studying the timing and amplitude of pressure and shear waves in the inter-string data, we know that shear waves are always lower amplitude than pressure waves.  Therefore we can use the maximum amplitude of the average waveform to determine the pressure wave time.  We then open a time window starting 5~ms before, and ending 26~ms after (chosen by visually inspecting all runs), this peak amplitude.  Because we only detect shear waves in the inter-string data for the nearest-neighbor string combinations (String A and to String B and vice versa), this windowing is sufficient to include all of the pressure signal while excluding all of the shear signal, for all transmitter-sensor distances.
\end{enumerate}

An example of the windowing algorithm applied to an average waveform is shown in Figure~\ref{waveform_example}.  For all average waveforms used in this analysis, the waveforms were examined by eye to verify that the windowing algorithm fully included the pressure wave while fully excluding the shear wave.

\subsection{Signal energy and effective amplitude}

Because the waveform shape varies from channel to channel and we combine nine different channels to fit for the attenuation length, the peak-to-peak amplitude is a bad choice of amplitude estimate: it is sensitive to variation in the resonant behavior of the sensor.  A better metric, which is not as sensitive to shape variation but is still dependent on an overall sensitivity normalization, is the energy in the signal.  Here we use \emph{energy} in the signal processing definition (sum of squared sample amplitudes, in arbitrary units), not in the physics sense (which would be proportional to the signal processing definition but normalized correctly to give the physical energy in the electrical or acoustic pulse).

First the total energy in the window-restricted waveform (including both signal and noise) was determined as

\begin{equation}
E_{S+N} = \sum_{i=1}^n V_i^2,
\end{equation}
where $n$ is the number of samples in the restricted time window, and $V_i$ is the mean voltage of the $i$th sample.  We include only the $n$ samples inside the restricted time window.  With the window of width 31~ms described above, and effective sampling frequency of the average waveform of 400~kHz, the number of samples in the restricted time window is $n =$~12,400 for all runs.

Next the amount of noise energy present in the average waveform (within the restricted time window) was calculated as
\begin{equation}
E_N = n (\mu^2 + \sigma^2).
\end{equation}
The first term in this equation is the noise contribution from the DC offset, and the second term is the contribution from the standard deviation of the noise.  The noise was subtracted to estimate the amount of pressure signal energy in the restricted time window:
\begin{equation}
E_S = E_{S+N} - E_N.
\end{equation}
Finally, the ``effective amplitude''
\begin{equation}
A = \sqrt{E_S}
\end{equation}
was introduced in order to directly fit for the \emph{amplitude} attenuation coefficient rather than the \emph{energy} attenuation coefficient.  The statistical uncertainty of the effective amplitude was determined with standard error propagation using the statistical uncertainty of each mean waveform sample and of $\mu$ and $\sigma$.

\subsection{Run selection}

The algorithm described above was applied to every inter-string run to determine an effective pressure signal amplitude $A$, along with the statistical uncertainty of it, $\delta A$.  The statistical uncertainty is dominated by the effect of the random Gaussian background noise varying from pulse to pulse.  Although the noise is reduced by a factor of $\sqrt{250}$ by the waveform averaging algorithm, for some transmitter-sensor combinations the signal amplitude is nevertheless small enough that the signal is swamped by noise.

For attenuation analysis, in order to only include transmitter-sensor combinations with signal-to-noise ratio sufficiently large to be sure that we are determining the transmitter amplitude correctly and not just measuring a noise fluctuation, we required that the effective amplitude signal-to-noise ratio (SNR) = $A/ \delta A$ be larger than 5.  This was chosen by visually inspecting waveforms: all waveforms with SNR larger than 5 had clear coherent waveform pulses, while some waveforms with SNR as large as e.g. 4.8 had waveforms that were difficult to distinguish from noise by visual inspection.  After applying this SNR cut, there were 93 unique transmitter-sensor combinations surviving.

For many transmitter-sensor combinations, more than one run was recorded as a consistency check.  These runs were used to verify that the average waveforms were reproducible in both amplitude and shape, as described above.  For such combinations with multiple runs taken, only the first run was used for attenuation analysis.

\subsection{Fitting for attenuation}

The waveform shape for each transmitter - sensor channel combination depends on both the transmitter and the sensor channel.  This is because both the transmitter and sensor resonate, and the resonance behavior of each transmitter and sensor is unique.  In addition to the shape variation from transmitter to transmitter and sensor to sensor, there is amplitude variation.  That is, each transmitter has different transmittivity, and each sensor has different sensitivity.  Here we use a single transmitter for each analysis, but multiple sensor channels.  Therefore in fitting for the attenuation length, the transmittivity is fit as part of an overall normalization factor that we do not care about, while the variation in sensitivity from channel to channel produces our dominant systematic uncertainty.

For each transmitter there are typically three sensor modules, each with three channels, at the same depth on other strings.  The acoustic attenuation was measured for each transmitter in the array, after applying the following two selection cuts to the data.  First, the SNR~$>$~5 cut described above was applied.  Second, there had to be at least one good channel at each of at least two distances from the transmitter, at the same depth as the transmitter, in order for a fit to be performed for a given transmitter.  Note that in the SPATS array geometry, only five of the nine instrumented depths are instrumented on all four strings.  12 of the 28 SPATS transmitters met these selection criteria.

For each good sensor channel, the quantity

\begin{equation}
y = \ln(A d)
\end{equation}

\noindent was calculated, where $A$ is the effective amplitude in Volts and $d$ is the transmitter-sensor distance in meters.  We then applied a linear fit for each of the 12 transmitters, using the model

\begin{equation}
y = -\alpha d + b,
\end{equation}

\noindent where $\alpha$ is the acoustic attenuation coefficient in m$^{-1}$ and b is a free normalization parameter related to the emission strength of the transmitter.  Linear regression was applied to directly determine the best fit and uncertainty on each of the two parameters $\alpha$ and $b$.

Both statistical and systematic uncertainty were included in the $y$ value for each channel recording each transmitter.  The statistical contribution was determined with error propagation from the statistical uncertainty of the effective amplitude.  The systematic uncertainty is dominated by the unknown relative sensitivity of the sensor channels.  The sensitivity of each channel is defined to be the response function converting acoustic pressure amplitude into electrical voltage amplitude.  We estimate the sensor channel-to-channel variation in sensitivity by treating the retrievable pinger results as an \emph{in situ} calibration of the sensor channels.  Because the pinger analysis fits for the $b$ parameter for each channel independently, $e^b$ can be taken as a measure of the sensitivity of each channel.

\subsection{Sensor sensitivity systematics}

\begin{figure}[tbp]
\begin{center}
\includegraphics[angle = 0, width = 0.7\textwidth]{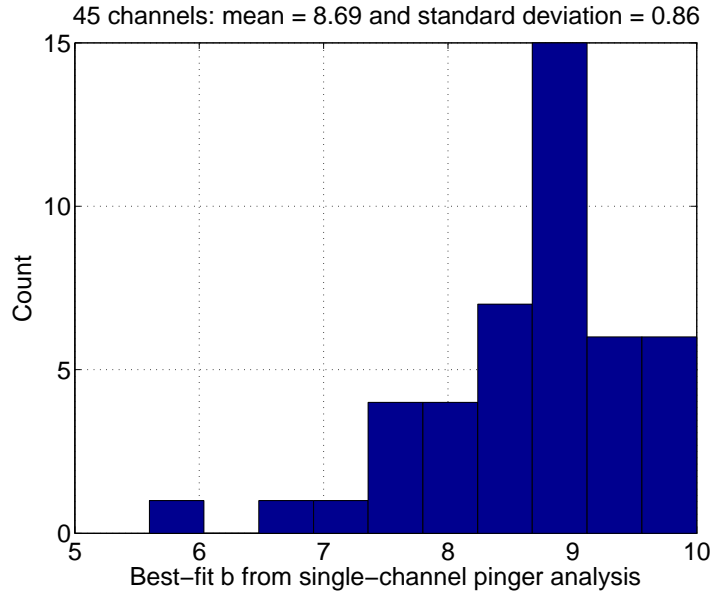}
\end{center}
\caption[Sensor sensitivity values from \emph{in situ} calibration with pinger ($b$ parameter)]{Sensor sensitivity values from \emph{in situ} calibration with pinger.  The histogram shows the best-fit $b$ values ($b$ is the offset in the linear attenuation fit) for the 45 attenuation analyses, one per channel, in the pinger analysis.}
\label{b_histogram}
\end{figure}

\begin{figure}[tbp]
\begin{center}
\includegraphics[angle = 0, width = 0.7\textwidth]{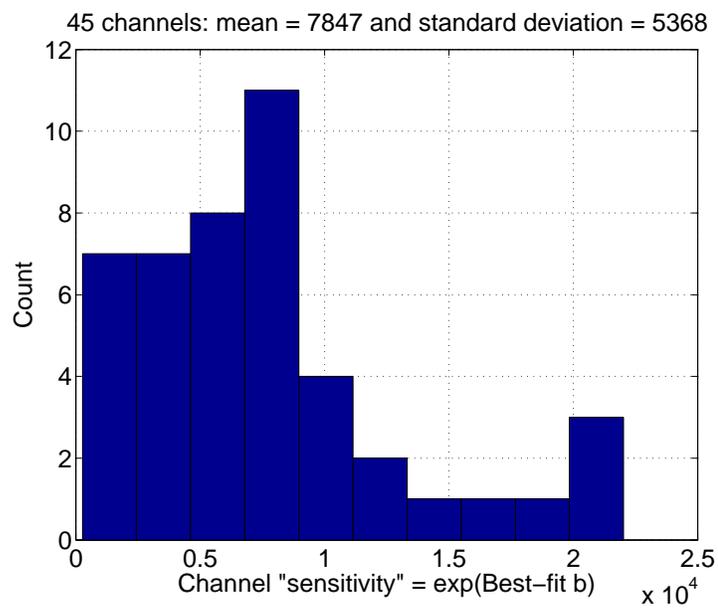}
\end{center}
\caption[Sensor sensitivity values from \emph{in situ} calibration with pinger ($e^b$ parameter)]{Sensor sensitivity values from \emph{in situ} calibration with pinger ($e^b$ parameter).  Histogram of $e^b$ for the 45 attenuation analyses, one per channel, in the pinger analysis.  $e^b$ is a measure of the relative sensitivity of each channel.}
\label{s_histogram}
\end{figure}

For each sensor channel recording, the effective amplitude $A$ and its statistical uncertainty $\delta A$ were determined as described above.  It is also necessary to estimate the systematic uncertainty of $A$.  As described above, we measure the attenuation length with the inter-string data by performing multiple fits, each using a single transmitter and multiple sensor channels at the same depth.  This method requires the sensitivity (conversion factor from pressure to voltage) of the different channels to be roughly equivalent, and the extent to which this is false must be included as a systematic uncertainty.

By taking the ensemble of all fitted $b$ parameters for all channels we can estimate the variation in the channel sensitivities.  Figure~\ref{b_histogram} shows a histogram of the 45 fitted $b$ parameters from the pinger analysis~\cite{Tosi09}.  To convert this into relative sensitivity, we need to consider $e^b$ for each channel, as shown in Figure~\ref{s_histogram}.  The standard deviation of the $e^b$ measurements is 68\% of the mean of the $e^b$ measurements.  More directly, because we are using the result of a fit in log space for another fit in log space, we can use the standard deviation of the best-fit b values measured in the pinger analysis, which is 0.86.

The sensitivity is in general direction-dependent for each sensor channel.  Because all retrievable pinger recordings were taken from nearly the same direction in azimuth, this is a reliable measurement of each channel sensitivity in the direction of the pinger.  The inter-string recordings, in contrast, are taken from a wide range of azimuths.  However, we do not need to know the absolute value of each channel sensitivity for the inter-string analyses; we only need to know the channel-to-channel variation which we take as a measure of the systematic uncertainty of each channel sensitivity.  We assume that the variation in sensitivities measured in the pinger direction is a good estimate for other directions also.  Therefore we take 0.86 as the systematic uncertainty on $y$ for each of the channels in the inter-string analysis.  This is added in quadrature to the statistical uncertainty of $y$ for each channel.

\subsection{Summary of inter-string attenuation results}

\begin{figure}[tbp]
\begin{center}
\includegraphics[angle = 0, width = 0.7\textwidth]{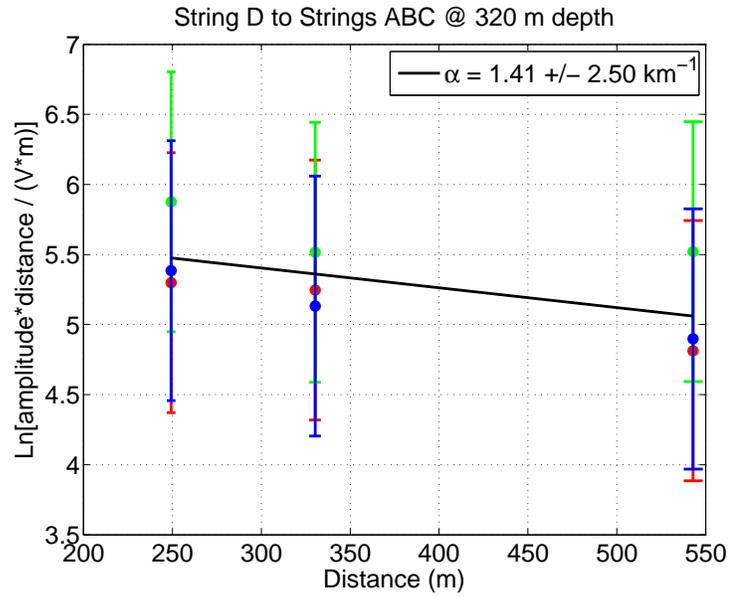}
\end{center}
\caption[Example attenuation fit from inter-string analysis]{Attenuation fit for one example transmitter in the single-transmitter, single-depth inter-string analysis.  The transmitter at 320~m depth on String D was transmitting, and the sensor modules at the same depth on Strings A, B, and C were recording.  In this example, the signal effective amplitude was determined with better than 33.3\% statistical precision on each of the 9 inter-string channels at the same depth as the transmitter.  The best fit attenuation coefficient is 0.0014~m$^{-1}$, and the one-sigma uncertainty is 0.0025~m$^{-1}$.  This single fit for a single transmitter is not sufficient to resolve non-zero attenuation.  Combining multiple independent analyses from different transmitters is necessary to do so.}
\label{interstringEnergyAnalysis_DT_320m}
\end{figure}

\begin{figure}[tbp]
\begin{center}
\includegraphics[angle = 0, width = 0.7\textwidth]{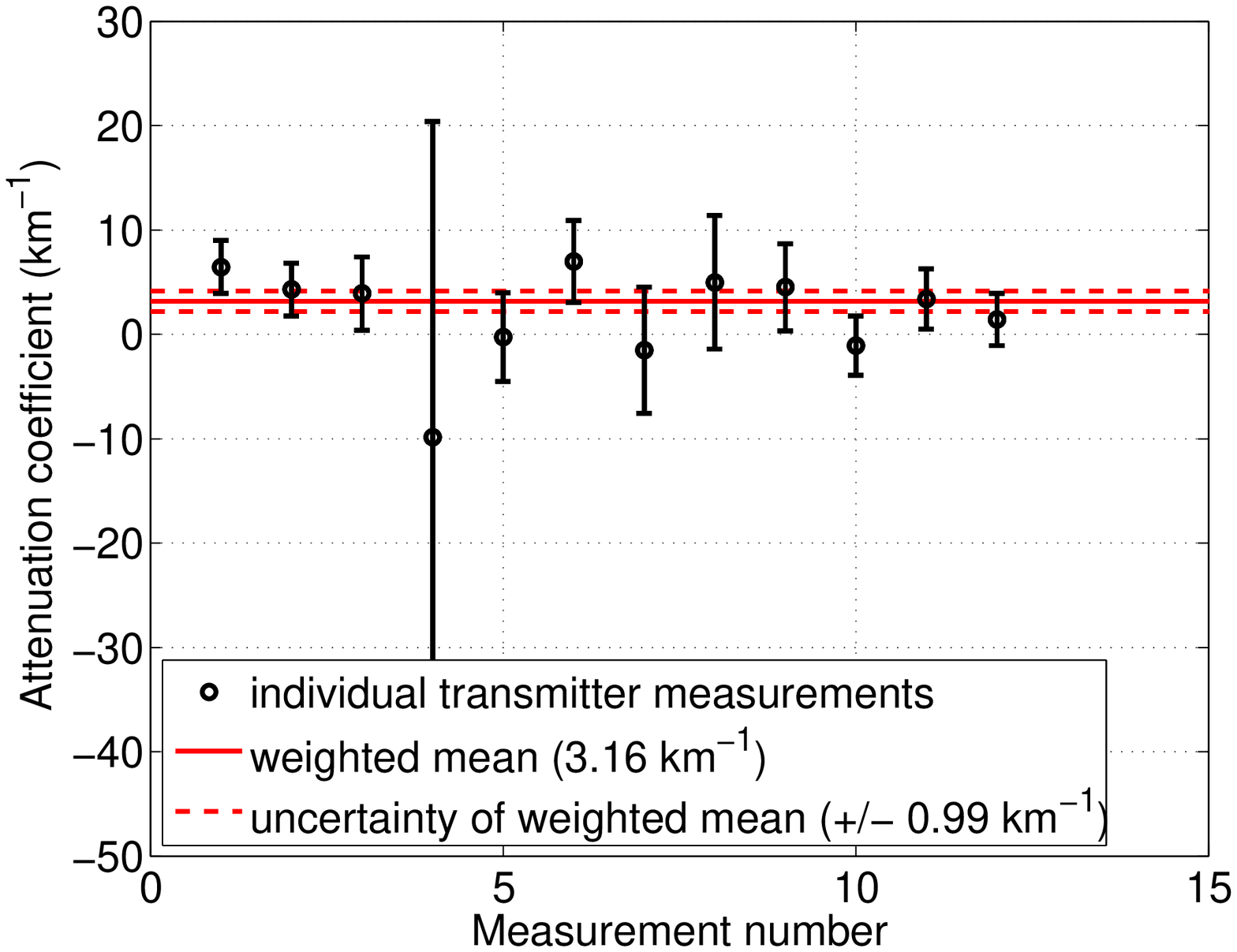}
\end{center}
\caption[Compilation of single-transmitter attenuation fits, with global fit]{Compilation of attenuation fits for the 12 transmitters used in the single-depth inter-string analysis.  The point with exceptionally large error bars is for the transmitter at 140~m depth on String B; all channels of the sensor module on String A at this depth are broken and so the only measurements are on Strings C and D, whose distance from B only differs by 28 m.  This anomalously small baseline results in a poorly constrained fit for this data point.  The weighted mean is shown as a solid horizontal line, and the one-sigma uncertainty of this global fit is indicated with dashed lines.  The result is $\alpha =$~3.16~$\pm$~1.05~km$^{-1}$.  The relative uncertainty of the $\alpha$ determination is 31\%.  Converted to attenuation length, this gives $\lambda =$~316~$\pm$~105~m.}
\label{attenuation_coefficient_SNR_greater_than_5}
\end{figure}

An example fit for the String D transmitter at 320~m depth is shown in Figure~\ref{interstringEnergyAnalysis_DT_320m}.  Results of the 12 fits for the 12 transmitters are compiled in Figure~\ref{attenuation_coefficient_SNR_greater_than_5}.  For each transmitter $i$, the best fit attenuation coefficient $\alpha_i$ and uncertainty $\sigma_i$ are shown.  Many of the single-transmitter fits individually give a result consistent with zero attenuation.  However, the 12 results can be combined to improve the precision of the measurement.  Determining the weighted mean of the measurements in the standard way (weighting by $1/\sigma_i^2$) gives a global best fit $\alpha =$3.16~$\pm$~1.05~km$^{-1}$.  Converting this to attenuation length, this analysis gives $\lambda =$~316~$\pm$~105~m.
\chapter{Conclusion and outlook}

\label{conclusionChapter}

\noindent\emph{In this chapter we give a brief summary of the history of SPATS, of our experimental achievements to date, and of what is next for the project.  We conclude with some thoughts on the outlook for acoustic neutrino detection in ice.}

\section{Achievements to date}

\subsection{The story of SPATS}
We started several years ago with sensor development neutrino sensitivity simulations, and a goal of hybrid GZK neutrino detection that required us to first determine the acoustic properties of South Pole ice, similar to AMANDA's need to determine the optical properties of the ice in its early days.  We started designing a project to make the necessary measurements, SPATS, on napkins at the Berkeley collaboration in the spring of 2005.  Successful deployment of SPATS followed 18 months later with the installation of Strings A, B, and C in January 2007.

While we learned a lot about noise conditions and made the first sound speed measurements with these strings, large systematic effects due to the unknown absolute, angular, and frequency responses of our sensors and transmitters, combined with the relatively short baselines available with the first three strings, prevented us from determining the attenuation length.  We installed a fourth string, String D, in the 2007-2008 South Pole season, and we also operated a retrievable pinger in six water-filled IceCube holes.  The new string gave us more and longer baselines as well as improved sensors and transmitters.  The retrievable pinger provided an interesting mystery: secondary waves detected under some but not all pinger-sensor geometries.  After thinking for some time that the secondary waves were reflections, we realized they were shear waves and simultaneously understood the smaller but also mysterious secondary waves we had seen in the inter-string data as shear waves.  The data from the first season of pinger operation (with pinger Version 1) were valuable for measuring the sound speed vs. depth of both shear waves and pressure waves.  However, both the first season of pinger data and the new four-string inter-string data were insufficient to determine the attenuation length.

In the 2008-2009 season we deployed an improved pinger in four IceCube holes.  The pinger had an order of magnitude larger repetition rate and mechanical centralizers to prevent the motion of the pinger, which had caused severe variation in the pinger signal during the first season.  This new pinger data allowed us to determine the attenuation length for the first time, and with good precision.  Simultaneously, progress in accounting for clock drift and optimizations in the data acquisition software allowed us to estimate the attenuation length with four-string inter-string data taken in the spring of 2009.  In the summer of 2008 we also upgraded the data acquisition software to acquire high-quality multi-channel background transient events, which have been successfully reconstructed.

By systematically addressing each of the challenges in our data from one year to the next, and responding to each of them in successive South Pole seasons, we have now completed nearly all of our measurement goals with SPATS.

\subsection{Sound speed}

We measured the sound speed as a function of depth, for both pressure and shear waves, between 80~m and 500~m depth, with better than 1\% precision.  Our measurements are compatible with others where they overlap, and we have measured the pressure wave speed for the first time beneath the firn and the shear wave speed for the first time at South Pole.  Most importantly for neutrino detection, we showed that at depths below the firn the level of refraction is small and therefore not a challenge to event reconstruction.  The results, our first complete analysis with SPATS, have been submitted to the Journal of Geophysical research.

\subsection{Attenuation length}

We have measured the attenuation length to be 300~m $\pm$ 100~m.  The amount of attenuation is more than an order of magnitude larger than theoretical predictions.  This should be understood both because it is an interesting scientific question and because it will help us determine whether acoustic neutrino detection is still feasible.  We emphasize that neutrino sensitivity does not scale directly with attenuation length, because $1/r$ divergence as well as the energy-dependent conversion of neutrino energy to acoustic radiation need to be considered.  Monte Carlo simulations are necessary to accurately estimate the neutrino sensitivity of a possible array.  These simulations can be performed with more confidence now that we have an experimental measurement of the attenuation length, rather than relying on theoretical predictions as we did previously.

Determining the attenuation dependence on depth and frequency will help distinguish different theoretical models of the attenuation mechanism, and will determine if there is a favorable part of the (depth, frequency) phase space for neutrino detection.  Although work is underway on both topics, so far we cannot distinguish any depth or frequency dependence with the existing data.

A paper on our attenuation length results is in preparation.

\subsection{Noise floor}

We have measured the noise floor to be very Gaussian, to be very stable with time, and to decrease with increasing depth.  The noise level in each channel has been stable over a $\sim$2~year period, and in particular is not correlated with weather or human activity on the surface.  Each of these features is encouraging for neutrino detection, and is in contrast to liquid water environments where other neutrino detection studies have been performed.  The Gaussianity of the noise causes our trigger rates to be stable and predictable with Gaussian statistics, where the Gaussian-tail events cause the bulk of triggers, and impulsive transient events contribute a small additional component.  Liquid water sites, in contradiction, typically have noise conditions that vary on multiple scales in both time and space, resulting in unstable trigger rates and/or the necessity of sophisticated adaptive threshold algorithms.

The absolute noise level decreases with increasing depth.  We predicted this before deploying SPATS, under the hypothesis that the effect is caused by the waveguide effect of surface noise being refracted back to the surface (toward lower wave propagation speed) due to the sounds speed gradient in the firn.  The SPATS sensors were initially calibrated only in water, and now they are deployed in ice at low temperature and high pressure.  The response to temperature and pressure have now been characterized in the lab, but it remains to perform an absolute calibration in ice.  Unfortunately it will not be possible to achieve a simultaneous calibration at high pressure, at low temperature, and in ice.  Instead it will be necessary to assume that these three properties affect the response independently.  We are working to calibrate the sensor response as well as possible in ice.  This will allow us to estimate the absolute noise level as a function of depth in South Pole ice.  The noise level determines the neutrino energy threshold, given a particular sensor array geometry.

\subsection{Transients}

The stability and Gaussianity of the noise have allowed stable and reliable acquisition of transient background events.  We have run with $\sim$70\% livetime since August~2008, accumulating more than 200 days of high-quality livetime reading out three sensor channels per string on each of our four strings.  We can reconstruct the emission time and location of these events and have found several interesting features after doing so.  Four of the five IceCube drill ``Rodriguez'' wells are present as strong and steady sources, as is an AMANDA Rod well last heated nine years ago.  Moreover, the individual IceCube holes drilled in December 2008 and January 2009 appeared as sources of transient acoustic noises, beginning emission $\sim$5~days after drilling.  The bulk of the emission is therefore associated with the re-freeze of the holes.  For one hole close to SPATS, we also see some events during the actual drilling of the hole.

We do not see any events (from IceCube holes or Rodriguez wells) deeper than $\sim$300~m.  This is a mystery because we are reading out sensors both above and below this depth.  This could be due to variation in the ice properties, but a simpler explanation is that the detected source distribution matches the intrinsic emission distribution.  If this is true then acoustic emission is strongest at shallow depths and decreases with increasing depth, perhaps due to the increasing ambient pressure.

The discovery and study of these sources has been a valuable verification of our full chain of hardware, online data acquisition software, and offline processing and reconstruction software.  The Rod wells and IceCube holes have served as useful calibration sources.  It is fortuitous that we upgraded the DAQ for this data in 2008, because the Rod wells have now quieted down considerably, and we could only hear the nearest of the IceCube holes that were drilled in December 2008 and January 2009.  Many of the 2008-2009 holes will be drilled farther from SPATS and it will likely be more difficult to hear them.

\section{Attenuation: reconciling theory and experiment}

The measured attenuation length of $\sim$300~m is shorter than the theoretical expectation by more than an order of magnitude.  We are considering several possible mechanisms for this:

\begin{enumerate}




\item The original theory presented in~\cite{Price06} evaluated attenuation as the sum of contributions from scattering and absorption. In the range of frequencies relevant to acoustic waves produced by ultrahigh-energy neutrino cascades (~10 to ~80 kHz), scattering was dominated by Rayleigh scattering off of ice grain boundaries, with a fourth power dependence on frequency and third power dependence on diameter. The use of mean diameter instead of the appropriate high moment led to a serious underestimate of the contribution of scattering to attenuation. An improved calculation using recent measurements of ice grain size distributions at depths down to nearly 300 m in ice near the South Pole suggests that scattering off the largest grains may account for the 300 m attenuation length.

\item Electrically charged linear dislocations in the ice crystals absorb energy from acoustic waves at a rate proportional to the concentration of dislocations and to the square of wave frequency. They also slow the acoustic waves by an amount that, though small, may be detectable by analyzing future pinger data at longer baselines with a larger signal-to-noise ratio made possible by the larger repetition rate. Data on dislocation concentrations in glacial ice, obtained since the evaluation in~\cite{Price06}, suggest that their contribution to absorption may account for the 300 m attenuation length.

\item There may be layered density fluctuations in the ice, with the thickness of the layers in the 1-10~cm range and the magnitude of the density fluctuations at the $\sim$0.1\% level.  Moreover, these layers undulate as a result of wind sculpting the snow surface\footnote{These undulations are called \emph{sastrugi}.} at the time of deposition.  The wavelength of these undulations is on the order of 1~m.  Therefore sound rays crossing propagating horizontally could cross hundreds of layers in propagating hundreds of meters.  Diagonal rays cross more layers than horizontal ones, but even horizontal rays can cross many layers due to the undulations.  Each layer crossing causes some propagation loss due to acoustic impedance ($Z = \rho v$ where $\rho$ is the density and $v$ is the sound speed variation.  Although the contribution of each layer crossing may be small, the accumulation over hundreds of layer crossings could be significant.

\end{enumerate}

The three attenuation mechanisms depend differently on frequency and depth. Future measurements could enable SPATS to determine their relative contributions by determining the frequency and depth dependence of the attenuation.

\section{What's next for SPATS}

While the sound speed precision achieved is already good (better than 1\%), it could be improved perhaps by a factor of 2-3.  The analysis used pinger data from the first pinger season, over a distance of 125~m.  The higher quality data from the second season could be used, and larger baselines could be used.  The inter-string data could also be used.  The limiting factors in the precision of this analysis are the precision with which we know the distance between holes, and the precision with which we can resolve the rising edge of each signal pulse as recorded by a sensor.  This resolution (on the rise time) is much smaller than width of the pulse.  So although the pulses can be seen clearly over large baselines, if the first oscillation of the signal cannot be clearly resolved above the noise then shorter baselines must be used.  This is why only the shortest baseline was used in the analysis of the 2007-2008 data.

It is a high priority to distinguish both the depth and frequency dependence of attenuation.  This is the primary goal of a new pinger data taking campaign planned for the 2009-2010 season.

Now that the transient detection and analysis chain has been verified with the IceCube holes and Rod wells, it is good that they are quieting down.  So far we have only detected emission from manmade sources.  Detecting emission from inside the bulk of the glacier itself (away from manmade holes) could provide evidence for stick-slip processes within the solid ice (in addition to the stick-slip motion thought to occur at the ice-bedrock interface).

We plan to search for neutrino-induced signals in addition to glaciological signals.  A null result could be used to set a limit on the neutrino flux.  The main challenge to performing a neutrino search is to understand the response of our sensors better, as they are known to be resonant sensors each with an unknown absolute response.  It will likely be necessary to incorporate the absolute calibration uncertainty as a systematic error band in the neutrino flux limit.

\section{Outlook for acoustic neutrino detection in ice}

We have determined that below the firn, both the amount of refraction and the noise level are low.  Moreover, the noise level is very stable which allows very stable acquisition of transient events with a simple threshold trigger.  All of these are beneficial for detecting high-energy neutrinos acoustically.  The amount of acoustic attenuation caused by the ice, on the other hand, is significantly larger than previously theorized.  It is important to understand why this is, and what it's mechanism is, in order to determine whether acoustic neutrino detection in ice is feasible.  In particular, if the attenuation is due to scattering rather than absorption, it may still be possible to use the scattered energy.

In any case, the outlook for acoustic neutrino detection in South Pole ice is not as good as that indicated by the simulations we performed several years ago.  At that time we used the only knowledge of the acoustic ice properties we had, which were theoretical.  Now that we have solid measurements of the various properties, it will be interesting to update the previous simulations.  We can expect that the GZK rate achieved with the previous array geometry will be reduced, or alternatively that to achieve the same sensitivity, greater instrumentation density is necessary.

It is now clear that the radio method can achieve superior sensitivity per cost than both the acoustic and the optical methods, in the GZK energy range.  While the prospects for multi-module, acoustic-only detection of GZK neutrinos by an autonomous acoustic array are reduced, individual hits from an acoustic array operated synchronously with a radio array could still be valuable.  This is true for both acoustic and optical hits.  The best configuration may be a large radio array that could trigger both an optical array (IceCube) and an acoustic array.  If and when detection of the first GZK neutrinos is claimed (likely either by the Pierre Auger Observatory or by a radio experiment), the events will likely be close to threshold and difficult to distinguish from both known and unknown backgrounds.  Even a single hit on a single module of an alternative type (optical or acoustic) could provide much-needed confirmation of the discovery.  Such information could also significantly improve the direction and energy reconstruction of the events, which will be the first quantities to determine once discovery is confirmed.

\appendix

\chapter{Simulation of a hybrid optical/radio/acoustic extension to IceCube for EeV neutrino detection}

\label{largeHybridAppendix}




\noindent\emph{The results of this simulation encouraged us to pursue the idea of hybrid neutrino detection of the South Pole and motivated the development of SPATS.  It appeared as~\cite{Besson05} and I reproduce it here for reference.}

\paragraph{Abstract}
Astrophysical neutrinos at $\sim$EeV energies promise to be an interesting source for 
astrophysics and particle physics. Detecting the predicted 
cosmogenic (``GZK'') neutrinos at 10$^{16}$ - 10$^{20}$ eV would test models of 
cosmic ray production at these energies and probe particle physics at $\sim$100~TeV
center-of-mass energy. While IceCube could detect $\sim$1 GZK event per year, it 
is necessary to detect 10 or more events per year in order to study temporal, 
angular, and spectral distributions. The IceCube observatory may be able to 
achieve such event rates with an extension including optical, radio, and 
acoustic receivers.  We present results from simulating such a hybrid detector. 


\section{Introduction}
Detecting and characterizing astrophysical neutrinos in the 10$^{16}$~eV to 10$^{20}$~eV range is a central 
goal of astro-particle
physics.  The more optimistic flux models in this range involve discovery physics including topological defects and
relic neutrinos.  Detecting the smaller flux of cosmogenic 
(or Greisen, Zatsepin, and Kuzmin, ``GZK'')
neutrinos produced via ultra-high energy cosmic ray interaction with the cosmic microwave background 
 would test models
of cosmic ray production and propagation and of particle physics at extreme energies.
  With $\sim$100 detected events, their angular distribution
would give a measurement of the total neutrino-nucleon cross section at $\sim$100 TeV center of mass,
probing an energy scale well beyond the reach of the LHC.  Hence, as a baseline, a detector capable 
of detecting $\sim$10 GZK events per year
has promising basic physics potential.  If any of the more exotic theories predicting greater EeV neutrino
fluxes is correct,
the argument in favor of such a detector is even stronger. 

To detect $\sim$10 GZK events per year,
a detector with an effective volume of $\sim$100~km$^3$ at EeV energies 
is necessary.
In addition to the possibility of identifying neutrino-induced air showers, there are three 
methods of ultra-high energy neutrino detection in solid media: optical, 
radio, and acoustic.  Optical Cherenkov detection is a well-established 
technique that has detected 
atmospheric neutrinos up to 10$^{14}$ eV and set limits up to
10$^{18}$~eV \cite{Ackermann05}.  Radio efforts 
have produced steadily improving upper limits on neutrino fluxes from 10$^{16}$ to
10$^{25}$~eV \cite{Kravchenko03,Gorham04,Lehtinen04}. Acoustic detection efforts are at an earlier stage, with one limit published thus far 
from 10$^{22}$ to 10$^{25}$~eV
 \cite{Vandenbroucke05}.

The currently planned 1~km$^3$ 
optical neutrino telescopes expect a GZK event rate of $\sim$1 per year.
It is possible to extend this by adding more optical strings for a modest
additional cost \cite{Halzen04}, but it's difficult to imagine achieving 10 or more events per year with
optical strings alone.  The radio and acoustic methods have potentially large effective volumes with relatively
few receivers, but the methods are unproven in that they have never detected a neutrino.  Indeed, if
radio experiments claim detection of a GZK signal, it may be difficult to confirm that it is really a
neutrino signal.
However, it may be possible to bootstrap the large effective volumes of radio and acoustic detection
with the optical method, by building a hybrid detector that can detect a large
rate of radio or acoustic events, a fraction of which are also detected by an optical detector.  
A signal seen
in coincidence between any two of the three methods
would be convincing. 
The information from multiple methods can be combined for hybrid reconstruction, yielding improved
angular and energy resolution.

We simulated the sensitivity of a detector that could be constructed by expanding the IceCube observatory 
currently under construction 
at the South Pole.  The ice at the South Pole is likely well-suited for all three methods:  Its optical clarity
has been established by the AMANDA experiment \cite{Ackermann05}, and its radio clarity and suitability for radio detection in the 
GZK energy range has been established by the RICE experiment \cite{Kravchenko03}.  
Acoustically, the signal in ice is ten times greater than that in water.
Theoretical estimates indicate 
low attenuation and noise
\cite{Price06}, and efforts are planned to measure both \cite{SPATS05} 
with sensitive transducers
developed for glacial ice \cite{Nahnhauer05}.
Here we estimate the sensitivity of such a detector by exposing all three components to a common
Monte Carlo event set and counting events
detected by each method alone and by each combination of multiple methods.

\section{Simulation}

IceCube will have 80 strings arrayed hexagonally with a horizontal spacing of 125~m.  In \cite{Halzen04},
the GZK sensitivity achieved by adding more optical strings at larger distances (``IceCube-Plus'')
was estimated, and the possibility of also adding radio and acoustic modules was mentioned. 
Here we consider an IceCube-Plus configuration consisting of 
a ``small'' optical array overlapped by a ``large'' acoustic/radio array
with a similar number of strings but larger horizontal spacing.  
The optimal string spacing for GZK detection was found to be $\sim$1~km
for both radio and acoustic strings.
This coincidence allows the two methods to share hole drilling and cable
costs, both of which are dominant costs of such arrays.

\begin{figure}
\centering
\noindent\includegraphics[width=30pc]{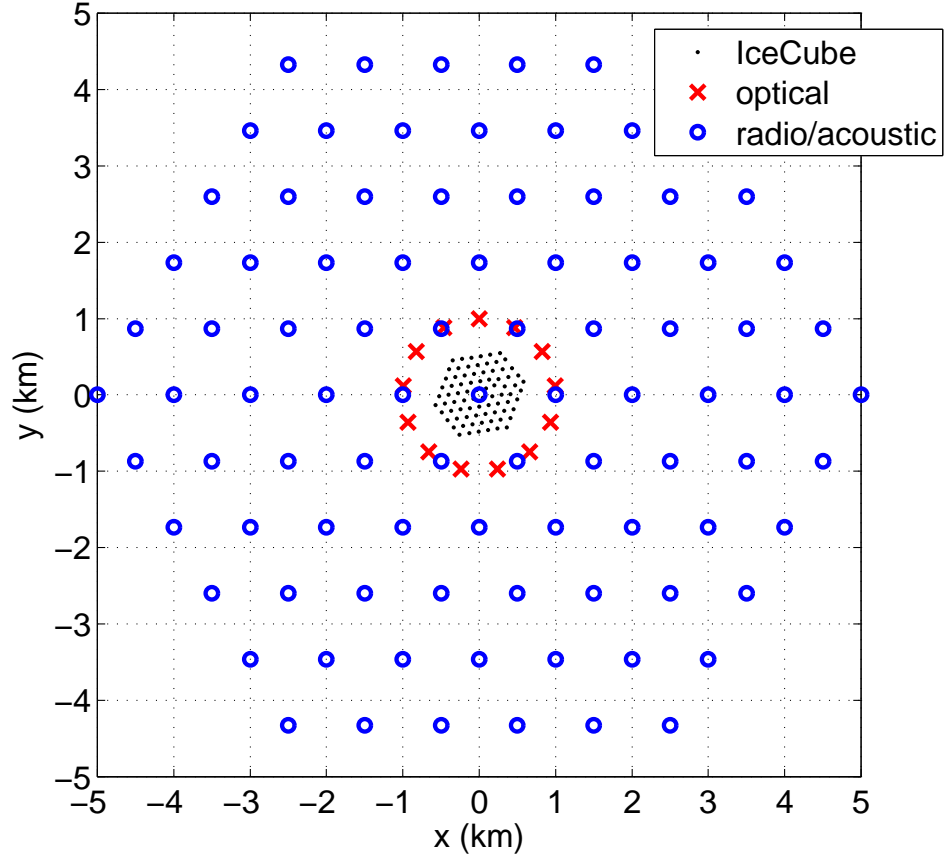}
\caption[Surface layout of large hybrid array]{Geometry of the simulated hybrid array.}
\label{fig1}
\end{figure}

The geometry of the simulated array is shown in Fig. \ref{fig1}.
We take the optical array to be IceCube as well as a ring
of 13 optical strings with a 1~km radius, surrounding IceCube.
All optical strings have standard IceCube geometry: 60 modules per
string, spaced every 17 m, from 1.4 to 2.4~km depth.
Encompassing this is a hexagonal array of 91 radio/acoustic strings
with 1~km spacing.  Each radio/acoustic hole has 5 radio receivers, spaced every 100~m from 
200~m to 600~m depth, and 300 acoustic receivers, spaced every 5~m from 5~m to 1500~m depth.  
At greater depths both methods suffer increased absorption due to the warmer ice.  The 
large acoustic density per string is necessary because the acoustic
radiation pattern is thin (only $\sim$10~m thick) in the direction along the shower.  
The array geometry was designed to seek
an event rate of $\sim$10 GZK events per year detectable with both radio and acoustic
independently.

\begin{figure}
\centering
\noindent\includegraphics[width=30pc]{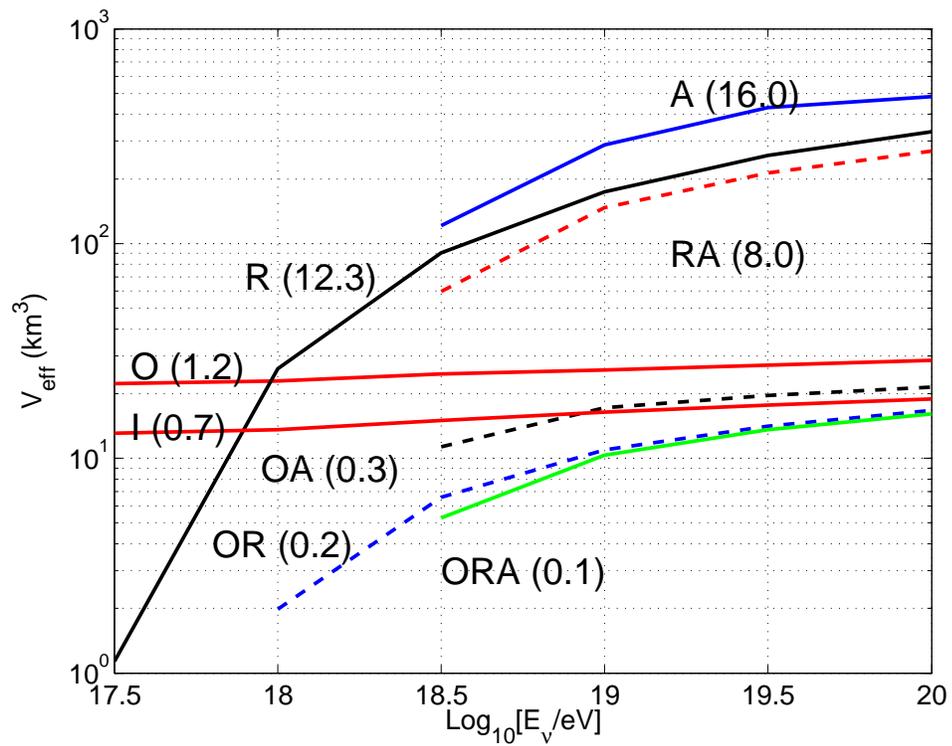}
\caption[Effective volume of large hybrid array]{Effective volume for each of the seven
combinations of detector components, as well as for IceCube alone (``I'').
GZK event rates per year are given in parentheses.  Note that different channels
were used for different combinations (see text).}
\label{fig2}
\end{figure}

To obtain rough event rate estimates, a very simple Monte Carlo generation scheme was chosen.  
Between 10$^{16}$ and 10$^{20}$~eV, the neutrino interaction length ranges between 6000 and 200~km
\cite{Gandhi}, so up-going
neutrinos are efficiently absorbed by the Earth and only down-going events are detectable.  A
full simulation would include the energy-dependent slow roll-off at the horizon.  Here we
assume all up-going neutrinos are absorbed before reaching the fiducial volume, and no
down-going neutrinos are; we generate incident neutrino directions isotropically in 2$\pi$~sr.
Vertices are also generated uniformly in a fiducial cylinder of radius 10 km, extending
from the surface to 3 km depth.  

The Bjorken parameter $y = E_{had}/E_{\nu}$ varies
somewhat with energy and from event to event, but we choose the mean value, $y = $ 0.2, for simplicity.
The optical method can detect both muons and showers, but here we only
consider the muon channel; simulation of the shower channel is in progress.  
The radio and acoustic methods cannot detect muon tracks but
can detect electromagnetic and hadronic showers. Under our assumptions of constant $y$
and no event-to-event fluctuations, 
all flavors interacting via both CC and NC produce the same hadronic shower.  Electron
neutrinos interacting via the charged current also produce  an 
electromagnetic cascade which produces radio and acoustic
signals superposed on the hadronic signals.  However, at the energies of interest here,
electromagnetic showers are lengthened to hundreds of meters by the Landau-Pomeranchuk-Migdal effect.  
This weakens their radio and acoustic signals significantly, and we assume
they are negligible.

For simulation of the optical response, the standard Monte Carlo chain 
used in current AMANDA-IceCube analyses \cite{Ackermann05} was performed.
After the primary trigger requiring any 5 hits in a 2.5~$\mu$s 
window, a
local coincidence trigger was applied: Ten local coincidences were required, where
a local coincidence is at least two hits on neighboring or next-to-neighboring 
modules within 1~$\mu$s. Compared with \cite{Halzen04}, we used an updated ice
model with increased absorption, which may account for our factor of $\sim$2 lower effective
volume.

Each simulated radio ``receiver'' consists of two vertical
half-wave dipole antennas separated vertically by 5~m to allow local rejection
of down-going anthropogenic noise. We assume an effective height at the
peak frequency (280~MHz in ice) equal to 10~cm, with
$\pm$20\% bandwidth to the --3 dB points. As currently under development
for RICE-II, we assume optical fiber transport of the signal to the
DAQ, with losses of 1~dB/km (measured) through the fiber.
The electric field strength $E(\omega)$ is calculated
from the shower according to the ZHS prescription \cite{ZHS91,Alvarez-Muniz98}.  Frequency-dependent 
ice attenuation effects are incorporated using measurements at
South Pole Station \cite{Barwick05}. The signal at the surface electronics 
is then transformed into the time domain, resulting in a waveform 10~ns long,
sampled at 0.5~ns intervals, at each antenna. Two
receivers with signals exceeding 3.5 times the
estimated root-mean-square noise temperature $\sigma_{kT}$ (thermal plus a system
temperature of 100~K) within a time window of 30~$\mu$s are required to trigger. 

The unattenuated acoustic pulse $P(t)$ produced 
at arbitrary position with
respect to a hadronic cascade is calculated by integrating over the cascade energy distribution.
The cascade is parametrized with the Nishimura-Kamata-Greisen parametrization, with 
$\lambda$ (longitudinal
tail length) parametrized from \cite{Alvarez-Muniz98}.
The dominant mechanism of acoustic wave absorption in South Pole ice is theorized \cite{Price06} to be 
molecular reorientation, which increases with ice temperature.  Using a temperature profile 
measured at the South Pole
along with laboratory absorption measurements, an absorption vs. depth profile was estimated.
The predicted absorption length ranges from 8.6~km at the surface to 4.8~km at 1~km depth to 0.7~km at 
2~km depth.
The frequency-independent absorption is integrated from source to receiver and applied in the time domain.

South Pole ice is predicted to be much quieter than
ocean water at the relevant frequencies ($\sim$10-60 kHz), because there are no waves, currents, or animals.  Anthropogenic
surface noise will largely be refracted back up to the surface due to the sound speed gradient
in the upper 200 m of snow that is not fully dense (``firn'').
For the current simulation we assume ambient noise
is negligible compared to transducer self-noise.  
Work is underway to produce transducers with self-noise at the 2-5~mPa level \cite{Nahnhauer05}. 
For comparison, ambient noise in the ocean is $\sim$100 mPa \cite{Vandenbroucke05}.
The acoustic trigger used in this simulation required that 3
receivers detect pressure pulses above a threshold of 9~mPa. 

\section{Results and conclusion}

Ten-thousand events were generated at each half-decade in neutrino energy 
in a cylinder of volume 942~km$^3$.  For each
method and combination of methods, the number of detected events was used to
calculate effective volume as a function of neutrino energy (Fig. \ref{fig2}).  This was folded with the
GZK flux model of \cite{Engel01,Seckel} and the cross-section
parametrizations of \cite{Gandhi} to estimate detectable
event rates (Fig. \ref{fig2}).  We use a flux model which assumes source evolution
according to $\Omega_{\Lambda}=$~0.7.  This model is a factor of $\sim$2
greater than that for $\Omega_{\Lambda}=$~0 evolution; it is unclear which model is
correct \cite{Seckel}. For radio and acoustic, and their combination,
all flavors and both interactions were included.  For those combinations including
the optical method, only the muon channel has been simulated thus far; including
showers will increase event rates for these combinations.  

It may be possible to build an extension like that considered here for a relatively small cost.
Holes for radio antennas and acoustic transducers can be narrow and shallow, and both
devices are simpler than photo-multiplier tubes.
The necessarily large acoustic channel multiplicity is partially offset by the fact that the 
acoustic signals are slower by five orders of magnitude, making data acquisition and processing easier.

The IceCube observatory will observe the neutrino universe from 10's of 
GeV to 100's of PeV.
Our simulations indicate that extending it with radio and acoustic strings could produce a neutrino detector
competitive with other projects optimized for high-statistics measurements of GZK neutrinos but with the unique 
advantage of cross-calibration via
coincident optical-radio, optical-acoustic, and radio-acoustic events. 












\bibliographystyle{unsrt_withcaps1.bst}



\bibliography{vandenbroucke_thesis}


\end{document}